# Roadmap for selected key measurements of LHCb

The LHCb Collaboration[1]

**Abstract**

Six of the key physics measurements that will be made by the LHCb experiment, concerning CP asymmetries and rare B decays, are discussed in detail. The "road map" towards the precision measurements is presented, including the use of control channels and other techniques to understand the performance of the detector with the first data from the LHC.

[1]Authors are listed on the following pages.

## The LHCb Collaboration

B. Adeva[36], M. Adinolfi[42], A. Affolder[48], Z. Ajaltouni[5], J. Albrecht[37], F. Alessio[6,37], M. Alexander[47], P. Alvarez Cartelle[36], A.A. Alves Jr[1], S. Amato[2], Y. Amhis[38], J. Amoraal[23], J. Anderson[39], O. Aquines Gutierrez[10], L. Arrabito[53], M. Artuso[52], E. Aslanides[6], G. Auriemma[22,k], S. Bachmann[11], Y. Bagaturia[11], D.S. Bailey[50], V. Balagura[30,37], W. Baldini[16], MdC. Barandela Pazos[37], R.J. Barlow[50], S. Barsuk[7], A. Bates[47], C. Bauer[10], Th. Bauer[23], A. Bay[38], I. Bediaga[1], K. Belous[34], I. Belyaev[23,30], M. Benayoun[8], G. Bencivenni[18], R. Bernet[39], M.-O. Bettler[38], M. van Beuzekom[23], A. Bizzeti[17,e], T. Blake[49], F. Blanc[38], C. Blanks[49], J. Blouw[11], S. Blusk[52], A. Bobrov[33], V. Bocci[22], A. Bondar[33], N. Bondar[29,37], W. Bonivento[15], S. Borghi[47], A. Borgia[52], E. Bos[23], T.J.V. Bowcock[48], C. Bozzi[16], J. van den Brand[24], J. Bressieux[38], S. Brisbane[51], M. Britsch[10], N.H. Brook[42], H. Brown[48], A. Büchler-Germann[39], J. Buytaert[37], J.-P. Cachemiche[6], S. Cadeddu[15], J.M. Caicedo Carvajal[37], O. Callot[7], M. Calvi[20,h], M. Calvo Gomez[35,l], A. Camboni[35], W. Cameron[49], P. Campana[18], A. Carbone[14], G. Carboni[21,i], A. Cardini[15], L. Carson[47], K. Carvalho Akiba[23], G. Casse[48], M. Cattaneo[37], M. Charles[51], Ph. Charpentier[37], A. Chlopik[27], P. Ciambrone[18], X. Cid Vidal[36], P.J. Clark[46], P.E.L. Clarke[46], M. Clemencic[37], H.V. Cliff[43], J. Closier[37], C. Coca[28], V. Coco[52], J. Cogan[6], P. Collins[37], F. Constantin[28], G. Conti[38], A. Contu[51], G. Corti[37], G.A. Cowan[46], B. D'Almagne[7], C. D'Ambrosio[37], W. Da Silva[8], P. David[8], I. De Bonis[4], S. De Capua[38], M. De Cian[39], F. De Lorenzi[12], J.M. De Miranda[1], L. De Paula[2], P. De Simone[18], D. Decamp[4], H. Degaudenzi[38,37], M. Deissenroth[11], L. Del Buono[8], C. Deplano[15], O. Deschamps[5], F. Dettori[15,c], J. Dickens[43], H. Dijkstra[37], M. Dima[28], S. Donleavy[48], A. Dovbnya[40], P.-Y. Duval[6], L. Dwyer[48], R. Dzhelyadin[34], C. Eames[49], S. Easo[45], U. Egede[49], V. Egorychev[30], D. van Eijk[23], F. Eisele[11], S. Eisenhardt[46], L. Eklund[47], D.G. d'Enterria[35,m], D. Esperante Pereira[36], L. Estève[43], S. Eydelman[33], E. Fanchini[20,h], C. Färber[11], G. Fardell[46], C. Farinelli[23], S. Farry[12], V. Fave[38], V. Fernandez Albor[36], M. Ferro-Luzzi[37], S. Filippov[32], C. Fitzpatrick[46], F. Fontanelli[19,g], R. Forty[37], M. Frank[37], C. Frei[37], M. Frosini[17,e], J.L. Fungueirino Pazos[36], S. Furcas[18], A. Gallas Torreira[36], D. Galli[14,b], M. Gandelman[2], Y. Gao[3], J-C. Garnier[37], L. Garrido[35], C. Gaspar[37], N. Gauvin[38], M. Gersabeck[37], T. Gershon[44], Ph. Ghez[4], V. Gibson[43], Yu. Gilitsky[34], V.V. Gligorov[37], C. Göbel[54,2], D. Golubkov[30], A. Golutvin[49,30,37], A. Gomes[1], M. Grabalosa Gándara[35], R. Graciani Diaz[35], L.A. Granado Cardoso[37], E. Graugés[35], G. Graziani[17], A. Grecu[28], G. Guerrer[1], B. Gui[52], E. Gushchin[32], Yu. Guz[34,37], Z. Guzik[27], T. Gys[37], F. Hachon[6], G. Haefeli[38], S.C. Haines[43], T. Hampson[42], S. Hansmann-Menzemer[11], R. Harji[49], N. Harnew[51], P.F. Harrison[44], J. He[7], K. Hennessy[48], P. Henrard[5], J.A. Hernando Morata[36], E. van Herwijnen[37], A. Hicheur[38], E. Hicks[48], W. Hofmann[10], K. Holubyev[11], P. Hopchev[4], W. Hulsbergen[23], P. Hunt[51], T. Huse[48], R.S. Huston[12], D. Hutchcroft[48], V. Iakovenko[7,41], C. Iglesias Escudero[36], J. Imong[42], R. Jacobsson[37], M. Jahjah Hussein[5], E. Jans[23], F. Jansen[23], P. Jaton[38], B. Jean-Marie[7], M. John[51], C.R. Jones[43], B. Jost[37], F. Kapusta[8], T.M. Karbach[9], J. Keaveney[12], U. Kerzel[43], T. Ketel[24], A. Keune[38],

S. Khalil[52], B. Khanji[6], Y.M. Kim[46], M. Knecht[38], J. Knopf[11], S. Koblitz[37], A. Konoplyannikov[30], P. Koppenburg[23], I. Korolko[30], A. Kozlinskiy[23], M. Krasowski[25], L. Kravchuk[32], P. Krokovny[11], K. Kruzelecki[37], M. Kucharczyk[52], I. Kudryashov[31], T. Kvaratskheliya[30], D. Lacarrere[37], A. Lai[15], R.W. Lambert[37], G. Lanfranchi[18], C. Langenbruch[11], T. Latham[44], R. Le Gac[6], R. Lefevre[5], A. Leflat[31], J. Lefrançois[7], O. Leroy[6], K. Lessnoff[42], L. Li[3], Y.Y. Li[43], J. Libby[51], M. Lieng[9], R. Lindner[37], S. Lindsey[48], C. Linn[11], B. Liu[3], G. Liu[16,d,37], J.H. Lopes[2], E. Lopez Asamar[35], J. Luisier[38], B. M'charek[24], F. Machefert[7], I. Machikhiliyan[30,37], F. Maciuc[10], O. Maev[29], J. Magnin[1], A. Maier[37], R.M.D. Mamunur[37], G. Manca[15,c], G. Mancinelli[6], N. Mangiafave[43], U. Marconi[14], F. Marin[6], J. Marks[11], G. Martellotti[22], A. Martens[7], L. Martin[51], D. Martinez Santos[36], Z. Mathe[12], C. Matteuzzi[20], V. Matveev[34], A. Mazurov[32,37], G. McGregor[50], R. McNulty[12], C. Mclean[46], M. Merk[23], J. Merkel[9], M. Merkin[31], R. Messi[21,i], F.C.D. Metlica[42], J. Michalowski[25], S. Miglioranzi[37], M.-N. Minard[4], S. Monteil[5], D. Moran[12], J.V. Morris[45], R. Mountain[52], I. Mous[23], F. Muheim[46], R. Muresan[38], F. Murtas[18], B. Muryn[26], M. Musy[35], J. Mylroie-Smith[48], P. Naik[42], T. Nakada[38], R. Nandakumar[45], J. Nardulli[45], Z. Natkaniec[25], M. Nedos[9], M. Needham[38], N. Neufeld[37], L. Nicolas[38], S. Nies[9], V. Niess[5], N. Nikitin[31], A. Noor[48], A. Oblakowska-Mucha[26], V. Obraztsov[34], S. Oggero[23], O. Okhrimenko[41], R. Oldeman[15,c], M. Orlandea[28], A. Ostankov[34], J. Palacios[23], M. Palutan[18], J. Panman[37], A. Papadelis[23], A. Papanestis[45], M. Pappagallo[13,a], C. Parkes[47], G. Passaleva[17], G.D. Patel[48], M. Patel[37], S.K. Paterson[39], G.N. Patrick[45], E. Pauna[28], C. Pauna (Chiojdeanu)[28], C. Pavel (Nicorescu)[28], A. Pazos Alvarez[36], A. Pellegrino[23], G. Penso[22,j], M. Pepe Altarelli[37], S. Perazzini[14,b], D.L. Perego[20,h], E. Perez Trigo[36], A. Pérez-Calero Yzquierdo[35], P. Perret[5], G. Pessina[20], A. Petrella[16,d], A. Petrolini[19,g], B. Pietrzyk[4], D. Pinci[22], S. Playfer[46], M. Plo Casasus[36], G. Polok[25], A. Poluektov[44,33], E. Polycarpo[2], D. Popov[10], B. Popovici[28], S. Poss[6], C. Potterat[38], A. Powell[51], S. Pozzi[16,d], T. du Pree[23], V. Pugatch[41], A. Puig Navarro[35], W. Qian[3,7], J.H. Rademacker[42], B. Rakotomiaramanana[47], I. Raniuk[40], G. Raven[24], S. Redford[51], W. Reece[49], A.C. dos Reis[1], S. Ricciardi[45], K. Rinnert[48], P. Robbe[7], E. Rodrigues[47], F. Rodrigues[2], C. Rodriguez Cobo[36], P. Rodriguez Perez[36], G.J. Rogers[43], V. Romanovsky[34], G. Rospabe[4], T. Ruf[37], H. Ruiz[35], G. Sabatino[18,i], J.J. Saborido Silva[36], N. Sagidova[29], B. Saitta[15,c], C. Salzmann[39], A. Sambade Varela[37], M. Sannino[19,g], R. Santacesaria[22], R. Santinelli[37], E. Santovetti[21,i], M. Sapunov[6], A. Sarti[18], C. Satriano[22,k], A. Satta[21], M. Savrie[16,d], D. Savrina[30], P. Schaack[49], M. Schiller[11], S. Schleich[9], M. Schmelling[10], B. Schmidt[37], O. Schneider[38], A. Schopper[37], M.-H. Schune[7], R. Schwemmer[37], A. Sciubba[18,j], M. Seco[36], A. Semennikov[30], K. Senderowska[26], N. Serra[23], J. Serrano[6], B. Shao[3], M. Shapkin[34], I. Shapoval[40], P. Shatalov[30], Y. Shcheglov[29], T. Shears[48], L. Shekhtman[33], V. Shevchenko[30], E. Simioni[24], H.P. Skottowe[43], T. Skwarnicki[52], A.C. Smith[37], K. Sobczak[5], F.J.P. Soler[47], A. Solomin[42], P. Somogy[37], F. Soomro[49], B. Souza De Paula[2], B. Spaan[9], A. Sparkes[46], E. Spiridenkov[29], P. Spradlin[51], F. Stagni[37], O. Steinkamp[39], S. Stoica[28], S. Stone[52], B. Storaci[23], U. Straumann[39], N. Styles[46], K. Syryczynski[27], M. Szczekowski[27], P. Szczypka[38], T. Szumlak[47,26], S. T'Jampens[4], H. Terrier[23],

F. Teubert[37], C. Thomas[51,45], E. Thomas[37], J. van Tilburg[39], M. Tobin[39], S. Topp-Joergensen[51], M.T. Tran[38], S. Traynor[12], A. Tsaregorodtsev[6], N. Tuning[23], A. Ukleja[27], O. Ullaland[37], U. Uwer[11], V. Vagnoni[14], G. Valenti[14], A. Van Lysebetten[23], R. Vazquez Gomez[35], P. Vazquez Regueiro[36], S. Vecchi[16], J.J. Velthuis[42], M. Veltri[17,f], K. Vervink[37], B. Viaud[7], I. Videau[7], X. Vilasis-Cardona[35,l], A. Vollhardt[39], A. Vorobyev[29], An. Vorobyev[29], H. Voss[10], H. de Vries[23], K. Wacker[9], S. Wandernoth[11], J. Wang[52], D.R. Ward[43], D. Websdale[49], M. Whitehead[44], D. Wiedner[11], L. Wiggers[23], G. Wilkinson[51], M. Williams[49], F.F. Wilson[45], M. Witek[25], W. Witzeling[37], S.A. Wotton[43], K. Wyllie[37], Y. Xie[46], F. Xing[51], Z. Yang[3], G. Ybeles Smit[23], R. Young[46], O. Yushchenko[34], L. Zhang[52], Y. Zhang[3], A. Zhelezov[11] and N. Zwahlen[38]

[1] *Centro Brasileiro de Pesquisas Físicas (CBPF), Rio de Janeiro, Brazil.*
[2] *Universidade Federal do Rio de Janeiro (UFRJ), Rio de Janeiro, Brazil.*
[3] *Center for High Energy Physics, Tsinghua University, Beijing, China.*
[4] *LAPP, Université de Savoie, CNRS/IN2P3, Annecy-Le-Vieux, France.*
[5] *Clermont Université, Université Blaise Pascal, CNRS/IN2P3, LPC, Clermont-Ferrand, France.*
[6] *CPPM, Aix-Marseille Université, CNRS/IN2P3, Marseille, France.*
[7] *LAL, Université Paris-Sud, CNRS/IN2P3, Orsay, France.*
[8] *LPNHE, Université Pierre et Marie Curie, Université Paris Diderot, CNRS/IN2P3, Paris, France.*
[9] *Fakultät Physik, Technische Universität Dortmund, Dortmund, Germany.*
[10] *Max-Planck-Institut für Kernphysik (MPIK), Heidelberg, Germany.*
[11] *Physikalisches Institut, Ruprecht-Karls-Universität Heidelberg, Heidelberg, Germany.*
[12] *School of Physics, University College Dublin, Dublin, Ireland.*
[13] *Sezione INFN di Bari, Bari, Italy.*
[14] *Sezione INFN di Bologna, Bologna, Italy.*
[15] *Sezione INFN di Cagliari, Cagliari, Italy.*
[16] *Sezione INFN di Ferrara, Ferrara, Italy.*
[17] *Sezione INFN di Firenze, Firenze, Italy.*
[18] *Laboratori Nazionali dell'INFN di Frascati, Frascati, Italy.*
[19] *Sezione INFN di Genova, Genova, Italy.*
[20] *Sezione INFN di Milano Bicocca, Milano, Italy.*
[21] *Sezione INFN di Roma Tor Vergata, Roma, Italy.*
[22] *Sezione INFN di Roma Sapienza, Roma, Italy.*
[23] *Nikhef, National Institute for Subatomic Physics, Amsterdam, Netherlands.*
[24] *Nikhef, National Institute for Subatomic Physics and Vrije Universiteit, Amsterdam, Netherlands.*
[25] *Henryk Niewodniczanski Institute of Nuclear Physics Polish Academy of Sciences, Cracow, Poland.*
[26] *Faculty of Physics & Applied Computer Science, Cracow, Poland.*
[27] *Soltan Institute for Nuclear Studies, Warsaw, Poland.*
[28] *Horia Hulubei National Institute of Physics and Nuclear Engineering, Bucharest-Magurele, Romania.*

[29]*Petersburg Nuclear Physics Institute (PNPI), Gatchina, Russia.*
[30]*Institute of Theoretical and Experimental Physics (ITEP), Moscow, Russia.*
[31]*Institute of Nuclear Physics, Moscow State University (SINP MSU), Moscow, Russia.*
[32]*Institute for Nuclear Research of the Russian Academy of Sciences (INR RAN), Moscow, Russia.*
[33]*Budker Institute of Nuclear Physics (BINP), Novosibirsk, Russia.*
[34]*Institute for High Energy Physics(IHEP), Protvino, Russia.*
[35]*Universitat de Barcelona, Barcelona, Spain.*
[36]*Universidad de Santiago de Compostela, Santiago de Compostela, Spain.*
[37]*European Organisation for Nuclear Research (CERN), Geneva, Switzerland.*
[38]*Ecole Polytechnique Fédérale de Lausanne (EPFL), Lausanne, Switzerland.*
[39]*Universität Zürich, Zürich, Switzerland.*
[40]*NSC Kharkiv Institute of Physics and Technology (NSC KIPT), Kharkiv, Ukraine.*
[41]*Institute for Nuclear Research of the National Academy of Sciences (KINR), Kyiv, Ukraine.*
[42]*H.H. Wills Physics Laboratory, University of Bristol, Bristol, United Kingdom.*
[43]*Cavendish Laboratory, University of Cambridge, Cambridge, United Kingdom.*
[44]*Department of Physics, University of Warwick, Coventry, United Kingdom.*
[45]*STFC Rutherford Appleton Laboratory, Didcot, United Kingdom.*
[46]*School of Physics and Astronomy, University of Edinburgh, Edinburgh, United Kingdom.*
[47]*Department of Physics and Astronomy, University of Glasgow, Glasgow, United Kingdom.*
[48]*Oliver Lodge Laboratory, University of Liverpool, Liverpool, United Kingdom.*
[49]*Imperial College London, London, United Kingdom.*
[50]*School of Physics and Astronomy, University of Manchester, Manchester, United Kingdom.*
[51]*Department of Physics, University of Oxford, Oxford, United Kingdom.*
[52]*Syracuse University, Syracuse, NY, United States of America.*
[53]*CC-IN2P3, CNRS/IN2P3, Lyon-Villeurbanne, France, associated member.*
[54]*Pontifícia Universidade Católica do Rio de Janeiro (PUC-Rio), Rio de Janeiro, Brazil, associated to* [2]*.*
[a]*Università di Bari, Bari, Italy.*
[b]*Università di Bologna, Bologna, Italy.*
[c]*Università di Cagliari, Cagliari, Italy.*
[d]*Università di Ferrara, Ferrara, Italy.*
[e]*Università di Firenze, Firenze, Italy.*
[f]*Università di Urbino, Urbino, Italy.*
[g]*Università di Genova, Genova, Italy.*
[h]*Università di Milano Bicocca, Milano, Italy.*
[i]*Università di Roma Tor Vergata, Roma, Italy.*
[j]*Università di Roma La Sapienza, Roma, Italy.*
[k]*Università della Basilicata, Potenza, Italy.*
[l]*LIFAELS, La Salle, Universitat Ramon Llull, Barcelona, Spain.*
[m]*Institució Catalana de Recerca i Estudis Avançats (ICREA), Barcelona, Spain.*

# Contents

# Chapter 1

# Introduction

The Technical Proposal of the LHCb experiment was submitted in February 1998 [1]. Construction of the two B Factories was already in progress and they were expected to produce physics results well before the LHCb experiment would become operational. The emphasis on physics described in the Technical Proposal was to test more precisely the flavour aspect of the Standard Model. The expected performance of the LHCb detector showed that the properties of the flavour-changing neutral currents and the consistency of the CKM Unitarity Triangle, determined by a set of four independent parameters (e.g. $\lambda$, $A$, $\rho$ and $\eta$ in the Wolfenstein parametrization), could be tested more precisely by LHCb. This was due to much higher statistics of $B^+$ and $B^0$ mesons than would be reachable by the B meson factories, as well as the unique availability of high statistics of $B_s^0$ decays at LHC energies. It was argued that:

- BABAR and Belle might discover signs of physics beyond the Standard Model. In such a case, higher precision on those measurements and results from the $B_s^0$ meson system would be essential to understand the nature of the new physics.
- Alternatively, the new physics could manifest itself in such a way that its existence would be concealed when only the $B^0$ and $B^+$ mesons were studied. By adding the $B_s^0$ meson the new physics would be revealed even with such a conspiracy.

The LHCb experiment was approved in December 1998.

When the design of the LHCb detector became mature and construction of some detector components started, it was noticed that the material budget of the entire tracking system had almost doubled compared to the original estimate. This would cause a deterioration of the momentum resolution due to the increased multiple scattering, a decrease of the tracking efficiency due to the hadronic interactions of particles with the detector material, and a loss of electrons and photons due to the electromagnetic interactions with the detector material. For these reasons, the set-up of the entire tracking system was reconsidered. As a result, the numbers of sensors in the vertex detector and the tracking stations were reduced. The most notable change was that the tracking stations inside the spectrometer dipole were removed and the track reconstruction algorithm was modified for this new set-up [2].

Another significant change in the experiment has concerned the trigger. An earlier design of the trigger consisted of three levels: the first level selection was based on the transverse momentum, $p_{\mathrm{T}}$, of hadrons, leptons and photons, and implemented with custom-made electronics. In the second level, data were read out from the vertex detector and the first tracking station into a processor farm, and used to reduce the event rate further. In the third level, all the detector information were read out to the processor farm for more sophisticated event selections based on the fully reconstructed events. After the third level, events would be written on a storage device with a rate of $\sim$ 200 Hz. This scheme has now been modified to have only two levels. After the first level, which remains essentially unchanged, all the detector information is readout with a rate of 1 MHz to a processor farm. Selection software running in the processor farm then reduces the event rate to 2 kHz to be recorded on a storage device for offline analysis. The selection of the events in the second level is performed in two stages. In the first stage, the decisions made by the first level of the trigger are confirmed with more accurate $p_{\mathrm{T}}$ determinations using the tracking system, in combination with the vertex detector information. For this, only the tracks under consideration are reconstructed. In the second stage, the events that passed the first stage are fully reconstructed and inclusive signatures of B-hadron decays are looked for. For a few decay modes, exclusive reconstruction of the final states can also be performed. The entire trigger scheme has been simulated and verified that it delivers the desired performance, as discussed below in Sect. 1.2.

Since the approval of the LHCb experiment, impressive progress has been made in B physics. The BABAR and Belle experiments produced many results testing the CKM mechanism of the Standard Model much more stringently than originally anticipated. There is now even strong evidence of $\mathrm{D}^0$-$\overline{\mathrm{D}}^0$ oscillations. The two Tevatron experiments, CDF and D0, have also produced several interesting measurements in the $\mathrm{B}_\mathrm{s}^0$ sector. At this moment, the data in general show a striking agreement with the Standard Model picture. Therefore, the emphasis of the LHCb experiment has shifted more to the search for physics *beyond* the Standard Model. Thanks to the large b-quark cross section expected in p-p collisions at the LHC energies, the high event reconstruction efficiency of the detector, and the flexible high-level trigger based on software, the LHCb experiment is ready to significantly extend the sensitivity to new physics, as the next generation heavy flavour experiment.

This document summarizes the expected physics performance of the LHCb detector obtained from a simulation set-up with a material budget describing the actually constructed detector components, and detector responses based on the prototype tests. The reconstruction and analysis programs are essentially identical to those which will be used soon on the real data. The physics programme of the LHCb experiment is broad, involving many channels of interest for CP asymmetries or rare decays in the B and charm sectors, as well as studies of lepton flavour violation, and more exotic searches. The physics topics selected for discussion in the following six chapters concern the subjects where LHCb should make a large impact, already during the early period of data taking, by constraining or, even better, discovering new physics. The chapters on the selected key measurements were independently produced, and no particular effort has been made to enforce uniformity of their treatment. The main purpose of this write-up is to demonstrate the readiness of the LHCb experiment to perform important physics analyses, after all the modifications of the detector and the changes in the physics landscape that have taken place since the Technical Proposal. In addition, although based on Monte Carlo simulation, the studies have prepared the way for analysis of the real data that will soon start flowing from the LHC, with a focus on extracting the results using techniques that are based, wherever possible, on the data themselves.

## 1.1 Key measurements

Particles associated to new physics would make additional contributions in the loop processes of the flavour-changing neutral current, such as penguin and box diagrams. Amplitudes of those processes could then:

1. change the phases of the couplings;
2. change their absolute values;
3. change the Lorentz structure.

For accurate amplitude predictions from the Standard Model, the four parameters of the CKM mass mixing matrix, in particular $\rho$ and $\eta$, must be measured precisely. Furthermore, this must be done in a way that the extracted parameters are not affected by possible new physics contributions, i.e. done only with the quantities related to tree diagrams. One of the most promising ways is to combine the decay width $\Gamma(\mathrm{b} \to \mathrm{uW})$, which depends on $\sqrt{\rho^2 + \eta^2}$, and CP violation in $\mathrm{B} \to \mathrm{DK}$ decays, which gives $\tan^{-1} \eta/\rho$ (the angle $\gamma$ of the Unitarity Triangle[1]). LHCb is expected to make a large improvement for the latter measurement, which is discussed in Chapter 2. It is particularly interesting to compare this with a measurement of $\gamma$ involving loop processes, since that could receive a contribution from new physics. This possibility is discussed in Chapter 3, where the extraction of $\gamma$ is made from the charmless two-body B-meson decays.

CP violation in the $\mathrm{B_s^0}$ system is still to be explored. CDF and D0 have put first constraints on the mixing-induced CP-violating phase in $\mathrm{B_s^0} \to \mathrm{J}/\psi\phi$ decays. Since the Standard Model effect is only $O(10^{-2})$ and the current measurement errors are still $O(10^{-1})$, a large contribution from new physics is not yet excluded. The study of this channel is presented in Chapter 4.

Absolute values of transition amplitudes are related to the branching fractions. One of the decay modes where there could be still a sizable effect from new physics is $\mathrm{B_s^0} \to \mu^+\mu^-$. The current limit on the branching fraction set by the Tevatron experiments is more than an order of magnitude above the Standard Model prediction of $O(10^{-9})$. This channel is discussed in Chapter 5.

The Lorentz structure of the flavour changing neutral current can be probed by studying the angular distribution of the decay final states in $\mathrm{B^0} \to \mathrm{K^{*0}}\mu^+\mu^-$, which is discussed in Chapter 6. Another way to probe the Lorentz structure is to study the time evolution of $B_s^0 \to \phi\gamma$ decays, presented in Chapter 7. A CP asymmetry can be present only if there exists CP violation in the decay amplitudes, or if there is an admixture of right-polarised photons, only possible due to the finite mass of the s-quark. Both effects are predicted to be very small in the Standard Model. It is worth noting that changes in the branching fraction or in the Lorentz structure could happen even for types of new physics where there is no extra phase.

[1]This angle is also referred to as $\phi_3$ in the literature.

## 1.2 The LHCb trigger

The LHCb trigger [3] is composed of two levels, called Level-0 (L0) and the high level trigger (HLT). L0 is synchronous, implemented in custom electronics and reduces the rate to 1 MHz. At this frequency the whole detector is read out, and the events are distributed over up to 2000 multicore computing nodes forming the Event Filter Farm (EFF). The HLT consists of a C++ application which is running on every CPU of the EFF. Each HLT application has access to all data in one event, and thus in principle could be executing the offline selection algorithms. But given the 1 MHz output rate of L0 and the limited CPU power available, the first part of the HLT (HLT1) aims at rejecting the bulk of the background events by using only part of the full information which is available, to reduce the rate to around 30 kHz. This rate is sufficiently low to allow full pattern recognition on the remaining events. The second part of the HLT (HLT2) uses a combination of inclusive and exclusive selections to reduce the rate to around 2 kHz, at which rate the data are written to storage.

L0 uses information from the Calorimeters and Muon Chambers. The $E_{\mathrm{T}}$ of hadron, electron and photon clusters, and the two highest $p_{\mathrm{T}}$ muons are reconstructed. L0 accepts an event when an L0-object with a transverse momentum above a threshold is found. A fast estimation of the number of p-p interactions per bunch crossing will be provided by two dedicated layers of the vertex detector and can be used in order to veto events with very high multiplicity.

HLT1 applies a progressive partial reconstruction through different sequences, called lines, each one seeded by an L0 candidate. The general strategy is to refine the $p_{\mathrm{T}}$ measurement by matching the L0 triggering objects to track segments from the Tracking Stations and to the vertex detector. Furthermore the impact parameter (IP) of the L0-confirmed candidate with respect to a reconstructed primary vertex (PV) is measured and can also be used in the decision. Finally it is possible to reconstruct secondary vertexes and take a decision based on their invariant mass. The main HLT1 lines are:

- Hadron Line [4]. Using a L0-Hadron candidate as seed, an event is confirmed if a reconstructed hadron has $p_{\mathrm{T}}$ and IP above a threshold. The event can also be accepted if a secondary vertex is identified from the seed and a reconstructed track.

- Muon Line [5]. HLT1 accepts an event if an L0-Muon or an L0-DiMuon seed is confirmed in a similar way as in the Hadron Line. In the muon line it is possible also to accept events without any IP requirement, only using more strict $p_{\mathrm{T}}$ and invariant mass thresholds.

- Muon and Track Line [6]. If it is possible to produce a secondary vertex from a confirmed L0-Muon candidate and a reconstructed track that fulfills $p_{\mathrm{T}}$, IP and invariant mass requirements this line accepts the event.

- Electron Line [7]. Starting from an L0-electron candidate, this line accepts an event if a confirmation of the electron or of an electron plus a track is obtained.

- Photon Line [7]. Using as seed an L0-electromagnetic candidate, the algorithm looks for a confirmed photon plus a track with IP, or a two track displaced vertex in order to accept the event.

In HLT2 the full event is reconstructed and particle identification is performed by the calorimeter and the muon system. Then a trigger decision is taken according to two categories of selections:

- Inclusive Selections.
  - Topological trigger. Triggers on a displaced vertex with invariant mass compatible with a b(c) decay.
  - Inclusive $\phi$. When it is possible to identify a $\phi \to \mathrm{K}^+\mathrm{K}^-$ decay not pointing to any PV the event is triggered. This line has a very low rate, which allows it to use the RICH particle ID.
  - Leptonic lines identify either one lepton, with $p_{\mathrm{T}}$ or IP requirements, or a two track vertex formed with at least one lepton, with invariant mass or IP requirements.

- Exclusive Selections. Full reconstruction of the event allows many interesting events to be selected at much lower trigger rates. Moreover, for some channels the inclusive selections cannot be very efficient and the exclusive approach is needed.

The total available bandwidth of the Trigger/DAQ is limited, and has to be shared by all trigger lines. The sharing of bandwidth between the different physics channels aims at minimizing the efficiency loss relative to its maximum efficiency possible if a particular channel would be allowed to absorb the whole bandwidth. The trigger efficiencies calculated with respect to offline-selected events for the channels which are discussed in the remainder of this paper are presented in Tables 1.1 and 1.2. These have been determined using high statistics samples of events from the full simulation[2] with a tuning of the thresholds optimized as described above, such that at nominal luminosity the output rate would be 2 kHz. Note that the trigger performance discussed for each of the key measurements in the following chapters may differ in detail from the results given in these tables, due to the recent evolution of the trigger selection criteria. It is beyond the scope of this document to list the trigger performance for a more complete list of channels which are under investigation in LHCb [8], but the trigger performance can be summarized with efficiencies in the range of 70–90% for leptonic channels, from 20–50% for hadronic channels, and of the order of 50% for radiative decays.

[2]DC06 Monte Carlo, at nominal LHC conditions of $\sqrt{s} = 14$ TeV, with spillover corresponding to 25 ns bunch crossing separation.

Table 1.1: L0 and L0×HLT1 efficiencies per trigger line, for offline-selected events.

| Channel | L0 | Hadron | Muon | | Eletromag. | | Total |
|---|---|---|---|---|---|---|---|
| | HLT1 | $h$ | $\mu$ | $\mu$+Tr. | $e$ | $\gamma$ | Efficiency |
| $B^+ \to D^0(K^0_S\pi\pi)K$ | L0 | 50% | 11% | | 7% | | 55% |
| | L0×HLT1 | 31% | 6% | 4% | 2% | 2% | 36% |
| $B \to hh$ | L0 | 60% | 10% | | 10% | | 65% |
| | L0×HLT1 | 49% | 4% | 2% | 4% | 1% | 52% |
| $B^0_s \to J/\psi\phi$ | L0 | 27% | 91% | | 8% | | 92% |
| | L0×HLT1 | 7% | 84% | 62% | 1% | 1% | 86% |
| $B^0_s \to \mu\mu$ | L0 | 17% | 98% | | 5% | | 98% |
| | L0×HLT1 | 3% | 97% | 91% | 0% | 2% | 97% |
| $B^0 \to K^*\mu\mu$ | L0 | 31% | 84% | | 7% | | 88% |
| | L0×HLT1 | 17% | 77% | 73% | 1% | 2% | 83% |
| $B^0_s \to \phi\gamma$ | L0 | 50% | 10% | | 72% | | 82% |
| | L0×HLT1 | 12% | 4% | 1% | 1% | 51% | 58% |

Table 1.2: L0×HLT1×HLT2 efficiencies per trigger line, for offline-selected events.

| Channel | b-topological | inclusive $\phi$ | leptonic | combined |
|---|---|---|---|---|
| $B^+ \to D^0(K^0_S\pi\pi)K$ | 18% | - | 4% | 21% |
| $B \to hh$ | 42% | - | 2% | 42% |
| $B^0_s \to J/\psi\phi$ | 34% | 38% | 82% | 84% |
| $B^0_s \to \mu\mu$ | 80% | - | 94% | 95% |
| $B^0 \to K^*\mu\mu$ | 59% | - | 70% | 74% |
| $B^0_s \to \phi\gamma$ | 1% | 50% | 1% | 51% |

# Chapter 2

# The tree-level determination of $\gamma$

M. Adinolfi, Y. Amhis, T. Bauer, S. Blusk, S. Brisbane, A.Contu, T. Gershon, V. Gibson, V. V. Gligorov, S. Haines, N. Harnew, P. Hunt, M. John, C. Jones, Y.-Y. Li, J. Libby, B. M'charek, L.Martin, M. Merk, J. Nardulli, M. Patel, A. Powell, J. Rademacker, S. Ricciardi, E. Rodrigues, O. Schneider, M.-H. Schüne, P. Spradlin, S. Stone, P. Szczypka, C. Thomas, G. Wilkinson and L. Zhang

**Abstract**

The current status of studies related to the measurement of the Standard Model $CP$-violating parameter $\gamma$ by LHCb is described. These measurements are made with decays that proceed by tree-level processes, which are expected to be unaffected by New Physics.

# Contents

# 1 Introduction

Results from the $e^+e^-$ $B$-factories and the Tevatron over the past decade have improved significantly the knowledge of the angles and sides of the Unitarity Triangle [1,2,3]. However, many of the tightest constraints are from processes that take place via virtual loops, which are sensitive to New Physics. Figure 1 shows the current constraints on the Unitarity Triangle from loop and tree-level (Standard Model) measurements; the tree-level determinations lead to significantly worse constraints. Therefore, a principal goal of future flavour experiments is to improve the Standard Model determinations to a level comparable to, or better than, those from loop processes. Any differences between the tree-level and loop Unitarity Triangles will indicate New Physics. In particular, comparison with the value of $\gamma$ measured in processes affected by penguin loops [4] will be a powerful probe of possible New Physics contributions; such measurements are another important goal of the LHCb experiment [5]. Furthermore, over constraining the Standard Model Unitarity Triangle is the most rigourous test of unitarity and the three-generation quark model.

The tree-level determination of $\gamma$ is the most promising route to a significant experimental improvement in the knowledge of the Standard Model Unitarity Triangle. Improved determinations of $|V_{ub}|/|V_{cb}|$ would, on the other hand, rely on more precise theoretical QCD-based calculations [6].

New Physics contributions to the tree-level processes sensitive to $\gamma$ are not excluded. For example an additional contribution could arise from a charged Higgs boson, $H^{\pm}$, mediating the decay, rather than a $W^{\pm}$ boson. However, given the large $H^{\pm}$ mass required to explain the observed rates of processes such as $b \to s\gamma$ and $B^+ \to \tau^+\nu$ (for example see Ref. [7]) no significant contribution is anticipated compared to those from Standard Model processes.

The LHCb experiment has been investigating several techniques, both time-dependent and time-independent, to determine $\gamma$ at tree-level. Studies have shown that with the expected total LHCb data set,[1] corresponding to an integrated luminosity of 10 fb$^{-1}$, a precision, including systematic uncertainties, of $1.9 - 2.7^{\circ}$ can be achieved [9]; this is around an order of magnitude improvement upon the current measurements and is better than the current precision from indirect determinations of $\gamma$ from CKM fits [2, 3]. Furthermore, these studies are by no means exhaustive; there are several additional channels that have either only been subject to preliminary study or have not yet been investigated.

This document or 'roadmap' presents the information related to the first measurements that will be performed to determine $\gamma$ at LHCb. By the end of 2010 it is foreseen that a data set corresponding to approximately 0.2 fb$^{-1}$ of integrated luminosity will be collected. The analyses that can be performed on these data are the focus of this document. However, most results are quoted with respect to larger data sets corresponding to 0.5 fb$^{-1}$ or 2 fb$^{-1}$ (one nominal year of LHCb data taking). With these data sets significant improvements of measurements of $\gamma$ over the $e^+e^-$ $B$ factories are anticipated.

[1]This does not include the twenty-fold increase in statistics for hadronic modes that the proposed LHCb upgrade [8] would collect.

The document is structured as follows. Sections 2 and 3 describe the benchmark techniques at LHCb to determine $\gamma$ and review the current status of these measurements at other experiments, respectively. Section 4 describes the LHCb triggering of these modes. Section 5 reviews the status of the offline-selection studies for these channels. Section 6 describes the control samples required for the analyses and the status of their selections. The various $\gamma$ extraction methods from the events selected in the different modes are briefly reviewed in Sec. 7. The systematic uncertainties related to these measurements and how they will be controlled are detailed in Sec. 8; this includes a discussion of the external inputs from other experiments that are required. Section 9 describes the important milestones that must be met before the first analyses can take place. The conclusions and outlook are given in Sec. 10.

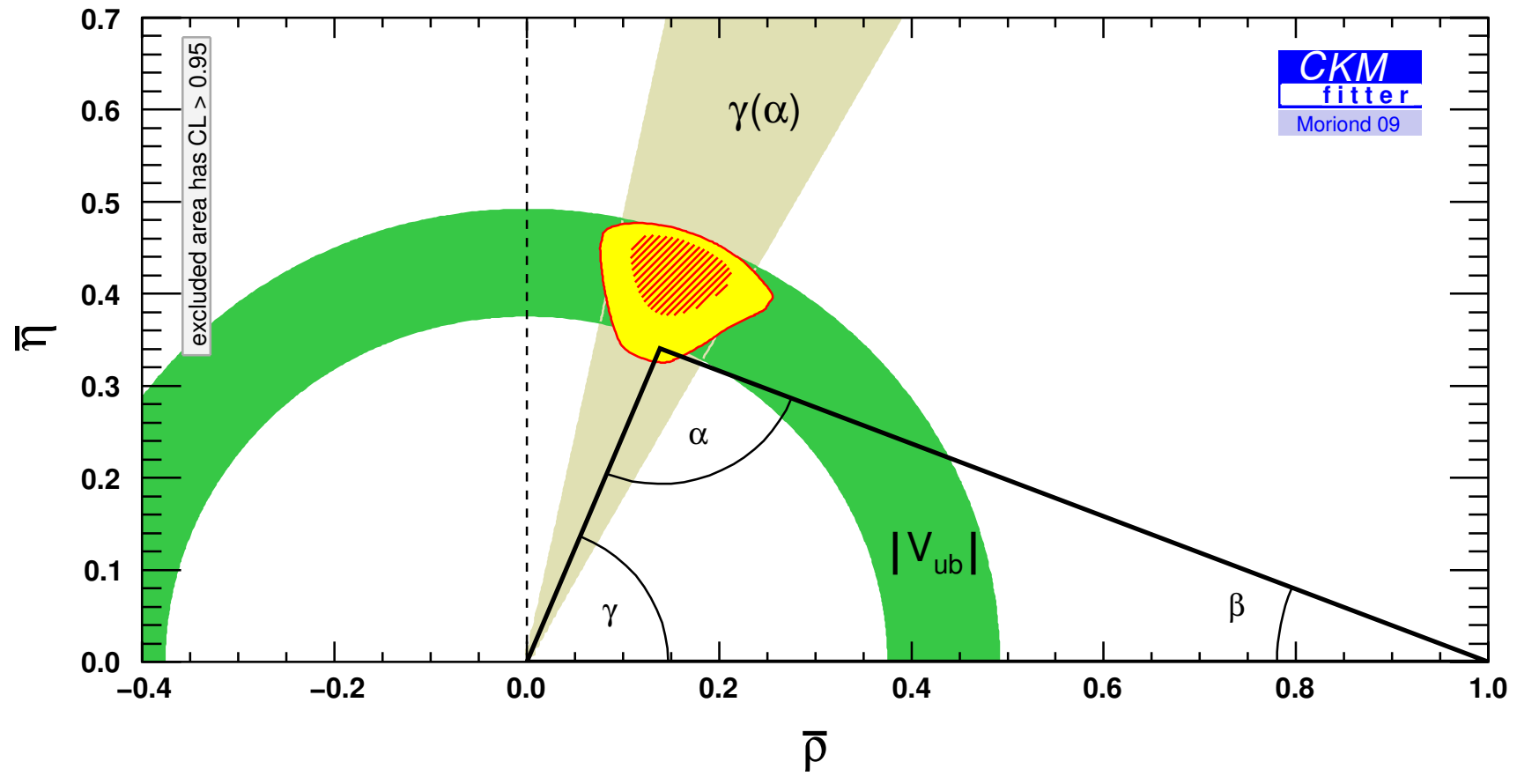

(a)

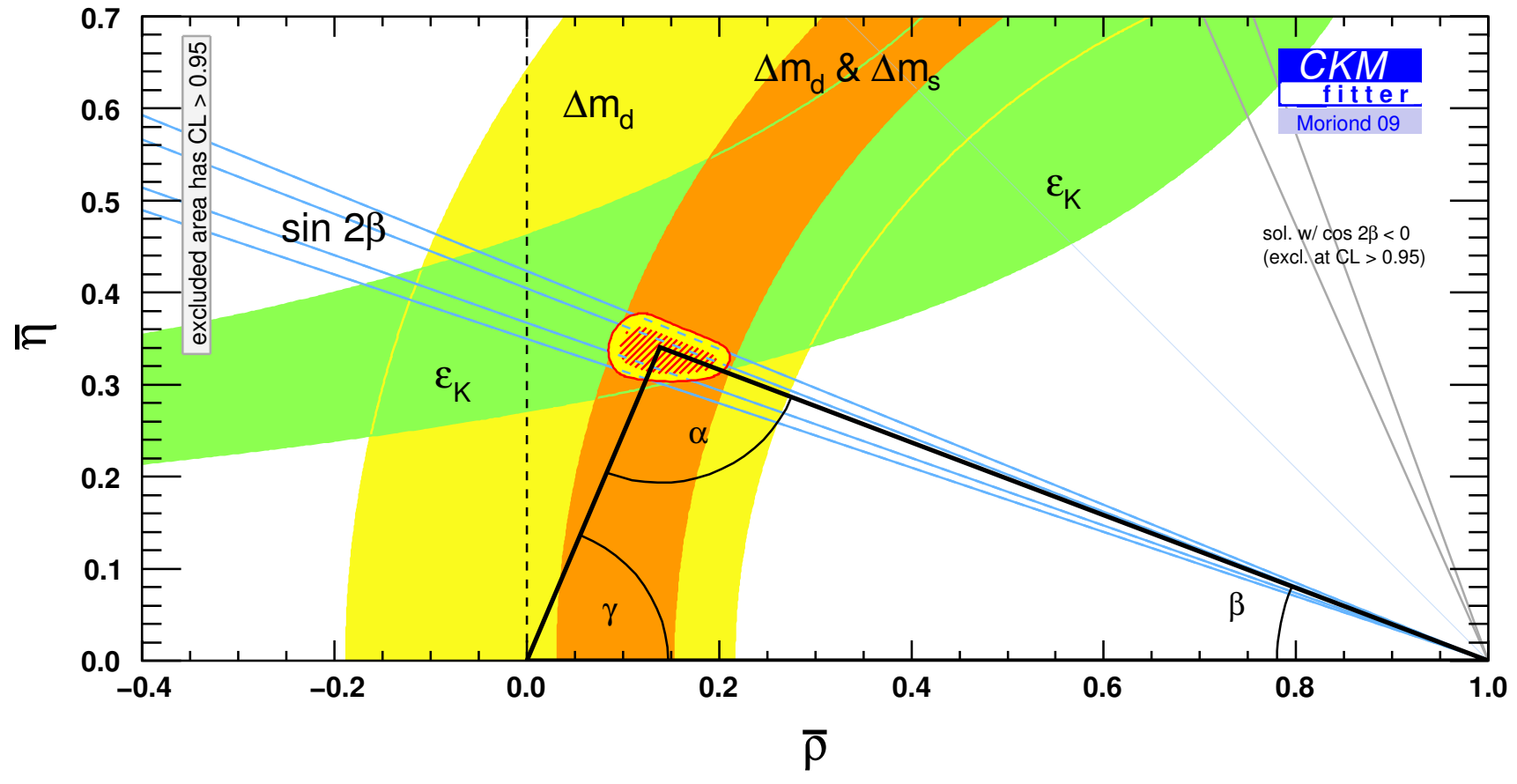

(b)

Figure 1: Constraints on the Unitarity Triangle from (a) tree and (b) loop processes. The constraints are the Winter 2009 results from the CKMfitter collaboration [3]. The tree-determination of $\gamma$ is dominated by constraints derived from measurements of $\alpha$ ($\gamma = 180^\circ - \beta - \alpha$) from $B \to \pi\pi$, $B \to \rho\rho$ and $B \to \rho\pi$ decays. However, the determination of $\alpha$ requires theoretical assumptions to be made related to isospin breaking and the contribution of electroweak penguin amplitudes.

# 2 Methods for measuring $\gamma$ at tree-level

The techniques for measuring $\gamma$ at LHCb fall into two categories: measurements of direct $CP$-violation in the decays $B^- \to DK^-$ and $\bar{B}^0 \to D\bar{K}^{*0}$,[2] where $D$ indicates a $D^0$ or $\bar{D}^0$ decaying into a common final state; and time-dependent measurements of $CP$-violation in $B^0 \to D^{(*)\mp}\pi^\pm$ and $B_s^0 \to D_s^\mp K^\pm$ decays. Sections 2.1 and 2.2 describe the $B^- \to DK^-/\bar{B}^0 \to D\bar{K}^{*0}$ and the time-dependent methods, respectively.

## 2.1 Time-integrated measurements of $B^- \to DK^-$ and $\bar{B}^0 \to D\bar{K}^{*0}$

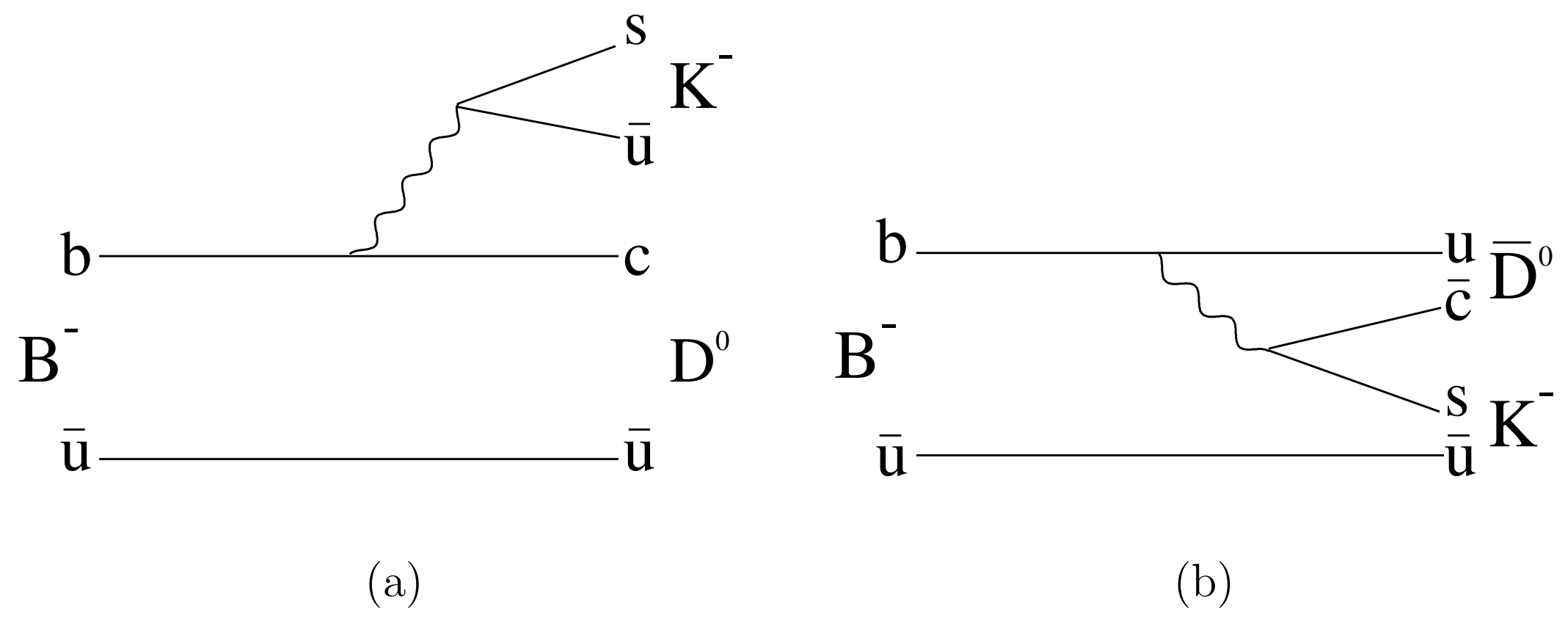

Figure 2: Feynman diagrams for (a) $B^- \to D^0K^-$ and (b) $B^- \to \bar{D}^0K^-$. There is a relative phase of $\delta_B - \gamma$ between the two amplitudes, and a relative magnitude of $r_B$.

Time-integrated measurements of the self-tagging modes $B^- \to DK^-$ and $\bar{B}^0 \to D\bar{K}^{*0}$ are sensitive to $\gamma$ at tree-level. Figure 2 shows the two tree-level processes for the charged $B$ case: diagram (a) depends on the CKM matrix element $V_{cb}$, whereas the diagram (b) depends on $V_{ub}$. Therefore it is CKM suppressed, in addition to colour suppression that occurs in internal $W$-emission diagrams. The weak-phase difference between $V_{ub}$ and $V_{cb}$ is $-\gamma$.[3] Therefore, interference between these two amplitudes when the $D^0$ or $\bar{D}^0$ decay to the same final state gives sensitivity to $\gamma$. In addition, there is a dependence on the ratio between the magnitude of the suppressed amplitude and the favoured amplitude, $r_B$, the strong-phase difference between these two amplitudes, $\delta_B$, and parameters of the specific $D$ decay; these parameters must also be determined from data or taken from other measurements of $D$ decay. The size of $r_B$ governs the amount of interference and hence the sensitivity to $\gamma$. The value of $r_B$ is expected to be $\frac{1}{3}|V_{ub}V_{cs}|/|V_{cb}V_{us}| \sim 0.1$, where the factor of $\frac{1}{3}$ is that due to colour suppression. The measured value of $r_B$ [2, 3] is in good agreement with this expectation.

[2]Charge-conjugate final states are implicit unless specifically stated otherwise.

[3]This is assuming that there is a negligible weak-phase difference between $V_{us}$ and $V_{cs}$, which is a good approximation in the three-generation quark model [10].

This method has no pollution from penguin loops, which mediate flavour changing neutral current processes such as $b \to sg$, that can be influenced by New Physics. The largest correction is due to $D^0\bar{D}^0$ mixing [11]. However, the observed rate of mixing [1] is such that a bias of $\ll 1^\circ$ on $\gamma$ is introduced [12]. In addition, $CP$ violation in $D$ decays is ignored; the current limits on $CP$ violation in the $D$ system [1] mean that any bias introduced by this assumption would have negligible impact on the extraction of $\gamma$.

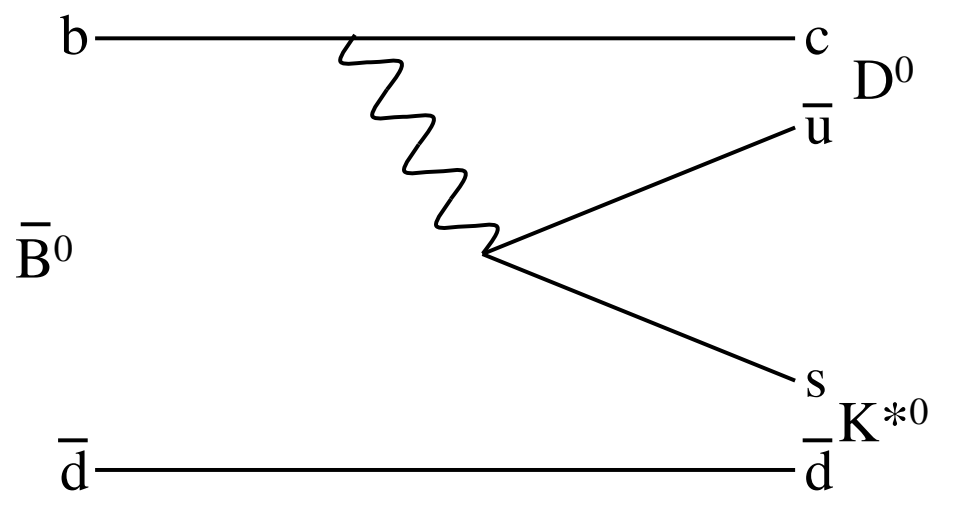

(a)

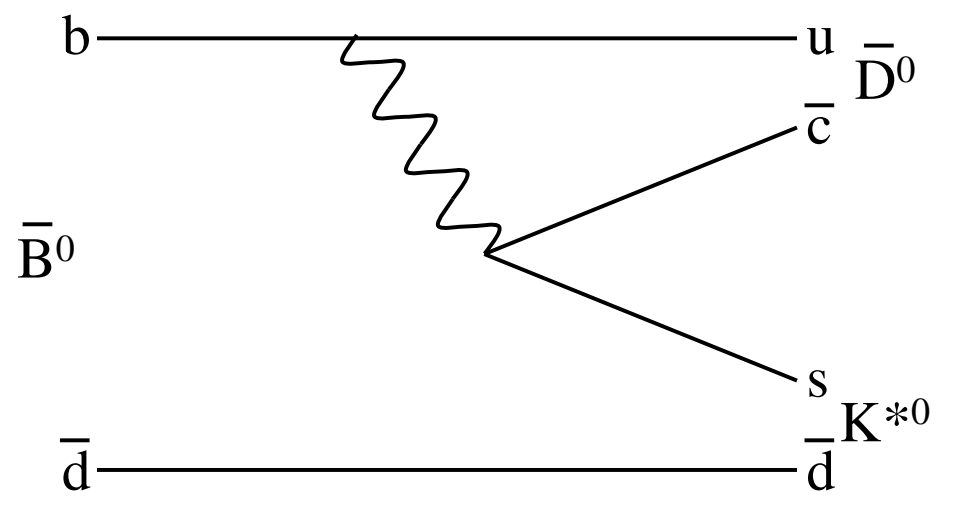

(b)

Figure 3: Feynman diagrams for (a) $\bar{B}^0 \to D^0K^{*0}$ and (b) $\bar{B}^0 \to \bar{D}^0K^{*0}$. There is a relative phase of $\delta_{B^0} - \gamma$ between the two amplitudes, and a relative magnitude of $r_{B^0}$.

For the neutral case, $\bar{B}^0 \to DK^{*0}$, the two diagrams are shown in Fig. 3. Both diagrams are colour suppressed which reduces the branching fraction but increases the size of the interference. There are also two new parameters introduced, $r_{B^0}$ and $\delta_{B^0}$, which are analogous to $r_B$ and $\delta_B$ in the charged case; the value of $r_{B^0}$ is expected to approximately 0.3 due to the lack of colour suppression between the two amplitudes compared to $B^- \to DK^-$.

The determination of $\gamma$ with different $D$ decay modes is discussed in Secs. 2.1.1 and 2.1.2 below.

### 2.1.1 ADS/GLW measurements

The first proposal to measure $\gamma$ with $B^- \to DK^-$ was with $D$ decay to $CP$ eigenstates [13], which is known as the GLW method after its authors Gronau, London and Wyler. LHCb has studied the $CP$-even final states $\pi^+\pi^-$ and $K^+K^-$. In addition, $D$ decays to the flavour specific final state $K^+\pi^-$ [14] can be used; this is known as the ADS method after its authors Atwood, Dunietz and Soni. The ADS technique involves measuring four separate rates; two of these rates are suppressed but the interference effects are very large leading to excellent sensitivity to $\gamma$. However, to determine $\gamma$, $\delta_B$ and $r_B$, the strong-phase difference between the doubly Cabibbo suppressed (DCS) decay $D^0 \to K^+\pi^-$ and the Cabibbo favoured (CF) decay $D^0 \to K^-\pi^+$, $\delta_{K\pi}$, and the normalisation, require further constraints. Therefore, to overconstrain the system, the ADS rates are fit together with the two GLW rates of $B^+ \to D(h^+h^-)K^+$ and $B^- \to D(h^+h^-)K^-$, where $h$ is a pion or a kaon, to determine all parameters and the normalisation [15]. A single normalisation parameter is required because those for the individual modes can be related together with knowledge of the $D^0$ branching fractions and the relative selection efficiencies. External constraints on $\delta_{K\pi}$ are also included [1, 16].

A similar analysis of the six ADS/GLW rates [17] has also been investigated for $\bar{B}^0 \to D\bar{K}^{*0}$ by LHCb [18]. Despite the reduced branching fraction the sensitivity to $\gamma$ can be better than the charged analogue due to the larger interference. However, there is a strong dependence on $\delta_{B^0}$ which leads to a factor 2.5 variation in the expected precision; at present there are no experimental constraints on $\delta_{B^0}$.

Other flavour specific final states, such as $D \to K^+\pi^-\pi^+\pi^-$, can be exploited in a similar manner to the two-body case. However, for multi-body final states different intermediate resonances can contribute, such as $K^{*0}\rho^0$ and $K^-a_1^+$, leading to the strong-phase difference varying over the phase space. This variation must be accounted for to allow multi-body modes to be used in an ADS analysis. The necessary changes to the formalism are described in Ref. [19], which leads to the introduction of a new parameter known as the coherence factor, $R_{K3\pi}$. The value of $R_{K3\pi}$ can be between zero and one; zero corresponds to many different interfering amplitudes contributing while one corresponds to a few non-overlapping intermediate states dominating. The value of $R_{K3\pi}$ can be determined from quantum correlated $D^0\bar{D}^0$ pairs produced in $e^+e^-$ collisions at a centre-of-mass energy equal to the mass of the $\psi(3770)$ [19, 20]. Such measurements are important for using additional flavour modes in an ADS analysis and are discussed further in Sec. 8.1.2.

LHCb has investigated measurements of $B^+ \to D(K^\pm\pi^\mp\pi^-\pi^+)K^+$ along with the two-body ADS measurements of $\gamma$ in a combined fit [9]; constraints from CLEO-c [20] are also included.

### 2.1.2 Dalitz plot analyses with $D \to K_S^0\pi^+\pi^-$

The analysis of three-body self-conjugate $D$ decays was proposed independently by Giri, Grossman, Soffer and Zupan [21] and Bondar [22] as a means to determine $\gamma$. The sensitivity to $\gamma$ arises from differences in the Dalitz plot of the $D$ decay coming from $B^+ \to DK^+$ and $B^- \to DK^-$. The decay $D \to K_S^0\pi^+\pi^-$ is considered the most favourable for this type of analysis because it has a relatively large branching fraction and a rich resonance structure.

Two methods have been investigated to determine $\gamma$ from $B^- \to D(K_S^0\pi^+\pi^-)K^-$: a likelihood fit to the Dalitz distributions, which requires a model of the $D^0 \to K_S^0\pi^+\pi^-$ decay, and a binned method, which relies on external knowledge of the strong-phase difference between the $D^0$ and $\bar{D}^0$ decays within those bins. The model-dependent method incurs a significant systematic uncertainty of between 6° and 15° [23, 24] due to the model assumptions.[4] The binned or model-independent method relies on measurements of the strong-phase differences from $\psi(3770)$ data; the uncertainty on strong-phase parameters replaces the model uncertainty and is estimated to be 2° (see Sec. 8.1.3). However, for a finite number of bins the model-independent method does not use the available information optimally, which leads to a decrease in the statistical precision compared to the model-dependent method.

[4]The uncertainties are inversely proportional to the measured value of $r_B$, therefore, for comparison both uncertainties have been scaled to $r_B = 0.1$.

## 2.2 Time-dependent measurements

### 2.2.1 $B_s^0 \to D_s^\pm K^\mp$

Measurements of the time-dependent $CP$ asymmetries in $B_s^0 \to D_s^\mp K^\pm$ allows $\gamma - \phi_M$ to be determined [25], where $\phi_M$ is the $B_s^0$ mixing phase. The sensitivity is improved further by including untagged events that are also sensitive to $\gamma$ via the non-zero width difference $\Delta\Gamma_s$ in the $B_s^0$ system. The tree-level sensitivity to $\gamma$ arises from the interference between the direct decay of $B_s^0$ and $\bar{B}_s^0$ to $D_s^+ K^-$ and decay after mixing; the Feynman diagrams for these decays are given in Fig. 4. The ratio of the magnitudes of the two interfering amplitudes, $x_s$, is expected to be approximately $\sim 0.4$, which is large enough for it to be determined from data. The value of $\gamma - \phi_M$ can be converted to a measurement of $\gamma$ because $\phi_M$ will be well constrained by measurements of $B_s^0 \to J/\psi\phi$ decays [26].

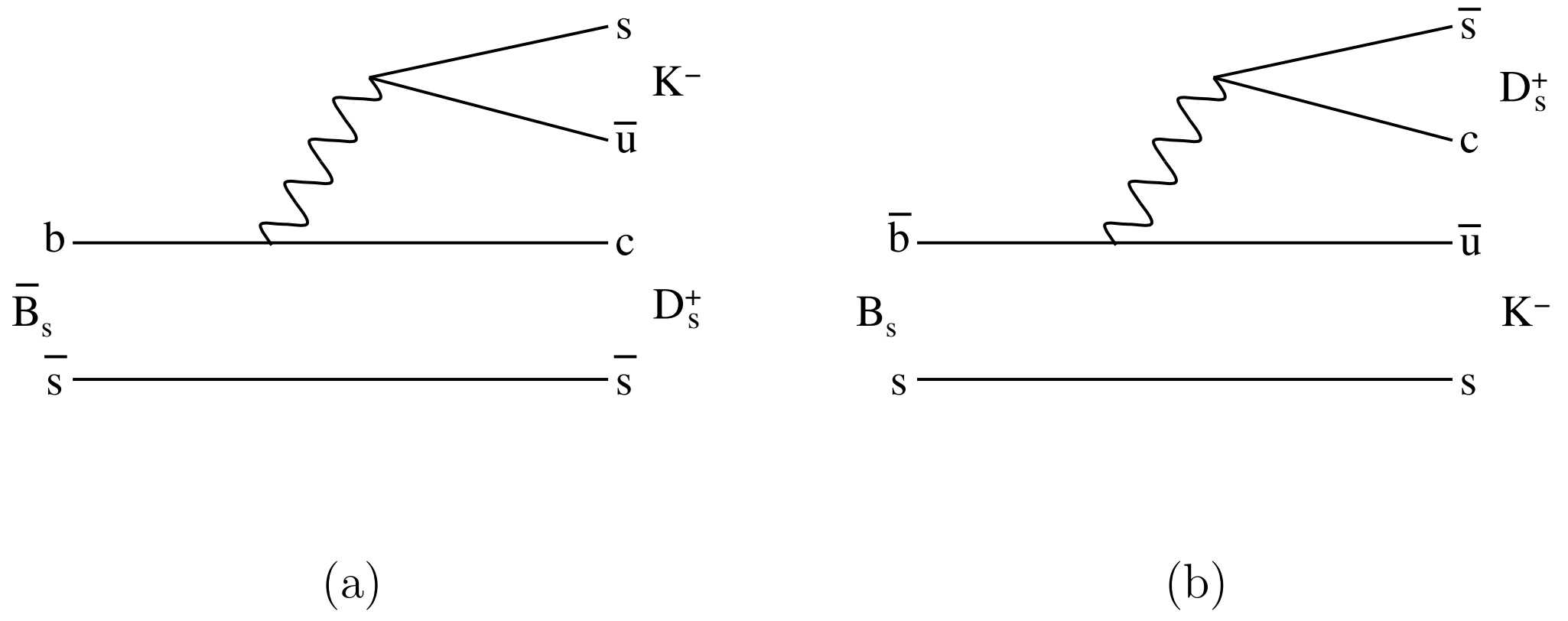

Figure 4: Feynman diagrams for $\bar{B}_s^0 \to D_s^+ K^-$ (left) and $B_s^0 \to D_s^+ K^-$ (right).

There have been several studies of $B_s \to D_s K$ at LHCb which describe the formalism in detail [27, 28, 29]. This measurement is likely to be unique to LHCb.

### 2.2.2 $B^0 \to D^\pm \pi^\mp$

The time-dependent $CP$ asymmetries in $B^0 \to D^\pm \pi^\mp$ allow $\gamma + 2\beta$ to be measured [30, 17]. Due to $\beta$ being already well measured [1], $\gamma$ can be determined from these asymmetries. The formalism is identical to $B_s^0 \to D_s^\pm K^\mp$. However, in practice, there are three significant differences:

Table 1: The relative weight (in percent) of each contributing analysis in the overall $\gamma$ determination for two values of $\delta_{B^0}$ and a dataset of 2 fb$^{-1}$. Table is taken from Ref. [9].

| Analysis | $\delta_{B^0} = 0°$ | $\delta_{B^0} = 45°$ |
|---|---|---|
| $B^- \to D^0(K^\pm\pi^\mp)K^-$, $D^0(h^+h^-)K^-$ and $D^0(K^\pm\pi^\mp\pi^+\pi^-)K^-$ | 25 | 38 |
| $B^- \to D^0(K^0_S\pi^+\pi^-)K^-$ | 12 | 25 |
| $B^0 \to D^0(K^\pm\pi^\mp)K^{*0}$ and $D^0(h^+h^-)K^{*0}$ | 44 | 8 |
| $B^0_s \to D^\mp_s K^\pm$ | 16 | 24 |
| $B^0 \to D^\mp\pi^\pm$ | 3 | 5 |

1. The absolute ratio between the two interfering amplitudes, $x_d$, is small $\sim 0.02$ and it cannot be determined from data. Instead it must be constrained from external measurements and theoretical arguments [31]; the uncertainties introduced and possible methods to improve them are discussed in Sec. 8.1.6.

2. The negligible lifetime difference in the $B^0$ system means that there is no sensitivity to $\gamma$ via untagged events.

3. The value of $\gamma$ is determined with an eight-fold ambiguity. Measurements from related channels with differing strong phases, such as $B^0 \to D^{*\pm}\pi^\mp$, or a joint U-spin based analysis [32] can resolve or reduce the ambiguities. An additional advantage of the U-spin analysis is that there is no dependence on $x_d$; however, uncertainties related to U-spin breaking are introduced.

Both the standalone and U-spin based determination have been investigated by LHCb [33].

### 2.3 Relative sensitivity of the modes at LHCb

A combined fit to all LHCb data will eventually be employed to obtain the best measurement of $\gamma$. Table 1 gives the weight of the each of the individual modes to the combined measurement of $\gamma$ for the most recent combination [9]. The results for two values of $\delta_{B^0}$ are shown as this parameter is currently unconstrained and has great influence on the weight of $B^0 \to K^{*0}$ decays. Most modes have considerable weight in both cases and no single mode has more than 50% of the statistical weight. Section 9 presents a discussion of which of these measurements are likely to be performed first once LHCb data-taking begins.

## 3 Summary of current measurements of $\gamma$

Before presenting the latest studies from LHCb, a brief review is presented of existing measurements from BABAR, Belle and CDF. This review puts into context the large improvements expected with only a limited amount of data collected by LHCb.

## 3.1 Current GLW/ADS measurements

Measurements of the $B^- \to DK^-$ GLW modes have been performed by BABAR [34], Belle [35] and CDF [36]. The results are presented in terms of the asymmetries

$$\begin{aligned} A_\pm &= \frac{\Gamma(B^- \to D_\pm K^-) - \Gamma(B^+ \to D_\pm K^+)}{\Gamma(B^- \to D_\pm K^-) + \Gamma(B^+ \to D_\pm K^+)} \\ &= \frac{\pm 2 r_B \sin\delta_B \sin\gamma}{1 + r_B^2 \pm 2 r_B \cos\delta_B \cos\gamma} \ , \end{aligned} \tag{1}$$

and the ratios

$$\begin{aligned} R_\pm &= 2\,\frac{\Gamma(B^- \to D_\pm K^-) + \Gamma(B^+ \to D_\pm K^+)}{\Gamma(B^- \to D^0 K^-) + \Gamma(B^+ \to D^0 K^+)} \\ &= 1 + r_B^2 \pm 2 r_B \cos\delta_B \cos\gamma \ , \end{aligned} \tag{2}$$

where $D_\pm$ denotes a decay of a $D^0$ or $\bar{D}^0$ to a $CP\pm$ eigenstate (we use the convention $D_\pm = \left(D^0 \pm \bar{D}^0\right)/2$). Only three of these parameters are independent because $A_+R_+ = -A_-R_-$. Therefore, the alternative parameters:

$$\begin{aligned} x_\pm &= r_B \cos(\delta_B \pm \gamma) \\ &= \frac{R_+(1 \mp A_+) - R_-(1 \mp A_-)}{4} \ , \end{aligned} \tag{3}$$

and

$$r_B^2 = \frac{R_+ + R_- - 2}{2} \ , \tag{4}$$

are sometimes used to present results. These three parameters allow easy comparison and combination with those found by some of the other determinations.

The average measured values of $A_\pm$ and $R_\pm$ [1] are:

$$\begin{aligned} A_+ &= 0.24 \pm 0.07 \ , \\ A_- &= -0.10 \pm 0.08 \ , \\ R_+ &= 1.10 \pm 0.09 \text{ and} \\ R_- &= 1.06 \pm 0.10 \ . \end{aligned}$$

There is evidence for direct $CP$ violation from the average value of $A_+$.

There have been additional measurements of the analogous asymmetries and ratios for $B^- \to D^* K^-$ [35,37] and $B^- \to DK^{*-}$ [38,39]; none of these measurements show any evidence of $CP$ violation given the current experimental uncertainties.

The measurements of the $B^- \to DK^-$ ADS modes have been made by both BABAR [40] and Belle [41]. Evidence for the suppressed modes has been found by BABAR [40]. Belle have not observed the suppressed decay and set a limit on the branching fraction at $< 2.8\times10^{-7}$ [41] at 95% confidence level (CL). The ADS asymmetry and ratio parameters related to $\gamma$, akin to those in the GLW analysis, are:

$$\begin{aligned} A_{\rm ADS} &= \frac{\Gamma(B^-\to D(K^+\pi^-)K^-)-\Gamma(B^+\to D(K^-\pi^+)K^+)}{\Gamma(B^-\to D(K^+\pi^-)K^-)+\Gamma(B^+\to D(K^-\pi^+)K^+)} \\ &= \frac{2r_Br_D\sin(\delta_B+\delta_{K\pi})\sin\gamma}{r_B^2+r_D^2+2r_Br_D\cos(\delta_B+\delta_{K\pi})\cos\gamma}\ , \end{aligned} \tag{5}$$

and

$$\begin{aligned} R_{\rm ADS} &= \frac{\Gamma(B^-\to D(K^+\pi^-)K^-)+\Gamma(B^+\to D(K^-\pi^+)K^+)}{\Gamma(B^-\to D(K^-\pi^+)K^-)+\Gamma(B^+\to D(K^+\pi^-)K^+)} \\ &= r_B^2+r_D^2+2r_Br_D\cos(\delta_B+\delta_{K\pi})\cos\gamma\ , \end{aligned} \tag{6}$$

where $r_D = |A(D^0\to K^+\pi^-)|/|A(D^0\to K^-\pi^+)| = 0.0579\pm0.0007$ [1]. The values of $R_{\rm ADS}$ and $A_{\rm ADS}$ reported by BABAR and Belle are given in Table 2.

Table 2: Measured values of $R_{\rm ADS}$ and $A_{\rm ADS}$ by BABAR [40] and Belle [41]. The first and second uncertainties are statistical and systematic, respectively.

| Experiment | $R_{\rm ADS}$ | $A_{\rm ADS}$ |
|---|---|---|
| BABAR | $(13.6\pm5.5\pm2.7)\times10^{-3}$ | $-0.70\pm0.35^{+0.09}_{-0.14}$ |
| Belle | $(7.8^{+6.2}_{-5.7}{}^{+2.0}_{-2.8})\times10^{-3}$ | $-0.13^{+0.97}_{-0.88}\pm0.26$ |

Analogous analyses have been performed by BABAR for $B^-\to DK^{*-}$ [42], $B^-\to D^*K^-$ [40] and $B^-\to D(K^\pm\pi^\mp\pi^0)K^-$ [43]; no significant evidence of non-zero values for these asymmetries and ratios has been observed. Both experiments have also measured $R_{ADS}$ and $A_{ADS}$ for $B^-\to D\pi^-$ decays [40,41]; no direct $CP$ asymmetry has been observed. Despite the reduced interference compared to $B^-\to DK^-$, because the ratio of the magnitudes of the two $B^-$ decay amplitudes is approximately 20 times smaller, $\mathcal{O}(10\%)$ differences in the suppressed rates are possible depending on the values of the phases.

The branching fraction for $\bar{B}^0\to D^0\bar{K}^{*0}$ has been measured and the decay $\bar{B}^0\to\bar{D}^0\bar{K}^{*0}$ has been searched for [44,45]. BABAR has recently presented an analysis of $\bar{B}^0\to D\bar{K}^{*0}$ with $D\to K^-\pi^+$, $D\to K^-\pi^+\pi^0$ and $D\to K^-\pi^+\pi^+\pi^-$ [46], which gives the constraint $0.07 < r_{B^0} < 0.41$ at the 95% CL.

## 3.2 Current Dalitz measurements

Both BABAR and Belle have recently updated their analyses of $B^-\to D(K_s^0\pi^+\pi^-)K^-$ using the model-dependent method [23,24]. The Belle analysis also includes the modes $B^-\to D^*(D\pi^0)K^-$ and $B^-\to D^*(D\gamma)K^-$; a previous analysis studied $B^-\to DK^{*-}$ [47]. The BABAR analysis also includes $B^-\to D^*(D\pi^0)K^-$, $B^-\to D^*(D\gamma)K^-$ and $B^- \to DK^{*-}$, but, in addition the self-conjugate decay $D \to K^0_S K^+ K^-$ was included. Another significant difference between the analyses is in the treatment of $\pi\pi$ and $K\pi$ $S$-wave in the amplitude models. Currently only averages of the $x_\pm$ and $y_\pm = r_B \sin(\delta_B \pm \gamma)$ measurements from BABAR and Belle has been prepared by the Heavy Flavour Averaging Group (HFAG). Table 3 summarises the most recent results from BABAR and Belle on $\gamma$, $r_B$ and $\delta_B$, combining all modes studied; the differing statistical precision between the two experiments is due to the uncertainty on $\gamma$ being inversely proportional to their measured value of $r_B$. The measurements of $B^- \to D(K^0_S\pi^+\pi^-)K^-$ dominate the precision on $\gamma$ obtained by the Unitarity Triangle fits discussed in Sec. 3.4.

Table 3: Latest results on $\gamma$, $\delta_B$ and $r_B$ from BABAR [23] and Belle [24] analyses of $B^+ \to D(K^0_S\pi^+\pi^-)^{(*)}K^{(*)+}$. The first uncertainty is statistical, the second is the experimental systematic and the third is related to the Dalitz model.

| Parameter | Belle | BABAR |
|---|---|---|
| $\gamma$ | $(78^{+11}_{-12} \pm 4 \pm 9)^\circ$ | $(76 \pm 22 \pm 5 \pm 5)^\circ$ |
| $r_B$ | $0.16 \pm 0.04 \pm 0.01^{+0.05}_{-0.01}$ | $0.086 \pm 0.032 \pm 0.010 \pm 0.011$ |
| $\delta_B$ | $(137^{+13}_{-16} \pm 4 \pm 23)^\circ$ | $(109^{+27}_{-30} \pm 4 \pm 7)^\circ$ |

The decay $\bar{B}^0 \to D(K^0_S\pi^+\pi^-)\bar{K}^{*0}$ has also been studied by BABAR [48]. The analysis measures $\gamma = (162 \pm 56)^\circ$ and $r_{B^0} < 0.55$ at the 90% CL.

## 3.3 Current time-dependent measurements

The decay $B^0_s \to D^\mp_s K^\pm$ has been observed by CDF [49] and Belle has seen evidence of this decay [50]. Both experiments quote their result with respect to the $B^0_s \to D^-_s\pi^+$ branching fraction $\mathcal{B}(B^0_s \to D^-_s\pi^+) = (3.0 \pm 0.7 \pm 0.1) \times 10^{-3}$ [51]. The results are shown in Table 4. At present the data size is too small for CDF to perform a measurement of the time-dependent $CP$ asymmetries.

Table 4: Measurements of the ratio $\mathcal{B}(B^0_s \to D^\mp_s K^\pm)/\mathcal{B}(B^0_s \to D^-_s\pi^+)$.

| Experiment | $\mathcal{B}(B^0_s \to D^\mp_s K^\pm)/\mathcal{B}(B^0_s \to D^-_s\pi^+)$ |
|---|---|
| CDF | $0.097 \pm 0.018(stat.) \pm 0.009(syst.)$ |
| Belle | $0.069^{+0.034}_{-0.028}$ |

The time-dependent $CP$ asymmetries in $B^0 \to D^\pm\pi^\mp$ have been measured by both BABAR [52] and Belle [53] using both inclusive and exclusive reconstruction. In addition, the $CP$ asymmetries in related modes $B^0 \to D^{*\pm}\pi^\mp$ have been measured by BABAR [54] and Belle [53], and $B^0 \to D^\pm\rho^\mp$ by BABAR [52]. The constraints on $2\beta + \gamma$ as a function of $x_d$ from these results are shown in Fig. 5; the current bounds are weak.

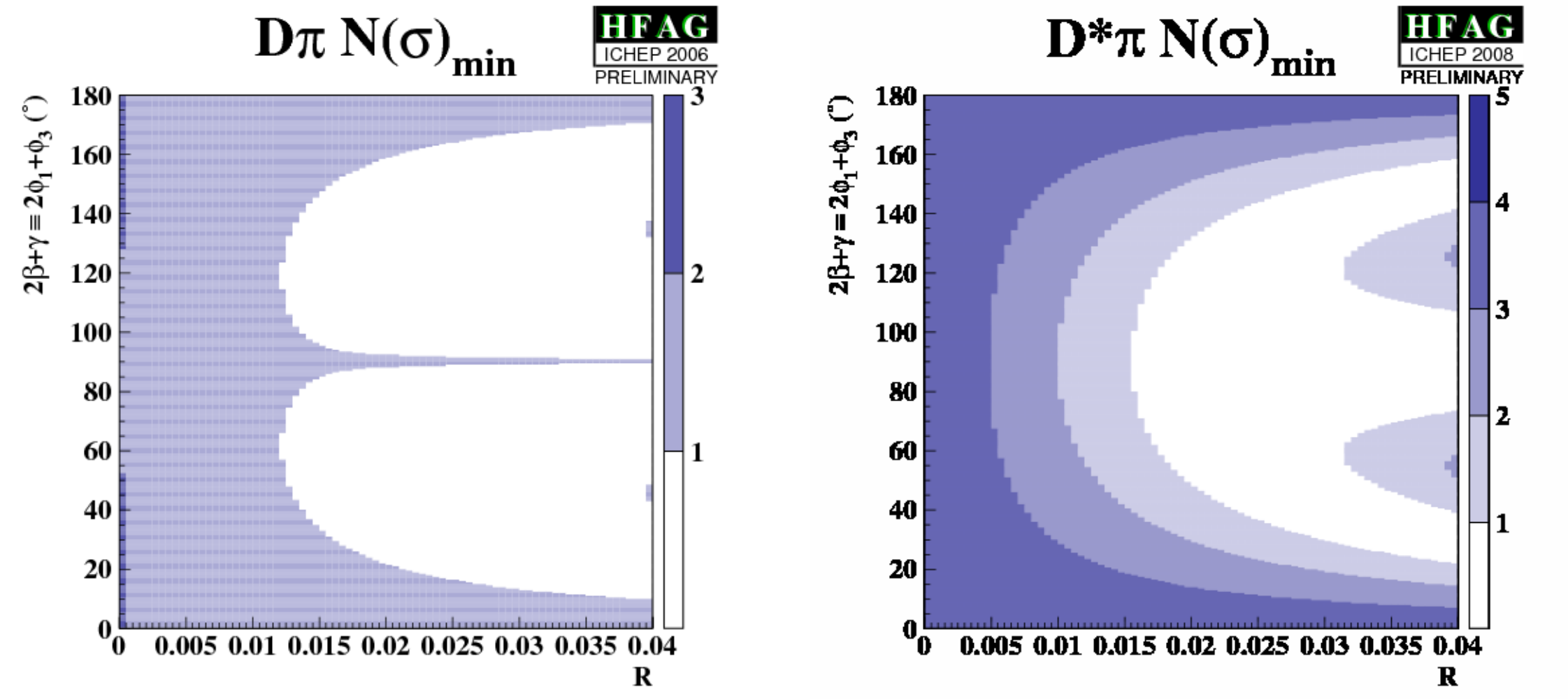

(a) (b)

Figure 5: Constraints on $2\beta+\gamma$ as a function of $x_d(=R)$ from a combination of BABAR and Belle measurements of (a) $B^0 \to D^{\pm}\pi^{\mp}$ and (b) $B^0 \to D^{*\pm}\pi^{\mp}$ [1]. The confidence levels are shown in terms of standard deviations.

A time-dependent analysis of the $B^0 \to D^{\pm}K^0_S\pi^{\mp}$ Dalitz plot to determine $\sin 2\beta+\gamma$ has been performed by BABAR experiment [55]. This measurement has the advantage that it resolves $2\beta+\gamma$ with a single ambiguity. The values favoured are $2\beta+\gamma = (83\pm53\pm20)^\circ$ and $(263\pm53\pm20)^\circ$, where the first uncertainty is statistical and the second is systematic.

### 3.4 Unitarity triangle fits to $\gamma$

Combinations of all these results have been performed by the UTfit [2] and CKMfitter [3] collaborations. The fit results for $\gamma$, $r_B$ and $\delta_B$ are given in Table 5. The precision on $\gamma$ varies between the two collaborations because of the Bayesian and frequentist statistical approaches adopted by the UTfit and CKMfitter collaborations, respectively. The average is dominated by the measurements of $\gamma$ from $B^- \to D(K^0_s\pi^+\pi^-)K^-$. The value of $r_B$ is reasonably well known with a relative uncertainty of approximately 20%.

### 3.5 Event yields at BABAR, Belle and CDF

Table 6 reports the signal yields and integrated luminosities of some of the analyses reported in this section that will also be performed at LHCb. The final $\Upsilon(4S)$ data set at BABAR corresponds to 426 fb$^{-1}$ of integrated luminosity. Belle have collected approximately 750 fb$^{-1}$ of integrated luminosity at the $\Upsilon(4S)$ and will continue taking data at the $\Upsilon(1S)$, $\Upsilon(2S)$ and $\Upsilon(5S)$ resonances until the end of 2009, before a major upgrade of the KEKB collider and the detector takes place [56]. CDF has collected 5 fb$^{-1}$ of integrated luminosity and will continue to accumulate data until at least the end of 2010 when $8-10$ fb$^{-1}$ will have been collected. It is also worth noting that the superior tagging performance of the $e^+e^-$ $B$ factories is irrelevant for the time-integrated measurements of $B^- \to DK^-$ and $\bar{B}^0 \to D\bar{K}^{*0}$, and that time-dependent measurements needed for $B_s^0 \to D_s^\pm K^\mp$ cannot be performed at the $\Upsilon(5S)$ with currently available detection technologies.

Table 5: Latest results of the UTfit [2] and CKMfitter [3] collaborations for averages of $\gamma$, $r_B$ and $\delta_B$. No result for $\delta_B$ is reported by the UTfit collaboration.

| Parameter | UTfit | CKMfitter |
|---|---|---|
| $\gamma$ | $(78 \pm 12)^\circ$ | $(70^{+27}_{-29})^\circ$ |
| $r_B$ | $0.102 \pm 0.017$ | $0.087^{+0.022}_{-0.018}$ |
| $\delta_B$ | – | $(110^{+22}_{-27})^\circ$ |

Table 6: Event yields and integrated luminosities, $\int \mathcal{L}dt$, used for some of the analyses by BABAR, Belle and CDF reported in this section.

| Mode | BABAR | | Belle | | CDF | |
|---|---|---|---|---|---|---|
| | Yield | $\int \mathcal{L}dt$ (fb$^{-1}$) | Yield | $\int \mathcal{L}dt$ (fb$^{-1}$) | Yield | $\int \mathcal{L}dt$ (fb$^{-1}$) |
| $B^- \to DK^-$ GLW | 240 | 351 | 143 | 252 | 91 | 1 |
| $B^- \to DK^-$ ADS | 1780 | 432 | 1220 | 602 | – | – |
| $B^- \to DK^-$ Dalitz | 610 | 351 | 756 | 602 | – | – |
| $B^0 \to D^\pm \pi^\mp$ | $15 \times 10^3$ | 212 | $26 \times 10^3$ | 353 | – | – |
| $B_s^0 \to D_s^\pm K^\mp$ | – | – | 7 | 22 (at $\Upsilon(5S)$) | 102 | 1.2 |

# 4 The LHCb trigger

The LHCb trigger is divided into two parts. The first part is a fixed latency hardware trigger (Level-0) which reduces the non-empty interaction rate of 10 MHz to 1 MHz by identifying high $p_T$ muons, hadrons, electrons and photons. The simulation studies presented in Sec. 5 find that the Level-0 trigger efficiency for fully hadronic modes is between 40 and 50%.

The second stage is a software trigger, which is referred to as the High Level Trigger (HLT). The HLT is split into two parts: HLT1, which confirms the Level-0 decision and includes the first use of impact parameter information, and HLT2, which runs exclusive and inclusive selections for the signal and control channels. The HLT1 will reduce the rate to ∼30 kHz and the HLT2 to ∼2 kHz, which is written to storage.

The combined Level-0 and HLT1 selection is currently being optimised using approximately 20 benchmark channels across the whole LHCb physics programme; these channels include $B_s^0 \to D\pm_s K^\mp$, $\bar{B}^0 \to D\bar{K}^{*0}$ and $B^- \to D(K_S^0\pi^+\pi^-)K^-$. The Level-0 efficiencies are unlikely to change significantly from those reported in Sec. 5. However, the HLT selections for hadronic modes are still undergoing significant development. The current status of the HLT for hadronic modes is summarised in Section **??**. Based on these studies an HLT efficiency of 65% for all hadronic modes is assumed in the sensitivity study presented in Sec. 7.

One mode that presents significant challenges to the HLT is $B^- \to D(K_S^0\pi^+\pi^-)K^-$ when the $K_S^0$ is reconstructed from downstream tracks without vertex detector information (DD $K_S^0$). As will be presented in Sec. 5.4, candidates reconstructed that include a DD $K_S^0$ comprise 75% of offline selected events. However, downstream tracking is too CPU intensive to be run on all events that pass the HLT1. Therefore, to prevent the loss of these events in HLT2, a staged selection has been developed where initially the long tracks with vertex detector information are used to identify topologies that are consistent with $B^- \to D(K_S^0\pi^+\pi^-)K^-$ decays. In particular, a dipion vertex and a large impact parameter bachelor kaon are identified. In addition, the dipion vertex and bachelor kaon are required to lie in the same plane as the primary vertex. The downstream reconstruction is run on this subset of events, which reduces the required CPU to a manageable level, and an exclusive selection is then performed. Preliminary studies show that this staged selection paradigm is an effective strategy for selecting these decays, but as is the case for the other HLT selections, no final efficiency numbers are available at present.

# 5 Offline selection studies

## 5.1 Introduction to the selection studies

This section presents the selection studies that have been performed with the latest production of simulated events of both signal and background. This production is referred to as DC06. Results are compared to the previous large production, referred to as DC04; yield and background estimates derived from the DC04 production were used to determine the expected precision on $\gamma$ [9] for differing data set sizes. The DC06 production contains many developments in the detector description, detector simulation and reconstruction, which make it more akin to the real LHCb environment. Therefore, the principal goal of the selection studies is to at least maintain the performance observed in the DC04 simulation and improve upon it if at all possible. Furthermore, these offline selections are an essential ingredient for development and optimisation of the LHCb trigger [57] and data stripping algorithms [58].

There are five selections described in this note:

- $B^- \to D(K^\pm\pi^\mp)K^+$ and $B^+ \to D(h^+h^-)K^-$,
- $B^- \to D(K^\pm\pi^\mp\pi^+\pi^-)K^-$,
- $B^- \to D(K^0_S\pi^+\pi^-)K^-$,
- $\bar{B}^0 \to D(K^\pm\pi^\mp)\bar{K}^{*0}$ and $\bar{B}^0 \to D(h^+h^-)\bar{K}^{*0}$, and
- $B^0_{(s)} \to D^\pm_{(s)}h^\mp$.

These selections cover all the channels that were used to determine the global precision on $\gamma$ at LHCb [9]. All the individual selections have been written up in supplementary notes [59, 60, 61, 62, 63], which contain the details of the simulation samples used, the selection optimisation procedures and background studies.

In the following subsections there is sometimes reference to a preselection. The preselection was applied to samples of signal and a sample of 22 million $b$-inclusive events within the LHCb acceptance from which background rates are determined.[5] The simultaneous application of preselections for many channels led to a *stripped* sample of events, which is approximately 30 times smaller than the full sample, and as such, the processing time required is much reduced. Other pieces of LHCb terminology are defined as they appear in the text.

All the signal, and some background, yield estimates require branching fractions as inputs. The branching fractions are all taken to be those reported in Ref. [64] apart from that for $B^0_s \to D^\pm_s K^\mp$ which is chosen to be the value reported in Ref. [49]. The values used are given in Table 7.

[5]This sample corresponds to an integrated luminosity of only 0.1 pb$^{-1}$, 0.005% of one nominal year of data taking.

Table 7: Branching fractions used to compute signal yields and some background estimates. All values taken to be those reported in Ref. [64] apart from that for $B_s^0 \to D_s^{\pm} K^{\mp}$ which is reported in Ref. [49].

| Process | Branching fraction |
|---|---|
| $B^- \to D^0 K^-$ | $(4.02 \pm 0.21) \times 10^{-4}$ |
| $B^- \to D^0 \pi^-$ | $(48.4 \pm 1.5) \times 10^{-4}$ |
| $\bar{B}^0 \to D^0 \bar{K}^{*0}$ | $(0.42 \pm 0.06) \times 10^{-4}$ |
| $\bar{B}^0 \to D^+ \pi^-$ | $(26.8 \pm 1.3) \times 10^{-4}$ |
| $\bar{B}_s^0 \to D_s^+ \pi^-$ | $(32 \pm 9) \times 10^{-4}$ |
| $\bar{B}_s^0 \to D_s^{\pm} K^{\mp}$ | $(3.4 \pm 1.2) \times 10^{-4}$ |
| $D^0 \to K^- \pi^+$ | $(3.89 \pm 0.05)\%$ |
| $D^0 \to K^- K^+$ | $(0.393 \pm 0.008)\%$ |
| $D^0 \to \pi^- \pi^+$ | $(0.140 \pm 0.003)\%$ |
| $D^0 \to K^- \pi^+ \pi^+ \pi^-$ | $(8.10 \pm 0.20)\%$ |
| $D^0 \to K_S^0 \pi^+ \pi^-$ | $(2.99 \pm 0.17)\%$ |
| $D^+ \to K^- \pi^+ \pi^+$ | $(9.22 \pm 0.21)\%$ |
| $D_s^+ \to K^- K^+ \pi^+$ | $(5.50 \pm 0.28)\%$ |
| $K_S^0 \to \pi^+ \pi^-$ | $(69.20 \pm 0.05)\%$ |
| $\bar{K}^{*0} \to K^- \pi^+$ | 2/3 |

## 5.2 Offline selection of $B^- \to D(K^{\pm}\pi^{\mp})K^-$ and $B^- \to D(h^+h^-)K^-$

### 5.2.1 Selection criteria

The background level observed applying the DC04 selection requirements [65] on the DC06 $b$-inclusive sample is substantially larger than that previously observed. In particular there was a significant increase in the number events reconstructed including a ghost track.[6] The selection requirements are therefore retuned using the algorithm described in Ref. [66] to optimise the ratio $S/\sqrt{S+B}$, where $S$ is the number of signal events observed in a sample corresponding to 2 fb$^{-1}$ of integrated luminosity, and $B$ is the equivalent number of background events. This algorithm automatically explored the multidimensional cut-space corresponding to the available variables, in order to maximise the above metric. At each iteration, the algorithm identified the cut with the greatest potential to increase the metric and then altered that cut by a small amount. The optimisation is done for the suppressed $B^- \to D(K^+\pi^-)K^-$ decay and identical selection criteria are then used for the favoured $B^- \to D(K^-\pi^+)K^-$ decay. Only the particle identification (PID) selection criteria are then adjusted to select $B^- \to D(h^+h^-)K^-$ decays. This common approach allows the normalisation factors amongst the six rates required to determine $\gamma$ from these modes [15] to be largely independent of knowledge of the individual modes ac-ceptance, apart from effects related to the PID. This approach will reduce the systematic uncertainties related to the normalisation.

[6] A ghost track is defined as one for which less than 70% of the hits associated to the track in the reconstruction come from a single charged particle.

In the DC04 studies the background was observed to be dominated by fragments of $B \to D^* X$ decays. This is also observed to be the case in the DC06 $b$-inclusive simulation sample. In order to increase the statistical power of the sample, such combinatoric events are selected in a mass window ten times larger than the so-called tight-mass window used to select signal events. The rate of events in the tight-mass window is estimated assuming a uniform distribution of events in the wide-mass window. Large inclusive samples of $B \to D^* X$ decays are available in DC06 and these are used to tune the selection. This tuning identifies significantly different requirements as optimal compared to the analysis of DC04. Tighter requirements are placed on the following topological variables: $B^-$ candidates Flight Significance (FS) criterion; $D^0$ daughter and bachelor $K^\pm$ Smallest Impact Parameter Significance (SIPS)[7]; the $B^\pm$ and $D^0$ candidate vertex $\chi^2$; and the separation between the $D^0$ and $B^\pm$ vertices in $z$ (the beam direction in the LHCb coordinate system). In addition, the criteria on the difference in log-likelihood ($\Delta LL$) of the kaon and pion hypotheses for the bachelor $K^\pm$ and the track fit $\chi^2/d.o.f.$ for all tracks are tightened. To compensate less strict requirements are applied to the $D^0$ $p_T$, FS and SIPS and no requirement is made on the angle between the momentum and the flight direction of the $D^0$.

The much stricter topological requirements on the $B^\pm$ candidate in the DC06 selection represents a shift in the selection strategy. While the DC04 selection accepted $B^\pm$ candidates much closer to the primary vertex, and used a range of additional criteria to control the background, the DC06 selection finds the optimal $S/\sqrt{S+B}$ removing such events. Tighter particle identification, impact parameter and track $\chi^2$ requirements are also needed to control the larger background seen in DC06. The $D^0$ to $B^\pm$ vertex separation requirement results in fewer additional requirements on the $D^0$. Even with the large $B \to D^* X$ samples, there are insufficient statistics to tune the vertex isolation requirement[8] that was used in DC04. This, or a similar, requirement will most likely be used in the final selection but it is not possible to determine its form with the samples of background presently available.

### 5.2.2 Annual signal event yield

Applying the selection requirements to the $B^- \to D(K^-\pi^+)K^-$ favoured signal sample results in a selection efficiency of $\epsilon_{sig} = (0.67 \pm 0.03)\%$. The value of $\epsilon_{sig}$ is the combination of the efficiency for decay products to be within the LHCb acceptance, the Level-0 trigger efficiency and the reconstruction efficiency.

Assuming an integrated luminosity of 2 fb$^{-1}$ and an inclusive $b\bar{b}$ production cross section of 500 $\mu$b, the event yield, $S$, of the favoured decay $B^- \to D(K^-\pi^+)K^-$ is given

[7]The value of SIPS is computed with the *smallest* impact parameter with respect to any primary vertex in the event. This impact parameter is then divided by its uncertainty to determine the significance.

[8]This requirement vetoes $B$ vertices that have more than a certain number of tracks with a minimum SIPS nearby; this removes background from poorly reconstructed primary vertices.

by:

$$S = 2N_{b\bar{b}} \times f(b \to B^+) \times \mathcal{B}(B^- \to D^0 K^-) \times \mathcal{B}(D^0 \to K^-\pi^+) \times \epsilon_{sig}, \quad (7)$$
$$= 83,800 \pm 5900,$$

where $N_{b\bar{b}} = 10^{12}$ is the number of $b\bar{b}$ pairs produced, $f(b \to B^\pm) = 0.4$ is the fraction of $b$ quarks that hadronise to $B^\pm$ and $\mathcal{B}$ are branching fractions. The uncertainty quoted is the statistical uncertainties on the selection efficiencies combined with those on the branching fractions.

The remaining expected 2 fb$^{-1}$ signal yields are shown in Table 8. The yield in the suppressed $B^+ \to D(K^-\pi^+)K^+$ signal channels is a strong function of the unknown parameters, $r_B$, $\delta_B$, $\delta_D^{K\pi}$ and $\gamma$. The number shown in the table assumes $r_B = 0.1$, $\delta_B = 130°$, $\delta_D^{K\pi} = -158°$ and $\gamma = 60°$. The values of $\epsilon_{sig}$ for the suppressed and $CP$ eigenstate modes are assumed to be the same as for the favoured modes; this was found to be a good assumption in previous studies [65].

Table 8: Summary of signal and background yields expected from 2 fb$^{-1}$ of data with the Level-0 trigger applied. The signal yield shown is for the sum of $B^+$ and $B^-$ decays. In the suppressed $K\pi$ decay these yields are highly asymmetric and strongly dependent on the underlying parameters. The value shown is an sum of the $B^+$ and $B^-$ signal yields and is computed assuming $r_B = 0.1$, $\gamma = 60°$, $\delta_B = 130°$, $\delta_D = -158°$ and $r_D = 0.0613$.

| Channel | S | B |
|---|---|---|
| $B^\pm \to D(K^\pm\pi^\mp)K^\pm$ | $83,800 \pm 5800$ | $50,900 \pm 3300$ |
| $B^\pm \to D(K^\mp\pi^\pm)K^\pm$ | $1600 \pm 100$ | $970 \pm 480$ |
| $B^\pm \to D(K^+K^-)K^\pm$ | $8400 \pm 600$ | $9760 \pm 2750$ |
| $B^\pm \to D(\pi^+\pi^-)K^\pm$ | $3000 \pm 200$ | $9520 \pm 3950$ |

### 5.2.3 Backgrounds and $B/S$ ratio

Having tuned the signal selection on half the available $B \to D^*X$ samples, the remaining half of the samples are used to make an unbiased estimate of the background. Three events are found of the type $B^\pm \to D^{*0}K^\pm$, where the $D^{*0} \to D^0(K\pi)\pi^0$ or $D^{*0} \to D^0(K\pi)\gamma$ and the $D^0$ daughters are both selected. However, the bachelor $K^\pm$ is not from the same decay. In two of the events the bachelor kaon candidate is a ghost track. In the other event the $K^\pm$ is from the other $B$ hadron produced in the event. These three events lead to an estimated $800 \pm 480$ background events per 2 fb$^{-1}$.[9] The background is divided equally between the favoured and suppressed modes because it is expected that the charge of the fake bachelor $K^\pm$ is independent of the $B$ decay that gives rise to the $D^0$ candidate. No events are selected from the $b$-inclusive sample which is consistent with the expectation from the higher significance $B \to D^* X$ sample.

[9]The $B \to D^*X$ sample only included around 50% of all possible decays, principally those with low multiplicity. The unsimulated topologies are assumed to yield the same number of background candidates. This is a conservative assumption because the higher multiplicity events contain tracks of lower momentum than the signal on average and are thus less likely to be reconstructed as signal candidates.

There is a significant background from the topologically identical decay $B^- \to D^0(K^-\pi^+)\pi^-$ where the $\pi$ is misidentified as a $K$; the branching fraction is twelve times larger than the signal mode. A sample of these events is used to estimate the background from $B^- \to D^0\pi^-$ events to the favoured mode of $50,100 \pm 3300$ per 2 fb$^{-1}$. The ratio between the favoured and suppressed modes of $B^- \to D(K + \pi^-)\pi^+$ events has been measured to be $(3.4 \pm 0.6) \times 10^{-3}$ [41]; this measurement is used to estimate a $B^- \to D\pi^-$ background of $170 \pm 30$ events for the suppressed modes.

The combinatoric and $B \to D\pi$ background estimates are summed to give the total backgrounds of $50,900 \pm 3300$ and $970 \pm 480$ for the favoured and suppressed $B^+ \to D(K\pi)K^+$ signals, respectively. The backgrounds for the GLW modes have not been studied in the DC06 simulation sample; the backgrounds measured in the DC04 studies [65] are taken as a conservative estimate since the combinatoric component is expected to be reduced by the reoptimised selection.

## 5.3 Offline selection of $B^- \to D(K^\pm\pi^\mp\pi^+\pi^-)K^-$

The preselection implemented is identical to that used for the DC04 selection studies of $B^- \to D(K^-\pi^\mp\pi^+\pi^-)K^-$ [67]. However, the offline selection reported in Ref. [67] applied to the DC06 $b$-inclusive background sample resulted in significantly higher background levels than the DC04 studies. Therefore, the selection is reoptimised on the DC06 signal and background samples [60].

The selection is optimised on the suppressed-signal mode, $B^- \to D(K^+\pi^-\pi^-\pi^+)K^-$, and is required to maximise the metric $S/\sqrt{S+B}$. Half the $b$-inclusive sample is used to optimise the selection and the other half is used to make an unbiased estimate of the background. To increase the statistical power of the $b$-inclusive sample a $\pm 500$ MeV$/c^2$ window about the nominal $B$ mass is used to estimate the background, even though the signal region is $\pm 50$ MeV$/c^2$. The combinatoric background is assumed to be uniformly distributed over the $\pm 500$ MeV$/c^2$ window to estimate the background in the $\pm 50$ MeV$/c^2$ window. Partially-reconstructed backgrounds within the $b$-inclusive sample, such as $B^- \to D^* K^-$, that peak away from the nominal $B$ mass are not considered in the optimisation.

The reoptimised selection favours stricter requirements on the SIPS of the bachelor $K$ and $D$ candidates than in the DC04 studies. In addition, two-dimensional linear criteria on correlated variables are found to be powerful discriminants between signal and background. An example of an optimised linear cut is shown in Fig. 6 for the $\chi^2$ of the $B$ flight distance and the $\chi^2$ of the smallest impact parameter of the bacheleor $K^\pm$.

No events from the second half of the $b$-inclusive event sample pass the optimised selection with the Level-0 trigger applied. The estimated signal yields to be reconstructed from a data sample corresponding to an integrated luminosity of 2 fb$^{-1}$ are $550^{+270}_{-180}$ and $53,000 \pm 3800$ events in each of the suppressed and favoured modes, respectively. The yield of in the suppressed mode depends on the value of $\gamma$, $\delta_B$, $r_B$, the CF and DCS branching fractions for $D \to K^-3\pi$, and the coherence parameter and average strong-phase difference for $D \to K^-3\pi$; the value assumed is based on the current measurements of these parameters. The $B/S$ ratios for the suppressed and favoured modes are $3.1^{+2.3}_{-2.0}$ and $0.20 \pm 0.06$, respectively. The background to the favoured mode is dominated by $B^- \to D^0\pi^-$.

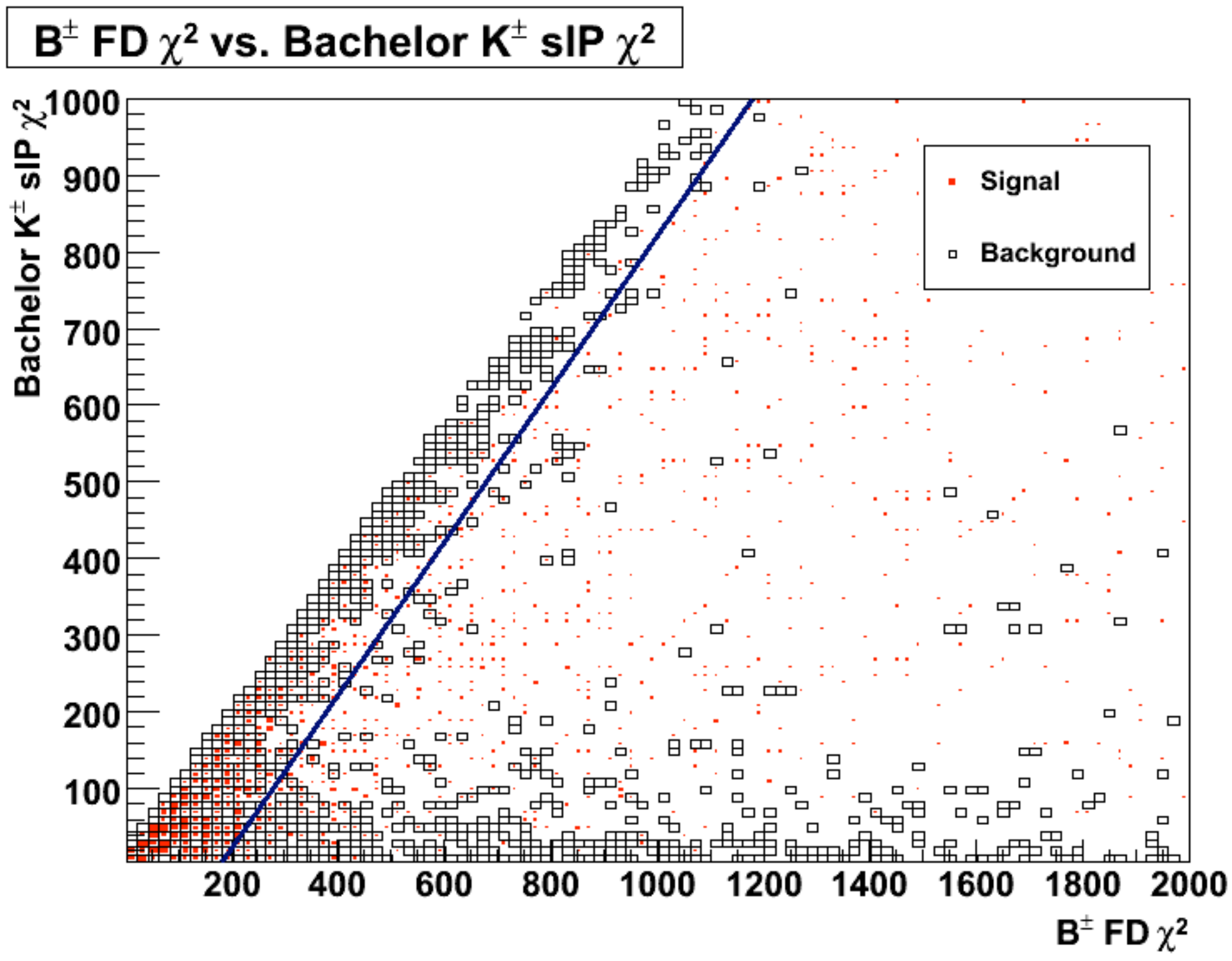

Figure 6: Two-dimensional distribution of the background (open box) and signal (solid box) of the $\chi^2$ of the $B$ flight distance (FD) and the $\chi^2$ of the bachelor $K$ smallest impact parameter (sIP). The optimised linear criteria on these variables is shown by the line; events are retained if they lie to the right of this line.

The yields are approximately a factor of two lower than those found in the DC04 studies. However, the reconstruction version used for the signal simulation in these studies, compared to all others presented in this document, is known to have significantly worse PID and tracking performance. Therefore, some of the signal loss is expected to be recuperated.

## 5.4 Offline selection of $B^- \to D(K_S^0\pi^+\pi^-)K^-$

### 5.4.1 Preselection

The largest change in the preselection between DC04 and DC06 is in the reoptimisation of the $K_S^0$ selection criteria. The changes consist of: a relaxation of the SIPS criterion with respect to the primary vertex (PV) for the LL $K_S^0$ pion daughters[10] to the same value as the DD $K_S^0$; the removal of requirements on both the vertex position of the $K_S^0$ along and transverse to the beam direction; and tighter requirements on track quality, the $K_S^0$ mass and vertex $\chi^2$. Overall, the reoptmisation of the $K_S^0$ preselection results in selection efficiencies of 9% (11%) and 11% (15%) for the LL and DD $K_S^0$ candidates, respectively, in DC06 (DC04). The $K_S^0$ mass resolutions are comparable between DC04, $(3.1 \pm 0.1)$ MeV/$c^2$, and DC06, $(3.5 \pm 0.3)$ MeV/$c^2$. In addition the tails of the $K_S^0$ mass distribution are significantly reduced with the DC06 reconstruction.

The preselection for the $D$ and $B$ candidates remains largely unchanged between DC04 and DC06, except for a tightening of the $D$-mass window and an additional $B$-vertex isolation requirement. An additional SIPS criterion on the bachelor kaon with respect to the PV is also applied to reduce background events where the candidate bachelor kaon originates from the PV.

After all preselection requirements, the $D$ selection efficiency is $(2.27 \pm 0.04)$% for DC06 compared to $(5.23 \pm 0.01)$% for DC04. The $B$ preselection efficiency for DC06 (DC04) is $(0.439 \pm 0.017)$% ($(0.484 \pm 0.004)$%).

### 5.4.2 Offline signal selection

The offline selection criteria consist of tighter requirements on all of the variables used in the preselection. The majority of selection criteria are unchanged from the DC04 optimisation study [68] with the exception of:

- an additional maximum momentum requirement ($<$ 100 GeV/$c$) on the bachelor kaon to reject tracks that are beyond the discriminating power of the RICH system,
- a tightening of the $K - \pi$ $\Delta LL$ PID criterion on the bachelor kaon from $> 2$ to $> 5$ [11] and
- a modification to the vertex isolation requirements on the $B$ candidate for events with a DD $K_S^0$.

[10] LL $K_S^0$ are composites of two long LHCb tracks which include silicon-vertex detector (VELO) information. DD $K_S^0$ are composites of two downstream tracks which are reconstructed by information from the tracking stations only. DD $K_S^0$ generally decay downstream of the VELO and upstream of the first tracking station (TT), which is after the first Ring Imaging Cherenkov (RICH) detector and before the magnet.

[11] The tightening of the $K - \pi$ $\Delta LL$ requirement means that a selection criterion on the $K - p$ $\Delta LL$ is no longer required.

After all offline selection requirements have been applied, $D$ and $B$ mass resolutions of $(9.6 \pm 0.2)$ MeV/$c^2$ and $(15.9 \pm 0.4)$ MeV/$c^2$ are obtained, respectively; these can be compared to the DC04 resolutions of 7 MeV/$c^2$ and 15 MeV/$c^2$. These mass resolutions are for the ensemble of events reconstructed with both DD and LL $K^0_S$.

### 5.4.3 Annual signal event yield

Applying the selection criteria to a $B^- \to D(K^0_S\pi^+\pi^-)K^-$ simulation sample the reconstruction efficiencies are $\epsilon_{sig} = (0.106 \pm 0.004)\,\%$ and $\epsilon_{sig} = (0.224 \pm 0.006)\,\%$ with and without the Level-0 trigger applied, respectively. Approximately 75% of candidates include a DD $K^0_S$.

Making the same assumptions for $N_{b\bar{b}}$ produced in 2 fb$^{-1}$ of integrated luminosity and $f(b \to B^\pm)$ as those made to estimate the $B^- \to D(K^\pm\pi^\mp)K^-$ signal yields reported in Sec. 5.2.2, the event yield, $S$, is given by:

$$\begin{aligned} S &= 2 \times N_{b\bar{b}} \times f(b \to B^\pm) \times \mathcal{B}\left(B^- \to D^0 K^-\right) \\ &\times \mathcal{B}\left(D^0 \to K^0_S\pi^+\pi^-\right) \times \mathcal{B}\left(K^0_S \to \pi^+\pi^-\right) \times \epsilon_{sig} \\ &= 6780 \pm 590\ , \end{aligned} \tag{8}$$

including the Level-0 trigger. The errors quoted are the combination of the statistical uncertainty on the selection efficiency and the uncertainties on the branching fractions.

### 5.4.4 Backgrounds and $B/S$

In the DC06 study, four categories of background have been identified:

- **combinatoric background** where the $D$ candidate has at least one final-state particle that is not from a $D$ combined with a reconstructed kaon from anywhere in the event;
- $D\pi$ **background** where a $\pi^-$ from a $B^- \to D^0(K^0_S\pi^+\pi^-)\pi^-$ event is misidentified as a $K^-$;
- $DK$**-random background** where a true $D \to K^0_S\pi^+\pi^-$ is combined with a reconstructed kaon from the *other* $b$-hadron decay or the underlying event; and
- $DK$**-signal background** a background arising from the combination of a true $D \to K^0_S\pi^+\pi^-$, originating either indirectly or directly from the decay of a $B$, and a reconstructed kaon from the *same* $B$ (excluding the $D\pi$ background). A subset of this background comes from the combination of a true $D$, originating from an intermediate $D^*$, and a real or fake kaon from the same $B$, which is referred to as $D^*$ **background**.

The first three sources of contamination were identified in the DC04 study. In all background studies presented here, a ten times larger $B$ mass window of $\pm 500$ MeV/$c^2$ ($\pm 350$ MeV/$c^2$) around the true $B$ mass is applied to the DD (LL) events to enlarge the sample with which to investigate these backgrounds.

The **combinatoric background** has been evaluated with the DC06 inclusive simulation sample. After all selection cuts zero events pass the offline selection in the wide mass window. An upper limit on the combinatoric background yield is estimated by assuming a uniform distribution across the wide mass window, which results in $B/S < 1.1$ at the 90% CL.

The $D\pi$ **background** is estimated using a dedicated $B^- \to D^0(K^0_S\pi^+\pi^-)\pi^-$ simulation sample. After all selection cuts, and without the Level-0 trigger applied to maximize statistics, two LL and ten DD events are observed in the wide-mass window; only two of the DD events fall within the signal-mass window. The two events are used to calculate a $B/S < 0.095$ at the 90% CL; this is a reduction of $\sim 60\%$ compared to the DC04 estimate, which is mostly due to the change in the $\Delta LL$ cut. The impact of the RICH on removing this background is illustrated in Fig. 7, where the signal and $B^- \to D^0\pi^-$ yields in 2 fb$^{-1}$ are shown with and without the PID criteria applied to the bachelor $K^-$ candidate.

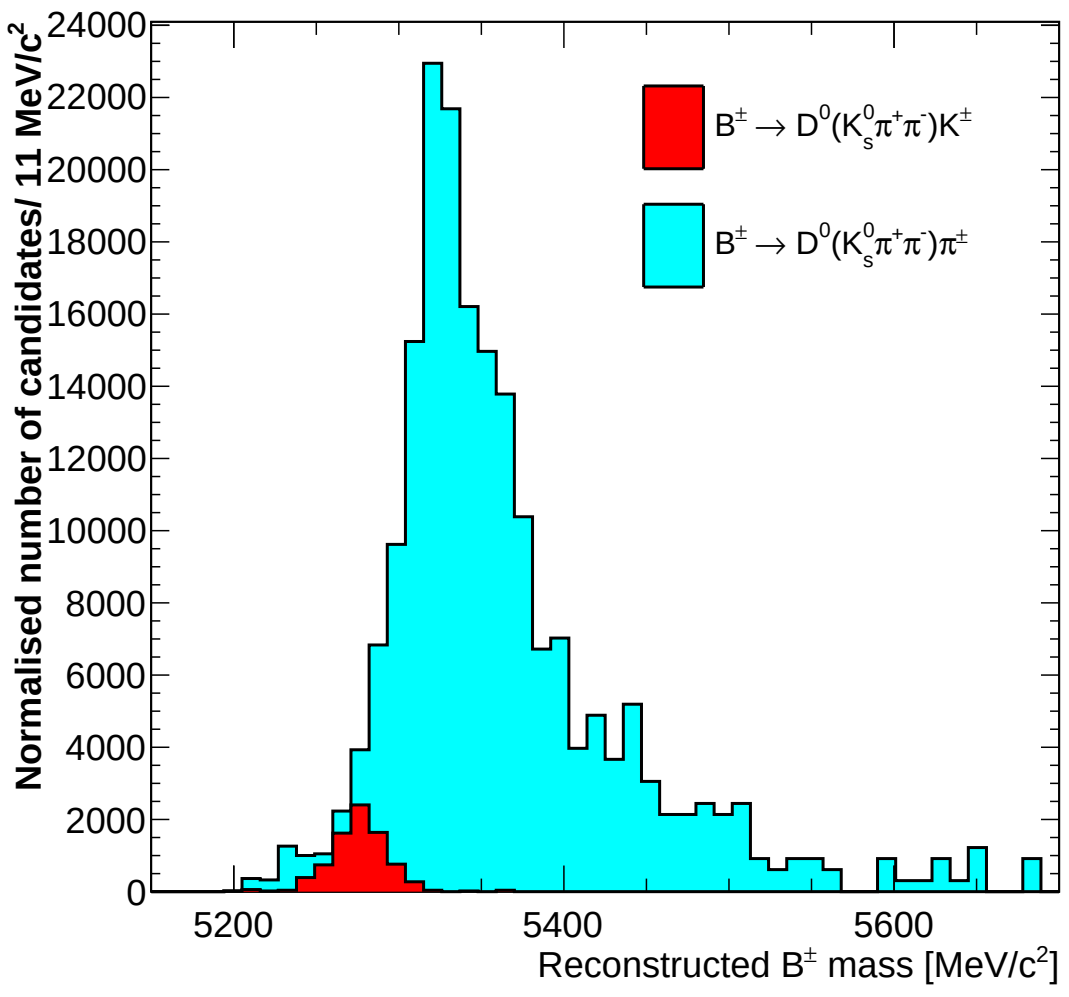

(a)

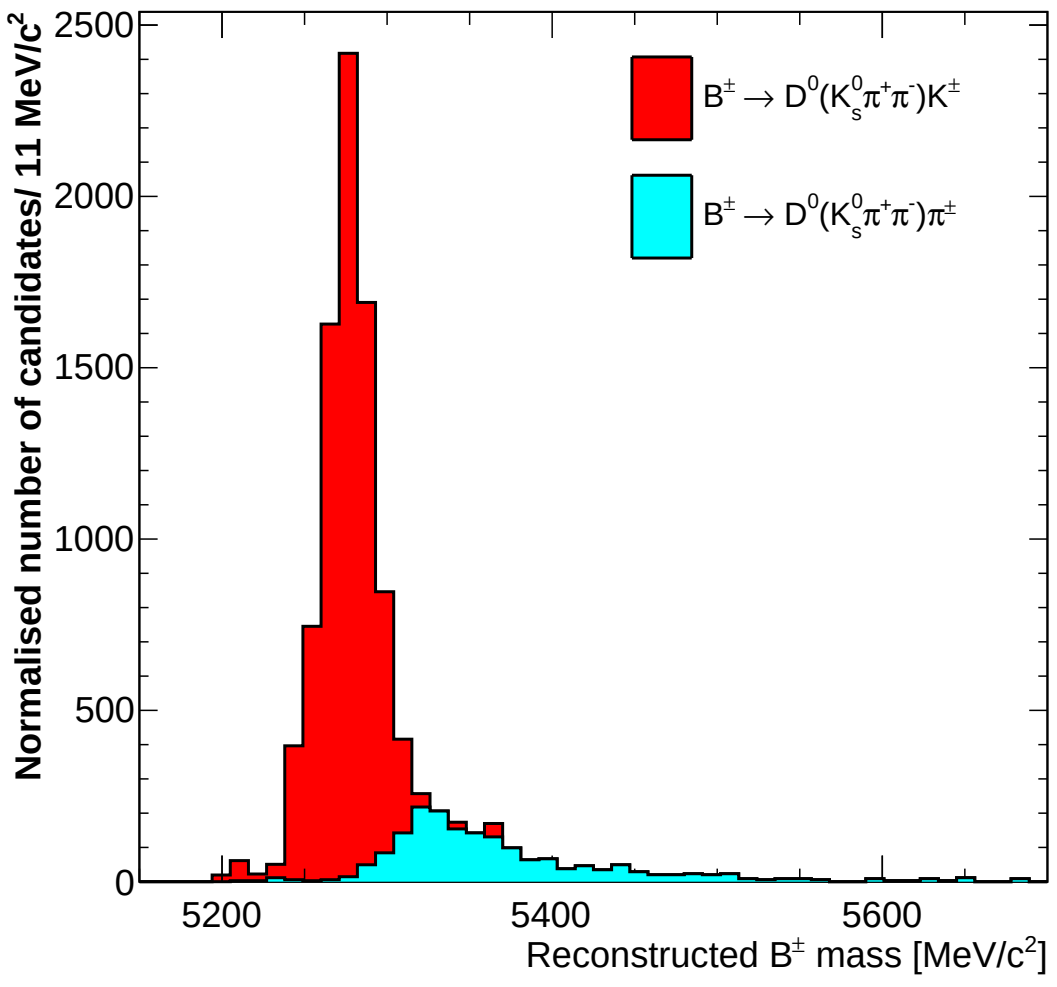

(b)

Figure 7: Distributions of reconstructed $B$ mass for $B^- \to DK^-$ and $B^- \to D\pi^-$ (a) without and (b) with PID criteria applied to the bachelor $K^-$.

The $DK$**-random background** is studied using the large statistics $B \to D^*X$ simulation samples, which are used as a source of true $D$ mesons. The background is estimated from the frequency that the true $D$ is paired with a reconstructed kaon which does not originate from the same $B$ and the combined mass lies within the wide $B$ mass window. Since the $D$ from the $D^*$ has been forced, at generation, to decay to the two-body final states, $K^\pm\pi^\mp$, $\pi^+\pi^-$ or $K^+K^-$, differences, such as branching fractions and relative selection efficiencies, between the two-body decays and three-body $D \to K^0_S\pi^+\pi^-$ decay are taken into account when calculating the background yield and $B/S$. After the offline selection cuts, excluding the $B$-vertex isolation cut and without the Level-0 trigger, a total of 160 $DK$-random background events are observed in the wide $B$ mass window; 13 events fall in the tight mass window. The majority (44%) of the reconstructed kaons originate from the PV, 34% from the other $B$ in the event, 18% are ghost and 5% come from an additional PV. Overall, it is estimated that the probability of picking up a reconstructed kaon, which does not come from the same $B$, to produce a $B$ candidate in the wide-mass window is $\sim$ 15%. The $DK$-random background level is estimated to be $4600 \pm 400$ events per 2 fb$^{-1}$ of integrated luminosity without the Level-0 trigger applied, leading to a $B/S = 0.35 \pm 0.03$.

Table 9: B/S estimates for the various sources of background to the $B^+ \to D(K^0_S\pi^+\pi^-)K^+$ described in the text. When an upper limit is given it is at the 90% CL.

| Type | $B/S$ |
|---|---|
| Combinatoric | $< 1.1$ |
| $DK$ random | $0.35 \pm 0.03$ |
| $DK$ signal | $< 0.09$ |
| $B \to D\pi$ | $< 0.095$ |

A limit on the **$DK$-signal background**, where the $D$ originates directly from the $B$, is estimated from the $b$-inclusive simulation sample. After the offline selection requirements no event remains. Hence, the $B/S$ ratio for the $DK$-signal background is $B/S < 0.09$ at the 90% CL. The **$D^*$ background** contribution to the $DK$-signal background is found to be $B/S < 0.05$ from a study of the $B \to D^*X$ simulation samples.

The background estimates from the various sources are summarised in Table 9. The $B^- \to D^0\pi^+$ background has been reduced significantly and the background from real $D^0 \to K^0_S\pi^+\pi^-$ decays paired with kaons from elsewhere in the event ($DK$-random) is better understood than in the DC04 studies. The expected distribution of the various backgrounds and signal in the reconstructed $B$ mass for 2 fb$^{-1}$ of data is shown in Fig. 8. The shapes of the background distributions have been obtained with full selection applied without the Level-0 trigger and with less stringent PID requirements. Fig. 8 is a conservative estimate of the background level because the combinatoric, $D\pi$ and $DK$-signal backgrounds have been normalised to their 90% CL.

### 5.5 Offline selection of $\bar{B}^0 \to D(K^\pm\pi^\mp)\bar{K}^{*0}$ and $\bar{B}^0 \to D(h^+h^-)\bar{K}^{*0}$

The $\bar{B}^0 \to D(K^\pm\pi^\mp)\bar{K}^{*0}$ and $\bar{B}^0 \to D(h^+h^-)\bar{K}^{*0}$ event selection relies on the high $p_T$ and large impact parameter of the $B$ decay products with respect to the PV and on the long $B$ lifetime; this is similar to other exclusive selections of hadronic $B$ decays. Furthermore, particle identification criteria are extremely important to suppress the background and to distinguish the different $D^0$ decay channels of interest. Many of these selection criteria were used in the DC04 selection [18] and optimised individually for the different decay modes. A different approach is followed in DC06, where a single set of kinematic and topological requirements has been used for all modes; these criteria are optimised on the ADS $D^0 \to K\pi$ mode. The modes are then separated using PID requirements on the $D^0$ daughters. This choice is motivated by the need to achieve a robust estimation of the relative selection efficiencies, which is an important ingredient of the analysis.

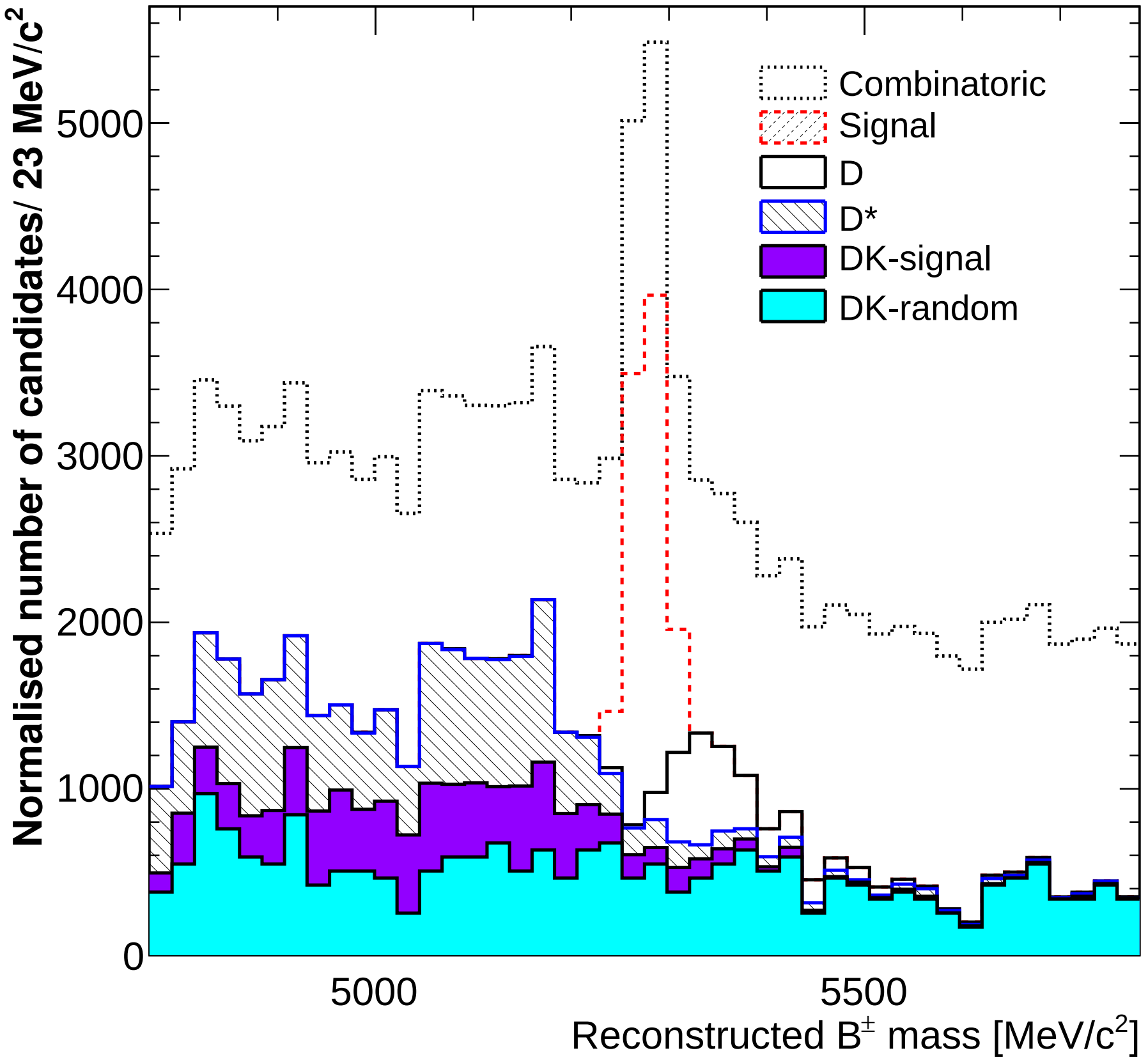

Figure 8: Expected $B$ mass signal and background distributions for $B^- \to D(K^0_S\pi^+\pi^-)K^-$ candidates in data corresponding to an integrated luminosity of 2 fb$^{-1}$. The plots are cumulative. For background contributions where only a 90% CL could be estimated, the level is set to this upper limit.

Significant differences of the DC06 selection with respect to the DC04 selection are:

- more restrictive FS criteria, compensated by less restrictive vertex-isolation requirements and
- the maximum $B$ and $D$ vertex $\chi^2$ requirements have been reduced.

Samples of signal events and $b$-inclusive events are used to optimise and benchmark the selection. The expected number of selected signal and background events, passing the Level-0 but without any requirement on passing the HLT, are shown, for a nominal year of data-taking, in Table 10. The corresponding selection efficiencies for the different signal channels are also listed. It should be noted that the number of events expected in the suppressed ADS mode depends heavily on the unknown parameters: $r_{B^0}$, $\delta_{B^0}$, $\gamma$ and $\delta_D^{K\pi}$. Therefore, we have neglected the interference term in the decay rate, so there is no dependence on the values of the phases, and assumed $r_{B^0} = 0.3$ to compute the numbers in Table 10.

Table 10: Summary of total efficiencies ($\epsilon_{sig}$), signal and background yields expected in 2 fb$^{-1}$, in a $\pm$ 50 MeV/$c^2$ $B$ mass window, computed with specific aseVsumptions on their composition and mass distributions, as described in the text. All upper limits are at 90% CL.

| Channel | $\epsilon_{sig}$ (%) | $S$ | $B_{b\bar{b}}$ | $B_{spec}$ |
|---|---|---|---|---|
| $\bar{B}^0 \to D(K^+\pi^-)\bar{K}^{*0}$ | $0.37 \pm 0.02$ | $3200 \pm 500$ | $780^{+1600}_{-500}$ | $2200^{+1900}_{-1200}$ |
| $\bar{B}^0 \to D(K^-\pi^+)\bar{K}^{*0}$ | $0.38 \pm 0.03$ | $290 \pm 50$ | $2850^{+1500}_{-1400}$ | $2100^{+1900}_{-1200}$ |
| $\bar{B}^0 \to D(K^+K^-)\bar{K}^{*0}$ | $0.31 \pm 0.02$ | $270 \pm 40$ | $<1650$ | $< 180$ |
| $\bar{B}^0 \to D(\pi^+\pi^-)\bar{K}^{*0}$ | $0.31 \pm 0.02$ | $100 \pm 20$ | $1600^{+1500}_{-950}$ | $80^{+75}_{-35}$ |

The $b$-inclusive sample is used to identify the following sources of background:

- combinatoric, where some of the final state kaons or pions are due to tracks from the rest of the event, which is assumed to follow a flat distribution in $B$ mass;
- different $B$ decays proceeding through a $D^*$, and a real $D^0$, where neutrals are lost and some of the charged pions in the true decay chain may be lost or misidentified; these events tend to peak at an invariant mass significantly less than the nominal $B$ mass;
- a peaking contribution in the $B$ mass from $\bar{B}^0 \to D^+(K^+\bar{K^{*0}})\pi^-$ with $\bar{K}^{*0} \to K^-\pi^+$. In the reconstruction of this event, the $K^{*0}$ has been properly reconstructed, while a fake $D^0$ is made with the pion coming from the $B$ and the kaon coming from the $D^+$.

Larger samples of specific background, which could mimic the signal, have been studied in order to evaluate better the background level [62]. Events passing the selection, in a window of $\pm$ 500 MeV/$c^2$ from the $B$ mass, have been subdivided into *low-mass*, *combinatorial* and *peaking background* events according to the truth information about the simulated decays [69]. On the basis of this categorisation we have estimated the number of background events in the signal window by simply assuming that the combinatorial background is uniformly distributed with respect to the $B$ mass, that the low-mass background can be neglected because it lies outside the signal-mass region and that the peaking background scales with its branching fraction. These are oversimplified assumptions, as the low-mass background will leak in to the signal window and the combinatorial background may not be uniformly distributed. The background values computed with these assumptions are reported in Table 10 for the sum of all the specific contributions and for the $b$-inclusive sample. The specific backgrounds are dominated by $B \to D^{(*)}X$ events, where the $D^0$ is correctly reconstructed in one of the signal decays. Unfortunately, the precision of these studies is limited by the size of the simulated samples.

## 5.6 Offline selection of $B^0_{(s)} \to D^{\pm}_{(s)} h^{\mp}$

This section describes a unified selection for the modes $B^0 \to D^{\pm}(K^{\mp}\pi^{\pm}\pi^{\pm})\pi^{\mp}$ and $B^0_s \to D^{\pm}_s(K^{\pm}K^{\mp}\pi^{\pm})K^{\mp}$; the decay $B^0_s \to D^{-}_s(K^{+}K^{-}\pi^{-})\pi^{+}$, which is an important control channel for these decays, is also selected. The motivation for this strategy is to allow an efficient common High Level Trigger (HLT) selection and to permit cancellation of some common systematic uncertainties when combining the channels to determine $\gamma$ [33]. The selection of the highest yield mode $B^0 \to D^{\pm}\pi^{\mp}$ is optimised first; then the criteria are adjusted to select $B^0_s \to D^{\pm}_s K^{\mp}$ while providing adequate background suppression.

### 5.6.1 $B^0 \to D^{\pm}\pi^{\mp}$

The starting point for the selection are the optimised selection criteria reported in Ref. [70]. Initially these cuts are reoptimised to reject all background events from a sample of $1.7 \times 10^6$ minimum bias events that pass the Level-0 trigger. This requirement ensures that offline selections used to optimise the Level-0 and HLT triggers did not have an unacceptably high rate for selecting minimum bias events. The results of this reoptimisation are slightly stricter requirements on the SIPS of the $B$, $D$ and bachelor $\pi$. The signal yield after the Level-0 trigger and the reoptimised offline selection is $(1.23 \pm 0.08) \times 10^6$ events in a sample corresponding to 2 fb$^{-1}$ of integrated luminosity. The background was estimated from a sample of $b$-inclusive events; the background is dominated by combinatoric events, in particular where one track is a ghost. A background of $(0.21 \pm 0.04) \times 10^6$ is estimated for 2 fb$^{-1}$ of data corresponding to a $S/B = 6.0 \pm 1.2$. The distribution of $B$ mass for events selected from the $b$-inclusive sample is shown in Fig. 9; the high purity mass peak is observed clearly.

### 5.6.2 $B^0_s \to D^{\pm}_s K^{\mp}$

The viability of applying a variant of the $B^0 \to D^{\pm}\pi^{\mp}$ selection to the decay mode $B^0_s \to D^{\pm}_s\,(K^{\pm}K^{\mp}\pi^{\pm})\,K^{\mp}$ has been studied [63]. Because of the approximately nine times lower branching ratio of $B^0_s \to D^{\pm}_s K^{\mp}$ as compared to $B^0 \to D^{\pm}\pi^{\mp}$, it is necessary to tighten certain requirements in order to reduce the background to an acceptable level. In addition to applying tight PID criteria on the bachelor kaon, the flight significance cut on the $B$ is increased from 2 to 12, and the impact parameter significance cut on the $B$ is tightened from 5 to 4. Furthermore, it is found that requiring the bachelor kaon to have a momentum of less than 100 GeV/$c$, so that it is within the momentum region with good $\pi - K$ separation, is valuable for this channel.

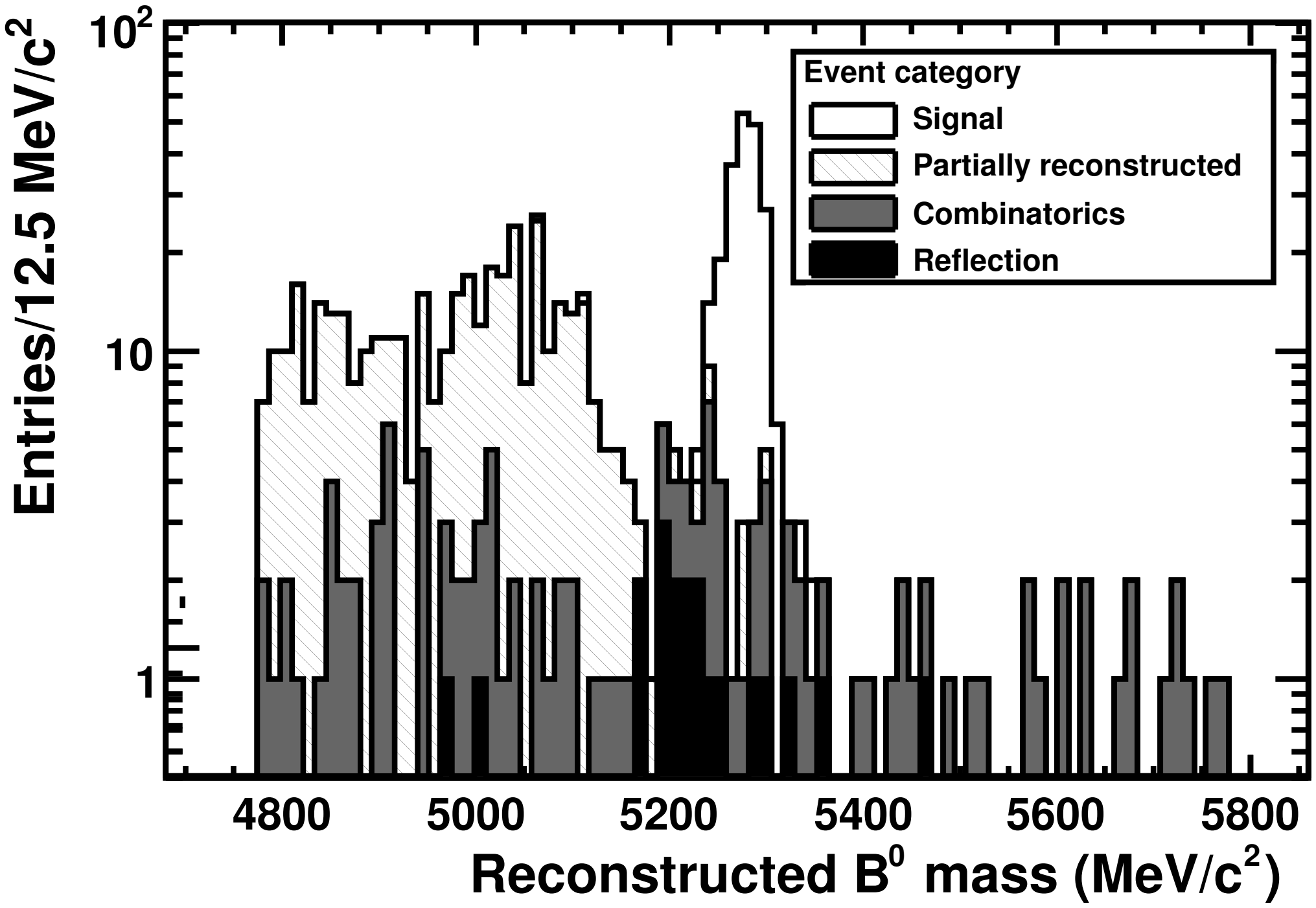

Figure 9: Distribution of $B^0 \to D^{\pm}(K^{\mp}\pi^{\pm}\pi^{\pm})\pi^{\mp}$ events selected from a sample of simulated $b$-inclusive decays.

In common with the DC06 study of $B^0 \to D^{\pm}\pi^{\mp}$, the cuts are optimised to reject all Level-0 stripped minimum bias events in a sample of $1.7 \times 10^6$ such events. The signal yield after the Level-0 trigger and the offline selection is $(14.0 \pm 4.8) \times 10^3$ events in 2 fb$^{-1}$ of data taking. The combinatoric background is estimated from the $b$-inclusive sample to be $(3.2^{+3.6}_{-2.0}) \times 10^3$ events. There is also a significant background from $B_s^0 \to D_s^- \pi^+$ events where the bachelor $\pi$ is misidentified as a $K$. The expected number of background $B_s^0 \to D_s^- \pi^+$ candidates is $(0.9 \pm 0.3) \times 10^3$ in each 2 fb$^{-1}$ of integrated luminosity collected.

### 5.6.3 $B^0 \to D^{*\pm}\pi^{\mp}$

No recent offline selection study has been performed for the channel $B^0 \to D^{\pm *}\pi^{\mp}$. Older studies estimate an untagged yield of $\sim 200k$ events after all triggers using an exclusive selection [71] and $\sim 800k$ events after all triggers using a combined inclusive and exclusive selection [72] in 2 fb$^{-1}$ of data taking. This mode will significantly aid the measurement of $\gamma$ from the related decay channel $B^0 \to D^{\pm}\pi^{\mp}$ [33] for two reasons: the increased statistics and the high probability of a different strong phase to $B^0 \to D^{\pm}\pi^{\mp}$, which will resolve ambiguous solutions.

# 6 Studies of sidebands and control samples

To perform systematically robust measurements of $\gamma$ in the modes discussed in this document several control channels must also be selected and studied. The status of these studies is outlined in the following sections. Many of these control samples are linked to evaluating the systematic uncertainties that affect these measurements, which are discussed in Sec. 8.

## 6.1 $B^- \to D(K^{\pm}\pi^{\mp})K^-$, $D(h^+h^-)K^-$ and $D(K^{\pm}\pi^{\mp}\pi^+\pi^-)K^-$

No dedicated studies have been performed of sidebands or control channels for the charged $B$ ADS and GLW modes. The principal control channel will be $B^- \to D^0\pi^+$. This decay is topologically identical to the signal modes but it has a branching fraction of $(4.84 \pm 0.15) \times 10^{-3}$ [64], which is twelve times larger. Furthermore, most $B^- \to D^0\pi^+$ decays are expected to exhibit negligible $CP$ violation in the Standard Model because $r_{B\to D\pi} \sim r_{B\to DK} \tan^2\theta_C \sim 5 \times 10^{-3}$, where $\theta_C = 13.0^{\circ}$ is the Cabibbo angle [64]. An exception is the suppressed ADS decays $B^- \to D(K^+\pi^-)\pi^-$ which can exhibit asymmetries of the $\mathcal{O}(10\%)$.

It is anticipated that there will be no PID criteria applied during the HLT selections so that a sample of $B^- \to D^0\pi^-$ events twelve times larger than the favoured $B^- \to D(K^-\pi^+)K^-$ and $B^- \to D(K^-\pi^+\pi^+\pi^-)K^-$, as well as $B^- \to D(h^+h^-)K^-$ modes, will be selected. The relative efficiency amongst the modes can be evaluated with these samples and bound the uncertainty on the normalisation technique employed to extract $\gamma$ from the combination of ADS and GLW modes. Furthermore, given the negligible $CP$ violation, first-order tests of the detection and production asymmetries can be made with the opposite charge decays reconstructed.

The PID criteria will be the only difference between the two selections. The efficiency and misidentification rates of pions and kaons will be calibrated with $D^{*+} \to D^0\pi^+$ decays, as discussed in Sec. 8.2.

## 6.2 $B^- \to D(K^0_S\pi^+\pi^-)K^-$

As for the ADS/GLW analysis, $B^- \to D^0\pi^-$ provides the primary control channel for the analysis of $B^- \to D(K^0_S\pi^+\pi^-)K^-$. This topologically identical decay allows the Dalitz plot acceptance and possible detection asymmetries to be determined from data. For the binned method the procedures to utilise this sample are described and the anticipated size of systematic uncertainties related to them are discussed in Ref. [73]. Also, in Ref. [73] the use of sidebands in $D$ and $B$ mass to determine the background level and its relative composition is discussed. Studies have been performed of this channel as a background (see Sec. 5.4). As long as no PID requirements are placed on the bachelor $K$ in the signal HLT2 selection this sample will be collected with the same efficiency as signal.

## 6.3 $B^0 \to D(K^{\pm}\pi^{\mp})K^{*0}$ and $B^0 \to D(h^+h^-)K^{*0}$

The main sources of systematic uncertainties for the extraction of $\gamma$ from $\bar{B}^0 \to D(K^{\pm}\pi^{\mp})\bar{K}^{*0}$ and $\bar{B}^0 \to D(h^+h^-)\bar{K}^{*0}$ are expected to be due to:

1. $K^+/K^-$and $\pi^+/\pi^-$charge detection asymmetries;
2. $B^0/\overline{B}^0$ production asymmetry;
3. $K^{*0} \to K^+\pi^-/\overline{K}^{*0} \to K^-\pi^+$ detection asymmetry;
4. relative selection efficiencies for $D^0 \to K^+K^+$ and $D^0 \to \pi^+\pi^-$, with respect to the $D^0 \to K^-\pi^+$ decay mode; and
5. background subtraction.

Suitable control samples and invariant mass sidebands will be used to control the main sources of systematic uncertainties on data. The evaluation of items 1-3 above will be discussed further in Sec. 8.

For item 4, as for the charged ADS/GLW and Dalitz analyses the mode $B^- \to D^0\pi^-$ is the most important control channel, which has a branching fraction two orders of magnitude larger than the $\bar{B}^0 \to D^0\bar{K}^{*0}$, and can be reconstructed with higher efficiency and purity, thanks to the lower track multiplicity in the final state. If identical $D^0 \to h^+h^-$ selections are used for the signal and the control sample, the ratio $R_{hh}$ between the reconstructed number ($N_{ev}$) of signal events in the GLW and in the ADS-favoured modes,

$$R_{hh} = \frac{N_{ev}(B^- \to D^0(hh)\pi^-)}{N_{ev}(B^- \to D^0(K^-\pi^+)\pi^-)},$$

is a good measurement of $N_{hh}/N_{K\pi}$, the product of the ratio of the total efficiencies ($\epsilon$) and the ratio of the $D^0$ branching fractions ($\mathcal{B}$) for the two corresponding neutral $B$ decay modes, *i.e.*,

$$R_{hh} \approx N_{hh}/N_{K\pi} = \frac{\epsilon(B^0 \to D(h^+h^-)K^{*0})}{\epsilon(B^0 \to \overline{D}^0(K^+\pi^-)K^{*0})} \times \frac{\mathcal{B}(D^0 \to h^+h^-)}{\mathcal{B}(\overline{D}^0 \to K^+\pi^-)} .$$

This relation assumes that efficiency for reconstructing the bachelor $\pi^{\pm}$ and $K^{*0}$ cancel and is just a measure of the relative $D$ reconstruction efficiencies. In the $\gamma$ sensitivity studies $N_{hh}/N_{K\pi}$ are taken to be exactly known and will be varied to evaluate the systematic uncertainty. We should be able to measure $R_{KK}$ and $R_{\pi\pi}$ from the control sample with a precision $< 1\%$; this is a smaller uncertainty than that on $N_{hh}/N_{K\pi}$ which is limited by the knowledge of the relative branching ratios to a precision of 3% [64].

The main potential sources of background have been identified on the limited DC06 $b$-inclusive sample (see Sec. 5.5). The dominant peaking background contribution for the ADS-suppressed channel is from $\bar{B}^0 \to D^+(K^-K^+\pi^+)\pi^-$ decays, where the $D^+$ decay can also proceed via quasi-two body decays, including $\bar{K}^{*0}$ resonances.[12] A similar analysis can be performed on $\bar{B}^0 \to D^+(K^+\pi^+\pi^-)\pi^-$ decay, which is a background to the GLW $D^0 \to \pi^+\pi^-$ channel. An inspection of the reconstructed $D^0$ mass in the $B$ mass window (Fig. 10(a)) shows that this is sufficiently uniform to be subtracted using the $D^0$ mass sidebands.

The distribution is used to evaluate the relative error on the expected number of events from these sources as a function of the $D^0$ sidebands width, which is shown in Fig. 10(b), for a data sample corresponding to an integrated luminosity of 2 fb$^{-1}$.[13] These plots show that we can achieve a 1% relative uncertainty on this background estimation with sidebands of $\pm 90$ MeV/$c^2$.[14]

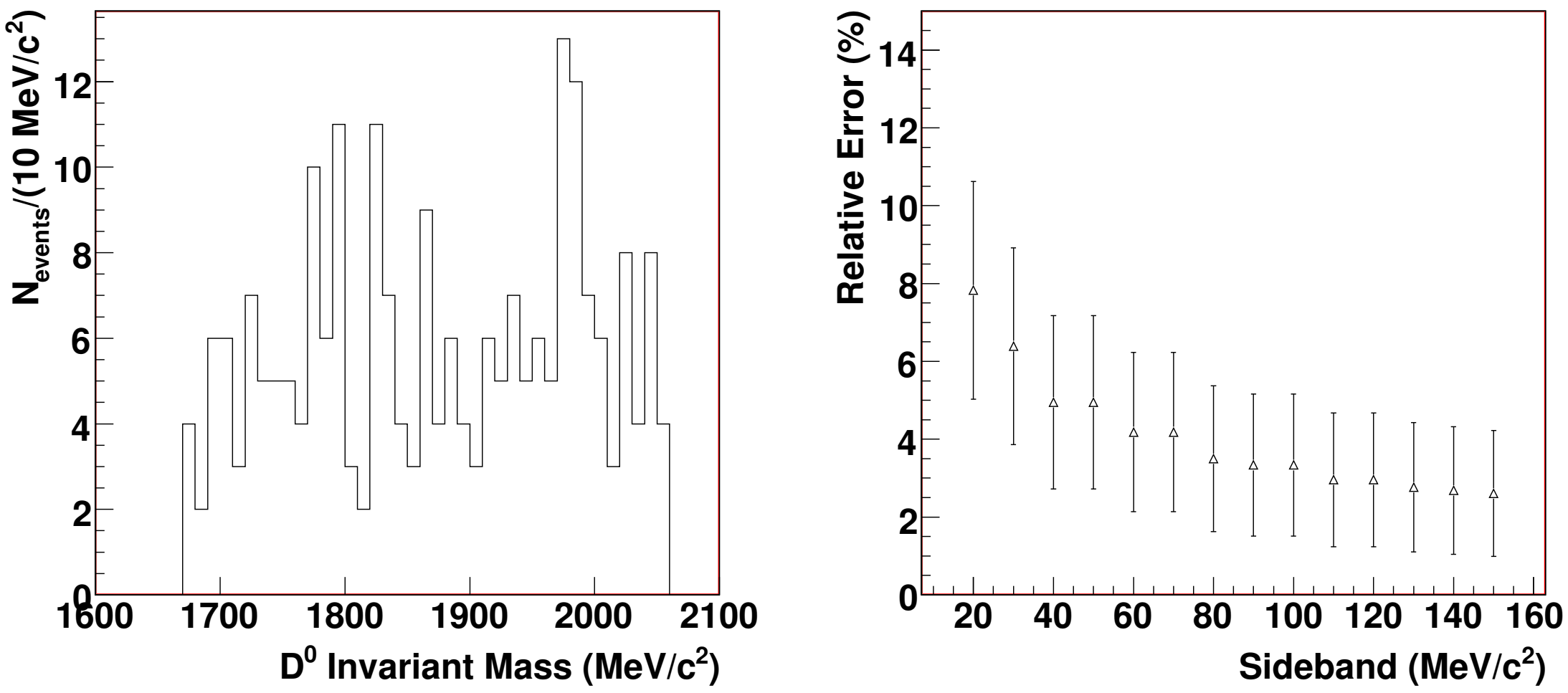

(a) (b)

Figure 10: (a) Reconstructed $D^0 \to \pi^+\pi^-$) mass distribution from $\bar{B}^0 \to D^+(K^+\pi^+\pi^-)\pi^-$; (b) corresponding relative error on the background evaluation as a function of the $D^0$ sidebands width.

## 6.4 $B^0_{(s)} \to D^{\pm}_{(s)} h^{\mp}$

Specific backgrounds to the decay mode $B^0_s \to D^{\pm}_s K^{\mp}$ have been studied [63], and the main source of background, aside from combinatorics, is expected to come from $B^0_s \to D^-_s \pi^+$ events. In order to study fully the combinatoric and specific backgrounds on real data, it will be necessary to examine the mass sidebands of the $D_s^+$ as well as those of the $B_s^0$. Given that the $B_s^0 \to D_s^- \pi^+$ events are topologically identical to the signal, PID requirements will be used to discriminate between them. Studies of the selection of $B_s^0 \to D_s^- \pi^+$ have been made [63]; these studies estimate a yield of $(1.7 \times 0.5) \times 10^5$ events passing the Level-0 trigger in a data sample corresponding to 2 $\text{fb}^{-1}$ of integrated luminosity. This is an important channel for calibrating the tagging, proper time resolution and proper time acceptance to make a time-dependent $CP$-asymmetry measurement (see Sec. 8.5). In fact a simultaneous fit to the $B_s^0 \to D_s^{\mp} K^{\pm}$ and $B_s^0 \to D_s^- \pi^+$ sample allows the mistag rates and proper-time resolutions to be extracted from data [27].

[12]This mode has not been generated in the DC06 simulation.

[13]In the estimate we use symmetric sidebands, centred around the $D^0$ mass, starting at a distance greater than $\pm 4\sigma$, where the $D^0 \to h^+h^-$ invariant mass resolution, $\sigma$, is 9 MeV/$c^2$ in DC06, and we assume Poisson fluctuations on the number of reconstructed events (scaled to 2 fb$^{-1}$).

[14]As well as the background source under discussion, the $D^0$ sideband in data will include several different contributions. These are harder to estimate with the given limited $b$-inclusive data samples.

No DC06 studies of specific background channels have been performed for the mode $B^0 \to D^{\pm}\pi^{\mp}$, but from DC04 studies [70] the only significant non-combinatoric background is expected to be $B^0 \to D^- K^+$ decays where the bachelor kaon is misidentified as a pion. However, given the branching fraction for $B^0 \to D^- K^+$ is more than an order of magnitude smaller than that for the signal and the $\pi$ to $K$ misidentification rate is approximately 5% there will be $< 1\%$ contamination from this decay. Therefore, with $B^0 \to D^{\pm}\pi^{\mp}$ having the largest branching ratio of all the $B^0_{(s)} \to D^{\pm}_{(s)} h^{\mp}$ family of modes by an order of magnitude, it can be used in calibrating the trigger and offline selections for all of its sister channels.

# 7 $\gamma$ extraction from data

The $\gamma$ sensitivity studies, as performed for the DC04 selection and reported in Ref. [9], have not been fully updated here due to final HLT efficiency estimates being unavailable. The global sensitivity expected from data samples corresponding to 0.5 $\text{fb}^{-1}$ and 2 $\text{fb}^{-1}$ of data presented in Ref. [9] are shown in Table 11. The expected sensitivity is shown for various values of $\delta_{B^0}$, which is currently unconstrained experimentally and greatly effects the sensitivity to $\gamma$ from $B^0 \to DK^{*0}$ decays.

A provisional estimate of the sensitivity expected from the signal yields and background estimates reported in Sec. 5 is found by assuming an HLT efficiency of 65% for all hadronic modes considered. This HLT efficiency is in line with current findings but will evolve as more studies are done for individual modes. There are also four other significant

Table 11: Expected combined sensitivity to $\gamma$ from $B \to DK$ and time-dependent measurements for data sets corresponding to integrated luminosities of 0.5 and 2 $\text{fb}^{-1}$. The table is taken from Ref. [9]. In these studies the Level-0 and Level-1, a precursor to HLT1, triggers were included. The HLT2 trigger was not included.

| $\delta_{B^0}$ (°) | 0 | 45 | 90 | 135 | 180 |
|---|---|---|---|---|---|
| $\sigma_\gamma$ for 0.5 $\text{fb}^{-1}$ (°) | 8.1 | 10.1 | 9.3 | 9.5 | 7.8 |
| $\sigma_\gamma$ for 2 $\text{fb}^{-1}$ (°) | 4.1 | 5.1 | 4.8 | 5.1 | 3.9 |

changes to the assumptions of the previous sensitivity study:

1. BABAR has presented the constraint $0.07 < r_{B^0} < 0.41$ at the 95% CL [46]; $r_{B^0}$ was assumed to be 0.4 in the DC04 sensitivity studies, which is at the edge of the confidence interval. Therefore, the updated study uses a value of 0.27. This new value will reduce the weight of the measurement of $\bar{B}^0 \to D\bar{K}^{*0}$ analysis in the global fit.

2. CLEO-c has now published measurements of the strong-phase difference in bins of the $D^0 \to K^0_S\pi^+\pi^-$ Dalitz plots [74]. The uncertainties on these measurements lead to a uncertainty of 1 to 2° (see Sec. 8.1.3), which is significantly smaller than the 5° uncertainty assumed in Ref. [9].

3. The constraint on $\delta_D^{K\pi}$ was that taken from the CLEO-c measurement [16] of $(-158^{+14}_{-16})^\circ$. However, the combination of this measurement with $D$-mixing measurements from BABAR, Belle and CDF [1] leads to a better constraint than the CLEO-c measurement alone of $\delta_D^{K\pi} = (-158.8^{+10.5}_{-10.9})^\circ$ (see Sec. 8.1.1).

4. The value of $r_D^{K\pi}$ used in these studies was 0.0616 which was the square root of the ratio of DCS to CF branching fractions. The current world-average value of $r_D^{K\pi}$ is $0.0579 \pm 0.0007$ [1], which is 6% lower due to the effect of $D$-mixing being accounted for. This will reduce the sensitivity to $\gamma$ from the suppressed ADS modes.

Incorporating all these changes gives an uncertainty on $\gamma$ of 4.8° for an integrated luminosity of 2 fb$^{-1}$and $\delta_{B^0} = 0^\circ$. This is worse than the DC04 performance principally as a result of the diminished weight of the $\bar{B}^0 \to D\bar{K}^{*0}$ modes because of the reduced signal yields, increased $B/S$ and lower value of $r_{B^0}$. However, the significant improvement in the expected yield and $B/S$ for $B^0_s \to D_s^\pm K^\mp$ partially compensates. Furthermore, the dependence on $\delta_{B^0}$ is greatly reduced with the performance actually improving over that of DC04 when $\delta_{B^0} = 45^\circ$ is assumed; the $\gamma$ sensitivity is of 4.9° and the improvement is again due to $B^0_s \to D_s^\pm K^\mp$. These studies will be updated once the mode-by-mode HLT efficiencies are known, but the preliminary conclusion is that there is no significant degradation in the overall sensitivity to $\gamma$.

# 8 Studies related to systematic uncertainties

This section describes the important systematic uncertainties related to the measurements. However, none of the measurements presented are likely to be systematically limited at LHCb given the low event yields in the modes most sensitive to $\gamma$ (suppressed ADS modes, GLW modes, Dalitz and $B^0_s \to D_s^\pm K^\mp$) or the low statistical sensitivity per event (favoured ADS modes and $B^0 \to D^\pm\pi^\mp$).

There are five general classes of uncertainties considered: those from external inputs, those from the reconstruction, those from production and detection asymmetries, those from the knowledge of backgrounds and those related to time-dependent $CP$-violation measurements. These are discussed in turn in the following subsections.

## 8.1 External inputs to the analyses

### 8.1.1 $\delta_D^{K\pi}$ from CLEO-c and charm-mixing measurements

External knowledge of the strong-phase $\delta_D^{K\pi}$ improves the determination of $\gamma$ from the ADS measurements. A Gaussian $\chi^2$ constraint on $\delta_D^{K\pi}$ is added to the global fit [9], therefore, no separate systematic uncertainty will be attributed to this source. The current combination of $D$-mixing measurements that are sensitive to $\delta_D^{K\pi}$ finds $\delta_D^{K\pi} = (-158.8^{+10.5}_{-10.9})^\circ$ [1]. This constraint is statistically limited and will improve once the full BABAR and CLEO-c data sets are analysed and additional data are added to the measurements from Belle and CDF. Furthermore, additional measurements of quantum-correlated $D\bar{D}$ production will be made by BES-III during the lifetime of LHCb.

### 8.1.2 $R_{K3\pi}$ and $\delta_D^{K3\pi}$ from CLEO-c

As discussed in Sec. 2.1.1, in order to include additional multi-body flavour-specific final states an additional parameter is introduced, the coherence factor $R$. The global fit of LHCb data includes the mode $B^- \to D(K^\pm\pi^\mp\pi^\mp\pi^\pm)K^-$. In addition, the constraints on $R_{K3\pi}$ and the average strong-phase difference, $\delta_D^{K3\pi}$, from CLEO-c [20] are included in the fit. The constraints are shown in Fig. 11. The measured value of the coherence factor is only $\sim 0.3$ therefore the sensitivity to $\gamma$ is diluted in this mode. However, this low coherence leads to enhanced sensitivity to $r_B$. In the global fit the improved knowledge of $r_B$ from including $B^- \to D(K^\pm\pi^\mp\pi^\mp\pi^\pm)K^-$ increases the sensitivity to $\gamma$ from the other modes that depend on $r_B$. The individual CLEO-c measurements sensitive to $R_{K3\pi}$ and $\delta_D^{K3\pi}$ are added as Gaussian $\chi^2$ constraints, as is the case for $\delta_D^{K\pi}$, so there is no separate systematic uncertainty on $\gamma$ coming from these measurements.

### 8.1.3 The determination of strong-phase differences over $D \to K_S^0\pi^+\pi^-$ Dalitz space from CLEO-c

The model-independent determination of $\gamma$ from $B^- \to D(K_S^0\pi^+\pi^-)K^-$ requires knowledge of the average sine and cosine of the strong-phase difference for the particular binning of the Dalitz plot used [21]. These parameters have been measured by the CLEO-c collaboration [74]. The impact of these measurements on the sensitivity to $\gamma$ has been studied using the model-independent fit described in Ref. [73] following a binning proposed in Ref. [75]. No background is included and 16 million events are generated for each toy experiment to make the statistical uncertainty negligible. For each toy experiment the events are generated with the strong-phase parameters selected randomly from Gaussian distributions with mean and width corresponding to the measured values and uncertainties reported in Ref. [74]; the fit is performed with the phase parameters fixed to the measured values. The other physics parameters are fixed to the following values: $\gamma = 60^\circ$, $\delta_B = 130^\circ$ and $r_B = 0.1$. (These are the same values assumed as the baseline sensitivities in Ref. [9].) The resulting distribution of the values of $\gamma$ returned by the fit for 10,000 toy experiments is shown in Fig. 12. The RMS of the distribution is $1.7^\circ$ and is indicative of the expected systematic uncertainty from the measurements of the strong phase difference. The RMS of the distribution is dependent on the central value of $\gamma$. For example, the RMS is found to decrease to $1^\circ$ if the assumed value of $\gamma$ is increased to $80^\circ$, which is within current experimental bounds.

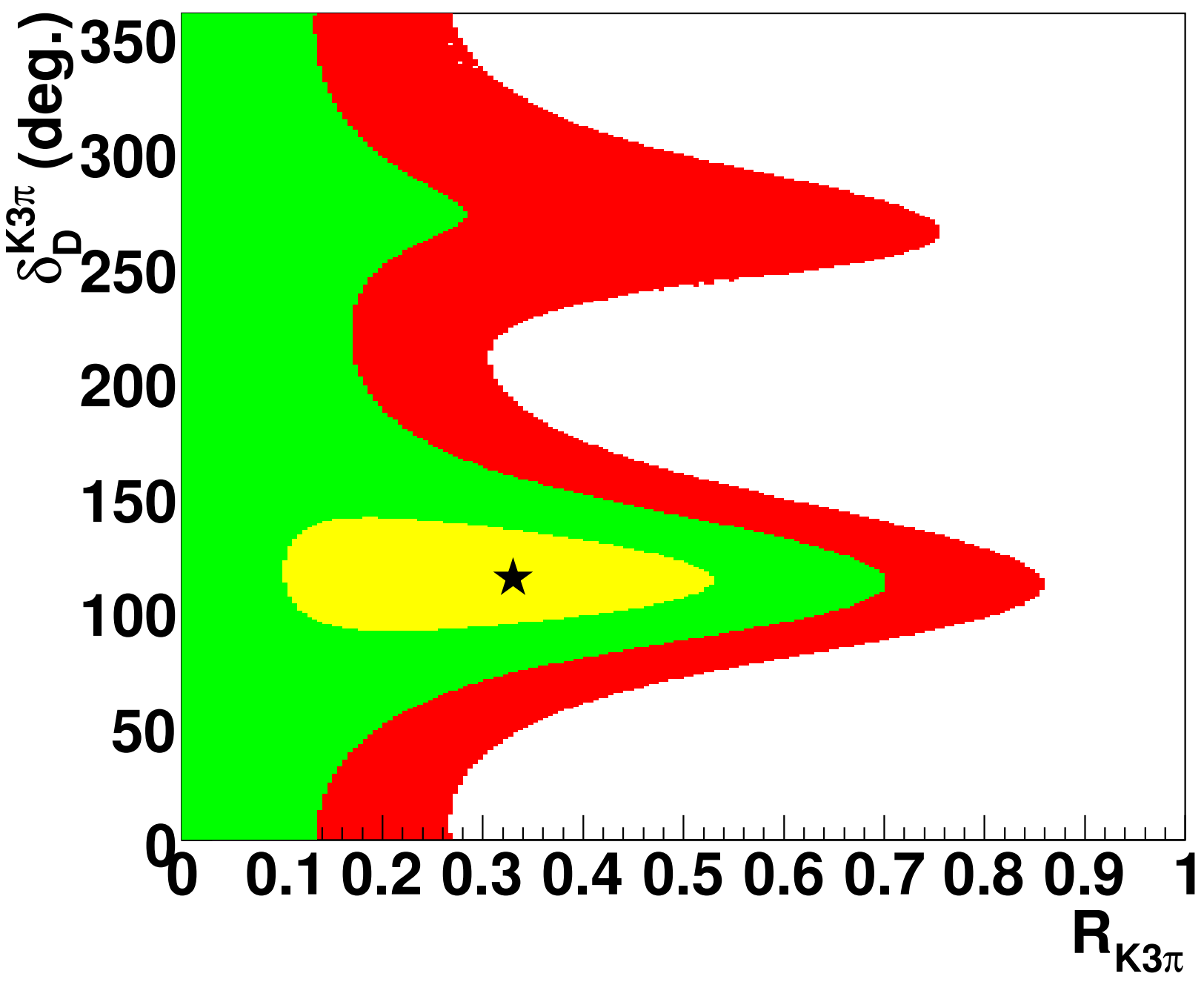

Figure 11: $1\sigma$ (yellow), $2\sigma$ (green) and $3\sigma$ (red) constraints on $R_{K3\pi}$ and $\delta_D^{K3\pi}$ from CLEO-c measurements [20]. The star indicates the best fit.

The systematic uncertainty related to the strong-phase measurements was assumed to be $5^\circ$ in the DC04-sensitivity studies. The reduction will increase the weight of this mode in the global fit particularly for large data sets ($> 5$ fb$^{-1}$).

### 8.1.4 The model of $D^0 \to K_S^0\pi^+\pi^-$

The model-dependent determination of $\gamma$ from $B^- \to D(K_S^0\pi^+\pi^-)K^-$ requires a model for the $D^0 \to K_S^0\pi^+\pi^-$ decay [68]. Both BABAR [23] and Belle [24] have published models determined from flavour-tagged samples of $D^0$ from $D^{*+} \to D^0(K_S^0\pi^+\pi^-)\pi^+$ decay. The systematic variation of these models when determining $\gamma$ lead to uncertainties of $6^\circ$ and $15^\circ$ for BABAR [23] and Belle [24], respectively. These models, or those derived from LHCb data, will be used by LHCb to determine $\gamma$ in a model-dependent fit and the minimum systematic uncertainty expected would be $6^\circ$.

The BABAR [23] model includes an improved description of the $\pi\pi$ $S$-wave using the $K$-matrix parameterisation [76]. This removes the need for unphysical resonances to be included in the model and provides a description of broad overlapping resonances that does not violate unitarity. The $K$-matrix parameterisation has been studied [77] for the model-dependent measurement of $\gamma$ at LHCb [68].

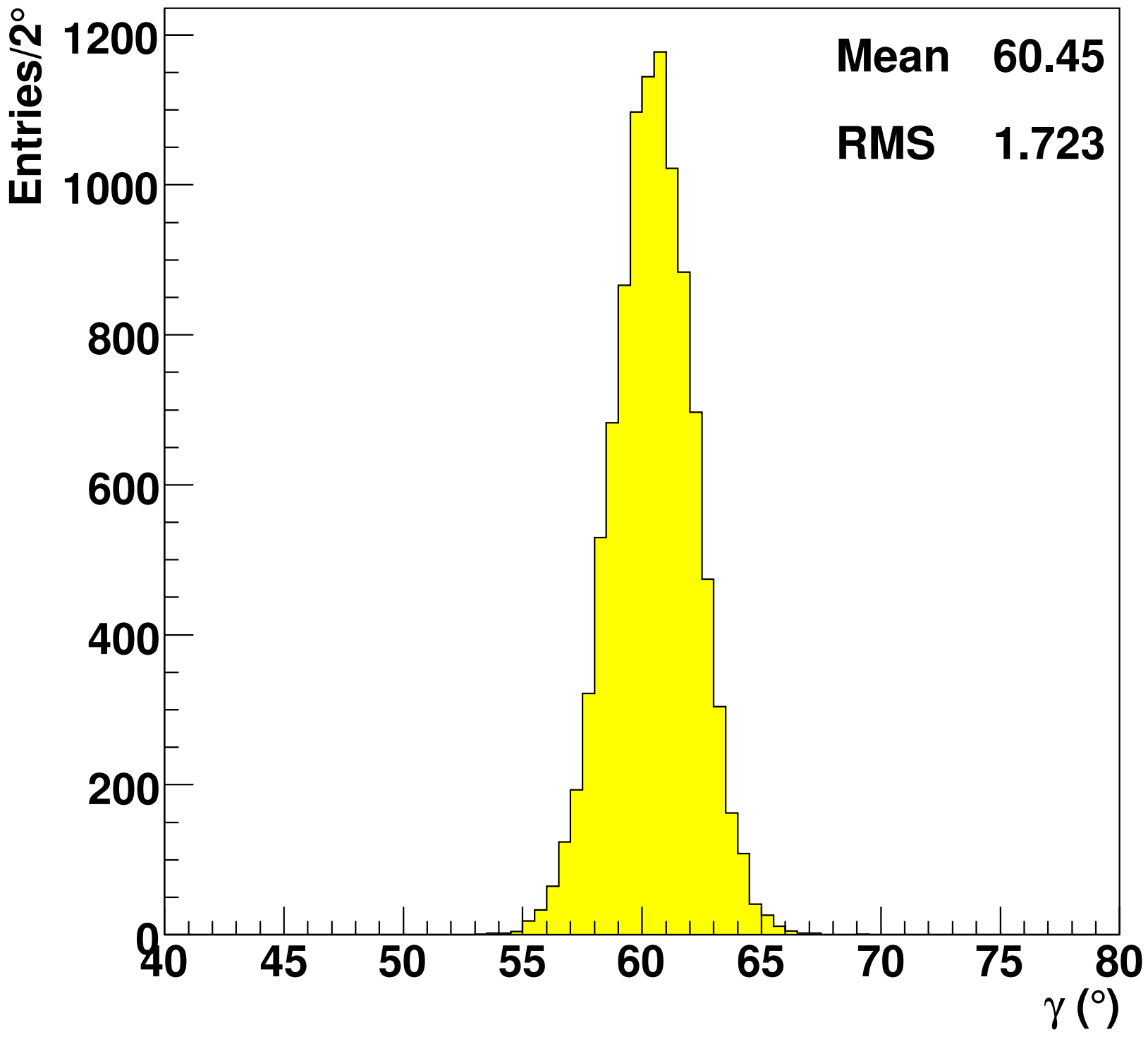

Figure 12: Distribution of $\gamma$ fit for 10,000 toy experiments where the values of strong-phase parameters measured by CLEO-c [74] have been varied by their experimental uncertainties.

#### 8.1.5 Non-resonant $B^0 \to DK^+\pi^-$

The ADS/GLW measurements using $\bar{B}^0 \to D\bar{K}^{*0}(K^-\pi^+)$ need to account for the presence of other $\bar{B}^0 \to DK^-\pi^+$ decays in the signal region. It has been shown that their presence will dilute the sensitivity to $\gamma$ by reducing the effective value of $r_{B^0}$ by a factor $\kappa$ [78]. The value of $\kappa$ has been estimated to be $0.95 \pm 0.03$ [79]. The uncertainty on $\kappa$ introduces a sub-degree uncertainty on $\gamma$ from the ADS/GLW fit to $\bar{B} \to D\bar{K}^{*0}$.

The need to correct for this dilution factor can be eliminated by fitting the whole $\bar{B}^0 \to DK^-\pi^+$ Dalitz space [80]. Furthermore, this technique exploits the extra sensitivity that is available from the additional events. This method is under investigation for use at LHCb.

### 8.1.6 $x_{D^{(*)}}$ from $e^+e^-$ $B$ factories and theory

The sensitivity to $\gamma$ in the decays $B^0 \to D^{\pm(*)}\pi^{\mp}$ depends on the expected interference between the tree-level diagrams for a $B^0$ or $\bar{B}^0$ meson to decay into the same final state, parameterized by $x_{D^{(*)}\pi}$. It is therefore important to determine the likely size of $x_{D^{(*)}\pi}$.

A recent BABAR analysis [31] of the decays $B^0 \to D^{(*)\mp}_{(s)}\pi^{\pm}$ has estimated $x_{D^{(*)}\pi}$ from the factorisation-based relation

$$x_{D^{(*)}\pi} = \tan(\theta_C)\sqrt{\frac{\mathcal{B}\left(B^0 \to D_s^{(*)+}\pi^-\right)}{\mathcal{B}\left(B^0 \to D^{(*)-}\pi^+\right)}}\frac{f_D}{f_{D_s}}, \tag{9}$$

where $\theta_C$ is the Cabibbo angle, and the term $\frac{f_D}{f_{D_s}}$ is the ratio of decay constants for the $D$ and $D_s$. The result is

$$x_{D\pi} = [1.78^{+0.14}_{-0.13}\,(\mathrm{stat}) \pm 0.08\,(\mathrm{syst}) \pm 0.10\,(\mathrm{th})]\%, \tag{10}$$

where the quoted theoretical uncertainty ($\approx 6\%$) comes from the estimate of $\frac{f_D}{f_{D_s}}$ from lattice QCD. The value of $\frac{f_D}{f_{D_s}}$ has been measured to similar precision [81] and could be used to determine $x_{D^{(*)}\pi}$. A similar estimate exists for $x_{D^*\pi}$

$$x_{D^*\pi} = [1.81^{+0.16}_{-0.15}\,(\mathrm{stat}) \pm 0.09\,(\mathrm{syst}) \pm 0.10\,(\mathrm{th})]\%. \tag{11}$$

The quoted uncertainty does not account for the entire theoretical uncertainty on $x_{D^{(*)}\pi}$, as SU(3) breaking effects are believed to affect $x_{D^{(*)}\pi}$ by up to 15%. As discussed in Ref. [82], the three major sources of theoretical uncertainty on $x_{D^{(*)}\pi}$ are rescattering diagrams, non-factorizable effects and W-exchange amplitudes. The contribution from rescattering diagrams can be parameterized as a multiplicative correction factor $R_i$,

$$x_{D^{(*)}\pi} = \tan(\theta_C)\sqrt{\frac{\mathcal{B}\left(B^0 \to D_s^{(*)+}\pi^-\right)}{\mathcal{B}\left(B^0 \to D^{(*)-}\pi^+\right)}}\frac{f_D}{f_{D_s}}R_i, \tag{12}$$

and the size of this factor can be calculated by fitting to the strong-interaction rescattering matrix. The uncertainty on this fit can be estimated by comparing the rescattered branching ratios for a variety of modes with the measured branching ratios. The theoretical uncertainty from the rescattering correction is 1%, and therefore negligible. The uncertainties due to non-factorizable effects and $W$-exchange amplitudes are estimated at 9% (Gaussian) and 5% (flat), respectively.

For the purpose of fitting to $\gamma$, it is necessary to calculate an overall uncertainty on $x_{D^{(*)}\pi}$ from the individual statistical, systematic and theoretical errors. The approach adopted is to add these uncertainties in quadrature, resulting in a total error of $\sigma_{x_{D^{(*)}\pi}} \approx 20\%$. This error can, however, be expected to improve in the future. The statistical uncertainties on the branching ratios will improve as more results are presented by the $B$-factories and LHCb, while the theoretical uncertainties may be better understood by repeating the analysis of [82] with more recent experimental inputs.

## 8.2 Particle identification

Understanding the particle identification performance of the RICH system is extremely important for determining the relative efficiency amongst the various ADS/GLW modes and for separating the topologically identical control samples described in Sec. 6. The PID performance will be calibrated from samples of $D^{*+} \to D^0(K^-\pi^+)\pi^+$ events which can be identified from kinematics alone, given that the mass difference between the $D^{*+}$ and $D^0$ is very close to the kinematic threshold. The determination of PID efficiencies and misidentification rates from these data samples as a function of $p$ and $p_T$ will be at the percent level with only a small sample of integrated luminosity, which is more than adequate for the ADS and GLW rate determinations from a few hundred signal events.

The PID calibration is even more crucial for separating $B^0_{(s)} \to h^+h'^-$ decays from one another [5] where the $\Delta LL$ distributions are parameterised as probability density functions to be used in maximum likelihood fits to determine signal yields and time-dependent $CP$-violating parameters. Such a procedure can be considered for the ADS/GLW rate determinations and exploitation of $B^- \to D^0\pi^-$ control samples if selection requirements on $\Delta LL$ are not adequate. Details of the PID calibration procedure can be found in Ref. [5] and the references therein.

## 8.3 Production and detection asymmetries

Any production and decay asymmetries must be understood when determining the direct $CP$ asymmetries that are sensitive to $\gamma$. Various methods have been proposed at LHCb and they fall in two categories: determinations of the combined production and detection asymmetry and determinations of the detection asymmetry alone. These are discussed in turn below.

### 8.3.1 Measurements of combined production and detection asymmetries

The $B^0/\overline{B}^0$ production asymmetry times $K^{*0}/\bar{K}^{*0}$ detection asymmetry can be measured using self-tagged $B^0 \to J/\psi K^{*0}(K^+\pi^-)$ decays. The asymmetry between these $B^0$ and $\overline{B}^0$ decay rates is expected to be negligible within the Standard Model.[15] So a potential difference in the measured decay rate at LHCb can be attributed to a $B^0/\overline{B}^0$ production or a $K^{*0}/\bar{K}^{*0}$ detection asymmetry effect. A large and clean sample of these decays will be available with the first LHCb data, given the large branching fraction of $(1.33\pm0.06)\times10^{-3}$ [64], which is three orders of magnitude larger than that for the signal. Furthermore, the presence of $J/\psi \to \mu^+\mu^-$ in the decay chain means it is triggered efficiently. Lifetime-unbiased selection studies [84] have shown that $650\times10^3$ events can be reconstructed and Level-0 triggered in a data sample corresponding to 2 fb$^{-1}$ of integrated luminosity; the $B/S$ is estimated to be 1.5 for candidates reconstructed with long lifetimes. Hence, the size and purity of this sample will be adequate to measure the asymmetries as a function of a few kinematic variables, which they are expected to depend upon, such as $p$, $p_T$, and whose distributions may be different for the selected signal and control samples.

[15] Some models of beyond-the-Standard-Model physics predict asymmetries of the $\mathcal{O}(1\%)$ [83].

In a similar fashion the $B^+/B^-$ production asymmetry times the $K^+/K^-$ detection asymmetry can be measured with $B^+ \to J/\psi K^+$. The lifetime-unbiased selection studies estimate [84] that $1.25 \times 10^6$ events can be reconstructed and Level-0 triggered in a data sample corresponding to 2 fb$^{-1}$ of integrated luminosity; the $B/S$ is estimated to be 0.3 for candidates reconstructed with large lifetimes.

### 8.3.2 Measurements of the detection asymmetry alone

It is desirable to have a determination of any detection asymmetry independent of the production asymmetry. It has been proposed that the $B \to D^{*+} X \mu\nu$, followed by $D^{*+} \to D^0(K^-\pi^+\pi^+\pi^-)\pi^+$, decays will be suitable for this because they are topologically and kinematically overconstrained such that one of the kaons or pions from the $D^0$ does not need to be found to reconstruct the event. The efficiency for kaons and pions of opposite charge can then be determined from the frequency with which the unused particle is reconstructed. The analysis is at an early stage and results are not ready to be included in this document; however, initial results are promising.

## 8.4 Background determinations

Given the restricted size of the simulation samples available, limited study has been made of the required sidebands for determining the background parameterisation apart from those reported in Sec. 6.3. Once the sidebands and control samples have been collected the effect of differing parameterisations of the background can be assessed. It is unlikely that this will be a significant source of uncertainty for most of the channels considered.

## 8.5 Systematic uncertainties related to the time-dependent measurements

No dedicated studies of the systematic uncertainties related to measurements of the time-dependent $CP$-violating parameters in $B^0_{(s)} \to D^{\pm}_{(s)} h^{\mp}$ have been performed. However, the more complicated time-dependent $CP$-violation measurements of $B^0_s \to J/\psi\phi$ have been studied in detail [26]. The principal additional systematic uncertainties for a time-dependent measurement are related to:

- proper-time acceptance,
- proper-time resolution and
- tagging.

For time-dependent asymmetry measurements any effects due to proper-time acceptance cancel to first-order. Any differences in acceptance can be bounded by the study of the control channel $B_s^0 \to D_s^- \pi^+$ which exhibits negligible $CP$-violation. The proper-time resolution can be parameterised and determined from a simultaneous fit to the $B_s^0 \to D_s^\pm K^\mp$ and $B_s^0 \to D^- \pi^+$ samples as is done for $B_s^0 \to J/\psi\phi$ [26].

The calibration of the flavour tagging is described in Ref. [85]. Once the offline selected yields, including the HLT, have been evaluated, the resulting systematic uncertainty on $\gamma$ can be determined by applying these calibrations to the fit $B_s \to D^\pm K^\mp$. It has been shown that the mistag rate for $B^0 \to D^\pm \pi^\mp$ can be determined directly from data [33].

# 9 Summary of steps before first measurements

There are several preliminary steps that need to be accomplished before the analyses presented in this document can be performed. This section gives a qualitative account of these. The qualitative nature is due to two factors. Firstly, the initial data-taking strategy has yet to be defined exactly. Secondly, simulation studies are ongoing of the detector and machine configuration in the first few months of operation.

Once the first time and spatial alignment of the sub-detectors has been performed, a large sample of minimum-bias events, and possibly a sample biased to contain tracks with large $p_T$, will be collected. From this sample one of the principal goals relevant to the $\gamma$ from trees analyses will be the reconstruction of long-lived neutral particles decaying to two charged tracks such as $K_S^0$; this will exercise the downstream tracking essential for the Dalitz measurements. Also, this sample will contain copious $D^0$, $D^\pm$ and $D_s^\pm$ mesons, and exclusive reconstruction of the final states of interest to the $\gamma$ analyses such $D^0 \to K^- \pi^+$ and $D_s^+ \to K^+ K^- \pi^+$ will be possible. These measurements will allow mass and proper-time resolutions to be evaluated as well as the combinatoric background to $D$ meson reconstruction. Studies such as these were made with the upgraded CDF detector at the beginning of Run II of the Tevatron (see for example Ref. [86]).

As the data samples are increased in size, exclusive reconstruction of $B$ final states will become possible. In particular, the high statistics $B^0 \to D^\pm \pi^\mp$ decay will be reconstructed. This mode will allow the mass resolution and combinatorial background levels to be assessed. Furthermore, this channel will be extremely important in evaluating the HLT and stripping performance as the instantaneous luminosity increases, which will require the implementation of more complex algorithms. The important control channel $B^- \to D^0 \pi^-$ will also be reconstructed with early data, as a first step toward the ADS, GLW and Dalitz analyses. Mass resolutions, backgrounds and any coarse estimates of charge asymmetries can be evaluated.

The most likely first measurement, amongst those presented here, is the two-body ADS analysis, where there is limited evidence for the suppressed modes [40,41]; the observation of these decays would be a significant first step toward the programme outlined in this document.

# 10 Conclusion and outlook

The modes that will be used to make the first measurements of $\gamma$ at tree-level with LHCb have been discussed. In particular, the selection efficiency and background studies have been updated to the latest simulated samples. Overall the offline performance is similar or slightly improved over previous studies. At this time the incomplete nature of the trigger studies for fully-hadronic modes means that the final yields are unknown and sensitivity studies have not been fully updated. Assuming a HLT efficiency of 65% for all the modes presented results in the sensitivity similar to previous studies [?] of around $10^\circ$ with a data set corresponding to an integrated luminosity of 0.5 $\text{fb}^{-1}$. The first work on control samples and systematic uncertainties has been summarised. Also, the significant advancements in the knowledge of external parameters measured at CLEO-c and the $B$-factories has been discussed. The most significant results reported in this note are summarised in Table 12.

Even if the HLT or selection performance is significantly worse than the estimates presented here there are many additional modes that are sensitive to $\gamma$ that are being studied and will improve the overall sensitivity. Some of these modes are listed below:

- It has been shown [87] that $B^- \to D^* K^-$ with $D^* \to D\gamma$ or $D^* \to D\pi^0$ has great sensitivity to $\gamma$ as long as the two $D^*$ decay modes can be distinguished. A first study [88] indicates that reconstruction of these final states is feasible with promising yields; however, investigations of the level of background are inconclusive given the restricted background simulation statistics.

- The channel $D \to K^+\pi^-\pi^0$ is a promising multi-body ADS mode because of its large branching fraction of $(13.9 \pm 0.5)\%$ [64] and its large coherence factor. CLEO-c has recently measured $R_{K\pi\pi^0} = 0.84 \pm 0.07$ [20].

- Dalitz analysis of the channel $D \to K^0_S K^+ K^-$ can be performed. (Such a measurement has been made by BABAR [23].) Although the branching fraction is only one fifth of that for $D \to K^0_S \pi^+\pi^-$ the background level is likely to be lower and the uncertainties related to the Dalitz models are uncorrelated.

- Sensitivity to $\gamma$ in the channel $B^- \to D(K^+K^-\pi^+\pi^-)K^-$ can be exploited in a four-body amplitude analysis [89].

- The ADS analysis of $B^- \to D(K^\pm\pi^\mp\pi^+\pi^-)K^-$ can be extended by considering separate bins of the $D$-decay phase space, each of which will be more coherent than integrating over the whole phase space [19]. The best sensitivity from this channel may come from a four-body amplitude analysis if a robust model of the DCS decay can be determined.

- The family of decays $B^- \to DK^{*-}(K^0_S\pi^-)$ can be included.

- Other $D$ decay modes can be added to the $\bar{B}^0 \to D\bar{K}^{*0}$ analysis.

- An amplitude analysis of the whole $\bar{B}^0 \to DK^-\pi^+$ Dalitz space can be performed, as discussed in Sec. 8.1.5.

- Time-integrated measurements of $\bar{B}^0_{(s)}$ decays, particularly the untagged strategy involving $\bar{B}^0_s \to D\phi$ and $\bar{B}^0 \to DK^0_S$ [90], can be made.

- The statistics in $B^0_s \to D^\mp_s K^\pm$ can be significantly improved by exploiting other $D_s$ decays than $K^+K^-\pi^+$, such as $\pi^+\pi^+\pi^-$, $K^+\pi^+\pi^-$ and $K^+K^-\pi^+\pi^0$.

- As discussed in Sec. 5.6.3 the mode $B^0 \to D^{*\pm}\pi^\mp$ is known to be feasible at LHCb either via inclusive or exclusive reconstruction. As the strong phase is likely to be different to that in $B^0 \to D^\pm\pi^\mp$, combining these two channels will reduce the number of ambiguous solutions, along with increasing the statistical precision. A $U$-spin combination [32] of $B^0 \to D^{*\pm}\pi^\mp$ and $B^0_s \to D^{*\mp}_s K^\pm$ is also an interesting analysis strategy.

- The $U$-spin pair of channels $B^0 \to D^\pm\rho^\mp$ and $B^0_s \to D^\mp_s K^{*\pm}$ is an experimentally challenging but interesting additional time-dependent measurement of $\gamma$. Decays involving other excited meson states can be investigated, such as $B^0_s \to D^\mp_s K_1(1270)^\pm$.

- A time-dependent Dalitz plot analysis of $B^0 \to D^\mp K^0\pi^\pm$ [91] has recently been reported by BABAR [48], which has the attractive feature that the interference effects accessible are substantially larger than those in $B^0 \to D^{(*)\mp}\pi^\pm$ and $B^0 \to D^{(*)\mp}\rho^\pm$ modes.

Most of the further measurements listed should not be considered as ones that will be performed with first data. However, once the measurements discussed in detail in this note have been executed these many additions will be made and enhance the sensitivity to $\gamma$ from the final LHCb data set.

Table 12: Summary of branching fractions, $\mathcal{B}$, expected yield in data corresponding to 0.5 fb$^{-1}$ of data, estimated $B/S$, external inputs and the anticipated largest systematic uncertainty. The yields do not include the HLT efficiency. All upper limits are at 90% CL.

| Mode | $\mathcal{B}$ ($\times 10^{-6}$) | 0.5 fb$^{-1}$ yield | $B/S$ | External inputs | Largest systematic uncertainty |
|---|---|---|---|---|---|
| $B^- \to D(K^-\pi^+)K^-$ | 16 | $21,000 \pm 1500$ | $0.6 \pm 0.1$ | $\delta_D^{K\pi}$ | Det./Prod. asymmetries |
| $B^- \to D(K^+\pi^-)K^-$ | 0.3 | $400 \pm 25$ | $0.6 \pm 0.3$ | $\delta_D^{K\pi}$ | Det./Prod. asymmetries |
| $B^- \to D(h^+h^-)K^-$ | 2.1 | $2850 \pm 160$ | $1.7 \pm 0.4$ | – | Det./Prod. asymmetries |
| $B^- \to D(K^-\pi^+\pi^+\pi^-)K^-$ | 32 | $13,300 \pm 1000$ | $0.20 \pm 0.06$ | $R_{K3\pi}$ and $\delta_D^{K3\pi}$ | Det./Prod. asymmetries |
| $B^- \to D(K^+\pi^-\pi^-\pi^+)K^-$ | 0.42 | $140^{+70}_{-50}$ | $3.1^{+2.3}_{-2.0}$ | $R_{K3\pi}$ and $\delta_D^{K3\pi}$ | Det./Prod. asymmetries |
| $\bar{B}^0 \to D(K^-\pi^+)\bar{K}^{*0}$ | 1.1 | $800 \pm 130$ | $0.24^{+0.49}_{-0.16}$ | $\delta_D^{K\pi}$ | Det./Prod. asymmetries |
| $\bar{B}^0 \to D(K^+\pi^-)\bar{K}^{*0}$ | 0.1 | $73 \pm 13$ | $9.8^{+5.4}_{-5.1}$ | $\delta_D^{K\pi}$ | Det./Prod. asymmetries |
| $\bar{B}^0 \to D(h^+h^-)\bar{K}^{*0}$ | 0.2 | $68 \pm 10$ | $< 6.1$ | – | Det./Prod. asymmetries |
| $B^- \to D(K_S^0\pi^+\pi^-)K^-$ | 8.3 | $1700 \pm 150$ | $< 1.1$ | Model or CLEO-c strong phases | Model or CLEO-c strong phases |
| $B^0 \to D^\pm\pi^\mp$ | 250 | $(310 \pm 20) \times 10^3$ | $0.17 \pm 0.06$ | $x_d$ | $x_d$ and background model |
| $B^0 \to D_s^\pm K^\mp$ | 19 | $3500 \pm 1200$ | $0.29^{+0.28}_{-0.18}$ | – | Background model and flavour tagging |

# Chapter 3

# Charmless charged two-body B decays

A. Bates, A. Carbone, L. Carson, H. Cliff, D. Galli, M. Gersabeck, V. Gibson, U. Kerzel, U. Marconi, R. Muresan, J. Nardulli, C. Parkes, S. Perazzini, E. Rodrigues, A. Sarti, V. Vagnoni, G. Valenti, S. Vecchi and G. Wilkinson

**Abstract**

The family of $B$ hadron decays into pairs of charmless charged mesons or baryons comprises a rich set of channels, each one characterized by charge or time dependent $\mathcal{CP}$ asymmetries, whose precise measurements play an important role in the quest for New Physics beyond the Standard Model. In particular, New Physics may show up as virtual contributions of new particles inside the loops of the strong and electroweak penguin graphs contributing to the amplitudes of such decays, altering in a subtle but observable way the Standard Model predictions of the $\mathcal{CP}$ asymmetries, provided that a sufficient experimental accuracy is achieved. LHCb has a great potential for triggering, reconstructing and selecting an unprecedented number of such decays, significantly increasing the statistics available today at the $B$ factories and the Tevatron. In this document we summarize the state-of-the-art studies performed at LHCb. The precision on the $\gamma$ angle of the Unitarity Triangle and on the $B_s^0$ mixing phase $\phi_s$, achievable by measuring the $\mathcal{CP}$ asymmetries of these decays with an integrated luminosity $L = 2\,\mathrm{fb}^{-1}$, is estimated to be 7° and 0.06 rad, respectively.

# Contents

# 1 Introduction

The family of charmless $H_b \to h^+h'^-$ decays, where $H_b$ can be either a $B^0$ meson, a $B_s^0$ meson or a $\Lambda_b$ baryon, while $h$ and $h'$ stand for $\pi$, $K$ or $p$, has been extensively studied and is matter of great interest at the $B$ factories and the Tevatron [1, 2, 3, 4, 5, 6, 7]. Such decays are sensitive probes of the Cabibbo-Kobayashi-Maskawa (CKM) [8, 9] matrix and have the potential to reveal the presence of New Physics (NP).

In contrast to the case of other theoretically clean measurements of $\mathcal{CP}$ violation in the $B$ sector, a simple interpretation of the $\mathcal{CP}$ violating observables of the charmless two-body $B$ decays in terms of CKM phases is not possible. For instance, if one considers the $B^0 \to \pi^+\pi^-$ decay, in addition to the $\bar{b} \to \bar{u} + W^+$ tree contribution, sizable $\bar{b} \to \bar{d} + g$ (and possibly $\bar{b} \to \bar{d} + \gamma$, $Z^0$) penguin contributions are expected to play a significant role. Such "penguin pollution" poses several problems for a clean measurement of CKM phases using these decays. On the other hand, the presence of loops inside the penguin diagrams has interesting implications, since they could be affected by sizable contributions from NP.

One promising way to exploit the presence of penguins for these decays, as a powerful resource rather than a limitation, was first suggested ten years ago in Ref. [10] (for the latest update of the analysis see Ref. [11]). In particular, it was shown how the combined measurement of the $B^0 \to \pi^+\pi^-$ and $B_s^0 \to K^+K^-$ time-dependent $\mathcal{CP}$ asymmetries, under the assumption of invariance of the strong interaction dynamics under the exchange of the $d \leftrightarrow s$ quarks (*U-spin* symmetry) in the decay graphs of these modes, provides an interesting way to determine the angle $\gamma$ of the Unitarity Triangle (UT), without the need of any dynamical assumption. Due to the possible presence of NP in the penguin loops, a measurement of $\gamma$ with these decays could differ appreciably from the one determined by using other $B$ decays governed by pure tree amplitudes [12].

The LHCb detector [13] can reconstruct some $10^5$ $H_b \to h^+h'^-$ events per year which can be exploited for $\mathcal{CP}$ violation measurements. These unprecedented sample sizes are a consequence of the large beauty production cross section at the LHC, which is expected to be around 500 $\mu$b, and to the excellent vertexing [14], triggering [15] and particle identification [16] capabilities of LHCb. Within a few months of commencing LHC operation at nominal luminosity, the collected sample will become larger than those available at the $B$ factories and the Tevatron, hence opening new possibilities for finding evidence of subtle inconsistencies between the Standard Model (SM) predictions and the measurements. This document summarizes the state-of-the-art studies performed in this sector at LHCb, following previous documents on the same subject [17, 18, 19].

Despite the large statistics we expect, it must be emphasized that the isolation of such decay modes and their exploitation for $\mathcal{CP}$ violation measurements is significantly more challenging at hadronic machines, with respect to the analyses at the $B$ factories. The much larger track multiplicity arising from the hadronic collisions makes the event selection more difficult due to a correspondingly larger combinatorial background. Moreover, the effective tagging power is estimated to be, from the current Monte Carlo (MC) studies, about five times smaller with respect to the $B$ factories, consequently reducing the effectively available statistics by the same amount. Furthermore, due to the simultaneous production of $B^0$, $B_s^0$ and $\Lambda_b$ hadrons, the mass peaks of many two-body decays overlap, giving rise to a single unresolved signal, if no particle identification information is used. Excellent control of the particle identification observables and an event-by-event fit technique are then required to fully exploit the statistical power of the sample.

The analyses carried out in this sector at the $B$ factories have been able to detect for the first time the presence of direct $\mathcal{CP}$ violation in the $B^0 \to K^+\pi^-$ decay and to measure the time dependent $\mathcal{CP}$ asymmetry in $B^0 \to \pi^+\pi^-$ [1,2,3]. While for the former the measurements of BaBar and Belle are in good agreement, in the latter case only the measurements of the mixing-induced $\mathcal{CP}$ asymmetry coefficient agree well, while those of the direct $\mathcal{CP}$ asymmetry coefficient are just consistent with a significance of 1.9 $\sigma$.

More recently, the CDF experiment at the Tevatron made impressive contributions by providing competitive measurements of the charge asymmetry of the $B^0 \to K^+\pi^-$ decay mode, and more interestingly by observing for the first time the decay modes $B_s^0 \to K^+K^-$, $B_s^0 \to \pi^+K^-$, $\Lambda_b \to p\pi^-$ and $\Lambda_b \to pK^-$ [4,5]. For these last three modes, CDF has also provided first measurements of the charge asymmetries. However, CDF has not yet published any time dependent measurement for the $B^0 \to \pi^+\pi^-$ and $B_s^0 \to K^+K^-$ modes, which may perhaps be attributed to the complexity of a time dependent analysis, requiring two additional ingredients with respect to the measurement of charge asymmetries, i.e. the tagging of the initial flavour of the $B$ mesons and the measurement of proper time of their decays. Nevertheless, the CDF results have successfully demonstrated that the different overlapping mass peaks of the $H_b \to h^+h'^-$ modes for measuring $\mathcal{CP}$ violation in a hadronic environment can be statistically disentangled, even in the case where the particle identification information coming from $dE/dx$ measurements is significantly inferior to that which will be available in the LHCb RICH system.

After a brief discussion of the physics interest in these decays in Sec. 2, we will describe the offline and online event selection algorithms in Sec. 3, estimating the expected signal and background event yields. Then, Sec. 4 will be dedicated to the measurement of the proper time of the decays, and Sec. 5 to the calibration of the RICH particle identification. A first attempt to understand the impact of misalignments of the vertex detector and of the tracking stations to the physics measurements will be described in Sec. 6, while in Sec. 7 we will discuss how to exploit the flavour tagging information and we will show the performance of the tagging algorithm for the decays under study. A further important aspect of the measurement is the control of the line shape of the invariant mass spectrum, and this is discussed in Sec. 8. By using all the ingredients defined in the previous sections, in Sec. 9 we will discuss the LHCb sensitivities of the measurements of $\mathcal{CP}$ violating quantities, while in Sec. 10 these will be translated to the corresponding sensitivity on measurement of the $\gamma$ angle and of the $B_s^0$ mixing phase. Finally, Sec. 11 will summarize the relevant results of the analysis.

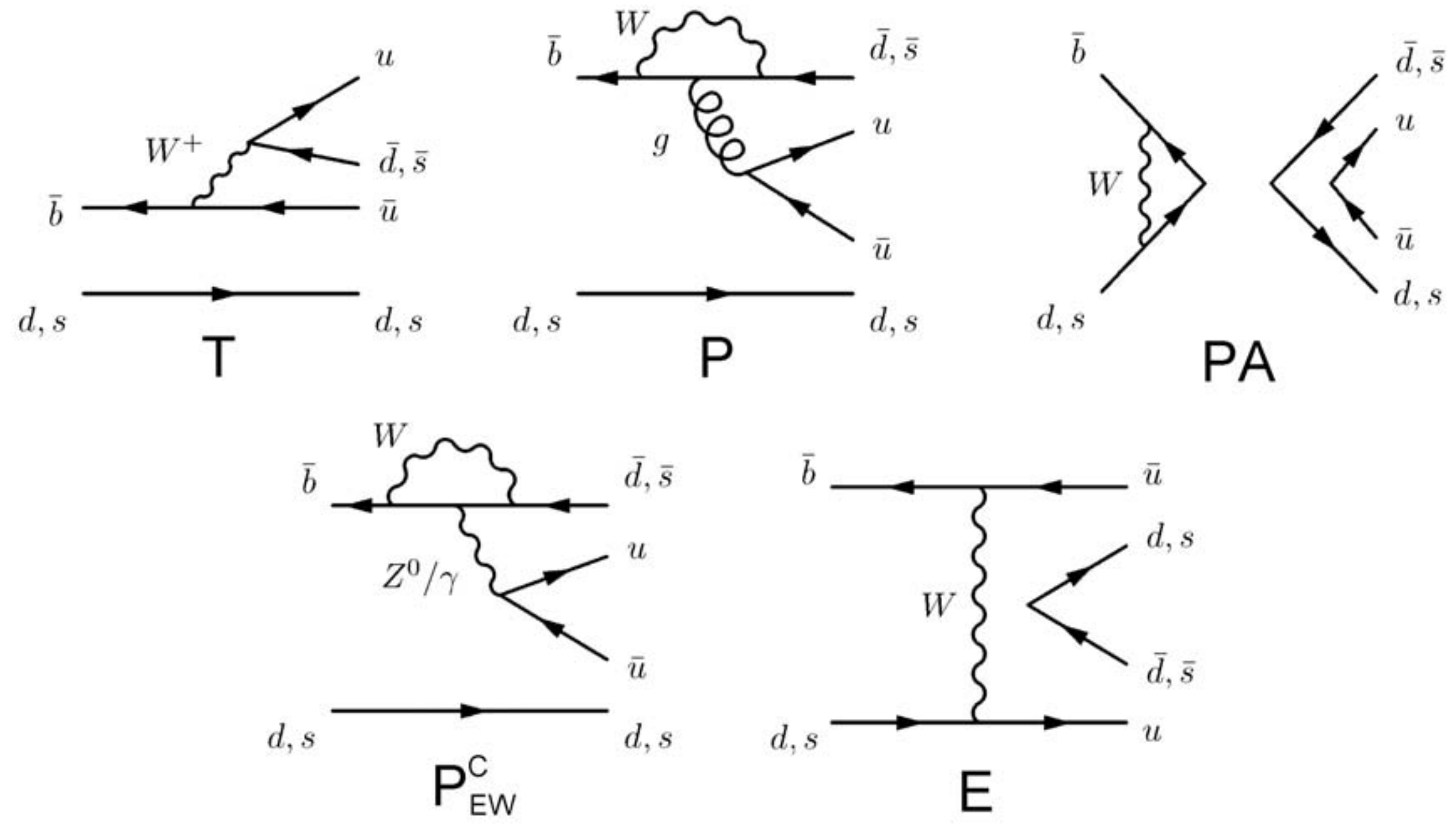

Figure 1: Diagrams contributing to the amplitudes of charmless $B^0_{(s)}$ decays to two charged mesons: Tree ($T$), Penguin ($P$), Penguin Annihilation ($PA$), Colour-suppressed Electroweak Penguin ($P^C_{EW}$) and Exchange ($E$).

# 2 Physics of $H_b \to h^+h'^-$ decays

The family of charmless two-body $B$ decays comprises several modes, providing many different ways for testing the SM picture of $\mathcal{CP}$ violation. In the studies presented in this document, we will take into account nine channels (not counting the $\mathcal{CP}$-conjugate ones): $B^0 \to \pi^+\pi^-$, $B^0 \to K^+\pi^-$, $B^0 \to K^+K^-$, $B^0_s \to K^+K^-$, $B^0_s \to \pi^+K^-$, $B^0_s \to \pi^+\pi^-$, $\Lambda_b \to p\pi^-$, $\Lambda_b \to pK^-$ and $B^0 \to p\bar{p}$. For each of these channels, relevant observables include branching ratios, charge $\mathcal{CP}$ asymmetries and, in the case of neutral $B$ mesons, time dependent $\mathcal{CP}$ asymmetries.

## 2.1 Decay diagrams and U-spin symmetry

Several topologies contribute to the decay amplitudes of the above decays in the SM. For the specific case of $B^0_{(s)}$ decays, all the diagrams are depicted in Fig. 1, and the ones contributing to each decay mode are summarized in Tab. 1.

Notably, the diagrams of the decays $B^0 \to \pi^+\pi^-$ and $B^0_s \to \pi^+K^-$ differ only by the interchange of the spectator quarks, which in the former case is a $d$ while in the latter is an $s$. For this reason, their strong interaction dynamics are connected by the so-called *U-spin* symmetry, i.e. a subgroup of SU(3) analogous to Isospin, but involving $d$ and $s$ quarks instead of $d$ and $u$ quarks. In fact, the $B^0 \to \pi^+\pi^-$ and $B^0_s \to \pi^+K^-$ decays are not fully connected by the U-spin symmetry, since two diagrams contributing to the amplitude of the former channel are not present in the latter mode, namely those referred to as $PA$ and $E$ in Fig. 1 and Tab. 1. Nevertheless, $PA$ and $E$ contributions are expected to be small and, as an example of the flexibility of such family of decays, their size can be probed by means of the $B^0 \to K^+K^-$ and $B_s^0 \to \pi^+\pi^-$ decays which proceed only through $PA$ and $E$ topologies. The situation is entirely analogous for the pair of decays $B_s^0 \to K^+K^-$ and $B^0 \to K^+\pi^-$.

| Decay mode | Contributing diagrams |
|---|---|
| $B^0 \to \pi^+\pi^-$ | $T$, $P$, $PA$, $P_{EW}^C$, $E$ |
| $B^0 \to K^+\pi^-$ | $T$, $P$, $P_{EW}^C$ |
| $B_s^0 \to \pi^+K^-$ | $T$, $P$, $P_{EW}^C$ |
| $B_s^0 \to K^+K^-$ | $T$, $P$, $PA$, $P_{EW}^C$, $E$ |
| $B^0 \to K^+K^-$ | $PA$, $E$ |
| $B_s^0 \to \pi^+\pi^-$ | $PA$, $E$ |

Table 1: Diagrams contributing to amplitude of each charmless $B^0_{(s)}$ decay to two charged mesons. See the caption of Fig. 1 for the definitions.

In contrast, the $B^0 \to \pi^+\pi^-$ and $B_s^0 \to K^+K^-$ decays are fully U-spin symmetric. There is no need of dynamical assumptions that some topologies do not contribute significantly as in the previous cases.

## 2.2 $B^0_{(s)} \to h^+h'^-$ decays

In this section we shall discuss in some detail the specific features of $B^0_{(s)} \to h^+h'^-$ decays, by recalling the main arguments outlined in Ref. [10]. In terms of the diagrammatic contributions reported in Tab. 1, the $B^0 \to \pi^+\pi^-$ instantaneous decay amplitude in the SM can be written as

$$A_{\pi^+\pi^-} = \mathcal{C}\left(e^{i\gamma} - de^{i\vartheta}\right), \tag{1}$$

with

$$\mathcal{C} = A\lambda^3 R_b\left(-T - \mathcal{P}^u + \mathcal{P}^t - E\right) \tag{2}$$

and

$$de^{i\vartheta} = \frac{1}{R_b}\frac{\mathcal{P}^c - \mathcal{P}^t}{T + \mathcal{P}^u - \mathcal{P}^t + E} \quad , \tag{3}$$

where $\lambda$ is the sine of the Cabibbo angle, $\gamma$ is the Unitarity Triangle (UT) angle and the penguin amplitudes $\mathcal{P}^q$ include strong, colour suppressed electroweak and annihilation penguin topologies:

$$\mathcal{P}^q = P^q + P_{EW}^{C\,q} + PA^q. \tag{4}$$

Note that the superscripts $u$, $c$ and $t$ stand for the corresponding quarks circulating inside the loops of the respective diagrams, and that all the CKM couplings have been factorized out explicitly. The factor $R_b$ is one of the sides of the UT, given by:

$$R_b = \frac{1}{\lambda}\left(1-\frac{\lambda^2}{2}\right)\left|\frac{V_{ub}}{V_{cb}}\right|. \tag{5}$$

The $\mathcal{CP}$-conjugate decay amplitude is obtained by conjugating the weak phase $\gamma$, but keeping the strong phases unchanged:

$$\overline{A}_{\pi^+\pi^-} = \eta_{\pi^+\pi^-}\mathcal{C}\left(e^{-i\gamma} - de^{i\vartheta}\right), \tag{6}$$

where $\eta_{\pi^+\pi^-} = +1$ is the $\mathcal{CP}$-parity of the $|\pi^+\pi^-\rangle$ $\mathcal{CP}$ eigenstate.

Analogously, the $B_s^0 \to K^+K^-$ instantaneous decay amplitude can be written as

$$A_{K^+K^-} = \frac{\lambda}{1-\lambda^2/2}\mathcal{C}'\left(e^{i\gamma} + \frac{1-\lambda^2}{\lambda^2}d'e^{i\vartheta'}\right), \tag{7}$$

with

$$\mathcal{C}' = A\lambda^3 R_b\left(-T' - \mathcal{P}'^u + \mathcal{P}'^t - E'\right) \tag{8}$$

and

$$d'e^{i\vartheta'} = \frac{1}{R_b}\frac{\mathcal{P}'^c - \mathcal{P}'^t}{T' + \mathcal{P}'^u - \mathcal{P}'^t + E'}, \tag{9}$$

where we have introduced primed quantities to distinguish them from the analogous ones of the $B^0 \to \pi^+\pi^-$ decay. The $\mathcal{CP}$-conjugate amplitude is then

$$\overline{A}_{K^+K^-} = \eta_{K^+K^-}\frac{\lambda}{1-\lambda^2/2}\mathcal{C}'\left(e^{-i\gamma} + \frac{1-\lambda^2}{\lambda^2}d'e^{i\vartheta'}\right), \tag{10}$$

where $\eta_{K^+K^-} = +1$ is the $\mathcal{CP}$-parity of the $|K^+K^-\rangle$ $\mathcal{CP}$ eigenstate.

By using the definition of the direct and mixing-induced $\mathcal{CP}$ asymmetry coefficients given in App. A, and introducing the mixing phases of the neutral $B$-meson systems, from Eqs. (1) and (6) it is straightforward to obtain [10]

$$\mathcal{A}^{dir}_{\pi^+\pi^-} = \frac{2d\sin(\vartheta)\sin(\gamma)}{1-2d\cos(\vartheta)\cos(\gamma)+d^2} \tag{11}$$

and

$$\mathcal{A}^{mix}_{\pi^+\pi^-} = -\frac{\sin(\phi_d+2\gamma) - 2d\cos(\vartheta)\sin(\phi_d+\gamma) + d^2\sin(\phi_d)}{1-2d\cos(\vartheta)\cos(\gamma)+d^2}, \tag{12}$$

where $\phi_d$ is the mixing phase of the $B^0$ meson. Analogously, from Eqs. (7) and (10) it follows that

$$\mathcal{A}^{dir}_{K^+K^-} = -\frac{2\tilde{d}'\sin(\vartheta')\sin(\gamma)}{1+2\tilde{d}'\cos(\vartheta')\cos(\gamma)+\tilde{d}'^2} \tag{13}$$

and

$$\mathcal{A}^{mix}_{K^+K^-} = -\frac{\sin(\phi_s + 2\gamma) + 2\tilde{d}' \cos(\vartheta') \sin(\phi_s + \gamma) + \tilde{d}'^2 \sin(\phi_s)}{1 + 2\tilde{d}' \cos(\vartheta') \cos(\gamma) + \tilde{d}'^2}, \tag{14}$$

where $\phi_s$ is the mixing phase of the $B^0_s$ meson and the parameter $\tilde{d}'$ is defined as

$$\tilde{d}' = \frac{1-\lambda^2}{\lambda^2} d'. \tag{15}$$

Furthermore, using the definition of $\mathcal{A}^{\Delta}_f$ again given in App. A, we have

$$\mathcal{A}^{\Delta}_{K^+K^-} = \frac{\cos(\phi_s + 2\gamma) + 2\tilde{d}' \cos(\vartheta') \cos(\phi_s + \gamma) + \tilde{d}'^2 \cos(\phi_s)}{1 + 2\tilde{d}' \cos(\vartheta') \cos(\gamma) + \tilde{d}'^2}. \tag{16}$$

It must be emphasized that Eq. (16) is not independent of Eqs. (13) and (14), due to the existence of the relation

$$\left(\mathcal{A}^{dir}_{K^+K^-}\right)^2 + \left(\mathcal{A}^{mix}_{K^+K^-}\right)^2 + \left(\mathcal{A}^{\Delta}_{K^+K^-}\right)^2 = 1. \tag{17}$$

It is of course possible to parameterize in the same way the complex terms $\lambda_{\pi^+\pi^-}$ and $\lambda_{K^+K^-}$, as in the following:

$$\mathrm{Re}\lambda_{\pi^+\pi^-} = \frac{\cos(\phi_d + 2\gamma) - 2d \cos(\vartheta) \cos(\phi_d + \gamma) + d^2 \cos(\phi_d)}{1 - 2d \cos(\gamma - \vartheta) + d^2}, \tag{18}$$

$$\mathrm{Im}\lambda_{\pi^+\pi^-} = -\frac{\sin(\phi_d + 2\gamma) - 2d \cos(\vartheta) \sin(\phi_d + \gamma) + d^2 \sin(\phi_d)}{1 - 2d \cos(\gamma - \vartheta) + d^2}, \tag{19}$$

$$\mathrm{Re}\lambda_{K^+K^-} = \frac{\cos(\phi_s + 2\gamma) + 2\tilde{d}' \cos(\vartheta') \cos(\phi_s + \gamma) + \tilde{d}'^2 \cos(\phi_s)}{1 + 2\tilde{d}' \cos(\gamma - \vartheta') + \tilde{d}'^2} \tag{20}$$

and

$$\mathrm{Im}\lambda_{K^+K^-} = -\frac{\sin(\phi_s + 2\gamma) + 2\tilde{d}' \cos(\vartheta') \sin(\phi_s + \gamma) + \tilde{d}'^2 \sin(\phi_s)}{1 + 2\tilde{d}' \cos(\gamma - \vartheta') + \tilde{d}'^2}. \tag{21}$$

By measuring the values of $\mathcal{A}^{dir}_{\pi^+\pi^-}$, $\mathcal{A}^{mix}_{\pi^+\pi^-}$, $\mathcal{A}^{dir}_{K^+K^-}$, $\mathcal{A}^{mix}_{K^+K^-}$ (or alternatively of $\mathrm{Re}\lambda_{\pi^+\pi^-}$, $\mathrm{Im}\lambda_{\pi^+\pi^-}$, $\mathrm{Re}\lambda_{K^+K^-}$ and $\mathrm{Im}\lambda_{K^+K^-}$) and taking the values of the two mixing phases $\phi_d$ and $\phi_s$ from measurements performed in other decays [21, 22], Eqs. (11), (12), (13) and (14) provide a system of four equations with five unknowns: $d$, $\vartheta$, $d'$, $\vartheta'$ and $\gamma$. In order to solve it one thus needs additional constraints.

It is at this point that the U-spin symmetry enters the scene. Within factorization, it can be demonstrated that the relations $d = d'$ and $\vartheta = \vartheta'$ do not receive U-spin breaking corrections [10].

A simple check of the validity of U-spin symmetry, concerning the exchange of the spectator quarks only, consists in comparing the direct $\mathcal{CP}$ asymmetries of the U-spin related decays $B^0_s \to K^+K^-$ and $B^0 \to K^+\pi^-$ (or $B^0 \to \pi^+\pi^-$ and $B^0_s \to \pi^+K^-$), although this must be done under the dynamical assumption that the contribution of penguin annihilation and exchange topologies are small enough, as already remarked. To this end, the measurement of the branching fractions of the $B^0 \to K^+K^-$ and $B_s^0 \to \pi^+\pi^-$ plays an important role. If measurements indicate that these $\mathcal{CP}$ asymmetries are very close to each other, i.e.[1]:

$$\mathcal{A}^{\mathcal{CP}}_{K^+\pi^-} \approx \mathcal{A}^{dir}_{K^+K^-} \tag{22}$$

and

$$\mathcal{A}^{\mathcal{CP}}_{\pi^+K^-} \approx \mathcal{A}^{dir}_{\pi^+\pi^-}, \tag{23}$$

it will be an important indication, even if not yet conclusive, that U-spin can be a good symmetry in the $B^0_{(s)} \to h^+h'^-$ sector.

Although the parameterization outlined in this section has been obtained assuming the validity of the SM, it must be emphasized that it remains valid also in extensions of the SM where NP affects the value of the neutral $B$ meson mixing phases. In particular, in case NP affects $b \leftrightarrow s$ $|\Delta F| = 2$ transitions, $\phi_s$ can be written as:

$$\phi_s = \phi_s^{SM} + \phi_s^{\Delta}, \tag{24}$$

where $\phi_s^{SM}$ is the SM value of the $B_s^0$ mixing phase while $\phi_s^{\Delta}$ is a correction due to NP.

## 2.3 $\Lambda_b \to ph'^-$ decays

Although $\Lambda_b$ decays to a proton and a charged pion or kaon have not yet received significant attention from a theoretical point of view, $\mathcal{CP}$ violation with these decays will be studied with high precision at LHCb. The author of Ref. [23] claims that the measurement of the $\mathcal{CP}$ charge asymmetry in the $\Lambda_b \to p\pi^-$ decay can be sensitive to NP effects in the Minimal Supersymmetric Standard Model (MSSM) with R-parity violation. While the SM predicts a charge asymmetry $\mathcal{A}^{\mathcal{CP}} \simeq 8\%$ and a branching fraction $\mathcal{BR} \simeq 10^{-6}$, in the R-parity violating model the charge asymmetry is predicted to be negligibly small, while the branching fraction as large as $1.6 \cdot 10^{-4}$. In other words, the presence of R-parity and lepton number violating couplings could significantly modify the SM predictions of the branching ratio and of the $\mathcal{CP}$ charge asymmetry, by enhancing the former and suppressing the latter. Although the recent measurements by CDF (see Tab. 5) exclude the possibility of a large branching ratio at the level of $10^{-4}$, a precise measurement of the charge asymmetry has not yet been made.

## 2.4 $B^0 \to p\bar{p}$ decay

Decays of $B$ mesons to baryons are difficult to describe theoretically, and the SM predictions for their branching ratios depend largely on the adopted theoretical framework. For the specific case of $B^0 \to p\bar{p}$, a series of calculations [24, 25, 26] in the early 1990's consistently predicted a branching ratio of $10^{-6}$ or higher. A more recent calculation, which uses the MIT bag model [27] to calculate the hadronic matrix elements, predicts a branching ratio of $1.1 \times 10^{-7}$, just at the edge of current experimental accessibility [6, 7]. While older predictions have been excluded experimentally, it will be interesting to see whether the more recent prediction will also be excluded if the experimental upper limit decreases with the latest $B$ factory datasets.

[1] In the remainder of this document, the charge asymmetries of the $B^0 \to K^+\pi^-$ and $B_s^0 \to \pi^+K^-$ decay modes will be denoted as $\mathcal{A}^{\mathcal{CP}}_{K^+\pi^-}$ and $\mathcal{A}^{\mathcal{CP}}_{\pi^+K^-}$ respectively.

| | $\mathcal{A}^{dir}_{\pi^+\pi^-}$ | $\mathcal{A}^{mix}_{\pi^+\pi^-}$ | $\mathcal{A}^{\mathcal{CP}}_{K^+\pi^-}$ |
|---|---|---|---|
| BaBar | $0.25 \pm 0.08 \pm 0.02$ | $-0.68 \pm 0.10 \pm 0.03$ | $-0.107 \pm 0.016^{+0.006}_{-0.004}$ |
| Belle | $0.55 \pm 0.08 \pm 0.05$ | $-0.61 \pm 0.10 \pm 0.04$ | $-0.094 \pm 0.018 \pm 0.008$ |
| CLEO | - | - | $-0.04 \pm 0.16 \pm 0.02$ |
| CDF | - | - | $-0.086 \pm 0.023 \pm 0.009$ |
| Average | $0.38 \pm 0.06$ | $-0.65 \pm 0.07$ | $-0.098^{+0.012}_{-0.011}$ |

Table 2: Current knowledge of $\mathcal{CP}$ violating parameters in the $B^0$ sector [1,2,3,4,28,29,30].

| | $\mathcal{BR}(B^0 \to \pi^+\pi^-)$ $\times 10^{-6}$ | $\mathcal{BR}(B^0 \to K^+\pi^-)$ $\times 10^{-6}$ | $\mathcal{BR}(B^0 \to K^+K^-)$ $\times 10^{-6}$ |
|---|---|---|---|
| BaBar | $5.5 \pm 0.4 \pm 0.3$ | $19.1 \pm 0.6 \pm 0.6$ | $0.04 \pm 0.15 \pm 0.08$ |
| Belle | $5.1 \pm 0.2 \pm 0.2$ | $19.9 \pm 0.4 \pm 0.8$ | $0.09^{+0.18}_{-0.13} \pm 0.01$ |
| CLEO | $4.5^{+1.4+0.5}_{-1.2-0.4}$ | $18.0^{+2.3+1.2}_{-2.1-0.9}$ | - |
| CDF | $5.10 \pm 0.33 \pm 0.36\,^*$ | - | $0.39 \pm 0.16 \pm 0.12\,^*$ |
| Average | $5.16 \pm 0.22$ | $19.4 \pm 0.6$ | $0.15^{+0.11}_{-0.10}$ |

Table 3: Measurements of the branching fractions of the three $B^0 \to h^+h'^-$ decay modes [2, 4, 31, 32, 33]. *The measurements of CDF are relative to the average branching fraction of the $B^0 \to K^+\pi^-$ decay measured by BaBar, Belle and CLEO.

Considering the inconsistency between the different theoretical predictions for the $B^0 \to p\bar{p}$ branching ratio and the latest experimental data, a search for this rare mode at LHCb would help to clarify the correct theoretical description of baryonic $B$ decays while providing the first observation of a charmless two-body baryonic $B$ decay.

The related decay $B^0_s \to p\bar{p}$ is expected to be significantly suppressed relative to $B^0 \to p\bar{p}$, hence we do not consider it for the moment.

## 2.5 Experimental status

The status of the $\mathcal{CP}$ asymmetry measurements in the $B^0$ sector is summarized in Tab. 2. It can be seen that the mixing-induced $\mathcal{CP}$ asymmetry terms of the $B^0 \to \pi^+\pi^-$ decay measured by BaBar [1] and Belle [3] agree well, while there exists tension at the level of 1.9 $\sigma$ for the direct $\mathcal{CP}$ terms, also apparent in Fig. 2. A further and more precise measurement will be needed in order to draw a definitive conclusion. The status of the $\mathcal{CP}$-averaged branching fraction measurements in the $B^0$ sector is reported in Tab. 3.

The topic of having NP in the $B \to K\pi$ sector has recently received considerable attention. For example, the value of the difference $\Delta\mathcal{A}^{\mathcal{CP}} = \mathcal{A}^{\mathcal{CP}}_{K^+\pi^0} - \mathcal{A}^{\mathcal{CP}}_{K^+\pi^-}$ between the direct $\mathcal{CP}$ asymmetries of the decay modes $B^+ \to K^+\pi^0$ and $B^0 \to K^+\pi^-$, expected to vanish in the SM, has been measured by the Belle Collaboration as $\Delta\mathcal{A}^{\mathcal{CP}} = 0.164 \pm 0.037$ [2]. Such a discrepancy has been argued to be a hint of NP, although alternative explanations within the SM have also been considered [34].

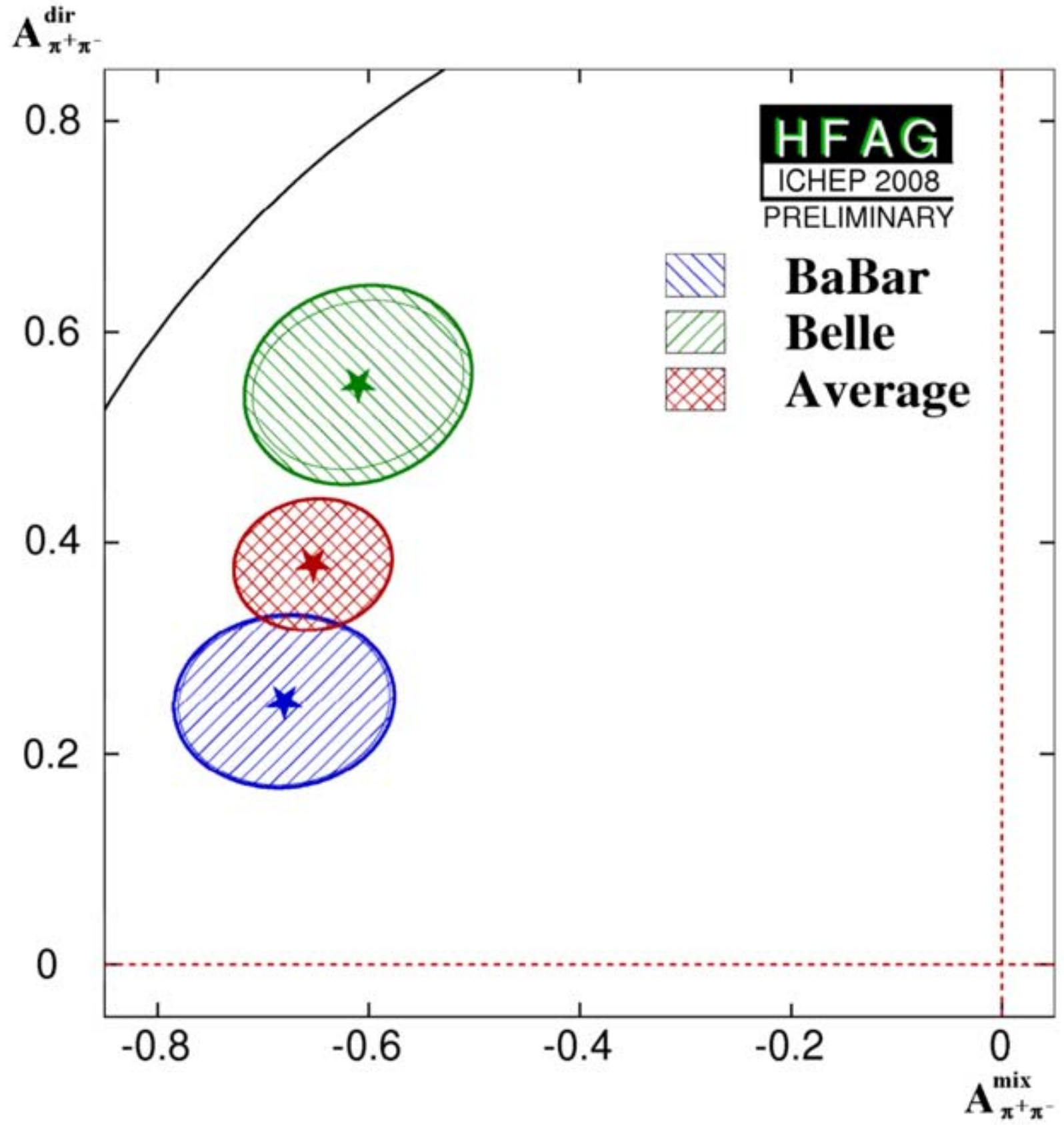

Figure 2: Representation of the measurements of the direct and mixing-induced $\mathcal{CP}$ violating coefficients for the $B^0 \to \pi^+\pi^-$ decay, performed by BaBar [1] and Belle [3]. The contours in the $\left(A^{dir}_{\pi^+\pi^-},\ A^{mix}_{\pi^+\pi^-}\right)$ plane correspond to $-2\Delta \log \mathcal{L} = \Delta\chi^2 = 1$, i.e. 60.7% C.L. for 2 degrees of freedom [29]. The curve on the top left of the plot delimits the allowed physical region.

Since the bulk of the luminosity at the $B$ factories has been delivered at the $\Upsilon(4S)$ energy, only CDF has been able to provide measurements of $B^0_s \to h^+h'^-$ decays so far [4]. CDF has also provided some preliminary results for $\Lambda_b$ charmless two-body decays [5]. The present status of the CDF analyses for the $\mathcal{CP}$ charge asymmetries of the modes $B^0_s \to \pi^+K^-$, $\Lambda_b \to p\pi^-$ and $\Lambda_b \to pK^-$ is summarized in Tab. 4. Interestingly, the value of $\mathcal{A}^{CP}_{\pi^+K^-}$ is compatible with that of $\mathcal{A}^{dir}_{\pi^+\pi^-}$, as predicted by U-spin, although the experimental uncertainty is still too large to draw any firm conclusion. CDF would need a statistics about 10 times larger in order to reach a sensitivity similar to the average of the $B$ factory results on $\mathcal{A}^{dir}_{\pi^+\pi^-}$.

| Quantity | Value |
|---|---|
| $\mathcal{A}^{\mathcal{CP}}_{\pi^+K^-}$ | $0.39 \pm 0.15 \pm 0.08$ |
| $\mathcal{A}^{\mathcal{CP}}_{p\pi^-}$ | $0.03 \pm 0.17 \pm 0.05$ |
| $\mathcal{A}^{\mathcal{CP}}_{pK^-}$ | $0.37 \pm 0.17 \pm 0.03$ |

Table 4: . Present knowledge of the $\mathcal{CP}$ charge asymmetries for the modes $B^0_s \to \pi^+K^-$, $\Lambda_b \to p\pi^-$ and $\Lambda_b \to pK^-$, as measured by CDF with an integrated luminosity $L = 1\,\text{fb}^{-1}$ [4, 5].

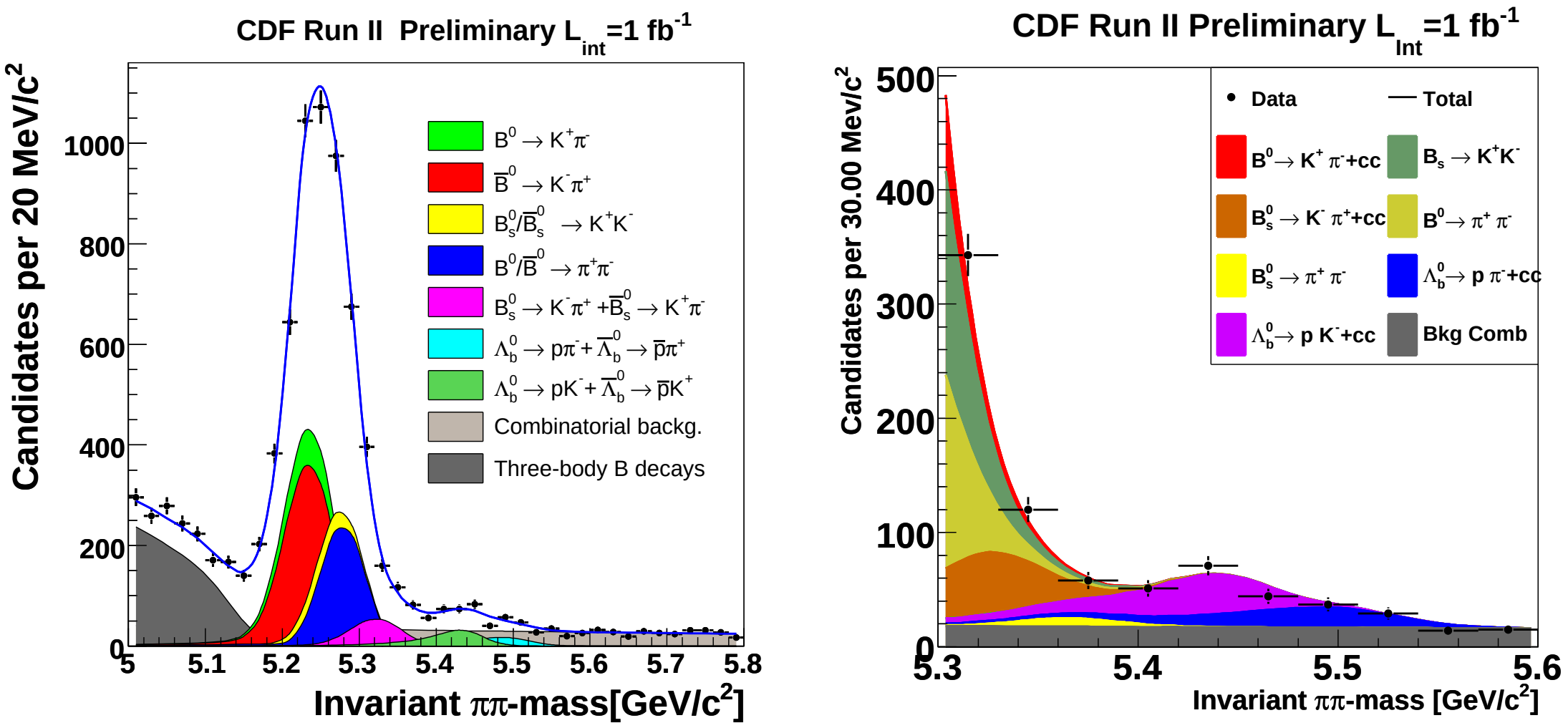

Figure 3: $H_b \to h^+h'^-$ invariant mass spectrum obtained by CDF using the $\pi^+\pi^-$ mass hypothesis for the final state: $B^0_{(s)} \to h^+h'^-$ analysis [4] (left) and $\Lambda_b \to ph^-$ analysis magnified around the $\Lambda_b$ mass region [5] (right). Note that different color conventions are used in the two plots.

Relatively more complex is the situation concerning the CDF measurements of the $B^0_s$ and $\Lambda_b$ branching fractions, which are all measured relative to the $B^0 \to K^+\pi^-$ branching ratio determined by the $B$ factories. All the relevant results are summarized in Tab. 5. Since they will be a very useful reference for the studies reported in the remainder of this document, we also show in Fig. 3 the invariant mass plots of the CDF $H_b \to h^+h'^-$ analyses [4, 5].

Finally, as far as charmless two-body baryonic $B$ decays are concerned, despite the searches at the $B$ factories and at the Tevatron, none of them has yet been observed. The current experimental 90% C.L. upper limit on the $B^0 \to p\bar{p}$ branching ratio is $1.1 \cdot 10^{-7}$, this value being dominated by the latest Belle search [7].

| Quantity | Value |
| --- | --- |
| $\frac{f_s\,\mathcal{BR}(B_s^0 \to K^+K^-)}{f_d\,\mathcal{BR}(B^0 \to K^+\pi^-)}$ | $0.347 \pm 0.020 \pm 0.021$ |
| $\mathcal{BR}(B_s^0 \to K^+K^-) \times 10^6$ | $25.8 \pm 1.5 \pm 3.9\,^*$ |
| $\frac{f_s\,\mathcal{BR}(B_s^0 \to \pi^+K^-)}{f_d\,\mathcal{BR}(B^0 \to K^+\pi^-)}$ | $0.071 \pm 0.010 \pm 0.007$ |
| $\mathcal{BR}(B_s^0 \to \pi^+K^-) \times 10^6$ | $5.27 \pm 0.74 \pm 0.90\,^*$ |
| $\frac{f_s\,\mathcal{BR}(B_s^0 \to \pi^+\pi^-)}{f_d\,\mathcal{BR}(B^0 \to K^+\pi^-)}$ | $0.007 \pm 0.004 \pm 0.005$ |
| $\mathcal{BR}(B_s^0 \to \pi^+\pi^-) \times 10^6$ | $0.52 \pm 0.29 \pm 0.38\,^*$ |
| $\frac{f_{\Lambda_b}\,\mathcal{BR}(\Lambda_b \to p\pi^-)}{f_d\,\mathcal{BR}(B^0 \to K^+\pi^-)}$ | $0.0415 \pm 0.0074 \pm 0.0058$ |
| $\mathcal{BR}(\Lambda_b \to p\pi^-) \times 10^6$ | $3.1 \pm 0.6 \pm 0.7\,^{**}$<br>$1.4 \pm 0.3^{+0.9}_{-0.5}\,^{***}$ |
| $\frac{f_{\Lambda_b}\,\mathcal{BR}(\Lambda_b \to pK^-)}{f_d\,\mathcal{BR}(B^0 \to K^+\pi^-)}$ | $0.0663 \pm 0.0089 \pm 0.0084$ |
| $\mathcal{BR}(\Lambda_b \to pK^-) \times 10^6$ | $5.0 \pm 0.7 \pm 1.0\,^{**}$<br>$2.2 \pm 0.3^{+1.4}_{-0.8}\,^{***}$ |

Table 5: Measurements of the branching fractions of the three $B_s^0 \to h^+h'^-$ and $\Lambda_b \to ph^-$ decay modes performed by CDF with an integrated luminosity $L = 1\,\text{fb}^{-1}$ [4,5]. *Relative measurement to the average $B^0 \to K^+\pi^-$ branching fraction measured by the $B$ factories, using in addition the hadronization fractions $f_s = 10.4 \pm 1.4\%$ and $f_d = 39.8 \pm 1.0\%$ [35]. **Relative measurement to the average $B^0 \to K^+\pi^-$ branching fraction measured by the $B$ factories, with in addition $f_{\Lambda_b}/f_d = 0.25 \pm 0.04$ [35]. ***Same measurement, but with $f_{\Lambda_b}/f_d = 0.56^{+0.28}_{-0.21}$ [36].

# 3 Event selection

In this section we present the latest $H_b \to h^+h'^-$ selection studies that have been performed using simulated LHCb event samples of signal and background events. In the following sub-sections we will first describe the set of MC samples employed for the studies, and then we will discuss structure and performance of the offline and online selection algorithms.

## 3.1 Monte Carlo samples

The online and offline event selections have been studied by means of a full GEANT4 [37] MC simulation of the detector response, implemented in the LHCb simulation program Gauss [38], fed by proton-proton primary collision events simulated by the PYTHIA event generator [39]. Different MC samples were produced during the so-called Data Challenge 2006 [40], and then analyzed offline by means of the LHCb analysis framework DaVinci [41]. Tab. 6 summarizes the event samples used in this analysis. In the same table we also report the current experimental values of the branching ratios for each decay mode of interest, together with the values of the probabilities that $b$ quarks hadronize to the relevant $B$ hadron species. In order to save computing time, MC events were generated with an acceptance cut at the generator level. For signal events, the cut consisted of requiring that the decay products of the $B$ hadrons had a polar angle comprised between 10 mrad and 400 mrad (with the exception of the $B_s^0 \to K^+K^-$ decay, see the caption of Tab. 6), i.e. discarding events with $B$ decay products produced outside the LHCb acceptance. Conversely, for $b\bar{b}$-inclusive events the generator level cut consisted in requiring that *at least* one of the two $B$ hadrons in the event lay inside a forward acceptance cone with a semi-aperture of 400 mrad. The corresponding generator level efficiencies are also reported in Tab. 6.

## 3.2 Offline selection

Assuming an inelastic cross section $\sigma_{inel} = 80\,\text{mb}$ and a beauty production cross section $\sigma_{b\bar{b}} = 500\,\mu\text{b}$ in proton-proton collisions at 14 TeV, and considering that the $H_b \to h^+h'^-$ decay modes of interest have branching fractions in the range $10^{-6}$-$10^{-5}$, very selective criteria must be used to reject the enormous background, in particular from other $B$ hadron decays. Furthermore, it is clear that a comprehensive understanding of the background by means of a full MC study, independently on any particular assumption concerning its nature, would require the simulation of several billions of LHC inelastic events, currently outside the computing capabilities of the experiment.

Both the CDF analysis [4] and the findings from the LHCb simulation studies indicate that there are two categories of background which need to be considered:

- *Physical background*, which is due to the partial reconstruction of three-body $B$ meson decays, like $B^0 \to \rho^{\pm}\pi^{\mp}$, in which only a $\pi^+\pi^-$ pair is reconstructed. Due to the missing $\pi^0$, the $\pi^+\pi^-$ invariant mass distribution is kinematically limited to the region below 5.14 GeV/c$^2$ (i.e. about $m_{B^0} - m_{\pi^0}$), apart from experimental resolution effects which might lead to higher values. For this reason, this background component mainly affects the left tail of the signal mass distribution. Other relevant decay modes belonging to this category are $B \to \rho K$ and $B \to K^*\pi$, as well as decays involving other intermediate resonances.

| Decay mode | MC events $\times 10^3$ | $\mathcal{BR}$ $\times 10^6$ | $f_{hadr}$ [%] | $\epsilon_{gen}$ [%] |
|---|---|---|---|---|
| $B^0 \to \pi^+\pi^-$ | 59.6 | $5.16 \pm 0.22$ | $40.3 \pm 0.9$ | $19.9 \pm 0.2$ |
| $B^0 \to K^+\pi^-$ | 218.2 | $19.4 \pm 0.6$ | $40.3 \pm 0.9$ | $20.1 \pm 0.2$ |
| $B_s^0 \to \pi^+K^-$ | 19.9 | $5.27 \pm 1.17$ | $10.1 \pm 0.9$ | $20.4 \pm 0.2$ |
| $B_s^0 \to K^+K^-$ | 149.6 | $25.8 \pm 4.2$ | $10.1 \pm 0.9$ | $34.5 \pm 0.3$ |
| $\Lambda_b \to p\pi^-$ | 19.9 | $3.1 \pm 0.9$ | $9.2 \pm 1.5$ | $20.8 \pm 0.1$ |
| $\Lambda_b \to pK^-$ | 49.7 | $5.0 \pm 1.2$ | $9.2 \pm 1.5$ | $21.1 \pm 0.2$ |
| $b\bar{b}$-inclusive | 22100 | – | – | $43.7 \pm 0.1$ |

Table 6: Number of analyzed MC events. The table contains also the branching fractions (see Tabs. 3 and 5) and the probabilities for a $b$ quark to form the various $B$ species during the hadronization phase [42], as well as the generator level cut efficiencies for each mode. The generator level cut efficiency corresponding to the $B_s^0 \to K^+K^-$ mode differs significantly from the other ones due to the fact that the cut applied in this case was different. It required that the $B$ hadron itself had to lie inside a forward cone with a semi-aperture of 400 mrad, while for all the other channels the requirement was that the polar angle of the $B$ daughter particles lay between 10 mrad and 400 mrad, hence resulting in a tighter cut. Nevertheless, all the MC samples are equivalent for our purposes.

- *Combinatorial background*, which is due to pairs of oppositely charged tracks not coming from a single $B$ hadron decay. Since, as we will see, the selection requires that the tracks must not originate from any of the primary vertices where the proton-proton collisions took place, one expects that the background tracks should mainly come from the following categories:

  – secondary interactions of primary particles within the detector material (e.g. in the RF foil);

  – decays of long-lived heavy flavoured hadrons, either charm or beauty, where the tracks come from different mother particles;

  – mis-reconstruction of physical tracks;

  – reconstruction of non existing (i.e. ghost) tracks, produced by the random association of detector hits.

  As tracks are randomly associated, this background component is not bound to a particular value of the invariant mass, hence it gives rise to a slowly falling spectrum around the mass region of the $B$ hadrons. The combinatorial component is the dominant source of background under the signal mass peaks as well as the sole source of candidates at masses above 5.6 GeV/c$^2$.

Whilst we could simulate a significant sample of physical background events, in order to study (at least) qualitatively their relevant properties, the generation of many billions of events for a full understanding of the combinatorial background would require at present a prohibitive amount of computing resources. Given this limitation, it is assumed, as in many other LHCb analyses, that the dominant source of combinatorial background comes from heavy flavour decays, in particular of $B$ hadrons. This is of course justified by the fact that the decay products of $B$ hadrons can better mimic the characteristics of particles from $H_b \to h^+h'^-$ signal decays. It is not possible to demonstrate the validity of such assumption on the basis of simple arguments. It may turn out that prompt $D$ hadrons, which have a production cross-section about 7 times larger than that of $B$ events, or even light-quark events, also contribute to the combinatorial background. However, it is expected that any cut which is effective against $B$ hadron decays will have much higher suppression against these other classes of events.

With these caveats, our working assumption that the dominant source of combinatorial background comes from events where a $b\bar{b}$ pair was produced in the primary collision (such events will be called $b\bar{b}$-inclusive events in the following) translates into a reduction of the number of events to be simulated of the order of $\sigma_{b\bar{b}}/\sigma_{inel} = 160$, i.e. from billions to tens of millions. During the DC06 LHCb data challenge [40], the mass production of some $10^7$ $b\bar{b}$-inclusive events was achieved in a few months of running on the WLCG Grid [43].

The optimization of the selection cuts was performed before simulating any trigger algorithm, in common with other simulation studies within LHCb. In this way the trigger algorithms and cuts can tuned in a second phase, only taking into account MC events which are selected offline and are useful for the final physics analysis. As the offline selection algorithm and cut values are well established, we merely report here the main aspects of the strategy, and refer to an older document for further details [19].

The first step of the selection consists of applying some filter criteria to each pair of oppositely charged tracks in the event, in particular cutting on (the small roman numerals in parentheses indicate the corresponding entries in Tab. 7):

- the impact parameter significances of the tracks, computed with respect to all the reconstructed primary vertices (i);
- the transverse momenta of the tracks (ii).

These two simple criteria provide a reduction of the initial sample. They reflect the basic signatures of the $B$ meson decays, i.e. the relatively long lifetime (an average $B$ meson reconstructed at LHCb is expected to fly order of 1 cm before decaying), and the high mass of the decaying $B$, leading to a large transverse momentum of the daughters with respect to minimum bias and underlying event primary particles. Then, for each pair of oppositely charged tracks, we perform cuts on:

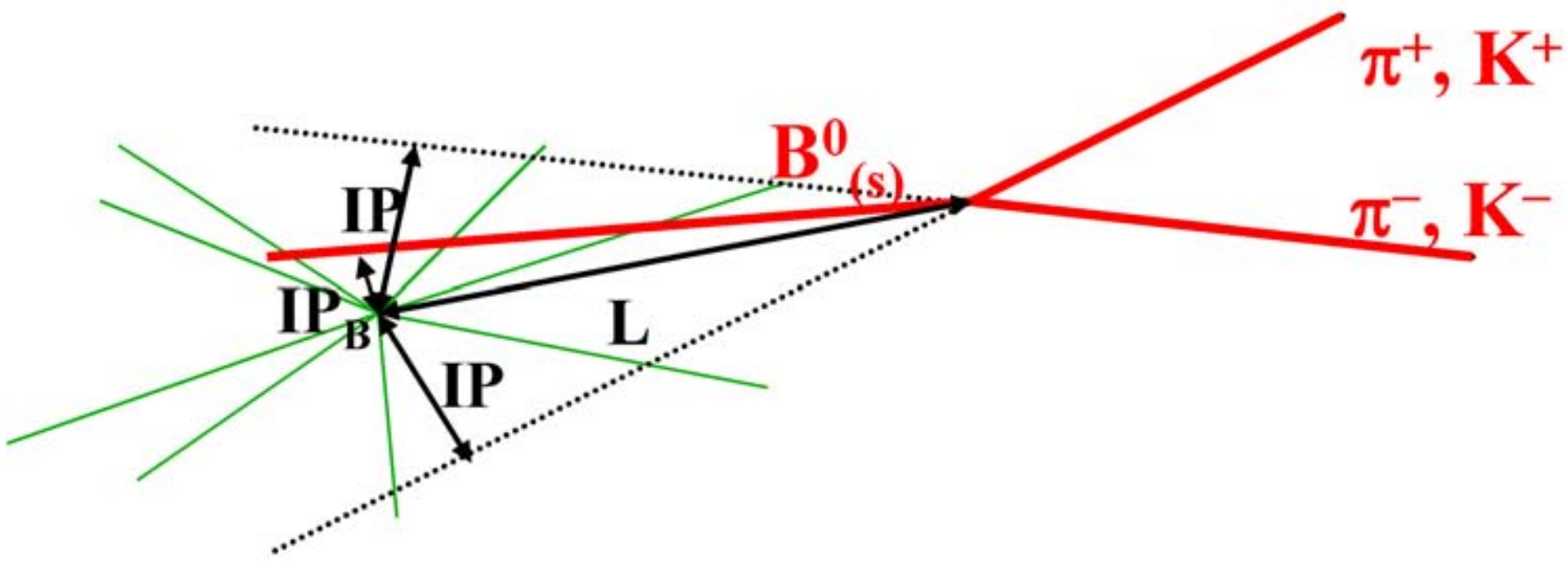

Figure 4: Sketch of the two-body $B$ meson decay topology.

- the invariant mass, calculated assuming the pion-mass hypothesis for each track (iii);
- the maximum of the impact parameter significances of the two tracks, computed with respect to all the reconstructed primary vertices (iv);
- the maximum of the transverse momenta of the two tracks (v).

Each pair surviving these cuts is then used to form the $B$ decay vertex by means of a common vertex fit, and then the momenta of the two tracks are summed up to form a $B$ candidate. In case the event contains more than one reconstructed primary vertex, the one giving the smallest value of the impact parameter significance of the $B$ candidate is chosen as the $B$ candidate primary vertex. The $B$ candidate selection is further refined by cutting on:

- the maximum value of the vertex fit $\chi^2$ (vi);
- the transverse momentum of the $B$ candidate (vii);
- the distance of flight, i.e. the distance between the primary and secondary vertices (viii);
- the impact parameter significance of the $B$ candidate, in order to constrain its direction of flight to point to the primary vertex (ix).

A graphical representation of the decay topology is shown in Fig. 4.

The values of the selection cuts are summarized in Tab. 7. In order not to introduce systematic differences between the various $H_b \to h^+h'^-$ channels through the selection, a single set of cuts has been chosen for all the decay modes, although the decay modes with the smaller branching fractions might generally benefit from a specific optimization employing tighter cuts, in order to reduce further the amount of combinatorial background below the signal mass peak.

| ID | Cut type | Accepted regions |
|---|---|---|
| i | $min[(IP/\sigma_{IP})^h, (IP/\sigma_{IP})^{h'}]$ | $> 6$ |
| ii | $min(p_T^h, p_T^{h'})$ [GeV/c] | $> 1$ |
| iii | $m$ [GeV/c$^2$] | [5.0, 5.8] |
| iv | $max[(IP/\sigma_{IP})^h, (IP/\sigma_{IP})^{h'}]$ | $> 12$ |
| v | $max(p_T^h, p_T^{h'})$ [GeV/c] | $> 3$ |
| vi | $\chi^2$ | $< 5$ |
| vii | $p_{T_B}$ [GeV/c] | $> 1.0$ |
| viii | $L_B/\sigma_{L_B}$ | $> 18$ |
| ix | $IP_B/\sigma_{IP_B}$ | $< 2.5$ |

Table 7: Summary of offline selection cuts. The index in the first column refers to the steps of the selection algorithm, as described in the text.

| Decay mode | $\epsilon_{gen}$ [%] | $\epsilon_{sel/gen}$ [%] | $\epsilon_{sel}$ [%] |
|---|---|---|---|
| $B^0 \to \pi^+\pi^-$ | $19.9 \pm 0.2$ | $19.8 \pm 0.2$ | $3.94 \pm 0.05$ |
| $B^0 \to K^+\pi^-$ | $20.1 \pm 0.2$ | $19.1 \pm 0.1$ | $3.84 \pm 0.04$ |
| $B_s^0 \to \pi^+K^-$ | $20.4 \pm 0.2$ | $18.8 \pm 0.3$ | $3.83 \pm 0.07$ |
| $B_s^0 \to K^+K^-$ | $34.6 \pm 0.3$ | $10.7 \pm 0.1$ | $3.69 \pm 0.05$ |
| $\Lambda_b \to p\pi^-$ | $20.8 \pm 0.1$ | $16.1 \pm 0.3$ | $3.36 \pm 0.06$ |
| $\Lambda_b \to pK^-$ | $21.1 \pm 0.2$ | $15.7 \pm 0.2$ | $3.32 \pm 0.05$ |

Table 8: Summary of offline selection efficiencies.

The total offline selection efficiencies for each of the $H_b \to h^+h'^-$ decays under study are shown in Tab. 8. The selection efficiency $\epsilon_{sel}$ can be factorized as:

$$\epsilon_{sel} = \epsilon_{gen} \cdot \epsilon_{sel/gen} \tag{25}$$

where $\epsilon_{gen}$ is the MC generator level cut efficiency, already shown in Tab. 6, and $\epsilon_{sel/gen}$ is the selection efficiency for generated events. The table clearly shows a decrease of efficiency going from the first two rows to the last two ones. This is due to the different lifetimes of the $B^0$, $B_s^0$ and $\Lambda_b$ hadrons, which in the MC simulation have been set to the values reported in Tab. 9. In fact, as will be seen later when discussing the acceptance as a function of the proper time, some of the cuts employed to isolate the signal tend to reject events with small proper time.

The invariant mass distribution of all the decay modes, under the pion mass hypothesis for both the charged tracks and after the offline event selection algorithm is applied to the MC samples, is shown in Fig. 5. The cumulative contributions of the various channels to the overall distribution are correctly normalized and kept distinct. All the contributions are clearly visible in the bottom plot which employs a logarithmic scale. Apart from the contribution of physical and combinatorial backgrounds, which are absent in this plot, the overall mass line shape is what LHCb will observe with an integrated luminosity

| $B$ hadron | $\tau_{MC}$ [ps] | $\tau_{HFAG}$ [ps] |
|---|---|---|
| $B^0$ | 1.536 | $1.530 \pm 0.008$ |
| $B_s^0$ | 1.461 | $1.478^{+0.020}_{-0.022}$ |
| $\Lambda_b$ | 1.229 | $1.379 \pm 0.051$ |

Table 9: Lifetimes of the $B$ hadrons under study: values used in the LHCb MC simulation ($\tau_{MC}$) and current experimental values ($\tau_{HFAG}$) [42]. The $\Lambda_b$ samples were generated with a proper lifetime value which is no longer compatible with current measurements.

$L = 0.36\,\mathrm{fb}^{-1}$.

## 3.3 Online selection

Although a full description of the LHCb trigger is beyond the scope of this note, it is useful to provide a brief overview in order to introduce the $H_b \to h^+h'^-$ online selection.

The LHCb trigger system is organized in two levels. The earliest level, called L0 (Level 0), is a hardware trigger implemented with custom electronic boards. Its goal is to select particles with high transverse energy and momentum using partial detector information, in particular from the calorimeter system and the muon chambers. This trigger level reduces the effective rate from about 10 MHz, i.e. the rate of bunch crossings with at least one visible interaction in the LHCb detector, to a maximum output rate of 1.1 MHz. At this rate, the full detector is read out into an Event Filter Farm (EFF). The second level, called HLT (High Level Trigger) is a software application running on the LHCb EFF. The HLT is further subdivided in two logical steps: the HLT first level (HLT1) which reduces the rate to a few tens of kHz, and the HLT second level (HLT2), which performs the last reduction to a rate of about 2 kHz.

The HLT1 applies different sequences of algorithms depending on the decision issued by the L0. It uses only a part of the full detector information available, in order to decrease the required processing power. The strategy is to confirm the L0 candidates by adding information from either the Vertex Locator (VELO) or the main tracker and applying cuts on the transverse momentum and the impact parameter with respect to the primary vertex.

After the HLT1, the events are fully reconstructed at a rate of a few tens of kHz, and a complete analysis of interesting $B$ decays is performed by the HLT2 selection algorithms. Tracks reconstructed at this level differ from the ones reconstructed offline since full covariance matrices are not available due to timing constraints. In general this means that, in contrast to what happens at the offline level, an online algorithm does not have complete estimates of the track parameter errors, although we must mention that the online tracking and vertexing algorithms achieve a precision just slightly worse than that obtained offline. These considerations have important consequences on how the online selection algorithm must be realized. After the full event reconstruction, two kinds of selections are applied, so-called inclusive and exclusive. Inclusive selections aim to collect events which are likely to contain $B$ decays or are useful for lifetime unbiased physics analyses and particle identification calibration. They include for example single and di-muon triggers, as well as a $D^*$ trigger, very useful for collecting large samples of kaons and pions for particle identification calibration purposes. The family of inclusive selections also include the so-called *topological* trigger, which is under development and aims to select decays on the basis of their topology (e.g. two-body and three-body modes). Instead, exclusive selections are specifically designed to provide high efficiencies within a limited bandwidth budget for fully reconstructed $B$ decays of interest, such as the ones of the $H_b \to h^+h'^-$ family. The order of 100 core physics channels are selected this way, sharing a bandwidth of about 200 Hz. The total output rate of the HLT trigger, whose global decision will be given by an $OR$ of all the decisions of the inclusive and exclusive algorithms, will be around 2 kHz. This is the final rate at which data will be sent to the mass storage.

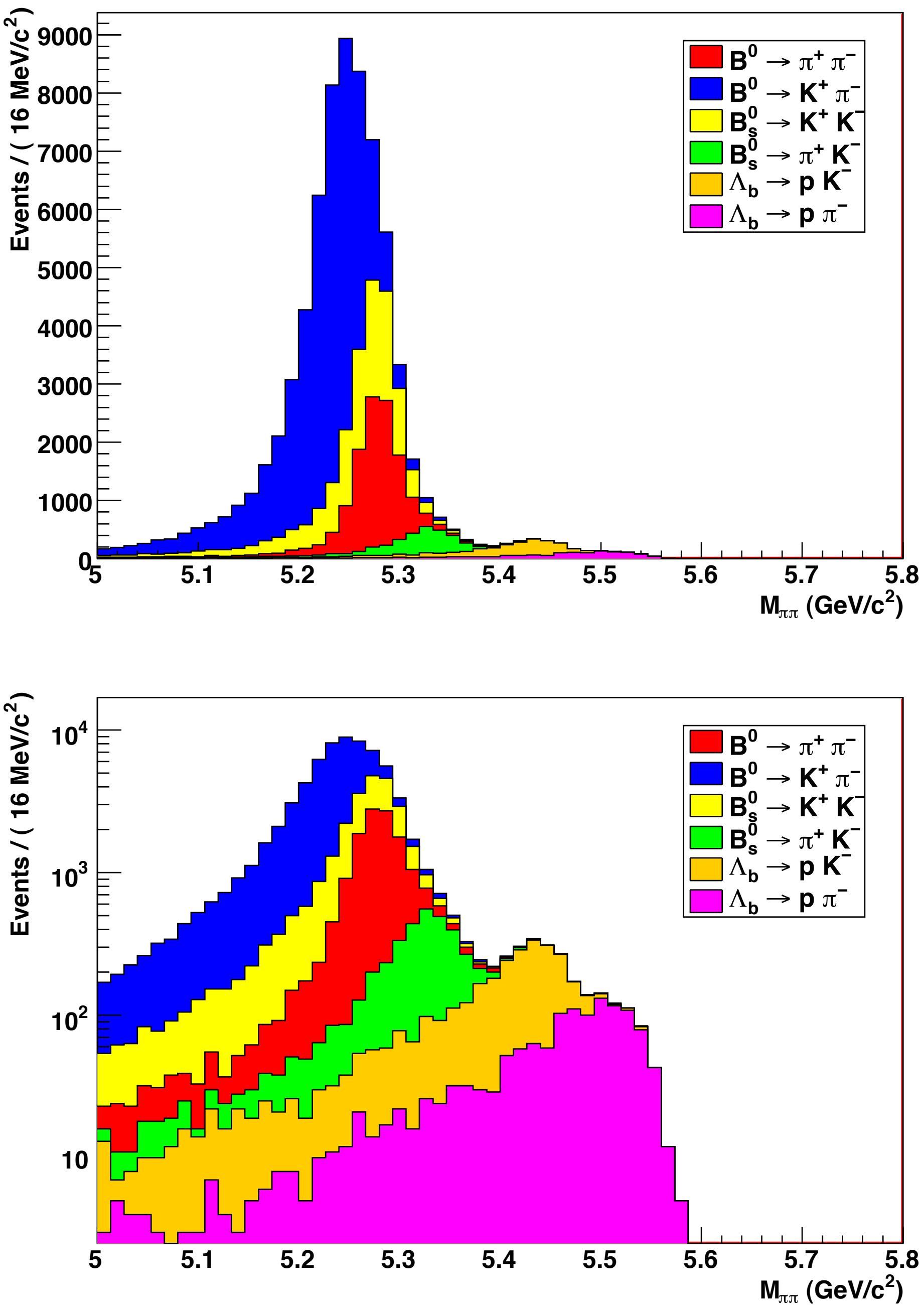

Figure 5: Invariant mass distribution, under the $\pi^+\pi^-$ hypothesis, for all the decay modes after the offline selection: linear scale (top) and logarithmic scale (bottom). The histograms of the various channels are cumulatively added up to form the overall mass line shape. The physical and combinatorial background components are not included.

Before going ahead with the description of the HLT2 $H_b \to h^+h'^-$ selection algorithm, it must be emphasized that while the L0 strategy is essentially frozen due to hardware constraints, the HLT software trigger can still be modified. In this sense, the description given above must be taken as the baseline at the time of writing, but specific numbers and even more, the overall strategy, might change significantly before the start of the physics data taking and even evolve further afterwards.

The HLT2 $H_b \to h^+h'^-$ selection algorithm is designed to have an output rate of a few Hz. In principle, the algorithm should not differ significantly from the one providing the offline selection. However, as we already said, for timing constraints the quality of the online tracking is in general slightly worse than the offline one and the error estimates are not based on a full covariance matrix calculation. For this reason it is important to place minimal reliance on cuts which involve the *significance* of certain observables, that is the measured value divided by the assigned error. In the offline selection, cuts are placed on the significance of the impact parameters and distance of flight. As in HLT2 these error assignments are less reliable, the strategy here is to base the selection on the absolute quantities.

Fig. 6 shows a comparison of the relevant distributions from the online and offline tracking for pions from the $B^0 \to \pi^+\pi^-$ decay, in events which have passed the offline event selection criteria. As can be seen, most of the distributions agree very well, demonstrating the good quality of the online tracking with respect to the offline. Nevertheless, significant differences are apparent for the invariant mass and the $B$ meson impact parameter with respect to the primary vertex distributions. Since for the $H_b \to h^+h'^-$ decays the invariant mass resolution is dominated by the momentum resolution, the slightly larger offline resolution is easily explained by the worsening of the momentum resolution of the $B$ meson daughters in the online tracking with respect to the offline. It can be demonstrated that the error on the mass measurement for the $H_b \to h^+h'^-$ decays is simply proportional to the relative error on the momentum of the daughters $\delta p/p$. Hence, the net effect of worsening the momentum resolution from the nominal value of approximately 0.4% to e.g. 0.8% both for the $\pi^+$ and for the $\pi^-$ causes a degradation of the invariant mass resolution from about 20 MeV/c$^2$ to about 40 MeV/c$^2$. As far as the $B$ meson impact parameter discrepancy is concerned, it is mostly due to the use in HLT2 of the primary vertex reconstructed with limited precision in HLT1. This can be easily corrected by using a more precise primary vertex reconstruction in HLT2, allowing to achieve a precision closer to the offline.

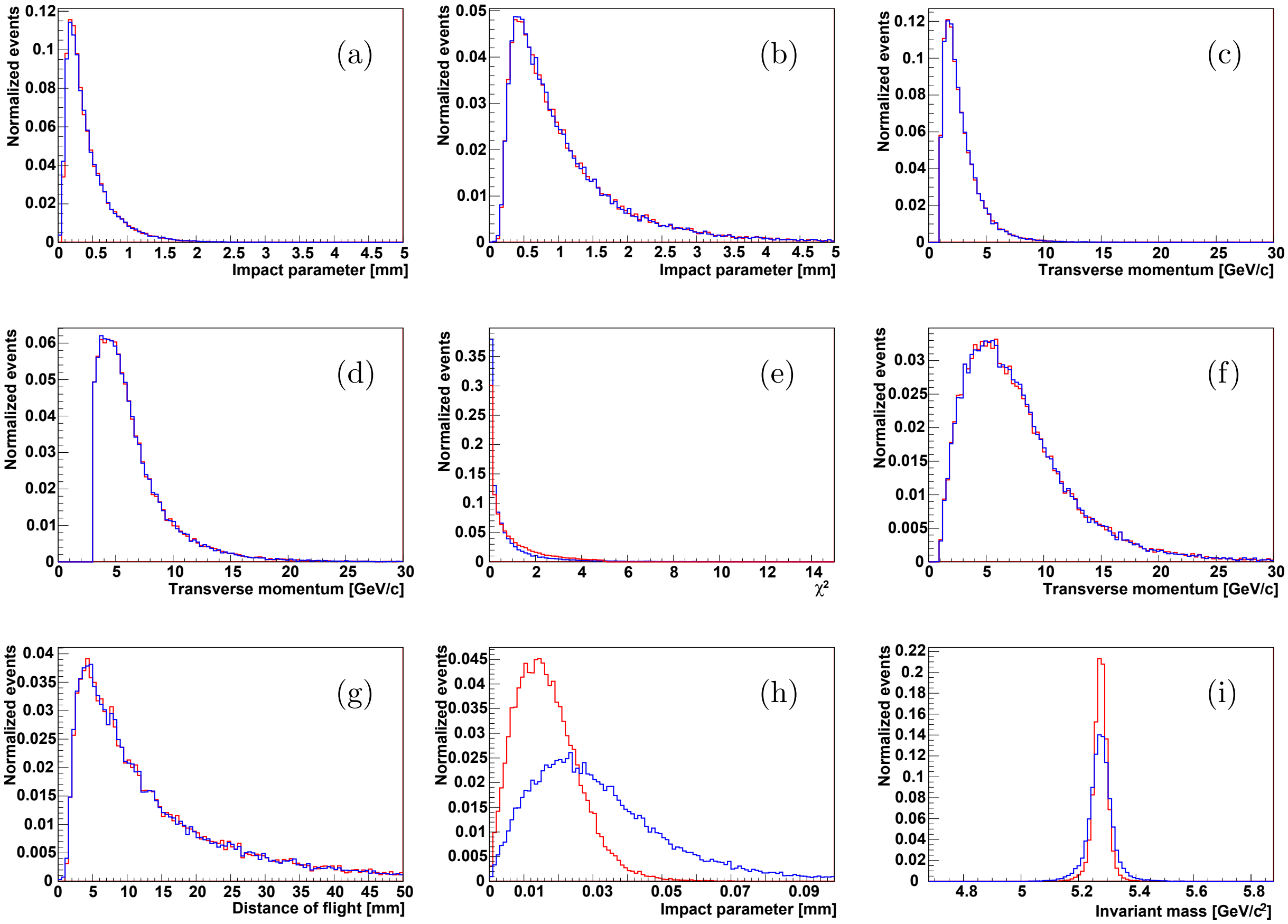

Figure 6: Comparison of distributions from the online (red curve) and offline (blue curve) tracking for pions from the $B^0 \rightarrow \pi^+\pi^-$ decay, for events which have passed the offline event selection: (a) minimum impact parameter of the two tracks with respect to all the primary vertices; (b) maximum impact parameter; (c) minimum transverse momentum; (d) maximum transverse momentum; (e) $\chi^2$ of common vertex fit of the two pions; (f) transverse momentum of the $B$ meson; (g) distance of flight of the $B$ meson; (h) impact parameter of the $B$ meson with respect to the closest primary vertex; (i) invariant mass.

The HLT2 selection algorithm follows the same lines as the offline selection outlined in the previous section. The only differences are that the cuts on the impact parameter and on the distance of flight significances are substituted with cuts on their absolute values. Tab. 10 shows the HLT2 cut values. They have been tuned in order to constrain the output rate to about 5 Hz, while keeping at the same time a large efficiency. The 5 Hz is an arbitrary number chosen so that the $H_b \rightarrow h^+h'^-$ output rate is small compared with the envisaged $\sim 200$ Hz of bandwidth presently foreseen for all exclusive HLT2 selections.

| ID | Cut type | Accepted regions |
|---|---|---|
| i | $min[IP^h, IP^{h'}]$ | $> 0.08\,\mathrm{mm}$ |
| ii | $min(p_T^h, p_T^{h'})$ | $> 1\,\mathrm{GeV/c}$ |
| iii | $m$ | $[5.0,\ 5.8]\,\mathrm{GeV/c^2}$ |
| iv | $max[IP^h, IP^{h'}]$ | $> 0.24\,\mathrm{mm}$ |
| v | $max(p_T^h, p_T^{h'})$ | $> 3\,\mathrm{GeV/c}$ |
| vi | $\chi^2$ | $< 8$ |
| vii | $p_{T_B}$ | $> 1.0\,\mathrm{GeV/c}$ |
| viii | $L_B$ | $> 2.2\,\mathrm{mm}$ |
| ix | $IP_B$ | $< 0.1\,\mathrm{mm}$ |

Table 10: Summary of the cut values employed by the $H_b \to h^+h'^-$ HLT2 exclusive selection.

| Decay mode | $\epsilon_{L0/sel}$ [%] | $\epsilon_{HLT1/L0}$ [%] | $\epsilon_{HLT2/HLT1}$ [%] | $\epsilon_{trig/sel}$ [%] |
|---|---|---|---|---|
| $B^0 \to \pi^+\pi^-$ | $52.4 \pm 0.7$ | $77.2 \pm 1.1$ | $88.6 \pm 1.4$ | $35.8 \pm 0.6$ |
| $B^0 \to K^+\pi^-$ | $52.2 \pm 0.4$ | $78.5 \pm 0.6$ | $88.1 \pm 0.7$ | $36.1 \pm 0.3$ |
| $B_s^0 \to \pi^+K^-$ | $53.4 \pm 1.2$ | $78.1 \pm 1.9$ | $88.8 \pm 2.4$ | $37.0 \pm 1.0$ |
| $B_s^0 \to K^+K^-$ | $53.4 \pm 0.6$ | $79.6 \pm 1.0$ | $88.0 \pm 1.1$ | $37.4 \pm 0.5$ |
| $\Lambda_b \to p\pi^-$ | $53.3 \pm 1.3$ | $80.2 \pm 2.2$ | $86.2 \pm 2.5$ | $36.8 \pm 1.1$ |
| $\Lambda_b \to pK^-$ | $51.2 \pm 0.8$ | $79.2 \pm 1.4$ | $88.1 \pm 1.7$ | $35.7 \pm 0.7$ |

Table 11: Summary of trigger efficiencies relative to offline-selected events.

Tab. 11 contains an overview of all the trigger efficiencies, starting from offline selected signal samples. The quantity $\epsilon_{L0/sel}$ represents the efficiency of the L0 hardware trigger on offline selected events. Similarly, $\epsilon_{HLT1/L0}$ is the efficiency of the HLT1 algorithm on events which passed the L0 and the offline selection, while $\epsilon_{HLT2/HLT1}$ is the efficiency of the HLT2 $H_b \to h^+h'^-$ exclusive selection algorithm on events which passed the HLT1, the L0 and the offline selection filter. These three efficiencies are seen to be very similar amongst the various decay modes, with values of approximately 53%, 79% and 88% respectively. The total trigger efficiency with respect to the offline-selected events can thus be written as the product of the three:

$$\epsilon_{trig/sel} = \epsilon_{L0/sel} \cdot \epsilon_{HLT1/L0} \cdot \epsilon_{HLT2/HLT1}, \tag{26}$$

yielding for all the modes a value approximately equal to 37%.

By running over a sample of $N_{L0}^{MB} = 5.186 \cdot 10^6$ MC minimum bias events, previously filtered with the L0 trigger, the $H_b \to h^+h'^-$ HLT selected $N_{HLT} = 25$ events in total. The output rate can then be calculated as:

$$R_{HLT} = \frac{N_{HLT}}{N_{L0}^{MB}} \cdot R_{L0}, \tag{27}$$

| Event type | Cross section [mb] | Fraction 68% C.L. | Fraction 95% C.L. |
|---|---|---|---|
| $b\bar{b}$ | 0.69 | [15%, 37%] | [9%, 51%] |
| $c\bar{c}$ | 3.54 | [21%, 45%] | [12%, 61%] |
| light flavour | 65 | [31%, 59%] | [21%, 77%] |

Table 12: Fractions of events containing heavy ($b\bar{b}$ or $c\bar{c}$) and light flavours after having applied the trigger filter to the minimum bias MC sample. Due to the limited statistics available, 68% C.L. and 95% C.L. Feldman-Cousins confidence intervals [44] are reported. For comparison, also the cross sections of the relative processes as predicted by the PYTHIA MC event generator are shown. The cross section quoted for the light flavour events is to the so-called visible cross section, i.e. the one corresponding to primary collisions which produce at least two reconstructible tracks in the whole detector. Although the ratio of light flavour and $b\bar{b}$ cross sections is almost a factor 100, the fraction of light flavour events after the trigger is comparable to that of $b\bar{b}$ events.

where $R_{L0} \simeq 1\,\mathrm{MHz}$ is the output event rate of the L0 trigger, yielding $R_{HLT} = (4.8 \pm 1.0)\,\mathrm{Hz}$.

It is also interesting to show the fractions of events containing heavy and light flavours after having applied the trigger filter. Tab. 12 shows the fraction of $b\bar{b}$, $c\bar{c}$ and light flavour events from the minimum bias sample after the trigger, compared with the cross sections of the respective processes predicted by the PYTHIA MC event generator, used in the DC06 LHCb simulation. As expected, the trigger enriches the amount of heavy flavoured events, depressing those containing only light flavours. However, the fraction of $b\bar{b}$ events is not yet dominating after the trigger, and we are actively working on it.

## 3.4 Event yields

In the previous sections we have estimated the offline and online selection efficiencies on the various $H_b \to h^+h'^-$ decay modes. The event yields can be calculated as:

$$Y = L \cdot \sigma_{b\bar{b}} \cdot f_{hadr} \cdot 2 \cdot \mathcal{BR} \cdot \epsilon_{sel} \cdot \epsilon_{trig/sel} \tag{28}$$

where $L = \int \mathcal{L} dt$ is the integrated luminosity at the LHCb interaction point, $\sigma_{b\bar{b}}$ is the beauty production cross section at the LHC, $f_{hadr}$ is the probability that the $b$ quark hadronizes to the $B$ hadron of interest, the factor 2 takes into account the presence of two $B$ hadrons per event, $\mathcal{BR}$, $\epsilon_{sel}$ and $\epsilon_{trig/sel}$ are respectively the branching ratio, the offline selection efficiency and the trigger efficiency for offline-selected events of the $B$ hadron decay of interest. Since the baseline instantaneous luminosity at the LHCb interaction point is foreseen to be $\mathcal{L} = 2 \cdot 10^{32}\,\mathrm{cm}^{-2}\mathrm{s}^{-1}$, assuming $10^7$ seconds of useful data taking per year, we obtain $L = 2\,\mathrm{fb}^{-1}$, that we conventionally assume as the baseline annual integrated luminosity. We assume a value of $\sigma_{b\bar{b}} = 500\,\mu\mathrm{b}$, which is consistent with the central value of the theoretical predictions [45], and use the values of $f_{hadr}$, $\mathcal{BR}$, $\epsilon_{sel}$ and $\epsilon_{trig/sel}$ reported in Tabs. 6, 8 and 11. The final expected annual yields are reported in Tab. 13, where we also report for convenience the relevant quantities needed for the computation. It can be seen that in a few years of LHCb operation the sample size of $H_b \to h^+h'^-$ decays will reach 1 million events, to be contrasted with the few thousands of events presently available from the $B$ factories and the Tevatron.

| Decay mode | $f_{hadr}$ [%] | $\mathcal{BR}$ $\times 10^6$ | $\epsilon_{sel}$ [%] | $\epsilon_{trig/sel}$ [%] | $\epsilon_{tot}$ [%] | Annual yield |
|---|---|---|---|---|---|---|
| $B^0 \to \pi^+\pi^-$ | $40.3 \pm 0.9$ | $5.16 \pm 0.22$ | $3.95 \pm 0.05$ | $35.8 \pm 0.6$ | $1.41 \pm 0.03$ | $58.8\,k$ |
| $B^0 \to K^+\pi^-$ | $40.3 \pm 0.9$ | $19.4 \pm 0.06$ | $3.84 \pm 0.04$ | $36.1 \pm 0.3$ | $1.39 \pm 0.02$ | $216.6\,k$ |
| $B_s^0 \to \pi^+K^-$ | $10.1 \pm 0.9$ | $5.27 \pm 1.17$ | $3.83 \pm 0.07$ | $37.0 \pm 1.0$ | $1.42 \pm 0.04$ | $15.1\,k$ |
| $B_s^0 \to K^+K^-$ | $10.1 \pm 0.9$ | $25.8 \pm 4.2$ | $3.69 \pm 0.05$ | $37.4 \pm 0.5$ | $1.38 \pm 0.02$ | $71.9\,k$ |
| $\Lambda_b \to p\pi^-$ | $9.2 \pm 1.5$ | $3.1 \pm 0.9$ | $3.36 \pm 0.06$ | $36.8 \pm 1.1$ | $1.24 \pm 0.04$ | $7.0\,k$ |
| $\Lambda_b \to pK^-$ | $9.2 \pm 1.5$ | $5.0 \pm 1.2$ | $3.32 \pm 0.05$ | $35.7 \pm 0.7$ | $1.18 \pm 0.03$ | $10.9\,k$ |

Table 13: Annual event yields for the various $H_b \to h^+h'^-$ decay modes. The table also contains the $b$ quark hadronization probabilities, the branching fractions, the offline, trigger and total efficiencies.

## 3.5 Backgrounds

As explained in Sec. 3.2, in order to have a quantitative estimate of the amount of combinatorial background, it is mandatory to make the working assumption that random tracks from $B$ hadron decays constitute the dominant contribution, i.e. that a reliable estimate of the amount of combinatorial background can be worked out by analyzing a $b\bar{b}$-inclusive MC sample.

The sample we have used for this study amounts to $N_{b\bar{b}} = 22.1 \cdot 10^6$ events with at least one $B$ hadron produced inside a forward cone with a semi-aperture of 400 mrad. The corresponding data taking time can be calculated as:

$$\Delta T = \frac{N_{b\bar{b}}}{\mathcal{L} \cdot \sigma_{b\bar{b}} \cdot \epsilon_{gen}^{b\bar{b}}} \tag{29}$$

where $\epsilon_{gen}^{b\bar{b}}$ is the generator level cut efficiency for $b\bar{b}$-inclusive MC events of Tab. 6, yielding $\Delta T \simeq 500\,s$ (i.e. about 8 minutes), corresponding to an integrated luminosity $L_{b\bar{b}} \simeq 0.1\,\mathrm{pb}^{-1}$. Nevertheless, if we apply the offline event selection to this sample, we are already able to see a signal peak sitting on top of a combinatorial background, having at lower mass values a shoulder due to partially reconstructed $B$ to three-body decays (physical background). Fig. 7 shows the invariant mass distribution under the $\pi^+\pi^-$ hypothesis for $b\bar{b}$-inclusive events passing the offline selection. Given the limited statistics, the trigger filter has not been applied. The plots show the various components (signal, physical background, combinatorial background) independently, as well as summed together. Tab. 14 summarizes the number of $b\bar{b}$-inclusive events passing the offline selection and the various trigger stages, subdivided into the various components. For all the three components, the trigger efficiencies on offline-selected events are statistically compatible with the signal trigger efficiencies reported in Tab. 11, although in the case of the combinatorial background this is only true within the 95% C.L..

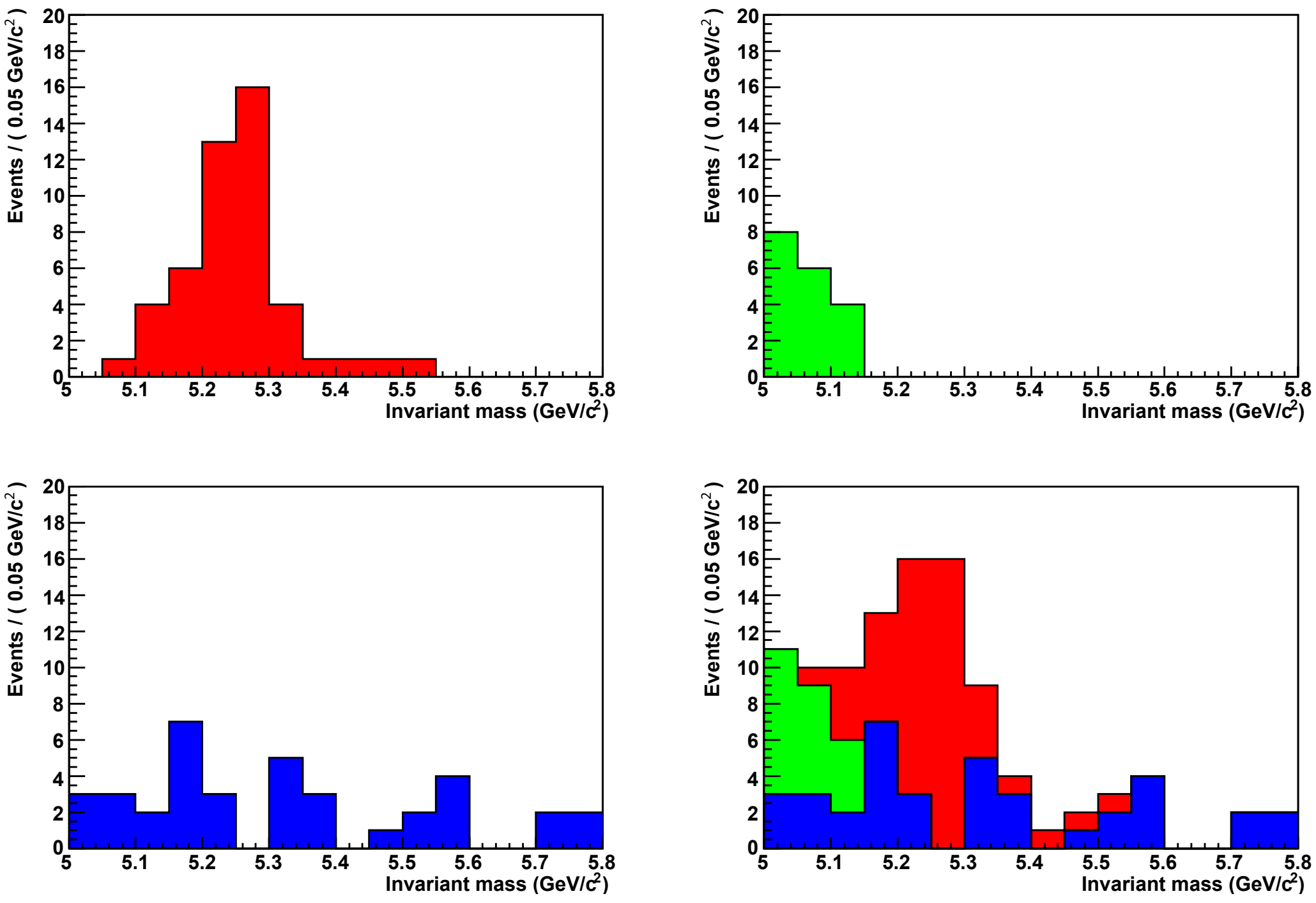

Figure 7: Invariant mass distributions under the $\pi^+\pi^-$ hypothesis for $b\bar{b}$-inclusive events passing the offline selection: $H_b \to h^+h'^-$ signals only (top left), physical background (top right), combinatorial background (bottom left) and all components summed together (bottom right).

Using the above results, we can now calculate the annual yields for the physical and combinatorial backgrounds. We can also calculate the overall signal yield from the $b\bar{b}$-inclusive sample, in order to check that it agrees with the one previously estimated by analyzing the specific signal samples. Such yields can be computed as

$$Y_{S,P,C} = \frac{N_{S,P,C}}{L_{b\bar{b}}} \cdot L_{year} \tag{30}$$

where $Y_{S,P,C}$ is the annual yield and $N_{S,P,C}$ is the number of triggered and offline-selected events reported in Tab. 14, with the subscripts $S$, $P$, and $C$ indicating the signal, physical background and combinatorial background components, respectively. Finally, $L_{b\bar{b}} = 0.1\,\text{pb}^{-1}$ and $L_{year} = 2\,\text{fb}^{-1}$ are respectively the integrated luminosity corresponding to the analyzed $b\bar{b}$-inclusive sample and the integrated luminosity corresponding to one nominal year of LHCb data taking. The 68% C.L. and 95% C.L. Feldman-Cousins unified confidence intervals [44] of the annual yields of the three components are shown in Tab. 15. Note that the 68% C.L. interval for the signal encompasses the sum of the signal modes reported in Tab. 13, showing the two calculations to be compatible.

| Component | $N_{sel}$ | $N_{L0}$ | $N_{HLT1}$ | $N_{HLT2}$ | $\epsilon_{trig/sel}$ 68% C.L. | $\epsilon_{trig/sel}$ 95% C.L. |
|---|---|---|---|---|---|---|
| Signal | 48 | 24 | 18 | 16 | [26%, 43%] | [19%, 53%] |
| Physical bkg | 18 | 8 | 6 | 6 | [21%, 52%] | [12%, 71%] |
| Combinatorial bkg | 37 | 17 | 14 | 8 | [14%, 30%] | [8%, 41%] |

Table 14: Number of $b\bar{b}$-inclusive events passing the offline selection and the various trigger stages, subdivided in the signal, physical background and combinatorial background components. $N_{sel}$ is the number of offline-selected events, $N_{L0}$ is the number of events which in addition pass the L0 trigger, $N_{HLT1}$ the HLT1 trigger and finally $N_{HLT2}$ the HLT2 trigger, i.e. the final result comprises the whole trigger chain and the offline selection. The last two columns show the 68% C.L. and 95% C.L. Feldman-Cousins unified confidence intervals [44] for the trigger efficiencies.

| Component | Annual yield $\times 10^3$ 68% C.L. | Annual Yield $\times 10^3$ 95% C.L. |
|---|---|---|
| Signal | [247, 416] | [187, 508] |
| Physical bkg | [76, 186] | [44, 255] |
| Combinatorial bkg | [106, 226] | [59, 306] |

Table 15: The 68% C.L. and 95% C.L. Feldman-Cousins unified confidence intervals [44] of the annual yields for signal, physical background and combinatorial background, as resulting from the analysis of the $b\bar{b}$-inclusive MC sample. The statistical uncertainties are just due to the number of simulated events passing the trigger and offline selection algorithms.

In the $\mathcal{CP}$ sensitivity studies discussed later in this note, as annual yields for the physical and combinatorial background components, we will conservatively adopt the upper limits of the 95% C.L. intervals of Tab. 15, i.e. 255,000 and 306,000 events respectively. In fact, as far as the physical background is concerned, only a few of the branching ratios of the $B$ to three-body modes are well known. Hence, it is not possible to make an accurate prediction of how many of these events we can expect. Furthermore, the modeling of these decays in the $b\bar{b}$-inclusive MC sample is not accurate. For example, the angular distribution of the decay products is incorrect because the polarization of the intermediate resonances has not been modelled. Concerning the combinatorial background, we have already mentioned that the dominance of the $b\bar{b}$-inclusive events is a working assumption we have to make. For these reasons, we feel that a conservative estimate of the corresponding yields is an appropriate choice.

While it is unrealistic to argue the shape of the invariant mass spectrum of the physical background from the few offline-selected events, it is instead reasonable to fit a decreasing exponential probability density function (p.d.f.) to the mass distribution of the combinatorial component. Fig. 8 shows the invariant mass spectrum of the combinatorial component only, superimposed with the result of an unbinned likelihood fit of an exponential p.d.f.. Although the uncertainty is large, the slope of the exponential can be determined, yielding:

$$\alpha = -1.5 \pm 0.7\,\mathrm{c}^2/\mathrm{GeV}. \tag{31}$$

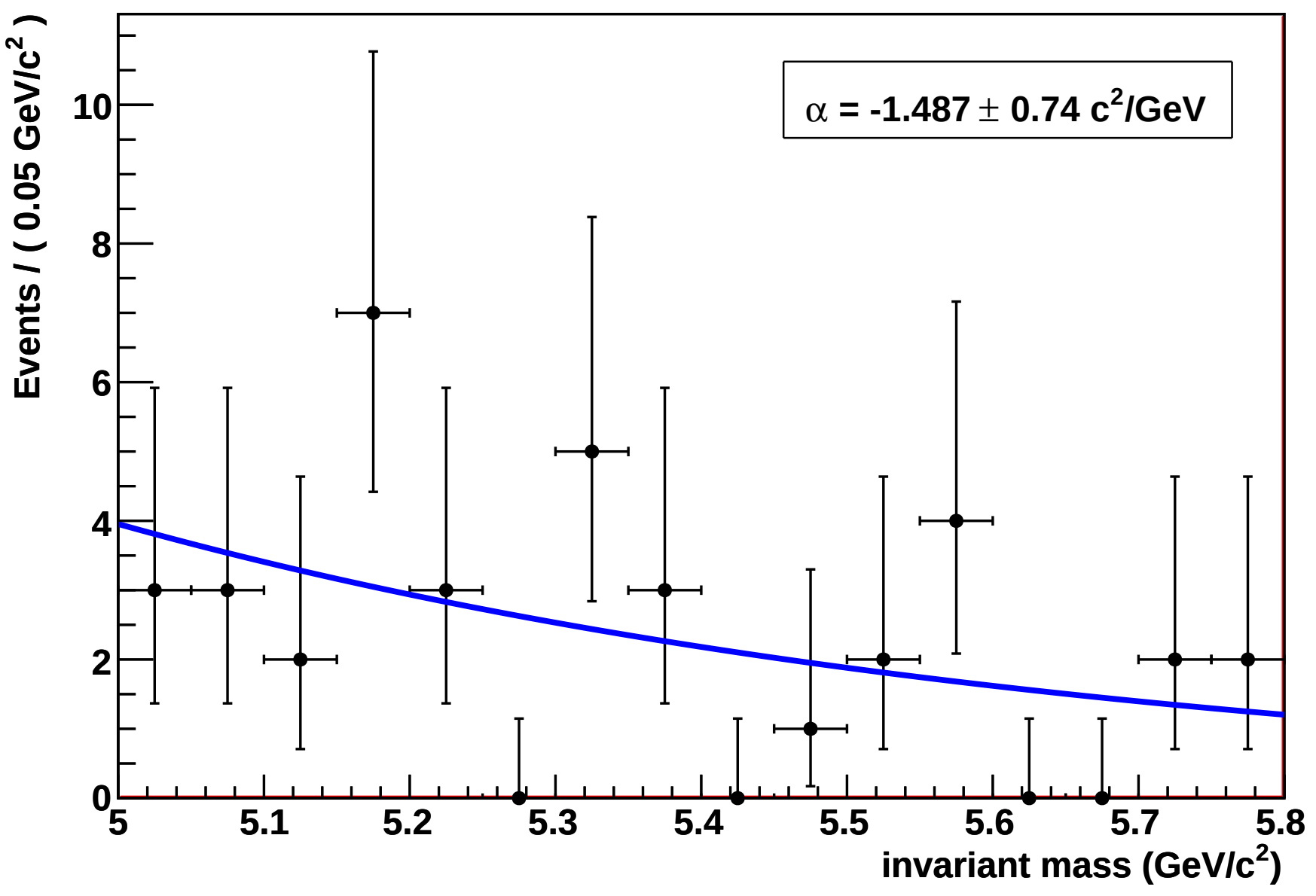

Figure 8: Invariant mass spectrum for combinatorial background events, superimposed with the result of an unbinned likelihood fit of an exponential p.d.f..

Since too few combinatorial events would remain after the trigger to perform a meaningful fit, we can make the reasonable assumption that the trigger would not have altered considerably the slope of the exponential, and use this value for further studies in the remainder of this note.

Further interesting information concerns the nature of the 74 tracks in the offline-selected combinatorial background sample, as summarized in Tab. 16. This information will be useful later for performing the $\mathcal{CP}$ sensitivity studies.

To conclude this section, we discuss the expected shape of the physical background invariant mass distribution. For this purpose, we analyzed a specific MC sample of $B^0 \rightarrow \rho\pi$ decays. Fig. 9 shows the $\pi^+\pi^-$ invariant mass spectrum for events passing the offline selection. The curve superimposed on the histogram is the result of an unbinned likelihood fit to the data of the following p.d.f.:

$$f_P(m) = A^{-1} \cdot m' \left(1 - \frac{m'^2}{m_0^2}\right) \Theta(m_0 - m')\, e^{-c_P \cdot m'} \otimes G(m - m';\, \sigma_P) \tag{32}$$

| Particle | Plus sign | Minus sign | Total |
|---|---|---|---|
| $e$ | 1 | 2 | 3 |
| $\mu$ | 3 | 5 | 8 |
| $\pi$ | 17 | 11 | 28 |
| $e$ or $\mu$ or $\pi$ | 21 | 18 | 39 |
| $K$ | 5 | 6 | 11 |
| $p$ | 2 | 2 | 4 |
| Ghost | 9 | 11 | 20 |
| Total | 37 | 37 | 74 |

Table 16: Nature of the 74 tracks in the offline-selected combinatorial background sample. Ghost tracks are defined as tracks with less than 70% of hits coming from a single MC truth particle.

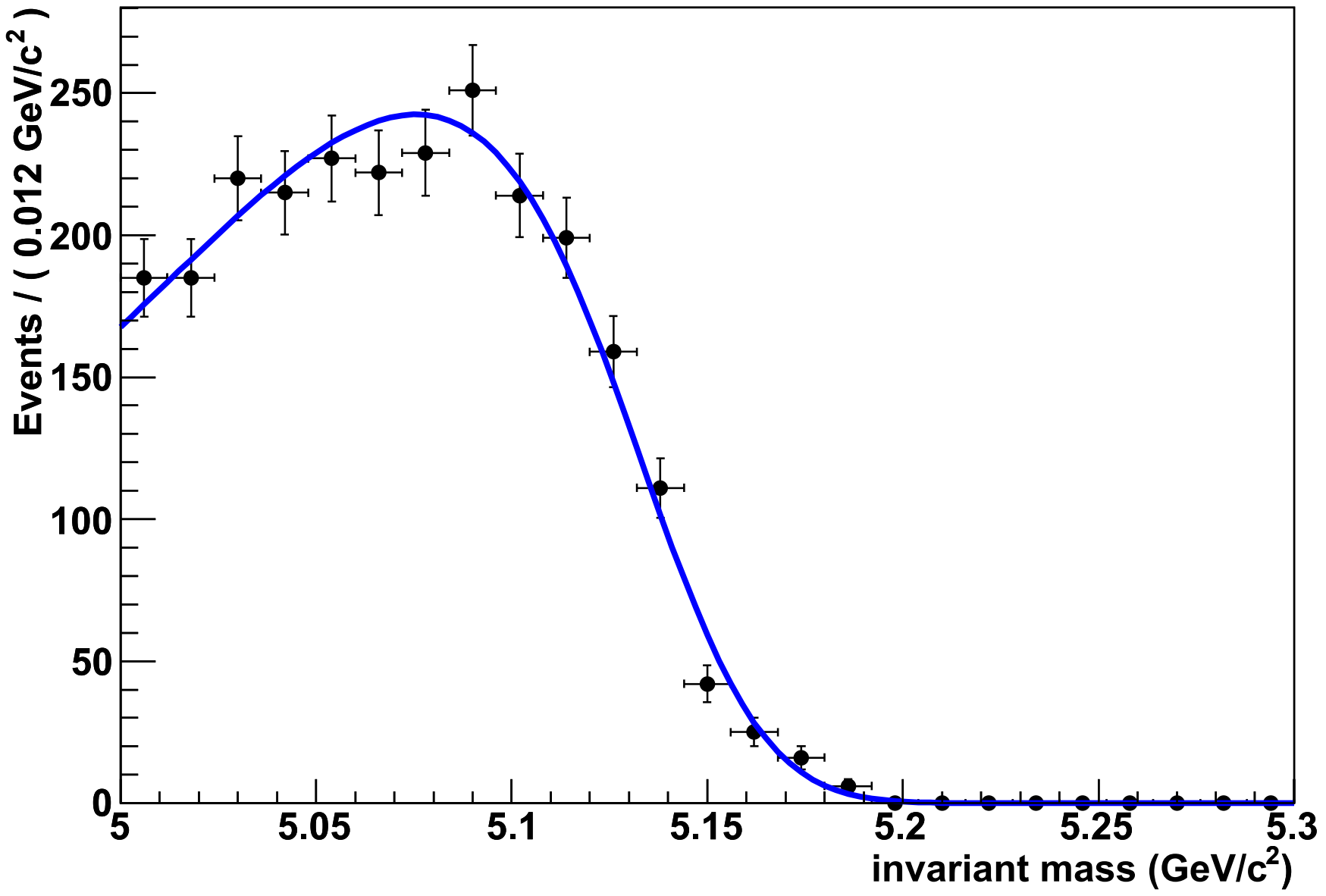

Figure 9: $\pi^+\pi^-$ invariant mass spectrum for $B^0 \to \rho\pi$ events passing the offline selection, with superimposed the result of an unbinned likelihood fit using the p.d.f. of Eq. (32).

where $\Theta$ is the Heaviside function (also commonly known as *step* function), $\otimes$ stands for convolution product ($m'$ is the variable over which the convolution integral is calculated), $G(m - m'; \sigma_P)$ is a Gaussian p.d.f. with standard deviation $\sigma_P$ representing the experimental resolution, $m_0$ is a free parameter and $A$ is a normalization factor. The results of the fit are reported in Tab. 17. The value of the parameter $m_0$ plays the role of the kinematical higher limit of the mass distribution in the absence of resolution effects, and turns out to be very close to the difference between the $B^0$ and the missing $\pi^0$ masses. Since the nominal mass resolution for the $H_b \to h^+h'^-$ decays is about 20 MeV/c$^2$, we expect that the kinematical limit can be overcome according to a Gaussian distribution with such a width, i.e. exactly what the fit shows. Finally, $c_P$ is a parameter governing the shape of the distribution, with no obvious physical meaning.

| Parameter | Fit result |
|---|---|
| $m_0$ [GeV/c$^2$] | $5.1495 \pm 0.0035$ |
| $c_P$ [c$^2$/GeV] | $14.6 \pm 1.3$ |
| $\sigma_P$ [MeV/c$^2$] | $21.6 \pm 2.0$ |

Table 17: Result of the fit of the function used for describing the physical background invariant mass line shape to the offline-selected $B^0 \to \rho\pi$ data.

## 3.6 Selection of $B^0 \to p\bar{p}$ decays

In this section we investigate the potential of LHCb to observe the rare decay $B^0 \to p\bar{p}$. Full details of the study are available in Ref. [46].

The offline event selection is identical to the one described in Sec. 3.2. In addition, in order to select protons and reject charged pions and kaons, particle identification cuts on the difference between the log likelihood of two particle hypotheses, $\Delta \log \mathcal{L}_{p\pi} > 5$ and $\Delta \log \mathcal{L}_{pK} > 0$ were used. Finally, a cut on the track quality, $\chi^2/n_{\mathrm{DoF}} < 3$, was applied to each of the daughter tracks in order to reduce the background due to ghost tracks.

Several sources of background were investigated: the background due to misidentified $H_b \to h^+h'^-$ final states or partially reconstructed three-body $B$ decays and the background from $b\bar{b}$-inclusive events. Given their similar signature and considerably higher branching ratios, the other $H_b \to h^+h'^-$ decays are expected to be a potentially dangerous source of background. We have considered the following decay modes: $B^0 \to K^+\pi^-$, $B^0_s \to K^+K^-$ and $\Lambda_b \to pK^-$.

Partially reconstructed three-body $B$ decays only affects the low mass tails of the other $H_b \to h^+h'^-$ signals, because their reconstructed mass tends to fall significantly below the signal mass peak when all daughter tracks are assigned the pion-mass hypothesis. However, three-body decays may populate the $B^0 \to p\bar{p}$ mass peak in the present analysis as their reconstructed mass will shift significantly under a proton mass hypothesis for the two reconstructed daughter particles. The three-body $B$ decays considered as specific backgrounds in the present study are: $B^0 \to \pi^+\pi^-\pi^0$, $B^0 \to K^0_S\pi^-\pi^+$, $B^+ \to \pi^+\pi^-\pi^+$, $B^+ \to \pi^+\pi^-K^+$, $B^+ \to \pi^+K^-K^+$, $B^+ \to p\bar{p}\pi^+$, $B^+ \to p\bar{p}K^+$ and $B^+ \to K^+K^-K^+$.

Tab. 18 gives the 90% C.L. upper limits on the number of events passing all selection cuts in $L = 2\,\mathrm{fb}^{-1}$ of data as well as the resulting background-to-signal ratios from each background category. The signal yield was calculated assuming the current experimental limit of $1.1 \cdot 10^{-7}$ for the branching ratio of $B^0 \to p\bar{p}$. The total background-to-signal ratio is found to be $B/S < 2.20$ at 90% C.L..

Taking the 90% C.L. upper limits for the background yields, the signal significance

| Parameter | $B^0 \to p\bar{p}$ | $H_b \to h^+h'^-$ | $H_b \to hhh$ | $b\bar{b}$-inclusive |
|---|---|---|---|---|
| Yield | 700 | $< 30$ | $< 270$ | $< 1.2k$ |
| B/S ratio | - | $< 0.05$ | $< 0.40$ | $< 1.75$ |

Table 18: Expected annual yield for the $B^0 \to p\bar{p}$ decay and 90% C.L. limits of the background yields and background-to-signal ratios for the background classes considered. Refer to the text for details.

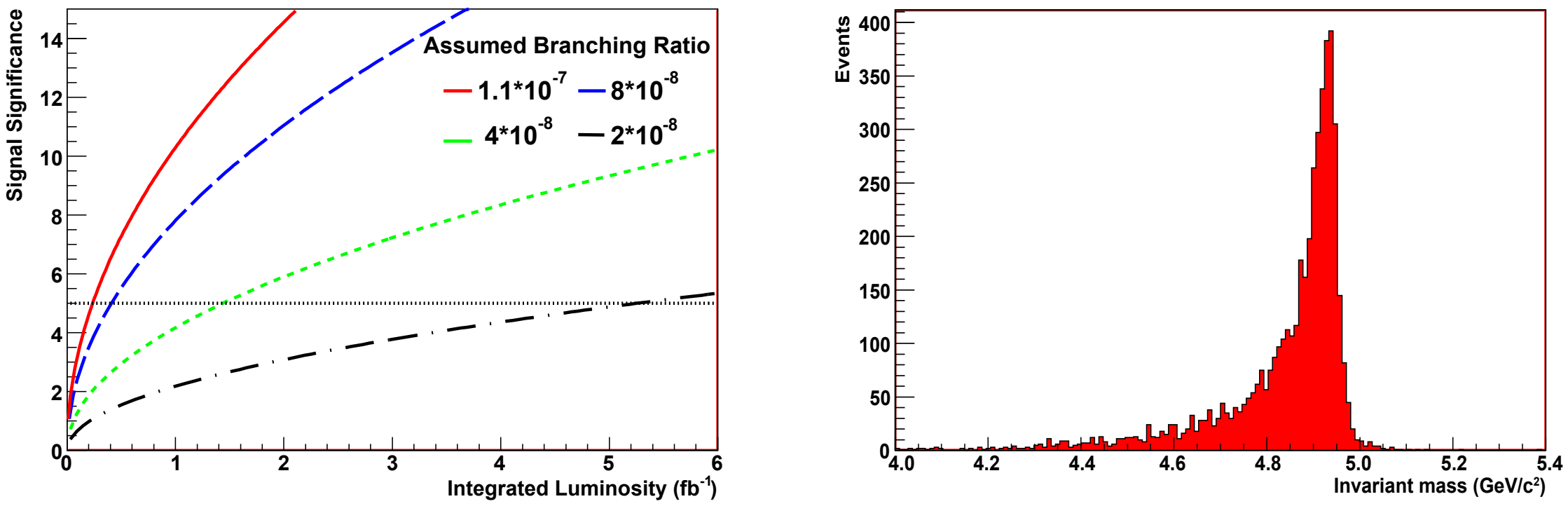

Figure 10: Left: significance of the $B^0 \to p\bar{p}$ signal as a function of integrated luminosity. Each curve assumes a different branching ratio for $B^0 \to p\bar{p}$; the values used are $1.1 \cdot 10^{-7}$ (red), $8 \cdot 10^{-8}$ (blue), $4 \cdot 10^{-8}$ (green) and $2 \cdot 10^{-8}$ (black). Right: reconstructed mass distribution for $B^0 \to p\bar{p}$ events when a pion mass hypothesis is assumed for the daughters.

$S/\sqrt{S+B}$ as a function of integrated luminosity and the branching ratio of $B^0 \to p\bar{p}$ can be calculated. Fig. 10 (left) shows, for several different branching ratio scenarios, how the signal significance improves as more data are collected. LHCb can expect to make an observation (i.e. achieve a $5\sigma$ significance) with $2\,\mathrm{fb}^{-1}$ of data, even if the true branching ratio of $B^0 \to p\bar{p}$ is significantly below the current experimental upper limit.

In the present trigger configuration for the selection of $H_b \to h^+h^{'-}$ modes, a pion hypothesis is assumed for the daughters. Under this assumption the reconstructed mass for a $B^0 \to p\bar{p}$ candidate will be far below the nominal $B^0$ mass. This is illustrated in Fig. 10 (right). For this trigger selection to incorporate efficiently this rare baryonic mode it will be necessary to extend the mass window at least $\simeq 500$ MeV/c$^2$ below the nominal $B^0$ mass; a mass window of $m_{B^0} \pm 600$ MeV$/c^2$ for example would cut away 15% of signal events. Provided that the mass window coverage is sufficient, the trigger efficiency for $B^0 \to p\bar{p}$ should be similar to those of the other $H_b \to h^+h'^-$ channels (see Tab. 11).

# 4 Proper time measurement

The accurate measurement of the proper decay time of $B$ hadrons is a crucial ingredient for determining time dependent $\mathcal{CP}$ violating observables and, obviously, for measuring average lifetimes. This is particularly true for $\mathcal{CP}$ measurements involving $B_s^0$ decays, where an inaccurate proper time resolution could dilute the $\mathcal{CP}$ sensitivity. This section presents the studies that have been specifically performed on $H_b \to h^+h'^-$ decays, although many issues here are in common with other LHCb analyses.

## 4.1 Acceptance and resolution

Since the offline selection and trigger filters for the $H_b \to h^+h'^-$ decays are based on tight cuts on the impact parameters of the $B$ decay products as well as on the $B$ distance of flight, the shape of the $B$ decay proper time distribution for selected events is distorted, in particular in the region of small proper time values. In fact, the $B$ decay proper time is related to the distance of flight by the expression:

$$\tau_B = m_B \frac{L_B}{p_B}, \tag{33}$$

where $m_B$, $L_B$ and $p_B$ are the mass, the distance of flight and the momentum of the $B$ hadron respectively. Hence, as already mentioned, any cut aimed to suppress events coming from the primary vertex has the effect of lowering the probability to accept events with low proper time. For this reason, the observed decay rate as a function of the proper time should be written as:

$$\Gamma_{exp}(t) \propto [\Gamma_{th}(t') \otimes R_t(t-t')]\, \epsilon_t(t) \tag{34}$$

where $\Gamma_{th}$ is the theoretical decay rate, $\epsilon_t(t)$ is an appropriate acceptance function of the proper time, and $R_t(t-t')$ describes the experimental resolution on the measured proper time.

The rate $\Gamma_{th}(t)$ cannot be described by a simple exponential for $B_s^0$ mesons decaying to $\mathcal{CP}$ eigenstates, given the non-negligible decay width difference of the $B_s^0$ system mass eigenstates. The general expression of the theoretical decay rates for the $B^0 \to \pi^+\pi^-$ and $B_s^0 \to K^+K^-$ decays is given by:

$$\Gamma_{th}(t) \propto \mathrm{e}^{-\frac{t}{\tau}} \left[ \left(1+|\lambda|^2\right) \cosh\left(\frac{\Delta\Gamma}{2}t\right) - 2\mathrm{Re}\lambda \sinh\left(\frac{\Delta\Gamma}{2}t\right)\right], \tag{35}$$

while for flavour specific decays like $B^0 \to K^+\pi^-$ and $B_s^0 \to \pi^+K^-$ it reduces to:

$$\Gamma_{th}(t) \propto \mathrm{e}^{-\frac{t}{\tau}} \cosh\left(\frac{\Delta\Gamma}{2}t\right). \tag{36}$$

It is clear that such decay rates become purely exponential only in the limit $\Delta\Gamma \to 0$, which is a good approximation for the $B^0$ meson, but not for the $B_s^0$ meson. As far as the $\Lambda_b$ decays are concerned, the theoretical rates are pure exponentials since the $\Lambda_b$ baryons are not subject to mixing.

For practical reasons, which will be clarified later when discussing the time dependent $\mathcal{CP}$ fits, instead of working with the proper time (whose calculation requires the corresponding $B$ hadron mass), it is convenient to work with a different observable, i.e.:

$$\xi = \frac{t}{m_B\, c^2} = \frac{L_B}{p_B\, c^2}, \tag{37}$$

which is by construction not dependent on the $B$ mass, and can then be used for all the $B$ decays under study. Introducing the average lifetime relative to the $B$ mass $\tau_\xi = \tau/(m_B c^2)$ and the function $\epsilon_\xi(\xi)$ representing the acceptance as a function of $\xi$, the p.d.f.'s for $\xi$ are:

$$f(\xi) = \frac{1}{C}\,\mathrm{e}^{-\frac{\xi'}{\tau_\xi}} \left[\left(1+|\lambda|^2\right)\cosh\left(\frac{\Delta\Gamma\, m_B\, c^2}{2}\xi'\right) - 2\mathrm{Re}\lambda \sinh\left(\frac{\Delta\Gamma\, m_B\, c^2}{2}\xi'\right)\right] \otimes R_\xi(\xi-\xi')\epsilon_\xi(\xi) \tag{38}$$

for the decays of neutral $B$ mesons to $\mathcal{CP}$ eigenstates,

$$f(\xi) = \frac{1}{C'}\,\mathrm{e}^{-\frac{\xi'}{\tau_\xi}} \cosh\left(\frac{\Delta\Gamma\, m_B\, c^2}{2}\xi'\right) \otimes R_\xi(\xi-\xi')\epsilon_\xi(\xi) \tag{39}$$

for the decays of neutral $B$ mesons to flavour specific states, and finally

$$f(\xi) = \frac{1}{C''}\,\mathrm{e}^{-\frac{\xi'}{\tau_\xi}} \otimes R_\xi(\xi-\xi')\epsilon_\xi(\xi) \tag{40}$$

for $\Lambda_b$ decays. The factors $C$, $C'$ and $C''$ provide the normalization of the p.d.f.'s to 1.

The resolution function $R_\xi(\Delta\xi)$ is simply defined as the difference between the measured value of the observable $\xi$ and its true value. By analyzing offline-selected events, it is shown that the distribution is well described by the sum of three Gaussian p.d.f.'s with common mean:

$$R_\xi(\Delta\xi) = \frac{f_1}{\sqrt{2\pi}\sigma_1}\mathrm{e}^{-\frac{(\Delta\xi-b_\xi)^2}{2\sigma_1^2}} + \frac{1-f_1-f_{tail}}{\sqrt{2\pi}\sigma_2}\mathrm{e}^{-\frac{(\Delta\xi-b_\xi)^2}{2\sigma_2^2}} + \frac{f_{tail}}{\sqrt{2\pi}\sigma_3}\mathrm{e}^{-\frac{(\Delta\xi-b_\xi)^2}{2\sigma_{tail}^2}}. \tag{41}$$

The distributions of the resolutions on $\xi$ with the results of an unbinned likelihood fit superimposed are shown in Fig. 11. The numerical results of the fits are summarized in Tab. 19.

It can be seen that the fit results for every decay mode are consistent within the statistical errors, but they also show a biased $\xi$ distribution of about 0.6 fs/GeV. By multiplying this value by the masses of the $B$ hadrons, it translates to an average bias on the proper time measurement of 3.2 fs. Although we do not expect such a small bias to have sizable impacts on $\mathcal{CP}$ violation measurements, its origin is still not understood and deserves investigations.

As far as the acceptance function $\epsilon_\xi(\xi)$ is concerned, its shape is not easily predictable from first principles. However, methods for extracting the acceptance function from data are being developed, as we will see in Sec. 4.2. Here we introduce an analytical expression that we have determined by means of full MC studies. We have found empirically that the following equation provides a very good parameterization:

$$\epsilon_\xi(\xi) \propto 1 - \mathrm{Erf}\left(\frac{c-\xi}{q\xi}\right), \tag{42}$$

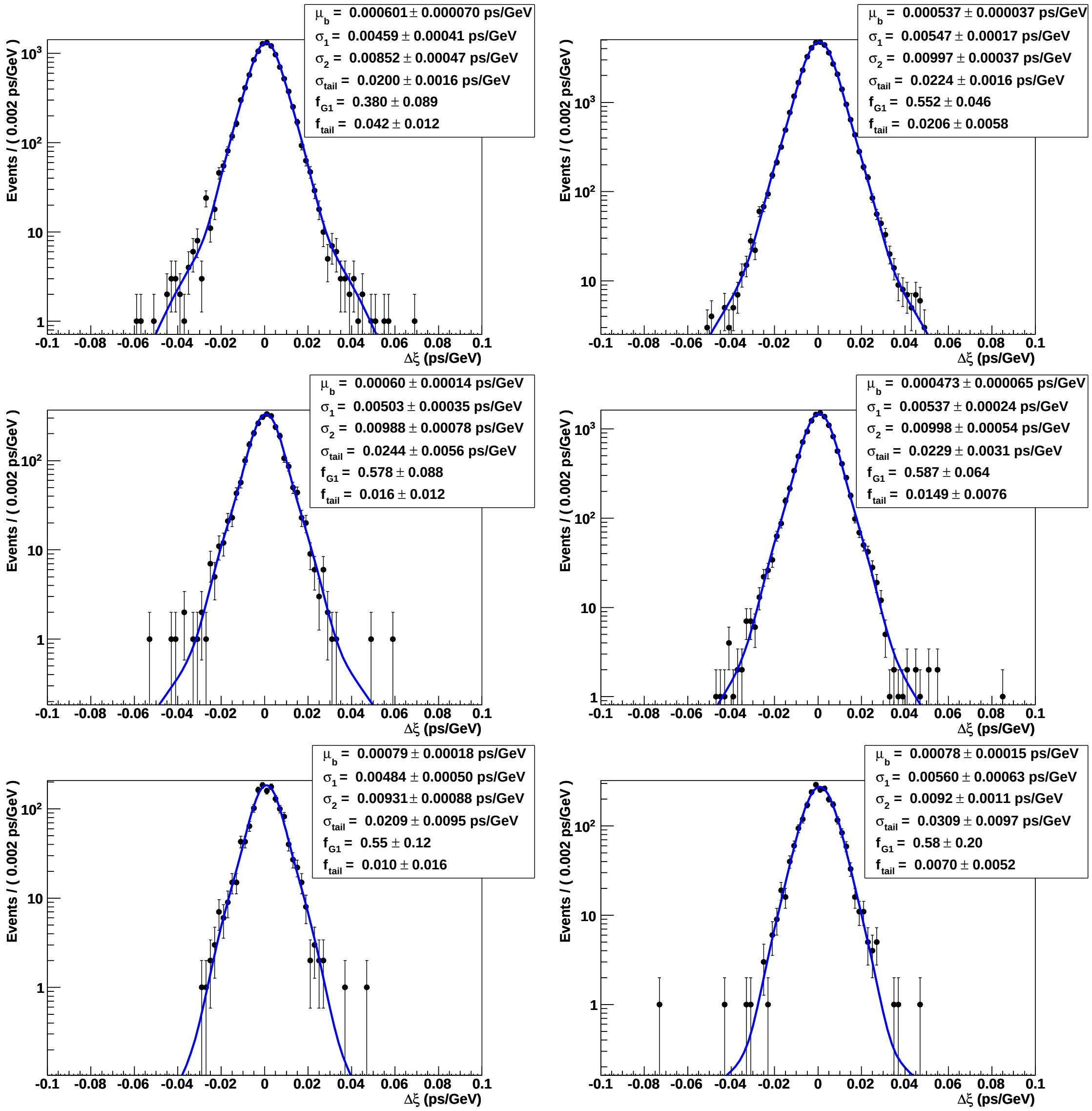

Figure 11: Distributions of the resolutions on $\xi$ with superimposed results of an unbinned likelihood fit to offline-selected events using a triple Gaussian p.d.f.: $B^0 \to \pi^+\pi^-$ (top left), $B^0 \to K^+\pi^-$ (top right), $B_s^0 \to \pi^+K^-$ (middle left), $B_s^0 \to K^+K^-$ (middle right), $\Lambda_b \to p\pi^-$ (bottom left) and $\Lambda_b \to pK^-$ (bottom right).

| Channel | $b_\xi$ [fs/GeV] | $f_1$ | $f_{tail}$ | $\sigma_1$ [fs/GeV] | $\sigma_2$ [fs/GeV] | $\sigma_{tail}$ [fs/GeV] |
|---|---|---|---|---|---|---|
| $B^0 \to \pi^+\pi^-$ | $0.60 \pm 0.07$ | $0.38 \pm 0.09$ | $0.042 \pm 0.012$ | $4.6 \pm 0.4$ | $8.5 \pm 0.5$ | $20 \pm 2$ |
| $B^0 \to K^+\pi^-$ | $0.54 \pm 0.04$ | $0.55 \pm 0.05$ | $0.021 \pm 0.006$ | $5.5 \pm 0.2$ | $10.0 \pm 0.4$ | $22 \pm 2$ |
| $B_s^0 \to \pi^+K^-$ | $0.60 \pm 0.14$ | $0.58 \pm 0.09$ | $0.016 \pm 0.012$ | $5.0 \pm 0.4$ | $9.9 \pm 0.8$ | $24 \pm 6$ |
| $B_s^0 \to K^+K^-$ | $0.47 \pm 0.07$ | $0.59 \pm 0.06$ | $0.015 \pm 0.008$ | $5.4 \pm 0.2$ | $10.0 \pm 0.5$ | $23 \pm 3$ |
| $\Lambda_b \to p\pi^-$ | $0.79 \pm 0.18$ | $0.54 \pm 0.12$ | $0.010 \pm 0.016$ | $4.8 \pm 0.5$ | $9.3 \pm 0.9$ | $21 \pm 10$ |
| $\Lambda_b \to pK^-$ | $0.78 \pm 0.15$ | $0.58 \pm 0.20$ | $0.007 \pm 0.005$ | $5.6 \pm 0.6$ | $9.2 \pm 1.1$ | $31 \pm 10$ |

Table 19: Numerical results of the likelihood fits of a triple Gaussian p.d.f. to the resolution on $\xi$ for offline-selected events.

| Channel | $\tau_\xi$ [ps/GeV] | $c$ [ps/GeV] | $q$ | $\tau_{\xi\,MC}$ [ps/GeV] |
|---|---|---|---|---|
| $B^0 \to \pi^+\pi^-$ | $0.286 \pm 0.004$ | $0.128 \pm 0.002$ | $0.47 \pm 0.03$ | 0.291 |
| $B^0 \to K^+\pi^-$ | $0.289 \pm 0.002$ | $0.131 \pm 0.001$ | $0.47 \pm 0.01$ | 0.291 |
| $B_s^0 \to \pi^+K^-$ | $0.284 \pm 0.008$ | $0.133 \pm 0.005$ | $0.48 \pm 0.06$ | 0.272 |
| $B_s^0 \to K^+K^-$ | $0.266 \pm 0.003$ | $0.130 \pm 0.002$ | $0.49 \pm 0.02$ | 0.272 |
| $\Lambda_b \to p\pi^-$ | $0.208 \pm 0.008$ | $0.129 \pm 0.007$ | $0.48 \pm 0.07$ | 0.219 |
| $\Lambda_b \to pK^-$ | $0.222 \pm 0.007$ | $0.130 \pm 0.006$ | $0.50 \pm 0.06$ | 0.219 |

Table 20: Numerical results of the likelihood fits of the p.d.f.'s given in Eqs. (38), (39) and (40) to the $\xi$ distributions for offline-selected events. For comparison, the last column reports the value of $\tau_\xi$ used in the MC simulation.

| Channel | $\tau_\xi$ [ps/GeV] | $c$ [ps/GeV] | $q$ | $\tau_{\xi\,MC}$ [ps/GeV] |
|---|---|---|---|---|
| $B^0 \to \pi^+\pi^-$ | $0.289 \pm 0.006$ | $0.144 \pm 0.003$ | $0.34 \pm 0.03$ | 0.291 |
| $B^0 \to K^+\pi^-$ | $0.287 \pm 0.003$ | $0.149 \pm 0.002$ | $0.38 \pm 0.02$ | 0.291 |
| $B_s^0 \to \pi^+K^-$ | $0.272 \pm 0.013$ | $0.159 \pm 0.011$ | $0.49 \pm 0.10$ | 0.272 |
| $B_s^0 \to K^+K^-$ | $0.273 \pm 0.005$ | $0.143 \pm 0.003$ | $0.34 \pm 0.03$ | 0.272 |
| $\Lambda_b \to p\pi^-$ | $0.217 \pm 0.012$ | $0.144 \pm 0.008$ | $0.37 \pm 0.07$ | 0.219 |
| $\Lambda_b \to pK^-$ | $0.222 \pm 0.011$ | $0.146 \pm 0.008$ | $0.38 \pm 0.08$ | 0.219 |

Table 21: Numerical results of the likelihood fits of the p.d.f. given in Eqs. (38), (39) and (40) to the $\xi$ distributions for triggered and offline-selected events. For comparison, the last column reports the value of $\tau_\xi$ used in the MC simulation.

where $c$ and $q$ are two free parameters, and Erf(x) is the error function given by:

$$\mathrm{Erf}(x) = \frac{2}{\sqrt{\pi}} \int_0^x \mathrm{e}^{-t^2} dt. \tag{43}$$

By using this functional form for the acceptance, we performed unbinned likelihood fits of the p.d.f.'s given by Eqs. (38), (39) and (40) to the $\xi$ distributions both for offline-selected events and for triggered and offline-selected events. We have left free to be fitted not only the parameters $c$ and $q$, which define the shape of the acceptance function, but also the parameter $\tau_\xi$, i.e. apart from a multiplicative factor, the average lifetime. We can then compare the fitted values of $\tau_\xi$ with the full MC values used to generate the events. The distributions of $\xi$, with the fit results superimposed, are shown in Figs. 12 and 13 for offline-selected and triggered and offline-selected events respectively. The numerical results of the fits are summarized in Tabs. 20 and 21.

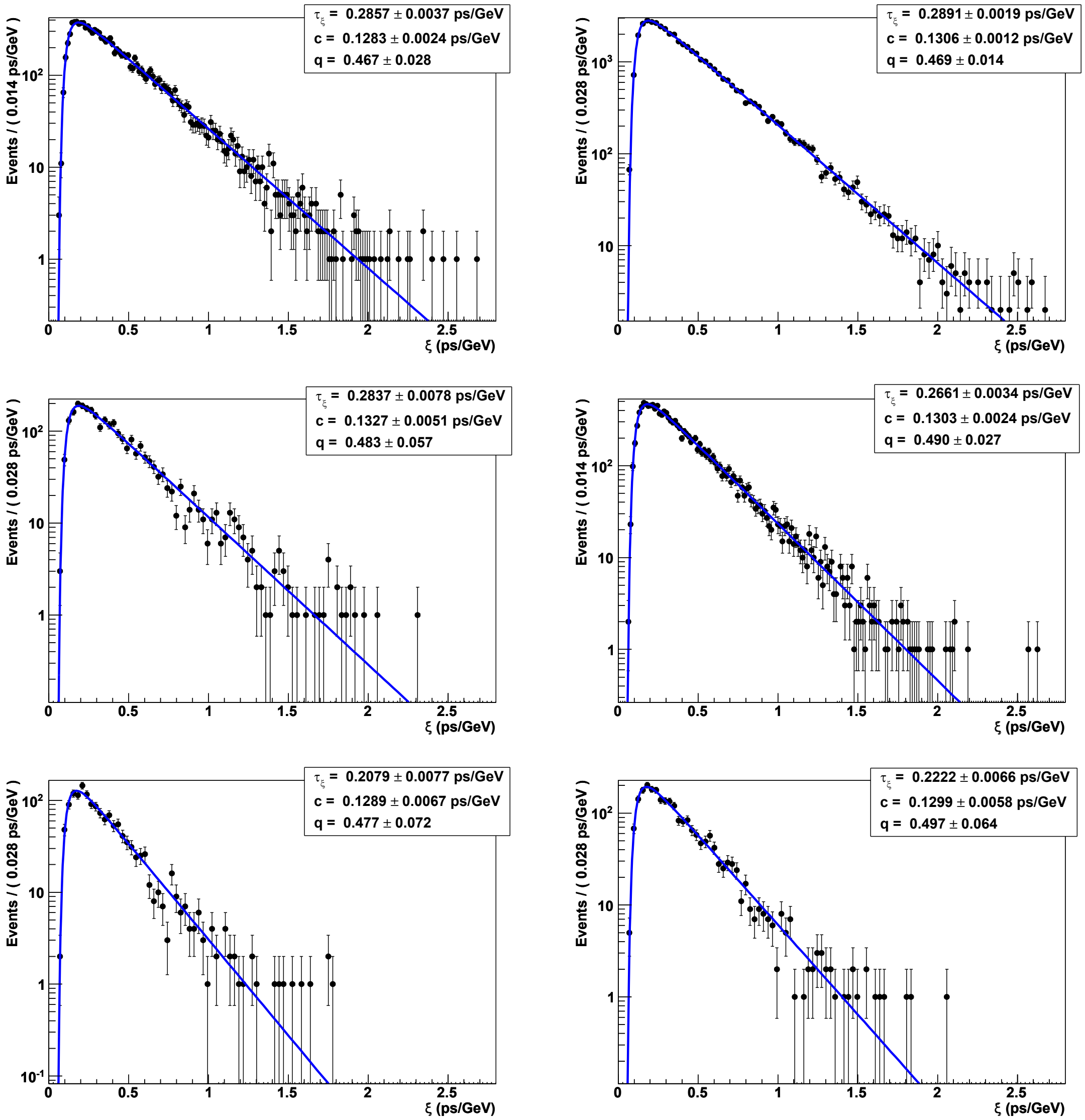

Figure 12: Distributions of $\xi$ for offline-selected events, with superimposed the result of an unbinned likelihood fit using the p.d.f.'s defined in Eqs. (38), (39) and (40): $B^0 \to \pi^+\pi^-$ (top left), $B^0 \to K^+\pi^-$ (top right), $B_s^0 \to \pi^+K^-$ (middle left), $B_s^0 \to K^+K^-$ (middle right), $\Lambda_b \to p\pi^-$ (bottom left) and $\Lambda_b \to pK^-$ (bottom right).

It can be seen that all the fitted values of the $\tau_\xi$ parameter are consistent within the statistical errors with the MC, a clear indication of the goodness of the acceptance function. However, viewed globally there is a tendency for the fit values to lie on the lower side. For each sample of offline-selected events or triggered and offline-selected events, the values of the $c$ and $q$ parameters are compatible amongst the various decay modes. Comparison between Tabs. 20 and 21 reveals that the values of $c$ and $q$ change as a result of applying the trigger, which modifies the shape of the acceptance function in a similar way mode to mode. As an example, the acceptance $\epsilon_\xi(\xi)$ with the values of the $c$ and $q$ parameters fixed to the central values determined from the fit, is plotted in Fig. 14 for offline-selected and for triggered and offline-selected $B^0 \to K^+\pi^-$ events.

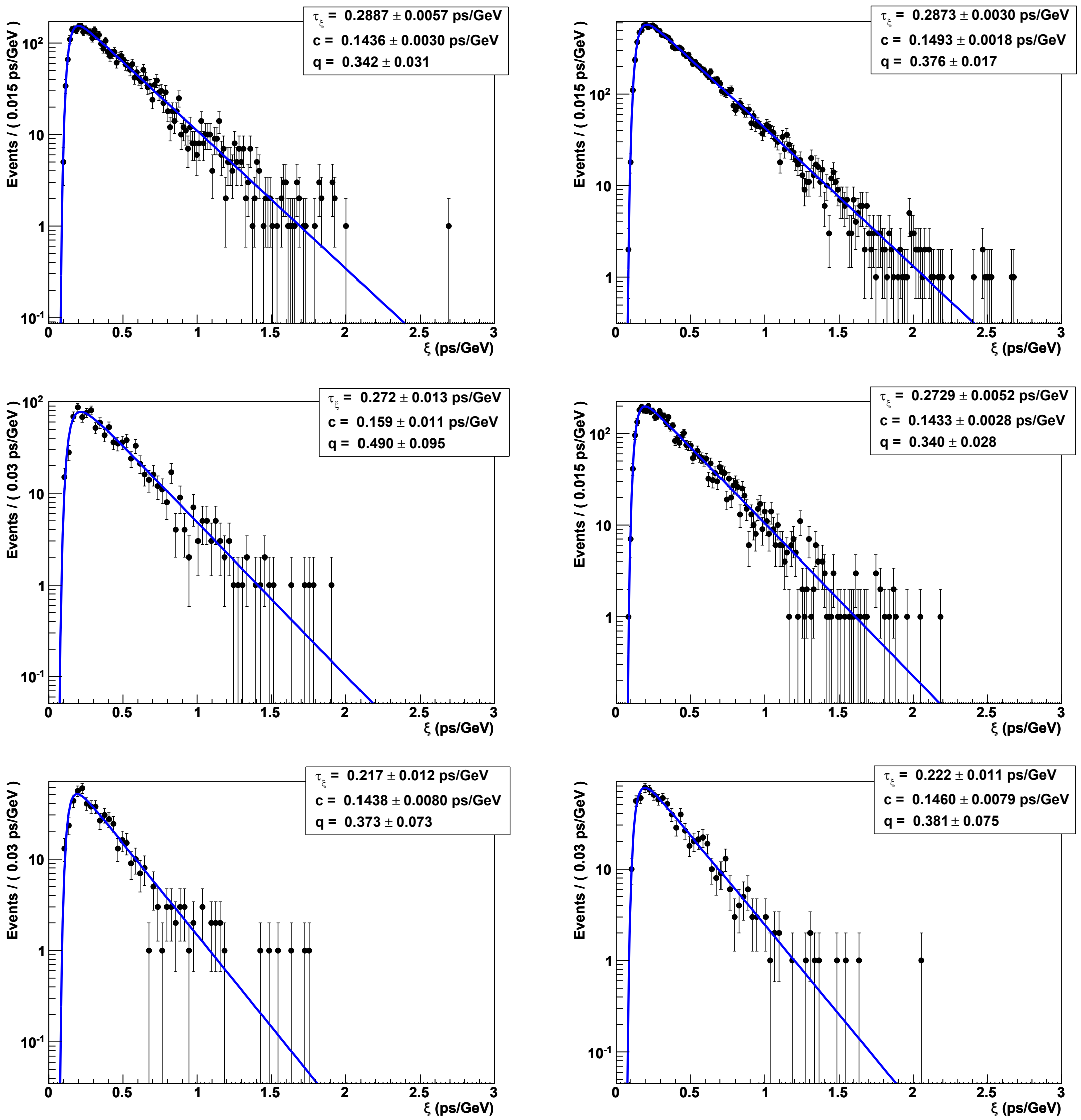

Figure 13: Distributions of $\xi$ for triggered and offline-selected events, with superimposed the result of an unbinned likelihood fit using the p.d.f.'s defined in Eqs. (38), (39) and (40): $B^0 \to \pi^+\pi^-$ (top left), $B^0 \to K^+\pi^-$ (top right), $B_s^0 \to \pi^+K^-$ (middle left), $B_s^0 \to K^+K^-$ (middle right), $\Lambda_b \to p\pi^-$ (bottom left) and $\Lambda_b \to pK^-$ (bottom right).

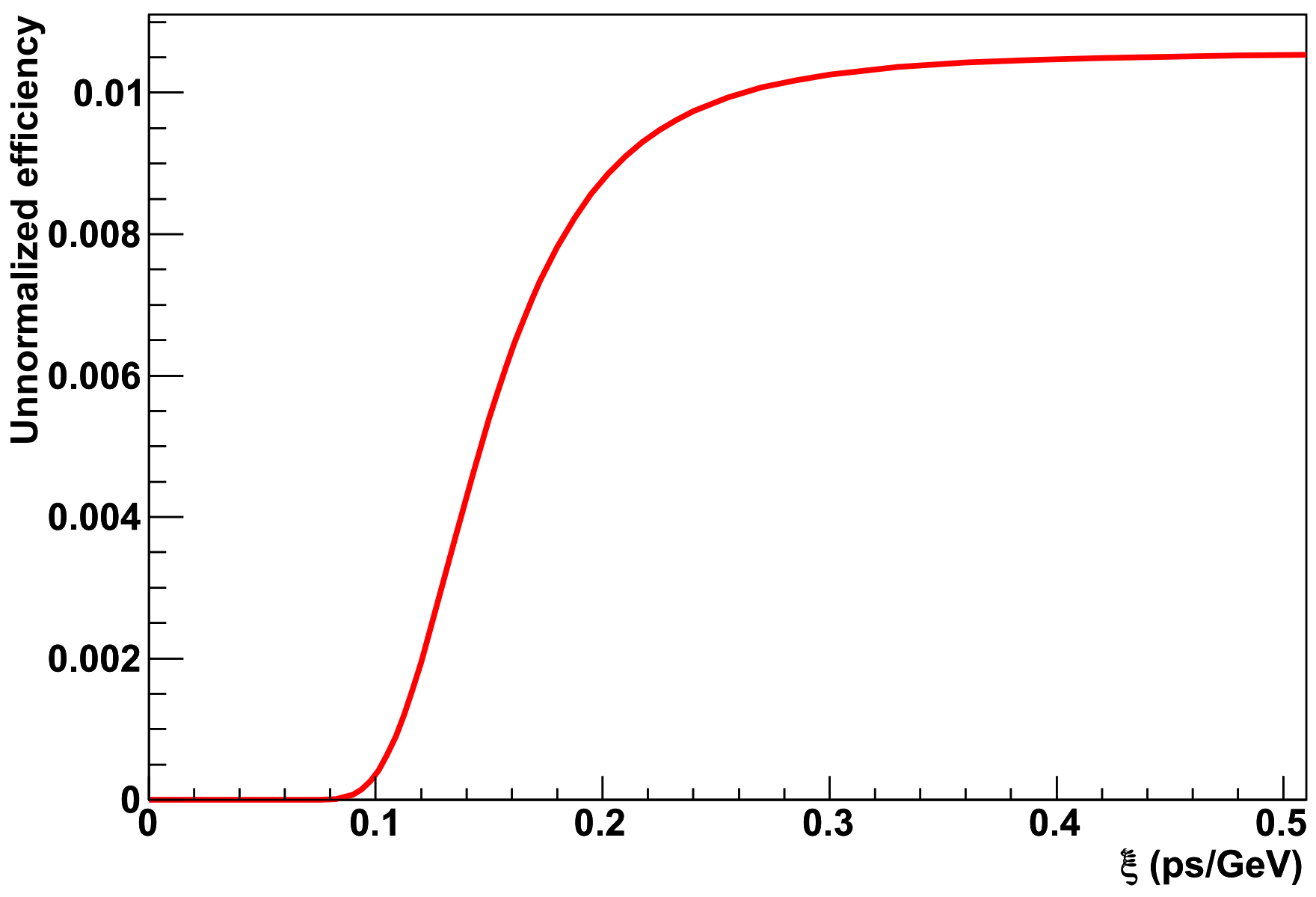

Figure 14: Shape of the function $\epsilon_\xi(\xi)$ for $B^0 \to K^+\pi^-$ triggered and offline-selected events, fixing the $c$ and $q$ parameters to the central values determined by the unbinned likelihood fit of the $\xi$ distribution, as described in the text. We have plotted just the low $\xi$ region, where the curve shows a non-trivial behaviour. For larger $\xi$ values it tends asymptotically to a constant value.

Although the proper time resolution model discussed in this section is solely based on MC studies, it will be also possible to obtain useful information from data. As detailed in Ref. [47], it is expected that the fit to the $B_s^0 \to \pi^+K^-$ proper time distribution will start providing constraints to the study of the proper time resolution model as the available statistics will exceed an integrated luminosity $L = 2\,\text{fb}^{-1}$.

## 4.2 Lifetime measurements

Measurement of the lifetime in the $\mathcal{CP}$ specific mode $B_s^0 \to K^+K^-$ provides important information about the $B_s^0$ system. If the behaviour of the SM is assumed, the measured lifetime can be used to extract information on $\Delta\Gamma_s$ and consequently to put constraints on NP contributions to the $B_s^0$ mixing phase [48].

While $\mathcal{CP}$ violation measurements are not severely affected by a limited knowledge of the acceptance as a function of the proper decay time, and consequently the modelization of the acceptance outlined in the previous section can suffice, a good control of the acceptance is a mandatory requirement for performing absolute lifetime measurements. To this aim, two independent approaches have been identified.

One approach is based on the standard $H_b \to h^+h'^-$ selection discussed in Sec. 3 and uses a MC free method to extract event-by-event acceptance functions from data [49, 50].

An important point in this context is the treatment of combinatorial background events. As it is not obvious which lifetime distribution events from combinatorial background will follow, no generic model can be used to calculate the background lifetime probability. Therefore, a method for modelling the lifetime distribution of combinatorial background events has been developed [51]. It consists of measuring the lifetime distribution of all events, and then of subtracting the contribution from signal and specific backgrounds.

A second approach relies instead on an event selection based on the application of particle identification cuts already in the HLT, in order to eliminate any distance of flight and impact parameter cut, thus avoiding any lifetime bias in the fit. Such a selection can basically rely on cuts on transverse momenta of the particles and on strong cuts on the $\Delta \log \mathcal{L}$ particle identification observables. A proof of principle of the validity of this approach can be found in Ref. [52], although the difficult problem of developing a trigger code able to process the RICH information at the HLT level within very limited time budget is still being worked on.

Complementary to the absolute lifetime measurements, also a method aiming to perform relative measurements has been studied in detail [53]. This method allows to remove the systematic bias introduced by the use of online and offline selections based on distance of flight and impact parameter cuts by fitting the ratio of the proper lifetime distributions of the decay modes $B_s^0 \to K^+K^-$ and $B^0 \to K^+\pi^-$, using the best current measurement of the mean $B^0$ lifetime. The technique is also capable of measuring the mean $B_s^0$ lifetime through the channel $B_s^0 \to \pi^+K^-$ and the lifetime of the $\Lambda_b$ baryon through the channels $\Lambda_b \to pK^-$ and $\Lambda_b \to p\pi^-$.

# 5 Particle Identification calibration

The particle identification (PID) at LHCb is based on the information provided by the RICH system [16], the electromagnetic and hadronic calorimeters [54], including preshower and scintillator pad detectors, and the muon chambers [55]. The variable used to discriminate between different particle hypotheses is $\Delta \log \mathcal{L}_{AB}$, that for each track is defined as:

$$\Delta \log \mathcal{L}_{AB} = \log \mathcal{L}_A - \log \mathcal{L}_B, \qquad (44)$$

where $\mathcal{L}_A$ and $\mathcal{L}_B$ are the likelihoods for particle hypotheses $A$ and $B$ respectively.

The high production rate of $D^*$ mesons at the LHC, both from $B$ decays and from prompt production, and the kinematical characteristics of the $D^* \to D\pi$ decay chain make $D^* \to D(K\pi)\pi$ events an ideal calibration sample for particle identification studies of kaons and pions. In this decay we designate the pion from the $D^*$ decay as the "slow pion", on account of its low energy in the $D^*$ rest frame.

An offline selection has been developed and optimized on a simulated sample of $B \to D^*X,\ D^* \to D^0\pi,\ D^0 \to hh'$ events, which makes no use of RICH information and therefore provides a sample which is suitable for calibration of the hadron identification. The selection cuts are shown in Tab. 22. When applying these cuts each candidate $D$ decay particle is attributed the pion and kaon masses in turn. The most important criterion is that which exploits the small mass difference between the $D^*$ and $D^0$ mesons.

| | | |
|---|---|---|
| $D^0$ | $p$ $\pi$, $K$ [GeV/c] | >2 |
| | $p_T$ $\pi$, $K$ [GeV/c] | >0.3 |
| | IP sig. $\pi$, $K$ | >3 |
| | $\chi^2$ $D^0$ vertex | <16 |
| | $D^0$ mass [GeV/c$^2$] | [1.84, 1.89] |
| | $p_T$ $D^0$ [GeV/c] | >1.250 |
| $D^*$ | IP sig. slow $\pi$ | >1. |
| | Dist. PV - $D^0$ DV sig. | >6 |
| | $\chi^2$ $D^*$ vertex | <16 |
| | $p_T$ $D^*$ [GeV/c] | >1.250 |
| | $D^*$ mass [GeV/c$^2$] | [1.985, 2.035] |
| | mass($D^* - D^0$) [GeV/c$^2$] | [0.1449, 0.1465] |

Table 22: Selection cuts for the $D^* \to D(K\pi)\pi$ pion and kaon calibration sample. No RICH information is used. All the particles are in turn attributed kaon mass and then pion mass.

For the studies presented here the candidate $D^*$ decays are obtained by running the selection cuts on 2 million $B^0 \to D^*X$ events. The background contamination in this sample after selection cuts is small ($\simeq 2\%$). However, when using real data it will be necessary to take into account the small but finite fraction of fake $D^*$ and $D^0$ candidates. This background correction will be made using sideband information. The sample obtained in real data will also receive a contribution from prompt $D^*$ events, which will lead to small differences in the kinematical properties of the candidate pions and kaons. In comparing with the physics signal, we have used 1.3 million simulated $B^0 \to \pi^+\pi^-$ events.

The RICH particle identification performance varies both as a function of momentum $p$ and of the transverse momentum $p_T$. As is apparent in Fig. 15, the $p$ and $p_T$ distributions of signal particles in the $H_b \to h^+h'^-$ sample are different from those of the particles in the calibration sample. This means that the $\Delta \log \mathcal{L}$ distributions are not expected to be identical for the two samples, as can be seen in Fig. 16 (top left) which presents the $\Delta \log \mathcal{L}_{\pi K}$ distribution for pions from $B$ and $D^*$ decays integrated over $p$ and $p_T$. However, when compared in bins of $p$ and $p_T$, the distributions look consistent. Therefore, by reweighting the $\Delta \log \mathcal{L}$ distributions in bins of $p$ and $p_T$, it becomes possible to use pions and kaons from the $D^*$ sample to calibrate the $H_b \to h^+h'^-$ analysis.

In the reweighting procedure we used 22 evenly spaced intervals in $p_T$ for the range $1\,\mathrm{GeV/c} < p_T < 12\,\mathrm{GeV/c}$, and 48 evenly spaced intervals in $p$ for the range $2\,\mathrm{GeV/c} < p < 98\,\mathrm{GeV/c}$. The entries in each bin for the calibration sample were summed with a weight which corresponds to the relative population of the bin in the physics signal sample. Fig. 17 (left) compares the $\Delta \log \mathcal{L}_{\pi K}$ distribution of pions from the $B^0 \to \pi^+\pi^-$ to that of pions from $D^*$, before and after reweighting. Fig. 17 (right) shows the relative difference between the calibration and signal samples before and after reweighting. It can be seen that, once reweighted, the calibration and signal distributions become more consistent, with a small difference around $\Delta \log \mathcal{L}_{\pi K} = -10$, which is still to be understood.

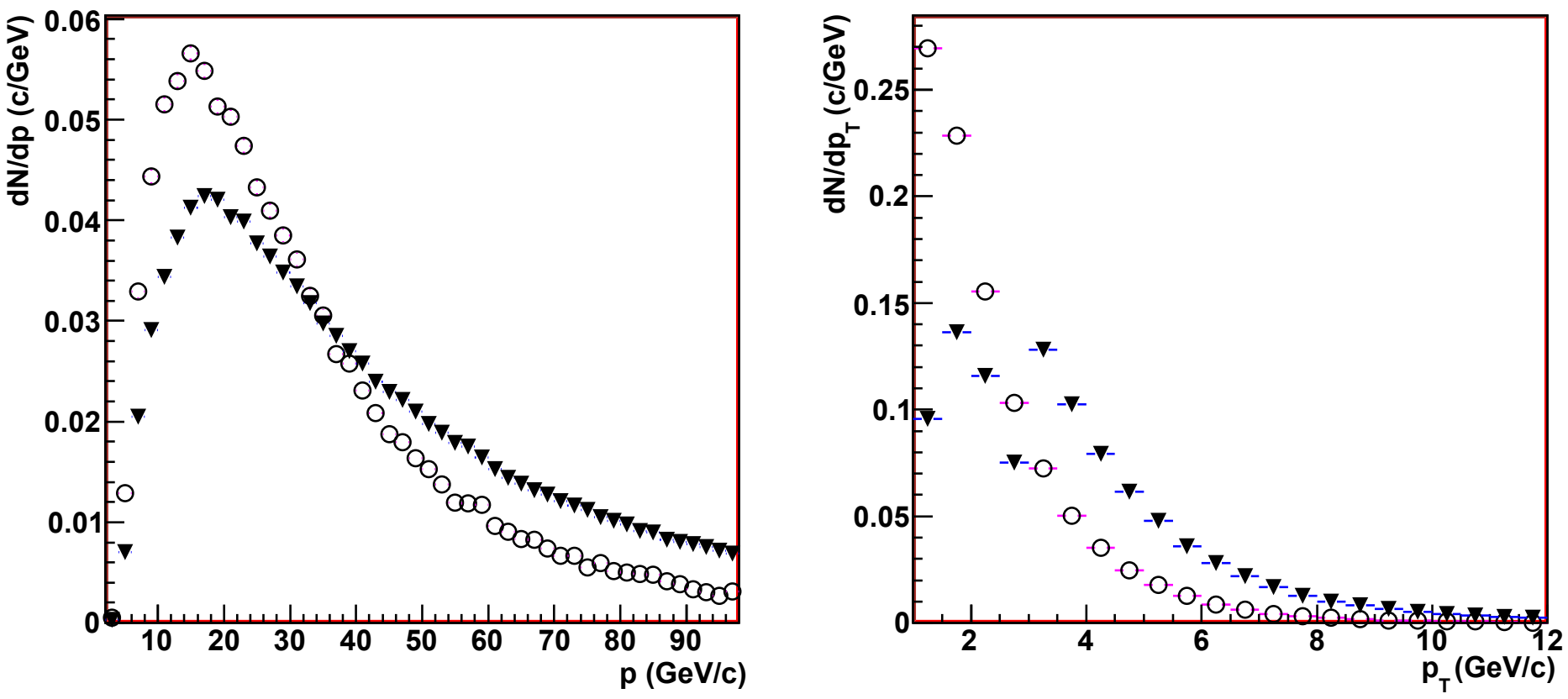

Figure 15: Momentum (left) and transverse momentum (right) distributions of pions from $B^0 \to \pi^+\pi^-$ (inverted triangles) and pions from $D^0$ (open circles).

It must be emphasized that such reweighting procedure requires the knowledge of the $p$ and $p_T$ distributions of the $B$ daughters, which will have to be studied with real data.

Following the method outlined in this section, it is also possible to calibrate the RICH system response for protons, information which is particularly important for studying the two $\Lambda_b \to ph^-$ decays. To this end, a sample of protons can be isolated by reconstructing large and pure samples of $\Lambda \to p\pi^-$ decays, selected on a kinematical basis without making use of any information coming from the particle identification systems.

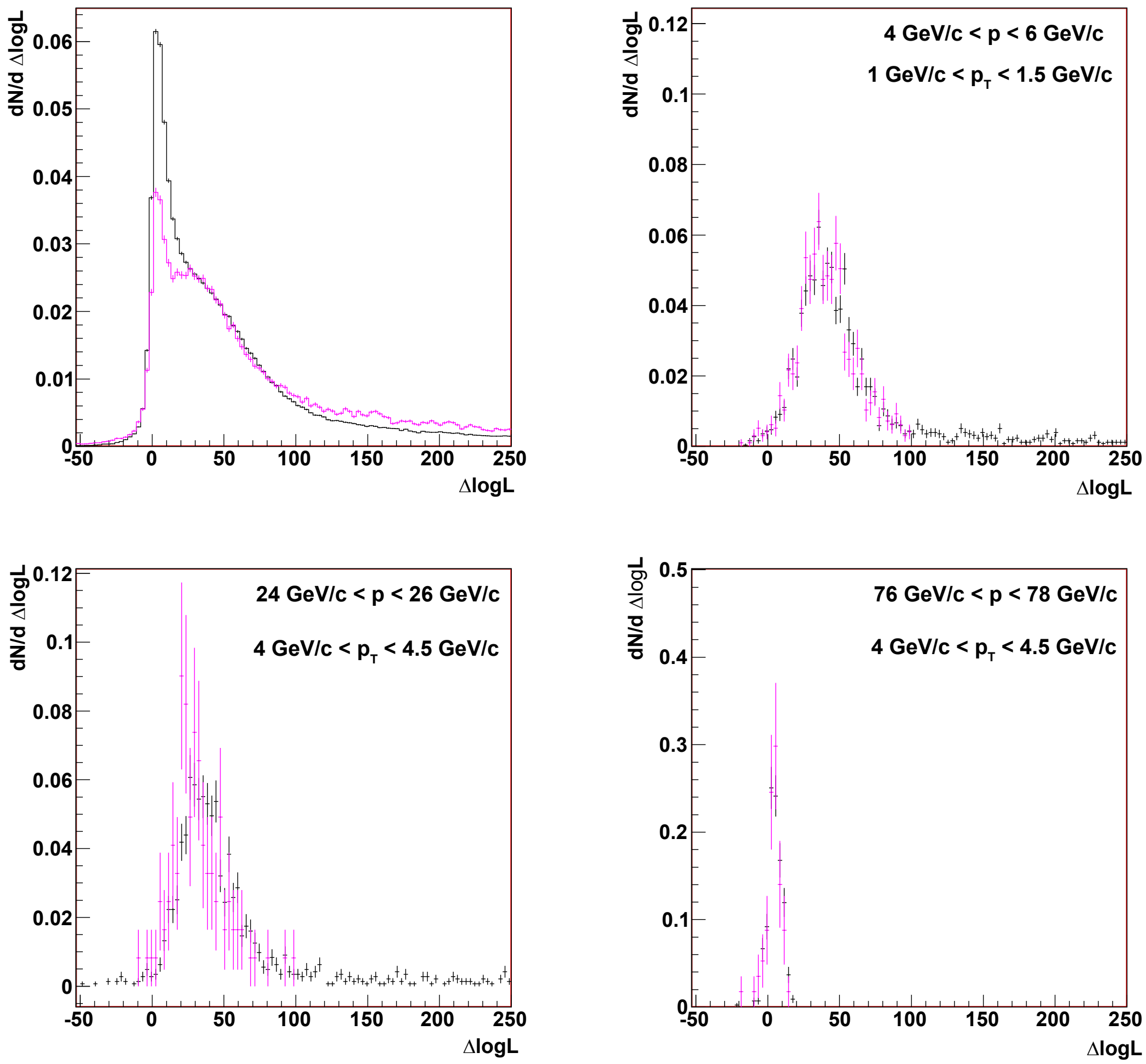

Figure 16: $\Delta \log \mathcal{L}_{\pi K}$ distributions for pions from $B^0 \to \pi^+\pi^-$ (black) and pions from $D^0$ decays (magenta) for all $p$ and $p_T$ (top-left) and selected bins of $p$ and $p_T$ (top right, bottom-left and bottom-right). The sizable difference between the two distributions, apparent in the top-left plot around $\Delta \log \mathcal{L} \simeq 0$, is mainly due to the longer tail of the momentum distribution for pions from $B^0 \to \pi^+\pi^-$ with respect to those from $D^0$ decays. In fact, for values of the momentum exceeding 100 GeV/c, the $\pi - K$ discriminating power of the RICH system starts dropping down.

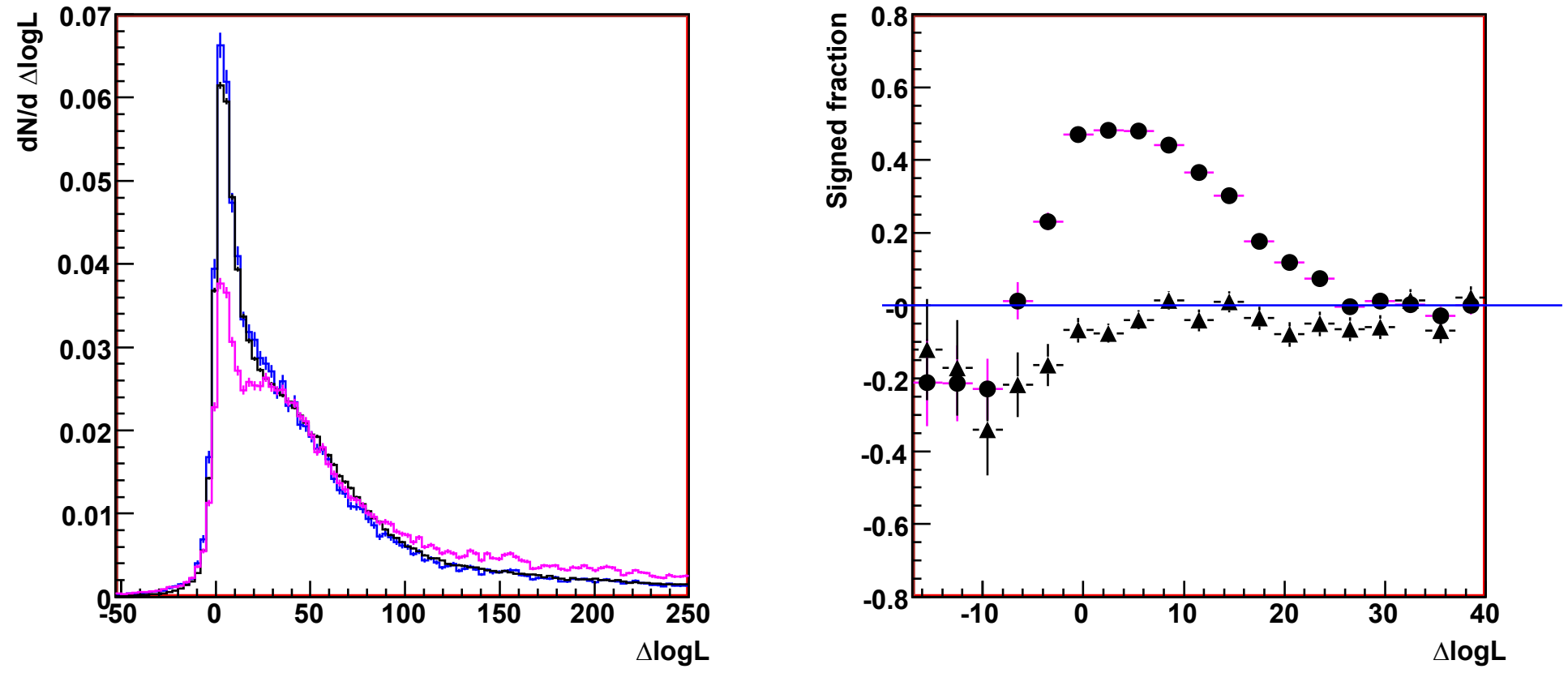

Figure 17: Left: $\Delta \log \mathcal{L}_{\pi K}$ distributions for pions from $B^0 \to \pi^+\pi^-$ (black) and pions from $D^*$, before (magenta) and after (blue) reweighting. Right: the relative difference between the $D^*$ and $B^0 \to \pi^+\pi^-$ $\Delta \log \mathcal{L}_{\pi K}$ distributions before (circles) and after (triangles) reweighting.

| Detector | Translations ($\mu$m) | | | Rotations (mrad) | | |
|---|---|---|---|---|---|---|
| | $\Delta_x$ | $\Delta_y$ | $\Delta_z$ | $R_x$ | $R_y$ | $R_z$ |
| VELO modules | 3 | 3 | 10 | 1.00 | 1.00 | 0.20 |
| VELO sensors | 3 | 3 | 10 | 1.00 | 1.00 | 0.20 |
| IT boxes | 15 | 15 | 50 | 0.10 | 0.10 | 0.10 |
| OT layers | 50 | 0 | 100 | 0.05 | 0.05 | 0.05 |

Table 23: Misalignment "$1\delta$" scales for the VELO modules and sensors, the IT boxes and OT layers.

# 6 Impact of misalignments

An accurate and efficient tracking system is of crucial importance to the success of the LHCb experiment. In this section misalignment effects of both the vertex locator and the inner and outer T-stations are discussed. A more detailed study can be found in Ref. [56].

Misalignment effects were examined as a function of a "misalignment scale". The scales were chosen to be roughly 1/3 of the detector single-hit resolution. This scale is termed "$1\delta$". Misalignments were then applied to each Vertex Locator (VELO) [14] module and sensor, each Inner Tracker (IT) [57] box and Outer Tracker (OT) [58] layer following a Gaussian distribution with a sigma corresponding to the $1\delta$ values. The list of these misalignment $1\delta$ scales is reported in Tab. 23. Ten sets of such $1\delta$ misalignments were generated. Likewise, this procedure was repeated with the creation of ten similar sets for each VELO module and sensor and each IT box and OT layer with misalignment scales increased by factors of 3 ($3\delta$) and 5 ($5\delta$). The study was performed with a sample of 20000 $B^0 \to \pi^+\pi^-$ perfectly aligned events (denoted "$0\delta$") and other three samples (corresponding to the $1\delta$, $3\delta$ and $5\delta$ scenarios) for the misalignments of both the VELO and the T-stations. Each sample consisted in reality of ten sub-samples of 2000 events, each of which was generated with a different set of the ten sets of a particular misalignment scenario. All the events were generated and digitized with a perfect geometry, and the misalignments were only introduced at the reconstruction level.

Tab. 24 summarizes how the various relevant resolutions deteriorate as the misalignment scale increases from $0\delta$ to $5\delta$. All the related studies in Ref. [56] indicate that the effects of the misalignments of the VELO and of the downstream IT and OT tracking stations are rather decoupled. In particular, the degradation in proper time resolution is driven by VELO misalignment, while it is the misalignment of the T-stations which is responsible for the worsening of the momentum and mass resolutions.

In addition to studying the effects of random misalignments, also the change of the VELO $z$-scale has been examined. This is of particular interest to lifetime measurements as it potentially introduces a direct bias in the measured proper time. A $z$-scaling effect could be expected from an expansion due to temperature variations of the VELO components, particularly the Aluminium base plate onto which the individual modules are screwed. However, the base plate is kept at a constant temperature of 20 °C by additional local heating. In addition, the scaling should be limited by the carbon-fibre constraint system that keeps the modules in place with a precision of 100 $\mu$m and which is less prone to temperature-induced expansion. A conservative estimate using a temperature change of 10 K yields a scaling in the $z$-direction of a factor $2 \cdot 10^{-5}$.

| Misalign. scenario | Resolutions | | | | | | | | | |
|---|---|---|---|---|---|---|---|---|---|---|
| | Momentum (%) | Mass (MeV/c$^2$) | Proper time (fs) | Primary vertex ($\mu$m) | | | $B^0$ vertex ($\mu$m) | | |
| | | | | $x$ | $y$ | $z$ | $x$ | $y$ | $z$ |
| $0\delta$ | 0.49 | 22.5 | 37.7 | 9 | 9 | 41 | 14 | 14 | 147 |
| $1\delta$ | 0.50 | 22.3 | 40.9 | 10 | 10 | 48 | 15 | 15 | 159 |
| $3\delta$ | 0.56 | 25.1 | 58.0 | 14 | 17 | 84 | 20 | 21 | 214 |
| $5\delta$ | 0.63 | 25.5 | 78.6 | 23 | 27 | 153 | 26 | 31 | 260 |

Table 24: Values of the resolutions on the $B^0$ daughter momentum, the $B^0$ mass, the $B^0$ proper time and on the position resolutions on the primary and the $B^0$ decay vertices for the different misalignment scenarios of both the VELO and the T-stations. The resolutions correspond to the sigmas of single-Gaussian fits.

To assess the influence of an incorrect knowledge of the VELO $z$-scale, four scenarios with different $z$-scales have been simulated and studied. For each scenario the $z$-position of each module has been changed according to the equation

$$z_{module} \rightarrow z_{module} \cdot (1 + h), \tag{45}$$

where $h$ takes the four values $\frac{1}{3} \cdot 10^{-4}$, $10^{-4}$, $\frac{1}{3} \cdot 10^{-3}$, and $10^{-3}$ for the four scenarios, respectively. For the first three $z$-scaling scenarios the observed changes in the resolutions of the various physics quantities of interest are minimal. Only for the largest $z$-scaling case a sizeable deterioration was observed, in particular of the proper time and vertex resolutions.

In conclusion of this section, we emphasize that in this study no assumptions based on the quality of the metrology or the expected performance of the alignment algorithms were made. However, the results here presented provide valuable information on how well alignment algorithms must behave in order to avoid sizable impacts of misalignments on the physics analysis.

# 7 Flavour tagging strategy and performance

An essential ingredient for performing $\mathcal{CP}$ violation measurements with $B$ decays to $\mathcal{CP}$ eigenstates is the tagging of the initial flavour of the reconstructed $B$ mesons. Flavour tagging in LHCb is performed by means of different algorithms, notably including opposite side (OS) and same side (SS) taggers [59, 60]. The terms "opposite" and "same" are a legacy of flavour tagging methods adopted at LEP and SLC, and merely refer to tracks associated with the signal $B$ and the other $B$ in the event, respectively.

The OS taggers determine the initial flavour of the signal $B$ meson by means of the charge of the lepton originating from semileptonic decays, either muon (OS muon tagger) or electron (OS electron tagger), and of the kaon from the $b \to c \to s$ transition (OS kaon tagger) of the opposite $B$. In addition, a fourth OS tagger exists, namely the vertex charge tagger, which is based on the inclusive reconstruction of the opposite $B$ decay vertex and on the computation of a weighted average of the charges of all tracks associated to that vertex [59].

The SS taggers instead are based on the determination of the charge of the pion (SS pion tagger) associated to the production of a $B^0$ meson during the hadronization phase of the $b\bar{b}$ pair, or of the kaon (SS kaon tagger) associated to the production of a $B^0_s$ meson. Contrary to the case of the OS taggers, that are equally applicable either if the signal is a $B^0$ or a $B^0_s$ meson, the SS pion tagger must be specifically applied to events where the signal is a $B^0$ while the SS kaon tagger to events where the signal is a $B^0_s$.

The decisions of all taggers must be properly combined in order to determine a unique tagging decision for each event. This is accomplished by estimating a probability for each tagger with the help of a *multi-layer perceptron* neural network technique [60]. For each tagger, the neural network combines several pieces of information such as momentum and transverse momentum, impact parameter significance and other kinematical variables of the tracks involved in the tagging procedure, and gives as output a quantity which is able to discriminate between the two possible tags. An event-by-event tagging probability is then obtained as a function of the neural network output, in general a polynomial of the first degree or higher, properly calibrated on a MC event sample. All the probabilities are finally combined into a single probability that the meson is a $B$ or a $\overline{B}$, within the approximation of independence of the various tagger responses. This way it is possible to take a unique decision and, furthermore, to estimate an event-by-event mistag rate. Although the training of the neural network, as well as the calibration of the tagging probability dependence on the neural network output, is performed on MC events, a proper recalibration on real data by means of control channels removes any residual dependence on the MC. Analogously, the approximation of independence of the various tagger responses employed to derive the combined decision and the event-by-event mistag rate should not affect the final results.

Furthermore, the knowledge of the event-by-event mistag allows for the categorization of events according to their tagging purity, hence improving the effective tagging efficiency $\epsilon_{eff}$. This is related to the tagging efficiencies $\epsilon_k$ and the mistag probabilities $\omega_k$ of the different exclusive tagging categories by:

$$\epsilon_{eff} = \sum_{k=1}^{n} \epsilon_k (1 - 2\omega_k)^2. \tag{46}$$

In the present implementation of the LHCb tagging strategy, $n = 5$ distinct categories is found to be an optimal choice. An increase beyond $n = 5$ would not lead to significant improvements in the effective tagging efficiency [60].

Ordered with increasing purity, the five categories are defined according to the following values of the estimated event-by-event mistag rate $\omega$:

- Category 1, if $\omega > 0.36$;
- Category 2, if $0.30 < \omega < 0.36$;
- Category 3, if $0.24 < \omega < 0.30$;
- Category 4, if $0.18 < \omega < 0.24$;
- Category 5, if $\omega < 0.18$.

Since we aim to make $\mathcal{CP}$ fits to the $H_b \to h^+h'^-$ sample as a whole, i.e. for a specific selected $B$ signal candidate we do not want to distinguish *a priori* whether it is a $B^0$, a $B_s^0$ or a $\Lambda_b$, we have to find a solution to the following problem: an appropriate SS tagger should be taken into account in the final tagging decision according to the signal $B$ hadron under consideration, i.e. for each event we cannot decide whether to use a SS pion tagger (for a $B^0$), a SS kaon tagger (for a $B_s^0$), or whether it makes no practical sense to use either of the two (for a $\Lambda_b$).

An advantageous feature of the $B^0_{(s)} \to h^+h'^-$ analyses is that the control channels needed to determine the mistag rates are selected by the same algorithm which is used for signal channels, i.e. the $B^0 \to K^+\pi^-$ is the control channel needed to measure the mistag rate to be used for the study of the time dependent $\mathcal{CP}$ asymmetry of the $B^0 \to \pi^+\pi^-$ decay, while the $B_s^0 \to \pi^+K^-$ decay is the natural control channel for the $B_s^0 \to K^+K^-$ decay. However, as is apparent in Tab. 13, while the $B^0 \to K^+\pi^-$ decay has a very rich event yield, the $B_s^0 \to \pi^+K^-$ decay is characterized by a much smaller statistics. However we expect, as the simulations clearly confirm, that the OS taggers give identical tagging efficiencies and mistag rates for $B^0$, $B_s^0$ and indeed also $\Lambda_b$ decays. It would be convenient then to determine the OS mistag rates for all the channels by exploiting the larger statistics of the $B^0 \to K^+\pi^-$ decay, differentiating amongst the various $B$ hadron species only for the SS taggers.

An effective way to deal with all these considerations consists of taking into account the responses of the OS taggers, the SS kaon tagger and the SS pion tagger in an exclusive way, i.e. first taking the OS tagger response, then the SS kaon tagger response only in the case the OS taggers did not give an answer, then the SS pion tagger response only if none of the previous steps gave an answer, obtaining in this way three mutually exclusive samples.

| $B^0 \to \pi^+\pi^-$ | | | |
|---|---|---|---|
| | $\epsilon_{eff}$ [%] | $\epsilon_{tag}$ [%] | $\omega_{tag}$ [%] |
| Opposite side taggers | | | |
| $\mu$ | $1.79 \pm 0.38$ | $9.09 \pm 0.44$ | $27.8 \pm 2.3$ |
| $e$ | $0.64 \pm 0.23$ | $3.05 \pm 0.26$ | $27.1 \pm 3.9$ |
| $K$ | $1.44 \pm 0.35$ | $15.21 \pm 0.55$ | $34.6 \pm 1.9$ |
| $Q_{vtx}$ | $1.33 \pm 0.35$ | $41.89 \pm 0.76$ | $41.1 \pm 1.2$ |
| Combination of opposite side taggers | | | |
| Cat. 1 | $0.21 \pm 0.14$ | $24.84 \pm 0.66$ | $45.4 \pm 1.5$ |
| Cat. 2 | $0.69 \pm 0.25$ | $7.18 \pm 0.40$ | $34.5 \pm 2.7$ |
| Cat. 3 | $0.41 \pm 0.19$ | $5.53 \pm 0.35$ | $36.3 \pm 3.1$ |
| Cat. 4 | $1.50 \pm 0.32$ | $4.09 \pm 0.30$ | $19.7 \pm 3.0$ |
| Cat. 5 | $1.64 \pm 0.32$ | $3.54 \pm 0.28$ | $16.0 \pm 3.0$ |
| Combined | $4.45 \pm 0.56$ | $45.17 \pm 0.76$ | $34.3 \pm 1.1$ |
| Same side kaon tagger | | | |
| | $0.05 \pm 0.07$ | $9.63 \pm 0.45$ | $53.7 \pm 2.5$ |
| Same side pion tagger | | | |
| | $0.67 \pm 0.24$ | $10.70 \pm 0.47$ | $37.5 \pm 2.3$ |
| All taggers | | | |
| | $5.17 \pm 0.61$ | $65.5 \pm 1.0$ | $36.0 \pm 0.8$ |

| $B^0 \to K^+\pi^-$ | | | |
|---|---|---|---|
| | $\epsilon_{eff}$ [%] | $\epsilon_{tag}$ [%] | $\omega_{tag}$ [%] |
| Opposite side taggers | | | |
| $\mu$ | $1.26 \pm 0.17$ | $8.54 \pm 0.23$ | $30.8 \pm 1.3$ |
| $e$ | $0.33 \pm 0.09$ | $2.99 \pm 0.14$ | $33.4 \pm 2.2$ |
| $K$ | $1.76 \pm 0.21$ | $15.53 \pm 0.30$ | $33.1 \pm 1.0$ |
| $Q_{vtx}$ | $1.29 \pm 0.18$ | $41.63 \pm 0.40$ | $41.2 \pm 0.6$ |
| Combination of opposite side taggers | | | |
| Cat. 1 | $0.17 \pm 0.07$ | $25.35 \pm 0.35$ | $46.0 \pm 0.8$ |
| Cat. 2 | $0.57 \pm 0.12$ | $7.02 \pm 0.21$ | $35.7 \pm 1.5$ |
| Cat. 3 | $1.03 \pm 0.15$ | $5.30 \pm 0.18$ | $28.0 \pm 1.6$ |
| Cat. 4 | $1.05 \pm 0.15$ | $4.20 \pm 0.16$ | $25.0 \pm 1.7$ |
| Cat. 5 | $1.28 \pm 0.15$ | $3.07 \pm 0.14$ | $17.7 \pm 1.8$ |
| Combined | $4.10 \pm 0.30$ | $44.95 \pm 0.41$ | $34.9 \pm 0.6$ |
| Same side kaon tagger | | | |
| | $0.03 \pm 0.03$ | $10.65 \pm 0.25$ | $52.8 \pm 1.2$ |
| Same side pion tagger | | | |
| | $0.34 \pm 0.09$ | $10.13 \pm 0.25$ | $40.8 \pm 1.3$ |
| All taggers | | | |
| | $4.47 \pm 0.31$ | $65.73 \pm 0.54$ | $37.0 \pm 0.5$ |

| $B_s^0 \to \pi^+K^-$ | | | |
|---|---|---|---|
| | $\epsilon_{eff}$ [%] | $\epsilon_{tag}$ [%] | $\omega_{tag}$ [%] |
| Opposite side taggers | | | |
| $\mu$ | $1.78 \pm 0.65$ | $7.95 \pm 0.73$ | $26.4 \pm 4.2$ |
| $e$ | $1.12 \pm 0.49$ | $3.40 \pm 0.49$ | $21.3 \pm 6.0$ |
| $K$ | $1.55 \pm 0.65$ | $16.99 \pm 1.01$ | $34.9 \pm 3.1$ |
| $Q_{vtx}$ | $1.31 \pm 0.61$ | $39.91 \pm 1.32$ | $40.9 \pm 2.1$ |
| Combination of opposite side taggers | | | |
| Cat. 1 | $0.43 \pm 0.35$ | $25.89 \pm 1.18$ | $43.6 \pm 2.6$ |
| Cat. 2 | $1.06 \pm 0.52$ | $6.72 \pm 0.67$ | $30.1 \pm 4.8$ |
| Cat. 3 | $0.83 \pm 0.46$ | $5.28 \pm 0.60$ | $30.1 \pm 5.4$ |
| Cat. 4 | $0.60 \pm 0.39$ | $4.19 \pm 0.54$ | $31.0 \pm 6.1$ |
| Cat. 5 | $1.88 \pm 0.56$ | $3.40 \pm 0.49$ | $12.8 \pm 4.9$ |
| Combined | $4.81 \pm 1.04$ | $45.48 \pm 1.34$ | $33.7 \pm 1.9$ |
| Same side kaon tagger | | | |
| | $1.41 \pm 0.61$ | $14.68 \pm 0.95$ | $34.5 \pm 3.3$ |
| Same side pion tagger | | | |
| | $0.09 \pm 0.16$ | $7.16 \pm 0.69$ | $44.4 \pm 5.0$ |
| All taggers | | | |
| | $6.31 \pm 1.21$ | $67.32 \pm 1.78$ | $34.7 \pm 1.5$ |

| $B_s^0 \to K^+K^-$ | | | |
|---|---|---|---|
| | $\epsilon_{eff}$ [%] | $\epsilon_{tag}$ [%] | $\omega_{tag}$ [%] |
| Opposite side taggers | | | |
| $\mu$ | $1.40 \pm 0.28$ | $7.66 \pm 0.34$ | $28.6 \pm 2.1$ |
| $e$ | $0.36 \pm 0.15$ | $2.69 \pm 0.21$ | $31.7 \pm 3.7$ |
| $K$ | $1.71 \pm 0.32$ | $15.32 \pm 0.47$ | $33.3 \pm 1.6$ |
| $Q_{vtx}$ | $1.57 \pm 0.32$ | $41.04 \pm 0.64$ | $40.2 \pm 1.0$ |
| Combination of opposite side taggers | | | |
| Cat. 1 | $0.53 \pm 0.19$ | $24.94 \pm 0.56$ | $42.7 \pm 1.3$ |
| Cat. 2 | $0.49 \pm 0.18$ | $6.72 \pm 0.32$ | $36.6 \pm 2.4$ |
| Cat. 3 | $1.07 \pm 0.25$ | $5.67 \pm 0.30$ | $28.3 \pm 2.4$ |
| Cat. 4 | $1.25 \pm 0.26$ | $4.33 \pm 0.26$ | $23.2 \pm 2.6$ |
| Cat. 5 | $1.00 \pm 0.21$ | $2.43 \pm 0.20$ | $17.9 \pm 3.2$ |
| Combined | $4.32 \pm 0.49$ | $44.09 \pm 0.64$ | $34.3 \pm 0.9$ |
| Same side kaon tagger | | | |
| | $1.83 \pm 0.33$ | $14.43 \pm 0.45$ | $32.2 \pm 1.6$ |
| Same side pion tagger | | | |
| | $0.00 \pm 0.00$ | $7.24 \pm 0.34$ | $49.9 \pm 2.4$ |
| All taggers | | | |
| | $6.15 \pm 0.59$ | $65.76 \pm 0.85$ | $34.7 \pm 0.7$ |

Table 25: Breakdown of effective tagging efficiencies, actual tagging efficiencies and mistag rates for each of the four $B^0_{(s)} \to h^+h'^-$ decay modes under study, calculated using triggered and offline-selected events.

The breakdown of tagging efficiencies for triggered and offline-selected events of each decay are reported in Tabs. 25 and 26. The tables show the tagging power, the tagging efficiency and the mistag probability for each OS tagger individually, as well as for the five OS tagging categories and the two SS tagging categories. The OS categories for all the channels, including the $\Lambda_b$ ones, are characterized by the same tagging efficiencies and mistag probabilities, within the statistical errors. The SS taggers, as expected, give the same tagging efficiencies and mistag probabilities for the two $B^0 \to h^+\pi^-$ decays, for the two $B_s^0 \to h^+K^-$ decays and for the two $\Lambda_b \to ph^-$ decays separately. While the SS pion tagging is just relevant for the $B^0 \to h^+\pi^-$ modes, giving a null effective tagging efficiency for all the other modes, the SS kaon tagging not only gives a large contribution to the effective tagging efficiency for the $B_s^0 \to h^+K^-$ decays, but also a small contribution by anti-tagging the $B^0 \to h^+\pi^-$ modes. In fact, the mistag probability for the SS kaon tagging turns out here to be slightly larger than 50%. This is due to the fact that the SS kaon tagger selects also a small fraction of OS kaons, which have opposite charge to that of a SS kaon, would it have existed. In contrast, the SS kaon tagger shows a mistag rate lower than 50% for the $\Lambda_b \to ph^-$ modes, and this can be interpreted as due to the misidentification of a proton produced nearby in phase space during the fragmentation process which gives rise to the $\Lambda_b$, hence mimicking a SS kaon tag.

| $\Lambda_b \to p\pi^-$ | | | |
|---|---|---|---|
| | $\epsilon_{eff}$ [%] | $\epsilon_{tag}$ [%] | $\omega_{tag}$ [%] |
| Opposite side taggers | | | |
| $\mu$ | $2.35 \pm 0.79$ | $7.33 \pm 0.77$ | $21.7 \pm 4.5$ |
| $e$ | $0.04 \pm 0.12$ | $3.18 \pm 0.52$ | $44.4 \pm 8.3$ |
| $K$ | $1.95 \pm 0.79$ | $14.93 \pm 1.06$ | $32.0 \pm 3.6$ |
| $Q_{vtx}$ | $2.39 \pm 0.90$ | $41.61 \pm 1.47$ | $38.0 \pm 2.2$ |
| Combination of opposite side taggers | | | |
| Cat. 1 | $0.80 \pm 0.53$ | $25.35 \pm 1.29$ | $41.1 \pm 2.9$ |
| Cat. 2 | $0.10 \pm 0.19$ | $6.27 \pm 0.72$ | $43.7 \pm 5.9$ |
| Cat. 3 | $0.74 \pm 0.49$ | $5.57 \pm 0.68$ | $31.7 \pm 5.9$ |
| Cat. 4 | $1.73 \pm 0.62$ | $3.53 \pm 0.55$ | $15.0 \pm 5.6$ |
| Cat. 5 | $2.47 \pm 0.63$ | $3.45 \pm 0.54$ | $7.7 \pm 4.3$ |
| Combined | $5.84 \pm 1.15$ | $44.17 \pm 1.48$ | $31.8 \pm 2.1$ |
| Same side kaon tagger | | | |
| | $0.32 \pm 0.33$ | $11.84 \pm 0.96$ | $41.8 \pm 4.3$ |
| Same side pion tagger | | | |
| | $0.36 \pm 0.35$ | $7.07 \pm 0.76$ | $61.2 \pm 5.4$ |
| All taggers | | | |
| | $6.52 \pm 1.25$ | $63.08 \pm 1.92$ | $33.9 \pm 1.6$ |

| $\Lambda_b \to pK^-$ | | | |
|---|---|---|---|
| | $\epsilon_{eff}$ [%] | $\epsilon_{tag}$ [%] | $\omega_{tag}$ [%] |
| Opposite side taggers | | | |
| $\mu$ | $2.26 \pm 0.52$ | $9.68 \pm 0.57$ | $25.9 \pm 2.7$ |
| $e$ | $0.63 \pm 0.28$ | $2.80 \pm 0.32$ | $26.3 \pm 5.1$ |
| $K$ | $1.77 \pm 0.49$ | $14.76 \pm 0.68$ | $32.7 \pm 2.3$ |
| $Q_{vtx}$ | $0.90 \pm 0.36$ | $43.03 \pm 0.95$ | $42.8 \pm 1.4$ |
| Combination of opposite side taggers | | | |
| Cat. 1 | $0.19 \pm 0.17$ | $26.09 \pm 0.84$ | $45.7 \pm 1.9$ |
| Cat. 2 | $0.63 \pm 0.30$ | $7.73 \pm 0.51$ | $35.7 \pm 3.3$ |
| Cat. 3 | $1.02 \pm 0.36$ | $5.26 \pm 0.43$ | $28.0 \pm 3.8$ |
| Cat. 4 | $1.19 \pm 0.37$ | $4.23 \pm 0.39$ | $23.5 \pm 4.0$ |
| Cat. 5 | $1.41 \pm 0.36$ | $2.91 \pm 0.32$ | $15.2 \pm 4.0$ |
| Combined | $4.45 \pm 0.72$ | $46.23 \pm 0.96$ | $34.5 \pm 1.4$ |
| Same side kaon tagger | | | |
| | $0.03 \pm 0.06$ | $10.16 \pm 0.58$ | $47.5 \pm 3.0$ |
| Same side pion tagger | | | |
| | $0.03 \pm 0.07$ | $7.77 \pm 0.51$ | $46.9 \pm 3.4$ |
| All taggers | | | |
| | $4.51 \pm 0.73$ | $64.16 \pm 1.23$ | $36.7 \pm 1.1$ |

Table 26: Breakdown of effective tagging efficiencies, actual tagging efficiencies and mistag rates for each of the two $\Lambda_b \to ph^-$ decay modes under study, calculated using triggered and offline-selected events.

In conclusion, the strategy described in this section consists of subdividing the sample of tagged events in seven categories, five for the OS tagging and two for the SS tagging. In the $\mathcal{CP}$ fits, described in the remainder of this document, the tagging efficiencies and mistag probabilities for the OS tagging categories will be assumed to be identical for all the $H_b \to h^+h'^-$ decay modes, while those of the SS pion and kaon tagging categories will be taken to be the same for the three pairs of channels $B^0 \to h^+\pi^-$, $B_s^0 \to h^+K^-$ and $\Lambda_b \to ph^-$ separately. As shown in Tab. 27, this corresponds to 11 tagging efficiencies and 11 mistag fractions, i.e. 22 parameters in total. The numbers in the table have been calculated over the union of the corresponding event samples. In the present analysis, the background components are assumed to be characterized by a 50% mistag probability, irrespectively of which tagging category the event belongs to.

| | $\epsilon_{eff}$ [%] | $\epsilon_{tag}$ [%] | $\omega_{tag}$ [%] |
|---|---|---|---|
| $H_b \to h^+h'^-$ Combination of opposite side taggers | | | |
| Cat. 1 | $0.26 \pm 0.06$ | $25.29 \pm 0.25$ | $44.9 \pm 0.6$ |
| Cat. 2 | $0.57 \pm 0.08$ | $7.01 \pm 0.15$ | $35.7 \pm 1.0$ |
| Cat. 3 | $0.91 \pm 0.10$ | $5.41 \pm 0.13$ | $29.5 \pm 1.1$ |
| Cat. 4 | $1.16 \pm 0.11$ | $4.19 \pm 0.11$ | $23.7 \pm 1.2$ |
| Cat. 5 | $1.35 \pm 0.11$ | $3.02 \pm 0.10$ | $16.6 \pm 1.2$ |
| Combined | $4.25 \pm 0.21$ | $44.92 \pm 0.28$ | $34.62 \pm 0.41$ |
| $B^0 \to h^+\pi^-$ Same side kaon tagger | | | |
| | $0.04 \pm 0.03$ | $10.42 \pm 0.22$ | $53.0 \pm 1.1$ |
| $B^0_s \to h^+K^-$ Same side kaon tagger | | | |
| | $1.74 \pm 0.29$ | $14.48 \pm 0.41$ | $32.6 \pm 1.4$ |
| $\Lambda_b \to ph^-$ Same side kaon tagger | | | |
| | $0.08 \pm 0.09$ | $10.65 \pm 0.50$ | $45.6 \pm 2.5$ |
| $B^0 \to h^+\pi^-$ Same side pion tagger | | | |
| | $0.41 \pm 0.09$ | $10.25 \pm 0.22$ | $40.0 \pm 1.1$ |
| $B^0_s \to h^+K^-$ Same side pion tagger | | | |
| | $0.00 \pm 0.01$ | $7.23 \pm 0.30$ | $48.9 \pm 2.2$ |
| $\Lambda_b \to ph^-$ Same side pion tagger | | | |
| | $0.00 \pm 0.02$ | $7.56 \pm 0.43$ | $50.9 \pm 2.9$ |

Table 27: Summary of the independent tagging efficiencies and mistag probabilities for the various decays under study, calculated for triggered and offline-selected events. The five OS tagging categories are characterized by a unique set of parameters for all the channels. The SS tagging categories take on a different value for each of the three pairs of modes $B^0 \to h^+\pi^-$, $B^0_s \to h^+K^-$ and $\Lambda_b \to ph^-$. The values are used as inputs to the fast MC $\mathcal{CP}$ sensitivity studies described in the remainder of this document.

# 8 Invariant mass line shape

An accurate description of the invariant mass distribution of each $H_b \to h^+h'^-$ decay mode is a key ingredient in order to disentagle the overlapped mass peaks of the various signal modes. In particular, this is extremely important for signals sharing the same final state signature, since they can only be separated on a kinematical basis. This is the case for $B^0 \to \pi^+\pi^-$ and $B_s^0 \to \pi^+\pi^-$, $B^0 \to K^+K^-$ and $B_s^0 \to K^+K^-$, and $B^0 \to K^+\pi^-$ and $B_s^0 \to \pi^+K^-$.

The question of the invariant mass spectrum parameterization is complicated by the fact that the decay products radiate photons due to QED final state radiation processes, hence leading to missing momentum that distorts the shape of the charged pair invariant mass. The net effect, as we shall see, will be the presence of a long tail on the lower side of the mass peak.

## 8.1 Parameterization under the correct mass hypothesis

The LHCb MC simulation includes the final state radiation by means of the PHOTOS generator [61], that is run on top of the EvtGen decay generator [62] and adds radiated photons to the decay tree. PHOTOS is a MC algorithm that simulates QED photon emissions in decays, by calculating $\mathcal{O}(\alpha)$ radiative corrections for charged particles using a leading log collinear approximation. Within the approximation, the program calculates the amount of bremsstrahlung radiation in the decay and modifies the final state according to the decay topology. A simplified algorithm like PHOTOS is commonly used since radiative corrections in heavy meson decays pose difficult theoretical problems, due to the lack of a universally valid effective theory.

However, it is possible to calculate and parameterize infra-red effects using the approximation of point-like hadrons. An important step in this direction can be found in Ref. [63], where the calculation has been performed for the non-leptonic decays of $B$ and $D$ mesons to two pseudo-scalar mesons. A successful attempt to employ the analytical QED calculation of Ref. [63] in order to obtain an accurate description of the invariant mass line shape for each $H_b \to h^+h'^-$ channel under study can be found in Ref. [64].

Although the parameterization discussed in Ref. [64] is well motivated on theoretical grounds and the fits to the invariant mass spectra give good results, its usage is not practical when performing high statistics toy MC studies of the sensitivity on $\mathcal{CP}$ violating observables, discussed in the remainder of this note, since the time needed for the numerical computation of convolution integrals would dominate the calculation of the likelihood function. As an alternative, we will make use of an approximated p.d.f. $f(m)$:

$$f(m) = f_E\, \Theta(m_B - m)\, \frac{1}{s} e^{-s(m_B-m)} + (1 - f_E)\, \delta(m_B - m), \tag{47}$$

where we have decomposed the mass distribution, in the absence of detector effects, into an exponential p.d.f. describing events where the emission of photons took place, plus a component with no emission given by a Dirac $\delta$ function. The factor $f_E$ weights the number of events where radiation occurred with respect to events with no radiation. From a physical point of view, such approximation corresponds to an underestimatation of the rate of soft photon emissions of low energies. Taking into account resolution effects, the mass p.d.f. $g(m)$ is given by:

$$g(m) = f_E\, E_d(m - m_B;\, f_1,\, \sigma_1,\, \sigma_2,\, s) + (1 - f_E)\, C\, G_d(m - m_B;\, f_1,\, \sigma_1,\, \sigma_2), \qquad (48)$$

where $C$ is a normalization factor, $G_d$ is a double Gaussian and $E_d$ describes the radiative tail of the distribution, defined as:

$$\begin{aligned} E_d(m - m_B;\, f_1,\, \sigma_1,\, \sigma_2,\, s) &= f_1\, K_1^{-1} e^{s(m-m_B)} \left[1 - \mathrm{Erf}\left(\frac{m - m_B + s\sigma_1^2}{\sqrt{2}\sigma_1}\right)\right] + \qquad (49)\\ &+ (1 - f_1)\, K_2^{-1} e^{s(m-m_B)} \left[1 - \mathrm{Erf}\left(\frac{m - m_B + s\sigma_2^2}{\sqrt{2}\sigma_2}\right)\right]. \end{aligned}$$

| Channel | $\mu$ [MeV/c$^2$] | $f_G$ | $\sigma_1$ [MeV/c$^2$] | $\sigma_2$ [MeV/c$^2$] | $f_E$ | $s$ [c$^2$/GeV] | $\mu_{MC}$ [MeV/c$^2$] |
|---|---|---|---|---|---|---|---|
| $B^0 \to \pi^+\pi^-$ | $5278.5 \pm 0.3$ | $0.832 \pm 0.018$ | $18.8 \pm 0.3$ | $47 \pm 2$ | $0.066 \pm 0.007$ | $9.6 \pm 1.2$ | 5279.4 |
| $B^0 \to K^+\pi^-$ | $5278.5 \pm 0.1$ | $0.835 \pm 0.010$ | $18.5 \pm 0.2$ | $46 \pm 1$ | $0.049 \pm 0.003$ | $9.1 \pm 0.7$ | 5279.4 |
| $B_s^0 \to \pi^+K^-$ | $5368.5 \pm 0.5$ | $0.86 \pm 0.02$ | $18.9 \pm 0.5$ | $54 \pm 4$ | $0.024 \pm 0.007$ | $3.6 \pm 2.6$ | 5369.6 |
| $B_s^0 \to K^+K^-$ | $5368.4 \pm 0.2$ | $0.838 \pm 0.017$ | $18.9 \pm 0.3$ | $45 \pm 2$ | $0.035 \pm 0.004$ | $8.4 \pm 1.1$ | 5369.6 |
| $\Lambda_b \to p\pi^-$ | $5623.5 \pm 0.6$ | $0.90 \pm 0.03$ | $19.1 \pm 0.7$ | $53 \pm 7$ | $0.053 \pm 0.014$ | $9.9 \pm 2.0$ | 5624.0 |
| $\Lambda_b \to pK^-$ | $5623.7 \pm 0.5$ | $0.83 \pm 0.03$ | $18.7 \pm 0.7$ | $47 \pm 4$ | $0.029 \pm 0.006$ | $5.7 \pm 1.3$ | 5624.0 |

Table 28: Results of the fits to the invariant mass distributions of the p.d.f. defined by Eq. (48). For comparison, the last column shows the mass value used in the MC simulation. The average width defined as $\sigma = \sqrt{f_1\sigma_1^2 + (1-f_1)\sigma_2^2}$ is about $25\,\mathrm{MeV/c^2}$.

The normalization factors $K_{1(2)}$ are given by:

$$K_{1(2)} = \int_{m_{min}}^{m_{max}} e^{s(m-m_B)} \left[1 - \mathrm{Erf}\left(\frac{m - m_B + s\sigma_{1(2)}^2}{\sqrt{2}\sigma_{1(2)}}\right)\right] dm, \qquad (50)$$

with $m_{min} = 5\,\mathrm{GeV/c^2}$ and $m_{max} = 5.8\,\mathrm{GeV/c^2}$, corresponding to the mass window accepted by the offline selection. Although we omit it for simplicity, the integral in Eq. (50) can be calculated analytically, hence the whole expression of the $g(m)$ p.d.f. is analytic and no longer involves lengthy numerical computations.

Fig. 18 shows the invariant mass distributions of each decay mode with the result of an unbinned likelihood fit using the p.d.f. defined by Eq. (48). The numerical results of the fits are shown in Tab. 28.

## 8.2 Parameterization under the $\pi^+\pi^-$ hypothesis

In the previous section we have assumed the correct mass hypotheses for the daughter particles. If instead one ignores any information on the masses of the daughter particles, and in particular makes the conventional assumption that they are pions for every decay mode, a different parameterization must be employed.

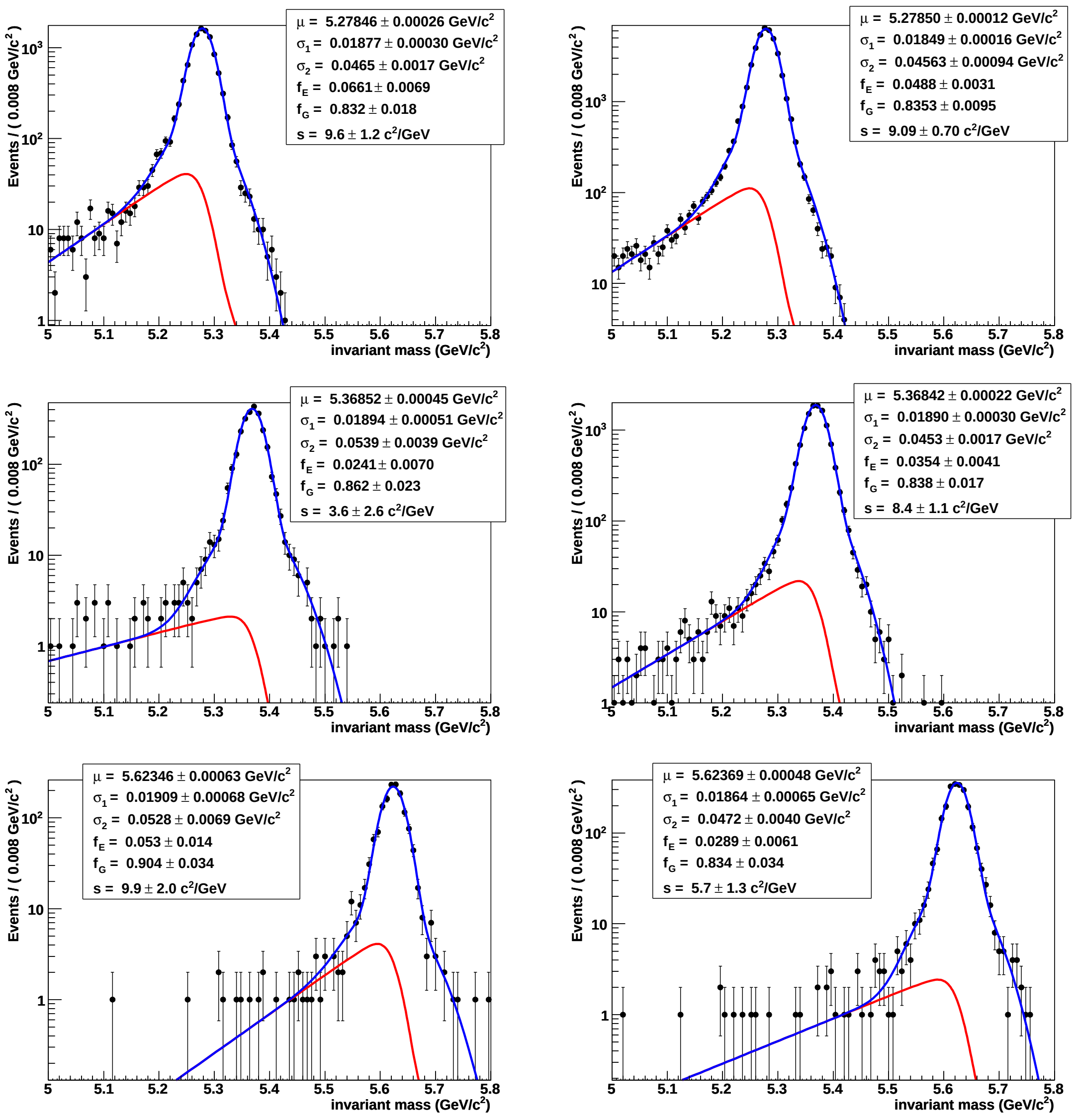

Figure 18: Invariant mass distributions of the various modes, with superimposed the result of the fit of the p.d.f. defined by Eq. (48): $B^0 \to \pi^+\pi^-$ (top left), $B^0 \to K^+\pi^-$ (top right), $B_s^0 \to \pi^+K^-$ (middle left), $B_s^0 \to K^+K^-$ (middle right), $\Lambda_b \to p\pi^-$ (bottom left) and $\Lambda_b \to pK^-$ (bottom right). As a reference, the lighter curve represents just the first component of Eq. (48), i.e. the one representing the radiation tail.

By skipping many mathematical details, which can be found Ref. [64], starting from the parameterization of the invariant mass under the correct mass hypothesis introduced in Eq. (48), we can write a joint p.d.f. $\tilde{f}(m_{\pi\pi}, \beta)$ as in the following:

$$\begin{aligned}\tilde{f}(m_{\pi\pi}, \beta) &= [f_E \, E_d(m - \mu(\beta);\, f_1,\, \sigma_1,\, \sigma_2,\, s) + \\ &+ (1 - f_E)\, C\, G_d(m - \mu(\beta);\, f_1,\, \sigma_1,\, \sigma_2)] \cdot \tilde{h}(\beta),\end{aligned} \tag{51}$$

| Channel | $\mu$ [MeV/c$^2$] | $f_G$ | $\sigma_1$ [MeV/c$^2$] | $\sigma_2$ [MeV/c$^2$] | $f_E$ | $s$ [c$^2$/GeV] | $\mu_{MC}$ [MeV/c$^2$] |
|---|---|---|---|---|---|---|---|
| $B^0 \to \pi^+\pi^-$ | $5278.5 \pm 0.3$ | $0.832 \pm 0.018$ | $18.8 \pm 0.3$ | $47 \pm 2$ | $0.066 \pm 0.007$ | $9.6 \pm 1.2$ | 5279.4 |
| $B^0 \to K^+\pi^-$ | $5278.6 \pm 0.1$ | $0.835 \pm 0.010$ | $18.8 \pm 0.2$ | $47 \pm 1$ | $0.049 \pm 0.004$ | $10.2 \pm 1.0$ | 5279.4 |
| $B_s^0 \to \pi^+K^-$ | $5368.6 \pm 0.5$ | $0.83 \pm 0.03$ | $18.6 \pm 0.6$ | $46 \pm 4$ | $0.028 \pm 0.008$ | $5.0 \pm 3.0$ | 5369.6 |
| $B_s^0 \to K^+K^-$ | $5368.4 \pm 0.2$ | $0.814 \pm 0.019$ | $19.0 \pm 0.3$ | $44 \pm 2$ | $0.031 \pm 0.005$ | $6.3 \pm 1.8$ | 5369.6 |
| $\Lambda_b \to p\pi^-$ | $5623.4 \pm 0.7$ | $0.85 \pm 0.06$ | $19.2 \pm 0.9$ | $43 \pm 5$ | $0.058 \pm 0.016$ | $10.6 \pm 2.4$ | 5624.0 |
| $\Lambda_b \to pK^-$ | $5623.7 \pm 0.5$ | $0.86 \pm 0.03$ | $20.0 \pm 0.7$ | $53 \pm 5$ | $0.023 \pm 0.007$ | $4.9 \pm 2.7$ | 5624.0 |

Table 29: Results of the fits to the distributions ($m_{\pi\pi}$, $\beta$) of the p.d.f. defined by Eq. (51). For comparison, the last column shows the mass value used in the MC simulation.

where $m_{\pi\pi}$ is the $\pi^+\pi^-$ invariant mass and $\beta$ is the asymmetry between the momenta $p_+$ and $p_-$ of the positive and negative particles, respectively:

$$\beta = \frac{p_+ - p_-}{p_+ + p_-}. \tag{52}$$

Here the function $\mu(\beta)$ is given by $\mu(\beta) = \sqrt{m_B^2 - F_{h^+h'^-}(\beta)}$, and in turn the function $F_{h^+h'^-}(\beta)$ is defined as:

$$F_{h^+h'^-}(\beta) = (m_{h^+}^2 - m_\pi^2)\left(1 + \frac{1-\beta}{1+\beta}\right) + (m_{h^-}^2 - m_\pi^2)\left(1 + \frac{1+\beta}{1-\beta}\right). \tag{53}$$

Finally, $\tilde{h}(\beta)$ is a p.d.f. for $\beta$, whose form depends on the specific decay channel, and the factor $C$ normalizes the double Gaussian $G_d$ to 1 inside the mass window.

Fig. 19 shows the invariant mass distributions under the $\pi^+\pi^-$ hypothesis for the various decay modes, with the projections on $m_{\pi\pi}$ of the results of unbinned likelihood fits of the p.d.f. defined in Eq. (51) to the spectra. The numerical results of the fits are summarized in Tab. 29. The fitted curves describe the data well, and the numerical results of Tab. 29 are in very good agreement with those obtained by studying the mass distributions under the correct mass hypothesis, summarized in Tab. 28.

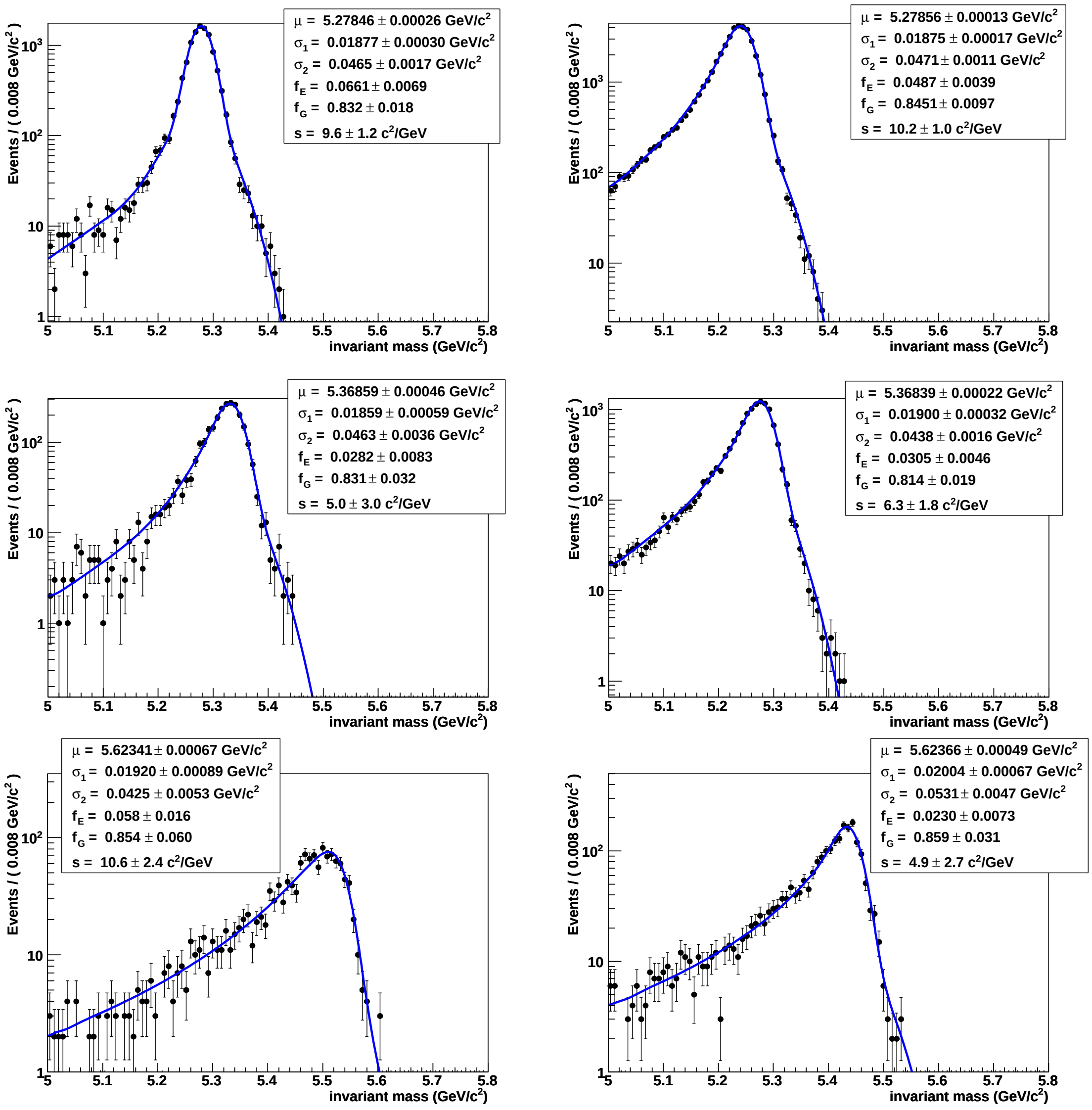

Figure 19: Invariant mass distributions of the various modes under the $\pi^+\pi^-$ hypothesis, with superimposed the projections on $m_{\pi\pi}$ of the results of unbinned likelihood fits of the p.d.f. defined by Eq. (51): $B^0 \to \pi^+\pi^-$ (top left), $B^0 \to K^+\pi^-$ (top right), $B_s^0 \to \pi^+K^-$ (middle left), $B_s^0 \to K^+K^-$ (middle right), $\Lambda_b \to p\pi^-$ (bottom left) and $\Lambda_b \to pK^-$ (bottom right).

# 9 $\mathcal{CP}$ sensitivity studies

In this section we present the method employed to extract the $\mathcal{CP}$ violating observables from the sample of selected $B$ candidates. First of all we will describe in detail how the likelihood function is built, and then we will show the results of unbinned maximum likelihood fits to the offline-selected full MC sample and to a series of toy experiments. We will conclude the section by quoting the sensitivity to various physics quantities of interest.

## 9.1 Experimental decay rates

The theoretical expressions of the decay rates for $B$ and $\overline{B}$ mesons can be found in App. A. From an experimental point of view these expressions need to be modified, by taking into account for example the possibility of mistagging the initial flavour of the $B$ meson, the presence of backgrounds, the acceptance as a function of the proper time and the resolution of the proper time measurement.

When introducing the effect of tagging, the observed decay rates for tagged neutral $B$ decays to $\mathcal{CP}$ eigenstates take the form:

$$\begin{aligned}\Gamma^q_{B\to f}(t) &= \epsilon_q\left[(1-\omega_q)\Gamma_{B\to f}(t)+\omega_q\Gamma_{\overline{B}\to f}(t)\right] = \\ &= \epsilon_q\frac{|A_f|^2}{2}e^{-\Gamma t}\left[I_+(t)+(1-2\omega_q)I_-(t)\right]\end{aligned} \tag{54}$$

and

$$\begin{aligned}\Gamma^q_{\overline{B}\to f}(t) &= \epsilon_q\left[\omega_q\Gamma_{B\to f}(t)+(1-\omega_q)\Gamma_{\overline{B}\to f}(t)\right] = \\ &= \epsilon_q\frac{|A_f|^2}{2}e^{-\Gamma t}\left[I_+(t)-(1-2\omega_q)I_-(t)\right],\end{aligned} \tag{55}$$

where $\epsilon_q$ and $\omega_q$ are the tagging efficiency and the mistag probability of the $q$-th tagging category respectively. For flavour specific decays, since $f \neq \overline{f}$, one has to consider the additional rates:

$$\begin{aligned}\Gamma^q_{\overline{B}\to \overline{f}}(t) &= \epsilon_q\left[\omega_q\Gamma_{B\to \overline{f}}(t)+(1-\omega_q)\Gamma_{\overline{B}\to \overline{f}}(t)\right] = \\ &= \epsilon_q\frac{|\overline{A}_{\overline{f}}|^2}{2}e^{-\Gamma t}\left[I_+(t)+(1-2\omega_q)I_-(t)\right],\end{aligned} \tag{56}$$

and

$$\begin{aligned}\Gamma^q_{B\to \overline{f}}(t) &= \epsilon_q\left[(1-\omega_q)\Gamma_{B\to \overline{f}}(t)+\omega_q\Gamma_{\overline{B}\to \overline{f}}(t)\right] = \\ &= \epsilon_q\frac{|\overline{A}_{\overline{f}}|^2}{2}e^{-\Gamma t}\left[I_+(t)-(1-2\omega_q)I_-(t)\right].\end{aligned} \tag{57}$$

## 9.2 Mass and time probability densities

A joint p.d.f. describing both tagged and untagged events of the continuous variable $t$ and the discrete variable $q$ can be written in the very compact form:

$$p(t,\ q) = f(q)\frac{e^{-\Gamma t}\left[I_+(t) + r\,\mathrm{sgn}(q)(1-2\omega_{|q|})I_-(t)\right]}{\int e^{-\Gamma t'}I_+(t')dt'}, \tag{58}$$

where the parameter $r$ has to be set to 1 if the final state is $f$ (including decays to $\mathcal{CP}$ eigenstates, where $f = \overline{f}$), and to -1 if it is $\overline{f}$. The variable $q$ can take the discrete value $+k$ for events tagged as $B$ and $-k$ for events tagged as $\overline{B}$ in the $k$-th tagging category, and 0 for untagged events. sgn($q$) stands for the sign of $q$, and the discrete function $f(q)$ is defined as:

$$f(q) = \frac{\epsilon_{|q|}}{2}(1-\delta_{0|q|}) + \left(1 - \sum_{k=1}^{n}\epsilon_k\right)\delta_{0|q|}, \tag{59}$$

where $\delta_{ij}$ is the Kronecker delta. The p.d.f. (58) is correctly normalized to unity by integrating over $t$ and summing over the $2n+1$ discrete values of $q$. It is remarkable that this joint p.d.f. is able to describe the whole sample of events tagged as $B$ and $\overline{B}$ and untagged events at the same time. Note that in the expressions above we are assuming that the tagging efficiencies and the mistag rates are identical for initial state $B$ and $\overline{B}$ mesons, indicated by the presence of the absolute value of the variable $q$ in the formulae. In reality, we expect small deviations from this assumption due to detector asymmetry effects, arising e.g. from the different interaction probability of final state particles and anti-particles within the detector material. Furthermore, in our treatment we are ignoring the possible existence of production asymmetries, arising from the different hadronization probabilities of $B$ and $\overline{B}$ mesons in a proton-proton collision. For example, by means of MC studies based on the PYTHIA event generator, a production asymmetry of the order of $10^{-3}$ for $B$ mesons has been predicted [65]. Future studies will investigate the impacts of these assumptions on the analysis.

The generalization of the p.d.f. to the case where also the proper time resolution and the proper time acceptance are taken into account is straightforward:

$$p(t,\ q) = f(q)\frac{\left\{e^{-\Gamma t'}\left[I_+(t') + r\,\mathrm{sgn}(q)(1-2\omega_{|q|})I_-(t')\right]\right\}\otimes R(t-t')\epsilon(t)}{\int e^{-\Gamma t''}I_+(t'')\otimes R(t'''-t'')\epsilon(t''')dt'''}, \tag{60}$$

where the symbol $\otimes$ stands for convolution product, $R(t-t')$ is a proper time resolution function and $\epsilon(t)$ is the proper time acceptance.

The decays considered so far comprise ten different final states in total, nominally:

- $\pi^+\pi^-$, $K^+\pi^-$ and $\pi^-K^+$ from the $B^0$ meson;
- $K^+K^-$, $\pi^+K^-$ and $K^-\pi^+$ from the $B^0_s$ meson;
- $p\pi^+$, $\pi^+\bar{p}$, $pK^+$ and $K^+\bar{p}$ from the $\Lambda_b$ baryon.

In addition, although we have not studied the corresponding full MC samples, we include the two rare modes $B^0 \to K^+K^-$ and $B_s^0 \to \pi^+\pi^-$, to give a total of twelve distinct final states. For the moment, the rare baryonic modes $B \to p\bar{p}$ are not included in this analysis.

As we have shown in Fig. 5, using the $\pi^+\pi^-$ hypothesis, all the invariant mass distributions largely overlap. The joint p.d.f. for $m_{\pi\pi}$ and the momentum asymmetry $\beta$ was introduced in Eq. (51). For each of the twelve final states, we can write the joint p.d.f. for $m_{\pi\pi}$, $\beta$, $t$ and $q$ as:

$$g_j(m_{\pi\pi},\, \beta,\, t,\, q) = f_j(m_{\pi\pi},\, \beta) \cdot p_j(t,\, q), \tag{61}$$

where the index $j$ identifies the final state.

However, since the observable $t$ depends on the mass of the $B$ hadron, as discussed in Sec. 4.1, we find it convenient to work with a different observable, i.e. the ratio $\xi$ between the reconstructed proper time and the $B$ mass, defined in Eq. (37), which is by definition no longer dependent on the $B$ mass. A p.d.f. for $\xi$ can be easily obtained by modifying Eq. (60) as in the following:

$$\tilde{p}(\xi,\, q) = f(q) \frac{\left\{ e^{-\Gamma m_B \xi'} \left[ \tilde{I}_+(\xi') + r\, \mathrm{sgn}(q)(1 - 2\omega_{|q|}) \tilde{I}_-(\xi') \right] \right\} \otimes \tilde{R}(\xi - \xi')\tilde{\epsilon}(\xi)}{\int e^{-\Gamma m_B \xi''} \tilde{I}_+(\xi'') \otimes \tilde{R}(\xi''' - \xi'')\tilde{\epsilon}(\xi''') d\xi'''}, \tag{62}$$

where $\tilde{R}(\xi - \xi')$ is now a resolution for the $\xi$ observable, $\tilde{\epsilon}(\xi)$ is the acceptance as a function of $\xi$, and the functions $\tilde{I}_+$ and $\tilde{I}_-$ are obtained from $I_+(t)$ and $I_-(t)$ by the substitutions $t \to \xi$, $\Delta m \to m_B\, c^2\, \Delta m$ and $\Delta\Gamma \to m_B\, c^2\, \Delta\Gamma$. Consequently, the p.d.f. defined in Eq. (61) becomes:

$$\tilde{g}_j(m_{\pi\pi},\, \beta,\, \xi,\, q) = f_j(m_{\pi\pi},\, \beta) \cdot \tilde{p}_j(\xi,\, q). \tag{63}$$

## 9.3 Particle identification probability densities

The final ingredient we need to introduce is the information coming from the PID system. As discussed in Sec. 5, the PID information in LHCb is described by so-called $\Delta \log \mathcal{L}$ observables. For each track, the PID system gives an answer in terms of the four observables $\Delta \log \mathcal{L}_{e\pi}$, $\Delta \log \mathcal{L}_{\mu\pi}$, $\Delta \log \mathcal{L}_{K\pi}$ and $\Delta \log \mathcal{L}_{p\pi}$, which are used to discriminate between the various hypotheses.

For the studies summarized in the remainder of this note, we only rely on the two observables $\Delta \log \mathcal{L}_{K\pi}$ and $\Delta \log \mathcal{L}_{p\pi}$. These observables are sufficient to discriminate particles coming from all the $H_b \to h^+h'^-$ decays, since signal decays contain neither electrons nor muons. It is observed, however, that the combinatorial background has a small contribution from lepton tracks, and hence the other $\Delta \log \mathcal{L}$ observables may be of use in suppressing this background.

Starting from Eq. (63), we could be tempted to write a joint p.d.f. for one of the twelve final states, including the PID observables, as:

$$\begin{aligned} & s_j(m_{\pi\pi},\, \beta,\, \xi,\, q,\, \Delta \log \mathcal{L}^+_{K\pi},\, \Delta \log \mathcal{L}^-_{K\pi},\, \Delta \log \mathcal{L}^+_{p\pi},\, \Delta \log \mathcal{L}^-_{p\pi}) = \\ & \quad = f_j(m_{\pi\pi},\, \beta)\, \tilde{p}_j(\xi,\, q)\, f^{K\pi}_{j^+}(\Delta \log \mathcal{L}^+_{K\pi})\, f^{K\pi}_{j^-}(\Delta \log \mathcal{L}^-_{K\pi})\, f^{p\pi}_{j^+}(\Delta \log \mathcal{L}^+_{p\pi})\, f^{p\pi}_{j^-}(\Delta \log \mathcal{L}^-_{p\pi}) \end{aligned} \tag{64}$$

where $f_{j^+}^{h\pi}$ and $f_{j^-}^{h\pi}$ stand for the $\Delta\log\mathcal{L}_{h\pi}$ distributions for the true positive particle and for the true negative particle of the final state $j$ respectively. In this last equation we have factorized the joint p.d.f. for the $\Delta\log\mathcal{L}$ observables into the product of one-dimensional p.d.f.'s, i.e. assuming for a given track the independence of the variables $\Delta\log\mathcal{L}_{K\pi}$ and $\Delta\log\mathcal{L}_{p\pi}$. However, such an assumption would be incorrect, as it can be shown that $\Delta\log\mathcal{L}_{K\pi}$ and $\Delta\log\mathcal{L}_{p\pi}$ are indeed strongly correlated. Hence the employment of such observables in order to discriminate between the various mass hypotheses would require the usage of 2-dimensional joint p.d.f.'s of $\Delta\log\mathcal{L}_{K\pi}$ and $\Delta\log\mathcal{L}_{p\pi}$, properly taking into account the correlation between the two. Although this is formally simple to state, it would considerably complicate the calibration of such distributions from data, since it would require a two-dimensional analysis.

As detailed in Ref. [66], in order to solve this issue, we can merge the information of the $\Delta\log\mathcal{L}_{K\pi}$ and $\Delta\log\mathcal{L}_{p\pi}$ observables into a single observable $\Delta$, defined as:

$$\Delta = 2 - P_\pi + P_p, \tag{65}$$

where $P_\pi$ and $P_p$ are given by:

$$P_\pi = \frac{1}{1 + e^{\Delta\log\mathcal{L}_{K\pi}} + e^{\Delta\log\mathcal{L}_{p\pi}}} \tag{66}$$

and

$$P_p = \frac{e^{\Delta\log\mathcal{L}_{p\pi}}}{1 + e^{\Delta\log\mathcal{L}_{K\pi}} + e^{\Delta\log\mathcal{L}_{p\pi}}}. \tag{67}$$

The observable $\Delta$ by construction lies between 1 and 3. As it is apparent in Fig. 20, the distribution of $\Delta$ is peaked at 1 for true pions, at 2 for true kaons, and at 3 for true protons.

By employing the PID observable $\Delta$, we can write a joint p.d.f. for one of the twelve final states as:

$$s_j(m_{\pi\pi}, \beta, \xi, q, \Delta^+, \Delta^-) = f_j(m_{\pi\pi}, \beta)\, \tilde{p}_j(\xi, q)\, d_{j^+}(\Delta^+)\, d_{j^-}(\Delta^-) \tag{68}$$

where $d_{j^+}$ and $d_{j^-}$ stand for the $\Delta$ p.d.f.'s for the true positive particle and for the true negative particle of the final state $j$ respectively. Note that this p.d.f. does not take into account the known momentum dependence of the PID observables, hence simplifying the problem at the price of loosing some information.

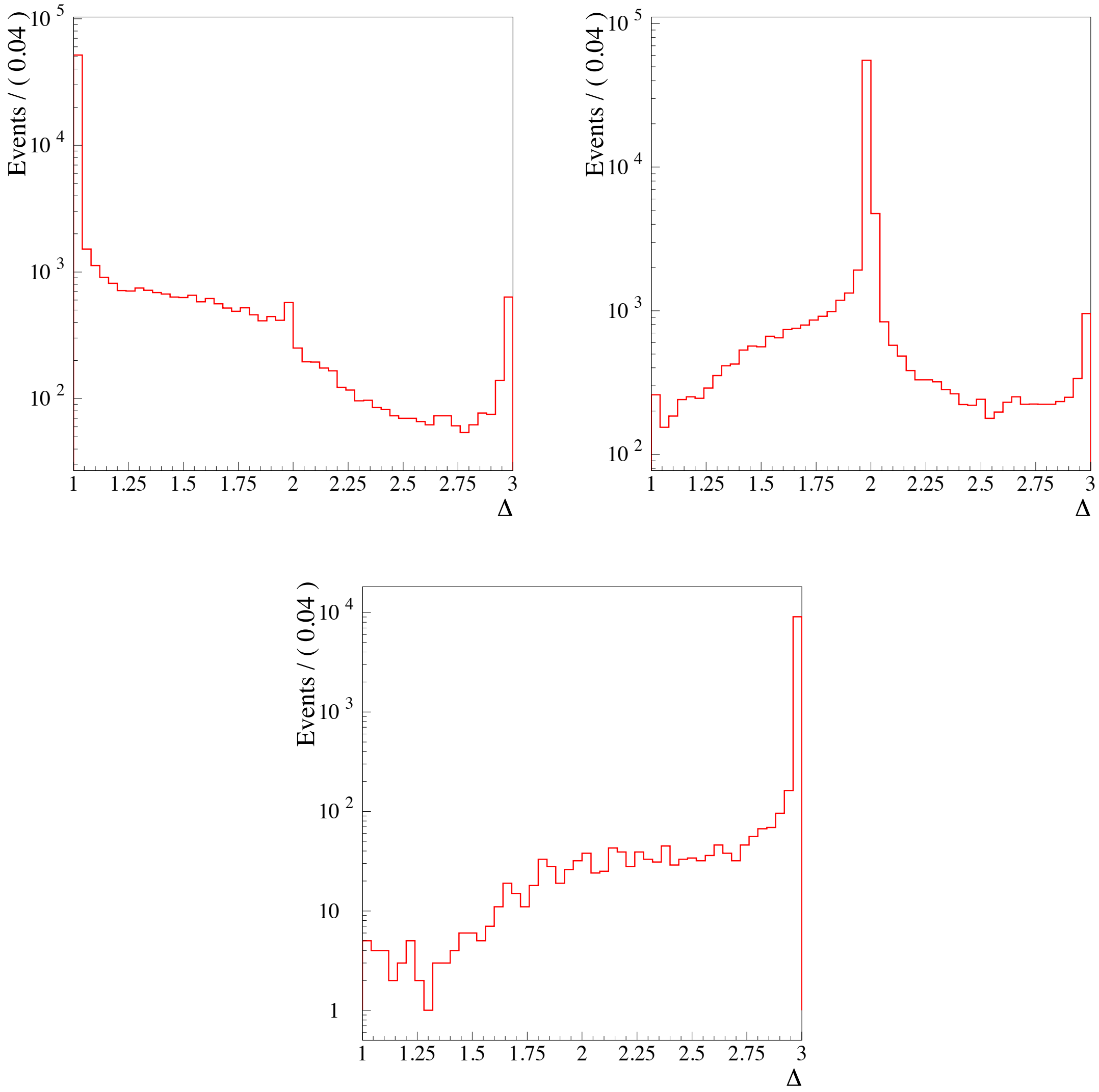

Figure 20: Distributions of $\Delta$ for pions (top left), kaons (top right) and protons (bottom) from offline-selected $H_b \to h^+h'^-$ events including all the different decay modes.

## 9.4 Final expression of the probability density

We are now able to write the full probability density including all the twelve final states:

$$s(m_{\pi\pi},\, \beta,\, \xi,\, q,\, \Delta^+,\, \Delta^-) = \sum_{j=1}^{12} w_j \cdot s_j(m_{\pi\pi},\, \beta,\, \xi,\, q,\, \Delta^+,\, \Delta^-), \tag{69}$$

where the weight $w_j$ is the fraction of events of the final state $j$, defined as:

$$w_j = \frac{N_j}{\sum_{k=1}^{12} N_k}, \tag{70}$$

with $N_j$ the number of events of the final state $j$.

In order to take into account the backgrounds, the p.d.f. must be modified as follows:

$$\begin{aligned}\tilde{s}(m_{\pi\pi},\, \beta,\, \xi,\, q,\, \Delta^+,\, \Delta^-) &= f_s \cdot s(m_{\pi\pi},\, \beta,\, \xi,\, q,\, \Delta^+,\, \Delta^-) + \\ &+ (1 - f_s) \cdot b(m_{\pi\pi},\, \beta,\, \xi,\, q,\, \Delta^+,\, \Delta^-),\end{aligned} \tag{71}$$

where $f_s$ represents the fraction of signal events:

$$f_s = \frac{\sum_{k=1}^{12} N_k}{N_{tot}}, \tag{72}$$

with $N_{tot}$ the total number of events in the sample.

As discussed in Sec. 3.5, two background components exist, i.e. combinatorial and physical due to partially reconstructed three-body modes. Hence, the background p.d.f. can be in turn decomposed as follows:

$$\begin{aligned}b(m_{\pi\pi},\, \beta,\, \xi,\, q,\, \Delta^+,\, \Delta^-) &= f_p \cdot b_p(m_{\pi\pi},\, \beta,\, \xi,\, q,\, \Delta^+,\, \Delta^-) + \\ &+ (1 - f_p) \cdot b_c(m_{\pi\pi},\, \beta,\, \xi,\, q,\, \Delta^+,\, \Delta^-),\end{aligned} \tag{73}$$

where $b_p$ and $b_c$ are the p.d.f.'s for the physical and combinatorial backgrounds respectively, and $f_p$ is the fraction of physical background events defined as:

$$f_p = \frac{N_p}{N_{tot} - \sum_{k=1}^{12} N_k}, \tag{74}$$

with $N_p$ the number of physical background events. We can factorize each of the two background components as follows:

$$b_k(m_{\pi\pi},\, \beta,\, \xi,\, q,\, \Delta^+,\, \Delta^-) = g_k(m_{\pi\pi})\, h_k(\beta)\, p_k(\xi)\, r_k(q)\, d_k(\Delta^+)\, d_k(\Delta^-), \tag{75}$$

where $k = \{p,\, c\}$, $g_k$ represents the p.d.f. of the background invariant mass, $h_k$ the p.d.f. of the background momentum asymmetry, $p_k$ the p.d.f. of $\xi$ for background events, $d_k$ the p.d.f. of $\Delta$ for background events, and $r_k(q)$ the background distribution for $q$ given by:

$$r_k(q) = \frac{\epsilon^k_{|q|}}{2}(1 - \delta_{0|q|}) + \left(1 - \sum_{j=1}^{n} \epsilon^k_j\right) \delta_{0|q|}, \tag{76}$$

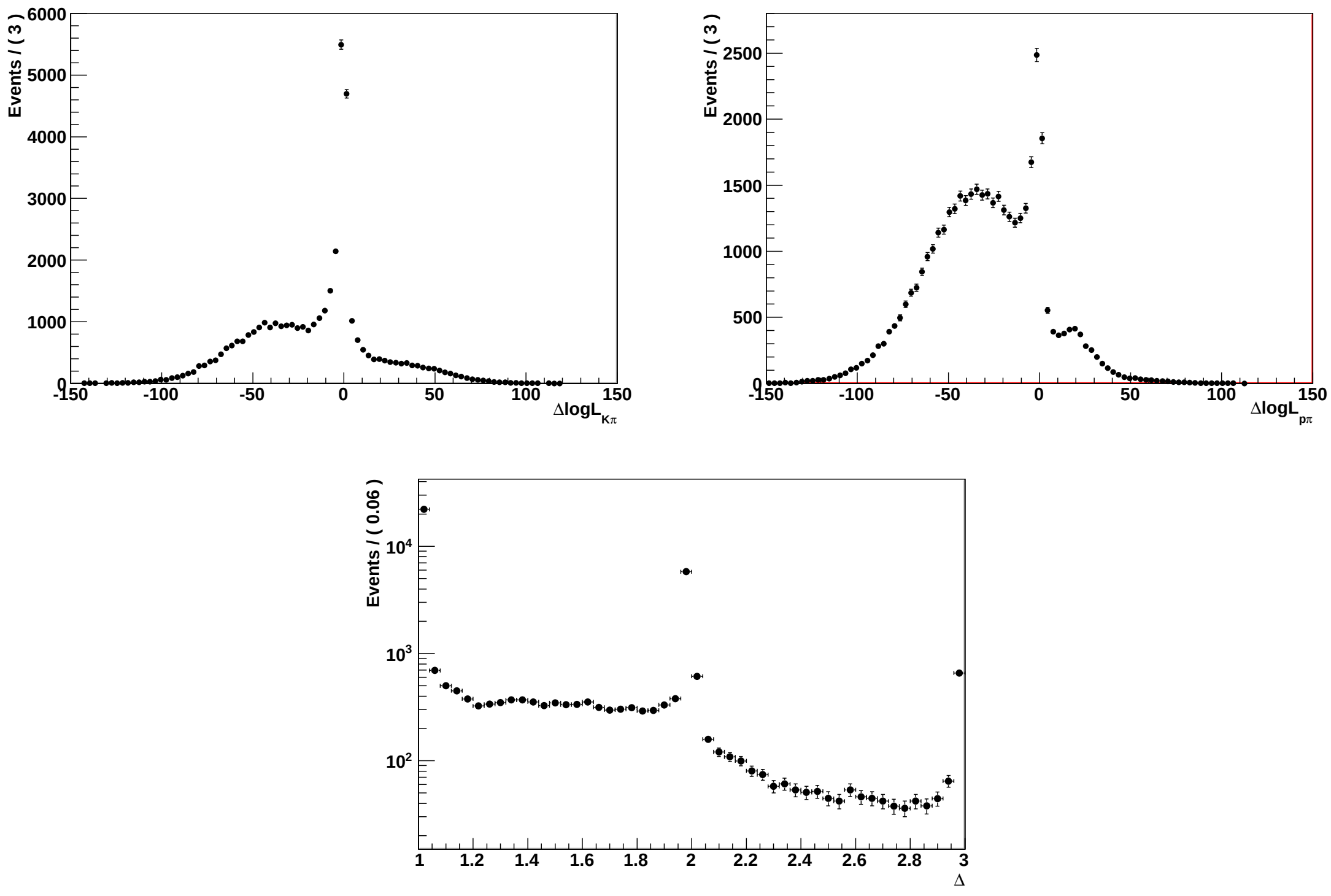

Figure 21: Effective distributions of $\Delta \log \mathcal{L}_{K\pi}$ (top left), $\Delta \log \mathcal{L}_{p\pi}$ (top right) and $\Delta$ (bottom) for combinatorial background events.

with $\epsilon_j^k$ equal to the background tagging efficiency for the $j$-th tagging category. Note that we have implicitly assumed that the background is flavour-tagging blind, i.e. that a background event can be tagged as a $B$ or $\overline{B}$ with identical probability.

As far as the combinatorial background component is concerned, we can assume, following the studies in Sec. 3.5, that the $g_c$ function is well described by a decreasing exponential function. However, with the limited MC statistics available after event selection it is not possible to make a meaningful study of the shape of the $h_c$ distribution. At present, we assume $h_c$ to be identical to the corresponding function for $B^0 \to \pi^+\pi^-$ events (see Ref. [64]). With real data it will be possible to characterize this distribution from a sample of sideband events.

Similarly, we assume that $p_c$ is a decaying exponential times an acceptance function for $\xi$ analogous to that of a signal decay. For the inverse of the slope of the exponential function $\tau_\xi$, according to fits performed on a handful of offline-selected MC $b\bar{b}$ events, we have chosen a value of 0.19 ps/GeV, i.e. about $\frac{2}{3}$ of the value corresponding to a $B$ meson decay (see Tab. 21). Finally, to build the p.d.f. $d_c$ we have summed up the MC $\Delta$ distributions for pions, kaons and protons from $H_b \to h^+h'^-$ decays, with weights equal to the fractions of the different particles present in the $b\bar{b}$-inclusive MC events which passed the offline selection (see Tab. 16). In practice, this corresponds to the assumption that the distribution of the momentum of combinatorial background tracks will be the same as the one of tracks coming from real two-body $B$ decays. This way we obtain the effective $\Delta$ distribution for combinatorial background events shown in Fig. 21. As a reference, in the same figure we also show the corresponding $\Delta \log \mathcal{L}_{h\pi}$ distributions.

Similar arguments hold for the physical background, i.e. we will have the possibility to study it from real data sideband events in the region $m_{\pi\pi} < 5.2\,\mathrm{GeV/c^2}$, a portion of the mass spectrum where, once the combinatorial background component is subtracted, it is expected to dominate. In the present studies, we assume that $h_p = h_c$ and $p_p = p_c$. For $d_p$ instead, we make the same assumption as for the combinatorial background case, but without the proton component, since the particles selected in the partial reconstruction of three-body $B$ decays will just be an admixture of charged pions and kaons. The $\Delta \log \mathcal{L}$ and $\Delta$ distributions corresponding to $d_p$ are not shown as they are almost identical to those of Fig. 21, since as is apparent in Tab. 16, the size of the proton component in the combinatorial background is tiny.

## 9.5 Monte Carlo studies of the $\mathcal{CP}$ sensitivity

Having introduced the expression of the global p.d.f. in Eq. (71), comprising all the twelve signal components and the two background components, we are now able to write the likelihood function as:

$$\mathcal{L}_{\mathcal{CP}} = \prod_{i=1}^{N_{tot}} \tilde{s}(m_{\pi\pi\,i},\, \beta_i,\, \xi_i,\, q_i,\, \Delta_i^+,\, \Delta_i^-), \tag{77}$$

where the index $i$ runs over the entire sample of selected events. In order to study the sensitivity to the $\mathcal{CP}$ violating parameters, we have performed a series of fast MC studies, by generating several samples of the 6 observables $m_{\pi\pi}$, $\beta$, $\xi$, $q$, $\Delta^+$ and $\Delta^-$, using the p.d.f. of Eq. (71), and then performing unbinned likelihood fits by maximizing the function given in Eq. (77) calculated over each generated sample. We have also performed a fit to the full MC sample surviving the offline selection. Although this sample is only composed of signal events, such a fit is a necessary check that the p.d.f. inside the fit correctly describes the characteristics found in the full simulation. All the results and relevant plots shown in the following are based on the RooFit statistical software toolkit [67].

### 9.5.1 Fit to the full Monte Carlo sample

In this study the data sample is composed of events which passed the offline selection algorithm, but with no trigger filter applied, in order not to reduce further the available statistics. The yields roughly correspond to an integrated luminosity $L \simeq 400\,\mathrm{pb}^{-1}$, although it is not possible to establish a strict relationship between the sample and the integrated luminosity, as the trigger algorithm was not applied.

The fit was performed by removing the background components from the p.d.f. given in Eq. (71), i.e. using Eq. (69), that was used to build the likelihood function, and it featured 53 free parameters in total. The maximum likelihood fit was realized by using the MIGRAD minimization engine of the MINUIT software library [68], configured with the so-called *Strategy 2*, followed upon convergence by the HESSE algorithm of the same library, in order to calculate with better precision the covariance matrix. The estimated distance to the minimum of the function $-\log \mathcal{L}_{\mathcal{CP}}$ at the fit solution, after HESSE, was about $2.3 \cdot 10^{-4}$.

Fit results for each parameter are summarized in Tab. 30, together with their expected values. Plots with the relevant distributions and with the results of the maximum likelihood fit superimposed are shown in Fig. 22. All the fitted parameters are consistent with expectations, with the exceptions of the masses and the average lifetimes, where there is a tendency for the fit to underestimate the input values. However, both of these effects are expected, the former mainly due to the choice of the approximate parameterization of the mass line shape in presence of QED radiation, as already discussed in Sec. 8.1, while the latter is mainly due to the fact that we are ignoring the small bias in the proper time resolution, as already seen in Sec. 4.1. Although both these effects deserve further studies, they are not expected to have any sizable impact on the measurements of the time dependent $\mathcal{CP}$ terms.

Note that the tagging power for this sample is significantly smaller than the one shown in Tab. 27. This can be attributed to the fact that the sample does not have the trigger applied. The trigger, and L0 in particular, selects with higher probability events which are likely to provide good OS tags, and hence enriches the tagging power.

Although no background components were involved in this fit, we conclude that the signal part of the model gives a satisfactory description of the full MC sample, at least within the available statistics. In view of the first analysis with real data, it will be important to perform a validation on a much larger full MC sample.

| Parameter | Expected value | Fit result | Parameter | Expected value | Fit result |
|---|---|---|---|---|---|
| Event Yields | | | | | |
| $N_{B^0\to K^+K^-}$ | 0 | $-6 \pm 32$ | $N_{B^0\to K^+\pi^-}$ | 40263 | $40345 \pm 219$ |
| $N_{B^0\to \pi^+\pi^-}$ | 11410 | $11217 \pm 121$ | $N_{B^0_s\to K^+K^-}$ | 15458 | $15407 \pm 138$ |
| $N_{B^0_s\to \pi^+K^-}$ | 3607 | $3640 \pm 85$ | $N_{B^0_s\to \pi^+\pi^-}$ | 0 | $-14 \pm 15$ |
| $N_{\Lambda_b\to pK^-}$ | 7341 | $7379 \pm 90$ | $N_{\Lambda_b\to p\pi^-}$ | 2959 | $3055 \pm 60$ |
| Charge asymmetries | | | | | |
| $\mathcal{A}^{\mathcal{CP}}_{K^+\pi^-}$ | -0.1 | $-0.096 \pm 0.005$ | $\mathcal{A}^{\mathcal{CP}}_{\pi^+K^-}$ | 0.39 | $0.38 \pm 0.02$ |
| $\mathcal{A}^{\mathcal{CP}}_{pK^-}$ | 0 | $0.017 \pm 0.012$ | $\mathcal{A}^{\mathcal{CP}}_{p\pi^-}$ | 0 | $0.029 \pm 0.020$ |
| Time dependent asymmetries | | | | | |
| $\mathrm{Re}\lambda_{\pi^+\pi^-}$ | -1.08 | $-1.10 \pm 0.29$ | $\mathrm{Re}\lambda_{K^+K^-}$ | 0.85 | $0.77 \pm 0.08$ |
| $\mathrm{Im}\lambda_{\pi^+\pi^-}$ | -1.04 | $-1.55 \pm 0.45$ | $\mathrm{Im}\lambda_{K^+K^-}$ | 0.21 | $0.29 \pm 0.15$ |
| Mixing | | | | | |
| $\Delta\Gamma_d$ [ps$^{-1}$] | 0 | $-0.070 \pm 0.034$ | $\Delta\Gamma_s$ [ps$^{-1}$] | 0.068 | $0.020 \pm 0.038$ |
| $\Delta M_d$ [ps$^{-1}$] | 0.502 | $0.494 \pm 0.014$ | $\Delta M_s$ [ps$^{-1}$] | 20 | $19.97 \pm 0.08$ |
| Lifetimes | | | | | |
| $\tau_d$ [ps] | 1.536 | $1.52 \pm 0.01$ | $\tau_s$ [ps] | 1.461 | $1.41 \pm 0.03$ |
| $\tau_{\Lambda_b}$ [ps] | 1.229 | $1.22 \pm 0.01$ | - | - | - |
| Masses | | | | | |
| $M_{B^0}$ [GeV/c$^2$] | 5.2794 | $5.2783 \pm 0.0001$ | $M_{B^0_s}$ [GeV/c$^2$] | 5.3696 | $5.3690 \pm 0.0002$ |
| $M_{\Lambda_b}$ [GeV/c$^2$] | 5.624 | $5.6238 \pm 0.0002$ | - | - | - |
| Mass resolution and radiative tail | | | | | |
| $f_G$ | 0.84 * | $0.847 \pm 0.013$ | $\sigma_1$ [MeV/c$^2$] | 19 * | $18.2 \pm 0.2$ |
| $\sigma_2$ [MeV/c$^2$] | 46 * | $42 \pm 2$ | $f_E$ | 0.042 * | $0.052 \pm 0.003$ |
| $s$ [c$^2$/GeV] | 9 * | $10.2 \pm 0.7$ | - | - | - |
| Opposite side tagging | | | | | |
| $\epsilon_1^{H_b}$ | 0.271 ** | $0.271 \pm 0.002$ | $\epsilon_2^{H_b}$ | 0.0676 ** | $0.0676 \pm 0.0009$ |
| $\epsilon_3^{H_b}$ | 0.0476 ** | $0.0476 \pm 0.0007$ | $\epsilon_4^{H_b}$ | 0.0334 ** | $0.0334 \pm 0.0006$ |
| $\epsilon_5^{H_b}$ | 0.0198 ** | $0.0198 \pm 0.0005$ | $w_1^{H_b}$ | 0.445 | $0.446 \pm 0.006$ |
| $w_2^{H_b}$ | 0.356 | $0.362 \pm 0.011$ | $w_3^{H_b}$ | 0.306 | $0.328 \pm 0.014$ |
| $w_4^{H_b}$ | 0.247 | $0.255 \pm 0.015$ | $w_5^{H_b}$ | 0.171 | $0.172 \pm 0.018$ |
| Same side kaon tagging | | | | | |
| $\epsilon_{SSK}^{B^0}$ | 0.1045 | $0.1050 \pm 0.0014$ | $\epsilon_{SSK}^{B^0_s}$ | 0.1416 | $0.1426 \pm 0.0026$ |
| $\epsilon_{SSK}^{\Lambda_b}$ | 0.1074 | $0.110 \pm 0.003$ | $w_{SSK}^{B^0}$ | 0.543 | $0.542 \pm 0.011$ |
| $w_{SSK}^{B^0_s}$ | 0.351 | $0.29 \pm 0.06$ | $w_{SSK}^{\Lambda_b}$ | 0.419 | $0.414 \pm 0.015$ |
| Same side pion tagging | | | | | |
| $\epsilon_{SS\pi}^{B^0}$ | 0.0868 | $0.0864 \pm 0.0013$ | $\epsilon_{SS\pi}^{B^0_s}$ | 0.0624 | $0.064 \pm 0.002$ |
| $\epsilon_{SS\pi}^{\Lambda_b}$ | 0.0684 | $0.066 \pm 0.002$ | $w_{SS\pi}^{B^0}$ | 0.406 | $0.408 \pm 0.013$ |
| $w_{SS\pi}^{B^0_s}$ | 0.469 | $0.47 \pm 0.06$ | $w_{SS\pi}^{\Lambda_b}$ | 0.506 | $0.51 \pm 0.02$ |

Table 30: Numerical results of the fit to the full MC sample.
*The expected values are calculated as the averages of independent fits to the mass distributions of single channel samples (see Tab. 29). Since they are not strictly identical amongst the various channels, as is apparent in Tab. 29, our choice of performing the fit using just one set of parameters is an approximation.
**Note that the central values resulting from the fit for these parameters agree perfectly with the expected values. This is due to the fact that the sample used to calculate the expected values is the same as the one used in the fit, and since the opposite side taggers are assumed to behave in the same way for all the signal channels, as far as the tagging efficiencies are concerned the fit is simply counting how many events are present in each tagging category.

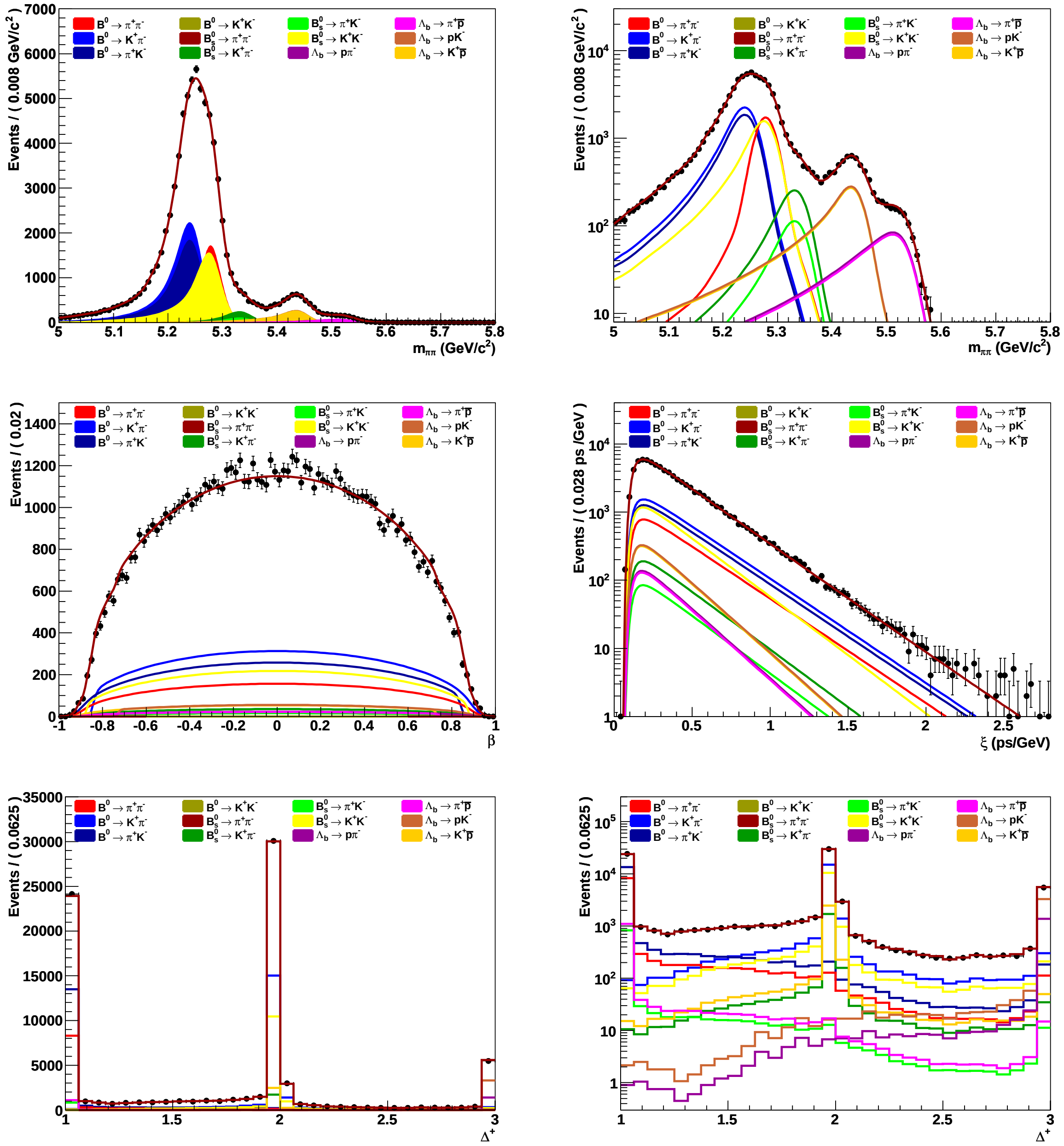

Figure 22: Distributions from the full MC sample with the result of the fit superimposed: $m_{\pi\pi}$ with linear scale (top left) and logarithmmic scale (top right), $\beta$ (middle left), $\xi$ (middle right), $\Delta^+$ with linear scale (bottom left) and logarithmic scale (bottom right). The various signal components are also shown.

### 9.5.2 Fast Monte Carlo studies and $\mathcal{CP}$ sensitivity

In order to estimate the LHCb sensitivity to the physics quantities of interest, a series of fast MC studies were performed by implementing the full model with background components included. By using the p.d.f. defined in Eq. (71), we generated 150 samples with different initial random seeds, each one with a statistics equivalent to an integrated luminosity $L = 200\,\mathrm{pb}^{-1}$, and then performed unbinned maximum likelhood fits in order to determine the best estimates and the sensitivities to the various relevant parameters. By generating a more limited number of larger samples (20 in total), corresponding to an integrated luminosity of $L = 0.8\,\mathrm{fb}^{-1}$, we also verified the scaling of the sensitivities as the inverse of the square root of the integrated luminosity.

The physics inputs to the fast MC simulations, i.e. the $B$ hadron masses and lifetimes, the mass and width differences of the $B^0$ and $B_s^0$ mass eigenstates, and the $\mathcal{CP}$ violating quantities were fixed according to the current experimental and theoretical knowledge. In particular, values of the yet unmeasured parameters $\mathrm{Re}\lambda_{K^+K^-}$ and $\mathrm{Im}\lambda_{K^+K^-}$ were estimated by taking the current values of $\mathrm{Re}\lambda_{\pi^+\pi^-}$, $\mathrm{Im}\lambda_{\pi^+\pi^-}$ and $\mathcal{A}^{\mathcal{CP}}_{K^+\pi^-}$, and assuming U-spin symmetry, as will be shown in Sec. 10. For setting the yields of the two yet unseen $B^0 \to K^+K^-$ and $B_s^0 \to \pi^+\pi^-$ decays, we have assumed the corresponding branching fractions to be $10^{-7}$ for both, slightly below the current limits set by the $B$ factories and the Tevatron (see Tabs. 3 and 5).

The maximum likelihood fits featured 68 free parameters in total, and they were performed by using the MIGRAD algorithm of the MINUIT library [68], configured with so-called *Strategy 1*. We could not employ *Strategy 2* followed by HESSE, as we did for the fit to the full MC sample described in the previous section, since so many fits would have required a very large aggregate computing time. Consider that, while MIGRAD with *Strategy 1* required about 2,500 likelihood function evaluations to converge to the minimum of $-\log\mathcal{L}_{\mathcal{CP}}$ (about 1.5 days on a modern E5420 Intel Xeon CPU running at 2.5 GHz, for a sample corresponding to an integrated luminosity $L = 200\,\mathrm{pb}^{-1}$), HESSE needed in addition 7,500 evaluations to calculate reliably the second derivative matrix by finite differences, and then inverting it to obtain the covariance matrix. With *Strategy 2* the situation would have been even worse, by roughly a factor 2. However, this is not a prohibitive amount of CPU time when performing a fit to a single sample of real data, also considering that RooFit allows the parallelization of the likelihood evaluation over several CPUs [67].

Without running HESSE after MIGRAD, the calculation of the second derivative matrix is approximate, in particular for what concerns the non-diagonal elements. This usually leads, especially when the number of parameters in the fit is large, to underestimate correlations amongst some parameters, and in turn to underestimate errors. In this case, the way to obtain reliable estimates of the sensitivities consists of taking for each parameter the root mean squares of the central values returned by the fits (procedure $a$), or equivalently in multiplying the average errors returned by the fits by scale factors equal to the widths of the corresponding pull distributions (procedure $b$). We checked that all the distributions of the central values returned by the fits were unbiased, and also that both the procedures $a$ and $b$ gave consistent results. Finally, on a sub-sample of only ten fits, we also verified that HESSE increased the average errors by the expected scale factors.

The sensitivities are summarized in Tab. 31, together with the corresponding full MC inputs. Plots with the relevant distributions taken from one of the generated samples with $L = 200\,\mathrm{pb}^{-1}$ and with the results of the maximum likelihood fit superimposed are shown in Fig. 23. The statistical sensitivities to the most relevant physics measurements are collected in Tab. 32.

Note that Tab. 32 reports the sensitivities for $\mathcal{A}^{dir}$ and $\mathcal{A}^{mix}$, while an alternative parameterization of time dependent $\mathcal{CP}$ violation using $\mathrm{Re}\lambda$ and $\mathrm{Im}\lambda$ as $\mathcal{CP}$ asymmetry coefficients was employed in the fits. In order to pass from $\mathrm{Re}\lambda$ and $\mathrm{Im}\lambda$ obtained from the $\mathcal{CP}$ fits to $\mathcal{A}^{dir}$ and $\mathcal{A}^{mix}$, a simple Gaussian error propagation using the functional relations reported in App. A can be adopted, of course also taking into account the correlation between $\mathrm{Re}\lambda$ and $\mathrm{Im}\lambda$ returned by the fit.

It can be seen that LHCb will surpass the current experimental knowledge involving ratios of branching fractions and charge asymmetries, either from the $B$ factories or the Tevatron, already with an integrated luminosity $L = 0.2\,\mathrm{fb}^{-1}$, i.e. 1/10 of a nominal year of LHCb data taking. This will hopefully be achievable quite early after the startup of the LHC physics run.

As far as the time dependent $\mathcal{CP}$ asymmetries are concerned, their measurement is complicated by the need for two additional ingredients, i.e. the proper time measurement and the tagging. As it is apparent in Tab. 32, LHCb will reach a better statistical sensitivity for the measurements of $\mathcal{A}^{dir}_{\pi^+\pi^-}$ and $\mathcal{A}^{mix}_{\pi^+\pi^-}$ than the current measurements with an integrated luminosity $L = 2\,\mathrm{fb}^{-1}$. This is mainly due to the fact that the larger LHCb statistics with respect to the $B$ factories is moderated in this case by the smaller tagging power, roughly amounting to 1/5 of that at the $B$ factories. For $\mathcal{A}^{dir}_{K^+K^-}$ and $\mathcal{A}^{mix}_{K^+K^-}$ no measurement yet exists, and only the CDF experiment will have an opportunity to make an earlier measurement.

| Parameter | Input value $0.2\,\text{fb}^{-1}$ | Sensitivity $0.2\,\text{fb}^{-1}$ | Sensitivity $0.8\,\text{fb}^{-1}$ | Parameter | Input value $0.2\,\text{fb}^{-1}$ | Sensitivity $0.2\,\text{fb}^{-1}$ | Sensitivity $0.8\,\text{fb}^{-1}$ |
|---|---|---|---|---|---|---|---|
| Event Yields | | | | | | | |
| $N_{B^0\to K^+K^-}$ | 110 | 40 | 60 | $N_{B^0\to K^+\pi^-}$ | 21660 | 200 | 400 |
| $N_{B^0\to\pi^+\pi^-}$ | 5880 | 120 | 240 | $N_{B_s^0\to K^+K^-}$ | 7190 | 100 | 200 |
| $N_{B_s^0\to\pi^+K^-}$ | 1510 | 90 | 200 | $N_{B_s^0\to\pi^+\pi^-}$ | 30 | 50 | 100 |
| $N_{\Lambda_b\to pK^-}$ | 1090 | 40 | 90 | $N_{\Lambda_b\to p\pi^-}$ | 700 | 30 | 50 |
| $N_{C\,bkg}$ | 30600 | 500 | 1100 | $N_{P\,bkg}$ | 25500 | 400 | 900 |
| Charge asymmetries | | | | | | | |
| $\mathcal{A}^{\mathcal{CP}}_{K^+\pi^-}$ | -0.098 | 0.008 | 0.004 | $\mathcal{A}^{\mathcal{CP}}_{\pi^+K^-}$ | 0.39 | 0.05 | 0.03 |
| $\mathcal{A}^{\mathcal{CP}}_{pK^-}$ | 0.37 | 0.03 | 0.016 | $\mathcal{A}^{\mathcal{CP}}_{p\pi^-}$ | 0.03 | 0.05 | 0.02 |
| Time dependent asymmetries | | | | | | | |
| $\mathrm{Re}\lambda_{\pi^+\pi^-}$ | -1.03 | 0.23 | 0.08 | $\mathrm{Re}\lambda_{K^+K^-}$ | 0.65 | 0.15 | 0.08 |
| $\mathrm{Im}\lambda_{\pi^+\pi^-}$ | -1.09 | 0.31 | 0.2 | $\mathrm{Im}\lambda_{K^+K^-}$ | 0.70 | 0.14 | 0.07 |
| Mixing | | | | | | | |
| $\Delta\Gamma_d$ [ps$^{-1}$] | 0 | 0.04 | 0.017 | $\Delta\Gamma_s$ [ps$^{-1}$] | 0.067 | 0.06 | 0.03 |
| $\Delta M_d$ [ps$^{-1}$] | 0.507 | 0.016 | 0.006 | $\Delta M_s$ [ps$^{-1}$] | 17.77 | 0.07 | 0.03 |
| Lifetimes | | | | | | | |
| $\tau_d$ [ps] | 1.530 | 0.013 | 0.006 | $\tau_s$ [ps] | 1.478 | 0.038 | 0.018 |
| $\tau_{\Lambda_b}$ [ps] | 1.379 | 0.035 | 0.015 | - | - | - | - |
| Masses | | | | | | | |
| $M_{B^0}$ [GeV/c$^2$] | 5.2795 | 0.0002 | 0.00009 | $M_{B_s^0}$ [GeV/c$^2$] | 5.3663 | 0.0003 | 0.00016 |
| $M_{\Lambda_b}$ [GeV/c$^2$] | 5.6202 | 0.0007 | 0.0003 | - | - | - | - |
| Mass resolution and radiative tail | | | | | | | |
| $f_G$ | 0.840 | 0.02 | 0.008 | $\sigma_1$ [GeV/c$^2$] | 0.0185 | 0.0003 | 0.0001 |
| $\sigma_2$ [GeV/c$^2$] | 0.045 | 0.004 | 0.0017 | $f_E$ | 0.050 | 0.009 | 0.005 |
| $s$ [c$^2$/GeV] | 10 | 4 | 2 | - | - | - | - |
| Signal tagging | | | | | | | |
| $\epsilon_1^{H_b}$ | 0.2530 | 0.0025 | 0.0015 | $\epsilon_2^{H_b}$ | 0.0700 | 0.0015 | 0.0007 |
| $\epsilon_3^{H_b}$ | 0.0540 | 0.0014 | 0.0007 | $\epsilon_4^{H_b}$ | 0.0420 | 0.0011 | 0.0004 |
| $\epsilon_5^{H_b}$ | 0.030 | 0.001 | 0.0005 | $\epsilon_{SS\pi}^{B^0}$ | 0.103 | 0.002 | 0.001 |
| $\epsilon_{SS\pi}^{B_s^0}$ | 0.072 | 0.003 | 0.0015 | $\epsilon_{SS\pi}^{\Lambda_b}$ | 0.076 | 0.007 | 0.004 |
| $\epsilon_{SSK}^{B^0}$ | 0.104 | 0.002 | 0.001 | $\epsilon_{SSK}^{B_s^0}$ | 0.145 | 0.004 | 0.002 |
| $\epsilon_{SSK}^{\Lambda_b}$ | 0.107 | 0.008 | 0.004 | $w_1^{H_b}$ | 0.449 | 0.010 | 0.004 |
| $w_2^{H_b}$ | 0.357 | 0.018 | 0.007 | $w_3^{H_b}$ | 0.295 | 0.021 | 0.011 |
| $w_4^{H_b}$ | 0.237 | 0.023 | 0.009 | $w_5^{H_b}$ | 0.166 | 0.026 | 0.012 |
| $w_{SS\pi}^{B^0}$ | 0.400 | 0.018 | 0.009 | $w_{SS\pi}^{B_s^0}$ | 0.489 | 0.060 | 0.029 |
| $w_{SS\pi}^{\Lambda_b}$ | 0.509 | 0.05 | 0.03 | $w_{SSK}^{B^0}$ | 0.530 | 0.019 | 0.009 |
| $w_{SSK}^{B_s^0}$ | 0.326 | 0.050 | 0.024 | $w_{SSK}^{\Lambda_b}$ | 0.456 | 0.044 | 0.023 |
| Background mass and lifetime parameters | | | | | | | |
| $m_0$ [GeV/c$^2$] | 5.1495 | 0.0016 | 0.0007 | $\sigma_P$ [GeV/c$^2$] | 0.0216 | 0.0015 | 0.0007 |
| $c_P$ | 14.6 | 0.6 | 0.2 | $\alpha$ [c$^2$/GeV] | -1.50 | 0.05 | 0.03 |
| $\tau_\xi^{P\,bkg}$ [ps/GeV] | 0.1900 | 0.0015 | 0.0007 | $\tau_\xi^{C\,bkg}$ [ps/GeV] | 0.1900 | 0.0016 | 0.0007 |
| Background tagging | | | | | | | |
| $\epsilon_1^{bkg}$ | 0.2530 | 0.0020 | 0.0011 | $\epsilon_2^{bkg}$ | 0.0700 | 0.0012 | 0.0005 |
| $\epsilon_3^{bkg}$ | 0.0540 | 0.0010 | 0.0004 | $\epsilon_4^{bkg}$ | 0.0420 | 0.0009 | 0.0004 |
| $\epsilon_5^{bkg}$ | 0.0300 | 0.0008 | 0.0003 | $\epsilon_{SS\pi}^{bkg}$ | 0.1030 | 0.0014 | 0.0008 |
| $\epsilon_{SSK}^{bkg}$ | 0.1040 | 0.0013 | 0.0007 | - | - | - | - |

Table 31: Sensitivities to the parameters of the final p.d.f. defined in Eq. (71), estimated as the root mean squares of the central values returned by the fits to 150 and 20 fast MC samples, corresponding to an integrated luminosity $L = 0.2\,\text{fb}^{-1}$ and $L = 0.8\,\text{fb}^{-1}$ respectively. Input values of the event yields are only indicated for $L = 0.2\,\text{fb}^{-1}$. For $L = 0.8\,\text{fb}^{-1}$ the sample sizes are a factor of four larger.

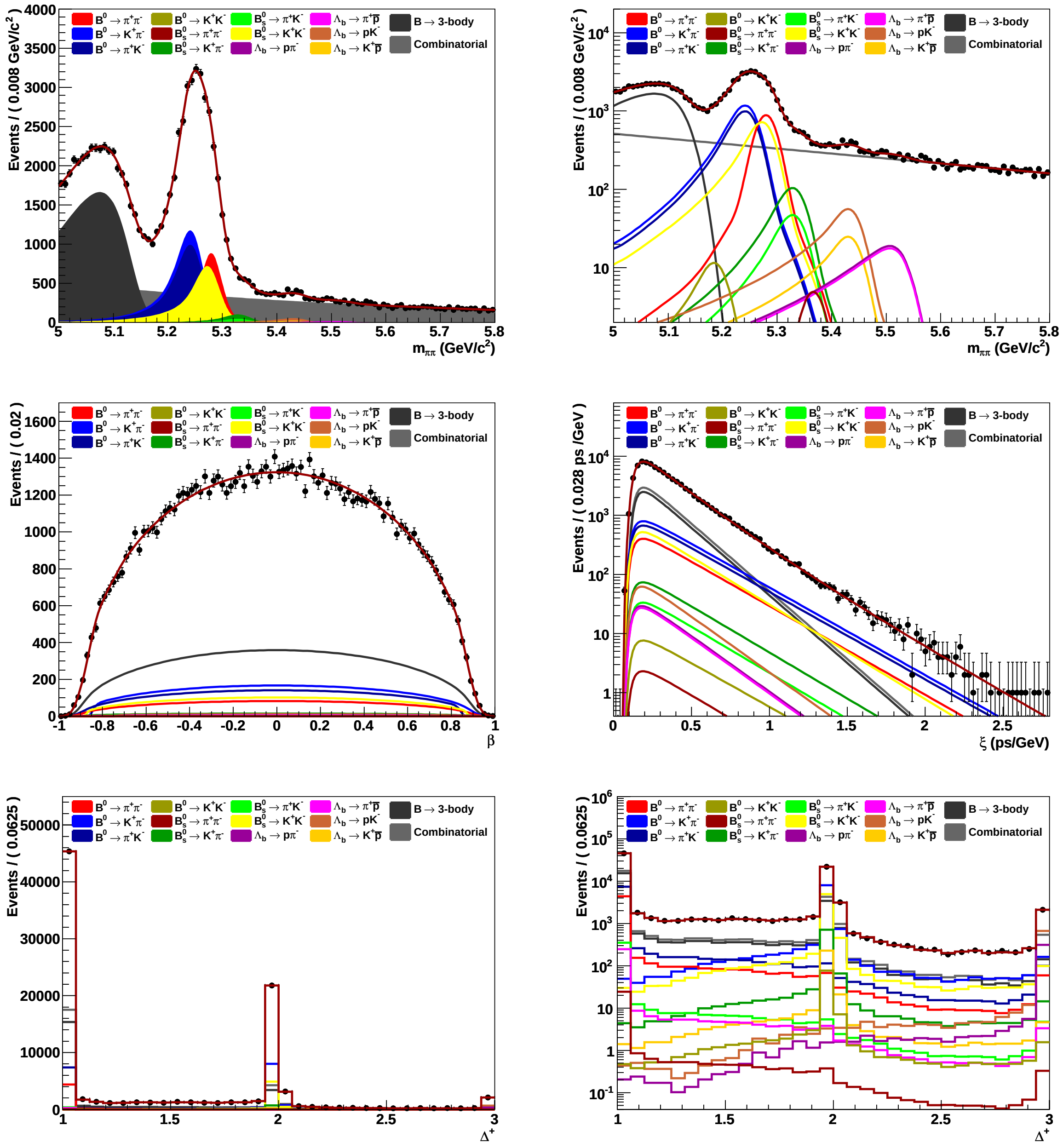

Figure 23: Distributions from a fast MC sample corresponding to an integrated luminosity $L = 200\,\text{pb}^{-1}$, with the result of the fit superimposed: $m_{\pi\pi}$ with linear scale (top left) and logarithmmic scale (top right), $\beta$ (middle left), $\xi$ (middle right), $\Delta^+$ with linear scale (bottom left) and logarithmic scale (bottom right). The various signal and background components are also shown.

| Quantity | Current experimental knowledge or prediction | LHCb stat. sensitivity | | |
|---|---|---|---|---|
| | | $0.2\,\text{fb}^{-1}$ | $2\,\text{fb}^{-1}$ | $10\,\text{fb}^{-1}$ |
| $\mathcal{A}^{\mathcal{CP}}_{K^+\pi^-}$ | $-0.098^{+0.012}_{-0.011}$ | 0.008 | 0.0025 | 0.001 |
| $\mathcal{A}^{\mathcal{CP}}_{\pi^+K^-}$ | $0.39 \pm 0.15 \pm 0.08$ | 0.05 | 0.016 | 0.007 |
| $\mathcal{A}^{\mathcal{CP}}_{p\pi^-}$ | $0.03 \pm 0.17 \pm 0.05$ | 0.05 | 0.016 | 0.007 |
| $\mathcal{A}^{\mathcal{CP}}_{pK^-}$ | $0.37 \pm 0.17 \pm 0.03$ | 0.03 | 0.009 | 0.004 |
| $\mathcal{A}^{dir}_{\pi^+\pi^-}$ | $0.38 \pm 0.06$ | 0.13 | 0.04 | 0.018 |
| $\mathcal{A}^{mix}_{\pi^+\pi^-}$ | $-0.65 \pm 0.07$ | 0.13 | 0.04 | 0.018 |
| $\text{Corr}(\mathcal{A}^{dir}_{\pi^+\pi^-}, \mathcal{A}^{mix}_{\pi^+\pi^-})$ | 0.08 | $-0.03$ | $-0.03$ | $-0.03$ |
| $\mathcal{A}^{dir}_{K^+K^-}$ | $-0.09 \pm 0.04\,^*$ | 0.15 | 0.047 | 0.021 |
| $\mathcal{A}^{mix}_{K^+K^-}$ | $0.77 \pm 0.18\,^*$ | 0.11 | 0.035 | 0.016 |
| $\text{Corr}(\mathcal{A}^{dir}_{K^+K^-}, \mathcal{A}^{mix}_{K^+K^-})$ | $-0.02\,^*$ | 0.02 | 0.02 | 0.02 |
| $\frac{\mathcal{BR}(B^0\to\pi^+\pi^-)}{\mathcal{BR}(B^0\to K^+\pi^-)}$ | $0.264 \pm 0.011$ | 0.006 | 0.0019 | 0.0008 |
| $\frac{\mathcal{BR}(B^0\to K^+K^-)}{\mathcal{BR}(B^0\to K^+\pi^-)}$ | $0.020 \pm 0.008 \pm 0.006$ | 0.005 | 0.0015 | 0.0007 |
| $\frac{f_s\,\mathcal{BR}(B^0_s\to K^+K^-)}{f_d\,\mathcal{BR}(B^0\to K^+\pi^-)}$ | $0.347 \pm 0.020 \pm 0.021$ | 0.006 | 0.0019 | 0.0008 |
| $\frac{f_s\,\mathcal{BR}(B^0_s\to\pi^+K^-)}{f_d\,\mathcal{BR}(B^0\to K^+\pi^-)}$ | $0.071 \pm 0.010 \pm 0.007$ | 0.004 | 0.0013 | 0.0006 |
| $\frac{f_s\,\mathcal{BR}(B^0_s\to\pi^+\pi^-)}{f_d\,\mathcal{BR}(B^0\to K^+\pi^-)}$ | $0.007 \pm 0.004 \pm 0.005$ | 0.002 | 0.0006 | 0.0003 |
| $\frac{f_{\Lambda_b}\,\mathcal{BR}(\Lambda_b\to p\pi^-)}{f_d\,\mathcal{BR}(B^0\to K^+\pi^-)}$ | $0.0415 \pm 0.0074 \pm 0.0058$ | 0.0016 | 0.0005 | 0.00023 |
| $\frac{f_{\Lambda_b}\,\mathcal{BR}(\Lambda_b\to pK^-)}{f_d\,\mathcal{BR}(B^0\to K^+\pi^-)}$ | $0.0663 \pm 0.0089 \pm 0.0084$ | 0.0018 | 0.0006 | 0.00025 |

Table 32: LHCb statistical sensitivities to the most relevant physics measurements under study. The sensitivities for $L = 0.2\,\text{fb}^{-1}$ have been scaled to those for $L = 2\,\text{fb}^{-1}$ and $L = 10\,\text{fb}^{-1}$ according to the inverse of the square root of the integrated luminosity. *These quantities are still unmeasured. The predictions reported in this table have been obtained employing the current $\phi_s$ measurement at the Tevatron, as discussed in App. B.

# 10 Extraction of $\gamma$ and $\phi_s$

As discussed in Sec. 2.2, once $\mathcal{A}^{dir}_{\pi^+\pi^-}$, $\mathcal{A}^{mix}_{\pi^+\pi^-}$, $\mathcal{A}^{dir}_{K^+K^-}$ and $\mathcal{A}^{mix}_{K^+K^-}$ are experimentally known, Eqs. (11), (12), (13) and (14) constitute a system of four equations with seven unknowns: $d$, $\vartheta$, $\phi_d$, $\gamma$, $d'$, $\vartheta'$ and $\phi_s$. However, $\phi_d$ is well measured by the $B$ factories, $\phi_d = (0.768 \pm 0.028)$ rad [69], and will be further refined by LHCb [21], while $\phi_s$ has been recently constrained by CDF and D0 although with a large uncertainty, $\phi_s = (-0.70 \pm 0.28)$ rad [69,70]. The phase $\phi_s$ will be eventually measured by LHCb with comparable precision to $\phi_d$, mainly by means of the $B^0_s \to J/\psi\phi$ decay [22]. By assuming the validity of the U-spin symmetry, which leads to the identities $d = d'$ and $\vartheta = \vartheta'$, just three unknowns are left, and the system is fully solvable. It is then possible to determine simultaneously $d$, $\vartheta$ and $\gamma$.

Furthermore, since we have more equations than unknowns, we can exploit the additional equation in order to make an alternative measurement of $\phi_s$ from these channels alone. In this section we will follow this approach. An alternative strategy, where $\phi_s$ is constrained to the results of the LHCb $B^0_s \to J/\psi\phi$ analysis, was instead adopted in previous studies, as reported in Ref. [19]. It is important to note, as we will show, that it is not necessary to rely on the U-spin symmetry validity *tout-court*, since meaningful results can be obtained also in the case when large non-factorizable U-spin breaking effects are taken into account.

As detailed in Ref. [11], first insights into U-spin breaking effects can be already obtained from present data using the measurements of the charge asymmetries and branching fractions of the U-spin related decay modes $B^0 \to K^+\pi^-$ and $B^0_s \to \pi^+K^-$. Their decay amplitudes can be also parameterized as:

$$A(B^0 \to K^+\pi^-) = -P\left(1 - re^{i\delta}e^{i\gamma}\right) \tag{78}$$

and

$$A(B^0_s \to \pi^+K^-) = P_s \frac{\lambda}{\sqrt{1-\lambda^2}}\left(1 + \frac{1-\lambda^2}{\lambda^2} r_s e^{i\delta_s} e^{i\gamma}\right), \tag{79}$$

where $P_{(s)}$ and $r_{(s)}e^{i\delta_{(s)}}$ are hadronic parameters which describe penguin amplitudes and tree-to-penguin ratios respectively. The validity of the U-spin symmetry leads to the identities $r = r_s$ and $\delta = \delta_s$, as well as to the relation:

$$\frac{\mathcal{A}^{\mathcal{CP}}_{\pi^+K^-}}{\mathcal{A}^{\mathcal{CP}}_{K^+\pi^-}} \simeq -\left|\frac{P_s}{P}\right|^2 \frac{\mathcal{BR}(B^0 \to K^+\pi^-)}{\mathcal{BR}(B^0_s \to \pi^+K^-)}. \tag{80}$$

It is then possible to obtain experimental insights into U-spin breaking effects by writing

$$\left|\frac{P_s}{P}\right|_{\text{exp}} = \left|\frac{P_s}{P}\right| \sqrt{\left[\frac{r_s}{r}\right]\left[\frac{\sin\delta_s}{\sin\delta}\right]} = 1.06 \pm 0.28, \tag{81}$$

that is in good agreement with the QCD sum-rule results of Ref. [71]

$$\left|\frac{P_s}{P}\right|^{\text{QCDSR}}_{\text{fact}} = 1.02^{+0.11}_{-0.10}, \tag{82}$$

although the experimental errors are still large. This quantity will be eventually measured with great precision at LHCb, thus providing a much more stringent test.

LHCb will measure all the four $\mathcal{CP}$ violating observables $\mathcal{A}^{dir}_{\pi^+\pi^-}$, $\mathcal{A}^{mix}_{\pi^+\pi^-}$, $\mathcal{A}^{dir}_{K^+K^-}$ and $\mathcal{A}^{mix}_{K^+K^-}$ with high precision (see Tab. 32). In particular, the determination of $\mathcal{A}^{mix}_{KK}$ will open a new avenue to the measurement of the $B^0_s$ mixing phase $\phi_s$ [11].

As U-spin symmetry predicts that $\mathcal{A}^{dir}_{K^+K^-}$ and $\mathcal{A}^{\mathcal{CP}}_{K^+\pi^-}$ shall assume the same value — neglecting exchange and annihilation topologies in the $\mathcal{A}^{dir}_{K^+K^-}$ decay amplitude, which are already quite constrained by current measurements of the branching fractions of the $B^0 \to K^+K^-$ and $B^0_s \to \pi^+\pi^-$ decays (see Tabs. 3 and 5) — we find that an optimal strategy consists of substituting the direct $\mathcal{CP}$ asymmetry term $\mathcal{A}^{dir}_{K^+K^-}$ with $\mathcal{A}^{\mathcal{CP}}_{K^+\pi^-}$, due to the much smaller statistical error achievable on $\mathcal{A}^{\mathcal{CP}}_{K^+\pi^-}$. LHCb will further constrain the size of the exchange and penguin annihilation amplitudes contributing to the $B^0_s \to K^+K^-$, by considerably improving the measurements of the $B^0 \to K^+K^-$ and $B^0_s \to \pi^+\pi^-$ branching ratios.

In order to estimate the LHCb sensitivity to $\gamma$ and $\phi_s$ we assume two alternative sets of hypothetical measurements corresponding to an integrated luminosity $L = 2\,\mathrm{fb}^{-1}$ in two different scenarios:

- *Scenario A*, large $B^0_s$ mixing phase implying NP in $b \leftrightarrow s$ $|\Delta F| = 2$ transitions;
- *Scenario B*, small $B^0_s$ mixing phase as expected in the SM.

By following the procedure outlined in App. B, but using the values of Tab. 33 as inputs, we obtain the central values and the sensitivities for $d$, $\vartheta$, $\gamma$ and $\phi_s$ reported in the same table, corresponding to an integrated luminosity $L = 2\,\mathrm{fb}^{-1}$. The corresponding p.d.f.'s are shown in Fig. 24. We emphasize that we are taking into account U-spin breaking effects as large as 20% for $d$ and $\pm 20°$ for $\vartheta$, independently varied in the expressions of $\mathcal{A}^{\mathcal{CP}}_{K^+\pi^-}$ and $\mathcal{A}^{dir}_{K^+K^-}$, in order to consider sizable U-spin breaking effects not only with respect to the $B^0 \to \pi^+\pi^-$ channel but also between $B^0 \to K^+\pi^-$ and $B^0_s \to K^+K^-$.

Note that we can achieve a good precision not only on $\gamma$, but also on $\phi_s$, either in *Scenario A* or in *Scenario B*. The $\phi_s$ sensitivity corresponding to $L = 2\,\mathrm{fb}^{-1}$ is only a factor 1.5 worse to the one obtainable from the LHCb $B^0_s \to J/\psi\phi$ analysis [22] in both scenarios.

The $\mathcal{CP}$ measurements from these channels will be important, first in order to give an independent confirmation of the baseline $\phi_s$ measurement from the $B^0_s \to J/\psi\phi$ decay, secondly to allow for an unambiguous determination of $\phi_s$ in conjuction with $B^0_s \to J/\psi\phi$. In particular, as pointed out in Ref. [11], we will be able to distinguish between the cases of $\phi_s = 0°$ and $\phi_s = 180°$, which is important for the search of NP.

We have also studied the dependence of the sensitivities to $\gamma$ and $\phi_s$ as functions of the size of non-factorizable U-spin breaking that is allowed. Fig. 25 shows the variations of the 68% and 95% probability intervals for $\gamma$ and $\phi_s$ in *Scenario A*, separately as functions of $\hat{\xi}$ and $\Delta\vartheta$. The variations of the 68% and 95% probability intervals for $\gamma$ and $\phi_s$ in *Scenario B* are not shown since they exhibit the same behaviour as in *Scenario A*. It is apparent that the dependence on the amount of U-spin breaking is significantly more pronounced for $\gamma$ than for $\phi_s$.

| Hypothetical LHCb measurements | | |
|---|---|---|
| | Scenario A | Scenario B |
| $\phi_d$ | $(0.768 \pm 0.020)$ rad | $(0.768 \pm 0.020)$ rad |
| $\mathcal{A}^{dir}_{\pi^+\pi^-}$ | $0.38 \pm 0.04$ | $0.38 \pm 0.04$ |
| $\mathcal{A}^{mix}_{\pi^+\pi^-}$ | $-0.65 \pm 0.04$ | $-0.65 \pm 0.04$ |
| $\mathcal{A}^{\mathcal{CP}}_{K^+\pi^-}$ | $-0.100 \pm 0.008\,^*$ | $-0.100 \pm 0.008\,^*$ |
| $\mathcal{A}^{mix}_{K^+K^-}$ | $0.700 \pm 0.035$ | $0.150 \pm 0.035\,^{**}$ |

| Sensitivities | | |
|---|---|---|
| | Scenario A | Scenario B |
| $d$ | $0.63 \pm 0.13$ | $0.63 \pm 0.13$ |
| $\vartheta$ | $(146 \pm 7)^\circ$ | $(146 \pm 7)^\circ$ |
| $\gamma$ | $(70 \pm 7)^\circ$ | $(70 \pm 7)^\circ$ |
| $\phi_s$ | $(-0.66 \pm 0.06)$ rad | $(-0.03 \pm 0.05)$ rad |

Table 33: Hypothetical LHCb measurements corresponding to an integrated luminosity $L = 2\,\mathrm{fb}^{-1}$, and consequent sensitivities on $d$, $\vartheta$, $\gamma$ and $\phi_s$, obtained by taking into account sizable U-spin breaking effects (see text). Two scenarios are considered: large NP contributions to the mixing phase of the $B^0_s$ system ($A$) and of the SM expectation ($B$).

$^*$The hypothetical measurement of $\mathcal{A}^{\mathcal{CP}}_{K^+\pi^-}$ here shown includes a conservative guess of the systematic error, about three times larger than the expected statistical one. No studies have yet been performed to make a realistic estimate of this uncertainty.

$^{**}$Note that the uncertainty on $\mathcal{A}^{mix}_{K^+K^-}$ remains unchanged while passing from *Scenario A* to *Scenario B*, as determined by means of maximum likelihood fits to specific toy MC data. The toy experiments producing these data were identical to those described in Sec. 9.5.2, apart from the settings of the true values of $\mathrm{Re}\lambda_{\mathrm{K^+K^-}}$ and $\mathrm{Im}\lambda_{\mathrm{K^+K^-}}$, which were tuned according to *Scenario B*.

Further information useful to constrain NP will come from the measurement of $\Delta\Gamma_s$, which can be also performed using $B^0_s \to K^+K^-$ decays. With an integrated luminosity $L = 2\,\mathrm{fb}^{-1}$, LHCb expects a statistical error on $\Delta\Gamma_s$ of about 0.02 $\mathrm{ps}^{-1}$ (obtained by rescaling to $L = 2\,\mathrm{fb}^{-1}$ the sensitivity reported in Tab. 31). In the presence of NP, $\Delta\Gamma_s$ is modified as follows [48]:

$$\Delta\Gamma_s \simeq \Delta\Gamma_s^{SM} \cos(\phi_s^\Delta), \tag{83}$$

where $\phi_s^\Delta$ has been defined in Eq. (24), i.e. NP effects can reduce the observed value of $\Delta\Gamma_s$ with respect to the SM expectation.

Before concluding this section, we emphasize that this study is based on the parameterization of the direct and mixing-induced $\mathcal{CP}$ terms outlined in Sec. 2.2. Such a parameterization is valid in the SM, as well as in extensions of the SM where NP only affects the values of the phases $\phi_d$ and $\phi_s$ via $b \leftrightarrow s$ $|\Delta F| = 2$ transitions. However, NP may also enter in other sectors, for example through $|\Delta F| = 1$ transitions in the penguin graphs contributing to the decay amplitudes. In such a case, the parameterization that we used would no longer be appropriate, and an analysis based on it could lead for example to a value of $\gamma$ in disagreement with the SM expectation, hence revealing the presence of NP effects.

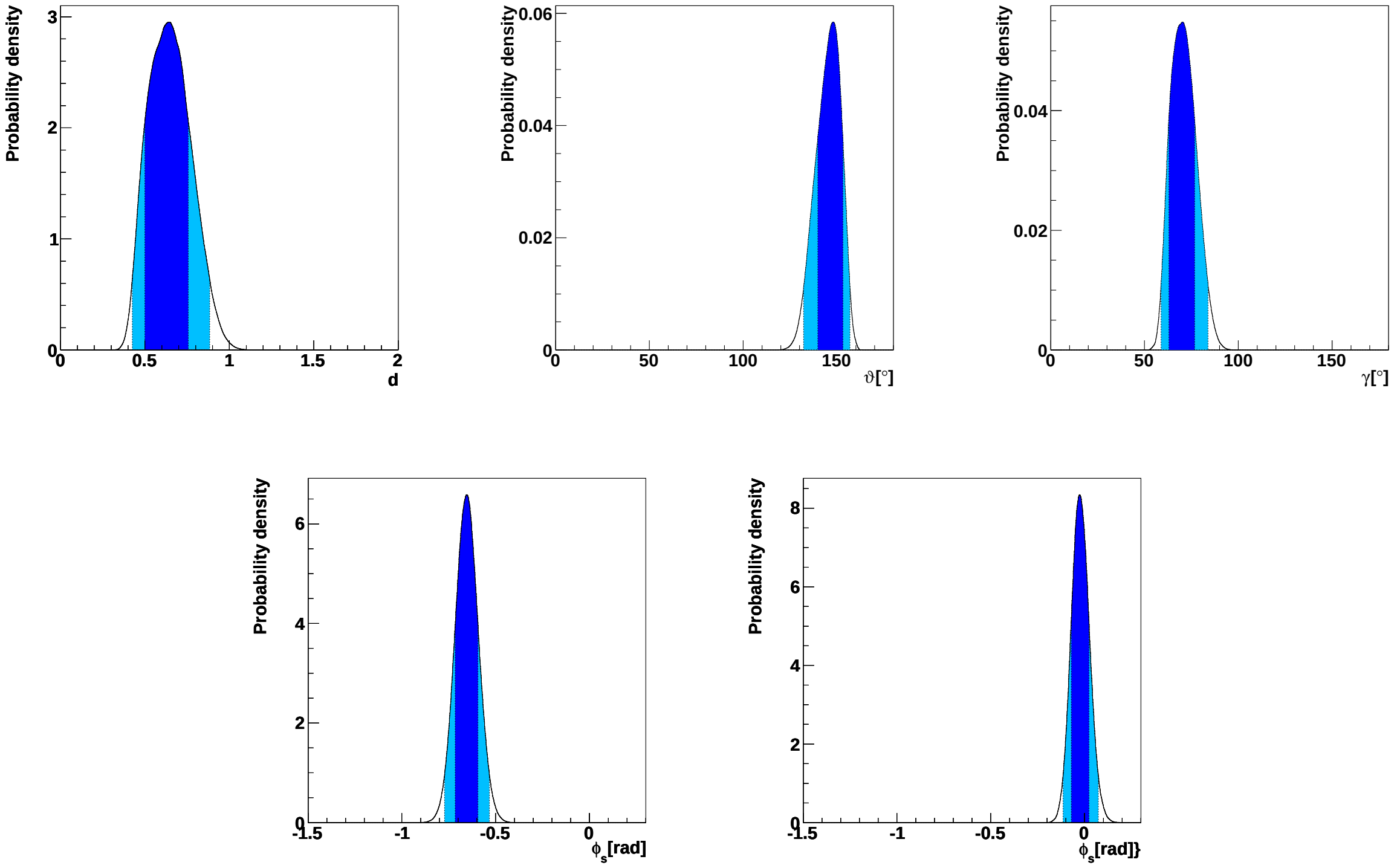

Figure 24: P.d.f.'s for $d$ (top left), $\vartheta$ (top middle), $\gamma$ (top right) and $\phi_s$ (bottom left) obtained by using the hypothetical LHCb measurements of *Scenario A* shown in Tab. 33, with sensitivities corresponding to an integrated luminosity $L = 2\,\mathrm{fb}^{-1}$. In the case of *Scenario B*, the p.d.f.'s for $d$, $\vartheta$ and $\gamma$ are identical to the ones of *Scenario A*. The p.d.f. for $\phi_s$ is instead different, and it is shown in the bottom right plot. The condition $\vartheta > 90^\circ$ was imposed in order to isolate the SM solution for $\gamma$ (see App. B for a discussion on this). 68% (dark) and 95% (light) probability intervals are also indicated.

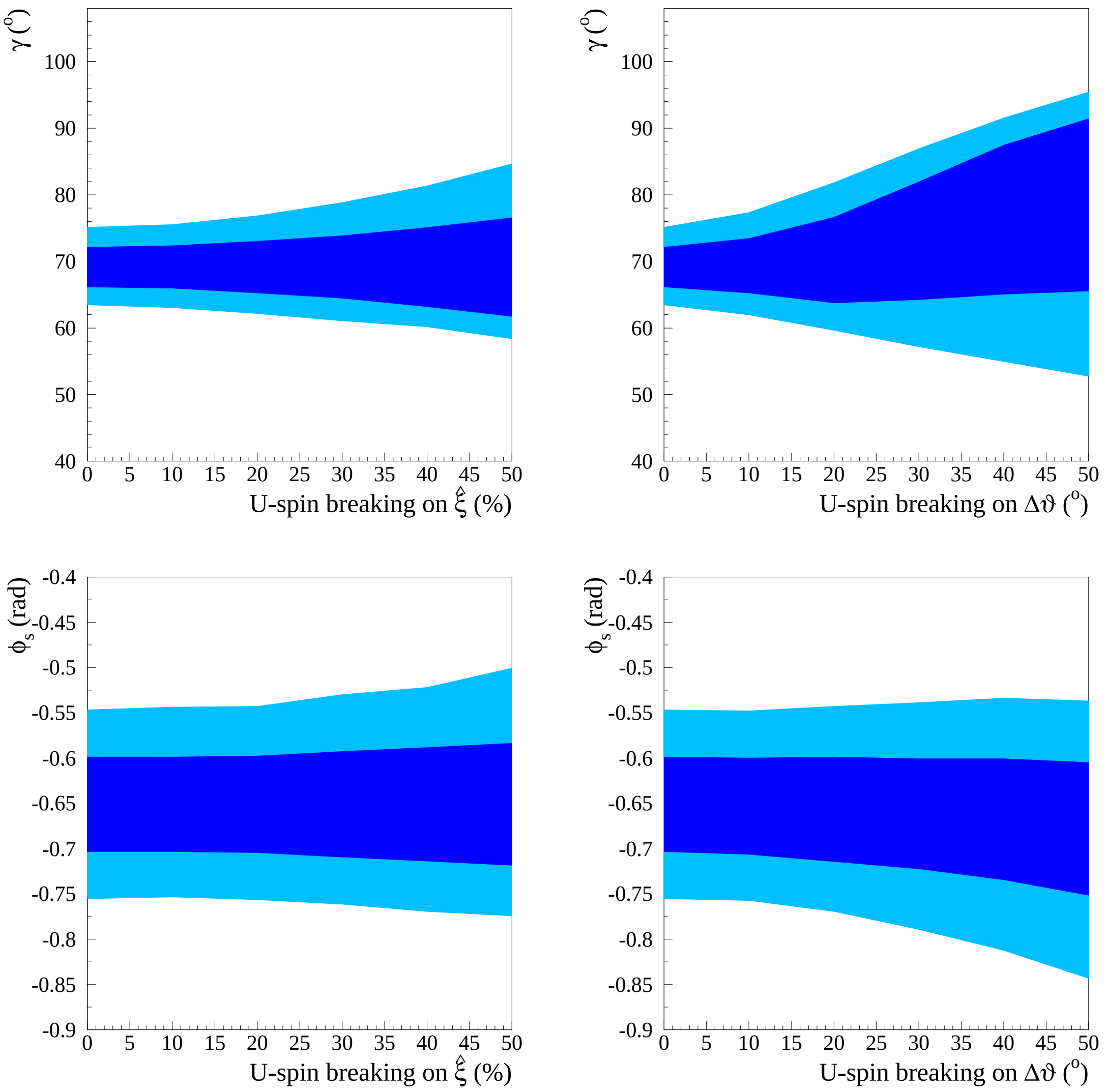

Figure 25: Variation of the 68% (dark) and 95% (light) probability intervals for $\gamma$ (top) and for $\phi_s$ (bottom) as a function of the size of U-spin breaking, separately in $\hat{\xi}$ (left) and $\Delta\vartheta$ (right). The plots are obtained by using the hypothetical LHCb measurements of *Scenario A* shown in Tab. 33, with sensitivities corresponding to an integrated luminosity $L = 2\,\mathrm{fb}^{-1}$. The condition $\vartheta > 90^\circ$ was imposed in order to isolate the SM solution for $\gamma$ (see App. B for a discussion on this).

# 11 Conclusions

The aim of this document is to make a comprehensive review of the readiness for the analysis of the $H_b \to h^+ h'^-$ family of decays at LHCb.

The theoretical interest in these channels is due to the fact that their $\mathcal{CP}$ asymmetries are sensitive to the presence of New Physics, either in the decay amplitudes (as penguin diagrams contribute to them) or in the mixing phases of neutral $B$ mesons (in the case of decays to $\mathcal{CP}$ eigenstates).

By studying in detail the efficiencies of the selection and trigger algorithms by means of full Monte Carlo simulations, we have shown that LHCb can collect about $2 \cdot 10^5$ such events per $\text{fb}^{-1}$ of integrated luminosity. This event yield is significantly larger than that presently available at the $B$ factories and the Tevatron.

Besides the event selection, our analysis covered a wide range of experimental aspects, notably including a first study on the effects of misalignments on the physics performance, the description of the technique employed to calibrate the particle identification observables, the measurement of the lifetime, the effect of QED radiation on the invariant mass line shape and the flavour tagging strategy and performance.

Furthermore, a complete model for extracting the physics parameters of interest from the data sample has been constructed and described in detail. The model consists of the definition of a p.d.f. for six different observables, namely the reconstructed invariant mass of the $B$ hadron candidate under the $\pi^+\pi^-$ hypothesis and its proper decay time, the momentum asymmetry between the two $B$ hadron daughters, the two particle identification observables of the $B$ hadron daughters and the response of the flavour tagging algorithm grouped into several categories of different tagging purities. Twelve distinct signal final states were included in the model, with in addition two background components.

By using such a model, first without including the background, we performed a fit featuring 53 free parameters to a sample of offline-selected full Monte Carlo events, obtaining results statistically consistent with the inputs of the full Monte Carlo simulation. In order to study the LHCb sensitivity we generated a series of fast Monte Carlo samples using the p.d.f. of our model, now including the background components, and then retrieving from the fits the set of input parameters used in the generation phase. These fits included 68 free parameters, and they were performed either to data samples corresponding to an integrated luminosity $L = 0.2\,\text{fb}^{-1}$ or $L = 0.8\,\text{fb}^{-1}$, the latter in order to verify the scaling of the sensitivities as the inverse of the square root of the integrated luminosity.

We have shown that LHCb will surpass the current experimental knowledge on the ratios of $\mathcal{CP}$-averaged branching fractions and charge asymmetries of the decays under study already with an integrated luminosity $L = 0.2\,\text{fb}^{-1}$, i.e. 1/10 of a nominal year of LHCb data taking. Such measurements will be performed very early in the experimental programme, as soon as an accurate calibration of the particle identification and a good tracking quality will be achieved. As the ratios of branching fractions involve kinematically similar decay modes, most systematic effects related to each individual mode cancel out in the ratios. For charge asymmetry measurements, systematic uncertainties can come from detector-induced charge asymmetries between positively and negatively charged particles, due to their different interaction probabilities with the detector material, and possibly from the presence of a so-called production asymmetry, arising during the hadronization phase of $b$ and $\bar{b}$ quarks into $B$ hadrons. Particle identification will be the crucial ingredient in this first phase, and the accuracy of its calibration will play a relevant role in the definition of systematic uncertainties. Another important aspect entering systematic uncertainties, particularly relevant for the discrimination of the rarest decay modes from the dominating ones, will involve the control of the invariant mass scale and line shape.

The measurements of the time dependent $\mathcal{CP}$ asymmetries will require some more time with respect to the measurements mentioned in the previous paragraph as, in addition, a good control of the proper time measurement and of the initial $B$ flavour tagging will be required. We have seen that LHCb will reach a better statistical sensitivity for the measurements of $\mathcal{A}^{dir}_{\pi^+\pi^-}$ and $\mathcal{A}^{mix}_{\pi^+\pi^-}$ than presently available with an integrated luminosity $L = 2\,\mathrm{fb}^{-1}$. For $\mathcal{A}^{dir}_{K^+K^-}$ and $\mathcal{A}^{mix}_{K^+K^-}$, no measurement yet exists, and LHCb will make a determination with a precision similar to that obtainable on $\mathcal{A}^{dir}_{\pi^+\pi^-}$ and $\mathcal{A}^{mix}_{\pi^+\pi^-}$. Additional sources of systematic uncertainties include here the proper time resolution model and parameters and the knowledge of the proper time acceptance function. As far as flavour tagging is concerned, the relevant parameters employed in the $\mathcal{CP}$ fitting model will be extracted from the $H_b \to h^+h'^-$ data sample itself, i.e. from kinematically similar control channels, without the need of external inputs from analyses of topologically different decay modes. For this reason, systematic effects due to the flavour tagging calibration are expected to be small.

Finally, we have seen how to extract from the measurements of the various $\mathcal{CP}$ violating terms the $\gamma$ angle of Unitarity Triangle and the $B^0_s$ mixing phase $\phi_s$. Even allowing for sizable U-spin symmetry breaking effects, we have shown that the sensitivities achievable on $\gamma$ and $\phi_s$, estimated to be $7^\circ$ and 0.06 rad respectively, will be very competitive with other measurements. As the method relies on the estimation of the size of U-spin breaking, theoretical progresses on this subject, driven by new LHCb experimental measurements, will play an important role.

# Appendix A

## Basic $\mathcal{CP}$ violation formalism

The decay rates for initial neutral $B$ and $\overline{B}$ mesons, decaying to a final state $f$ at proper time $t$, can be written as:

$$\Gamma_{B\to f}(t) = \frac{|A_f|^2}{2} e^{-\Gamma t} \left[I_+(t) + I_-(t)\right] \tag{A.1}$$

and

$$\Gamma_{\overline{B}\to f}(t) = \frac{|A_f|^2}{2} \left|\frac{p}{q}\right|^2 e^{-\Gamma t} \left[I_+(t) - I_-(t)\right]. \tag{A.2}$$

Analogously, for the $\mathcal{CP}$-conjugate final state $\overline{f}$, one has:

$$\Gamma_{\overline{B}\to \overline{f}}(t) = \frac{|\overline{A}_{\overline{f}}|^2}{2} e^{-\Gamma t} \left[\overline{I}_+(t) + \overline{I}_-(t)\right] \tag{A.3}$$

and

$$\Gamma_{B\to \overline{f}}(t) = \frac{|\overline{A}_{\overline{f}}|^2}{2} \left|\frac{q}{p}\right|^2 e^{-\Gamma t} \left[\overline{I}_+(t) - \overline{I}_-(t)\right]. \tag{A.4}$$

$A_f$ and $\overline{A}_{\overline{f}}$ are the instantaneous decay amplitudes for $B \to f$ and $\overline{B} \to \overline{f}$ respectively, and $\Gamma$ is the average decay width for the two mass eigenstates $|B_L\rangle$ and $|B_H\rangle$ — $L$ and $H$ denoting the lightest and the heaviest of the two states respectively — which can be expressed in the $|B\rangle$ and $\left|\overline{B}\right\rangle$ basis as:

$$|B_L\rangle = \frac{1}{\sqrt{|p|^2 + |q|^2}} \left(p\,|B\rangle + q\left|\overline{B}\right\rangle\right) \tag{A.5}$$

and

$$|B_H\rangle = \frac{1}{\sqrt{|p|^2 + |q|^2}} \left(p\,|B\rangle - q\left|\overline{B}\right\rangle\right). \tag{A.6}$$

The functions $I_+(t)$, $I_-(t)$, $\overline{I}_+(t)$ and $\overline{I}_-(t)$ are:

$$I_+(t) = \left(1 + |\lambda_f|^2\right) \cosh\left(\frac{\Delta\Gamma}{2}t\right) - 2Re(\lambda_f) \sinh\left(\frac{\Delta\Gamma}{2}t\right), \tag{A.7}$$

$$I_-(t) = \left(1 - |\lambda_f|^2\right) \cos(\Delta m\, t) - 2Im(\lambda_f) \sin(\Delta m\, t), \tag{A.8}$$

$$\overline{I}_+(t) = \left(1 + |\overline{\lambda}_{\overline{f}}|^2\right) \cosh\left(\frac{\Delta\Gamma}{2}t\right) - 2Re(\overline{\lambda}_{\overline{f}}) \sinh\left(\frac{\Delta\Gamma}{2}t\right) \tag{A.9}$$

and

$$\overline{I}_{-}(t) = \left(1 - \left|\overline{\lambda}_{\overline{f}}\right|^2\right) \cos(\Delta m\, t) - 2Im(\overline{\lambda}_{\overline{f}}) \sin(\Delta m\, t), \tag{A.10}$$

where $\Delta m$ is the mass difference, positive by definition:

$$\Delta m = m_H - m_L,$$

while $\Delta\Gamma$ is the difference of the decay widths of the mass eigenstates, whose sign is positive in the SM:

$$\Delta\Gamma = \Gamma_L - \Gamma_H.$$

The complex numbers $\lambda_f$ and $\overline{\lambda}_{\overline{f}}$ are defined by the following equations:

$$\lambda_f = \frac{q}{p}\frac{\overline{A}_f}{A_f} \tag{A.11}$$

and

$$\bar{\lambda}_{\overline{f}} = \frac{p}{q}\frac{A_{\overline{f}}}{\overline{A}_{\overline{f}}}. \tag{A.12}$$

If $f$ is a $\mathcal{CP}$ eigenstate, one has $f = \overline{f}$, and the four decay rates reduce to two. We can define the time dependent $\mathcal{CP}$ asymmetry as:

$$\mathcal{A}_f^{\mathcal{CP}}(t) = \frac{\Gamma_{\overline{B}\to f}(t) - \Gamma_{B\to f}(t)}{\Gamma_{\overline{B}\to f}(t) + \Gamma_{B\to f}(t)} = \frac{\left(|\lambda_f|^2 - 1\right)\cos(\Delta m\, t) + 2\mathrm{Im}\lambda_f \sin(\Delta m\, t)}{\left(|\lambda_f|^2 + 1\right)\cosh\left(\frac{\Delta\Gamma}{2}t\right) - 2\mathrm{Re}\lambda_f \sinh\left(\frac{\Delta\Gamma}{2}t\right)}. \tag{A.13}$$

In the derivation of Eq. (A.13) we have used $\left|\frac{q}{p}\right| \simeq 1$, that is a very good approximation for both the $B^0 - \overline{B}^0$ and $B_s^0 - \overline{B}_s^0$ systems [20].

Introducing the quantities

$$\mathcal{A}_f^{dir} = \frac{|\lambda_f|^2 - 1}{|\lambda_f|^2 + 1}, \tag{A.14}$$

$$\mathcal{A}_f^{mix} = \frac{2\mathrm{Im}\lambda_f}{|\lambda_f|^2 + 1} \tag{A.15}$$

and

$$\mathcal{A}_f^{\Delta} = \frac{2\mathrm{Re}\lambda_f}{|\lambda_f|^2 + 1}, \tag{A.16}$$

the time-dependent $\mathcal{CP}$ asymmetry (A.13) can be rewritten as

$$\mathcal{A}_f^{\mathcal{CP}}(t) = \frac{\mathcal{A}_f^{dir}\cos(\Delta m\, t) + \mathcal{A}_f^{mix}\sin(\Delta m\, t)}{\cosh\left(\frac{\Delta\Gamma}{2}t\right) - \mathcal{A}_f^{\Delta}\sinh\left(\frac{\Delta\Gamma}{2}t\right)}. \tag{A.17}$$

$\mathcal{A}_f^{dir}$, $\mathcal{A}_f^{mix}$ and $\mathcal{A}_f^{\Delta}$ are by definition related by the following equation:

$$\left(\mathcal{A}_f^{dir}\right)^2 + \left(\mathcal{A}_f^{mix}\right)^2 + \left(\mathcal{A}_f^{\Delta}\right)^2 = 1. \tag{A.18}$$

The quantities $\mathcal{A}_f^{dir}$ and $\mathcal{A}_f^{mix}$ parametrize direct and mixing-induced $\mathcal{CP}$ violation respectively.

Another interesting case is when $f$ is a flavour specific final state, i.e. $f \neq \overline{f}$. One can write:

$$\lambda_f = \overline{\lambda}_{\overline{f}} = 0, \tag{A.19}$$

since only the $B$ has instantaneous access to the decay channel $f$, while only the $\overline{B}$ has instantaneous access to the decay channel $\overline{f}$. In this case, the functions $I_+(t)$, $I_-(t)$, $\overline{I}_+(t)$ and $\overline{I}_-(t)$ reduce to:

$$I_+(t) = \overline{I}_+(t) = \cosh\left(\frac{\Delta\Gamma}{2}t\right) \tag{A.20}$$

and

$$I_-(t) = \overline{I}_-(t) = \cos(\Delta m\, t). \tag{A.21}$$

By using the four decay rates, it is possible to define the following $\mathcal{CP}$ asymmetry:

$$\mathcal{A}^{\mathcal{CP}}(t) = \frac{\left[\Gamma_{\overline{B}\to\overline{f}}(t) + \Gamma_{B\to\overline{f}}(t)\right] - \left[\Gamma_{\overline{B}\to f}(t) + \Gamma_{B\to f}(t)\right]}{\left[\Gamma_{\overline{B}\to\overline{f}}(t) + \Gamma_{B\to\overline{f}}(t)\right] + \left[\Gamma_{\overline{B}\to f}(t) + \Gamma_{B\to f}(t)\right]}, \tag{A.22}$$

that is not dependent on time, and is identically equal to the charge asymmetry defined as:

$$\mathcal{A}^{\mathcal{CP}} = \frac{\left|\overline{A}_{\overline{f}}\right|^2 - |A_f|^2}{\left|\overline{A}_{\overline{f}}\right|^2 + |A_f|^2} = \frac{\left|\frac{\overline{A}_{\overline{f}}}{A_f}\right|^2 - 1}{\left|\frac{\overline{A}_{\overline{f}}}{A_f}\right|^2 + 1}. \tag{A.23}$$

The charge asymmetry differs from zero in the presence of direct $\mathcal{CP}$ violation and parameterizes its magnitude.

A further relevant case is when no mixing can occur, i.e. for charged $B$ mesons and $B$ baryons, due to charge and baryon number conservation respectively. We have in this case just two decay rates, $\Gamma_{B\to f}$ and $\Gamma_{\overline{B}\to\overline{f}}$, which can be obtained by observing that:

$$I_+(t) = \overline{I}_+(t) = I_-(t) = \overline{I}_-(t) = 1, \tag{A.24}$$

thus leading to purely exponential rates. As in the flavour specific decays of neutral $B$ mesons, $\mathcal{CP}$ violation may enter through a non-zero charge asymmetry.

# Appendix B

## Angle $\gamma$ from present $B^0 \to h^+h'^-$ measurements

The direct and mixing-induced $\mathcal{CP}$ coefficients $\mathcal{A}^{dir}_{\pi^+\pi^-}$ and $\mathcal{A}^{mix}_{\pi^+\pi^-}$ are experimentally known, while $\mathcal{A}^{dir}_{K^+K^-}$ and $\mathcal{A}^{mix}_{K^+K^-}$ are still unmeasured. However, still employing the U-spin symmetry, the value of $\mathcal{A}^{dir}_{K^+K^-}$ can be estimated to be equal to that of $\mathcal{A}^{\mathcal{CP}}_{K^+\pi^-}$, which is already well measured. This requires that penguin annihilation and exchange topologies do not give significant contributions to the decay amplitudes, a fact that can be eventually probed by measuring the branching fractions of the $B^0 \to K^+K^-$ and $B^0_s \to \pi^+\pi^-$ decays [10]. In this case we have a closed system of three equations and three unknowns. It is then possible to determine $d$, $\vartheta$ and $\gamma$ by using the currently available measurements.

In order to infer a joint p.d.f. for $d$, $\vartheta$ and $\gamma$ we will make use of a Bayesian approach implemented in the software packages developed by the UTfit Collaboration [69, 72]. The problem is in fact analogous to the one of inferring a p.d.f. for the CKM parameters $\bar{\rho}$ and $\bar{\eta}$, given a set of measurements and theoretical predictions related to them by analytical constraints. We do not need to rely on the full validity of the U-spin symmetry, and so we allow for a non-factorizable breaking of the U-spin relations $d = d'$ and $\vartheta = \vartheta'$ of up to 20% and $\pm 20°$ respectively, similar to what is done in Ref. [11], i.e. $\hat{\xi}$ and $\Delta\vartheta$ are varied uniformly in the ranges:

$$\hat{\xi} = d'/d \in [0.8,\, 1.2], \tag{B.1}$$

and

$$\Delta\vartheta = \vartheta' - \vartheta \in [-20°,\, 20°]. \tag{B.2}$$

By integrating out two variables in turn from the joint p.d.f. we obtain the one dimensional p.d.f.'s for $d$, $\vartheta$ and $\gamma$ shown in Fig. B.1. Due to the non-linearity of the system of equations, it is apparent in the plots that two solutions are present. However, it can be shown that the two solutions are well separated in the (d, $\vartheta$, $\gamma$) space, and it turns out that the solution which has the central value of $\gamma \simeq 70°$ corresponding to the SM expectation could be easily isolated by imposing a requirement like $\vartheta > 90°$. Although non-factorizable effects might have a relevant impact on $\vartheta$, it is not expected that they will change the sign of the cosine of this strong phase, which is predicted to be negative by factorization [11]. Consequently, the solution with $\vartheta \simeq 40°$ can be excluded through this argument, and thus we are justified to employ the constraint $\vartheta > 90°$. In this way we get the p.d.f.'s shown in Fig. B.2. The corresponding 68% probability intervals for $d$, $\vartheta$ and $\gamma$ are reported in Tab. B.1.

The 68% probability interval $\gamma = (70 \pm 8)°$ that we obtain is fully consistent with the current average $\gamma = (78 \pm 12)°$ [69], determined from pure tree decays at the $B$ factories, or with the indirect prediction from UT fits assuming the SM validity $\gamma = (65.6 \pm 3.3)°$ [69]. Furthermore, we have also obtained probability intervals for the hadronic parameters $d$ and $\vartheta$, that can be used to drive theory in correctly modelling the underlying hadron dynamics involved in the decays under study.

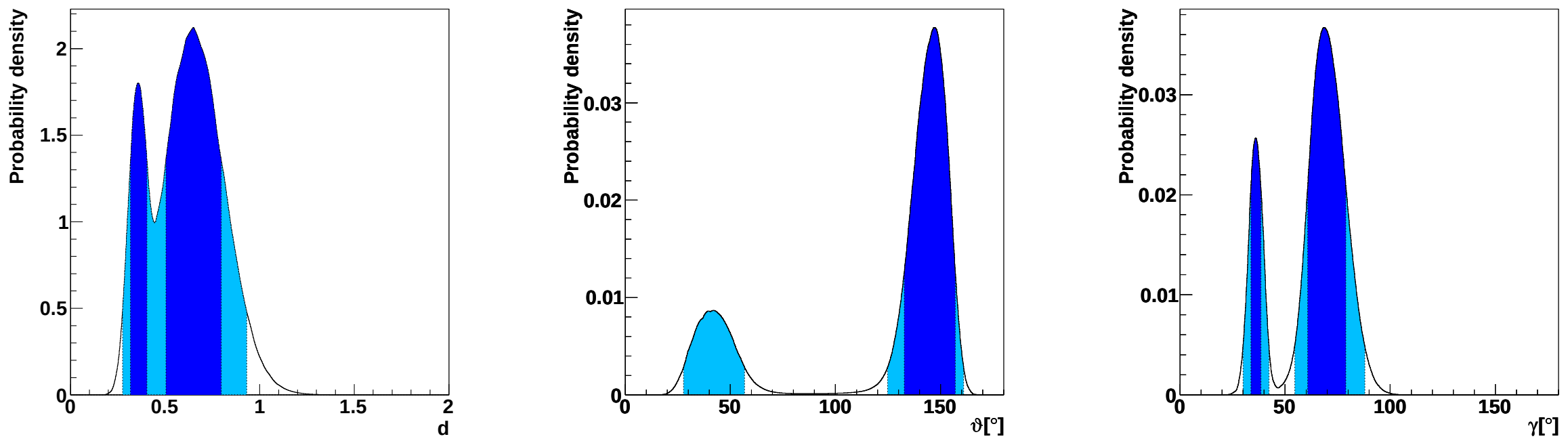

Figure B.1: From left to right: p.d.f.'s for $d$, $\vartheta$ and $\gamma$ obtained by using the current experimental measurements. 68% (dark) and 95% (light) probability intervals are also indicated.

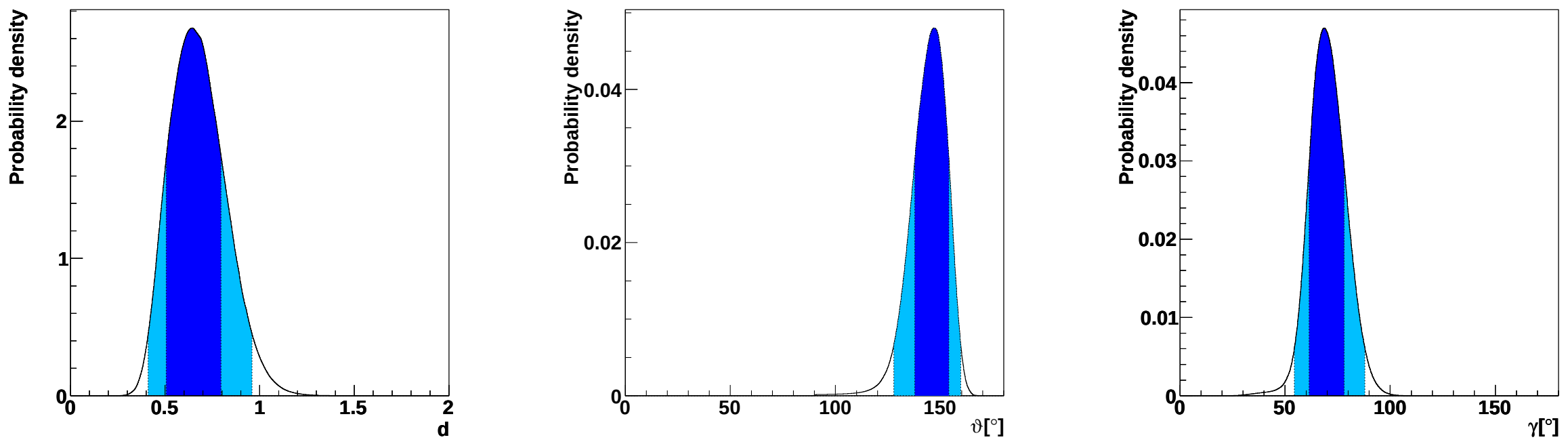

Figure B.2: From left to right: p.d.f.'s for $d$, $\vartheta$ and $\gamma$ obtained by using the current experimental measurements, and imposing $\vartheta > 90^\circ$ in order to isolate the SM solution for $\gamma$. 68% (dark) and 95% (light) probability intervals are also indicated.

It is also interesting to see how the sensitivity depends on the size of U-spin breaking that is allowed. Fig. B.3 shows the variations of the 68% and 95% probability intervals for $\gamma$ as a function of the size of U-spin breaking, separately in $\hat{\xi}$ and $\Delta\vartheta$. The half-width of the 68% (95%) probability interval increases from about $5^\circ$ ($11^\circ$) in the case of zero non-factorizable U-spin breaking effects, up to $9^\circ$ ($18^\circ$) and $14^\circ$ ($25^\circ$) for breakings of 50% on $\hat{\xi}$ and $\pm 50^\circ$ on $\Delta\vartheta$ respectively.

We can also predict probability intervals for $\mathcal{A}^{dir}_{K^+K^-}$ and also, by relying on the current knowledge of the $B^0_s$ mixing phase $\phi_s$, for $\mathcal{A}^{mix}_{K^+K^-}$. Of course, using the alternative parameterization of the time dependent $\mathcal{CP}$ asymmetry terms, we can also make a prediction for $\mathrm{Re}\lambda_{K^+K^-}$ and $\mathrm{Im}\lambda_{K^+K^-}$. Such predictions have been obtained by assuming an additional independent U-spin breaking of 20% and ±20° respectively on the hadronic parameters entering the expressions of $\mathcal{A}^{dir}_{K^+K^-}$, $\mathcal{A}^{mix}_{K^+K^-}$, $\mathrm{Re}\lambda_{K^+K^-}$ and $\mathrm{Im}\lambda_{K^+K^-}$, in order to take into account U-spin breaking effects also while exchanging the values of $\mathcal{A}^{mix}_{K^+K^-}$ and $\mathcal{A}^{\mathcal{CP}}_{K^+\pi^-}$. For this reason, the prediction of $\mathcal{A}^{dir}_{K^+K^-}$ is not simply equal to the measurement of $\mathcal{A}^{\mathcal{CP}}_{K^+\pi^-}$, but has a larger uncertainty.

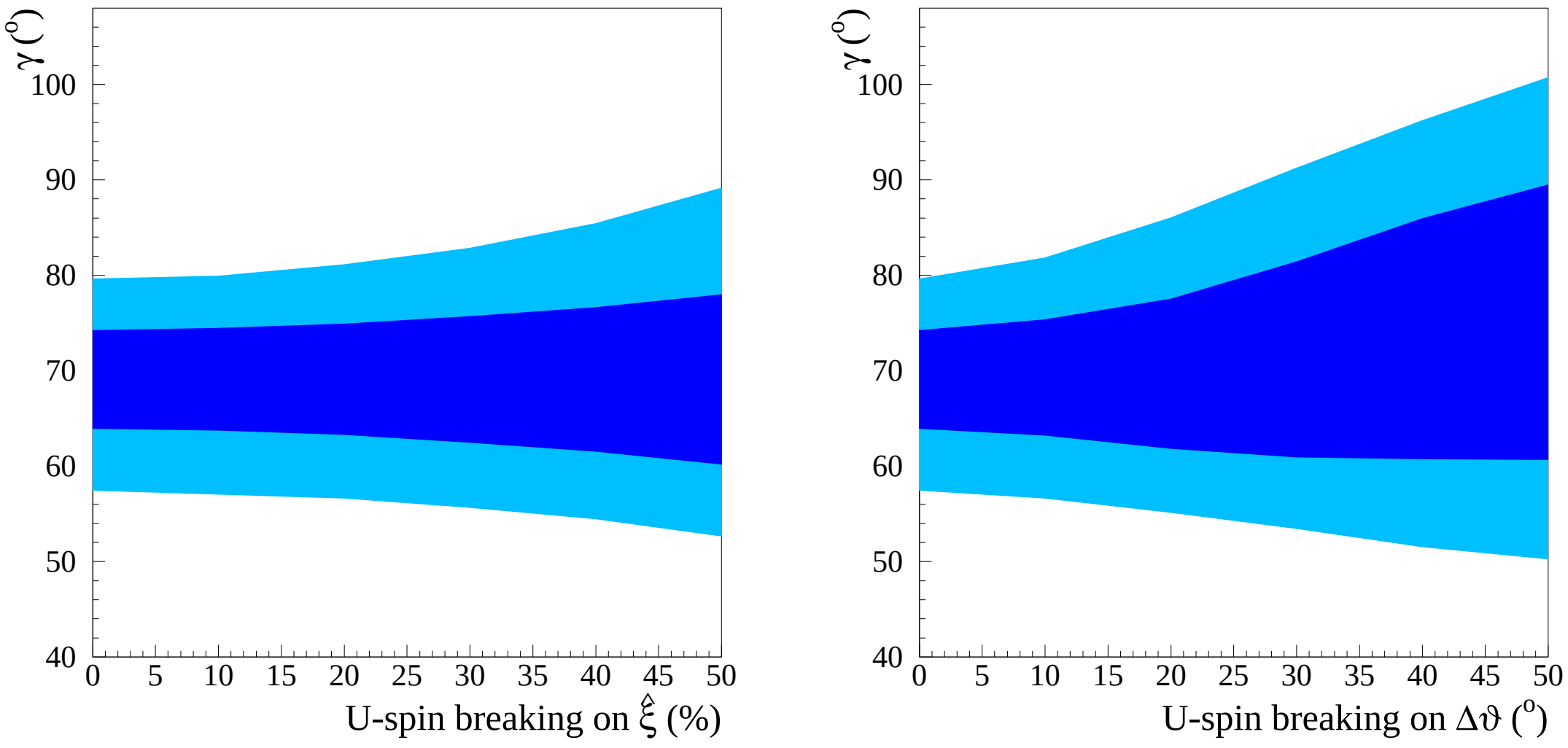

Figure B.3: Variation of the 68% (dark) and 95% (light) probability intervals for $\gamma$ as a function of the size of U-spin breaking that is allowed, separately in $\hat{\xi}$ and $\Delta\vartheta$. The plots are obtained by using the current experimental measurements, and imposing $\vartheta > 90°$ in order to isolate the SM solution for $\gamma$.

All the relevant inputs and the predictions for the parameters of interest are summarized in Tab. B.1. The p.d.f.'s for $\mathcal{A}^{dir}_{K^+K^-}$, $\mathcal{A}^{mix}_{K^+K^-}$, $\mathrm{Re}\lambda_{K^+K^-}$ and $\mathrm{Im}\lambda_{K^+K^-}$ are shown in Fig. B.4. While the central values and the uncertainties of $d$, $\vartheta$, $\gamma$ and $\mathcal{A}^{dir}_{K^+K^-}$ are not affected by the measurement of $\phi_s$, those of $\mathcal{A}^{mix}_{K^+K^-}$ are strongly affected. In particular, the large central value predicted for $\mathcal{A}^{mix}_{K^+K^-}$ is a direct consequence of the large measured value of $\phi_s$.

| Inputs | |
|---|---|
| $\phi_d$ | $(0.768 \pm 0.028)$ rad |
| $\phi_s$ | $(-0.70 \pm 0.28)$ rad |
| $\mathcal{A}^{dir}_{\pi^+\pi^-}$ | $0.38 \pm 0.06$ |
| $\mathcal{A}^{mix}_{\pi^+\pi^-}$ | $-0.65 \pm 0.07$ |
| Corr($\mathcal{A}^{dir}_{\pi^+\pi^-}$, $\mathcal{A}^{mix}_{\pi^+\pi^-}$) | 0.08 |
| $\mathcal{A}^{\mathcal{CP}}_{K^+\pi^-}$ | $-0.098 \pm 0.012$ |

| Predictions | |
|---|---|
| $d$ | $0.65 \pm 0.15$ |
| $\vartheta$ | $(146 \pm 8)^\circ$ |
| $\gamma$ | $(70 \pm 8)^\circ$ |
| $\mathcal{A}^{dir}_{K^+K^-}$ | $-0.09 \pm 0.04$ |
| $\mathcal{A}^{mix}_{K^+K^-}$ | $0.77 \pm 0.18$ |
| Corr($\mathcal{A}^{dir}_{K^+K^-}$, $\mathcal{A}^{mix}_{K^+K^-}$) | $-0.02$ |
| Re$\lambda_{K^+K^-}$ | $0.65 \pm 0.18$ |
| Im$\lambda_{K^+K^-}$ | $0.70 \pm 0.17$ |
| Corr(Re$\lambda_{K^+K^-}$, Im$\lambda_{K^+K^-}$) | $-0.88$ |

Table B.1: Experimental inputs and predictions in terms of 68% probability intervals for the various parameters of interest. Note that the large central value of $\mathcal{A}^{mix}_{K^+K^-}$ follows from the large central value of the $\phi_s$ measurement. By assuming instead that $\phi_s$ is equal to the SM expectation from UT fits $\phi_s = (-0.0366 \pm 0.0015)$ rad [69], one would obtain the prediction $\mathcal{A}^{mix}_{K^+K^-} = 0.15 \pm 0.03$.

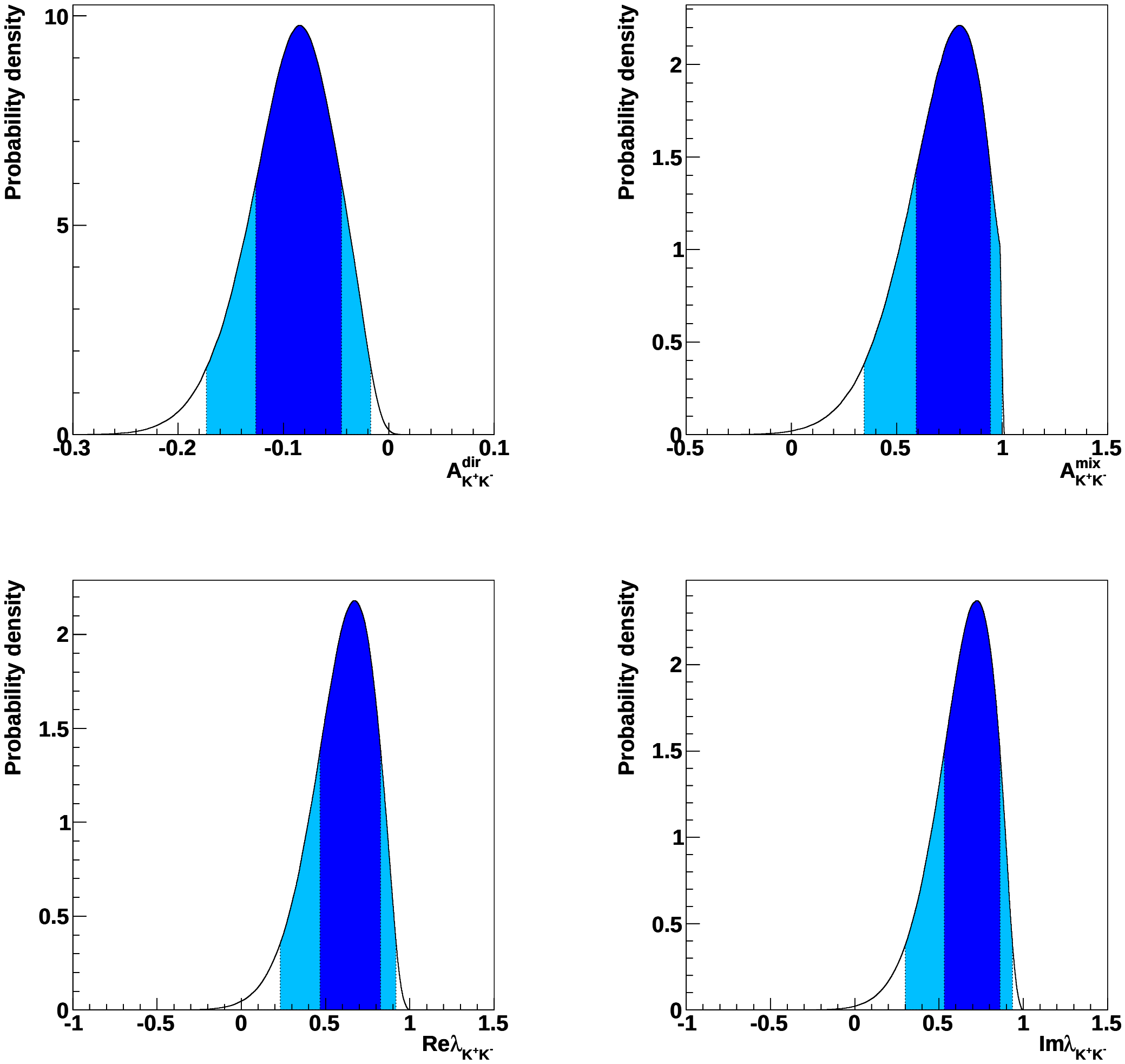

Figure B.4: P.d.f.'s for $\mathcal{A}^{dir}_{K^+K^-}$ (top left), $\mathcal{A}^{mix}_{K^+K^-}$ (top right), $\mathrm{Re}\lambda_{K^+K^-}$ (bottom left) and $\mathrm{Im}\lambda_{K^+K^-}$ (bottom right), obtained by using the current experimental measurements and imposing $\vartheta > 90^\circ$ in order to isolate the SM solution for $\gamma$. 68% (dark) and 95% (light) probability intervals are also indicated.

# Chapter 4

# Measurement of mixing-induced CP violation in $B_s^0 \to J/\psi\phi$

J. Albrecht, S. Amato, M. Calvi, A. Carbone, P. Clarke, G. Conti, G. A. Cowan, L. de Paula, T. du Pree, M. Gandelman, C. Göbel, B. Khanji, C. Langenbruch, G. Lanfranchi, C. Linn, S. Hansmann-Menzemer, A. Hicheur, O. Leroy, F. Muheim, M. Musy, N. Mangiafave, S. Poss, G. Raven, F. Rodrigues, A. Satta, B. Souza de Paula, U. Uwer, S. Vecchi and Y. Xie

**Abstract**

For the $B_s^0 \to J/\psi\phi$ decay, the interference between the direct amplitude and the amplitude via $B_s^0$–$\overline{B}_s^0$ oscillation gives rise to a CP violating phase $\phi_s^{J/\psi\phi}$. In the Standard Model, this phase is predicted to be $-2\beta_s = (-0.0360^{+0.0020}_{-0.0016})$ rad, where $\beta_s = \arg\left(-V_{ts}V_{tb}^*/V_{cs}V_{cb}^*\right)$. The measurement of this phase is one of the key goals of the LHCb experiment. Indeed, $\phi_s^{J/\psi\phi}$ is one of the CP observables with the smallest theoretical uncertainty in the Standard Model, and New Physics could significantly modify this prediction, if, for example, new particles contribute to the $B_s^0$–$\overline{B}_s^0$ box diagram. In this document, we explain the steps for measuring $\phi_s^{J/\psi\phi}$ at LHCb.

# Contents

# 1 Introduction

The interference between $B_s^0$ decays to $J/\psi\phi$ either directly or via $B_s^0$–$\overline{B}_s^0$ oscillation gives rise to a CP violating phase $\phi_s^{J/\psi\phi}$. In the Standard Model, this phase is predicted to be $-2\beta_s$, where $\beta_s = \arg\left(-V_{ts}V_{tb}^*/V_{cs}V_{cb}^*\right)$ is the smallest angle of the "b−s unitarity triangle" (Figure 1). The indirect determination via global fits to experimental data gives $2\beta_s = (0.0360^{+0.0020}_{-0.0016})$ rad [1]. The direct measurement of this phase is one of the key goals of the LHCb experiment. Indeed, $\phi_s^{J/\psi\phi}$ is one of the CP observables with the smallest theoretical uncertainty in the Standard Model, and New Physics could significantly modify this prediction, if new particles contribute with a new phase to the $B_s^0$–$\overline{B}_s^0$ box diagram [3, 4]. In this document, we explain the steps necessary to measure $\phi_s^{J/\psi\phi}$. It is organised as follows.

In Section 2, we remind the reader of the essential phenomenological aspects linked to the $B_s^0 \to J/\psi\phi$ decays and define the notation that will be used. In Section 3, we summarize the existing results on $\phi_s^{J/\psi\phi}$, obtained by the Tevatron experiments. In Section 4, we present the analysis strategy and give the steps required before the $\phi_s^{J/\psi\phi}$ measurement itself can be made. In Section 5, we describe the Monte Carlo simulations used in this document. Sections 6 and 7 are dedicated respectively to the selection and triggering of $B_s^0 \to J/\psi(\mu\mu)\phi(KK)$ and related control channels. We also give the distributions of mass, proper time and angles of signal and backgrounds, and we discuss the resolution and acceptance of proper time and angles. Section 8 is devoted to flavour tagging and Section 9 to the fitting procedure. In Section 10, we validate the fitting procedure on Monte Carlo. In Section 11, we discuss the systematic uncertainties.

More details on technical aspects can be found in the appendices and in the referenced notes.

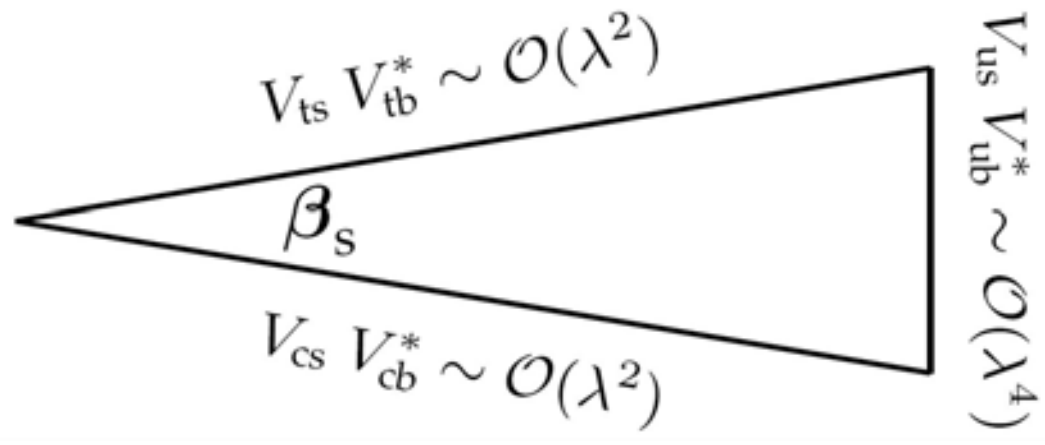

Figure 1: The "b−s unitarity triangle", corresponding to the relation: $V_{us}V_{ub}^* + V_{cs}V_{cb}^* + V_{ts}V_{tb}^* = 0$. $\lambda$ is the sine of the Cabibbo angle [2].

# 2 Phenomenology

## 2.1 Notation, conventions

The phenomenological aspects linked to $B_s^0 \to J/\psi\phi$ decays are described in many articles, e.g. in [5, 6]. Here we give a brief summary and introduce our notation.

We adopt the convention that flavour eigenstates $|B_q\rangle$ and $|\overline{B_q}\rangle$ are associated with the particles $B_q \sim \overline{b}q$ and $\overline{B}_q \sim b\overline{q}$, respectively (q=s,d). These states are linked through

the CP symmetry:

$$\mathrm{CP}|\mathrm{B_q}\rangle = \mathrm{e}^{\mathrm{i}\xi}|\overline{\mathrm{B_q}}\rangle, \qquad \mathrm{CP}|\overline{\mathrm{B_q}}\rangle = \mathrm{e}^{-\mathrm{i}\xi}|\mathrm{B_q}\rangle\,, \tag{1}$$

where $\xi$ is an arbitrary phase. An arbitrary combination of flavour eigenstates

$$a|\mathrm{B_q}\rangle + b|\overline{\mathrm{B_q}}\rangle\,, \tag{2}$$

has a time evolution described by an effective Schrödinger equation:

$$i\frac{\partial}{\partial t}\left(\begin{array}{c} a(t) \\ b(t) \end{array}\right) = \mathcal{H}\left(\begin{array}{c} a(t) \\ b(t) \end{array}\right), \tag{3}$$

where $\mathcal{H}$ is the effective Hamiltonian describing the system.

$$\mathcal{H} = \mathbf{M} - i\frac{\mathbf{\Gamma}}{2}\,, \tag{4}$$

where $\mathbf{M}$ and $\mathbf{\Gamma}$ are $2\times 2$ hermitian matrices. Due to the CPT theorem, their diagonal elements are equal; they represent the mass $M_{\mathrm{B_q}}$ and the width $\Gamma_{\mathrm{q}}$ respectively of the $\mathrm{B_q}$ and $\overline{\mathrm{B}}_{\mathrm{q}}$ mesons. The mass eigenstates are:

$$|\mathrm{B_L}\rangle = p|\mathrm{B_q}\rangle + q|\overline{\mathrm{B_q}}\rangle\,, \tag{5}$$

$$|\mathrm{B_H}\rangle = p|\mathrm{B_q}\rangle - q|\overline{\mathrm{B_q}}\rangle\,. \tag{6}$$

The complex coefficients $p$ and $q$ obey the normalisation condition:

$$|p|^2 + |q|^2 = 1\,. \tag{7}$$

The mass difference, $\Delta m_{\mathrm{q}}$ and the width difference, $\Delta\Gamma_{\mathrm{q}}$, between the mass eigenstates are defined by:

$$\Delta m_{\mathrm{q}} = M_{\mathrm{H}} - M_{\mathrm{L}}, \qquad \Delta\Gamma_{\mathrm{q}} = \Gamma_{\mathrm{L}} - \Gamma_{\mathrm{H}}\,. \tag{8}$$

Hence, the average mass and width can be written:

$$M_{\mathrm{B_q}} = \frac{M_{\mathrm{H}} + M_{\mathrm{L}}}{2}, \qquad \Gamma_{\mathrm{q}} = \frac{\Gamma_{\mathrm{L}} + \Gamma_{\mathrm{H}}}{2}\,. \tag{9}$$

The eigenvalues of $\mathcal{H}$ allow one to link $\Delta m_{\mathrm{q}}$, $\Delta\Gamma_{\mathrm{q}}$ and $q/p$ to the matrix elements $M_{12}$ and $\Gamma_{12}$, and to describe the time evolution of the $\mathrm{B_q}$ and $\overline{\mathrm{B}}_{\mathrm{q}}$ mesons.

We define the instantaneous decay amplitudes:

$$A_f = \langle f|\mathcal{H}|\mathrm{B_q}\rangle\,, \qquad \overline{A}_{\overline{f}} = \langle \overline{f}|\mathcal{H}|\overline{\mathrm{B_q}}\rangle\,, \tag{10}$$

$$A_{\overline{f}} = \langle \overline{f}|\mathcal{H}|\mathrm{B_q}\rangle\,, \qquad \overline{A}_{f} = \langle f|\mathcal{H}|\overline{\mathrm{B_q}}\rangle\,. \tag{11}$$

The decay rates of $\mathrm{B_q}$ or $\overline{\mathrm{B}}_{\mathrm{q}}$ to a final state $f$ or its CP conjugate $\overline{f}$, as a function of its proper time $t$, can be written (omitting a normalisation factor):

$$\Gamma(\mathrm{B_q}(t)\to f) = \frac{|A_f|^2}{2}\mathrm{e}^{-\Gamma_{\mathrm{q}}t}\left[g_+(t) + g_-(t)\right]\,, \tag{12}$$

$$\Gamma(\overline{\mathrm{B}}_{\mathrm{q}}(t)\to f) = \frac{|A_f|^2}{2}\left|\frac{p}{q}\right|^2\mathrm{e}^{-\Gamma_{\mathrm{q}}t}\left[g_+(t) - g_-(t)\right]\,, \tag{13}$$

$$\Gamma(\overline{\mathrm{B}}_{\mathrm{q}}(t)\to \overline{f}) = \frac{|\overline{A}_{\overline{f}}|^2}{2}\mathrm{e}^{-\Gamma_{\mathrm{q}}t}\left[\overline{g}_+(t) + \overline{g}_-(t)\right]\,, \tag{14}$$

$$\Gamma(\mathrm{B_q}(t)\to \overline{f}) = \frac{|\overline{A}_{\overline{f}}|^2}{2}\left|\frac{q}{p}\right|^2\mathrm{e}^{-\Gamma_{\mathrm{q}}t}\left[\overline{g}_+(t) - \overline{g}_-(t)\right]\,. \tag{15}$$

The functions $g_\pm(t)$ are:

$$g_+(t) = (1+|\lambda_f|^2)\cosh(\Delta\Gamma_{\rm q}\,t/2) - 2\,\Re(\lambda_f)\sinh(\Delta\Gamma_{\rm q}\,t/2)\,, \tag{16}$$
$$g_-(t) = (1-|\lambda_f|^2)\cos(\Delta m_{\rm q}\,t) - 2\,\Im(\lambda_f)\sin(\Delta m_{\rm q}\,t)\,, \tag{17}$$
$$\overline{g}_+(t) = (1+|\overline{\lambda}_f|^2)\cosh(\Delta\Gamma_{\rm q}\,t/2) - 2\,\Re(\overline{\lambda}_f)\sinh(\Delta\Gamma_{\rm q}\,t/2)\,, \tag{18}$$
$$\overline{g}_-(t) = (1-|\overline{\lambda}_f|^2)\cos(\Delta m_{\rm q}\,t) - 2\,\Im(\overline{\lambda}_f)\sin(\Delta m_{\rm q}\,t)\,, \tag{19}$$

where we have defined the complex quantities:

$$\lambda_f = \frac{q}{p}\frac{\overline{A}_f}{A_f} \quad\text{and}\quad \overline{\lambda}_f = \frac{p}{q}\frac{A_{\overline{f}}}{\overline{A}_{\overline{f}}}\,. \tag{20}$$

Assuming $f=\bar{f}$ and no CP violation in either the mixing or in the decay, the decay rates can be re-written:

$$\Gamma(\mathrm{B_q}\to f) \propto |A_f|^2 \mathrm{e}^{-\Gamma_{\rm q}t}\Big[\cosh\left(\frac{\Delta\Gamma_{\rm q}t}{2}\right) - \Re\lambda_f \sinh\left(\frac{\Delta\Gamma_{\rm q}t}{2}\right) - \Im\lambda_f \sin(\Delta m_{\rm q}t)\Big]\,, \tag{21}$$
$$\Gamma(\overline{\mathrm{B}}_{\rm q}\to f) \propto |A_f|^2 \mathrm{e}^{-\Gamma_{\rm q}t}\Big[\cosh\left(\frac{\Delta\Gamma_{\rm q}t}{2}\right) - \Re\lambda_f \sinh\left(\frac{\Delta\Gamma_{\rm q}t}{2}\right) + \Im\lambda_f \sin(\Delta m_{\rm q}t)\Big]\,. \tag{22}$$

Let us now consider the case $\mathrm{B_q} = \mathrm{B^0_s}$ and $f = \mathrm{J}/\psi\phi$. Within the Standard Model, the decay $\mathrm{B^0_s}\to \mathrm{J}/\psi\phi$ is dominated by $\overline{\rm b}\to\overline{\rm c}{\rm c}\overline{\rm s}$ quark level transitions, as represented in Figure 2. Following [7], the $\overline{\rm b}\to\overline{\rm c}{\rm c}\overline{\rm s}$ decay amplitude can be expressed as a combination of tree ($A_T$), electroweak and QCD penguin amplitudes:

$$\begin{aligned} A(\overline{\rm b}\to\overline{\rm c}{\rm c}\overline{\rm s}) &= V_{\rm cs}V^*_{\rm cb}(A_T+P_{\rm c}) + V_{\rm us}V^*_{\rm ub}P_{\rm u} + V_{\rm ts}V^*_{\rm tb}P_{\rm t} \\ &= V_{\rm cs}V^*_{\rm cb}(A_T+P_{\rm c}-P_{\rm t}) + V_{\rm us}V^*_{\rm ub}(P_{\rm u}-P_{\rm t})\,, \end{aligned} \tag{23}$$

because $V_{\rm ts}V^*_{\rm tb} = -V_{\rm us}V^*_{\rm ub} - V_{\rm cs}V^*_{\rm cb}$, with three generations. $P_i$ denotes the penguin amplitude with internal $i$-quark, $i\in\{\rm u,c,t\}$.

Since $V_{\rm cs}V^*_{\rm cb}\sim A\lambda^2(1-\lambda^2/2)$ and $V_{\rm us}V^*_{\rm ub}\sim A\lambda^4(\rho+i\eta)$, the contribution of $(P_{\rm u}-P_{\rm t})$ is suppressed by a factor $\sim\lambda^2\simeq 0.05$ with respect to $(A_T+P_{\rm c}-P_{\rm t})$. Therefore in the following, we assume that the $\mathrm{B^0_s}\to\mathrm{J}/\psi\phi$ decay amplitudes are dominated by a single weak phase, $\Phi_{\rm D} = \arg(V_{\rm cs}V^*_{\rm cb})$. The violation of this assumption will be treated as theoretical uncertainty. For discussion of this issue and proposed methods to control this uncertainty see [8,9,10].

Let us denote $\eta_f$ the CP-eigenvalue of the final state, i.e. $\mathrm{CP}|f\rangle = \eta_f|f\rangle$. The amplitude ratio is given, by:

$$\frac{\overline{A}_f}{A_f} = \eta_f \exp^{i(2\Phi_{\rm D}-\xi)}\,. \tag{24}$$

The ratio $q/p$ is given by:

$$\frac{q}{p} = \exp^{-i(\Phi_{\rm M}-\xi)}\,, \tag{25}$$

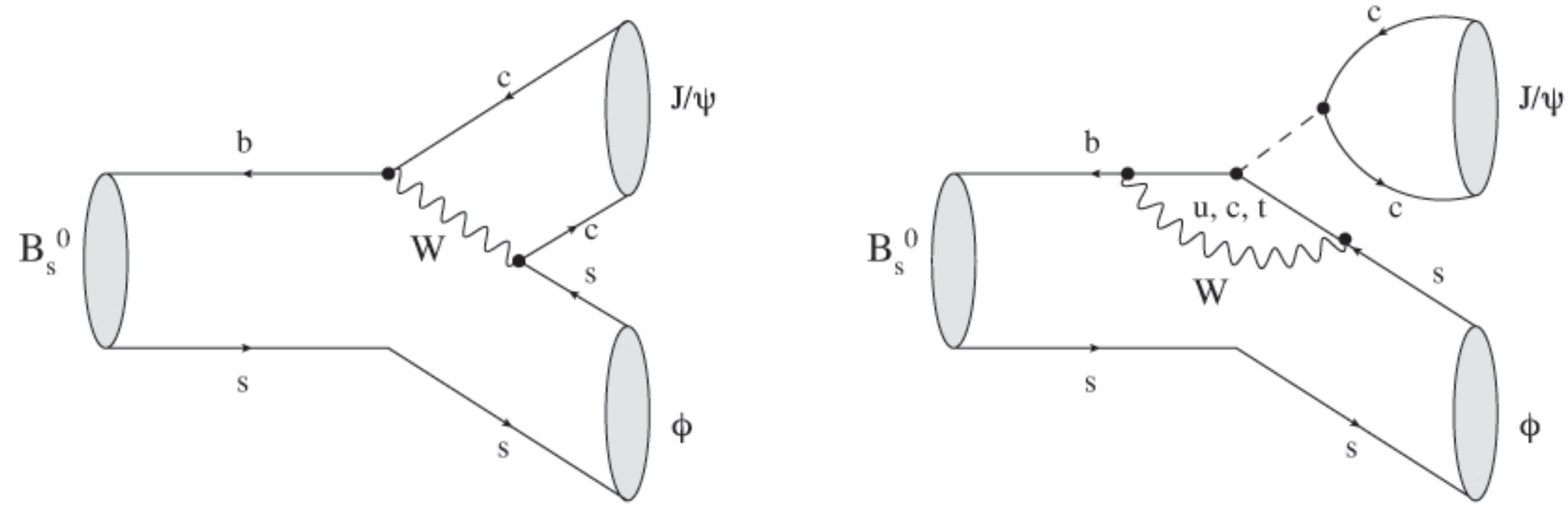

Figure 2: Feynman diagrams contributing to the decay $B_s^0 \to J/\psi\phi$, within the Standard Model. Left: tree; right: penguins.

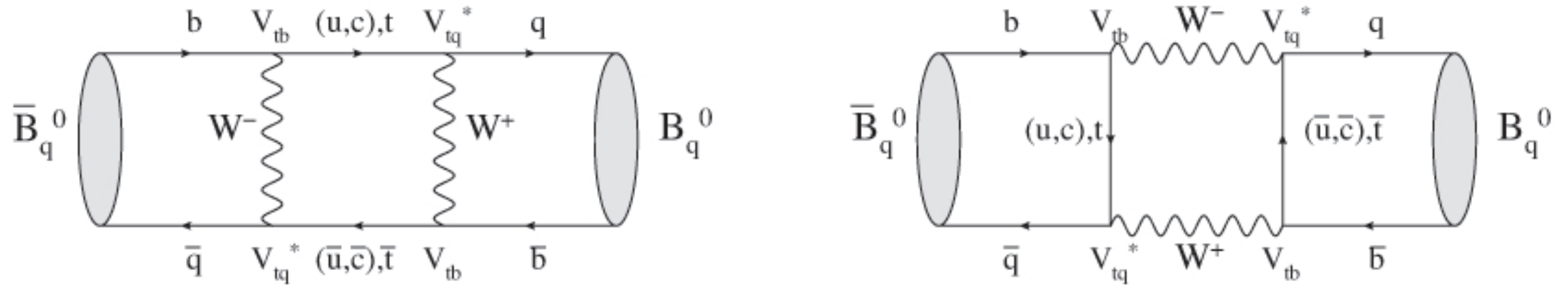

Figure 3: Feynam diagrams responsible for $B_q$–$\overline{B}_q$ mixing, within the Standard Model (q=s,d).

where $\Phi_M = \arg(M_{12})$.

In the Standard Model, $B_q$ mesons oscillate through box diagrams represented in Figure 3. The phase $\Phi_M$, sometimes called "the $B_s^0$ mixing phase", is equal to $2\arg(V_{ts}V_{tb}^*)$. Note that $\overline{A}_f/A_f$, $q/p$, $\Phi_D$ and $\Phi_M$ depend on phase convention. However $\lambda_f$ does not. The weak phase

$$\phi_s^{J/\psi\phi} \equiv -\arg(\eta_f\lambda_f) = \Phi_M - 2\Phi_D \tag{26}$$

is the observable phase we will measure. In the rest of the document, we simply call it $\phi_s$ to simplify the equations and because there is no ambiguity with other decay modes.

Noting that:

$$\Im\lambda_f = -\eta_f \sin\phi_s \quad \text{and } \Re\lambda_f = \eta_f\cos\phi_s\,, \tag{27}$$

we can re-write the decay rates (21-22):

$$\begin{aligned}\Gamma(B_q \to f) &= |A_f|^2 e^{-\Gamma_s t}\Big[\cosh\left(\frac{\Delta\Gamma_s t}{2}\right) - \eta_f\cos\phi_s\sinh\left(\frac{\Delta\Gamma_s t}{2}\right) \\ &\quad +\eta_f\sin\phi_s\sin(\Delta m_s t)\Big]\,, \qquad (28)\\ \Gamma(\overline{B}_q \to f) &= |A_f|^2 e^{-\Gamma_s t}\Big[\cosh\left(\frac{\Delta\Gamma_s t}{2}\right) - \eta_f\cos\phi_s\sinh\left(\frac{\Delta\Gamma_s t}{2}\right) \\ &\quad -\eta_f\sin\phi_s\sin(\Delta m_s t)\Big]\,. \qquad (29)\end{aligned}$$

If we consider that New Physics affects only $M_{12}$ and not $\Gamma_{12}$, its contribution to $\Delta B = 2$ transitions can be parametrised in a model independent manner by introducing the complex parameter $\Delta_s$ [11]:

$$M_{12}^{\mathrm{tot}} = M_{12}^{\mathrm{SM}} \Delta_s = M_{12}^{\mathrm{SM}} |\Delta_s| e^{i\phi_s^{\Delta}} . \quad (30)$$

The phase $\phi_s$ can then be expressed as a function of its Standard Model value and the phase $\phi_s^{\Delta}$ of the New Physics parameter $\Delta_s$:

$$\phi_s = \phi_s^{\mathrm{SM}} + \phi_s^{\Delta} , \quad (31)$$

where:

$$\phi_s^{\mathrm{SM}} = 2 \arg(V_{ts}^* V_{tb}) - 2 \arg(V_{cb} V_{cs}^*) + \delta^{\mathrm{Penguins}} . \quad (32)$$

When the Standard Model penguins are neglected, $\delta^{\mathrm{Penguins}} = 0$, as discussed above, we have:

$$\phi_s^{\mathrm{SM}} = -2\beta_s . \quad (33)$$

The angle $\beta_s$ is defined to be the (positive) smallest angle of the b−s unitarity triangle (Figure 1), corresponding to the relation:

$$V_{us} V_{ub}^* + V_{cs} V_{cb}^* + V_{ts} V_{tb}^* = 0 , \quad (34)$$

$$\beta_s = \arg \left( -\frac{V_{ts} V_{tb}^*}{V_{cs} V_{cb}^*} \right) . \quad (35)$$

Using the Wolfenstein parametrization:

$$\beta_s = \eta \lambda^2 + \mathcal{O}(\lambda^4) . \quad (36)$$

It is important to note that the same New Physics phase, $\phi_s^{\Delta}$, is expected to modify other independent quantities:

$$\Delta\Gamma_s = 2|\Gamma_{12}^{\mathrm{SM}}| \cos(\phi_{s,\mathrm{SM}}^{M/\Gamma} + \phi_s^{\Delta}) , \quad (37)$$

$$a_{fs} = \frac{|\Gamma_{12}^{\mathrm{SM}}|}{|M_{12}^{\mathrm{SM}}|} \frac{\sin(\phi_{s,\mathrm{SM}}^{M/\Gamma} + \phi_s^{\Delta})}{|\Delta_s|} . \quad (38)$$

In the above equations, $a_{fs}$ is the flavour specific asymmetry [12] and

$$\phi_{s,\mathrm{SM}}^{M/\Gamma} = \arg \left( -\frac{M_{12}^{\mathrm{SM}}}{\Gamma_{12}^{\mathrm{SM}}} \right) . \quad (39)$$

The phases $\phi_{s,\mathrm{SM}}^{M/\Gamma}$ and $\phi_s^{\mathrm{SM}}$ (Eq. 33) are two different quantities, as is explained at the end of Ref. [11]. There is no trivial relation between these two observables. In the Standard Model, $\phi_{s,\mathrm{SM}}^{M/\Gamma} = (3.40^{+1.32}_{-0.77}) \times 10^{-3}$ rad [1] while $2\beta_s = (3.6 \pm 0.2) \times 10^{-2}$ rad.

## 2.2 Differential $B_s^0 \to J/\psi\phi$ decay rates

The decay $B_s^0 \to J/\psi\phi$ is a pseudo-scalar to vector-vector decay. Due to total angular momentum conservation, the final state is a superposition of three possible states with relative orbital momentum between the vector mesons $\ell = 0, 1, 2$. The CP-eigenvalue of the final state, $\eta_f$, is given by:

$$\begin{aligned} \mathrm{CP}|\mathrm{J}/\psi\phi\rangle_\ell &= \eta_f|\mathrm{J}/\psi\phi\rangle_\ell \\ &= (-1)^\ell|\mathrm{J}/\psi\phi\rangle_\ell\,. \end{aligned} \tag{40}$$

In the transversity formalism [5], the amplitudes at $t = 0$, $A_0(0)$ and $A_\parallel(0)$, are CP-even ($\ell = 0, 2$), while $A_\perp(0)$ is CP-odd ($\ell = 1$)[1]. The argument of these amplitudes are strong phases denoted $\delta_0$, $\delta_\parallel$ and $\delta_\perp$. Only two amplitudes and two strong phases are independent. We choose the convention:

- $\delta_0 = 0$, $\delta_\parallel = \arg(A_\parallel(0)A_0^*(0))$ and $\delta_\perp = \arg(A_\perp(0)A_0^*(0))$ ;
- $|A_\perp(0)|^2 + |A_\parallel(0)|^2 + |A_0(0)|^2 = 1$, so that $|A_\perp(0)|^2$ represents the fraction of CP-odd at time $t = 0$. To simplify the notation, we write:

$$R_0 = \frac{|A_0(0)|^2}{|A_\perp(0)|^2 + |A_\parallel(0)|^2 + |A_0(0)|^2} = |A_0(0)|^2\,, \tag{41}$$

and

$$R_\perp = \frac{|A_\perp(0)|^2}{|A_\perp(0)|^2 + |A_\parallel(0)|^2 + |A_0(0)|^2} = |A_\perp(0)|^2\,. \tag{42}$$

An angular analysis of the decay products is required to disentangle, *statistically*, the three components of the final state. The three decay product angles are $\Omega = \{\theta, \varphi, \psi\}$ as shown in Figure 4. In the coordinate system of the J/$\psi$ rest frame (where the $\phi$ and $\mathrm{B_s^0}$ meson move in the $x$ direction, the $z$ axis is perpendicular to the decay plane of $\phi \to \mathrm{K^+K^-}$, and $p_y(\mathrm{K^+}) \geq 0$), the transversity polar and azimuthal angles $(\theta, \varphi)$ describe the direction of the $\mu^+$. In the rest frame of the $\phi$ meson, the angle $\psi$ is the angle between $\vec{p}(\mathrm{K^+})$ and $-\vec{p}(\mathrm{J}/\psi)$. The differential decay rates for $\mathrm{B_s^0}$ and $\overline{\mathrm{B}}\mathrm{_s^0}$ mesons produced as flavour eigenstates at $t = 0$ are given by the general expressions:

$$\frac{\mathrm{d}^4\Gamma(\mathrm{B_s^0} \to \mathrm{J}/\psi\phi)}{\mathrm{d}t\;\mathrm{d}\cos\theta\;\mathrm{d}\varphi\;\mathrm{d}\cos\psi} \equiv \frac{\mathrm{d}^4\Gamma}{\mathrm{d}t\;\mathrm{d}\Omega} \propto \sum_{k=1}^{6} h_k(t) f_k(\Omega)\,, \tag{43}$$

and:

$$\frac{\mathrm{d}^4\Gamma(\overline{\mathrm{B}}\mathrm{_s^0} \to \mathrm{J}/\psi\phi)}{\mathrm{d}t\;\mathrm{d}\cos\theta\;\mathrm{d}\varphi\;\mathrm{d}\cos\psi} \equiv \frac{\mathrm{d}^4\overline{\Gamma}}{\mathrm{d}t\;\mathrm{d}\Omega} \propto \sum_{k=1}^{6} \bar{h}_k(t) f_k(\Omega)\,. \tag{44}$$

The functions $h_k(t)$, ,$\bar{h}_k(t)$ and $f_k(\Omega)$ are defined in Table 1.

The full expression for each of the time dependent $\mathrm{B_s^0}$ amplitudes are:

$$|A_0(t)|^2 \;=\; |A_0(0)|^2 \mathrm{e}^{-\Gamma_\mathrm{s} t}\Big[\cosh\left(\frac{\Delta\Gamma_\mathrm{s} t}{2}\right) - \cos\phi_\mathrm{s}\sinh\left(\frac{\Delta\Gamma_\mathrm{s} t}{2}\right) + \sin\phi_\mathrm{s}\sin(\Delta m_\mathrm{s} t)\Big]\,, \tag{45}$$

[1]Using Eqs 10 and 11, we can write:

$$A_0(0) = \langle \mathrm{J}/\psi\phi|_{\ell=0}\mathcal{H}|\mathrm{B_s^0}\rangle\,, \quad A_\parallel(0) = \langle \mathrm{J}/\psi\phi|_{\ell=2}\mathcal{H}|\mathrm{B_s^0}\rangle\,, \quad A_\perp(0) = \langle \mathrm{J}/\psi\phi|_{\ell=1}\mathcal{H}|\mathrm{B_s^0}\rangle\,.$$

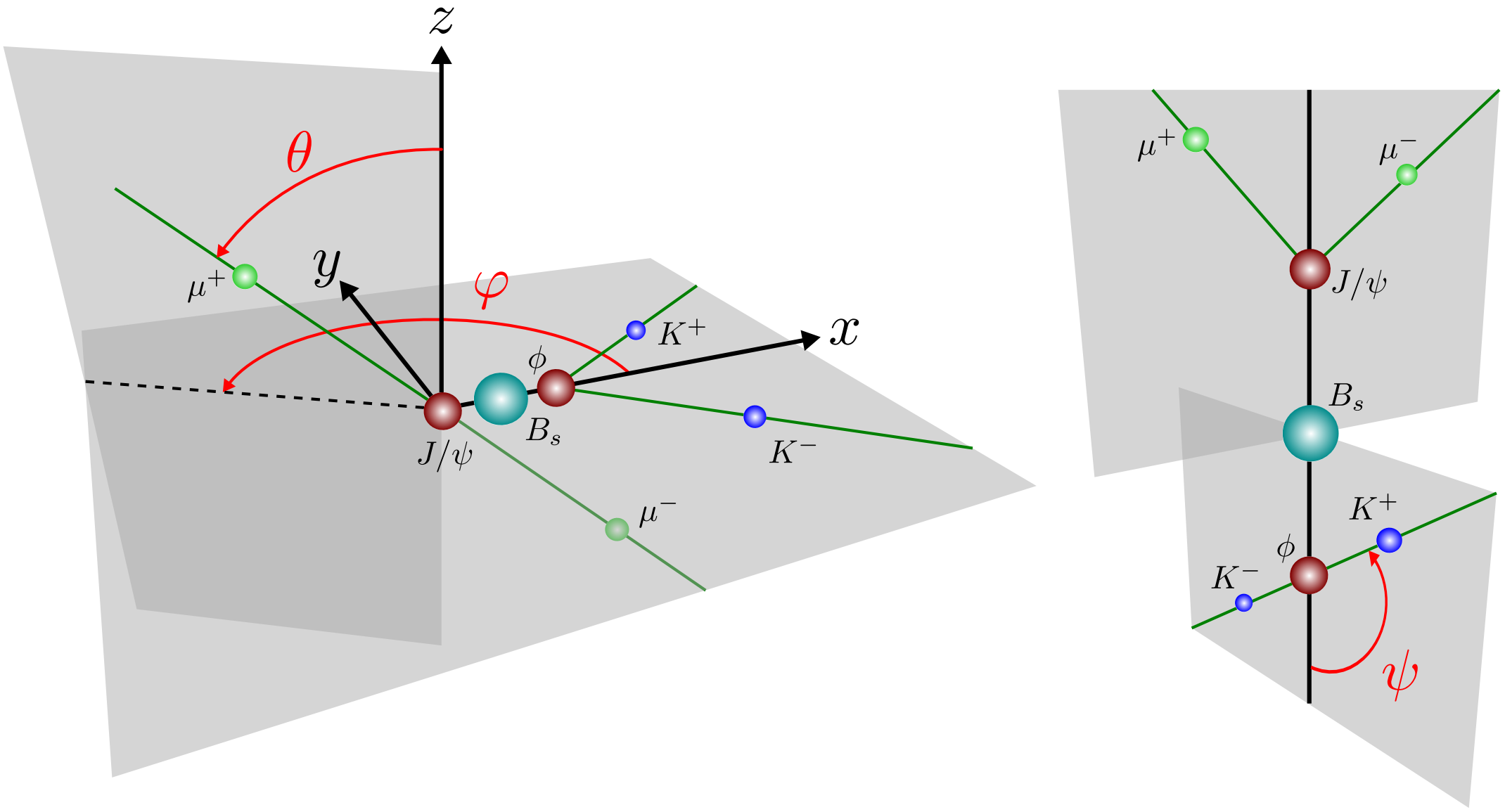

Figure 4: Angle definition: $\theta$ is the angle formed by the positive lepton ($\ell^+$) and the $z$ axis, in the J/$\psi$ rest frame. The angle $\varphi$ is the azimuthal angle of $\ell^+$ in the same frame. In the $\phi$ meson rest frame, $\psi$ is the angle between $\vec{p}(\mathrm{K}^+)$ and $-\vec{p}(\mathrm{J}/\psi)$. The definition is the same whether a $\mathrm{B_s^0}$ or a $\overline{\mathrm{B}}_\mathrm{s}^0$ decays.

| $k$ | $h_k(t)$ | $\bar{h}_k(t)$ | $f_k(\theta,\psi,\varphi)$ |
|---|---|---|---|
| 1 | $\lvert A_0(t)\rvert^2$ | $\lvert \bar{A}_0(t)\rvert^2$ | $2\cos^2\psi(1-\sin^2\theta\cos^2\varphi)$ |
| 2 | $\lvert A_{\vert\vert}(t)\rvert^2$ | $\lvert \bar{A}_{\vert\vert}(t)\rvert^2$ | $\sin^2\psi(1-\sin^2\theta\sin^2\varphi)$ |
| 3 | $\lvert A_{\perp}(t)\rvert^2$ | $\lvert \bar{A}_{\perp}(t)\rvert^2$ | $\sin^2\psi\sin^2\theta$ |
| 4 | $\Im\{A^*_{\vert\vert}(t)A_{\perp}(t)\}$ | $\Im\{\bar{A}^*_{\vert\vert}(t)\bar{A}_{\perp}(t)\}$ | $-\sin^2\psi\sin 2\theta\sin\varphi$ |
| 5 | $\Re\{A^*_0(t)A_{\vert\vert}(t)\}$ | $\Re\{\bar{A}^*_0(t)\bar{A}_{\vert\vert}(t)\}$ | $\frac{1}{\sqrt{2}}\sin 2\psi\sin^2\theta\sin 2\varphi$ |
| 6 | $\Im\{A^*_0(t)A_{\perp}(t)\}$ | $\Im\{\bar{A}^*_0(t)\bar{A}_{\perp}(t)\}$ | $\frac{1}{\sqrt{2}}\sin 2\psi\sin 2\theta\cos\varphi$ |

Table 1: Definition of the functions $h_k(t)$, $\bar{h}_k(t)$ and $f_k(\theta,\psi,\varphi)$ of Eq. 43 and 44.

$$|A_{||}(t)|^2 = |A_{||}(0)|^2 e^{-\Gamma_s t}\left[\cosh\left(\frac{\Delta\Gamma_s t}{2}\right) - \cos\phi_s \sinh\left(\frac{\Delta\Gamma_s t}{2}\right) + \sin\phi_s\sin(\Delta m_s t)\right], \quad (46)$$

$$|A_{\perp}(t)|^2 = |A_{\perp}(0)|^2 e^{-\Gamma_s t}\left[\cosh\left(\frac{\Delta\Gamma_s t}{2}\right) + \cos\phi_s \sinh\left(\frac{\Delta\Gamma_s t}{2}\right) - \sin\phi_s\sin(\Delta m_s t)\right], \quad (47)$$

$$\begin{aligned}\Im\{A^*_{||}(t)A_{\perp}(t)\} &= |A_{||}(0)||A_{\perp}(0)|e^{-\Gamma_s t}\Big[-\cos(\delta_\perp-\delta_{||})\sin\phi_s\sinh\left(\frac{\Delta\Gamma_s t}{2}\right)\\ &+ \sin(\delta_\perp-\delta_{||})\cos(\Delta m_s t) - \cos(\delta_\perp-\delta_{||})\cos\phi_s\sin(\Delta m_s t)\Big], \quad (48)\end{aligned}$$

$$\begin{aligned}\Re\{A^*_0(t)A_{||}(t)\} &= |A_0(0)||A_{||}(0)|e^{-\Gamma_s t}\cos\delta_{||}\Big[\cosh\left(\frac{\Delta\Gamma_s t}{2}\right) - \cos\phi_s\sinh\left(\frac{\Delta\Gamma_s t}{2}\right)\\ &+ \sin\phi_s\sin(\Delta m_s t)\Big], \quad (49)\end{aligned}$$

$$\begin{aligned}\Im\{A^*_0(t)A_{\perp}(t)\} &= |A_0(0)||A_{\perp}(0)|e^{-\Gamma_s t}\Big[-\cos\delta_\perp\sin\phi_s\sinh\left(\frac{\Delta\Gamma_s t}{2}\right)\\ &+ \sin\delta_\perp\cos(\Delta m_s t) - \cos\delta_\perp\cos\phi_s\sin(\Delta m_s t)\Big]. \quad (50)\end{aligned}$$

In the case of the $\overline{B}^0_s$, the time evolution is given by the conjugated functions $\bar{h}_k(t)$ and the angular dependence remains the same. We have highlighted the signs that change:

$$|\bar{A}_0(t)|^2 = |\bar{A}_0(0)|^2 e^{-\Gamma_s t}\left[\cosh\left(\frac{\Delta\Gamma_s t}{2}\right) - \cos\phi_s \sinh\left(\frac{\Delta\Gamma_s t}{2}\right) - \sin\phi_s\sin(\Delta m_s t)\right], \quad (51)$$

$$|\bar{A}_{||}(t)|^2 = |\bar{A}_{||}(0)|^2 e^{-\Gamma_s t}\left[\cosh\left(\frac{\Delta\Gamma_s t}{2}\right) - \cos\phi_s \sinh\left(\frac{\Delta\Gamma_s t}{2}\right) - \sin\phi_s\sin(\Delta m_s t)\right], \quad (52)$$

$$|\bar{A}_{\perp}(t)|^2 = |\bar{A}_{\perp}(0)|^2 e^{-\Gamma_s t}\left[\cosh\left(\frac{\Delta\Gamma_s t}{2}\right) + \cos\phi_s \sinh\left(\frac{\Delta\Gamma_s t}{2}\right) + \sin\phi_s\sin(\Delta m_s t)\right], \quad (53)$$

$$\begin{aligned}\Im\{\bar{A}^*_{||}(t)\bar{A}_{\perp}(t)\} &= |\bar{A}_{||}(0)||\bar{A}_{\perp}(0)|e^{-\Gamma_s t}\Big[-\cos(\delta_\perp-\delta_{||})\sin\phi_s\sinh\left(\frac{\Delta\Gamma_s t}{2}\right)\\ &- \sin(\delta_\perp-\delta_{||})\cos(\Delta m_s t) + \cos(\delta_\perp-\delta_{||})\cos\phi_s\sin(\Delta m_s t)\Big], \quad (54)\end{aligned}$$

$$\begin{aligned}\Re\{\bar{A}^*_0(t)\bar{A}_{||}(t)\} &= |\bar{A}_0(0)||\bar{A}_{||}(0)|e^{-\Gamma_s t}\cos\delta_{||}\Big[\cosh\left(\frac{\Delta\Gamma_s t}{2}\right) - \cos\phi_s\sinh\left(\frac{\Delta\Gamma_s t}{2}\right)\\ &- \sin\phi_s\sin(\Delta m_s t)\Big], \quad (55)\end{aligned}$$

$$\begin{aligned}\Im\{\bar{A}^*_0(t)\bar{A}_{\perp}(t)\} &= |\bar{A}_0(0)||\bar{A}_{\perp}(0)|e^{-\Gamma_s t}\Big[-\cos\delta_\perp\sin\phi_s\sinh\left(\frac{\Delta\Gamma_s t}{2}\right)\\ &- \sin\delta_\perp\cos(\Delta m_s t) + \cos\delta_\perp\cos\phi_s\sin(\Delta m_s t)\Big]. \quad (56)\end{aligned}$$

There is an exact symmetry in the decay rates which are invariant under the simultaneous transformations:

$$\begin{aligned} \Phi &\longleftrightarrow \pi - \Phi\,, \\ \Delta\Gamma_{\rm s} &\longleftrightarrow -\Delta\Gamma_{\rm s}\,, \\ \delta_{\parallel} &\longleftrightarrow -\delta_{\parallel}\,, \\ \delta_{\perp} &\longleftrightarrow \pi - \delta_{\perp}\,. \end{aligned} \tag{57}$$

It may be possible to resolve this two-fold ambiguity as discussed in Appendix A.

To reduce the decay rates (Eqs. 43 and 44) to the case where only the single transversity angle $\theta$ is used, one integrates over $\varphi$ and $\cos\psi$. The form of the differential decay rates are in this case simple enough that they can be written as single expressions:

$$\begin{aligned} \frac{{\rm d}\Gamma}{{\rm d}t\;{\rm d}\cos\theta} &\propto (1-R_{\perp}){\rm e}^{-\Gamma_{\rm s}t}\Big[\cosh\left(\frac{\Delta\Gamma_{\rm s}t}{2}\right) - \cos\phi_{\rm s}\sinh\left(\frac{\Delta\Gamma_{\rm s}t}{2}\right) \\ &+ \sin\phi_{\rm s}\sin(\Delta m_{\rm s}t)\Big]\frac{1}{2}(1+\cos^2\theta) \\ &+ R_{\perp}\,{\rm e}^{-\Gamma_{\rm s}t}\Big[\cosh\left(\frac{\Delta\Gamma_{\rm s}t}{2}\right) + \cos\phi_{\rm s}\sinh\left(\frac{\Delta\Gamma_{\rm s}t}{2}\right) \\ &- \sin\phi_{\rm s}\sin(\Delta m_{\rm s}t)\Big]\sin^2\theta \end{aligned} \tag{58}$$

and

$$\begin{aligned} \frac{{\rm d}\bar\Gamma}{{\rm d}t\;{\rm d}\cos\theta} &\propto (1-R_{\perp}){\rm e}^{-\Gamma_{\rm s}t}\Big[\cosh\left(\frac{\Delta\Gamma_{\rm s}t}{2}\right) - \cos\phi_{\rm s}\sinh\left(\frac{\Delta\Gamma_{\rm s}t}{2}\right) \\ &- \sin\phi_{\rm s}\sin(\Delta m_{\rm s}t)\Big]\frac{1}{2}(1+\cos^2\theta) \\ &+ R_{\perp}\,{\rm e}^{-\Gamma_{\rm s}t}\Big[\cosh\left(\frac{\Delta\Gamma_{\rm s}t}{2}\right) + \cos\phi_{\rm s}\sinh\left(\frac{\Delta\Gamma_{\rm s}t}{2}\right) \\ &+ \sin\phi_{\rm s}\sin(\Delta m_{\rm s}t)\Big]\sin^2\theta\,. \end{aligned} \tag{59}$$

All of the terms contain information concerning the mass eigenstates. To the extent that the angular information separates the eigenstates, then each of $\Gamma_{\rm L}$ and $\Gamma_{\rm H}$ are determined and hence $\Delta\Gamma$. In the specific case of the Standard Model where $\cos\phi_{\rm s} \approx 1$ then the differential decay rates take a particularly simple form where (considering the diagonal $|A_i|^2$ and the real interference terms) $\Gamma_{\rm L}$ is determined by the CP-even component and $\Gamma_{\rm H}$ by the CP-odd component.

Evidently $\phi_{\rm s}$ can be extracted from several of the terms, but around the Standard Model value the sensitivity from the $\cos\phi_{\rm s}$ terms is very poor since $|\cos\phi_{\rm s}| \sim 1$ and all information comes from the $\sin\phi_{\rm s}$ terms. The diagonal and real interference terms are all multiplied by $\sin(\Delta m_{\rm s}t)$ and hence information on $\phi_{\rm s}$ is mainly obtained from observation of the amplitude of the sinusoid in the time distribution. However, these terms have opposite sign between $\mathrm{B_s^0}$ and $\mathrm{\overline{B}{}_s^0}$ and if untagged events are used then almost complete cancellation of the $\sin\phi_{\rm s}$ information occurs. The analysis, therefore, benefits significantly from flavour tagging.

# 3 Current Tevatron results

Both the CDF and DØ collaborations have recently presented results of tagged analyses of $B_s^0 \to J/\psi\phi$ decays [14, 15, 16]. Their measurements are summarized in Table 2. CDF uses a neural network for the selection, while DØ uses a cut-based selection. The tagging power, $\varepsilon_{\rm tag}(1-2\omega)^2$ (see Section 8) is similar in both experiments, while the signal over background ratio $S/B$ is better for CDF. The combined result is [17][2]:

$$\phi_s \in [-1.18;\, -0.54] \,\cup\, [-2.60;\, -1.94]\,\text{rad at 68\% CL}\,,$$

Note that they do not present a central value and its uncertainty, due to the highly non-Gaussian shape of the likelihood function. Instead, they present a confidence region in the $\phi_s - \Delta\Gamma_s$ plane.

By the end of Run 2, assuming $9\,\text{fb}^{-1}$ for each CDF and DØ and simple scaling with $1/\sqrt{\mathcal{L}_{\rm int}}$, the combined Tevatron sensitivity to $\phi_s$ is expected to be $\sim 0.18\,$rad.

| | CDF [15] | DØ [16] |
|---|---|---|
| Integrated luminosity | $2.8\,\text{fb}^{-1}$ | $2.8\,\text{fb}^{-1}$ |
| Selection | NNet | Cut based |
| Signal candidates | $3\,166 \pm 56$ | $1\,967 \pm 65$ |
| $S/B$ | $\simeq 1$ | $\simeq 1/4$ |
| $\varepsilon_{\rm tag}(1-2\omega)^2$ | 4.81%(*) | 4.68% |
| $\phi_s$ [rad] | $[-2.58;\, -0.56]$ at 68%CL | $-0.57^{-0.30}_{+0.24}$ |

Table 2: Comparison of the CDF and DØ results of tagged $B_s^0 \to J/\psi\phi$ analyses from [15, 16]. Only the $\phi_s$ solution closest to the Standard Model is shown. The DØ analysis [16] imposes constraints on the strong phases $\delta_\parallel$ and $\delta_\perp$ while the CDF [15] analyis does not. (*) The tagging power obtained by CDF was 4.81% on the first $1.4\,\text{fb}^{-1}$ and only 1.2% on the last $1.4\,\text{fb}^{-1}$ because in the latter case, the same side tagging was not yet used.

# 4 Analysis strategy

In this section we describe the analysis strategy that we intend to follow in LHCb. The phase $\phi_s$ is fitted from differential decay rates of $B_s^0$ and $\overline{B}_s^0$ mesons to $J/\psi\phi$, as shown in Eqs. 43 and 44. Therefore we need to:

- trigger and select $B_s^0 \to J/\psi\phi$ events;
- measure their proper time;
- measure the transversity angles of their decay products;
- tag their initial flavour;
- fit $\phi_s$, together with the other unknown physics parameters.

[2] Note that, in Ref. [17], the observable $\phi_s \equiv \phi_s^{J/\psi\phi}$ of Eq. 26 is called "$-2\beta_s$", while the CKM phase $\beta_s$ of Eq. 35 is called "$\beta_s^{\rm SM}$".

The fit depends on 8 physics parameters: $\{\phi_s, \Gamma_s, \Delta\Gamma_s, R_\perp, R_0, \delta_\perp, \delta_\parallel, \Delta m_s\}$. The parameter $\Delta m_s$ is already known [2]. Constraints exist on $\Gamma_s$ and $\Delta\Gamma_s$ [2]. The 5 others are unknown, so that they have to be measured. However, several steps can be envisaged. For example, the parameters $\{\Gamma_s,\ \Delta\Gamma_s,\ R_\perp,\ R_0\}$ can be extracted from an untagged three-angle and time-dependent analysis. The correlations between all physics parameters are not trivial. They are discussed in Section 10.2 and Appendix C.

We have seen in Section 2 that information on $\phi_s$ is mainly obtained from the measurement of the amplitude of the $\sin(\Delta m_s t)$ term in the time-dependent distribution of $B_s^0 \to J/\psi\phi$, particularly in the case of the Standard Model. In case $\phi_s$ is large, the $\cos(\Delta m_s t)$ is also important. The amplitude of the oscillation can be affected by several experimental factors: imperfect flavour tagging, proper time resolution, angular and proper time acceptances and background contamination. Thus the important ingredients for reaching high sensitivity on $\phi_s$ are the following:

- large signal yield and tagging power to decrease the statistical uncertainty;
- good proper time resolution to resolve the fast $B_s^0$ oscillations;
- good control of proper time and angular acceptances: those can lead to systematic biases in the cases where they are not properly evaluated. In this regard, the flatter the acceptances are, the smaller are the corrections to be applied and the smaller is the systematic error associated to them.

LHCb plans to run at an average luminosity of $\mathcal{L} \sim 2 \times 10^{32}\,\mathrm{cm^{-2}s^{-1}}$ which is smaller than the nominal LHC luminosity at the ATLAS and CMS interaction points[3], and will collect an integrated luminosity of $\sim 2\,\mathrm{fb^{-1}}$ per nominal year ($10^7$ s). At the LHC centre-of-mass energy the $b\bar{b}$ quarks are produced with a large cross section ($\sigma_{b\bar{b}} \sim 500\mu$b) and mostly in the same forward or backward cone [18]. The LHCb detector is a forward spectrometer covering a pseudo-rapidity range from 1.9 to 5.9. This forward geometry allows the LHCb first level trigger to collect events with one or two muons with $p_T$ value as low as 1.3 GeV/$c$ for a single muon and $|p_{T,1}(\mu)| + |p_{T,2}(\mu)| > 1.5$ GeV/$c$ for di-muon events (Section 7).

The selection of the $B_s^0 \to J/\psi\phi$ channel is described in Section 6. One of the optimization criteria of this selection is to minimize any possible bias on proper time and angular distribution. The advantages of this kind of selection are the following.

1. We avoid proper time and angular acceptance corrections. Since it is not trivial to determine these acceptances, we remove this constraint, at least for the beginning.
2. Since the proper time distribution follows a decreasing exponential, there are more events at small proper times than at larger ones, so the less we cut on these low proper time events, the more statistical power we keep.
3. Using a lifetime unbiased selection allows us to use the "negative tail" of the proper time distribution to estimate the proper time resolution.

[3] The luminosity delivered at ATLAS and CMS interaction points is expected to be $\mathcal{L} \sim 2 \times 10^{33}\,\mathrm{cm^{-2}s^{-1}}$ in the low luminosity regime and $\mathcal{L} \sim 10^{34}\,\mathrm{cm^{-2}s^{-1}}$ in the high luminosity regime.

4. Cutting away part of the proper time spectrum affects in different way the CP-odd and CP-even components that have different lifetimes. So, indirectly one would affect also in a non trivial way the angular distributions, which are given by the superposition of the two CP-components.

Two drawbacks of the lifetime unbiased selection are:

1. Since no cut on impact parameter of any track is allowed, a huge prompt background survives the selection. This background is not a problem for the $\phi_s$ measurement, since it is well identified by the fitting procedure (very small lifetime, flat mass), but it will take a big part of the trigger bandwidth, if not pre-scaled.

2. The way to trigger "lifetime unbiased" $B_s^0 \to J/\psi\phi$ events relies mainly on a unique trigger alley, as we will see in Section 7. In addition, the HLT1 efficiency is only $\sim 80\%$, for these unbiased events.

In this document we present an analysis strategy based on a "lifetime unbiased" selection. However we plan also to study a "lifetime biased" selection in order to be ready in case the prompt background rate will be too high. Moreover a lifetime biased selection would allow us to use also trigger lines with cuts on the impact parameter of the decay products, increasing the efficiency and the robustness of the trigger on this channel. Studies of this subject are ongoing.

In order to control systematic effects we need to extract from data as much information as possible, such as mistag rates and proper time resolution and to find a way to study angular and proper time acceptances. Therefore the use of control channels in this analysis is of fundamental importance. In particular the two modes $B^0 \to J/\psi K^{*0}(K^+\pi^-)$ and $B^+ \to J/\psi K^+$ are flavour-specific channels with yields that are 4 and 8 times respectively more abundant than $B_s^0 \to J/\psi\phi$. They will be used to estimate the opposite-side (OS) mistag rate (see Section 8), while the same-side (SS) mistag rate will be evaluated by using the $B_s^0 \to D_s^-\pi^+$ and $B_s^0 \to D_s^-\mu^+\nu$ decays. Since the tagging properties of OS taggers depend on the phase space of the selected b-hadron due to the $b\overline{b}$ correlation, the selection of $B_s^0$, $B^0$ and $B^+$ designed for this analysis is such that the selected b-hadrons have the same $p$ and $p_T$ distributions (Section 6).

Moreover the decay $B^0 \to J/\psi K^{*0}$ is a pseudo-scalar to vector-vector ($P \to VV$) transition with amplitudes known from several measurements [19, 20, 21, 22] and with kinematics that are similar to $B_s^0 \to J/\psi\phi$, so it is a good control channel to study the angular acceptances and validate the fit procedure. The study of the $B^0 \to J/\psi K^{*0}$ decay is a very important step before the measurement of $\phi_s$ and it is discussed in [60].

Other necessary steps before the measurement of $\phi_s$ are the measurement of $\Delta m_d$ (briefly discussed in Section 8.3), the measurement of $\sin(2\beta)$ in the $B^0 \to J/\psi K_S^0$ decay (see [61]) and the measurement of the $B_s^0$ oscillation frequency $\Delta m_s$ [23]. The latter three measurements, will be the first "stress-tests" of our flavour tagging calibration procedure and the $\Delta m_s$ analysis will validate that we can satisfactorily resolve $B_s^0$ oscillations.

The full three-angle tagged time-dependent analysis of $B_s^0 \to J/\psi\phi$, including treatment of resonant and non-resonant contribution to the KK final state, is a long term plan. Before this "ultimate" measurement, the analysis will proceed through several intermediate steps of increasing complexity, e.g.:

- We will perform an analysis of the $B_s^0 \to J/\psi\phi$ decay without tagging the flavour of the $B_s^0$ (untagged analysis). This analysis is mainly sensitive to $\Gamma_s$, $\Delta\Gamma_s$, $R_\perp$ and $R_0$, while it has very limited sensitivity to $\phi_s$ [24]. The requirement on the proper time resolution is less demanding than in the tagged analysis, since we do not need to resolve the fast $B_s^0$–$\overline{B}_s^0$ oscillations.

- The time-dependent tagged analysis requires the calibration of the tagging procedure: we will start by calibrating opposite-side taggers with the two flavour-specific control channels $B^0 \to J/\psi K^{*0}$ and $B^+ \to J/\psi K^+$ (see Section 8) and we will continue by measuring the tagging performance of the same-side tagger. Furthermore, it requires a precise knowledge and control of the proper time resolution. A detailed evaluation of the proper time resolution model is discussed in [25]. A first estimate of the average proper time resolution can be extracted in several ways from the channels with $J/\psi \to \mu\mu$ in the final state (see Sections 6.3.2 and 6.4).

- The angular analysis of a tagged and time-dependent sample requires a complex fit procedure that must be validated: we can simplify by using only one angle out of three (i.e. integrating over the other two angles). We will validate the three angle fit using $B^0 \to J/\psi K^{*0}(K^+\pi^-)$, which does not require flavour tagging [60], then we will include tagging. As we will see in Section 10.2, the complete fit has 26 free parameters. We will start with a reduced number of parameters, keeping constant in the fit the "detector parameters". These can first be taken from Monte Carlo, then measured in control channels and, eventually, fitted from the $B_s^0 \to J/\psi\phi$ channel alone as a cross-check. In particular, the three-angle time-dependent analysis provides an independent way to extract the mistag rate from the $B_s^0 \to J/\psi\phi$ sample itself [24].

- The possible resonant or non-resonant contributions to the KK final state, in $B_s^0 \to J/\psi KK$, is a non-trivial complication to the analysis that can eventually be included in the angular fit, as discussed in Appendix A. At first however we would neglect this possible contribution when calculating our central value and account for any resulting bias by assigning a conservative systematic uncertainty.

- In the longer term, we will perform the analysis to allow for direct CP violation and penguin contributions following e.g. [9].

In the rest of this document, we describe all the steps required to perform a tagged three-angles time-dependent analysis. The focus is mainly on the measurement with $2\,\mathrm{fb}^{-1}$, rather than with very first data.

# 5 Monte Carlo simulations

The following studies are based on Monte Carlo events generated using Pythia [26] and fully simulated using GEANT4 programs [27] in the LHCb Data Challenge 2006 (DC06). The b-hadrons are decayed with the EvtGen program [28]. In particular, the $B_s^0 \to J/\psi\phi$ decays are treated using the the PVV_CPLH model (see Appendix B). The final state radiation is included by means of the PHOTOS generator [29], which adds radiated photons to the decay trees generated by EvtGen.

We have studied in detail the selection of $B_s^0 \to J/\psi\phi$ and related control channels with $J/\psi \to \mu\mu$ in the final states, $B^0 \to J/\psi K^{*0}(K^+\pi^-)$ and $B^+ \to J/\psi K^+$.

Six different background samples are used: minimum bias events, inclusive $b\bar{b}$ events, inclusive $J/\psi$ events, $B^+ \to J/\psi X$, $B^0 \to J/\psi X$ and $B_s^0 \to J/\psi X$ events. In the following, the sum of the latter three samples, with natural proportions 40%/40%/10%, will be called $B_{u,d,s} \to J/\psi X$.

To save CPU time, cuts are applied at the generation level to enforce particles of interest to be within the LHCb acceptance. For the signal samples, all decays products should be between 10 mrad and 400 mrad; for the $B_{u,d,s} \to J/\psi X$ samples, the two muons should be between 10 mrad and 400 mrad. For the $b\bar{b}$ sample, at least one b-hadron should be below 400 mrad.

A second filtering step is applied on the data after the generation, simulation and reconstruction, known as the *stripping preselection.* It consists of applying a set of loose preselection cuts to suppress events poorly reconstructed or with some of the decay products outside the acceptance. This stripping phase is the OR of all physics channel preselections available so far.

For the minimum bias sample, no generator-level cuts nor stripping are applied, but the events must pass the L0-trigger. The sample sizes, after those requirements, are given in Table 3, together with the generator level cut efficiencies and the stripping efficiencies.

The branching fractions are given in Table 4, the cross sections at 14 TeV are given in Table 5. The $B_s^0$ and $B^0$ physical properties are given in Tables 6 and 7.

| Channel | $\varepsilon_{\rm gen}$ % | $\varepsilon_{\rm strip}$ % | Statistics | $\mathcal{L}_{\rm int}$ |
|---|---|---|---|---|
| Signals | | | | |
| $B^+ \to J/\psi(\mu\mu)K^+$ | $17.89 \pm 0.03$ | $63.12 \pm 0.04$ | 1 420 737 | $0.52\,{\rm fb}^{-1}$ |
| $B^0 \to J/\psi(\mu\mu)K^{*0}(K\pi)$ | $17.16 \pm 0.04$ | $62.66 \pm 0.03$ | 1 640 673 | $0.75\,{\rm fb}^{-1}$ |
| $B_s^0 \to J/\psi(\mu\mu)\phi(KK)$ | $18.14 \pm 0.04$ | no stripping | 9 218 605 | $17.0\,{\rm fb}^{-1}$ |
| Backgrounds | | | | |
| $b\bar{b}$ | $43.7 \pm 0.1$ | $3.643 \pm 0.004$ | 924 407 | $5.8 \times 10^{-5}\,{\rm fb}^{-1}$ |
| Inclusive $J/\psi(\mu\mu)$ | $19.70 \pm 0.04$ | $48.37 \pm 0.03$ | 1 259 878 | $0.00078\,{\rm fb}^{-1}$ |
| $B^+ \to J/\psi(\mu\mu)X$ | $20.37 \pm 0.03$ | $57.19 \pm 0.02$ | 2 913 243 | $0.22\,{\rm fb}^{-1}$ |
| $B^0 \to J/\psi(\mu\mu)X$ | $20.36 \pm 0.03$ | $55.85 \pm 0.02$ | 2 644 263 | $0.19\,{\rm fb}^{-1}$ |
| $B_s^0 \to J/\psi(\mu\mu)X$ | $20.24 \pm 0.04$ | $54.53 \pm 0.07$ | 310 280 | $0.08\,{\rm fb}^{-1}$ |
| Minimum bias | 100 | $\varepsilon_{\rm L0} = 5.934 \pm 0.004$ | 5 535 650 | $9.1 \times 10^{-7}\,{\rm fb}^{-1}$ |

Table 3: Generator-level cut efficiency, stripping efficiency and available statistics after the stripping for the decay modes used in this note. $\mathcal{L}_{\rm int}$ is the equivalent integrated luminosity. For minimum bias, the "stripping efficiency" given is the L0-trigger efficiency.

# 6 Selection of $B_s^0 \to J/\psi\phi$ and related control channels $B_d^0 \to J/\psi K^{*0}$ and $B^+ \to J/\psi K^+$

In this section we describe the selection of the $B_s^0 \to J/\psi\phi$ mode together with the two control channels with $J/\psi \to \mu\mu$ in the final state, $B^0 \to J/\psi K^{*0}$ and $B^+ \to J/\psi K^+$. The guidelines followed in designing the selections can be summarized as follows:

1. maximize the signal yield while keeping the background at a reasonable level;

| Branching fraction | PDG [2] | DC06 |
|---|---|---|
| $\mathcal{B}(J/\psi \to \mu\mu)$ | $(5.93 \pm 0.06) \times 10^{-2}$ | $5.933 \times 10^{-2}$ |
| $\mathcal{B}(J/\psi \to e^+e^-)$ | $(5.94 \pm 0.06) \times 10^{-2}$ | $5.933 \times 10^{-2}$ |
| $\mathcal{B}(K^{*0} \to K^+\pi^-)$ | – | $66.57 \times 10^{-2}$ |
| $\mathcal{B}(\phi \to K^+K^-)$ | $(49.2 \pm 0.6) \times 10^{-2}$ | $49.10 \times 10^{-2}$ |
| $\mathcal{B}(B^0 \to J/\psi K^{*0})$ | $(1.33 \pm 0.06) \times 10^{-3}$ | $1.291 \times 10^{-3}$ |
| $\mathcal{B}(B^0_s \to J/\psi \phi)$ | $(9.3 \pm 3.3) \times 10^{-4}$ | $13.5 \times 10^{-4}$ |
| $\mathcal{B}(B^+ \to J/\psi K^+)$ | $(1.007 \pm 0.035) \times 10^{-3}$ | $1.008 \times 10^{-3}$ |
| $\mathcal{B}_{vis}(B^0_s \to J/\psi(\mu\mu)\phi(KK))$ | $(2.71 \pm 0.96) \times 10^{-5}$ | $3.93 \times 10^{-5}$ |
| $\mathcal{B}_{vis}(B^0_s \to J/\psi(ee)\phi(KK))$ | $(2.72 \pm 0.96) \times 10^{-5}$ | $3.93 \times 10^{-5}$ |
| $\mathcal{B}_{vis}(B^0 \to J/\psi(\mu\mu)K^{*0}(K\pi))$ | $(5.25 \pm 0.24) \times 10^{-5}$ | $5.10 \times 10^{-5}$ |
| $\mathcal{B}_{vis}(B^+ \to J/\psi(\mu\mu)K^+)$ | $(5.9 \pm 0.2) \times 10^{-5}$ | $5.98 \times 10^{-5}$ |
| $f_u$ (%) | $39.9 \pm 1.1$ | 40.5 |
| $f_d$ (%) | $39.9 \pm 1.1$ | 40.5 |
| $f_s$ (%) | $11.0 \pm 1.2$ | 10.0 |

Table 4: Individual branching fractions for all decays and B meson production fractions. Values from PDG and from the Monte Carlo simulation (DC06) are indicated.

| Cross section | Value |
|---|---|
| $\sigma_{pp}$ | 102.9 mb |
| $\sigma_{pp \to b\bar{b}}$ | 0.698 mb |
| $\sigma_{pp \Rightarrow J/\psi X}$ | 0.286 mb |
| $\sigma_{pp \to b \Rightarrow J/\psi X}$ | 0.0204 mb |

Table 5: Production cross sections at 14 TeV predicted by the Monte Carlo simulation (DC06) [26, 28]. The double arrow (i.e. $\Rightarrow$) indicates the existence of possible intermediate states, e.g. $pp \Rightarrow J/\psi X$ includes $pp \to b \to J/\psi X$. The "prompt $J/\psi$" cross section is $\sigma_{pp \to J/\psi X} \equiv \sigma_{Pr} = \sigma_{pp \Rightarrow J/\psi X} - \sigma_{pp \to b \Rightarrow J/\psi X}$ =0.266 mb. The $b\bar{b}$ cross-section used for the yield estimates is 500 mb.

| Parameter | Value |
|---|---|
| $B_s^0$ mass | $5.3696\,\mathrm{GeV}/c^2$ |
| $B_s^0$ lifetime | $1.461\,\mathrm{ps}$ |
| $\Delta m_s$ | $20\,\mathrm{ps}^{-1}$ |
| $\Delta\Gamma_s$ | $0.06852\,\mathrm{ps}^{-1}$ |
| $\lvert A_{\parallel}(0)\rvert$ | 0.49 |
| $\delta_{\parallel}$ | $2.50\,\mathrm{rad}$ |
| $\lvert A_0(0)\rvert$ | 0.775 |
| $\delta_0$ | $0.0\,\mathrm{rad}$ |
| $\lvert A_{\perp}(0)\rvert$ | 0.4 |
| $\delta_{\perp}$ | $-0.17\,\mathrm{rad}$ |
| $\phi_s$ | $-0.04\,\mathrm{rad}$ |

Table 6: $B_s^0$ parameters used in the generation of the $B_s^0 \to J/\psi(\mu\mu)\phi(KK)$ sample. $|A_{\parallel}(0)|$, $|A_0(0)|$ and $|A_{\perp}(0)|$ are the absolute values of three decay amplitudes, in the transversity basis; $\delta_{\parallel}$, $\delta_0$ and $\delta_{\perp}$ the corresponding strong phases (as defined in Section 2). From the values of the $B_s^0$ lifetime ($\tau_{B_s^0}$) and $\Delta\Gamma_s$, one can derive the lifetime of the $B_s^0$ mass eigenstates: $\tau_H$=1.538 ps and $\tau_L$=1.391 ps. $\Gamma_s \equiv 1/\tau_{B_s^0} = 0.684\,\mathrm{ps}^{-1}$.

| Parameter | Value |
|---|---|
| $B^0$ mass | $5.2794\,\mathrm{GeV}/c^2$ |
| $B^0$ lifetime | $1.536\,\mathrm{ps}$ |
| $\Delta m_d$ | $0.502\,\mathrm{ps}^{-1}$ |
| $\Delta\Gamma_d$ | 0 |

Table 7: $B^0$ parameters used in the generation of the $B^0 \to J/\psi(\mu\mu)K^{*0}(K\pi)$ sample. The amplitudes and strong phases, expressed in the transversity basis, are the same as for $B_s^0 \to J/\psi\phi$ (Table 6).

2. minimize lifetime and angular acceptance distortions: the selection has been designed in such a way that the distortions are small and as similar as possible among signal and control channels;

3. select the b-hadrons in such a way that their momentum distributions are similar: this is important in order to allow the tagging performance determined on control channels to be applied on the signal with minimal correction.

The selection is designed to achieve reasonably small statistical error while keeping systematic errors under control. The chosen set of cuts and the corresponding event yield and background contamination is the result of a compromise among the three points listed above.

## 6.1 Selections

The selection cuts for the three channels are detailed in a separate note [30]. Here we summarize the results. The selections developed for this analysis are *cut-based* selections: each cut is chosen sequentially, with a maximization of the number of signal event for a given reasonable background level. This approach is well suited for first data analysis since it allows the study of the impact of each single variable on the analysis.

The starting point of the analysis are events reconstructed and preselected with very loose cuts. This allows us to separate the effects of acceptance, reconstruction and particle identification, from the selection itself. It also allows to have a reference point for evaluating the efficiencies of each single cut then used in the selection. The preselection of J/$\psi$ is performed at the stripping level. It requires two oppositely-charged tracks reconstructed and identified as muons to come from the same vertex and to have an invariant mass compatible with the J/$\psi$. The preselection of the $\phi$ and the $K^{*0}$ is done requiring the decays products to be reconstructed and identified (as $K^+K^-$ and $K^+\pi^-$ respectively), to come from the same vertex and to have an invariant mass compatible with the one of a $\phi$ and $K^{*0}$, respectively. The combined efficiency of these preselections is 24.7%, 21.1% and 35.8% for the $B_s^0 \to J/\psi\phi$, $B^0 \to J/\psi K^{*0}$ and $B^+ \to J/\psi K^+$ channels, respectively.

The $J/\psi \to \mu\mu$ decay is selected in a similar way for the three channels as a consequence of the fact that J/$\psi$ kinematics is similar in the three cases. The cuts are summarized in Table 8. For the $B^0 \to J/\psi K^{*0}$ and $B^+ \to J/\psi K^+$ cases an additional cut on the particle identification likelihood of the muons ($\Delta \ln \mathcal{L}_{K\pi}$), estimated using the RICH systems, is required to reduce the contribution of muons coming from decays in flight that can fake a J/$\psi$ candidate[4]. This cut is not required for the $B_s^0 \to J/\psi\phi$ selection since a cut around the narrow $\phi$ meson resonance is enough to suppress this background. No J/$\psi$ mass constraint is used, since this would bias the $B_s^0$ proper time in a non-trivial way.

The effect of a given cut on the $J/\psi \to \mu\mu$ decay is equal for the three channels and therefore we expect the same to be true for a trigger based on a $J/\psi \to \mu\mu$ inclusive selection. This issue is a strong argument in favour of using an inclusive trigger line based on the $J/\psi \to \mu\mu$ selection (see Section 7).

[4] As far as the RICH particle identification is concerned we neglect the difference between $\pi$ and $\mu$ so $\Delta \ln \mathcal{L}_{K\pi}$ is equivalent to $\Delta \ln \mathcal{L}_{K\mu}$. A fake J/$\psi$ can be made adding a real muon with a pion or kaon decaying in flight. If it is a pion, we cannot detect it. Instead if it is a kaon, it will likely pass through at least RICH1 before decaying. With the $\Delta \ln \mathcal{L}_{K\pi}$ cut, we remove some of these kaons decaying in flight.

| Cut | $\varepsilon(\mathrm{B_s^0} \to \mathrm{J}/\psi\phi)$ | $\varepsilon(\mathrm{B^0} \to \mathrm{J}/\psi\mathrm{K^{*0}})$ | $\varepsilon(\mathrm{B^+} \to \mathrm{J}/\psi\mathrm{K^+})$ |
|---|---|---|---|
| $\mu$ minimum $\Delta\ln\mathcal{L}_{\mu\pi} > -5$ | $99.49 \pm 0.01$ | $99.48 \pm 0.01$ | $98.93 \pm 0.01$ |
| $\mu$ track max $\chi^2_{\mathrm{track}}/\mathrm{nDoF} < 3$ (5 for $\mathrm{B_s^0} \to \mathrm{J}/\psi\phi$) | $99.10 \pm 0.02$ | $97.39 \pm 0.02$ | $97.57 \pm 0.02$ |
| $\mu$ minimum $p_{\mathrm{T}} > 500\,\mathrm{MeV}/c$ | $96.22 \pm 0.02$ | $96.61 \pm 0.02$ | $96.4 \pm 0.02$ |
| $\mu$ minimum $\Delta\ln\mathcal{L}_{\mathrm{K}\pi} < 5$ | — | $96.5 \pm 0.02$ | $97.04 \pm 0.02$ |
| $\mathrm{J}/\psi$ $\chi^2_{\mathrm{vtx}}/\mathrm{nDoF} < 6$ | $97.3 \pm 0.04$ | $97.3 \pm 0.02$ | $97.3 \pm 0.02$ |
| $\mathrm{J}/\psi$ $p_{\mathrm{T}} > 1\,\mathrm{GeV}/c$ | $92.41 \pm 0.07$ | $92.45 \pm 0.02$ | $93.12 \pm 0.03$ |
| $\lvert M(\mu\mu) - M(\mathrm{J}/\psi)\rvert < 3\sigma$ | $94.47 \pm 0.06$ | $94.45 \pm 0.03$ | $94.2 \pm 0.03$ |
| total | $80.47 \pm 0.04$ | $76.77 \pm 0.05$ | $77.13 \pm 0.05$ |

Table 8: J/$\psi$ selection: cuts and efficiencies (%). The efficiencies are calculated with respect to the previous cut. The efficiency of the first cut is evaluated with respect to the preselected sample.

Only the events which survive the J/$\psi$ selection are used as the input of the selection of $\phi$, $\mathrm{K^{*0}}$ and $\mathrm{K^{\pm}}$. The J/$\psi$ selection is a necessary filter to remove most of the huge combinatorics coming mostly from prompt pions and kaons that can easily fake a $\mathrm{K^{*0}}$, $\phi$ or $\mathrm{K^{\pm}}$ signal. The detail of the selections of $\phi$, $\mathrm{K^{*0}}$ and $\mathrm{K^{\pm}}$ can be found in [30]. Finally the $\mathrm{B^0}$, $\mathrm{B_s^0}$ and $\mathrm{B^+}$ mesons are reconstructed using the same basic criteria, which are a vertex quality cut $\chi^2_{\mathrm{vtx}}/\mathrm{nDoF} < 5$, and an impact parameter significance (IPS) cut with respect to the primary vertex IPS$< 5$. If more than one primary vertex is found, the one that corresponds to the smallest IPS of the $\mathrm{B_s^0}$ candidate is chosen as its origin vertex. In the $\mathrm{B^0} \to \mathrm{J}/\psi\mathrm{K^{*0}}$ case, an additional cut on the $\mathrm{B^0}$ transverse momentum $p_{\mathrm{T}} > 2\,\mathrm{GeV}/c$ is used, to further reduce prompt background that has, on average, a lower $p_{\mathrm{T}}$. Moreover, for $\mathrm{B^0} \to \mathrm{J}/\psi\mathrm{K^{*0}}$ candidates, we reject those events where the system formed by the J/$\psi$ and the $\mathrm{K^+}$ alone has an invariant mass compatible with the $\mathrm{B^+}$ mass within $\pm 60\,\mathrm{MeV}/c^2$. Those events are mostly true $\mathrm{B^+} \to \mathrm{J}/\psi\mathrm{K^+}$ events where a prompt pion is picked up from the underlying event to form the $\mathrm{B^0} \to \mathrm{J}/\psi\mathrm{K^{*0}}$ candidate: in general they do not affect the invariant mass distribution under the peak but populate the upper sideband of the spectrum. Eliminating this peaking background allows for the use of the sidebands to infer the background properties under the peak [30]. In case several B candidates are found within one event, only the one with the smallest $\chi^2_{\mathrm{vtx}}/\mathrm{nDoF}$ is kept. Table 9 summarizes the total efficiencies for the three channels after applying the selections and the L0-trigger.

| $B_s^0 \to J/\psi\phi$ | |
|---|---|
| Cut | Efficiency [%] |
| Preselection | $\varepsilon_{presel/gen}$=24.75 ± 0.05 |
| $J/\psi \to \mu^+\mu^-$ selection | 80.47 ± 0.04 |
| $\phi \to K^+K^-$ selection | 80.9 ± 0.1 |
| $B_s^0$ $\chi^2_{vtx}$/nDoF < 5 | 97.15 ± 0.07 |
| IPS($B_s^0$) < 5 | 98.65 ± 0.03 |
| $\varepsilon_{sel/presel}$ | 62.3 ± 0.1 |
| $\varepsilon_{L0/sel}$ | 93.9 ± 0.1 |
| $\varepsilon_{sel/presel} \times \varepsilon_{L0/sel}$ | 58.5 ± 0.1 |
| $\varepsilon_{presel/gen} \times \varepsilon_{sel/presel} \times \varepsilon_{L0/sel}$ | 14.5 ± 0.1 |
| $\varepsilon_{tot} = \varepsilon_{presel/gen} \times \varepsilon_{sel/presel} \times \varepsilon_{L0/sel} \times \varepsilon_{gen}$ | 2.61 ± 0.01 |

| $B^0 \to J/\psi K^{*0}$ | |
|---|---|
| Cut | Efficiency [%] |
| Preselection | $\varepsilon_{presel/gen}$=21.05 ± 0.10 |
| $J/\psi \to \mu^+\mu^-$ selection | 76.77 ± 0.04 |
| $K^{*0} \to K\pi$ selection | 65.16 ± 0.06 |
| $B^0$ $\chi^2_{vtx}$/nDoF < 5 | 97.78 ± 0.06 |
| IPS($B^0$) < 5 | 98.27 ± 0.02 |
| $p_T(B^0) > 2\,GeV/c$ | 92.98 ± 0.06 |
| $\lvert M(\mu\mu K) - M(B^+)\rvert > 60$ MeV/$c^2$ | 100 |
| $\varepsilon_{sel/presel}$ | 44.35 ± 0.07 |
| $\varepsilon_{L0/sel}$ | 94.88 ± 0.04 |
| $\varepsilon_{sel/presel} \times \varepsilon_{L0/sel}$ | 42.08 ± 0.04 |
| $\varepsilon_{presel/gen} \times \varepsilon_{sel/presel} \times \varepsilon_{L0/sel}$ | 8.87 ± 0.02 |
| $\varepsilon_{tot} = \varepsilon_{presel/gen} \times \varepsilon_{sel/presel} \times \varepsilon_{L0/sel} \times \varepsilon_{gen}$ | 1.54 ± 0.01 |

| $B^+ \to J/\psi K^+$ | |
|---|---|
| Cut | Efficiency [%] |
| Preselection | $\varepsilon_{presel/gen}$=35.78 ± 0.04 |
| $J/\psi \to \mu^+\mu^-$ selection | 77.13 ± 0.05 |
| Kaon selection | 59.02 ± 0.06 |
| $B^+$ $\chi^2_{vtx}$/nDoF < 5 | 97.34 ± 0.03 |
| IPS($B^+$) < 5 | 99.44 ± 0.01 |
| $\varepsilon_{sel/presel}$ | 43.34 ± 0.06 |
| $\varepsilon_{L0/sel}$ | 94.45 ± 0.04 |
| $\varepsilon_{sel/presel} \times \varepsilon_{L0/sel}$ | 40.94 ± 0.05 |
| $\varepsilon_{presel/gen} \times \varepsilon_{sel/presel} \times \varepsilon_{L0/sel}$ | 14.65 ± 0.04 |
| $\varepsilon_{tot} = \varepsilon_{presel/gen} \times \varepsilon_{sel/presel} \times \varepsilon_{L0/sel} \times \varepsilon_{gen}$ | 2.61 ± 0.01 |

Table 9: Efficiencies for $B_s^0 \to J/\psi\phi$, $B^0 \to J/\psi K^{*0}$ and $B^+ \to J/\psi K^+$ selections. Each efficiency is defined with respect to the previous cut. The preselection efficiency, $\varepsilon_{presel/gen}$, is defined with respect to the number of events produced, after the generator-level cuts. It includes acceptance, reconstruction and some loose selection cuts described in the text.

## 6.2 Annual event yields

For a given channel the annual yield is expressed as follows:

$$\mathcal{S} = \mathcal{L}_{\rm int} \times \sigma_{\rm b\overline{b}} \times 2 \times f_{\rm u,d,s} \times \mathcal{B}_{\rm vis} \times \varepsilon_{\rm tot}\,, \qquad (60)$$

where $\mathcal{L}_{\rm int} = 2\,{\rm fb}^{-1}$ is the integrated luminosity in one nominal year, $\sigma_{\rm b\overline{b}}$ =500 $\mu$b is the $\rm b\overline{b}$ production cross section at 14 TeV[5], the factor 2 accounts for the pair production of b quarks, $f_{\rm u,d,s}$ are the B meson production fractions; $\mathcal{B}_{\rm vis}$ is the total visible branching fraction, and $\varepsilon_{\rm tot}$ is the total efficiency. The total efficiency is given by:

$$\varepsilon_{\rm tot} = \varepsilon_{\rm gen} \times \varepsilon_{\rm presel/gen} \times \varepsilon_{\rm sel/presel} \times \varepsilon_{\rm L0/sel}\,, \qquad (61)$$

where $\varepsilon_{\rm gen}$ is the generator level cut efficiency, $\varepsilon_{\rm presel/gen}$ is the combination of acceptance, reconstruction and preselection cuts efficiencies, $\varepsilon_{\rm sel/presel}$ the selection efficiency normalized to preselected events and finally $\varepsilon_{\rm L0/sel}$ is the L0-trigger efficiency on selected events.

In Table 10 we show the total selection efficiencies and the annual yields for the three channels. The annual event yields are given for selected and L0-triggered events without/with including the High Level Trigger (HLT) that will be discussed in Section 7.

The dominant uncertainty in the yield is due to the uncertainty ($\sim$ 40%) in the value of the $\sigma_{\rm b\overline{b}}$ at 14 TeV [26]. For $\rm B^0_s \to J/\psi\phi$ there is also the contribution from $\mathcal{B}(\rm B^0_s \to J/\psi\phi)$ which is known with an accuracy of $\sim$ 35% [2].

| Channel | $\varepsilon_{\rm tot}$ [%] | Event yield after L0 | Event yield after L0+HLT |
|---|---|---|---|
| $\rm B^0_s \to J/\psi(\mu\mu)\phi(KK)$ | $2.61 \pm 0.01$ | 156 k | 117 k |
| $\rm B^0 \to J/\psi(\mu\mu)K^{*0}(K\pi)$ | $1.54 \pm 0.01$ | 648 k | 489 k |
| $\rm B^+ \to J/\psi(\mu\mu)K^+$ | $2.61 \pm 0.01$ | 1 248 k | 942 k |

Table 10: Total selection efficiency (before HLT), untagged event yield for selected and L0-triggered events, and including HLT in the last column. The event yields correspond to an integrated luminosity of $2\,{\rm fb}^{-1}$.

## 6.3 Signal distributions

The main goals are to minimize biases in proper time and angular acceptances and to select the $\rm B^0_s \to J/\psi\phi$ signal and control channels such as they share the same phase space. These aspects are discussed in the following sections.

### 6.3.1 Invariant mass distribution

The mass distribution of the $\rm B^0_s$ candidates passing the selection is shown in Figure 5. The whole mass range is well fitted using a triple Gaussian function:

$$\begin{aligned} &F(m) = f_{m,1} \times G_1 + f_{m,2} \times G_2 + (1 - f_{m,1} - f_{m,2}) \times G_3\,, \\ &\text{where} \quad G_i = \frac{1}{\sqrt{2\pi}\sigma_{m,i}} \exp\left(-\frac{1}{2}\left(\frac{m-\mu_{m,i}}{\sigma_{m,i}}\right)^2\right) \qquad i = 1,2,3\,, \end{aligned} \qquad (62)$$

[5] For the yields computation we use the LHCb convention of $\sigma_{\rm b\overline{b}} = 500\,\mu$b and the branching fractions from Ref. [2].

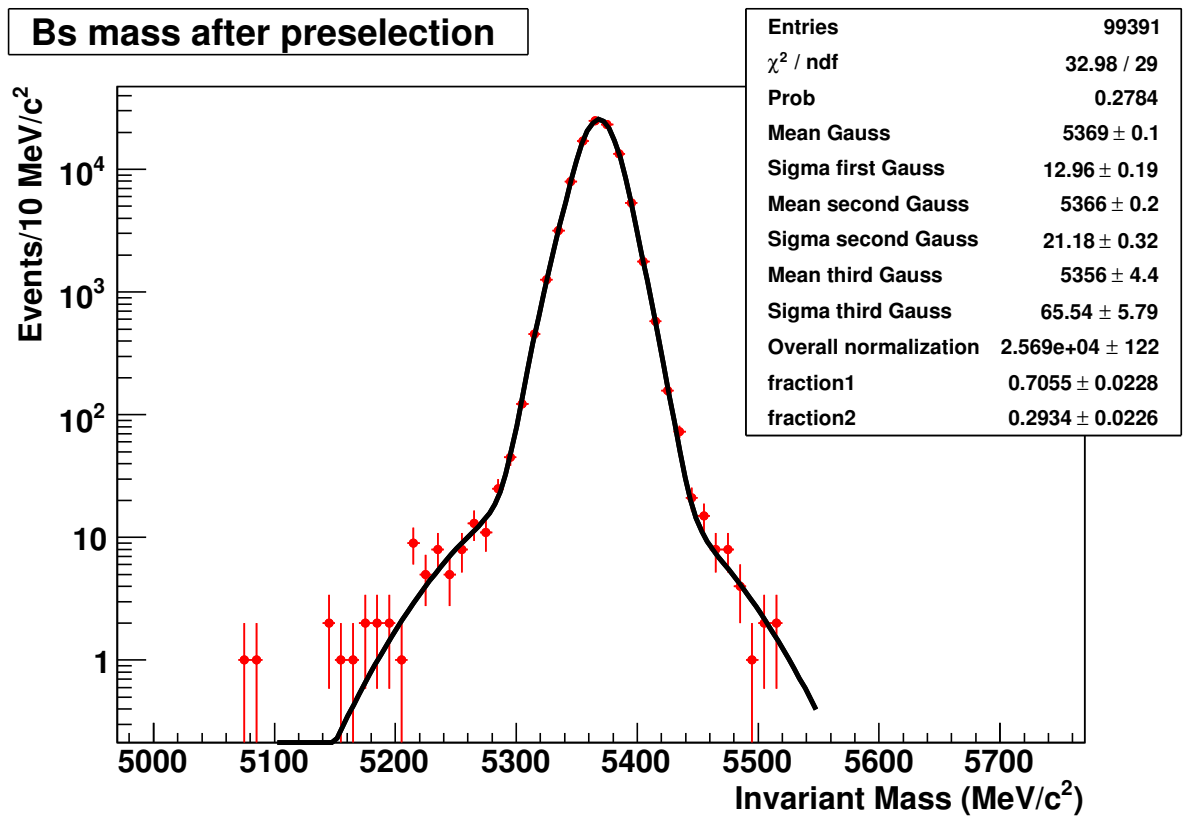

Figure 5: $B_s^0 \to J/\psi\phi$ selection: $B_s^0$ mass for signal events passing all cuts and L0-trigger. No $J/\psi$ mass constraint is used.

with $\sigma_{m,i}$ the individual resolution terms and $f_{m,i}$ the relative weights. The corresponding average resolution is:

$$\langle\sigma_m\rangle = \sqrt{f_{m,1} \times \sigma_{m,1}^2 + f_{m,2} \times \sigma_{m,2}^2 + (1 - f_{m,1} - f_{m,2}) \times \sigma_{m,3}^2} \simeq 16\,\mathrm{MeV}/c^2\,. \tag{63}$$

The result of the fit of the invariant mass distribution is shown in Table 11 (first column); for comparison the results for the control channels $B^0 \to J/\psi K^{*0}$ and $B^+ \to J/\psi K^+$ are also reported.

| Channel | $\langle\sigma_m\rangle$ [MeV/$c^2$] | $\langle\sigma_t^{rec}\rangle$ [fs] | $\tau_{input}$ [ps] | $\tau_{true}^{fitted}$ [ps] | $\tau_{rec}^{fitted}$ [ps] |
|---|---|---|---|---|---|
| $B_s^0 \to J/\psi\phi$ | 16.0 | $38 \pm 1$ | $\tau_L = 1.391$ | – | – |
| | | | $\tau_H = 1.538$ | – | – |
| | | | $\tau_{eq} = 1.415$ | $1.407 \pm 0.005$ | $1.413 \pm 0.005$ |
| $B^0 \to J/\psi K^{*0}$ | 16.5 | $38 \pm 1$ | 1.536 | $1.516 \pm 0.003$ | $1.524 \pm 0.003$ |
| $B^+ \to J/\psi K^+$ | 18.3 | $36 \pm 1$ | 1.671 | $1.641 \pm 0.003$ | $1.638 \pm 0.003$ |

Table 11: Average mass resolution, average proper time resolution $\langle\sigma_t^{rec}\rangle$ from a fit of $\Delta t = t^{rec} - t^{true}$ distribution, lifetime input value, lifetime obtained from a fit on the true proper times of selected and L0-triggered events and lifetime obtained from a fit of reconstructed proper times of selected and L0-triggered events.

### 6.3.2 Proper time distributions: acceptance and resolution

The $B_s^0$ proper time, $t^{rec}$, and its uncertainty, $\delta_t^{rec}$, are obtained from a kinematic fit, using Lagrange multipliers under the constraint that the $B_s^0$ momentum points to the primary vertex (PV). The secondary vertex is solely defined by the decay products, while the primary vertex is determined by using all the reconstructed tracks or segments with a dedicated algorithm [31]. To limit systematic effects in the PV and the proper time measurements a refit of the PV without the $B_s^0$ products needs to be done. This correction is particularly important in this analysis since a large fraction of the $B_s^0$ products contribute to the PV determination due to the absence of IP cuts in the applied selection.

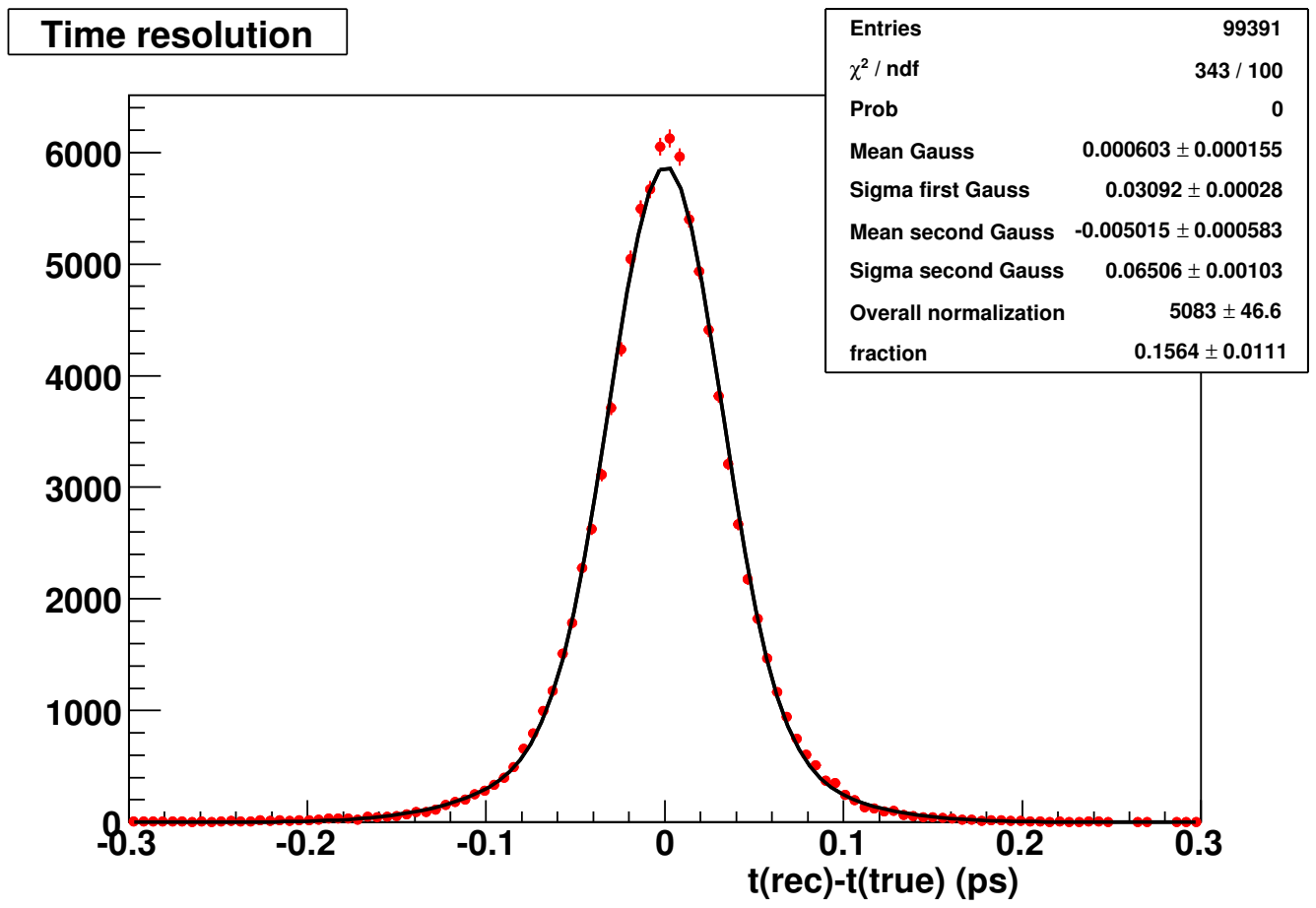

Figure 6: Proper time resolution: $t_{\rm rec} - t_{\rm true}$ distribution for $\rm B_s^0 \to J/\psi\phi$ selected and L0-triggered events.

The proper time resolution has been evaluated by fitting with a double Gaussian function the $\Delta t = t^{\rm rec} - t^{\rm true}$ distribution represented in Figure 6, where $t^{\rm true}$ is the true proper time. The results for the fit parameters are: $f_{t,1} = 0.844 \pm 0.011$, $\sigma_{t,1} = (30.9 \pm 0.3)$ fs and $\sigma_{t,2} = (65 \pm 1)$ fs corresponding to a mean resolution $\langle\sigma_t\rangle \sim 38$ fs.
The results of the fit for $\rm B_s^0 \to J/\psi\phi$, $\rm B^0 \to J/\psi K^{*0}$ and $\rm B^+ \to J/\psi K^+$ are summarized in Table 11 (second column) and are in agreement with each other within the errors.

The distribution of the event-by-event proper time error is represented in Figure 7 (left). The pull distribution, $(t^{\rm rec} - t^{\rm true})/\delta_t^{\rm rec}$, is shown in Figure 7 (right). It is fitted with a single Gaussian, the standard deviation of which shows that the time errors are underestimated by a factor $SF$=$1.112 \pm 0.001$. This behaviour can be attributed to an underestimation of the track parameter and vertex errors. Therefore, in order to use event-by-event errors in the fit it will be necessary to use a calibration procedure to correct for this underestimation. Studies indicate, however, that for sample sizes corresponding to $2\,{\rm fb}^{-1}$, the assignment of per-event errors in the fit does not significantly increase the sensitivity to $\phi_{\rm s}$ (Section 11.2). Therefore the baseline proper time resolution model for toy MC study is a fixed resolution, extracted from the double-Gaussian of Figure 6. The natural evolution of this model will be to use the per event proper time estimate (Figure 7). A detailed study of a proper time resolution model is described in [25].

For the $\rm B_s^0 \to J/\psi\phi$ we show in Figure 8 the proper time acceptance defined as the true proper time distribution divided by the theoretical expectation. The acceptance slightly decreases with proper time, with a slope of $-0.0036 \pm 0.0003\,{\rm ps}^{-1}$. A similar behaviour is observed in the two control channels. The origin of this very small effect is being investigated.

The true and reconstructed proper time distributions for selected and L0-triggered $\rm B_s^0 \to J/\psi\phi$ events are shown in Figure 9. The $\rm B_s^0$ is a superposition of the two states evolving in time with $\tau_{\rm L}$ and $\tau_{\rm H}$ whose values are shown in Table 6. For the study here, we instead fit using a single exponential with an *average* lifetime of $\tau_{\rm eq} = 1.415\,{\rm ps}$:

$$\tau_{\rm eq} = \left((1-\cos\phi_{\rm s})\frac{|A_0(0)|^2}{2} + (1-\cos\phi_{\rm s})\frac{|A_\parallel(0)|^2}{2} + (1+\cos\phi_{\rm s})\frac{|A_\perp(0)|^2}{2}\right)\tau_{\rm H} +$$

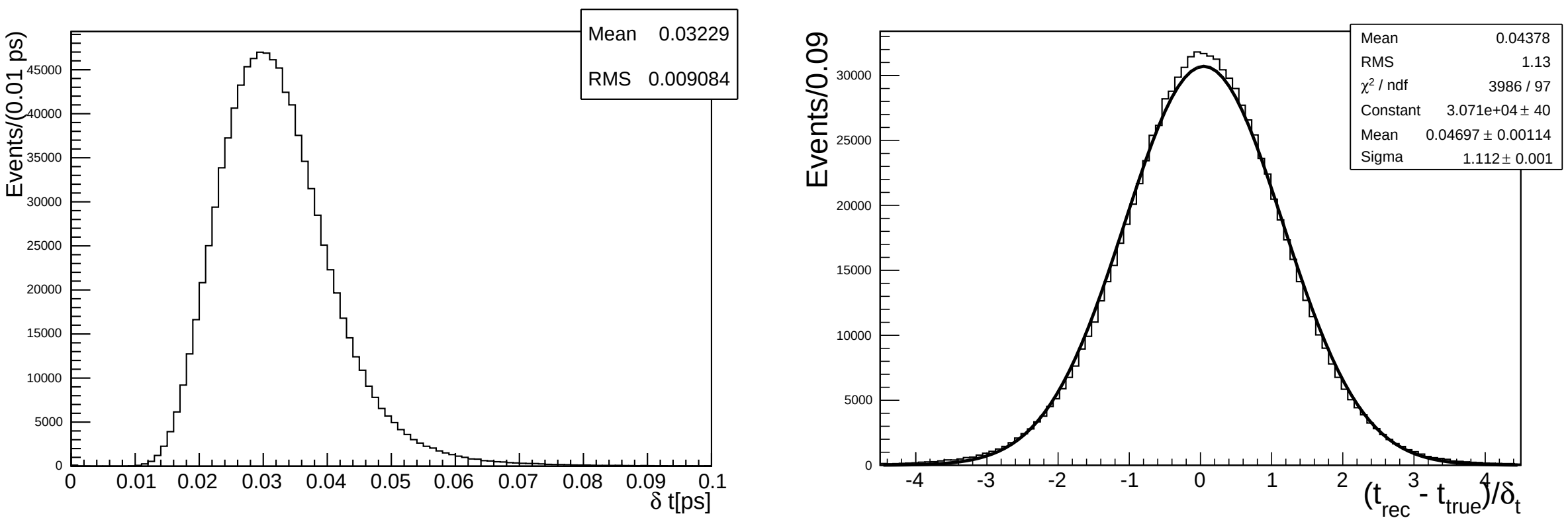

Figure 7: For selected $B_s^0 \to J/\psi\phi$ events, event-by-event proper time error (left) and pull distribution (right).

$$\left((1+\cos\phi_s)\frac{|A_0(0)|^2}{2} + (1+\cos\phi_s)\frac{|A_\parallel(0)|^2}{2} + (1-\cos\phi_s)\frac{|A_\perp(0)|^2}{2}\right)\tau_L\,. \quad (64)$$

The fit of the proper time distribution is performed with a simple exponential if the true proper time is considered and with an exponential convoluted with a Gaussian if the reconstructed proper time is considered. The same study has been performed for the other two channels $B^0 \to J/\psi K^{*0}$ and $B^+ \to J/\psi K^+$ [30]. The results are summarized in the last three columns of Table 11: in all cases the fitted true and reconstructed lifetimes agree within the errors. For the $B_s^0$, $B^0$ and $B^+$ channels the result of the fit are respectively 0.8%, 1% and 2% below the input values, because the proper time acceptance has not been taken into account at this level.

For the three channels the proper time resolution has been extracted from a fit to the proper time distribution convoluted with a Gaussian with free parameters. The results are compatible with the values extracted from a fit to $\Delta t$ (Table 11, second column). These results show that a fit of the proper time distribution of events in the signal mass region, properly background subtracted, can give a first estimate of the average proper time resolution. Other samples can be used to extract from data a first average estimate of the proper time resolution and will be described in Section 6.4.

### 6.3.3 Angular resolutions and acceptances

In Figures 10−12 (left) we show the distributions of the three angles in the transversity basis for selected and L0-triggered signal events compared with the theoretical predictions. In the same figures, we show the acceptances obtained by dividing the true angular distributions by the the theoretical expectation. Although distortions are observed, the measurements of the three angles are characterized by acceptance curves which deviate from uniformity by less than 10%. The distortions come from the requirement of having the four decay tracks reconstructed within the LHCb geometrical acceptance. The effect on the physics parameters of these residual distortions is not negligible if it is not properly taken into account (see Section 10.1).

The angular resolutions, obtained from a weighted average from a triple-Gaussian fit, are shown in Table 12. Their effect on the extraction on physics parameters is discussed in Sections 9.5 and 11.1.2.

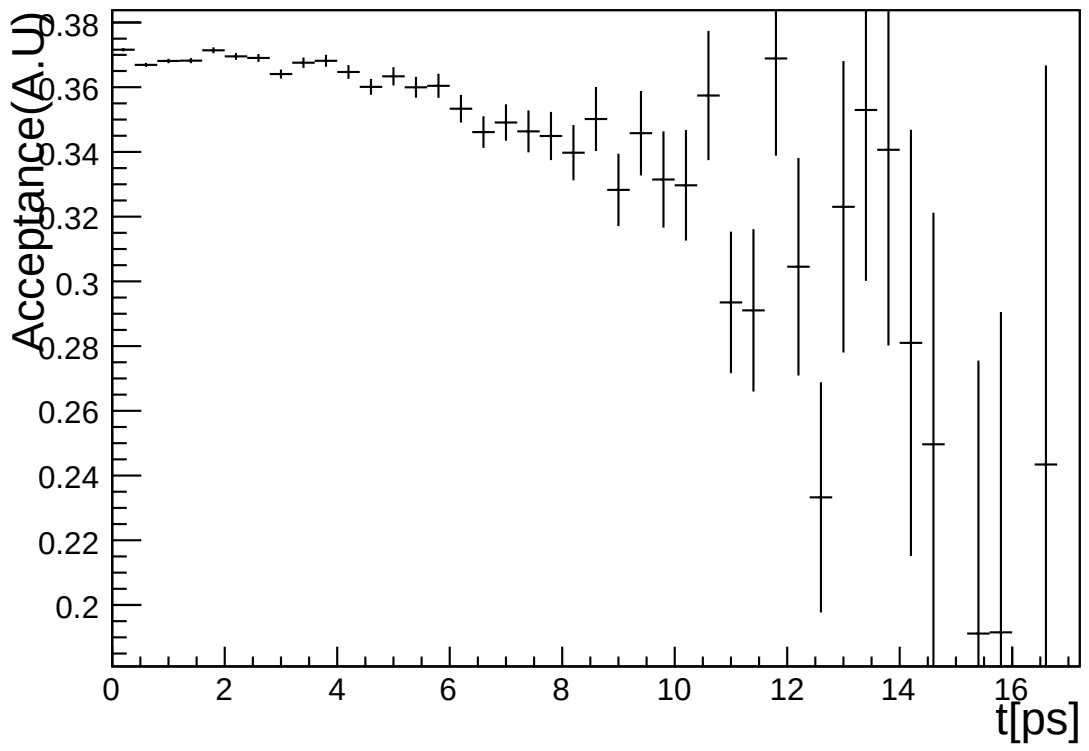

Figure 8: Proper time acceptance for $B_s^0 \to J/\psi\phi$ events passing all the selection cuts and L0-trigger. Note that the plot is zero suppressed and made with a sample size equivalent to $\sim 17\,\mathrm{fb}^{-1}$ of data.

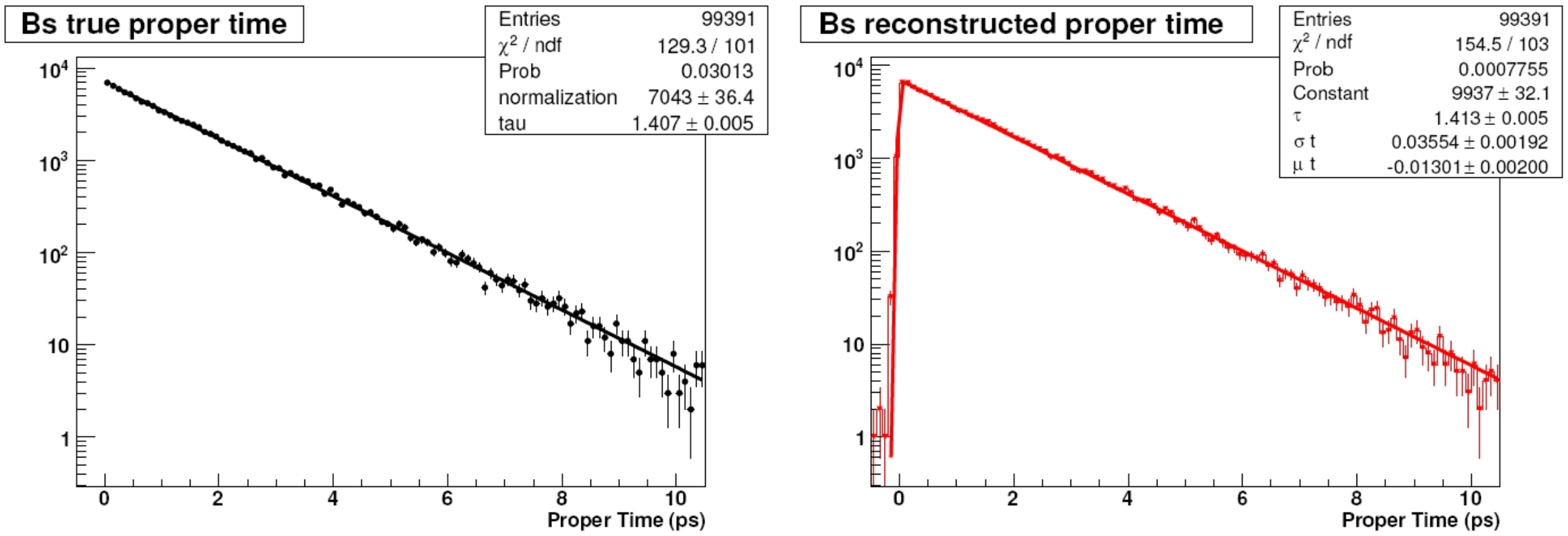

Figure 9: Proper time distribution for $B_s^0 \to J/\psi\phi$ events passing all the selection cuts and L0-trigger: true proper time (left) and reconstructed proper time (right).

| | Average resolution |
|---|---|
| $\theta$ | 27 mrad |
| $\varphi$ | 27 mrad |
| $\psi$ | 20 mrad |

Table 12: Average resolution on the three transversity angles.

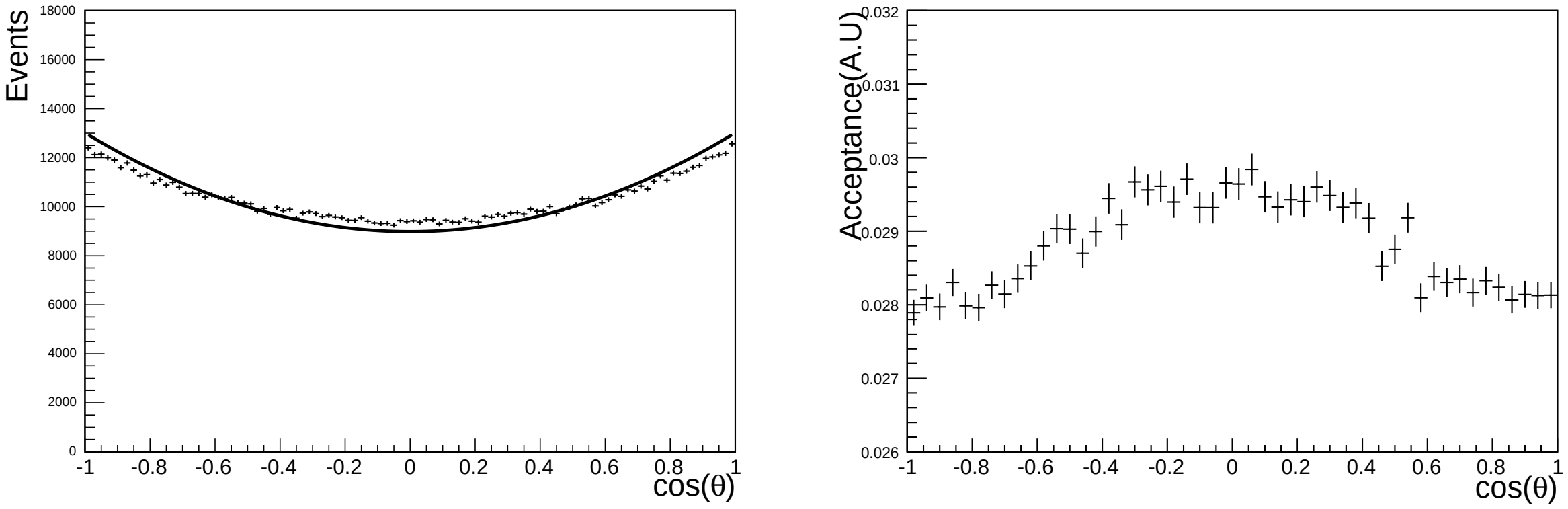

Figure 10: $B_s^0 \to J/\psi\phi$: $\cos\theta$ distribution (left) and acceptance (right) for reconstructed, selected and L0-triggered events. The superimposed curve is the expectation from theory.

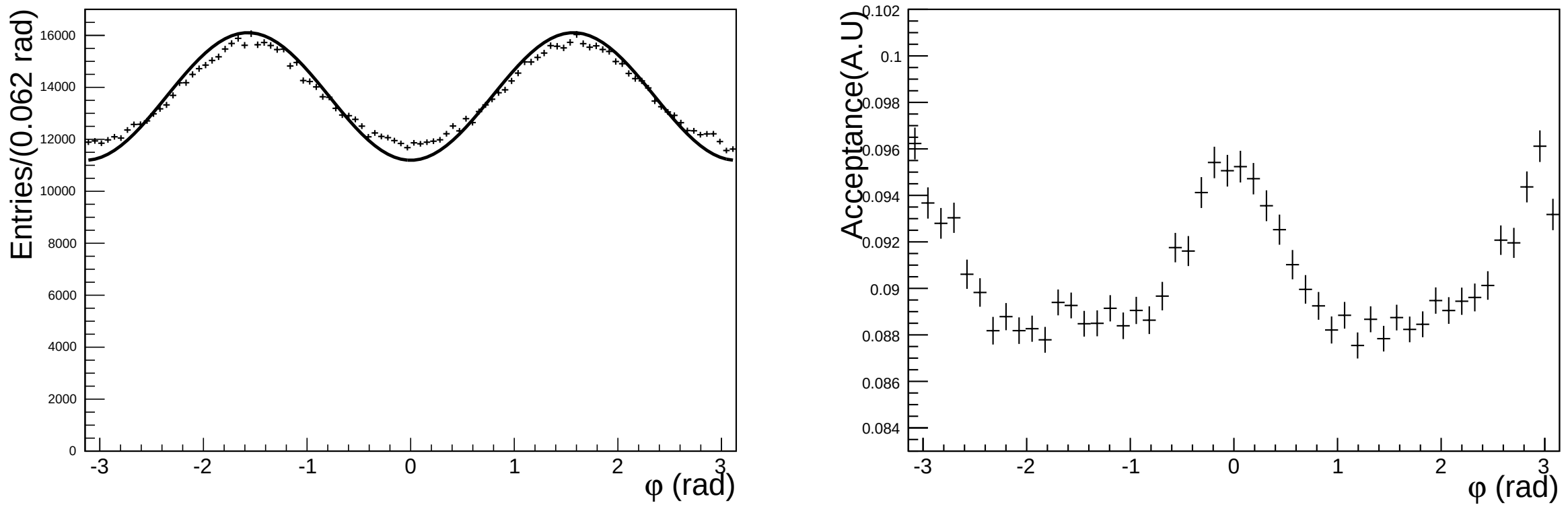

Figure 11: $B_s^0 \to J/\psi\phi$: $\varphi$ distribution (left) and acceptance (right) for reconstructed, selected and L0-triggered events. The superimposed curve is the expectation from theory.

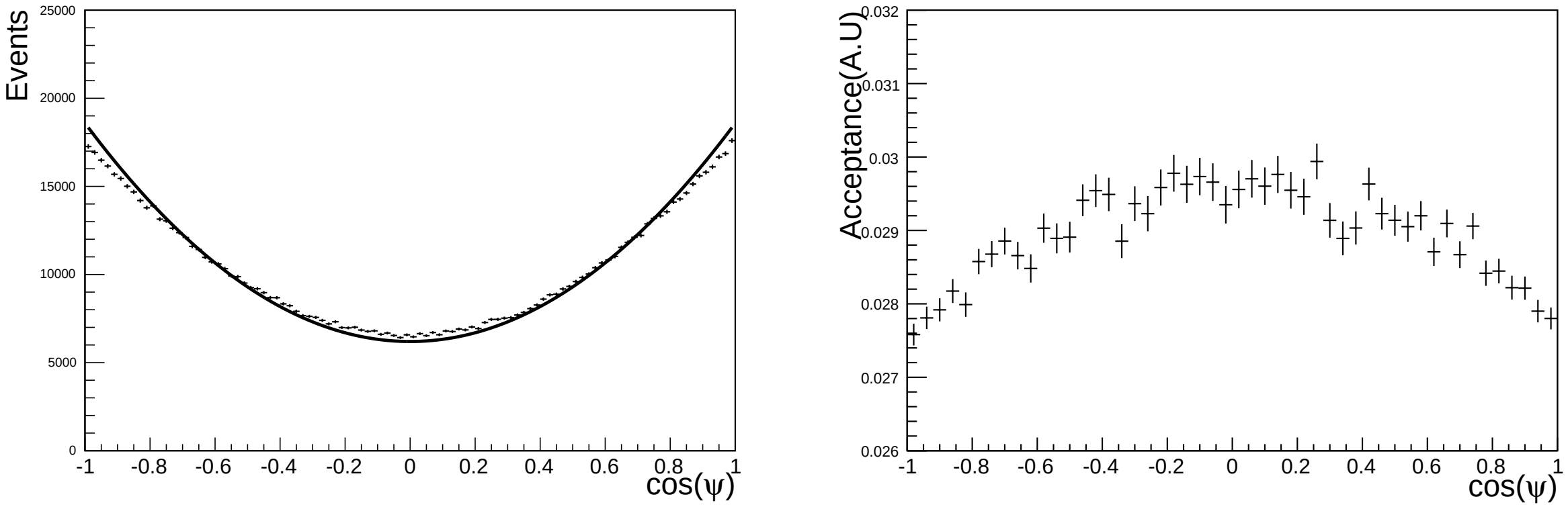

Figure 12: $B_s^0 \to J/\psi\phi$: $\cos\psi$ distribution (left) and acceptance (right) for reconstructed, selected and L0-triggered events. The superimposed curve is the expectation from theory.

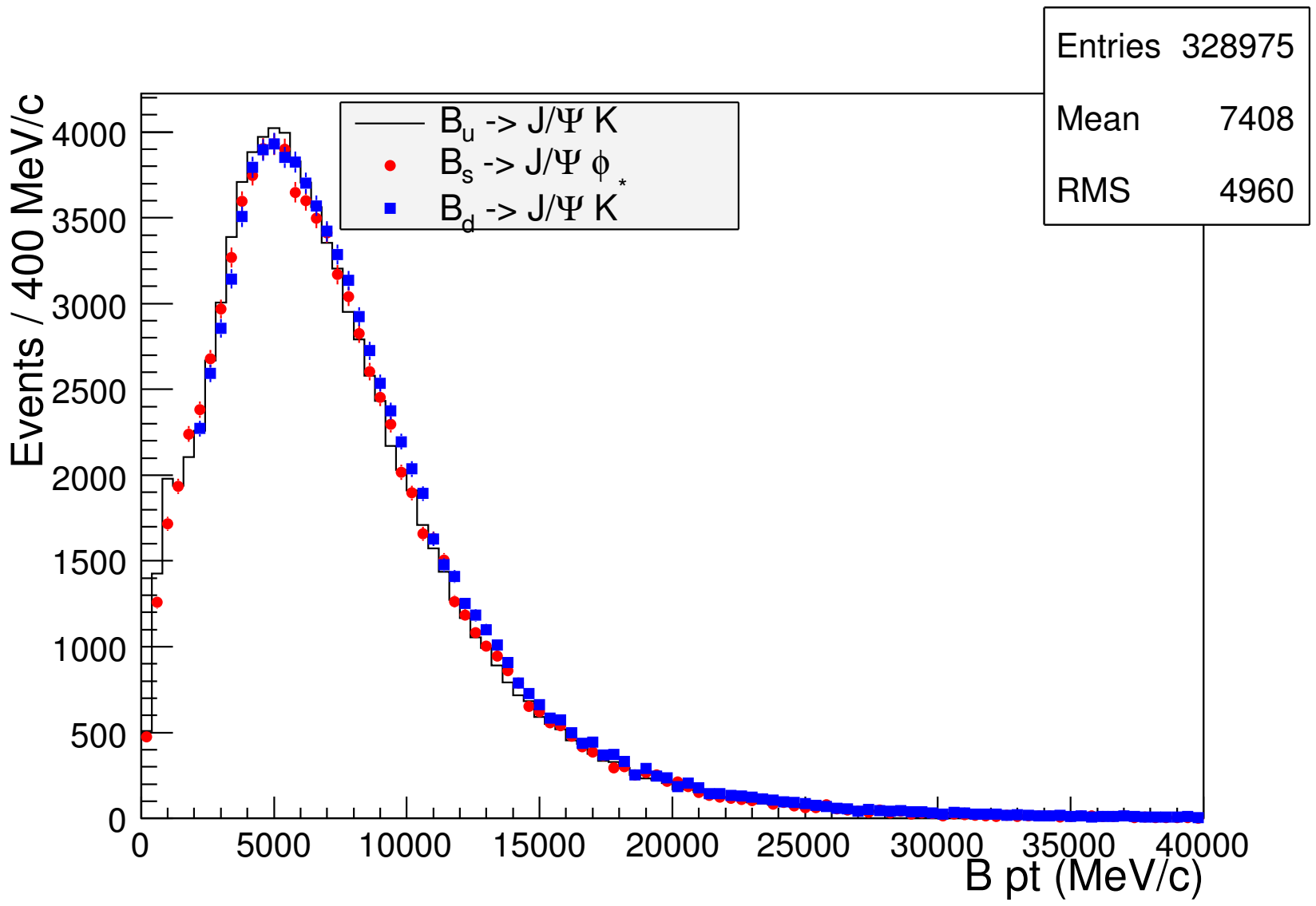

Figure 13: Transverse momentum distributions for the $B_s^0$, $B^0$ and $B^+$ candidates after selection and L0-trigger.

### 6.3.4 $p_T$ spectrum of the selected B mesons

In Figure 13 we show the transverse momentum spectrum for selected signal channels after L0: the selected $B_s^0$ signal candidates share the same phase space as the two flavor specific control channels and therefore we expect that opposite-side tagging properties can be safely determined from control channels, without large corrections. This will be discussed in Section 8.

## 6.4 Backgrounds

The $\phi_s$ phase is extracted from a fit to the tagged time-dependent angular distributions. Therefore the impact of the background on the final accuracy on $\phi_s$ depends not only on the total contamination in the signal region but also on its proper time and angular distributions and on its tagging properties.

We expect, for example, that the prompt background component will have a smaller impact on the final sensitivity of $\phi_s$ with respect to the long-lived one, since most of the events have very low proper times ($t < 0.2\,\mathrm{ps}$). Moreover the prompt background proper time distribution can be easily fitted and parametrized. Since the prompt component is mainly given by combinatorics, we expect that the tagging efficiency for those events will be lower with respect to the one obtained for long-lived partially reconstructed B decays.

In data the background will be extrapolated from the sidebands of the invariant mass distribution and its proper time, angular distributions and tagging efficiency will be parametrized.

The background study is complicated by the fact that the available inclusive Monte Carlo samples (minimum bias and $b\bar{b}$ inclusive) have limited statistics. More abundant specific background samples describe only a fraction (even if the dominant part) of the total background. Therefore the use of the available Monte Carlo samples in order to extract background properties is rather limited and some approximations must be made.

Here we describe the Monte Carlo samples that have been analyzed and how each sample has been used to evaluate a specific background property.

Six different samples have been considered: minimum bias events, inclusive $b\overline{b}$ events, inclusive $J/\psi(\mu\mu)$ events, $B^+ \to J/\psi X$, $B^0 \to J/\psi X$ and $B^0_s \to J/\psi X$ events. For each sample we divide the backgrounds events in two categories: prompt and long-lived. The prompt component is defined as a sample where all tracks used to build the candidate come from prompt processes. The long-lived component is all the rest and is mainly composed of events where at least one of the tracks used to build the candidate comes from the decay of a b-hadron.

For prompt and long-lived background components we have studied the $B/S$ ratio, the composition, the proper time and angular distributions and the tagging properties. We use the six Monte Carlo samples in the following way:

1. the minimum bias sample is used to get a rough estimate of the total background contamination in an enlarged mass window ($\pm 300\,\mathrm{MeV}/c^2$) around the nominal $B_{u,d,s}$ mass; this broad mass window is necessary in order to study the background in the sidebands of the signal mass distribution;

2. $J/\psi \to \mu\mu$ inclusive events are used to study the properties of the prompt background component after the removal of the long-lived $b \to J/\psi X$ events;

3. the $b\overline{b}$ inclusive sample is used to estimate the total amount of the long-lived background. The dominant contribution of this background comes from $B_{u,d,s} \to J/\psi X$ events so we use the dedicated $B_{u,d,s} \to J/\psi X$ sample to study the proper time, mass and angular distribution of this contamination.

A detailed study of all the background components is performed in [30]. In Table 13 we summarize the main results.

- The first column shows the ratio of prompt background over signal in a $\pm 50\,\mathrm{MeV}/c^2$ mass window around the nominal B meson mass. It is evaluated from the $J/\psi \to \mu\mu$ prompt sample using the formula:

$$\frac{B_{\mathrm{Pr}}}{S} = \frac{\sigma_{\mathrm{Pr}} \times \mathcal{B}(J/\psi \to \mu\mu) \times \varepsilon_{\mathrm{Pr}} \times f_{\mathrm{MW}}}{2 \times \sigma_{b\overline{b}} \times f_{u,d,s} \times \mathcal{B}_{\mathrm{vis}} \times \varepsilon_{\mathrm{tot}}}\,, \quad (65)$$

  where $\sigma_{\mathrm{Pr}}$ is the cross-section $pp \to J/\psi X$ given in Table 5, $\varepsilon_{\mathrm{Pr}}$ is the selection efficiency for the prompt component in the $\pm 300\,\mathrm{MeV}/c^2$ mass window, $f_{\mathrm{MW}}$ is the scale factor needed to pass from $\pm 300\,\mathrm{MeV}/c^2$ to the $\pm 50\,\mathrm{MeV}/c^2$ mass window[6] and $\varepsilon_{\mathrm{tot}}$ is the total efficiency for the signals (Table 10). The $f_{u,d,s}$ are the production fractions for $B_{u,d,s}$, $\mathcal{B}_{\mathrm{vis}}$ are the visible branching fractions (Table 4, DC06 column). The $b\overline{b}$ cross-section value is the one from DC06 (Table 5).

- The second column shows the ratio of long-lived background over signal in $\pm 50\,\mathrm{MeV}/c^2$ mass window around the nominal B meson mass. It is evaluated from the inclusive $b\overline{b}$ sample using the formula:

$$B_{\mathrm{LL}}/S = \frac{\varepsilon_{\mathrm{sel},b\overline{b}}}{2 \times f_{u,d,s} \times \mathcal{B}_{\mathrm{vis}} \times \varepsilon_{\mathrm{tot}}}\,, \quad (66)$$

[6] We assume a linear mass distribution.

where $\varepsilon_{\mathrm{sel,b\bar{b}}}$ is the selection efficiency in the $\mathrm{b\bar{b}}$ sample after removing the signal events.

- the third column shows the rate of minimum bias events in the $\pm 300\,\mathrm{MeV}/c^2$ mass window surviving the selections. It has been calculated assuming a minimum bias rate of 0.9 MHz after the L0-trigger. Two $\mathrm{B_s^0} \to \mathrm{J}/\psi\phi$ candidates pass the selection and the L0-trigger among 5.9 M events. The first is a prompt $\mathrm{J}/\psi \to \mu\mu$, with a proton and a kaon, so it is already accounted for in the "prompt" background (Eq. 65). Its invariant mass is $5.084\,\mathrm{GeV}/c^2$. The second is composed of two prompt kaons decaying in flight and one prompt $\phi \to \mathrm{K^+K^-}$. Its invariant mass is $5.077\,\mathrm{GeV}/c^2$. This additional background is not accounted for in the sensitivity studies (Section 10.2), but will be carefully monitored when a larger background sample will be available. Indeed, with only one event, one cannot extrapolate any proper time or angular distribution.

| Channel | $B_{\mathrm{Pr}}/S$ | $B_{\mathrm{LL}}/S$ | Minimum bias rate |
|---|---|---|---|
| $\mathrm{B_s^0} \to \mathrm{J}/\psi(\mu\mu)\phi(\mathrm{KK})$ | $1.6 \pm 0.6$ | $0.51 \pm 0.08$ | $\sim 0.3$ Hz |
| $\mathrm{B^0} \to \mathrm{J}/\psi(\mu\mu)\mathrm{K^{*0}}(\mathrm{K}\pi)$ | $5.2 \pm 0.3$ | $1.53 \pm 0.08$ | $\sim 8.1$ Hz |
| $\mathrm{B^+} \to \mathrm{J}/\psi(\mu\mu)\mathrm{K^+}$ | $1.6 \pm 0.2$ | $0.29 \pm 0.06$ | $\sim 1.4$ Hz |

Table 13: $B/S$ ratio estimated in a $\pm 50\,\mathrm{MeV}/c^2$ mass window for prompt and long-lived backgrounds and minimum bias rate in a $\pm 300\,\mathrm{MeV}/c^2$ mass window, after selection and L0-trigger.

In Figure 14 we show the invariant mass distribution of $\mathrm{B_s^0} \to \mathrm{J}/\psi\phi$ candidates from the $\mathrm{J}/\psi$ inclusive sample (left) and the proper time distribution for the prompt component only (right). The fit of the proper time distribution of the prompt background component is performed with a single Gaussian function with a standard deviation of $\sigma_t \sim 43$ fs which is in agreement within $3\sigma$ with that found for the signal distribution. Similar results have been found for the other two channels which is to be expected since the events have similar topology and share the same phase space [30].

In Figure 15 we show the invariant mass distribution of $\mathrm{B_s^0} \to \mathrm{J}/\psi\phi$ candidates from the $\mathrm{B_{u,d,s}} \to \mathrm{J}/\psi\mathrm{X}$ inclusive samples (left) and proper time distribution for the long-lived component only (right).

The proper time distribution for the long-lived background is fitted with a double exponential convoluted with a double Gaussian function: $\sim 72\%$ of the distribution is described by an exponential with a short decay constant, $\tau = 0.18\,\mathrm{ps}$, while a longer one ($\tau = 1\,\mathrm{ps}$) accounts for the remainder of the distribution. The core of the proper time resolution distribution is described again by a Gaussian with $\sigma_t = 39\,\mathrm{fs}$, which is in agreement with the other determinations.

In Figures 16-17 we show the angular distributions for prompt background events and in Figures 18-19 the same for long-lived background events extracted from the sidebands after selection and L0-trigger. For the prompt background events, the distributions are flat, as these events are mostly combinatorial. Some structure is observed in the long-lived sample because the fraction of partially reconstructed signal events is not negligible.

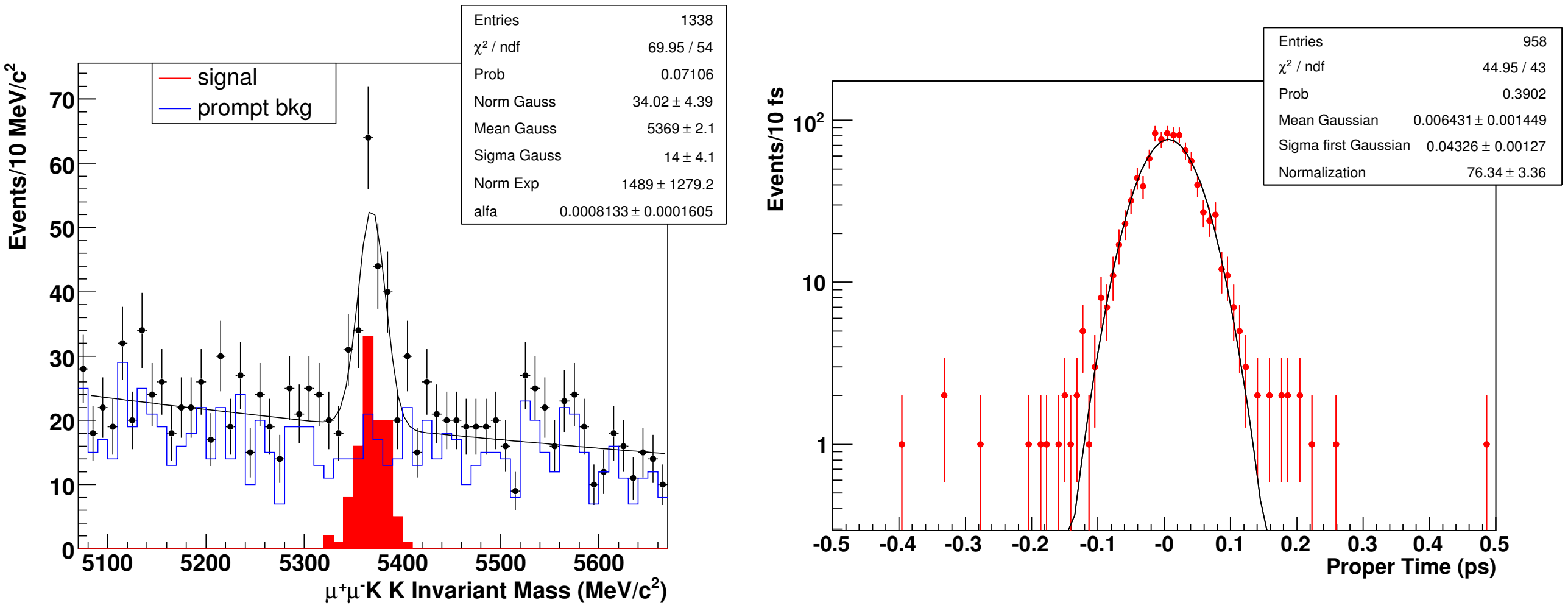

Figure 14: $B_s^0 \to J/\psi\phi$ candidates in the $J/\psi(\mu\mu)$ inclusive sample: invariant mass (left) and proper time distribution for the prompt component, where the true signal events have been removed (right).

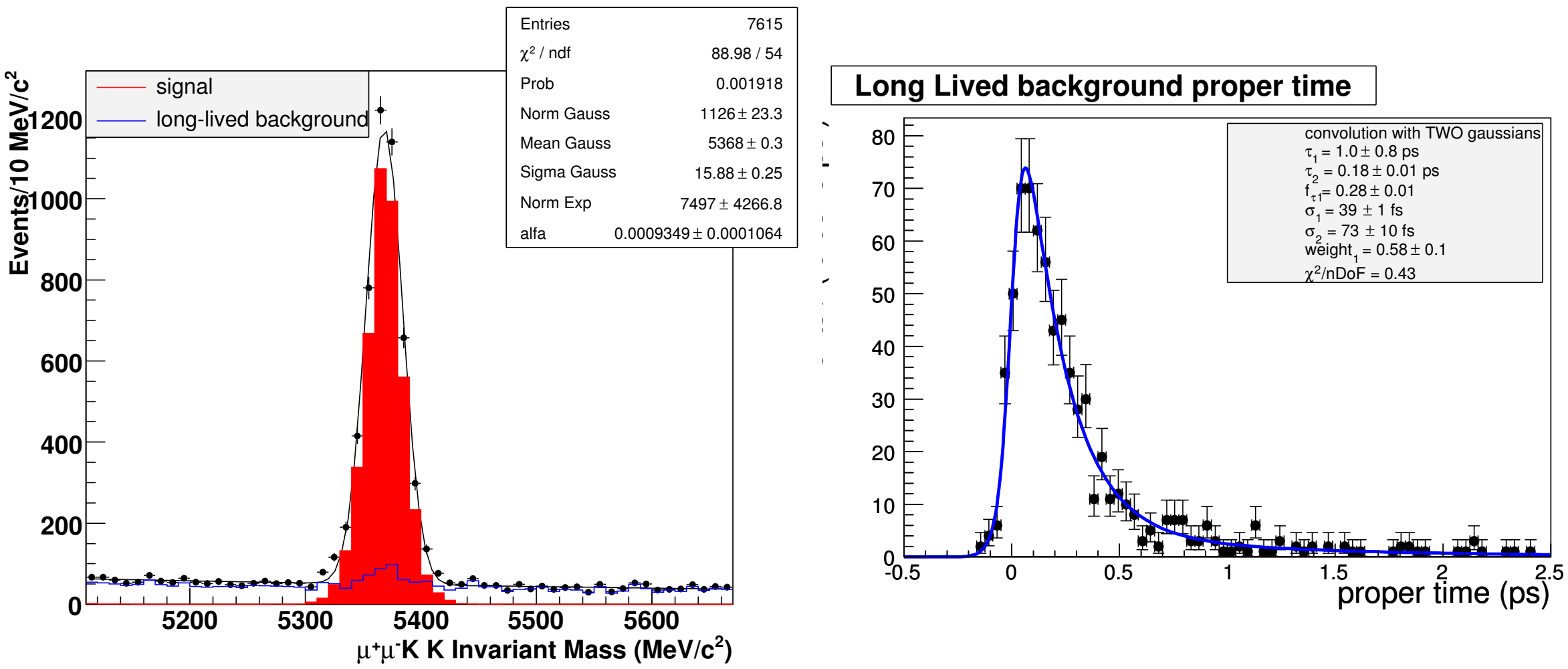

Figure 15: $B_s^0 \to J/\psi\phi$ candidates in the $B_{u,d,s} \to J/\psi X$ inclusive sample: invariant mass (left) and proper time distribution for the long-lived background component (right).

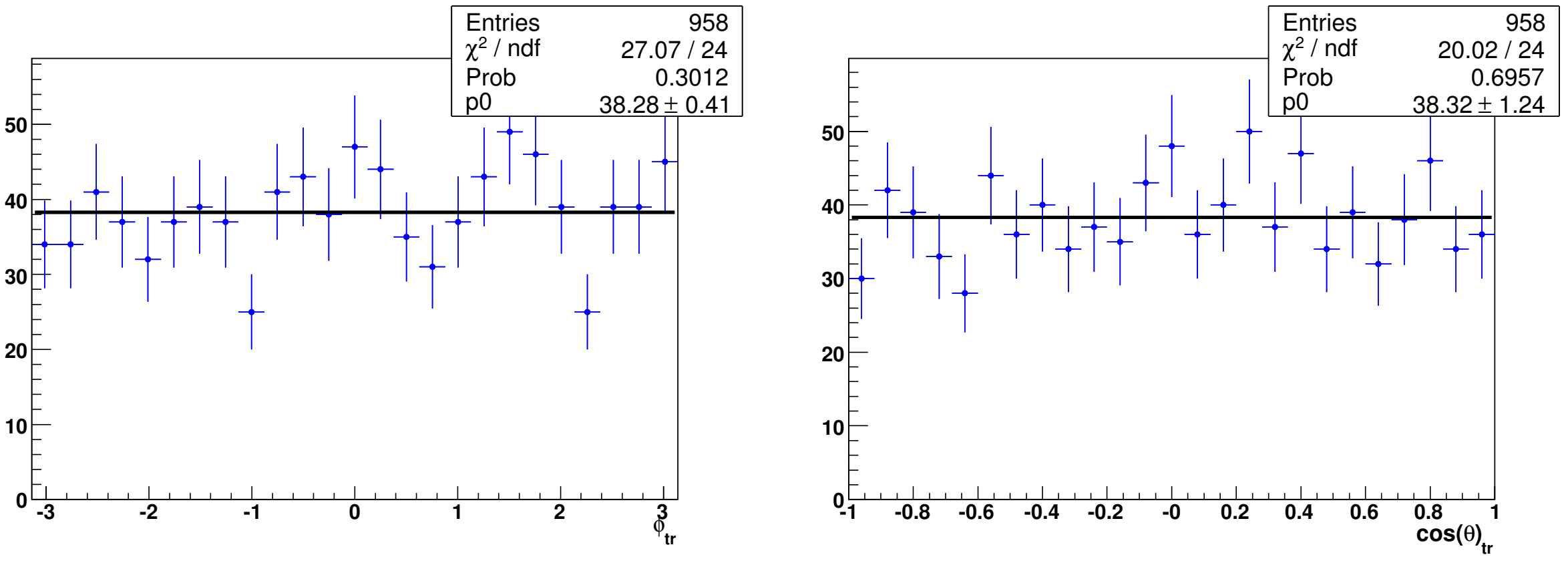

Figure 16: Prompt background for $B_s^0 \to J/\psi\phi$: angular distributions for $\varphi$ and $\cos\theta$ in the transversity basis.

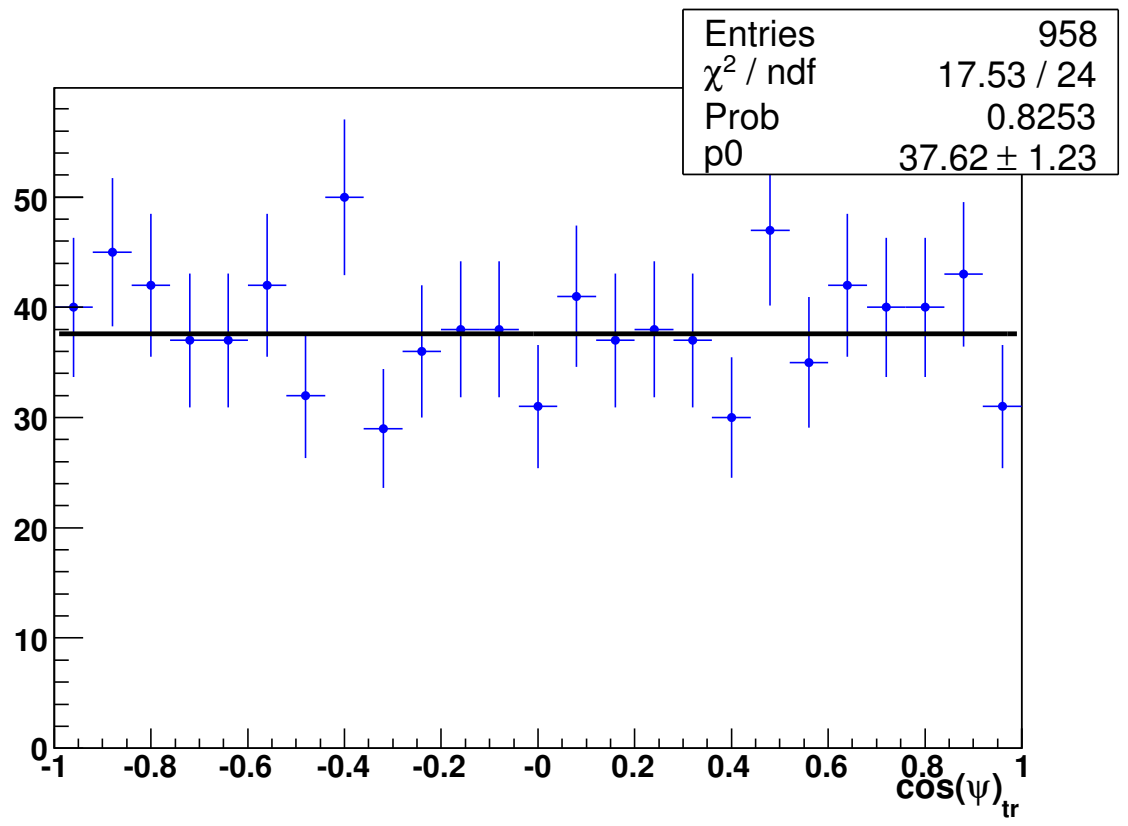

Figure 17: Prompt background for $B_s^0 \to J/\psi\phi$: angular distribution for $\cos\psi$ in the transversity basis.

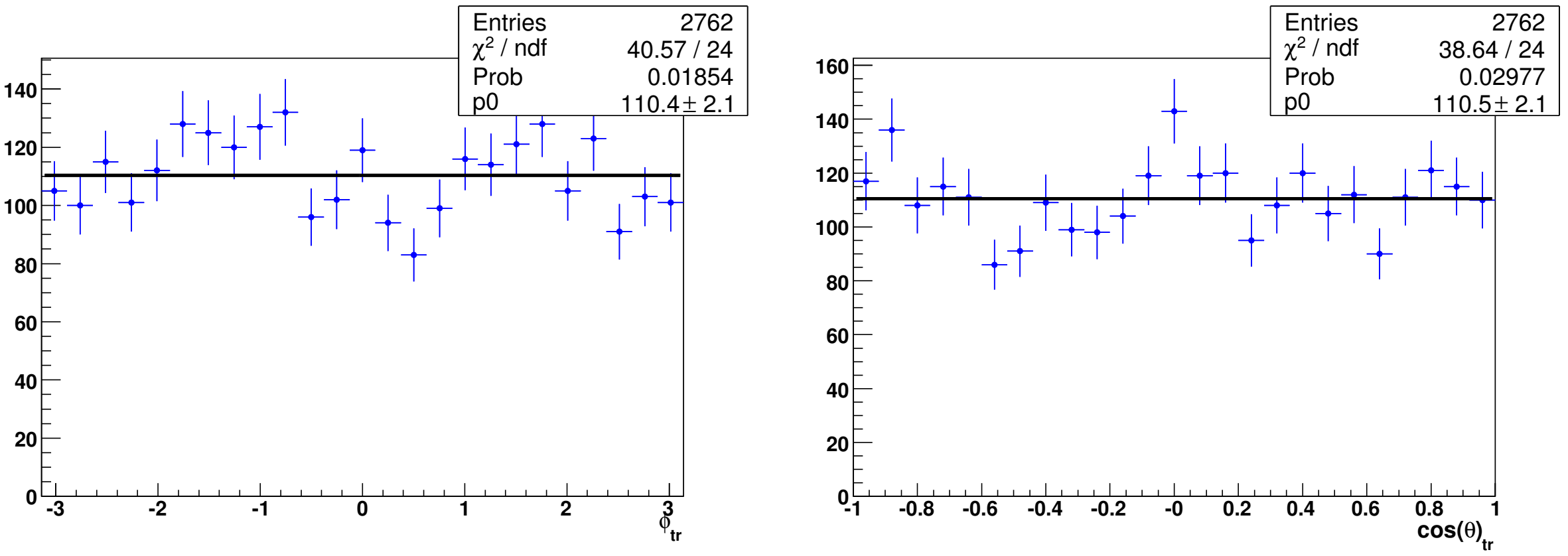

Figure 18: Long-lived background for $B_s^0 \to J/\psi\phi$: angular distributions for $\varphi$ and $\cos\theta$ in the transversity basis.

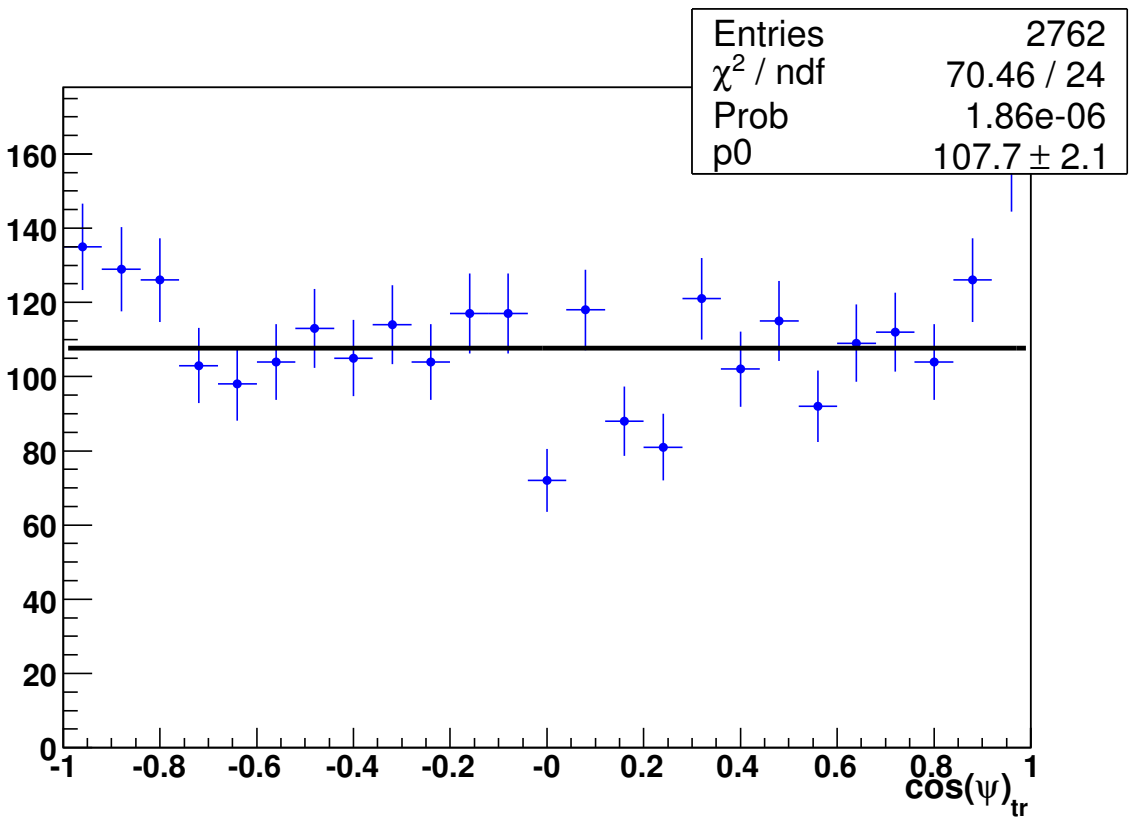

Figure 19: Long-lived background for $B_s^0 \to J/\psi\phi$: angular distribution for $\cos\psi$ in the transversity basis.

# 7 Trigger

A detailed discussion concerning the trigger is beyond the scope of this note but it is useful to review here the general LHCb trigger structure and to discuss which trigger lines are most suitable for the present selection.

As discussed in the previous sections, we have designed a selection which is both proper time unbiased and as similar as possible between the signal $B_s^0 \rightarrow J/\psi\phi$ and its related control channels with $J/\psi$ in the final state, $B^+ \rightarrow J/\psi K^+$ and $B^0 \rightarrow J/\psi K^{*0}$.

In order to maintain this philosophy at trigger level, we need a trigger which does not apply any cut that could bias the lifetime and that treats in the same way all the channels under study. The ideal trigger would be an inclusive and lifetime unbiased trigger on the $J/\psi \rightarrow \mu\mu$.

The LHCb trigger is divided in two levels: Level-0 (L0) and High Level Trigger (HLT). The L0 selects events with high $p_T$ muon or calorimeter objects (hadrons, pions, $\gamma$ and electrons). It reduces the input 40 MHz rate to 1 MHz and it is implemented using custom made electronics. The HLT reduces the L0 output rate to 2 kHz. The HLT selected events are saved on permanent storage (see [33] for details). The High Level Trigger is divided in two parts named HLT1 and HLT2. The purpose of HLT1 is to confirm the Level-0 triggering objects by reconstructing them in the tracking system. HLT1 allows to reduce the rate to about 30 kHz, such that on the remaining events the full pattern recognition, which is time consuming, can be performed. The final reduction is performed by HLT2 which selects event by means of inclusive and exclusive B decay trigger algorithms.

We briefly describe the key features of these three trigger levels, relevant for $B_s^0 \rightarrow J/\psi\phi$ and related control channels.

## 7.1 Level-0 trigger

With the present tuning the L0-trigger accepts events which satisfy any of the following conditions:

- a muon candidate detected in the muon chambers with a $p_T > 1.5\,\text{GeV}/c$ regardless of whether the GEC[7] have fired or not (L0-$\mu$, no GEC);
- a muon candidate detected in the muon chambers with a $p_T > 1.3\,\text{GeV}/c$ and satisfying the GEC (L0-$\mu$, GEC);
- two muon candidates detected in the muon chambers with $|p_{T,1}| + |p_{T,2}| > 1.5\,\text{GeV}/c$ (L0-di$\mu$);
- a hadron detected in the Hadron Calorimeter with $E_T > 3.5\,\text{GeV}$ (L0-hadron);
- an electron or a photon detected in the Electromagnetic Calorimeter (ECAL) with $E_T > 3.5\,\text{GeV}$. The clusters in ECAL are identified as photons (L0-$\gamma$) or electrons (L0-$e^{\pm}$) depending on the information from the SPD.

[7] The Global Event Cuts (GEC) are defined in [32]: the total transverse energy measured in the HCAL should be above 5 GeV, the pile-up and SPD multiplicity smaller than 112 and 280 hits respectively and the number of tracks in the second vertex found by the pile-up system less than 3.

As we have seen in Table 9, the L0-trigger is very efficient ($\varepsilon_{\rm trg} \sim 94\%$) in the phase space covered by selected events. In Table 14 we summarize again the L0 efficiency for the channels under study and we show the relative contributions of each sub-trigger or *line*. The OR of the different lines is used to compute the total efficiency. The column (L0-$\mu$, tot) corresponds to the OR of the muon lines.

We notice that $\sim$99% of the L0-trigger is provided by one of the muon lines, in particular by the L0-$\mu$, no GEC, which provides alone almost 97% of the L0-triggers. This observation is relevant since the HLT1 algorithms are separated into alleys according to the nature of the L0-trigger.

| L0-trigger | $B_s^0 \to J/\psi\phi$ | $B^0 \to J/\psi K^{*0}$ |
|---|---|---|
| $\varepsilon_{\rm L0}/\varepsilon_{\rm sel}$ | $93.9 \pm 0.1$ | $94.8 \pm 0.04$ |
| L0-$\mu$, tot | $98.75 \pm 0.04$ | $98.69 \pm 0.3$ |
| L0-$\mu$, no GEC | $96.70 \pm 0.06$ | $97.07 \pm 0.04$ |
| L0-$\mu$, GEC | $71.7 \pm 0.2$ | $72.5 \pm 0.1$ |
| L0-di$\mu$, no GEC | $69.6 \pm 0.2$ | $70.3 \pm 0.1$ |
| L0-hadron | $4.5 \pm 0.7$ | $5.1 \pm 0.5$ |
| L0-$\gamma$/$e^{\pm}$ | $11.5 \pm 0.5$ | $12.6 \pm 0.2$ |

Table 14: L0-trigger efficiency (%) and relative contribution of each sub-trigger or *lines* for selected $B_s^0 \to J/\psi\phi$ and $B^0 \to J/\psi K^{*0}$ events.

## 7.2 High-Level-Trigger 1

The HLT1 is composed of several trigger paths, named "alleys", which are invoked depending on the nature of the object that fired the L0-trigger: muon, hadron, electron $\gamma$ and $\pi^0$. This section describes the ideas and algorithms used in the HLT1 Muon Alley, and its performance using a sample of about 1 k simulated $B_s^0 \to J/\psi(\mu\mu)\phi(KK)$ and 200 k minimum bias events. More details can be found in [34].

The purpose of the HLT1 Muon Alley is to confirm a L0-$\mu$ candidates using the tracking system. As the L0 can give a decision of a single muon or a dimuon we will also have these two distinct trigger lines in the alley. The HLT1 single-$\mu$ line starts from the confirmation of one L0-$\mu$ candidate. The HLT1 di-$\mu$ line starts from the confirmation either of two L0-$\mu$ candidates or of one L0-di$\mu$ candidate. In addition to those, the HLT1 tries to recover some other di-$\mu$ events made by one L0-$\mu$ candidate plus one additional muon. Indeed, the dimuon events failing the L0-di$\mu$ selection criteria can still be selected by the L0-$\mu$ line.

For confirmed events, the HLT1 muon alley accepts events satisfying any of the following conditions:

- the confirmed L0-$\mu$ candidate has an impact parameter (IP) IP>0.08 mm and a $p_{\rm T}$ >1.3 GeV/$c$ (*lifetime biased single-$\mu$ line*);
- the confirmed L0-$\mu$ candidate has $p_{\rm T}$ >6 GeV/$c$ (*lifetime unbiased single-$\mu$ line*);
- two muons are found with a distance of closest approach (DOCA) DOCA< 0.5 mm and with a $\mu\mu$ invariant mass larger than 2.5 GeV/$c^2$ (*lifetime unbiased di-$\mu$ line*);

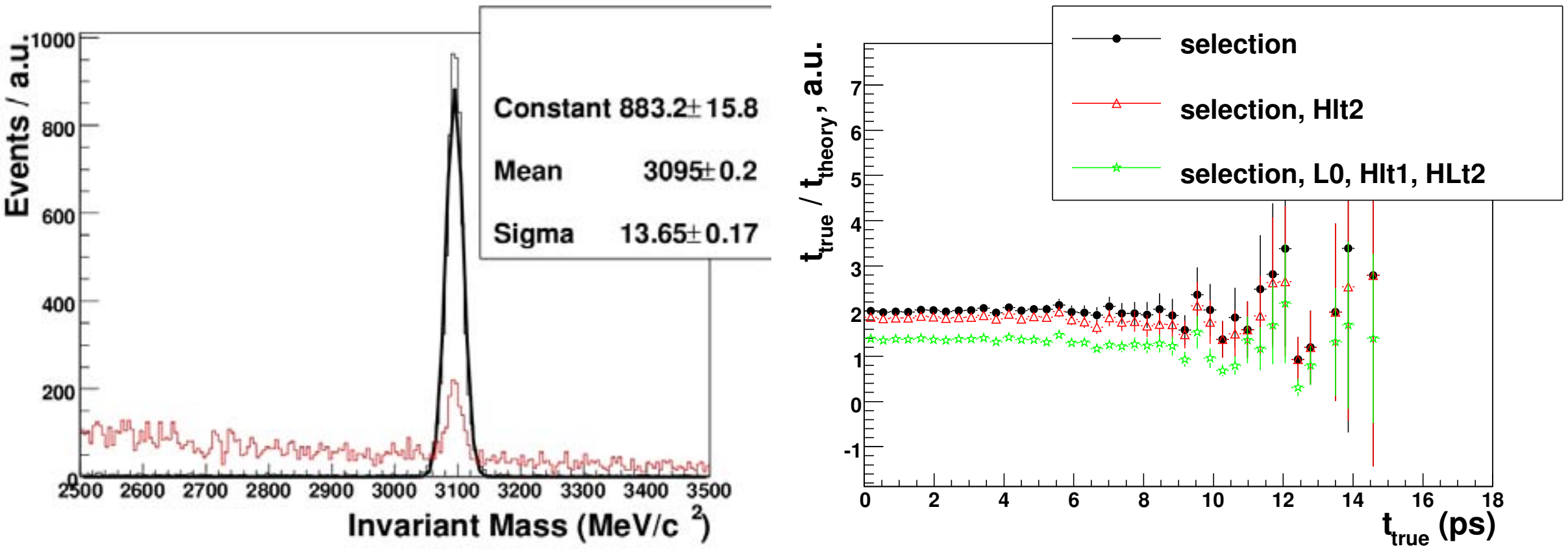

Figure 20: Invariant mass of the J/$\psi$ events for selected signal (black) and minimum bias (red); the normalization is arbitrary. A single Gaussian is fitted to the signal curve (left). The MC proper times of the B normalized to theory (right).

- two muons are found with an impact parameter IP greater than 0.15 mm and an invariant mass larger than 0.5 GeV/$c^2$ (*lifetime biased di-$\mu$ line*).

This set of cuts gives a minimum bias retention of $\sim$17 kHz and an efficiency of 91% on $B_s^0 \to J/\psi\phi$ selected and L0-triggered signal events. This 91% is composed of 78% of lifetime biased single-$\mu$ line, 79.4% of lifetime unbiased di-$\mu$ line and 48.2% of lifetime biased di-$\mu$ line. The baseline HLT1 selection of $B \to J/\psi X$ events considered in this document is performed by the lifetime unbiased di-$\mu$ line. The single-$\mu$ lifetime unbiased line slightly improves the selection efficiency from 79.4% to 81.4%. However, it requires a strong $p_T$ cut on one of the muons, which distorts the angular distributions more than the smaller $p_T$ cuts of the di-$\mu$ line.

## 7.3 High-Level-Trigger 2

At the second level of the HLT, the rate is sufficiently low to reconstruct fully the relevant part of the event (in our case the muons) in the tracking stations, using tracks from the VELO as seeds. We foresee two possible lines for triggering $B_s^0 \to J/\psi(\mu\mu)\phi(KK)$, $B^0 \to J/\psi(\mu\mu)K^{*0}(K\pi)$ and $B^+ \to J/\psi(\mu\mu)K^+$events at the HLT2 level: an inclusive $J/\psi \to \mu\mu$ line and three exclusive trigger selections. The bandwidth given to inclusive and exclusive selections will depend on the running conditions.

### 7.3.1 Inclusive $J/\psi \to \mu\mu$ HLT2 trigger line

The muon track candidates are fitted with a simplified Kalman track fit; both muon candidates must form a common vertex with a sufficiently low $\chi^2$. In addition, a minimum transverse momentum of both tracks is required. The invariant mass of signal events is shown in Figure 20, the resolution is 14 MeV/$c^2$. The background in minimum bias events is shown in red; it consists predominantly of prompt J/$\psi$'s, in the narrow mass window. A set of cuts requiring the same minimum $p_T$ of 500 MeV/$c$ as the offline selection, a narrow mass window ($\pm 3\sigma$) around the nominal J/$\psi$ mass[8] and a rather hard cut on the vertex

[8] A larger mass window is also possible, e.g. pre-scaling the sidebands.

$\chi^2$ (<7) gives a minimum bias rate of about 200 Hz on events which passed the previous two trigger levels. With this, all J/$\psi$ channels under study are selected with an efficiency above 95%, measured on offline selected events which pass the previous trigger levels.

These events are triggered without applying cuts on impact parameter with respect to the primary vertex. The true MC proper time distribution of the reconstructed $B^0_s$, normalized to theory, is shown in Figure 20. The black histogram shows the normalized proper times for selected events. In the red histogram, the HLT2 unbiased-J/$\psi(\mu\mu)$ selection is additionally required, and in the green histogram the full trigger chain is required (L0, HLT1 (unbiased di-muon), HLT2 (unbiased-J/$\psi(\mu\mu)$)). All three curves are flat, so there is no bias in the lifetime introduced by the trigger cuts.

With the present tuning the total trigger efficiency for selected $B^0_s \to J/\psi\phi$ signal events at the end of the HLT2 $J/\psi \to \mu\mu$ inclusive trigger line is given by:

$$\varepsilon_{\rm trig} = \varepsilon_{\rm L0} \times \varepsilon_{\rm HLT1} \times \varepsilon_{\rm HLT2,inclusive} = 93.9\% \times 80\% \times 95\% \sim 70\%$$

where $\varepsilon_{\rm L0}$ is the efficiency of the L0-trigger with respect to offline selected events, $\varepsilon_{\rm HLT1}$ is the efficiency of the HLT1 lifetime unbiased dimuon alley with respect to offline selected and L0-triggered events and $\varepsilon_{\rm HLT2,inclusive}$ is the efficiency of the inclusive $J/\psi \to \mu\mu$ trigger line with respect to offline selected, L0 and HLT1 accepted events. These are the baseline HLT efficiencies assumed throughout this document.

#### 7.3.2 Exclusive $B^0_s \to J/\psi\phi$, $B^0 \to J/\psi K^{*0}$ and $B^+ \to J/\psi K^+$ HLT2 trigger lines

In case the minimum bias rate is much higher than expected from simulation we might be forced to use exclusive trigger selections for $B^0_s \to J/\psi\phi$, $B^0 \to J/\psi K^{*0}$ and $B^+ \to J/\psi K^+$ modes by reconstructing all the tracks in the final state and using cuts on the intermediate resonance masses and the final B meson mass. A detailed study is described in [35]. The conclusions of this work is that we can have a total rate of minimum bias events for the three channels of $\sim$33 Hz ($\sim$18 Hz) without prescaling (with prescaling) events in the sidebands of the B meson mass spectra while keeping the efficiency on signal events $\varepsilon_{\rm HLT2,exclusive} \sim 77\%$.

With the present tuning the total trigger efficiency for $B^0_s \to J/\psi\phi$ channel at the end of the HLT2 $B^0_s \to J/\psi\phi$ exclusive trigger line is:

$$\varepsilon_{\rm trig} = \varepsilon_{\rm L0} \times \varepsilon_{\rm HLT1} \times \varepsilon_{\rm HLT2,exclusive} = 93.9\% \times 80\% \times 77\% \sim 57.8\%.$$

# 8 Flavour tagging

## 8.1 Flavour tagging procedure and performance

The identification of the initial flavour of the reconstructed B mesons, which is necessary for the CP asymmetry measurement, is performed at LHCb by several flavour tagging algorithms as described in [36] and [37]. Opposite-side tags (muon, electron, kaon and inclusive secondary vertex) and same-side pion or kaon tags can be used.

The tuning of these algorithms and the measurement of the associated mistag rate is performed on control channels with flavour-specific final states. For the opposite-side tags, and for the same-side pion tag, $\mathrm{B^+}$ and $\mathrm{B^0}$ control channels can be used, while only $\mathrm{B_s^0}$ channels can be considered for the same-side kaon tag. Since in $\mathrm{B_s^0}$ channels there will be less events available than in the $\mathrm{B^+}$ and $\mathrm{B^0}$ channels, and the proper time resolution will play a more important role, the first tags to be tuned will be the opposite-side ones.

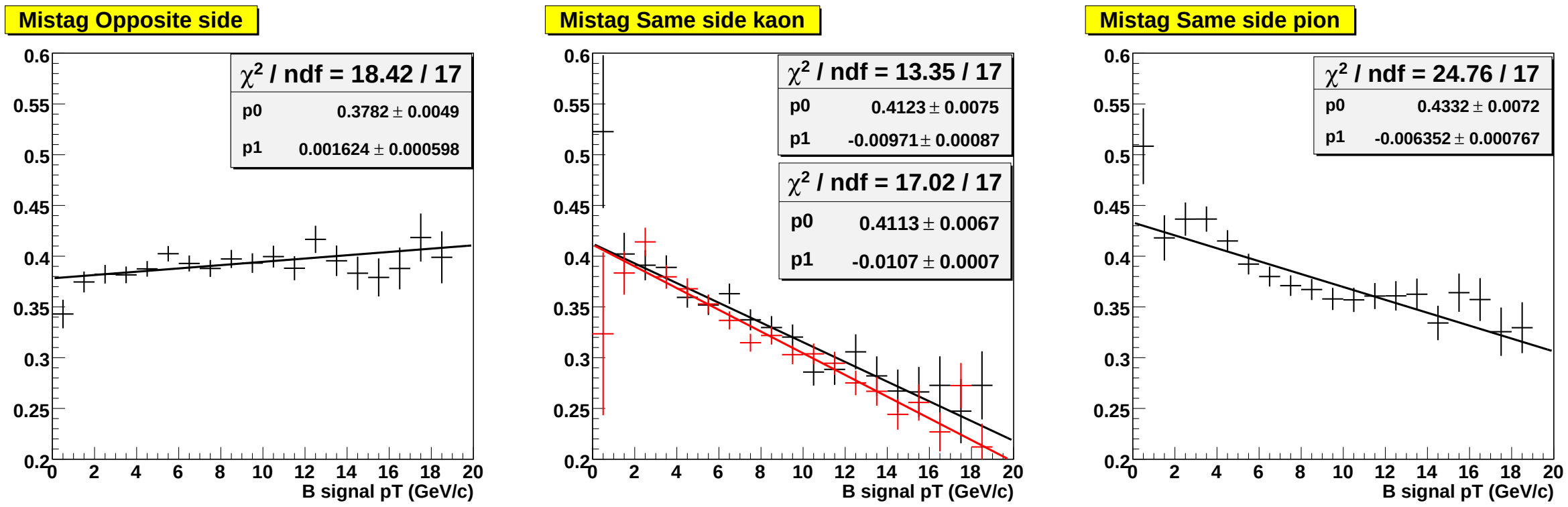

Figure 21: Mistag rate as a function of the signal B transverse momentum. Left: combination of all opposite-side tags in $\mathrm{B_s^0} \to \mathrm{J}/\psi\phi$ events. Centre: same-side kaon tag in $\mathrm{B_s^0} \to \mathrm{J}/\psi\phi$ events (black, upper parameter box) and superimposed $\mathrm{B_s^0} \to \mathrm{D_s^-}\pi^+$ events (red, lower parameter box). Right: same-side pion tag in $\mathrm{B^+} \to \mathrm{J}/\psi\mathrm{K^+}$ events.

First results on flavour tagging will probably come from the observation of flavour oscillation in $\mathrm{B^0} \to \mathrm{D^{*-}}\mu^+\nu$, which is the exclusive B channel with the highest yield in LHCb. However, opposite-side tagging performance is not completely independent of the signal channel, since trigger and event selection can affect the opposite b-hadron kinematic distributions. As an example, in Figure 21 the dependence of the mistag rate on the signal B transverse momentum is shown, for the combination of all opposite-side tags and for same-side kaon tag, in $\mathrm{B_s^0} \to \mathrm{J}/\psi\phi$ events, and for same-side pion tag, in $\mathrm{B^+} \to \mathrm{J}/\psi\mathrm{K^+}$ events. A linear fit is superimposed, to show the trend.

Therefore the final calibration of the opposite-side tags performance for the $\mathrm{B_s^0} \to \mathrm{J}/\psi(\mu\mu)\phi(\mathrm{KK})$ channel and for $\mathrm{B^0} \to \mathrm{J}/\psi(\mu\mu)\mathrm{K_S^0}(\pi\pi)$ channel will be performed by using two similar control channels: $\mathrm{B^+} \to \mathrm{J}/\psi(\mu\mu)\mathrm{K^+}$ and $\mathrm{B^0} \to \mathrm{J}/\psi(\mu\mu)\mathrm{K^{*0}}(\mathrm{K}\pi)$ . All these channels are triggered mainly by the $\mathrm{J}/\psi(\mu\mu)$ part, and the event selection criteria described in Section 6 have been tuned in order to minimize the differences among the three channels.

In a sample of events, the tagging efficiency, the mistag rate and the effective efficiency are defined as:

$$\varepsilon = \frac{W+R}{R+W+U}, \qquad \omega = \frac{W}{R+W}\,, \qquad \varepsilon_{\mathrm{eff}} = \varepsilon(1-2\omega)^2\,, \tag{67}$$

where $R$, $W$ and $U$ are the number of correctly tagged, incorrectly tagged and untagged events, respectively. For $N$ exclusive samples, the tagging efficiencies and the tagging effective efficiencies can be summed to obtain the total *combined* values:

$$\varepsilon_{\mathrm{tag}}^{\mathrm{comb}} = \sum_{k=1}^{N} \varepsilon_k, \qquad \varepsilon_{\mathrm{eff}}^{\mathrm{comb}} = \sum_{k=1}^{N} \varepsilon_k(1-2\omega_k)^2\,. \tag{68}$$

A *combined* mistag rate can be calculated *a posteriori* as:

$$\omega^{\text{comb}} = \left(1 - \sqrt{\varepsilon_{\text{eff}}^{\text{comb}} / \varepsilon_{\text{tag}}^{\text{comb}}}\right) / 2 \,. \tag{69}$$

This quantity is a figure of merit that corresponds to the mistag rate of a single sample of events having the same effective efficiency. When the samples are obtained by grouping events with similar mistag rate, higher *combined* effective efficiency, and lower *combined* mistag rate, are expected with respect to having all events in a single sample.

Performance of flavour tagging on $B_s^0 \to J/\psi(\mu\mu)\phi(KK)$, $B^+ \to J/\psi(\mu\mu)K^+$ and $B^0 \to J/\psi(\mu\mu)K^{*0}(K\pi)$ events, selected as described in Section 6, are shown in Tables 15, 16 and 17, respectively. The L0 trigger is required in all cases. Only reconstructed signal events with tracks matching to the generated Monte Carlo ones are used. Values in these tables are calculated comparing the tagging results with the b-flavour determined from MC truth. The first part of the tables contains the performance for each tag alone. The second and third parts of the tables contain the performance after sorting all events, as explained in detail below, into five exclusive samples of increasing tagging purity. The *combined* performance shown in the tables are obtained by summing over these five exclusive samples. The *average* performance reported in the tables are those obtained for all events not sorted into samples.

Numbers in the tables show that there is indeed a good general agreement in the tagging performance among control channels and signal channel, when one isolates the contribution of the OS tags.

In the tables, b and $\overline{\text{b}}$ events are merged. Differences between flavour tagging performance for b and $\overline{\text{b}}$ mesons can arise from production or detection asymmetries. The production asymmetry between b and $\overline{\text{b}}$ meson is estimated to be $\sim$1% [38]. It is discussed in [12]. Detection asymmetries are a consequence, for example, of the different interaction cross section for positive and negative kaons and charge asymmetries in the track curvature produced by the magnetic field. The observed difference in the mistag rate for b and $\overline{\text{b}}$ in the Monte Carlo data is not significant with the available sample sizes [39] and will be below the statistical error on the mistag rate for 2 $\text{fb}^{-1}$. The mistag rates will be measured separately on b and $\overline{\text{b}}$ events in the data in flavour-specific control channels, and can be explicitly accounted for in CP fits. In Table 18 we show the tagging efficiency for the prompt and long-lived background components: the long-lived component has a tagging efficiency comparable with the signal while the prompt one, dominated by combinatorics, has a much lower tagging efficiency, of the order of $\sim$30%.

As discussed in [37], the combination of the tagging information can be performed in different ways. We describe only one of them here.

As described in [37], a mistag probability $\eta_i$ can be assigned to each tag $i$ as a function of several kinematic properties of the tag itself and of the full event[9], merged together into a single variable by using a neural net. The training of the neural net is performed on simulated data. In the learning phase the true flavour of the B meson is used to define the combination of variables which maximize the separation between correctly tagged and incorrectly tagged events. The probability of mistag $\eta_i$ is then determined as a linear function of the neural net output. When more than one tag is available per event, the tag probabilities are combined into a global probability of mistagging the event, as explained in [37]. This event probability can be used to sort events into sub-samples of increasing tagging purity, or directly in the CP asymmetry fits. Performances shown in the second and third parts of Tables 15, 16 and 17 correspond to sorting into five exclusive samples of increasing tagging purity and to their sum.

[9]For example, for the opposite-side muon tag, the neural net combines: the muon momentum, the muon transverse momentum, the significance of the muon impact parameter with respect to the primary vertex, the track multiplicity in the event and the signal B transverse momentum.

| $B_s^0 \to J/\psi(\mu\mu)\phi(KK)$ | | | |
|---|---|---|---|
| | $\varepsilon_{tag}(1-2\omega)^2$ % | $\varepsilon_{tag}$ % | $\omega$ % |
| Individual tags | | | |
| $\mu$ | $0.76 \pm 0.05$ | $5.77 \pm 0.08$ | $31.9 \pm 0.6$ |
| e | $0.38 \pm 0.04$ | $2.91 \pm 0.06$ | $32.0 \pm 0.9$ |
| $K_{opp}$ | $1.25 \pm 0.07$ | $15.06 \pm 0.12$ | $35.6 \pm 0.4$ |
| $K_{same}$ | $2.39 \pm 0.10$ | $26.37 \pm 0.15$ | $34.9 \pm 0.3$ |
| $Q_{vtx}$ | $1.09 \pm 0.07$ | $44.35 \pm 0.17$ | $42.1 \pm 0.2$ |
| Combination of opposite-side tags only | | | |
| Average | $2.18 \pm 0.10$ | $45.61 \pm 0.17$ | $39.07 \pm 0.24$ |
| **Combined** | $\varepsilon_{eff}^{comb}$=**3.32 ± 0.11** | $\varepsilon_{tag}^{comb}$=**45.61 ± 0.17** | $\omega^{comb}$=**36.51 ± 0.24** |
| Combination of all tags | | | |
| Average | $4.45 \pm 0.14$ | $55.71 \pm 0.17$ | $35.88 \pm 0.21$ |
| **Combined** | $\varepsilon_{eff}^{comb}$=**6.23 ± 0.15** | $\varepsilon_{tag}^{comb}$=**55.71 ± 0.17** | $\omega^{comb}$=**33.27 ± 0.21** |

Table 15: Results of flavour tagging obtained for $B_s^0 \to J/\psi\phi$ events passing Level-0, for the individual tags and for their combination. Average: result from the global tagging decision for all events together. Combined: results after splitting into the 5 categories and summing the effective efficiencies. Uncertainties are due to the Monte Carlo statistics.

In order to check the estimation of the probability of mistag on real data, its distributions are studied on a control channel with a full model which takes into account signal and background events. Assuming for simplicity a linear dependence of the mistag rate in real data with respect to the mistag rate calculated using the coefficients from simulation, two correction factors are defined for each tag as:

$$\omega_i(\eta_i) = p0_i + p1_i \times (\eta_i - \bar{\eta}_i)\,. \tag{70}$$

In order to decrease the correlation between each couple of parameters $p0_i$ and $p1_i$ the linear dependence on $\eta_i$ has been centered around the average value of $\bar{\eta}_i$. These correction factors $p0_i, p1_i$ determined on the control channel, can be used in other channels to estimate correctly the probability of mistag for that tag.

The $B^+ \to J/\psi K^+$ control channel is used to fit the correction factors from data, as described in Section 8.2; more details are available in [39].

In the next step, these correction factors are used in the $B^0 \to J/\psi K^{*0}$ channel to sort events into five samples, in each sub-sample a fit to mixing oscillation as a function of proper-time is performed. The results of these fits provide a check of the tagging procedure and a direct measurement of the mistag rate for $B^0 \to J/\psi K^{*0}$ events in each of the five categories.

The same factors can also be used to provide an opposite-side probability of mistag per event in the $B_s^0 \to J/\psi\phi$ channel to sort events into five samples of increasing tagging purity. In each sub-sample the mistag rate will be either measured directly on data (as can be done in the complete three angle fit) or considered as an input parameter. In this case the value is fixed to the one measured either in the $B^+ \to J/\psi K^+$ or in the $B^0 \to J/\psi K^{*0}$ channel. A fit using the probability of mistag per event can also be envisaged.

| $B^+ \to J/\psi(\mu\mu)K^+$ | | | |
|---|---|---|---|
| | $\varepsilon_{tag}(1-2\omega)^2$ % | $\varepsilon_{tag}$ % | $\omega$ % |
| Individual tags | | | |
| $\mu$ | $0.76 \pm 0.04$ | $5.42 \pm 0.06$ | $31.3 \pm 0.5$ |
| e | $0.37 \pm 0.03$ | $2.71 \pm 0.04$ | $31.5 \pm 0.8$ |
| $K_{opp}$ | $1.61 \pm 0.07$ | $14.15 \pm 0.09$ | $33.1 \pm 0.3$ |
| $\pi_{same}$ | $1.14 \pm 0.05$ | $19.14 \pm 0.11$ | $37.8 \pm 0.3$ |
| $Q_{vtx}$ | $0.97 \pm 0.05$ | $42.36 \pm 0.13$ | $42.4 \pm 0.2$ |
| Combination of opposite-side tags only | | | |
| Average | $2.23 \pm 0.08$ | $43.63 \pm 0.13$ | $38.69 \pm 0.20$ |
| **Combined** | $\varepsilon_{eff}^{comb}$=$\mathbf{3.35 \pm 0.09}$ | $\varepsilon_{tag}^{comb}$=$\mathbf{43.63 \pm 0.13}$ | $\omega^{comb}$=$\mathbf{36.15 \pm 0.20}$ |
| Combination of all tags | | | |
| Average | $3.21 \pm 0.09$ | $52.76 \pm 0.14$ | $37.67 \pm 0.18$ |
| **Combined** | $\varepsilon_{eff}^{comb}$=$\mathbf{4.45 \pm 0.10}$ | $\varepsilon_{tag}^{comb}$=$\mathbf{52.76 \pm 0.14}$ | $\omega^{comb}$=$\mathbf{35.48 \pm 0.18}$ |

Table 16: Results of flavour tagging obtained for $B^+ \to J/\psi K^+$ events passing Level-0, for the individual tags, for categories and for their combination. Average: result from the global tagging decision for all events together. Combined: results after splitting into the 5 categories and and summing the effective efficiencies. Uncertainties are due to the Monte Carlo statistics.

The flavour of $B^0$ and $B^+$ mesons can also be determined by a same-side pion tag. This tag is calibrated from the channels $B^+ \to J/\psi K^+$ and $B^0 \to J/\psi K^{*0}$, in the same manner as the opposite-side tags. The combined probability of mistag including same-side pion tag will be used for example in the flavour tagging of $B^0 \to J/\psi(\mu\mu)K^0_S(\pi\pi)$ events for the sin2$\beta$ measurement. Flavour tagging performances, as determined from MC truth, for the $B^0 \to J/\psi(\mu\mu)K^0_S(\pi\pi)$ channel are shown in [61].

On the other hand, the same-side kaon tag, which is specific to $B^0_s$ mesons, will be tuned using the $B^0_s \to D^-_s\pi^+$ control channel. The possibility to use double tag events with the high statistics $B^0_s \to D^-_s\mu^+\nu$ channel has also been considered in [40]. The tagging performance in the $B^0_s \to D^-_s\pi^+$ channel, as determined from MC truth, is shown in Appendix D of [39]. A cheated selection has been used for these events, which requires all the signal tracks being reconstructed, without imposing additional cuts. The tagging performance is improved by the trigger requirements and the mistag rate in $B^0_s \to D^-_s\pi^+$ L0-triggered events is smaller than in $B^0_s \to J/\psi\phi$ events. However, for the same-side kaon tag, the main difference in mistag rate is related to the different momentum spectra of the reconstructed B meson. In Figure 21 the two channels show the same dependence. In order to get mistag values compatible with $B^0_s \to J/\psi\phi$ to be used as inputs parameters in CP fits, the mistag rate measured in $B^0_s \to D^-_s\pi^+$ events have to be re-weighted, as discussed in [41]. $B^0_s \to D^-_s\pi^+$ events can also be used to determine the probability of mistag $\eta_i$ for the same-side kaon tag, as a function of the event variables, in a similar way as for the other tags. A combined mistag per event including all tags can be calculated [42].

| $B^0 \to J/\psi(\mu\mu)K^{*0}(K\pi)$ | | | |
|---|---|---|---|
| | $\varepsilon_{\text{tag}}(1-2\omega)^2$ % | $\varepsilon_{\text{tag}}$ % | $\omega$ % |
| Individual tags | | | |
| $\mu$ | $0.78 \pm 0.05$ | $5.54 \pm 0.07$ | $31.2 \pm 0.6$ |
| e | $0.39 \pm 0.03$ | $2.69 \pm 0.05$ | $30.9 \pm 0.8$ |
| $K_{\text{opp}}$ | $1.63 \pm 0.07$ | $14.11 \pm 0.10$ | $33.0 \pm 0.4$ |
| $\pi_{\text{same}}$ | $1.16 \pm 0.06$ | $20.35 \pm 0.12$ | $38.1 \pm 0.3$ |
| $Q_{\text{vtx}}$ | $1.07 \pm 0.06$ | $42.73 \pm 0.14$ | $42.1 \pm 0.2$ |
| Combination of opposite-side tags only | | | |
| Average | $2.28 \pm 0.09$ | $43.59 \pm 0.14$ | $38.62 \pm 0.21$ |
| **Combined** | $\varepsilon_{\text{eff}}^{\text{comb}}$=**3.45 ± 0.10** | $\varepsilon_{\text{tag}}^{\text{comb}}$=**43.95 ± 0.14** | $\omega^{\text{comb}}$=**36.00 ± 0.21** |
| Combination of all tags | | | |
| Average | $3.25 \pm 0.10$ | $53.60 \pm 0.15$ | $37.69 \pm 0.19$ |
| **Combined** | $\varepsilon_{\text{eff}}^{\text{comb}}$=**4.52 ± 0.11** | $\varepsilon_{\text{tag}}^{\text{comb}}$=**53.60 ± 0.15** | $\omega^{\text{comb}}$=**35.48 ± 0.19** |

Table 17: Results of flavour tagging obtained for $B^0 \to J/\psi K^{*0}$ events passing Level-0, for the individual tags, for categories and for their combination. Average: result from the global tagging decision for all events together. Combined: results after splitting into the 5 categories and summing the effective efficiencies. Uncertainties are due to the Monte Carlo statistics.

| Channel | Prompt Background $\varepsilon_{\text{tag}}^{\text{Pr}}$ [%] | Long-lived background $\varepsilon_{\text{tag}}^{\text{LL}}$ [%] |
|---|---|---|
| $B_s^0 \to J/\psi\phi$ | $27.0 \pm 0.2$ | $50.0 \pm 0.6$ |
| $B^0 \to J/\psi K^{*0}$ | $33.5 \pm 0.6$ | $46.1 \pm 0.2$ |
| $B^+ \to J/\psi K^+$ | $26.1 \pm 0.6$ | $56.9 \pm 0.3$ |

Table 18: Tagging efficiency for $B_s^0 \to J/\psi\phi$, $B^0 \to J/\psi K^{*0}$ and $B^+ \to J/\psi K^+$candidates in prompt and long-lived background, after L0-trigger and selection.

## 8.2 Flavour tagging calibration with $B^+ \to J/\psi(\mu\mu)K^+$ events

The mistag rate can be measured in real data in the $B^+ \to J/\psi K^+$ channel by comparing the charge of the reconstructed B meson, as given by the kaon charge, and the decision of the tagging algorithms. One relevant issue is the subtraction of the background contribution, in particular when background events consist of partially reconstructed B mesons whose reconstructed charge is correlated to the true b flavour. For these events the tagging answer is not random and the associated mistag rate is different from 50%. In order to separate signal from background events both the mass and the time distributions are used. In the model developed the total PDF is the sum of three PDFs for signal, prompt background and non-prompt background, respectively. The observables are the B candidate mass $m$, the B candidate proper-time $t$ and the tagging decision $d_i$ of each considered tag $i$. The tagging decision is: $d_i = +1$ for a right tag, $d_i = -1$ for a wrong tag, and $d_i = 0$ for no tag, respectively. The parametrization of the PDF is inspired by results on simulated data. For the mass distribution a double Gaussian is used for signal and an exponential for background. The time distribution is described by an exponential con-voluted with a Gaussian resolution for signal and the long-lived background component, while a simple Gaussian is used for the prompt background component.

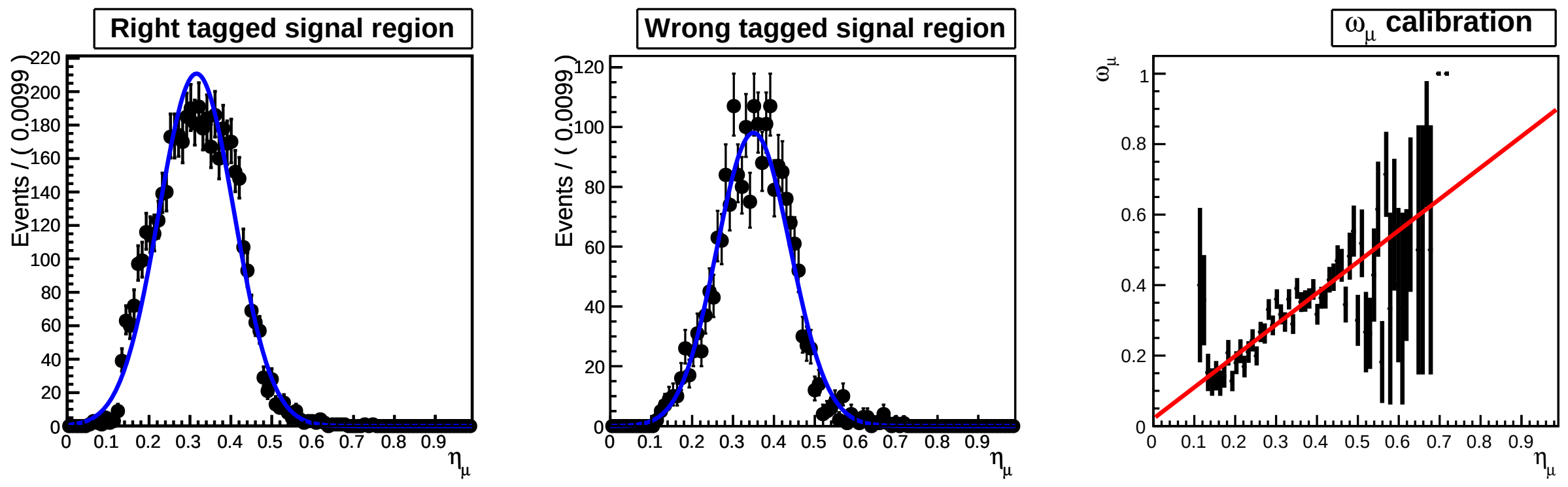

Figure 22: Mistag distribution for correctly (left) and wrongly (centre) muon tagged $B^+ \to J/\psi(\mu\mu)K^+$ signal events. On the right the dependence of the measured mistag on the probability of mistag is represented. The best fit calibration curve $\omega_\mu(\eta_\mu) = p0_\mu + p1_\mu(\eta_\mu - \bar{\eta}_\mu)$ is superimposed.

For the signal, for each tag, the tagging part of the PDF is written, in case of procedure B), as:

$$\mathcal{S}_i^{\text{tag}}(d_i, \eta_i) = \begin{cases} \epsilon_i^s \times \{1 - (p0_i + p1_i \times (\eta_i - \overline{\eta_i}))\} \times \mathcal{N}(\eta_i) & \text{if } d_i = 1, \text{ right tag,} \\ \epsilon_i^s \times \{p0_i + p1_i \times (\eta_i - \overline{\eta_i})\} \times \mathcal{N}(\eta_i) & \text{if } d_i = -1, \text{ wrong tag,} \\ \{1 - \epsilon_i^s\} & \text{if } d_i = 0, \text{ untag,} \end{cases} \tag{71}$$

where $\mathcal{N}(\eta_i)$ is the distribution of the probability of mistag, which is modeled with a Gaussian for the electron and the muon tags, and with a histogram PDF for the kaon, the vertex charge and the pion tags, on account of these variables having distributions which cannot be conveniently described by standard functions. $\epsilon_i^s$ is the signal tagging efficiency for tag $i$. In an early stage, we can simplify the procedure by setting $p1_i = 0$.

For the background, the PDF is an expression similar to the signal one, but the mistag rates $\omega_i^{\text{LL/Pr}}$ are assumed to be constant, set to 0.50 for the prompt component and left as a free parameter for the non-prompt one. The tagging efficiencies are different in the prompt and non prompt background components.

An unbinned extended likelihood fit is performed. The fit procedure has been checked first on signal Monte Carlo data, using up to five tags together. The results on the calibration parameters $p0_i$ and $p1_i$ are in agreement with the ones that can be directly calculated from the Monte Carlo truth information. An example of the fit to the mistag distribution of Monte Carlo signal events is shown in Figure 22.

Given the relatively small number of background events available from full simulation, a toy Monte Carlo has been used to fully test the model. We generated and fitted about 300 samples of signal and background events corresponding to a statistics of about $0.1\,\text{fb}^{-1}$. The input value of the parameters for the toy for the signal coincide with the best fit parameters of the Monte Carlo events, while for the background they are close to the parameters that fit the Monte Carlo events in [30]. The pull distributions are as expected for all parameters, except a few cases where the fit errors are overestimated. For the muon tag the statistical sensitivity on $p0$ and $p1$ is 0.8% and 2.4%, respectively, for $2\,\text{fb}^{-1}$.

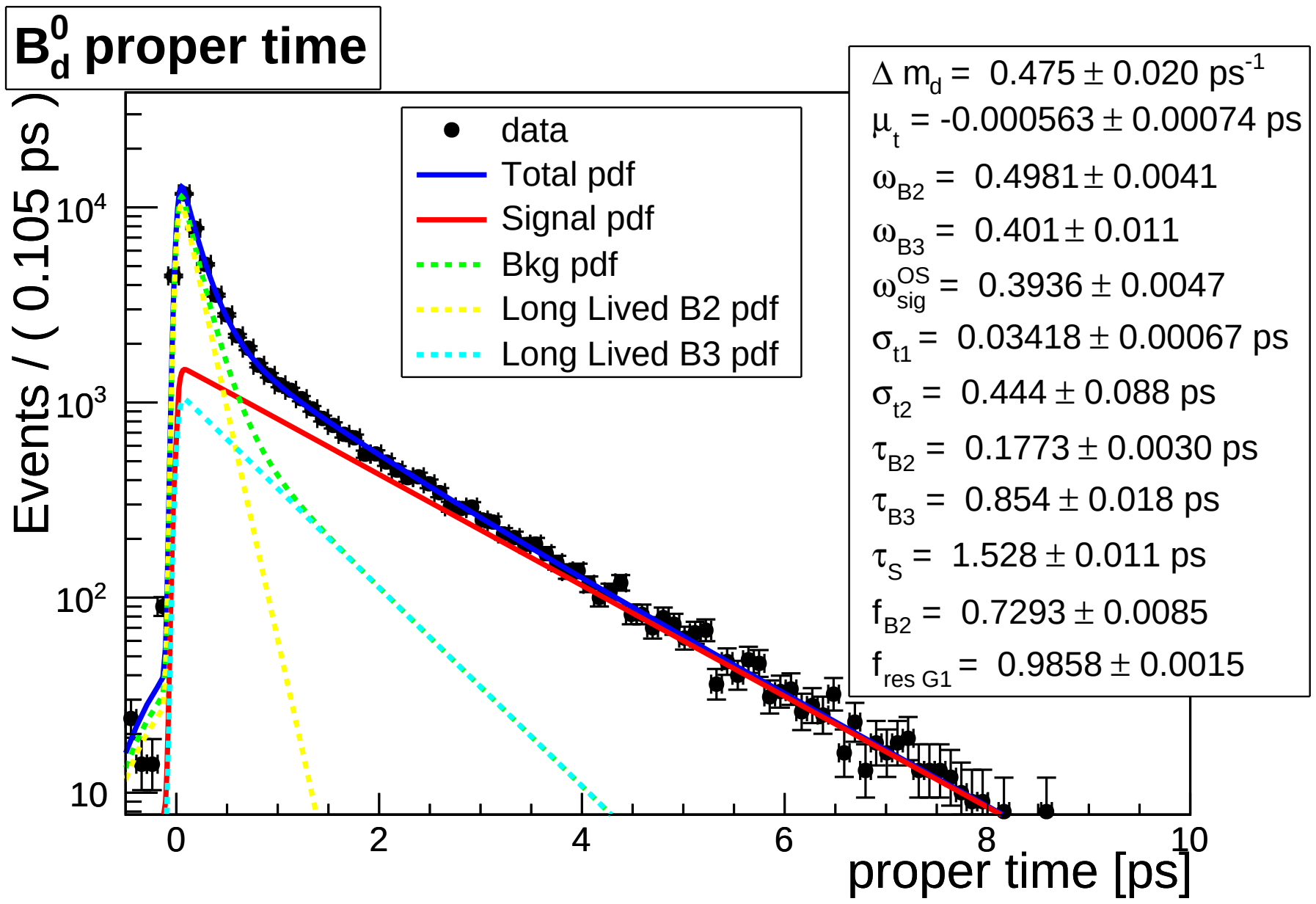

Figure 23: Proper time fit of the $B_{u,d} \to J/\psi X$ selected sample. In the box the output parameters of the fit are given.

## 8.3 Flavour tagging calibration with $B^0 \to J/\psi(\mu\mu)K^{*0}(K\pi)$ events

The mistag rate can be measured in real data in the $B^0 \to J/\psi(\mu\mu)K^{*0}(K\pi)$ channel through a fit of the flavour oscillations of the $B^0$ mesons as a function of proper time. The flavour of the $B^0$ meson at production is determined from the tagging algorithms, while the flavour at decay is determined by the $K^{*0}$ flavour, which is in turn defined by the kaon charge.

The event selection described in Section 6 is used for this study. In this case a large fraction of the background events is due to prompt $J/\psi$ production where no dependence of the reconstructed B flavour on the reconstructed proper time is expected. However part of the background is due to mis-reconstructed $B^0$ events, where a dependence on proper time is expected. The mistag rate is also expected to be different in the various background components. In order to separate signal from background events both the mass and the time distribution are used. Only flavour tagged events are used. The PDF used to describe the $B^0 \to J/\psi K^{*0}$ data is the sum of signal and background PDFs.

The observables are the $B^0$ candidate mass $m$, which is the invariant mass of the $\mu\mu K\pi$ system (in $\mathrm{GeV}/c^2$), the $B^0$ candidate reconstructed proper-time $t$ (in ps) and the measured mixing state $q$ ($q = +1$ for unmixed and $q = -1$ for mixed state). The signal PDF is given by:

$$\mathcal{S}(m,t,q) = \left\{ \frac{1}{2\tau_{B^0}} e^{-\frac{t}{\tau_{B^0}}} \left[ 1 + q(1-2\omega_{OS}^{Sig}) \cos(\Delta m_d t) \right] \right\} \otimes R(t;\mu_t,\sigma_{t1},\sigma_{t2},f_{res\ G1}) \times G(m;\mu_{G1},\sigma_{G1}) \,, \quad (72)$$

where $\tau_{B^0}$ is the $B^0$ lifetime, $\omega_{OS}^{Sig}$ is the mistag fraction and $\Delta m_d$ is the oscillation frequency. The proper-time resolution function is given by a double Gaussian while the mass

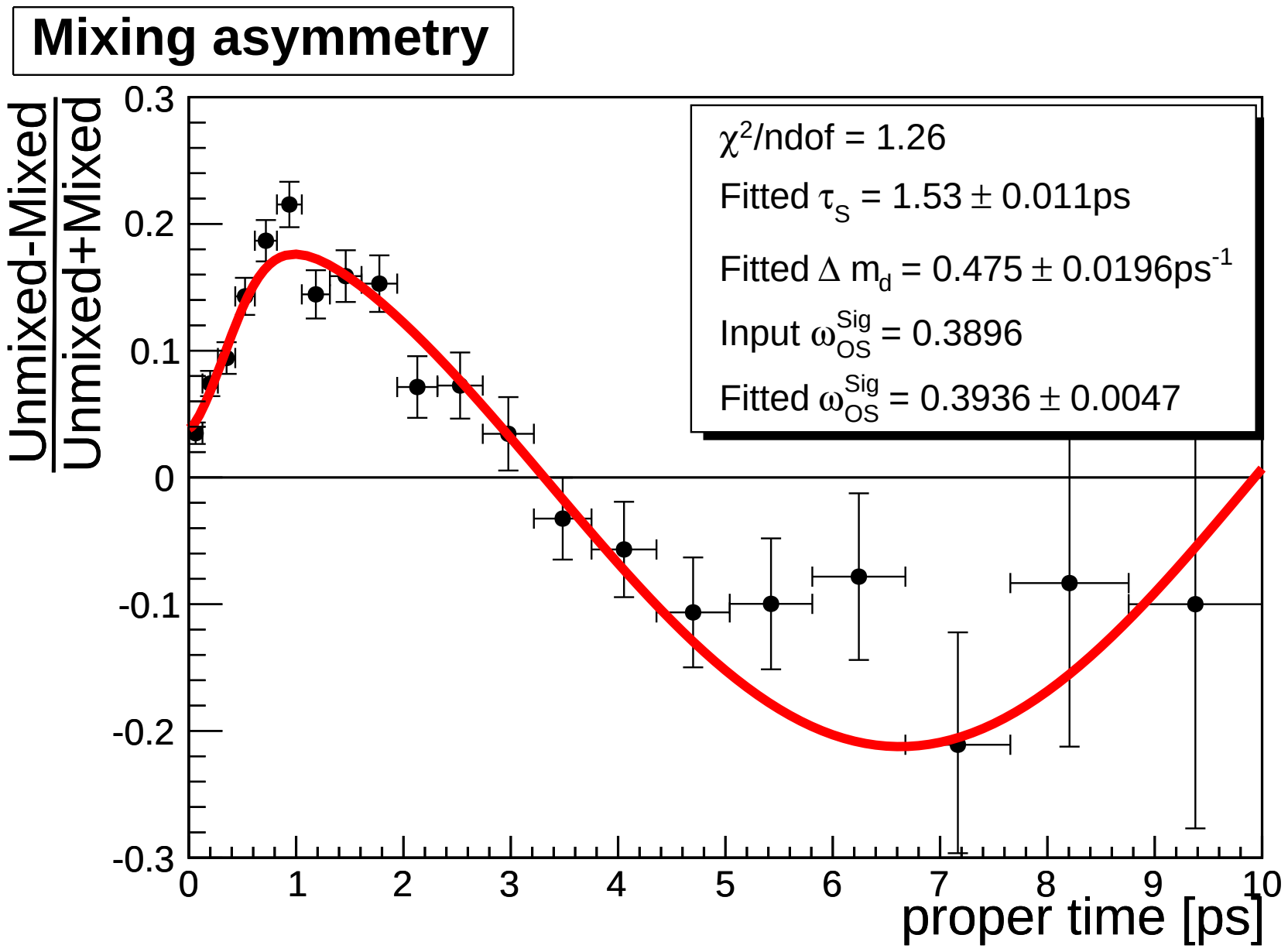

Figure 24: Mixing asymmetry of the $B_{u,d} \to J/\psi X$ selected sample. In the box: the $\chi^2$ between the best fitted curve in red and the data distribution and the input $\omega_{OS}^{sig}$ for the opposite-side mistag, measured using the Monte Carlo truth.

is described by a single Gaussian.

Three types of background are considered in this study: a prompt component, where all final state tracks come directly from the primary vertex; events where the $J/\psi$ really comes from a long-lived particle, but the $K^{*0}$ is made out of prompt tracks, and events where three out of four tracks come from a long-lived particle, namely a B meson. For all types of backgrounds the B mass distribution is described as a decreasing exponential, but with different slope parameters. In the prompt component the reconstructed proper time is described by a double Gaussian distribution, with the same parameters as the resolution function used for signal events. In the long-lived backgrounds the proper time distribution has similar properties to those of the signal, but with different values of lifetime and mistag rates.

To have a sufficient number of signal and background events to test the model on full MC data, $B_{u,d} \to J/\psi(\mu\mu)X$ events are used. There is no prompt component in this sample, but it is added in toy studies. As the main interest of this analysis is to measure the mistag fraction for opposite-side tags, only opposite-side tagged events are used. A first test is performed, using events all together and measuring the average mistag rate. Results for the unbinned likelihood fit are reported in Figure 23 showing the proper time distribution and in Figure 24, showing the mixing asymmetry. The mixing asymmetry is defined as:

$$\mathcal{A}(t) = \frac{\text{Unmixed}(t) - \text{Mixed}(t)}{\text{Unmixed}(t) + \text{Mixed}(t)}\,, \tag{73}$$

where $t$ is the proper time of the B candidate and Mixed (Unmixed) is the number of events with a final state flavour different (identical) to the initial one. The fit has been repeated splitting events into five samples of increasing tagging purity, according to the probability of mistag, measured in each event as explained in the previous sections. A simultaneous fit to five samples has been performed and the measured mistag rates all are in agreement with those calculated from MC truth.

In order to check fully the model, toy experiments were performed generating all events according to the expected PDFs, including the background component of prompt events. A total of 899 toys have been generated, with a statistics of about $0.03\,\mathrm{fb}^{-1}$ each. All fitted parameters show canonical Gaussian behaviour. The relative statistical error obtained from these fits is $\frac{\sigma(\omega_{\mathrm{OS}}^{\mathrm{Sig}})}{\omega_{\mathrm{OS}}^{\mathrm{Sig}}} = 2.5\%$ which corresponds to a sensitivity of 0.3% in $2\,\mathrm{fb}^{-1}$ of data. For $\Delta m_{\mathrm{d}}$ the resulting sensitivity scaled to $2\,\mathrm{fb}^{-1}$ of data is $\frac{\sigma(\Delta m_{\mathrm{d}})}{\Delta m_{\mathrm{d}}} = 0.7\%$.

# 9 Fit procedures

The procedure for determining physics parameters from the data is based upon an unbinned likelihood method. Construction and normalisation of the PDF in detail depends upon exactly which detector parameters one wishes to use on an event-by-event basis and how one wishes to handle acceptance effects. Here we mainly describe the features, principles and constraints.

The likelihood function for $N$ events can be written generically as:

$$\mathcal{L} = \prod_{e}^{N} \mathcal{P}(X_e; \lambda)\,, \tag{74}$$

where

- $X_e$ are the measured event physics attributes. These are: proper time $t$, decay angles $\Omega$, B mass $m$ and initial B flavour tag $q$.
- $\lambda$ are the physics parameters, $\lambda_{\mathrm{phys}}$, and detector parameters, $\lambda_{\mathrm{det}}$, where:
- $\lambda_{\mathrm{phys}} = \{\Gamma_{\mathrm{s}}, \Delta\Gamma_{\mathrm{s}}, R_{\perp}, R_0, \delta_{\perp}, \delta_{\parallel}, \Delta m_{\mathrm{s}}, \phi_{\mathrm{s}}\}$;
- $\lambda_{\mathrm{det}}$ are the detector parameters: mass resolution $\sigma_m$, proper time resolution $\sigma_t$, mistag rate $\omega$, background properties, ...

The PDF consists of a signal PDF, $\mathcal{S}$, and a background PDF, $\mathcal{B}$:

$$\mathcal{P} = f_{\mathrm{sig}}\mathcal{S} + (1 - f_{\mathrm{sig}})\mathcal{B}\,, \tag{75}$$

where $f_{\mathrm{sig}}$ is the expected overall signal fraction.

The signal PDF can be factorised as:

$$\mathcal{S}(X_e; \lambda) = \mathcal{S}_1(t, \Omega, q; \lambda)\, \mathcal{S}_2(m; \sigma_m)\,, \tag{76}$$

where:

- $\mathcal{S}_1(t, \Omega, q; \lambda)$ is the core PDF discussed below;
- $\mathcal{S}_2(m; \sigma_m)$ is the mass PDF.

In the following, we describe step by step all the ingredients of the fitting procedure.

## 9.1 Including flavour tagging of signal

$\mathcal{S}_1(t,\Omega,q;\lambda)$ is constructed from the differential decay rate for $B_s^0$ and $\overline{B}_s^0$ described in Section 2, $\frac{d^4\Gamma}{dt\,d\Omega}$, $\frac{d^4\bar{\Gamma}}{dt\,d\Omega}$. To simplify the equations, we write:

$$\mathcal{P}_s = \frac{d^4\Gamma}{dt\,d\Omega}\,, \quad \bar{\mathcal{P}}_s = \frac{d^4\bar{\Gamma}}{dt\,d\Omega}\,. \tag{77}$$

Using these we may write the joint PDF of $t, \Omega$ and $q$, (assuming perfect resolution and without background or acceptance) as:

$$\mathcal{S}_1(t,\Omega,q;\lambda) \propto \left(\frac{1+q}{2}\mathcal{P}_s + \frac{1-q}{2}\bar{\mathcal{P}}_s\right)\,. \tag{78}$$

where the normalization condition is:

$$\sum_{q=0,-1,+1} \int dt d\Omega\, \mathcal{S}_1(t,\Omega,q;\lambda) = 1\,. \tag{79}$$

When the detector dependent mistag-fraction parameter $\omega$ is introduced the PDF can then be re-written:

$$\mathcal{S}_1(t,\Omega,q;\lambda) \propto \left(\frac{1+qD}{2}\mathcal{P}_s + \frac{1-qD}{2}\bar{\mathcal{P}}_s\right)\,, \tag{80}$$

where $D = (1-2\omega)$.

It is instructive to write down the components which make up the numerator of Eq. 78 including the mistagging rate $\omega$. These are obtained by adding terms of Eqs. 45−56 from Section 2. For events tagged as $B_s^0$ the result is:

$$\begin{aligned}(1-\omega)|A_0(t)|^2 + \omega|\bar{A}_0(t)|^2 &= |A_0(0)|^2 e^{-\Gamma_s t} \times \\ &\left[\cosh\left(\frac{\Delta\Gamma_s t}{2}\right) - \cos\phi_s \sinh\left(\frac{\Delta\Gamma_s t}{2}\right) + (1-2\omega)\,\sin\phi_s \sin(\Delta m_s t)\right],\end{aligned} \tag{81}$$

$$\begin{aligned}(1-\omega)|A_\parallel(t)|^2 + \omega|\bar{A}_\parallel(t)|^2 &= |A_\parallel(0)|^2 e^{-\Gamma_s t} \times \\ &\left[\cosh\left(\frac{\Delta\Gamma_s t}{2}\right) - \cos\phi_s \sinh\left(\frac{\Delta\Gamma_s t}{2}\right) + (1-2\omega)\,\sin\phi_s \sin(\Delta m_s t)\right],\end{aligned} \tag{82}$$

$$\begin{aligned}(1-\omega)|A_\perp(t)|^2 + \omega|\bar{A}_\perp(t)|^2 &= |A_\perp(0)|^2 e^{-\Gamma_s t} \times \\ &\left[\cosh\left(\frac{\Delta\Gamma_s t}{2}\right) + \cos\phi_s \sinh\left(\frac{\Delta\Gamma_s t}{2}\right) - (1-2\omega)\,\sin\phi_s \sin(\Delta m_s t)\right],\end{aligned} \tag{83}$$

$$\begin{aligned}(1-\omega)\Im\{A_\parallel^*(t)A_\perp(t)\} + \omega\Im\{\bar{A}_\parallel^*(t)\bar{A}_\perp(t)\} &= |A_\parallel(0)||A_\perp(0)|\; e^{-\Gamma_s t}\Big[-\cos(\delta_\perp - \delta_\parallel)\sin\phi_s \sinh\left(\frac{\Delta\Gamma_s t}{2}\right) \\ &+ (1-2\omega)\;\{\sin(\delta_\perp - \delta_\parallel)\cos(\Delta m_s t) - \cos(\delta_\perp - \delta_\parallel)\cos\phi_s \sin(\Delta m_s t)\}\Big],\end{aligned} \tag{84}$$

$$\begin{aligned}(1-\omega)\Re\{A_0^*(t)A_\parallel(t)\} + \omega\Re\{\bar{A}_0^*(t)\bar{A}_\parallel(t)\} &= |A_0(0)||A_\parallel(0)|\; e^{-\Gamma_s t}\cos(\delta_\parallel) \times \\ &\left[\cosh\left(\frac{\Delta\Gamma_s t}{2}\right) - \cos\phi_s \sinh\left(\frac{\Delta\Gamma_s t}{2}\right) + (1-2\omega)\,\sin\phi_s \sin(\Delta m_s t)\right],\end{aligned} \tag{85}$$

$$\begin{aligned}(1-\omega)\Im\{A_0^*(t)A_\perp(t)\} + \omega\Im\{\bar{A}_0^*(t)\bar{A}_\perp(t)\} &= |A_0(0)||A_\perp(0)|\; e^{-\Gamma_s t}\Big[-\cos\delta_\perp \sin\phi_s \sinh\left(\frac{\Delta\Gamma_s t}{2}\right) \\ &+ (1-2\omega)\;\{\sin(\delta_\perp)\cos(\Delta m_s t) - \cos(\delta_\perp)\cos\phi_s \sin(\Delta m_s t)\}\Big],\end{aligned} \tag{86}$$

where the full differential decay rate is obtained by multiplying each component by the relevant angular terms and adding them together as described in Section 2.

Several features can be observed.

- The diagonal and real interference terms contain the product $(1-2\omega)\sin\phi_{\rm s}$ multiplying the sinusoid $\sin(\Delta m_{\rm s}t)$. These are the terms which primarily determine $\phi_{\rm s}$ if it is small, as expected in the Standard Model, provided that $\omega$ is determined elsewhere.

- As noted earlier the $\cos\phi_{\rm s}$ terms are in principle accessible using untagged events, but in practice they do not contribute much sensitivity if $\phi_{\rm s}$ is small.

- The imaginary interference terms are very interesting. Inspection shows that if $\Delta\Gamma_{\rm s}\neq 0$ they separate the $\sin\phi_{\rm s}$ term from the $(1-2\omega)$ term providing orthogonal information such that both $\phi_{\rm s}$ and $\omega$ can be simultaneously determined from the data. This property is an important feature of this analysis and details can be found in [24]. Naturally the question of dependence upon the strong phases arises and this is fully discussed in the reference. This issue is discussed further in the fitting section.

In the reduced one-angle analysis (using only $\theta$) the full differential decay rate for $\mathrm{B_s^0}$ tagged events is:

$$\begin{aligned}(1-\omega)\frac{\mathrm{d}\Gamma}{\mathrm{d}t\;\mathrm{d}\cos\theta}+\omega\frac{\mathrm{d}\bar{\Gamma}}{\mathrm{d}t\;\mathrm{d}\cos\theta}\propto&\\ (1-R_\perp)\mathrm{e}^{-\Gamma_{\rm s}t}\Big[\cosh\left(\frac{\Delta\Gamma_{\rm s}t}{2}\right)-\cos\phi_{\rm s}\sinh\left(\frac{\Delta\Gamma_{\rm s}t}{2}\right)&\\ +(1-2\omega)\sin\phi_{\rm s}\sin(\Delta m_{\rm s}t)\Big]\frac{1}{2}(1+\cos^2\theta)&\\ +R_\perp\mathrm{e}^{-\Gamma_{\rm s}t}\Big[\cosh\left(\frac{\Delta\Gamma_{\rm s}t}{2}\right)+\cos\phi_{\rm s}\sinh\left(\frac{\Delta\Gamma_{\rm s}t}{2}\right)&\\ -(1-2\omega)\sin\phi_{\rm s}\sin(\Delta m_{\rm s}t)\Big]\sin^2\theta\,.&\qquad(87)\end{aligned}$$

## 9.2 Including backgrounds and their flavour tagging

In this section, we introduce both the two sources of backgrounds presented in Section 6: prompt (Pr) and long-lived (LL). Their tagging efficiencies are denoted $\varepsilon_{\rm tag}^{\rm Pr}$ and $\varepsilon_{\rm tag}^{\rm LL}$. The signal tagging efficiency is simply called $\varepsilon_{\rm tag}$.

Taking this tagging efficiency into account, the signal PDF is re-written:

$$\mathcal{S}_1(t,\Omega,q;\lambda)\propto\varepsilon_{\rm tag}|q|\left(\frac{1+qD}{2}\mathcal{P}_s+\frac{1-qD}{2}\bar{\mathcal{P}}_s\right)+(1-\varepsilon_{\rm tag})(1-|q|)\left(\mathcal{P}_s+\bar{\mathcal{P}}_s\right)\,.\quad(88)$$

The two first terms correspond to the tagged part of the signal, while the last term corresponds to the untagged part.

The total background PDF is:

$$\mathcal{B}(t,\Omega,q)=f_{\rm Pr}\mathcal{B}_{\rm Pr}(t,\Omega,q)+(1-f_{\rm Pr})\mathcal{B}_{\rm LL}(t,\Omega,q)\,.\quad(89)$$

The prompt background PDF is:

$$\mathcal{B}_{\rm Pr}(t,\Omega,q)=\frac{1}{2}\varepsilon_{\rm tag}^{\rm Pr}|q|\mathcal{P}_{\rm Pr}(t,\Omega)+(1-\varepsilon_{\rm tag}^{\rm Pr})(1-|q|)\mathcal{P}_{\rm Pr}(t,\Omega)\,.\quad(90)$$

Again, the first part corresponds to tagged prompt background, while the second is the untagged one. $\mathcal{P}_{\rm Pr}(t,\Omega)$ is the joint PDF of the proper time and the angles for the prompt component.

The long-lived background PDF is:

$$\mathcal{B}_{\rm LL}(t,\Omega,q) = \frac{1}{2}\varepsilon_{\rm tag}^{\rm LL}|q|\mathcal{P}_{\rm LL}(t,\Omega) + (1-\varepsilon_{\rm tag}^{\rm LL})(1-|q|)\mathcal{P}_{\rm LL}(t,\Omega)\,. \tag{91}$$

## 9.3 Including the proper time resolution

The proper time resolution model is discussed in Section 6.3.2 leading to the choice of a double Gaussian $\mathcal{G}$ for the signal, with parameters $R$, given in Table 22. For fit stability a third Gaussian can be added to account for additional outlying events. We convolve $\mathcal{S}$ with this function:

$$\begin{aligned}\mathcal{S}_1'(t,\Omega,q;\lambda,R) &= \int_{t'=0}^{\infty} \mathrm{d}t'\,\mathcal{S}_1(t,\Omega,q;\lambda)G(t'-t;R) \\ &\equiv \mathcal{S}_1(t,\Omega,q;\lambda)\otimes G(t'-t;R)\,,\end{aligned} \tag{92}$$

where $t'$ denotes the true proper time. The parameters $R$ describe the proper time resolution model and are defined in time units.

If an event-by-event resolution is used the PDF becomes conditional to the additional observable $\delta_t$ which is the proper time error, whose distribution $\mathcal{S}_3(\delta_t)$ for the signal events can be obtained on real data from the signal mass region after subtracting the background distribution from the side-bands.

$$\mathcal{S}_1'(t,\delta_t,\Omega,q;\lambda) = \mathcal{S}_1'(t,\Omega,q|\delta_t;\lambda)\mathcal{S}_3(\delta_t) = \mathcal{S}_1(t,\Omega,q;\lambda)\otimes G(t'-t|\delta_t;R)\mathcal{S}_3(\delta_t)\,. \tag{93}$$

In this form the parameters $R$ that describe the proper time resolution model become unit-less variables like the bias and error scale factors.

## 9.4 Including proper time and angular acceptance

It has been shown in Section 6 that the angular acceptance in the baseline selection in the current MC is close to being flat. In reality this might be different. Moreover the $\rm B^0 \to J/\psi K^{*0}$ control channel does show a non trivial angular acceptance (see [60]). In general a non flat angular acceptance will lead to a biased estimate of the amplitudes, the strong phases and of $\phi_{\rm s}$. The difficulty in incorporating this is that the acceptance is a non-factorisable 4-dimensional function of $\Omega$ and $t$.

Introducing a proper-time and angular dependent acceptance $\varepsilon(t,\Omega)$ while neglecting for the moment the angular resolution (which will be discussed in Section 9.5) the PDF including acceptance and resolution (denoted by $''$) can be written in the following way:

$$\mathcal{S}''(t,\Omega,q;\lambda) \propto \mathcal{S}_1'(t,\Omega,q;\lambda)\varepsilon(t,\Omega)\,, \tag{94}$$

where the normalization is ensured by Eq. 79. To estimate the parameters $\lambda$ one finds the stationary point of the joint log likelihood, i.e.

$$\frac{\partial}{\partial\lambda}\sum_e \ln \mathcal{S}_e''(t_e,\Omega_e,q_e;\lambda) = 0\,, \tag{95}$$

where the sum is over events $e$. Using the fact that the angular dependent part and the time and amplitude dependent part factorize (into $f_i(\Omega)$ and $h_i(t,q;\lambda)$ respectively; see Eqs 43 and 44), the likelihood maximization can be written as:

$$\frac{\partial}{\partial\lambda}\sum_e \ln \frac{h_i(t_e,q_e;\lambda)f_i(\Omega_e)}{\int h_j(t,q;\lambda)\int f_j(\Omega)\varepsilon(t,\Omega)\mathrm{d}\Omega\mathrm{d}t} = 0\,, \tag{96}$$

where the sums over the $i,j$ (from 1 to 6) are omitted (here and in the following). The acceptance $\varepsilon(t,\Omega)$ has disappeared from the numerator because it does not depend on $\lambda$ and $\ln AB = \ln A + \ln B$. The second integral in the denominator can be determined from MC. Denoting this normalization integral $\xi_i(t)$, it can be written as:

$$\xi_i(t) \equiv \int f_i(\Omega)\varepsilon(t,\Omega)\mathrm{d}\Omega\,. \tag{97}$$

To correct for the acceptance it suffices to determine this integral. The log-likelihood maximization becomes:

$$\frac{\partial}{\partial\lambda}\sum_e \ln \frac{h_i(t_e,q_e;\lambda)f_i(\Omega_e)}{\int h_j(t,q;\lambda)\xi_j(t)\mathrm{d}t}\,. \tag{98}$$

If the proper time acceptance and the angular acceptance factorize the acceptance weights can be written as:

$$\xi_j(t) \rightarrow \varepsilon(t)\xi_j\,. \tag{99}$$

The terms $\xi_j$ can be pre-calculated before the fit, per bin of $t$.

The results presented in Section 10.1 are obtained using a four dimensional binned histogram for reconstructed proper time and angular acceptance. It is defined as:

$$\begin{aligned}\varepsilon'\left(t_i,\cos\theta_j,\phi_k,\cos\psi_l\right) = {} & \mathrm{Accepted}\left(t_i,\cos\theta_j,\phi_k,\cos\psi_l\right)/ \\ & \Big[\mathrm{Theory}(t_i,\cos\theta_j,\phi_k,\cos\psi_l)\otimes \\ & \left(f^{\mathrm{s}}_{t,1}G(t'-t_i;\sigma^{\mathrm{s}}_{t,1}) + (1-f^{\mathrm{s}}_{t,1})G(t'-t_i;\sigma^{\mathrm{s}}_{t,2})\right)\Big]\,,\end{aligned} \tag{100}$$

where the indices $i,j,k,l$ denote the bins of the histogram. The integral in Eq. 97 can be approximated by performing a numerical integration. The disadvantage of this method is that systematic binning effects have to be studied. In this approach, both the input sample or theoretical distribution of the simulation and the output of the selection and reconstruction are needed. An alternative method using normalization weights is presented in Ref. [43].

## 9.5 Including the angular resolution

Applying an angular resolution, $R(\Omega,\Omega')$, before the acceptance correction the likelihood equation changes to:

$$\begin{aligned}\frac{\partial}{\partial\lambda}\ln\mathcal{L}(\lambda) &= \frac{\partial}{\partial\lambda}\sum_e \ln\frac{h_i(t_e,q;\lambda)f_i(\Omega_e)\varepsilon(t_e,\Omega_e)}{\int h_j(t,q;\lambda)\xi'_j(t)\mathrm{d}t} \\ &= \frac{\partial}{\partial\lambda}\sum_e \ln\frac{h_i(t_e,q;\lambda)f_i(\Omega_e)}{\int h_j(t,q;\lambda)\xi'_j(t)\mathrm{d}t}\,,\end{aligned} \tag{101}$$

with

$$\xi'_j(t) = \int\int f_j(\Omega) R(\Omega, \Omega')\, d\Omega\, \varepsilon(\Omega', t)\, d\Omega' . \quad (102)$$

As in Eq. 96, the resolution $R(\Omega, \Omega')$ has disappeared from the numerator because it does not depend on $\lambda$ and $\ln AB = \ln A + \ln B$. In order to be able to neglect the angular resolution it has to be small in comparison with the scale on which $f_i$ and $f_i\varepsilon$ change. Then $f'_i \to f_i$ and $\xi'_i \to \xi_i$ and the resolutions in the numerator and denominator can just be neglected. The effect of the angular resolution on fitted parameters is discussed in Section 11.1.2.

## 9.6 Including proper time and angular distributions of background

Background will in general be included in the fit by using the sidebands of the $B_s^0$ mass peak to describe the background in the signal region.

If a full 4D proper time and angular acceptance is used to describe the signal a fit to or a histogram of the angular and proper time distribution of the sidebands will be used to describe the background.

## 9.7 Parametrization of the strong phase terms

As explained in [43], phases in general can lead to various problems in the fit. The fit can become unstable when close to the boundary of the parameters. Also the estimates of the parabolic errors appear to be underestimated due to non-Gaussian fit distributions. Multiple closely spaced solutions can also occur close to the maximum of the likelihood. These problematic properties are exacerbated when the strong phases have values close to a multiple of $\pi$, which is the theoretically preferred situation.

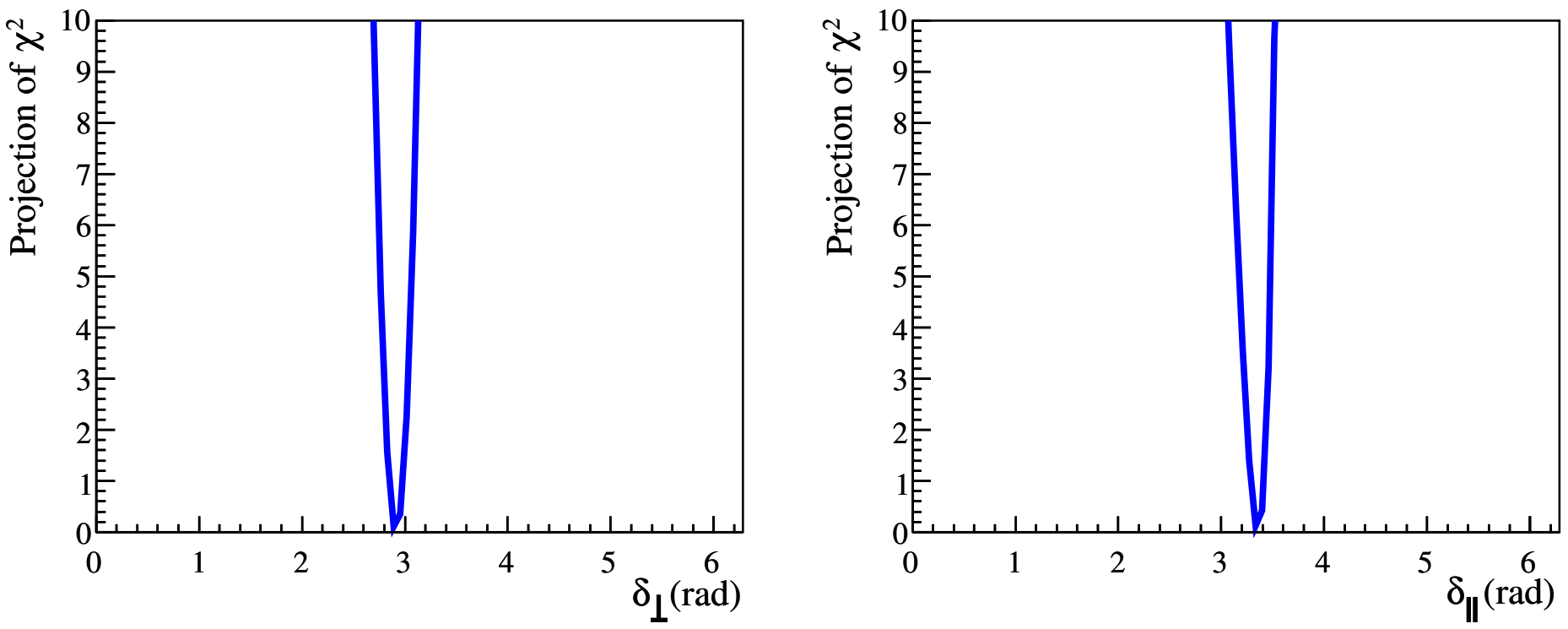

Figure 25: The $\chi^2$ of the fit for $\delta_i$. The minimization returns $\delta_\perp = 2.90^{+0.22}_{-0.27}$ rad and $\delta_\parallel = 3.35^{+0.17}_{-0.40}$ rad. The parabolic errors are 0.23 and 0.22 respectively.

To solve the above problems the following parametrization is proposed:

$$C_{\perp,0} = \cos(\delta_\perp - \delta_0) ,$$
$$S_{\perp,0} = \sin(\delta_\perp - \delta_0) ,$$

$$
\begin{aligned}
C_{\parallel,0} &= \cos(\delta_0 - \delta_\parallel)\,, \\
C_{\perp,\parallel} &= \cos(\delta_\perp - \delta_\parallel)\,, \\
S_{\perp,\parallel} &= \sin(\delta_\perp - \delta_\parallel)\,,
\end{aligned} \tag{103}
$$

which makes the fit more stable and reliable since the parametrization is linear. As shown in [43] this parametrization leads to correctly determined Gaussian errors. The sensitivity to the strong phases and other signal parameters does not decrease. In order to retrieve the physical values of the strong phases one proceeds as follows. Writing the three strong phases as $\delta_i$, let us introduce the residuals:

$$
\begin{aligned}
\mathbf{R}_1 &\equiv C_{\perp,0} - \cos(\delta_\perp - \delta_0)\,, \\
\mathbf{R}_2 &\equiv S_{\perp,0} - \sin(\delta_\perp - \delta_0)\,, \\
\mathbf{R}_3 &\equiv C_{\parallel,0} - \cos(\delta_\parallel - \delta_0)\,, \\
\mathbf{R}_4 &\equiv C_{\perp,\parallel} - \cos(\delta_\perp - \delta_\parallel)\,, \\
\mathbf{R}_5 &\equiv S_{\perp,\parallel} - \sin(\delta_\perp - \delta_\parallel)\,.
\end{aligned} \tag{104}
$$

Then, fixing $\delta_0$ to zero, we may write $\chi^2(\delta_\perp, \delta_\parallel)$ ,

$$
\chi^2(\delta_\perp, \delta_\parallel) = \mathbf{R}^T V^{-1} \mathbf{R}\,, \tag{105}
$$

with $V$ the covariance matrix of $C_{\perp,0}$, $S_{\perp,0}$, $C_{\parallel,0}$, $C_{\perp,\parallel}$ and $S_{\perp,\parallel}$.

Figure 25 shows the typical shape of the $\chi^2(\delta_i)$ after minimizing it. The results are in accordance with the input values and the sensitivity has not decreased.

In case similar fit problems occur for the weak phase, this approach might also be useful to parametrize the sine and cosine of $\phi_s$. This approach might be particularly interesting for the early measurement with a small number of events. However, it is not used in the rest of this document which assumes $2\,\text{fb}^{-1}$ of data.

# 10 Fit validations and sensitivity studies

## 10.1 Fit to signal events in the full Monte Carlo

### 10.1.1 Data sample

The fit has been checked using the Monte-Carlo sample of $B_s^0 \to J/\psi(\mu\mu)\phi(KK)$ signal events to verify that it will retrieve the known physics parameters used in the generation (see Table 6) [44]. After the L0-trigger and the offline selection described in Section 6, the data set contains 1.4 million $B_s^0$ or $\overline{B}_s^0$.

### 10.1.2 Determination of the acceptances

To compensate for detector inefficiencies and geometrical effects, acceptances in the transversity angles and the proper time have to be included in the fit. The implementation adopted in this case uses a three-dimensional acceptance histogram for the transversity angles $\cos\theta$, $\varphi$ and $\cos\psi$, so that the strong correlations between the variables are correctly accounted for. In the case of both angular and proper time acceptance, a four-dimensional histogram is used. The histograms consists of 20 bins in each reconstructed transversity angle and 10 bins in the reconstructed proper time. This gives a total of 8 000 bins when using angular acceptance and 80 000 bins when using both angular and time acceptances.

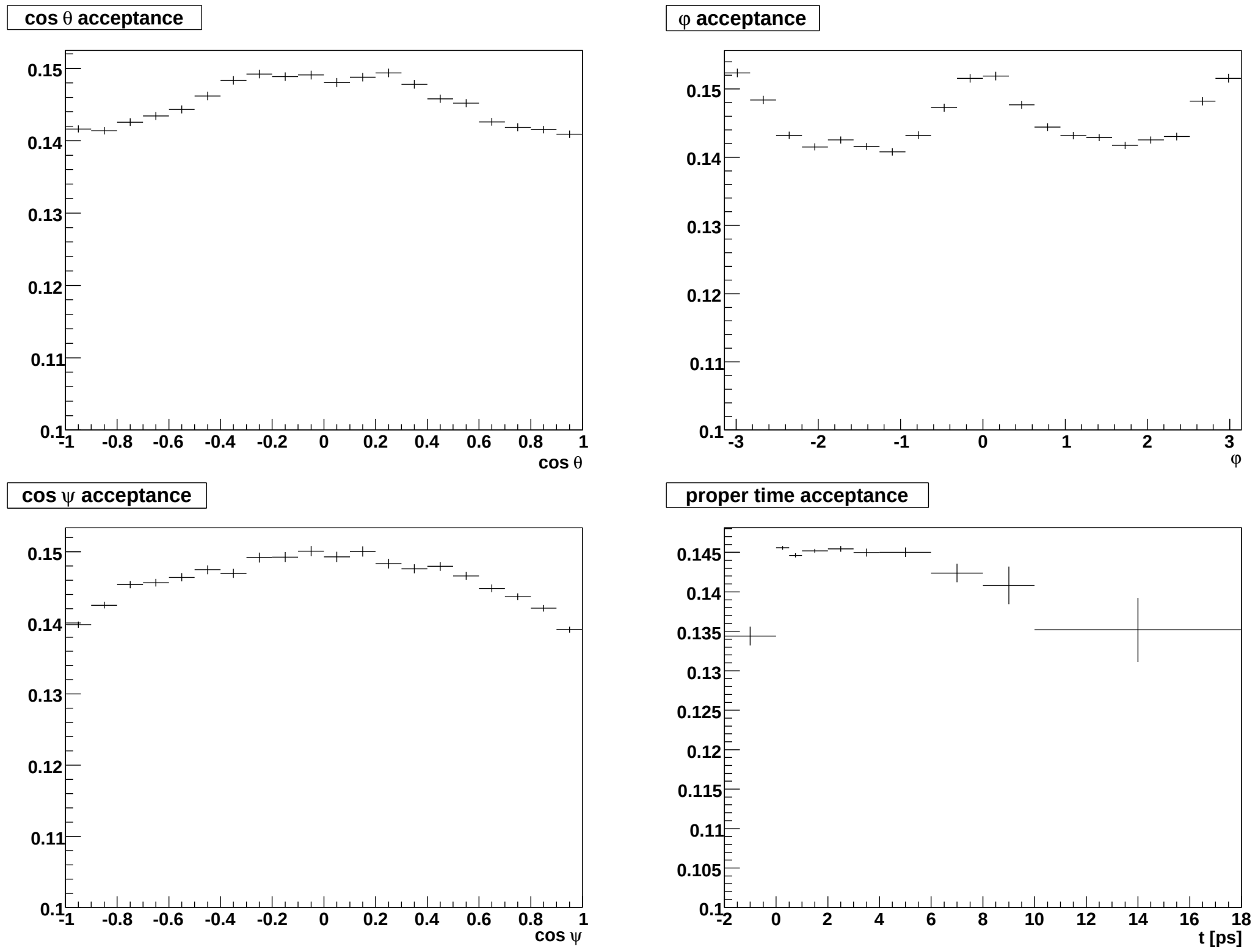

Figure 26: The projection of the determined acceptances used in the fit to Monte Carlo signal data. The deviations of the acceptances from being flat are less than 10%.

Each bin content is determined by dividing the number of reconstructed events in that particular bin by the corresponding expected number of events taken from theory for the parameters used in the Monte Carlo generation, convoluted with a double Gaussian resolution in $t$ (see Eq. 100). The full statistics of the MC signal data sample has been used to determine the acceptances; their projections can be seen in Figure 26. The effect of statistical fluctuations in the bins is discussed in Section 11.1. Effects due to angular resolution are considered in Section 10.4.

### 10.1.3 Results of the fit

The fit to the simulated data is performed using 115 k signal events, corresponding to about 1 year at a nominal luminosity of $2\,\mathrm{fb}^{-1}$. Fitting the signal Monte Carlo data neglecting acceptance effects gives the results shown in the left columns of Table 19. The mistag fraction was kept fixed in the fit to its true value (33.3%). Large deviations can be seen in the amplitudes $|A_\perp(0)|^2$ and $|A_0(0)|^2$.

The results including angular and proper time acceptances are shown in the right columns of Table 19. In addition to the eight physics parameters, also the widths of the double Gaussian describing the mass and the proper time resolution parameters are successfully fitted. The mistag fraction was kept fixed. The fit converge and determine the physics parameters with less than $2\sigma$ deviation from the generated value. The error on $\phi_\mathrm{s}$ is 0.027. The bias in the amplitudes observed when neglecting acceptance effects disappears. From these results we conclude that the angular acceptance effects, even when the acceptance are rather flat, can not be neglected in the fit. The fit including only the acceptance angular is presented in [44]. We do not observe significant differences between the results for angular and time acceptance and angular acceptance only and can thus conclude that including the proper time acceptance in the fit might not be necessary for determining $\phi_{\mathrm{s}}$ and the amplitudes.

| | Acceptance neglected | | Acceptance included | |
|---|---|---|---|---|
| Parameter | Result | $\sigma$ from nominal | Result | $\sigma$ from nominal |
| $m_{\mathrm{B_s}}$ (MeV/$c^2$) | $5368.01 \pm 0.05$ | | $5368.01 \pm 0.05$ | |
| $f^{\mathrm{s}}_{m,1}$ | $0.47 \pm 0.13$ | | $0.47 \pm 0.13$ | |
| $\sigma^{\mathrm{s}}_{m,1}$ (MeV/$c^2$) | $12.0 \pm 0.7$ | | $12.0 \pm 0.7$ | |
| $\sigma^{\mathrm{s}}_{m,2}$ (MeV/$c^2$) | $19.0 \pm 1.3$ | | $19.0 \pm 1.3$ | |
| $\|A_0(0)\|^2$ | $0.578 \pm 0.003$ | **−9.1** | $0.599 \pm 0.002$ | −0.6 |
| $\|A_\perp(0)\|^2$ | $0.173 \pm 0.004$ | **3.7** | $0.162 \pm 0.004$ | 0.5 |
| $\delta_\parallel$ (rad) | $2.50 \pm 0.02$ | −0.1 | $2.49 \pm 0.02$ | −0.4 |
| $\delta_\perp$ (rad) | $-0.28 \pm 0.10$ | −1.0 | $-0.28 \pm 0.10$ | −1.1 |
| $\phi_{\mathrm{s}}$ (rad) | $-0.0385 \pm 0.0273$ | 0.6 | $-0.0399 \pm 0.0272$ | 0.0 |
| $\Gamma_{\mathrm{s}}$ (ps$^{-1}$) | $0.686 \pm 0.003$ | 0.5 | $0.686 \pm 0.004$ | 0.4 |
| $\Delta\Gamma_{\mathrm{s}}$ (ps$^{-1}$) | $0.060 \pm 0.010$ | −0.9 | $0.061 \pm 0.010$ | −0.8 |
| $f_{t,1}$ | $0.96 \pm 0.01$ | | $0.96 \pm 0.01$ | |
| $\sigma^{\mathrm{s}}_{t,1}$ (ps) | $0.031 \pm 0.001$ | | $0.032 \pm 0.001$ | |
| $\sigma^{\mathrm{s}}_{t,2}$ (ps) | $0.113 \pm 0.011$ | | $0.12 \pm 0.01$ | |
| $\Delta m_{\mathrm{s}}$ (MeV/$c^2$) | $19.95 \pm 0.04$ | −1.0 | $19.96 \pm 0.04$ | |

Table 19: The results for the fitted parameters for a dataset equivalent to $2\,\mathrm{fb}^{-1}$, when the acceptances effects are neglected (left) or taken into account (right). The columns "$\sigma$ from nominal" give the fitted value minus the input value divided by the fitted uncertainty.

The projections of data and fitted probability density function on the transversity angles and the proper time can be found in Figure 27. The fitted PDF and data show very good agreement.

The correlations between all 15 signal parameters, including the mass and resolution parameters can be found in Table 20. The correlations between $\phi_{\mathrm{s}}$ and the other parameters are small which makes the measurement of $\phi_{\mathrm{s}}$ robust against systematic effects.

## 10.2 Fit signal and background on toy Monte Carlo

Since there are not enough events in the full Monte Carlo sample to perform a realistic fit of signal and background, toy MC samples are used to complete the study. In this way it is possible to estimate the sensitivity to the parameters for different input parameter values, both to test the robustness of the fit and to study specific systematic effects.

As a reference, the physics input parameters used for the "default" toy MC studies are listed in Table 21. No error is given for most of the parameters; this implies that only the central values were used for the purposes of generating toy MC data. Although it is possible to measure $\Delta m_s$ simultaneously in this channel with quite good precision, it will be measured externally with much better precision and so we consider it as a fit parameter constrained by an external input with a Gaussian error applied to the PDF.

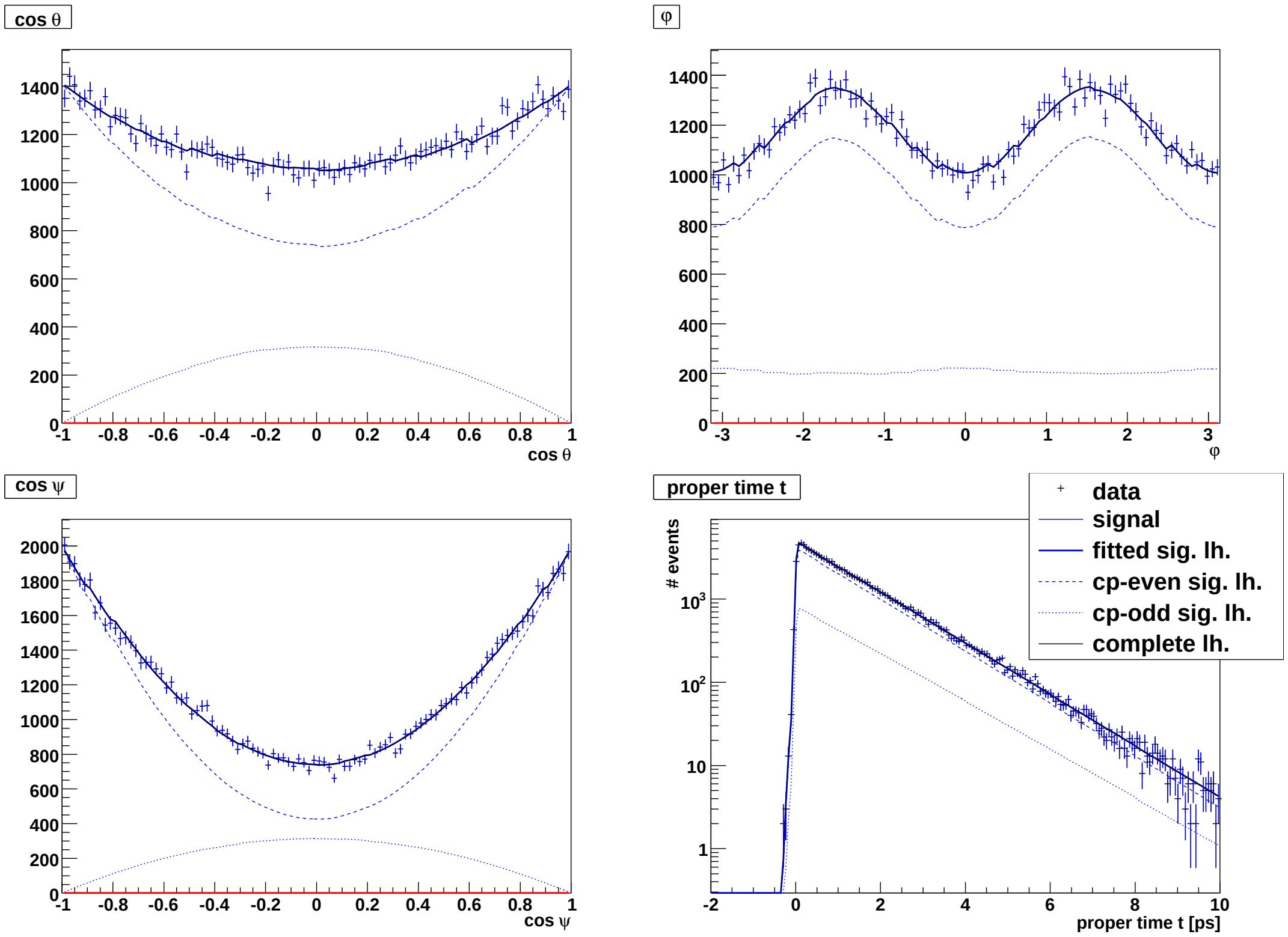

Figure 27: The projections of data and fitted signal PDF including both the angular and proper time acceptance effects, in a sample of fully simulated signal events. Also shown are the CP-even (dashed) and the CP-odd (dotted) components. The angular and the proper time functions show slight binning effects.

| Parameter | Global | $m_{\mathrm{B_s}}$ | $f^{\mathrm{s}}_{m,1}$ | $\sigma^{\mathrm{s}}_{m,1}$ | $\sigma^{\mathrm{s}}_{m,2}$ | $\lvert A_0(0)\rvert^2$ | $\lvert A_\perp(0)\rvert^2$ | $\delta_\parallel$ | $\delta_\perp$ | $\phi_{\mathrm{s}}$ | $\Gamma_{\mathrm{s}}$ | $\Delta\Gamma_{\mathrm{s}}$ | $f^{\mathrm{s}}_{t,1}$ | $\sigma^{\mathrm{s}}_{t,1}$ | $\sigma^{\mathrm{s}}_{t,2}$ | $\Delta m_{\mathrm{s}}$ |
|---|---|---|---|---|---|---|---|---|---|---|---|---|---|---|---|---|
| $m_{\mathrm{B_s}}$ | 0.04 | 1.00 | −0.02 | −0.02 | −0.02 | 0 | 0 | 0 | 0 | 0 | 0 | 0 | 0 | 0 | 0 | 0 |
| $f^{\mathrm{s}}_{m,1}$ | 0.999 | | 1.00 | 0.99 | 0.99 | 0 | 0 | 0 | 0 | 0 | 0 | 0 | 0 | 0 | 0 | 0 |
| $\sigma^{\mathrm{s}}_{m,1}$ | 0.996 | | | 1.00 | 0.97 | 0 | 0 | 0 | 0 | 0 | 0 | 0 | 0 | 0 | 0 | 0 |
| $\sigma^{\mathrm{s}}_{m,2}$ | 0.997 | | | | 1.00 | 0 | 0 | 0 | 0 | 0 | 0 | 0 | 0 | 0 | 0 | 0 |
| $\lvert A_0(0)\rvert^2$ | 0.62 | | | | | 1.00 | −0.51 | 0 | 0.01 | −0.03 | −0.48 | 0.58 | 0 | 0 | 0 | 0.01 |
| $\lvert A_\perp(0)\rvert^2$ | 0.72 | | | | | | 1.00 | −0.36 | 0.04 | 0.03 | 0.54 | −0.61 | 0 | 0.01 | 0 | 0 |
| $\delta_\parallel$ | 0.46 | | | | | | | 1.00 | 0.13 | 0 | 0.05 | −0.04 | 0 | 0.01 | 0 | 0.02 |
| $\delta_\perp$ | 0.68 | | | | | | | | 1.00 | −0.14 | 0 | 0.01 | 0 | 0.01 | 0 | 0.67 |
| $\phi_{\mathrm{s}}$ | 0.20 | | | | | | | | | 1.00 | 0.04 | −0.05 | −0.01 | −0.02 | 0 | −0.20 |
| $\Gamma_{\mathrm{s}}$ | 0.81 | | | | | | | | | | 1.00 | −0.81 | 0 | 0 | 0 | −0.02 |
| $\Delta\Gamma_{\mathrm{s}}$ | 0.85 | | | | | | | | | | | 1.00 | 0 | 0 | 0 | 0.02 |
| $f^{\mathrm{s}}_{t,1}$ | 0.88 | | | | | | | | | | | | 1.00 | 0.65 | 0.84 | 0 |
| $\sigma^{\mathrm{s}}_{t,1}$ | 0.65 | | | | | | | | | | | | | 1.00 | 0.49 | 0.02 |
| $\sigma^{\mathrm{s}}_{t,2}$ | 0.84 | | | | | | | | | | | | | | 1.00 | 0 |
| $\Delta m_{\mathrm{s}}$ | 0.68 | | | | | | | | | | | | | | | 1.00 |

Table 20: The correlations of the 15 fitted parameters in the fit using angular acceptance only, in a sample of pure $\mathrm{B^0_s} \to \mathrm{J}/\psi\phi$ fully simulated Monte Carlo events, corresponding to $2\,\mathrm{fb}^{-1}$. The inputs parameters are Standard-Model like. They are given in Tables 4, 5 and 6. Absolute correlations less than 0.01 are displayed as 0. The correlations between proper time resolution parameters amongst themselves are high. However, the correlations of $\phi_{\mathrm{s}}$ with the other variables are small. The high correlations between mass-related parameters are a consequence of the tight cut on the $\mathrm{B^0_s}$ mass chosen to be compatible with the toy MC results given later. As a consequence, the tails are missing and the broader component is hard to fit.

| Parameter | Input | Unit | Source |
|---|---|---|---|
| $\phi_s$ | $-0.0368$ | rad | [1](*) |
| $\Gamma_s$ | 0.680 | $fs^{-1}$ | [2] |
| $\Delta\Gamma_s$ | 0.049 | $fs^{-1}$ | [2] |
| $\lvert A_0(0)\rvert^2$ | 0.556 | | [19] |
| $\lvert A_\parallel(0)\rvert^2$ | 0.211 | | [19] |
| $\lvert A_\perp(0)\rvert^2$ | 0.233 | | [19] |
| $\delta_\perp$ | 2.91 | rad | [19] |
| $\delta_\parallel$ | $-2.93$ | rad | [19] |
| $\Delta m_s$ | $17.77 \pm 0.012$ | $fs^{-1}$ | [2] |
| $m_{B_s^0}$ | 5366.4 | $MeV/c^2$ | [2] |

Table 21: Summary of physics input parameters to fits. (*) The $\phi_s$ value comes from the early 2008 estimate made by [1]. Since then, it has been updated to $-(0.0360^{+0.0020}_{-0.0016})$ rad.

The parameters that describe the detector performance (signal resolutions) and define the background parametrisation are listed in Table 22[10].

In the following we will first present the results of toy MC studies where only the physics parameters are free in the fit and the detector parameters are fixed. In a second step, we let free all the physics and detector parameters.

Toy studies are based on 300 experiments, each corresponding to $2\,fb^{-1}$ of signal and background data. A full three-angular analysis of the tagged and untagged data is applied. The results were obtained by running different fit programs developed within the collaboration, that have been demonstrated to be compatible among themselves.

For each fitted parameter, the sensitivity is defined as the width of the Gaussian that best fits the value distribution. The pull distributions are also examined in order to search for systematic effects in the parameter determination.

The fitted parameters mean and error are reported on Table 23. The two columns correspond to the cases in which different sets of parameters are free: only the physics parameters (first) and all the parameters free (second).

With the baseline setting (first column), the sensitivity to $\phi_s$ is $(0.030 \pm 0.002)$ rad.

We note that in the least constrained case (last column) the fit converges successfully demonstrating that the full angular analysis is capable of determining all parameters simultaneously. Moreover, the sensitivity on $\phi_s$ does not degrade significantly if more parameters are left free, being largely uncorrelated to all the other parameters as one can see in Table 26 of Appendix C. In all the cases the pull distributions are compatible with normal Gaussian, with the exception of the parameter $f^s_{t,1}$, which being correlated to $\sigma^s_{t,1}$ and $\sigma^s_{t,2}$ and limited between 0 and 1 has a non-Gaussian shape.

It should be noted that the correlation matrix changes significantly with the input value of the physics parameters. As an example, the correlation matrix obtained with

[10] They are nearly identical to the parameters extracted from the selection of Section 6, with small differences arising from the fact that the selection analyses kept evolving after the parameters were frozen for the fit studies.

| Parameter | Signal | Prompt background (Pr) | Long-lived background (LL) |
|---|---|---|---|
| Events in $B_s^0$ mass window $\pm 50\,\mathrm{MeV}/c^2$ (fractions, $B/S$) | 117 k (30.3%) | 211 k (54.4%, $B_{\mathrm{Pr}}/S = 1.8$) | 59.6 k (15.4%, $B_{\mathrm{LL}}/S = 0.5$ ) |
| Mass $m$ ( $\mathrm{MeV}/c^2$) | 2 Gauss; $f^{\mathrm{s}}_{m,1} = 0.74$; $\sigma^{\mathrm{s}}_{m,1} = 13.2$; $\sigma^{\mathrm{s}}_{m,2} = 22.5$ | $\mathrm{Exp}(-\alpha^{\mathrm{Pr}}_m m)$; $\alpha^{\mathrm{Pr}}_m = 0.0006$ | $\mathrm{Exp}(-\alpha^{\mathrm{LL}}_m m)$; $\alpha^{\mathrm{LL}}_m = 0.001$ |
| Proper time $t$ (fs) | Signal PDF $\otimes$ 2 Gauss; $f^{\mathrm{s}}_{t,1} = 0.85$; $\mu^{\mathrm{s}}_{t,1} = 0$; $\sigma^{\mathrm{s}}_{t,1} = 31.5$; $\mu^{\mathrm{s}}_{t,2} = 0$; $\sigma^{\mathrm{s}}_{t,2} = 66.7$ | $\delta(t) \otimes$ 1 Gauss; $\mu^{\mathrm{Pr}}_t = 0$; $\sigma^{\mathrm{Pr}}_t = 44$ | 2 Exponentials $\otimes$ 1 Gauss; $f^{\mathrm{LL}}_{\tau_1} = 0.22$; $\tau^{\mathrm{LL}}_1 = 1114$, $\tau^{\mathrm{LL}}_{t,2} = 161$; $\mu^{\mathrm{LL}}_t = 0$; $\sigma^{\mathrm{LL}}_t = 66$ |
| Angles | no acceptance flat background | no acceptance flat background | no acceptance flat background |
| Flavour tagging | $\varepsilon_{\mathrm{tag}} = 0.564$; $\omega = 0.334$ | $\varepsilon^{\mathrm{Pr}}_{\mathrm{tag}} = 0.30$ | $\varepsilon^{\mathrm{LL}}_{\mathrm{tag}} = 0.62$ |

Table 22: Summary of baseline detector input parameters for sensitivity studies, extracted from full Monte Carlo.

$-\phi_s = 0.736$ is given in Table 27 of Appendix C. In this case, $\phi_s$ is correlated with the mistag fraction, $R_\perp$, $R_0$ and $\Gamma_s$.

| Parameter | Units | Sensitivity $\times 10^3$ : $\sigma \pm \delta_\sigma$ | |
|---|---|---|---|
| | | Only phys. parameters free | All parameters free |
| $\phi_s$ | rad | $30 \pm 2$ | $31 \pm 2$ |
| $\Gamma_s$ | ps$^{-1}$ | $3.1 \pm 0.1$ | $2.9 \pm 0.1$ |
| $\Delta\Gamma_s$ | ps$^{-1}$ | $9.1 \pm 0.4$ | $9.1 \pm 0.5$ |
| $R_\perp$ | | $4.2 \pm 0.2$ | $4.2 \pm 0.2$ |
| $R_0$ | | $3.1 \pm 0.1$ | $3.0 \pm 0.2$ |
| $\delta_\parallel$ | rad | $74 \pm 3$ | $72 \pm 3$ |
| $\delta_\perp$ | rad | $89 \pm 4$ | $130 \pm 7$ |
| $\Delta m_s$ | ps$^{-1}$ | – | $44 \pm 2$ |
| $M_{B_s}$ | MeV/$c^2$ | – | $52 \pm 3$ |
| $\omega$ | | – | $13 \pm 1$ |
| $f^s_{m,1}$ | | – | $32 \pm 2$ |
| $\sigma_{m,1}$ | MeV/$c^2$ | – | $180 \pm 10$ |
| $\sigma_{m,2}$ | MeV/$c^2$ | – | $840 \pm 40$ |
| $f^s_{t,1}$ | | – | $20 \pm 10$(*) |
| $\mu^s_t$ | ps | – | $3.6 \pm 0.2$ |
| $\sigma^s_{t,1}$ | ps | – | $8.6 \pm 0.4$ |
| $\sigma^s_{t,2}$ | ps | – | $14 \pm 1$ |
| $\alpha^{Pr}_m$ | (MeV/$c^2$)$^{-1}$ | – | $0.016 \pm 0.001$ |
| $\mu^{Pr}_t$ | ps | – | $0.082 \pm 0.004$ |
| $\sigma^{Pr}_t$ | ps | – | $0.081 \pm 0.004$ |
| $\alpha^{LL}_m$ | (MeV/$c^2$)$^{-1}$ | – | $0.032 \pm 0.002$ |
| $f^{LL}_{\tau_1}$ | | – | $3.8 \pm 0.2$ |
| $\tau^{LL}_{t,1}$ | ps$^{-1}$ | – | $9.7 \pm 0.5$ |
| $\tau^{LL}_{t,2}$ | ps$^{-1}$ | – | $1.5 \pm 0.1$ |
| $\mu^{LL}_t$ | ps | – | $3.0 \pm 0.2$ |
| $\sigma^{LL}_t$ | ps | – | $1.16 \pm 0.06$ |

Table 23: Parameter precisions and corresponding error obtained from simultaneous three-angle fit to different set of parameters using tagged and untagged events. Each value is the mean (and its uncertainty) of a single-Gaussian fit to the parameter distribution obtained in 300 toy Monte Carlo experiments, each corresponding to a dataset equivalent to 2 fb$^{-1}$. The value marked with (*) represent the error mean and RMS, since the distribution is not Gaussian.

## 10.3 Sensitivity versus physics parameters

A study of the $\phi_s$ sensitivity versus different central values chosen for physics parameters has been performed in [24] and [47]. The statistical uncertainty on $\phi_s$ decreases when $\Delta\Gamma_s$ grows and increases when the CP-odd fraction ($R_\perp$) grows. No significant dependence on $\Gamma_s$ is observed. When $\phi_s$ is small then $\phi_s$ is mainly determined through the $\sin\phi_s$ terms measured through the tagged analysis of the $\sin(\Delta m_s t)$ time dependence of the cross section. In this case the uncertainty on $\phi_s$ and $\sin\phi_s$ are the same and the $\cos\phi_s$ term is insensitive to $\phi_s$. The uncertainly on $\sin\phi_s$ measured in this way is approximately constant and therefore as $\sin\phi_s$ becomes larger then the uncertainly on $\phi_s$ itself shows a $1/|\cos\phi_s|$ dependency. This is illustrated with values of $\phi_s = -0.0368\,\mathrm{rad}$ and $\phi_s = -0.736\,\mathrm{rad}$ in Table 24. When $\sin\phi_s$ approaches one, the uncertainty on $\phi_s$ from the tagged analysis increases sharply, and the extraction of $\cos\phi_s$ from untagged proper time distribution becomes the major contribution in determining $\phi_s$.

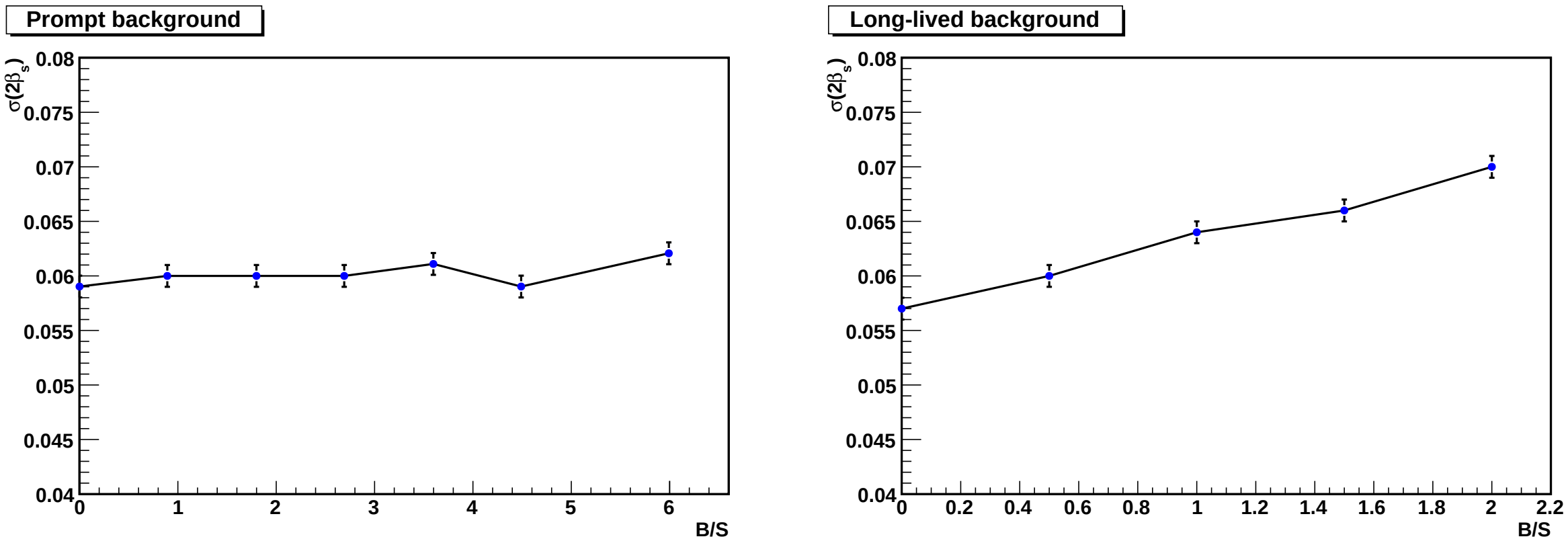

Figure 28: Sensitivity to $\phi_s$ versus $B/S$ for prompt background (left) and long-lived background (right). The sample size is equivalent to $0.5\,\mathrm{fb}^{-1}$ of data and the default fitting procedure is used (Section 10.2), with only the physics parameters left free.

As noted above, the correlation matrix changes significantly with the input value of the physics parameters.

## 10.4 Sensitivity versus detector parameters

The study of $\phi_s$ sensitivity versus proper time resolution is presented in [24, 48]. It decreases as the resolution degrades. It also decreases when the mistag fraction increases [24].

The sensitivity on $\phi_s$, as a function of $B/S$ for prompt and long-lived backgrounds, are shown in Figure 28. Each point has been obtained with 30 toy MC experiments corresponding to $0.5\,\mathrm{fb}^{-1}$ each. On the left plot, the background over signal ratio of the long-lived component is fixed to its default value, i.e. $B_{\mathrm{LL}}/S = 0.5$. On the right plot, the background over signal ratio of the prompt component is fixed to its default value, i.e. $B_{\mathrm{Pr}}/S = 1.8$. The prompt background has very limited effect on $\phi_s$. The long-lived background is naturally more dangerous since it is more signal like. If both background levels are set to zero, the sensitivity, rescaled to $2\,\mathrm{fb}^{-1}$ is $(0.027 \pm 0.002)\,\mathrm{rad}$. This is compatible with the result of the single fit made to signal only, in the full Monte Carlo (see Section 10.1).

## 10.5 Sensitivity versus integrated luminosity

Figure 29 shows the statistical uncertainty on $\phi_s$ versus the integrated luminosity. The sensitivity has been estimated only at 0.5 and $2\,\mathrm{fb}^{-1}$, with the baseline set of parameters, as in Section 10.2. The values are respectively $0.060 \pm 0.005$ and $0.030 \pm 0.002$. The red line is an extrapolation from these two values, assuming the errors scale like $1/\sqrt{\mathcal{L}_{\mathrm{int}}}$.

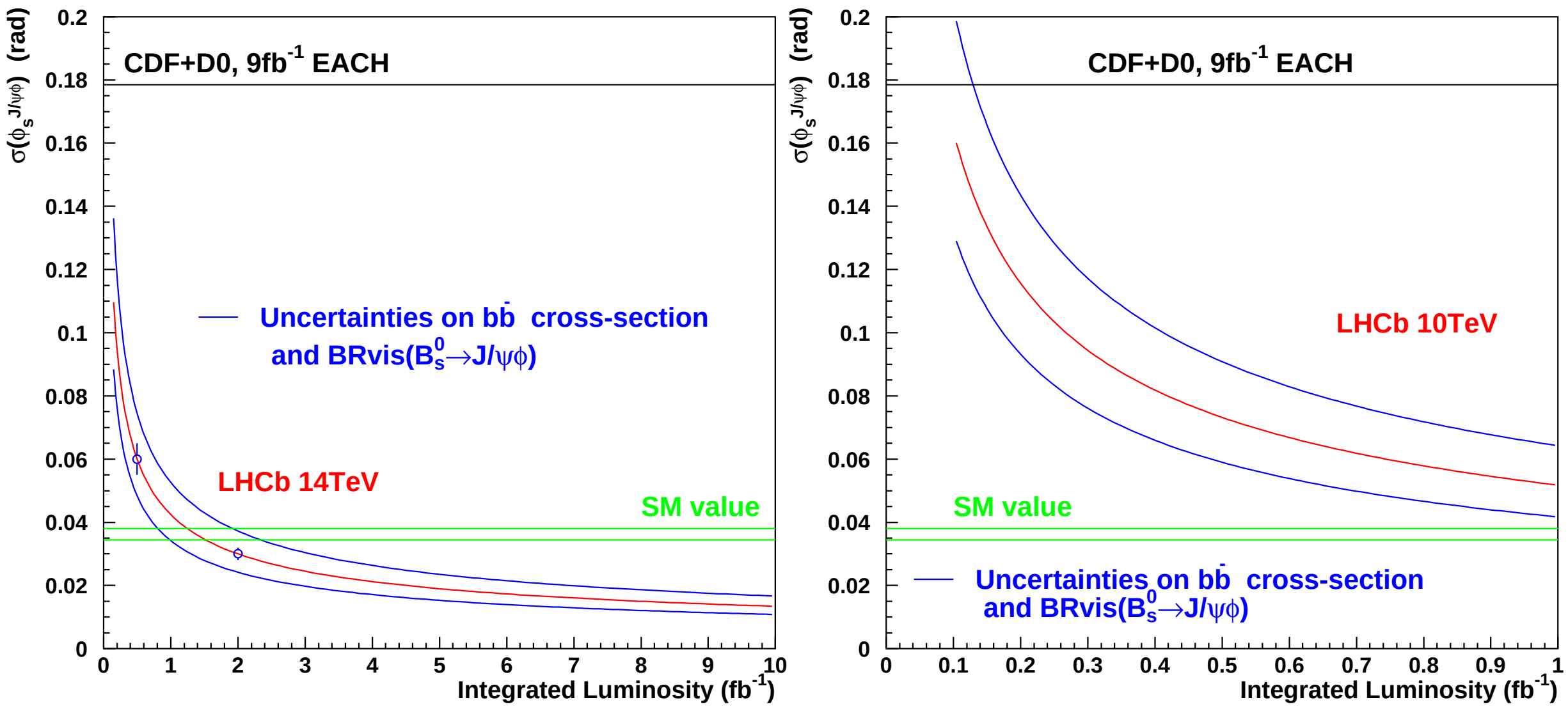

Figure 29: Red line: Statistical uncertainty on $\phi_\mathrm{s}$ versus the integrated luminosity. Blue line: uncertainties coming from the $\mathrm{b\overline{b}}$ cross-section and the visible branching ratio on $\mathrm{B_s^0} \to \mathrm{J}/\psi(\mu\mu)\phi(\mathrm{KK})$. The green band is the Standard Model value: $2\beta_\mathrm{s} = (0.0360^{+0.0020}_{-0.0016})\,\mathrm{rad}$ [1]. Left: from 0 to $10\,\mathrm{fb}^{-1}$, assuming a centre-of-mass energy of 14 TeV. Right: zoom between 0 and $1\,\mathrm{fb}^{-1}$, assuming a centre-of-mass energy of 10 TeV.

The blue lines show the uncertainties coming from the $\mathrm{b\overline{b}}$ cross-section and the visible branching ratio of $\mathrm{B_s^0} \to \mathrm{J}/\psi(\mu\mu)\phi(\mathrm{KK})$. No estimate is given for integrated luminosity lower than $0.1\,\mathrm{fb}^{-1}$. Indeed, with such a low statistics, we observe some fit instabilities, when trying to fit all detector parameter together. Work is still ongoing of this subject. The Tevatron line is the combined CDF/DØ uncertainty in 2009 scaled to $18\,\mathrm{fb}^{-1}$, as expected at the Tevatron by the end of Run 2 and discussed in Section 3.

It should be noted that for the first data, the LHC centre-of-mass energy is expected to be lower than 14 TeV, so that the $\mathrm{b\overline{b}}$ cross-section will also be smaller. A centre-of-mass energy of 10 TeV is assumed on the right plot of Figure 29.

### 10.6 Sensitivity with simplified analysis

The sensitivity to $\phi_\mathrm{s}$ has been studied with alternative and simplified analysis strategies [24].

The one-angle analysis gives a sensitivity $\sim$20% lower than the three-angle analysis [24].

The untagged analysis is not sensitive to small $\phi_\mathrm{s}$ values. It only provides useful measurement of $R_\perp$, $R_0$, $\Delta\Gamma_\mathrm{s}$ and $\Gamma_\mathrm{s}$.

# 11 Systematic uncertainties

In this section, we report some results on possible systematic effects due angular acceptance modeling, proper time and angular resolutions and the flavour tagging performance.

## 11.1 Angles

### 11.1.1 Angular acceptance

Systematic over- or underestimations of the acceptances in certain areas of the angles in the transversity base can lead to biases. To estimate the effect we use a simple model to distort the acceptances:

$$\epsilon_{\cos\theta}(\cos\theta, \varphi, \cos\psi) = \epsilon(\cos\theta, \varphi, \cos\psi)\,(1 \pm 0.05\cos((1+\cos\theta)\,\pi))\ , \quad (106)$$

$$\epsilon_{\varphi}(\cos\theta, \varphi, \cos\psi) = \epsilon(\cos\theta, \varphi, \cos\psi)\,(1 \pm 0.05\cos(\pi+\varphi))\ , \quad (107)$$

$$\epsilon_{\cos\psi}(\cos\theta, \varphi, \cos\psi) = \epsilon(\cos\theta, \varphi, \cos\psi)\,(1 \pm 0.05\cos((1+\cos\psi)\,\pi))\ . \quad (108)$$

The variations of $\pm 5\%$ are conservative upper bound on the distortions. Indeed, if the acceptances extracted from Monte Carlo would be wrong by more than $\pm 5\%$, one would detect it through the $B^0 \to J/\psi K^{*0}$ study, where the amplitudes and strong phases are already measured with high precision [60]. Table 28 in Appendix D shows that the results for the amplitudes and the strong phases differ significantly from the nominal values in the case of a systematic change in $\cos\theta$ and $\cos\psi$. The generally smaller effect for systematic changes dependent on the angle $\varphi$ can be explained by the less pronounced separation of CP-even and CP-odd eigenstates in this transversity angle. For the phase $\phi_s$ we find relative deviations of up to 3% in the New Physics scenario. The largest deviations when using the Standard Model value of $\phi_s$ is $(-7 \pm 4)\%$.

### 11.1.2 Angular resolution

Toy Monte Carlo studies have been performed to quantify the importance of the angular resolution, using $0.2\,\mathrm{fb}^{-1}$ of signal data [43]. Including resolutions which are slightly overestimated compared to the resolutions shown in Section 6, we do not observe biases. However, for angular resolutions twice as large as the current values systematic biases *are* observed. This means that the assumption to neglect the angular resolutions should be monitored. The systematic bias for 1 M $B^0_s \to J/\psi\phi$ signal events due to neglecting the angular resolutions while fixing the strong phases is significant for $|A_\perp|^2$ and equals $(0.23 \pm 0.07) \times 10^{-3}$. With $2\,\mathrm{fb}^{-1}$ of data the sensitivity to $|A_\perp|^2$ is $(4.6 \pm 0.2) \times 10^{-3}$ and hence the systematic bias can be neglected.

## 11.2 Proper time

The analysis of Monte Carlo data has shown that the signal proper time resolution models can be rather complicated. A double Gaussian resolution model can be a reasonable approximation if a fixed resolution model is assumed, but if a more precise per-event resolution model is considered and more precision is requested, Monte Carlo studies indicate that one additional Gaussian is needed and that the resolution parameters depend on the time itself.

| $\phi_s$ [rad] | $\mathcal{L}_{int}$ [$fb^{-1}$] | Sensitivity to $\phi_s$ : $\sigma \pm \delta_\sigma$ | | | |
|---|---|---|---|---|---|
| | | fit (a) | fit (b) | fit (c) | fit (d) |
| $-0.0368$ | 0.5 | $0.067 \pm 0.003$ | $0.064 \pm 0.003$ | $0.071 \pm 0.004$ | $0.065 \pm 0.003$ |
| $-0.0368$ | 2.0 | $0.0320 \pm 0.002$ | $0.030 \pm 0.002$ | $0.032 \pm 0.002$ | $0.029 \pm 0.002$ |
| $-0.736$ | 0.5 | $0.080 \pm 0.004$ | $0.077 \pm 0.004$ | $0.086 \pm 0.004$ | $0.078 \pm 0.004$ |
| $-0.736$ | 2.0 | $0.041 \pm 0.002$ | $0.039 \pm 0.002$ | $0.042 \pm 0.002$ | $0.039 \pm 0.002$ |

Table 24: Sensitivity to $\phi_s$ from the analysis of 300 toy experiments corresponding to different simulated statistics ($0.5\,fb^{-1}$and $2\,fb^{-1}$) and $\phi_s$ input values. The quoted results correspond to different approximated fit models of the same events generated with a per-event time resolution model. Fit model (a) is the same model (double Gaussian) used to generate the events (the baseline); fit (b) is based on a per-event time resolution model that neglects the tail contribution (single Gaussian); fit model (c) is based on a double Gaussian fixed time resolution model, while fit (d) is based on a single Gaussian fixed time resolution model. More details are described in Ref. [48].

Systematic uncertainties can appear in the B proper time measurement and error for different reasons, for example the imperfect alignment of the VELO detector. Studies on the $B^0_{(s)} \to h^+h^-$ channel [49] have shown that the proper time resolution deteriorates as the alignment accuracy decreases. A systematic effect on the VELO length scale [23] or in the $B$-field can also introduce a systematic effect on the track reconstruction (slope and momentum), and consequently on the time measurement. However, the global effect is small [49].

A possible method to reveal systematic effects in the track or vertex reconstruction was studied in Refs. [50, 51].

Systematic errors due to an incorrect track momentum scale calibration can also be checked on the $J/\psi \to \mu\mu$ channel, as discussed in Ref. [52].

In Ref. [25] it was demonstrated how to determine the parameters of the time resolution model through analyzing the time distribution of the control channels $B^0 \to J/\psi K^{*0}$ and $B^+ \to J/\psi K^+$.

It is also important to understand the effects of an imprecise resolution model or systematic effects in the time measurement on the physics parameters determination. This is discussed in Ref. [48]. In the following we briefly summarize the tests performed and the main conclusions of these studies.

Within the assumptions of the chosen parameters for the proper time resolution model of the signal and the parametrization of the background given in Table 22, we did not find any difference in the determination of $\phi_s$ if an event-by-event resolution model is considered (according to the formalism described in Section 9.3) rather than a fixed one. No systematic bias or differences in the sensitivity estimate are found. These conclusions are valid both for statistics corresponding to $0.5\,fb^{-1}$ and $2\,fb^{-1}$, with either an input value of $\phi_s = -0.0368\,rad$ or $-0.786\,rad$. In the cases where we approximated the signal time resolution model by neglecting the contribution of the tails, the value of $\phi_s$ is correctly determined without any biases or differences in sensitivity. The main results are summarized in Table 24. In this case, though, the determination of $\omega$ turns out to be significantly biased, with an effect which grows the greater the contribution of the tails to the time resolution. These results are valid even in the case of a worse signal mean time resolution, up to a value of 80 fs. We considered also the case where no scale factor for the resolution is included in the fit.

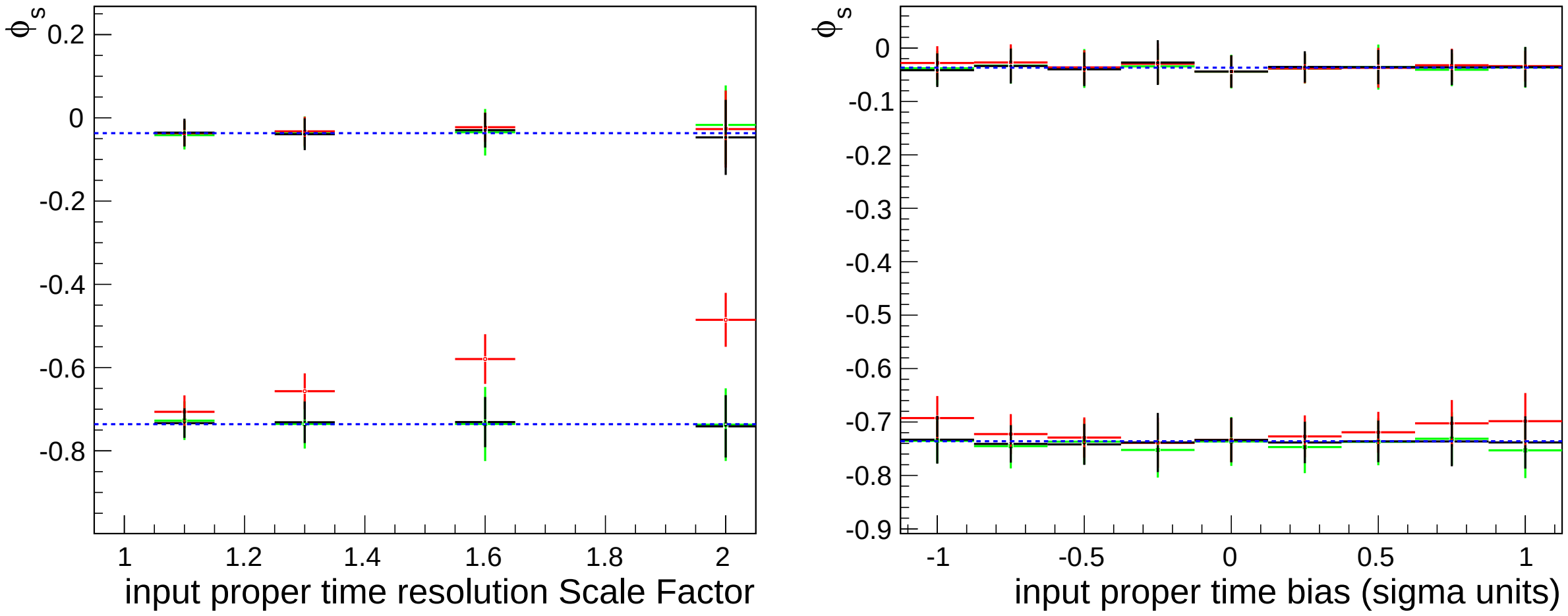

Figure 30: The plots show the dependence of the fitted mean of $\phi_s$ on the input scale factor (left) and bias (right) to the signal proper time resolution. The results are based on 150 toy experiments corresponding to $2\,\text{fb}^{-1}$ of data each. Both the input values $\phi_s = -0.0368$ rad and $-0.736$ rad are indicated. For the black data points, the proper time resolution parameters were fixed to the input values. For the red points, the proper time resolution parameters were fixed to the nominal values (bias=0 and scale factor=1). For the green points, the proper time resolution parameters were free to adjust in the fit. In all cases the mistag parameter is fixed to the input value.

By varying the input proper time resolution scale factor in the range [1.1, 2.0] no systematic deviation in the mean value of $\phi_s$ is found in the case of $\phi_s = -0.0368$ rad. For the case of $\phi_s = -0.736$ rad a large systematic error arises if the scale factor is ignored in the fit, but can be recovered leaving the time resolution parameters free to adjust in the fit. In a similar way, in case we introduced a bias in the proper time, the determination of $\phi_s$ is correct in case of $\phi_s = -0.0368$ rad, while it shows a small systematic deviation in case of $\phi_s = -0.736$ rad for large bias values. Also in this case the systematic bias cancels leaving the time resolution parameters free to adjust in the fit. These results are shown in Figure 30.

In conclusion, the $\phi_s$ measurement turns out to be robust with respect to the proper time systematic uncertainties. The results obtained are in agreement with the fact that the correlation of $\phi_s$ with all the other variables is low.

## 11.3 Flavour tagging

It is possible to extract the tagging parameters from the likelihood fit itself, but it is desirable to determine these quantities in other signal channels (see Section 8) to reduce the number of free parameters in the analysis. The important quantity to be determined is the mistag probability $\omega_{\text{tag}}$. The danger lies in an incorrect determination of this quantity. The effect of this can be seen in Figure 31, where three different studies of toy data are compared. In one of the data sets, represented by the black data points, $\omega_{\text{tag}}$ is changed in the generation and the fit. The other data sets, namely the red and green data points, simulate the effect of a wrongly determined $\omega_{\rm tag}$. In this case $\omega_{\rm tag}$ is changed in the generation of the toy data but not in the fit, where it is always fixed at $\omega_{\rm tag} = 0.334$. For the red data points the SM value of $\phi_{\rm s} = -0.0368$ rad has been used while for the data set with the green data points a value of $\phi_{\rm s} = -0.736$ rad was used. The fitted value of $\phi_{\rm s}$ for large $-\phi_{\rm s}$ is clearly affected by the wrongly determined $\omega_{\rm tag}$. For the small SM value of $\phi_{\rm s}$ this effect is not visible. The error on the measured $\phi_{\rm s}$ is affected in both cases. It is systematically over- or underestimated. The expected precision on $\omega_{\rm tag}$ leads to a small systematic error on $\phi_{\rm s}$ (if taken from an external source) or an equivalent small increase in the statistical error on $\phi_{\rm s}$ (if fitted in the same fit).

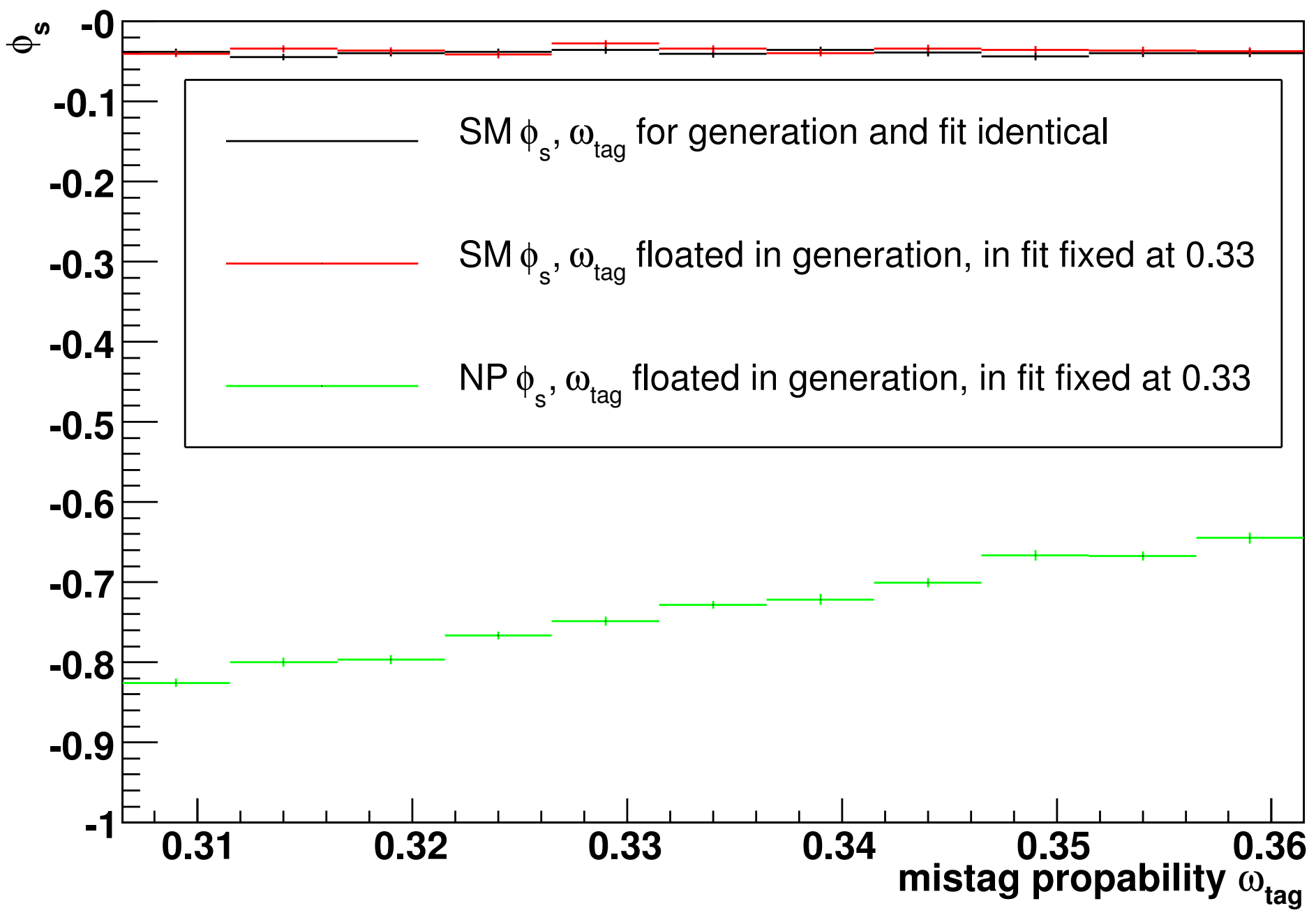

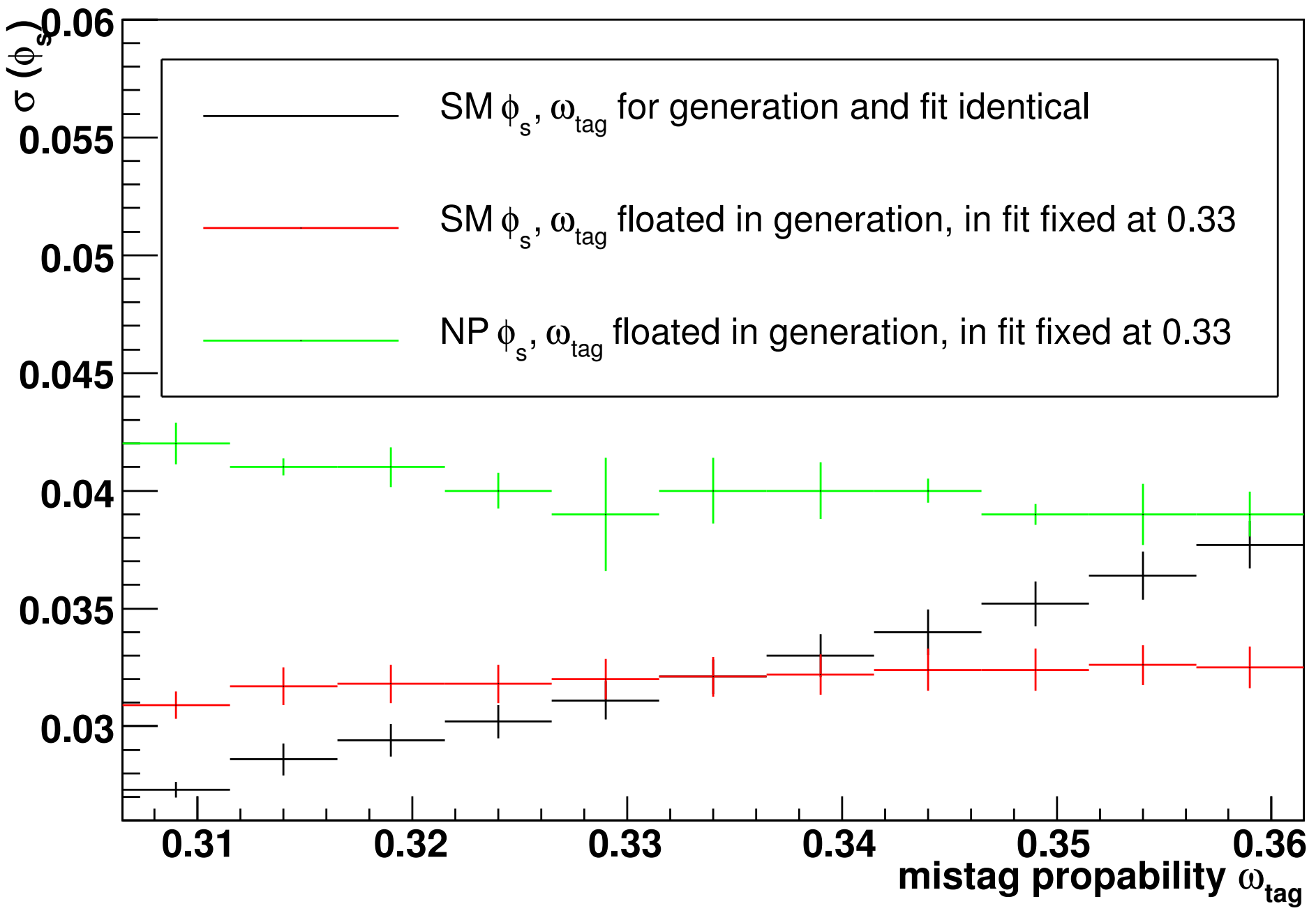

Figure 31: The upper plot shows the fitted mean of $\phi_s$ and its error for a scan over the mistag probability $\omega_{tag}$. For the black data points, $\omega_{tag}$ was varied for the generation and then fixed in the fit to the varied value. The red data points have been obtained by varying $\omega_{tag}$ in the generation but then fixing it to 0.334 in the fit. Every data point corresponds to 50 toy experiments using the configuration detailed in Table 22. No bias can be seen for the value of $\phi_s$ if the SM value for $\phi_s$ is used. In contrast to that a sizable dependence on $\phi_s$ is visible in the case of large $\phi_s = -0.736$ rad. The lower plot shows the effect that the wrongly determined $\omega_{tag}$ has on the error on $\phi_s$. If the data were generated with smaller $\omega_{tag}$ than 0.334 the error on $\phi_s$ is overestimated, if the $\omega_{tag}$ used in the generation is larger than 0.334 the fit underestimates the error.

The present observed difference on mistag for $\mathrm{B_s^0}$ and $\overline{\mathrm{B}}{}_\mathrm{s}^0$ in the Monte Carlo is below statistical error for $2\,\mathrm{fb}^{-1}$ [39]. It will be measured in data from flavour-specific control channels. It can be explicitly accounted for in the fit.

## 11.4 Background

The possible systematic effects due to the background description remain to be studied. We plan to measure all background properties in real data, using the $\mathrm{B_s^0}$ mass side-bands. According to [53], we expect this systematic uncertainty to be small, with respect to the statistical uncertainty on $\phi_{\rm s}$.

## 11.5 Summary on systematic uncertainties

A summary of the systematic effects presented in the previous sections is given in Table 25. No irreducible systematic uncertainty has been identified so far.

| Parameter | Variation | $\lvert\phi_{\rm s}^{\rm wrong} - \phi_{\rm s}^{\rm true}\rvert/\phi_{\rm s}^{\rm true}$ |
|---|---|---|
| Angular distortions | $\pm 5\%$ | 7% |
| Proper time resolution | $(38 \pm 5)$ fs | 6% |
| Mistag | $(34 \pm 1)\%$ | 7% |

Table 25: Relative systematic variation on $\phi_{\rm s}$ (column 3), due to parameter variations (columns 1 and 2).

The justification of the $\pm 5\%$ variations of the angular distribution is given in Section 11.1.1. For the systematic biases on $\phi_{\rm s}$ due to proper time resolution and the mistag, an approximate formula is used [54]:

$$\frac{\lvert\phi_{\rm s}^{\rm wrong} - \phi_{\rm s}^{\rm true}\rvert}{\phi_{\rm s}^{\rm true}} = \frac{D^{\rm wrong}}{D^{\rm true}} - 1 \tag{109}$$

where $D = \exp(-\frac{1}{2}(\sigma_t \Delta m_{\rm s})^2)$ for the proper time resolution and $D = (1 - 2\omega)$ for the mistag. We have checked with toy Monte Carlo experiment that the above formula gives a conservative estimate of the systematic bias. The proper time resolution can be measured thanks to the prompt background component. The difference between the proper time resolution measured in signal (38 fs) and in prompt events (43 fs) is taken as a systematic uncertainty. We have seen in Section 8.3 that the opposite side mistag can be extracted with a relative precision of 0.3% using only $B^0 \to J/\psi K^{*0}$ events. The total mistag can be extracted from $B_s^0 \to D_s^- \pi^+$ with a relative precision of 1.1% [55]. Taking into account possible differences when exporting the mistag from control channels to the signal channel, a conservative variation of 2.9% is considered in Table 25.

# 12 Conclusions

We have presented the basic steps necessary to perform a measurement of $\phi_s$ at LHCb. This measurement is one of the most important, but also one of the most challenging at LHCb, due to various experimental techniques that have to be under control: flavour tagging, time and angular dependent analysis, background rejection, multi-parameter fit (including low statistics instabilities, resonant and non-resonant contributions to the $K^+K^-$ final state). A strategy has been defined to start from an analysis "as simple as possible" and increase sensitivity and complexity step by step.

Based on our latest Monte Carlo samples, we expect to select $\sim 117\,000$ $B_s^0 \to J/\psi(\mu\mu)\phi(KK)$ events per $2\,fb^{-1}$ after full trigger (L0 and HLT). The background has been estimated using various event samples: $b\overline{b}$, $B_{u,d,s} \to J/\psi X$, minimum bias, inclusive $J/\psi$. The background over signal ratio is $B/S$ ~2.1, dominated by the prompt component. This prompt component should be easy to isolate in real data, since its average proper time is zero. The typical proper time resolution is $\sim 40\,fs$. The effective tagging power is $\varepsilon_{tag}(1-2\omega)^2$ ~6.2%. The mistag rate can be extracted from the control channels like $B^+ \to J/\psi K^+$ and $B^0 \to J/\psi K^{*0}$ for opposite-side and $B_s^0 \to D_s^- \pi^+$ and $B_s^0 \to D_s^- \mu^+ \nu$ for same-side tagging. It can also be measured, though with less precision, on the $B_s^0 \to J/\psi\phi$ sample itself. The angular and proper time acceptances are flat within $\sim 10\%$. The small distortions come from the requirement to have the four decay tracks reconstructed within the LHCb detector. They can be ignored without producing bias on $\phi_s$ during the first $2\,fb^{-1}$, if $\phi_s$ has the Standard Model value. However, they bias the amplitudes by $\sim 8\%$.

From toy Monte Carlo studies, with a three-angle time-dependent tagged analysis, including background, proper time resolution and mistag, we expect a statistical uncertainty of $\sigma(\phi_s) \simeq 0.03$ rad for a data set of $2\,fb^{-1}$. This uncertainty would increase strongly were the flavour tagging performance to degrade, decrease with $\Delta\Gamma_s$, increase less strongly with the CP-odd fraction ($R_\perp$), increase slightly with $-\phi_s$, increase with the fraction of long-lived background and increase with the proper time resolution. The dependence on $\Gamma_s$ and on the fraction of prompt background is negligible. The relative uncertainty on the expected $\sigma(\phi_s)$ is greater than 25%, due to the large uncertainties on the $b\overline{b}$ cross-section at the LHC and on the branching ratio of $B_s^0 \to J/\psi\phi$.

Control channels are used to measure the flavour tagging performance and proper time resolution. As a cross-check, these parameters can be extracted from the $B_s^0 \to J/\psi\phi$ channel alone. The extraction of the angular acceptance relies on Monte Carlo. It can be cross-checked with real data using $B^0 \to J/\psi K^{*0}$ events.

The sensitivity to systematic uncertainties due to proper time and angular resolutions, angular acceptance and flavour tagging have been studied and found to be smaller than the statistical uncertainty, for $2\,fb^{-1}$ and for the Standard Model value of $\phi_s$. Some of these systematic uncertainties are expected to increase with the absolute value of $\phi_s$.

In addition to the golden mode $B_s^0 \to J/\psi(\mu\mu)\phi(KK)$, several other $B_s^0$ decay channels dominated by the $\overline{b} \to \overline{c}c\overline{s}$ tree-level transition are interesting to search for New

Physics using mixing-induced CP violation. The following pure CP-even modes have been studied with an older version of the simulation: $B_s^0 \to J/\psi\eta(\gamma\gamma)$, $B_s^0 \to J/\psi\eta(\pi^+\pi^-\pi^0)$, $B_s^0 \to \eta_c(4h)\phi$, $B_s^0 \to D_s^- D_s^+$ [56], $B_s^0 \to J/\psi\eta'(\rho^0\gamma)$ [57] and $B_s^0 \to J/\psi\eta'(\eta\pi^+\pi^-)$ [58]. The decay $B_s^0 \to J/\psi f_0$, with $f_0 \to \pi^+\pi^-$ has been recently studied [59]. In the future, we may also consider $B_s^0 \to J/\psi(ee)\phi$, $B_s^0 \to D_s{}^{(*)-} D_s{}^{(*)+}$ and higher $\psi$ resonances like $B_s^0 \to \psi(2S)\phi$.

Should $\phi_s$ be as large as the value currently preferred by the Tevatron data, then we will be able to discover New Physics in $B_s^0$–$\overline{B}_s^0$ mixing (i.e. establish $\phi_s \neq -2\beta_s$ with $5\sigma$ significance) with $0.2\,\mathrm{fb}^{-1}$ of data.

# A KK S-wave contributions

The quasi-two-body treatment of the decay $B_s^0 \to J/\psi\phi$ considers only the KK P-wave amplitudes in the vicinity of the $\phi(1020)$. In principle, the KK system can have contributions from any partial wave. However, the BaBar experiment showed that in both $B^0 \to K^+K^-K_S^0$ and $D^0 \to K^+K^-\pi^0$ the S-wave and P-wave contributions dominate in the mass range above threshold and below $1.1\,\mathrm{GeV}/c^2$ [62, 63]. In both cases the $f_0(980)$ contribution in this region is found to be comparable to and significantly interfere with that from $\phi(1020)$. A smaller non-resonant S-wave component is also found to be necessary to describe the data. These results lead us to investigate how large is the S-wave contribution to $B_s^0 \to J/\psi K^+K^-$ in the $\phi(1020)$ region. We notice that the production mechanism of the $K^+K^-$ pair is mode-dependent. In $B_s^0 \to J/\psi K^+K^-$ the $K^+K^-$ pair can only arise from $s\bar{s}$ while in $B^0 \to K^+K^-K_S^0$ and $D^0 \to K^+K^-\pi^0$ it can have contributions from both $s\bar{s}$ and $d\bar{d}$. This makes it difficult to give any quantitative argument for the S-wave component based on analogy between these decay modes. In [59] the S-wave $K^+K^-$ contribution under the $\phi(1020)$ peak is estimated to be $5-10\%$ for decay modes in which the $K^+K^-$ pair must come from $s\bar{s}$.

The S wave has a CP-eigenvalue of $-1$. Its angular distribution, $1 - \sin^2\theta\cos^2\varphi$ , is different from any of the three P-wave components. Including the S-wave contribution and its interference with the P-wave contributions in the time-dependent angular distribution of $B_s^0 \to J/\psi K^+K^-$ is necessary for correct extraction of $\phi_s$. Initial studies show that ignoring a 10% S-wave contribution can lead to a 15% bias on $\phi_s$ towards zero, and including it increases the error on $\phi_s$ by 20% but also removes any biases on the central value [64]. Further understanding of the effect of possible S-wave contribution on the measurement of $\phi_s$ requires a more complete model to simulate the dependence of the magnitude and phase of each partial wave amplitude on the invariant mass of the KK.

The description of the mass line shapes of $f_0(980)$ and $\phi(1020)$ are model dependent. In order to reduce model dependence we can choose not to use the KK mass but rely on angular distributions to separate S and P waves.

If the interference between the S-wave and P-wave amplitudes turns out to be significant, we can further use it to measure $\cos\phi_s$ and resolve the ambiguity in the determination of $\phi_s$, in the same way as BaBar measured $\cos 2\beta$ in $B^0 \to J/\psi K_S^0\pi^0$ [62]. This requires measuring $\delta_{S0}$, the strong phase difference between the S wave and the longitudinal P wave, as a function of the KK mass. Two branches are expected when plotting this function, each corresponding to a different solution for the strong phases and weak phase. It is straightforward to choose the physical solution since the P-wave phase is expected to rise rapidly through the $\phi(1020)$ mass region, while the S-wave is expected to vary relatively slowly, resulting in a rapidly falling $\delta_{S0}$ [62].

# B Specific aspects of EvtGen for P2VV decays

The decay of a B meson into a final state is governed by the EvtGen package, which generates per event the decay, given certain amplitudes. For $B_s^0 \to J/\psi\phi$ the calculation of the angular dependence is done using the EvtSVVHelAmp model, the same as used for $B^0 \to J/\psi K^{*0}$. However for $B_s^0 \to J/\psi\phi$ in general $\Delta\Gamma \neq 0$. Since the angular dependence in that case evolves with proper time, a new model has been developed for the calculation of the proper time-dependent amplitudes.

EvtPVVCPLH takes care of this calculation, distributing the correct time-dependent transversity amplitudes for the angular distribution to EvtSVVHelAmp. The model allows for CP violating time asymmetries including different lifetimes for the light and heavy mass eigenstates. Either of the two decay widths can be chosen to be the largest. The model is particularly intended for decays like $B_s^0 \to J/\psi\phi$. The code is based on SVV_CPLH by Anders Ryd [28].

## B.1 Functionality

The decay model works as follows. First of all the flavour of the *other* B at production is determined. This is used to determine the relevant flavour of the decaying B meson itself at production. Since the meson decays to a CP-eigenstate the flavour mixing of the produced B is irrelevant. Hence the proper time-dependent mixing behaviour generated in EvtIncoherentMixing is subsequently overruled, by generating a proper time distribution with an envelope lifetime equal to $\tau_{\mathrm{envelope}} = 1/(\Gamma - |\Delta\Gamma|/2)$. Here $\Gamma - |\Delta\Gamma|/2$ is the smallest eigenvalue of the decay width, leading to the longest possible lifetime. This is chosen such that the generation of the proper time distribution is most efficient.

Finally the transversity amplitudes are calculated *with respect to* this envelope lifetime probability. Defining

$$g_\pm = \frac{1}{2}\left(\mathrm{e}^{-(im_{\mathrm{L}}+\Gamma_{\mathrm{L}}/2)t} \pm \mathrm{e}^{-(im_{\mathrm{H}}+\Gamma_{\mathrm{H}}/2)t}\right) , \tag{110}$$

with $\Gamma_{\mathrm{H}} = \Gamma - \Delta\Gamma/2, \Gamma_{\mathrm{L}} = \Gamma + \Delta\Gamma/2$, the proper time-dependent transversity amplitudes are

$$h_i(t) = \langle f_{\mathrm{CP}}|H_i|\mathrm{B}(t)\rangle = h_i[g_+(t) + \lambda_f g_-(t)] , \tag{111}$$

with

$$\lambda_f = \eta \mathrm{e}^{-i\phi_{\mathrm{s}}} . \tag{112}$$

Here $\eta = 1$ for the CP-even amplitudes $A_{0,\parallel}$ and $\eta = -1$ for the CP-odd amplitude $A_\perp$. Since the envelope lifetime has already been generated, it now suffices to calculate

$$h_i(t)/\sqrt{\mathrm{e}^{-t/\tau_{\mathrm{envelope}}}} = \mathrm{e}^{(\Gamma-|\Delta\Gamma|/2)t} h_i(t) \tag{113}$$

and transfer these amplitudes to EvtSVVHelAmp to generate the proper time-dependent angular distribution.

## B.2 Usage in the decay file

The usage in the decay file is the following:

BrFr V1 V2 PVV_CPLH $-\phi_s/2$ $\eta$ $|A_\parallel(0)|$ $\delta_\parallel$ $|A_0(0)|$ $\delta_0$ $|A_\perp(0)|$ $\delta_\perp$

Here the different arguments represent the following parameters:

- The first term is the branching fraction.
- The second and third term are the names of the two decay daughter vectors.
- The fourth argument is the name of the decay mode.
- The following argument is $-\phi_s/2$, the relevant CKM angle in radians.
- The next argument, $\eta$, is the CP-eigenvalue of the final state. Since the final state is a mixture of CP-eigenstates this argument is not used.
- The last six arguments are the absolute values and phases of the polarization amplitudes in the transversity basis $A_\parallel$, $A_0$, and $A_\perp$, and their respective strong phases $\delta_i$, at proper time equal to zero. Only the differences between the strong phases are physical.

The width difference $\Delta\Gamma_s$ and the mass difference $\Delta m_s$ are not input parameters to the model. These are called technically deltaGamma and deltaMs in a separate file called DECAY.DEC. The model can be used for negative deltaGamma.

The example below decays the $B_s^0$ meson to the two vector mesons J/$\psi$ and $\phi$. Input values are the SM value of $-\phi_s/2 = 0.02$ rad and the approximate world averages of the polarization amplitudes (as taken from Ref. [19]): $|A_\parallel(0)|^2 = 0.24$, $|A_0(0)|^2 = 0.60$ and $|A_\perp(0)|^2 = 0.16$, with the phases $\delta_\parallel = 2.50$ rad, $\delta_\perp = -0.17$ rad and $\delta_0 \equiv 0$.

```
Decay B_s0
   1.000  J/psi  phi  PVV_CPLH  0.02  1  0.49  2.50  0.775  0.0  0.4  -0.17
Enddecay
```

These are the values used in the LHCb event generation.

## B.3 Versions

The above decay is used in Bs_Jpsiphi,mm=CPV.dec since DecFiles/dkfiles version v13r9. The Gen/EvtGen code of EvtPVVCPLH is used since version v8r13 (also in DC06). Higher order terms have been corrected in November 2007 and will be used in versions > v8r16.

# C Full correlation matrix with 26 free parameters

The two following tables give the mean and RMS of the correlation coefficient distributions over ∼300 toy MC experiments for simultaneous fit to all 26 parameters using tagged and untagged events, as in Section 10.2. The number of events in each toy experiment is equivalent to $2\,\mathrm{fb}^{-1}$ of data. Absolute correlations greater than 0.20 are highlighted in yellow, while those ones less than 0.01 are displayed as 0. The first row gives the global correlation coefficient, the second row gives the uncertainty on this coefficient. In Table 26, $\phi_\mathrm{s} = -0.0368\,\mathrm{rad}$ while in Table 27, $\phi_\mathrm{s} = -0.736\,\mathrm{rad}$.

# D Sytematic effects due to angular acceptance distortions

See Table 28.

| | $\Gamma_s$ | $\Delta\Gamma_s$ | $R_\perp$ | $R_0$ | $\delta_\parallel$ | $\delta_\perp$ | $\Delta m_s$ | $M_{B_s^0}$ | $\phi_s$ | $\omega$ | $f_{m,1}^s$ | $\sigma_{m,1}$ | $\sigma_{m,2}$ | $f_{t,1}^s$ | $\mu_t^s$ | $\sigma_{t,1}^s$ | $\sigma_{t,2}^s$ | $\alpha_m^{Pr}$ | $\mu_t^{Pr}$ | $\sigma_t^{Pr}$ | $\alpha_m^{LL}$ | $f_{\tau_1}^{LL}$ | $\tau_{t,1}^{LL}$ | $\tau_{t,2}^{LL}$ | $\mu_t^{LL}$ | $\sigma_t^{LL}$ |
|---|---|---|---|---|---|---|---|---|---|---|---|---|---|---|---|---|---|---|---|---|---|---|---|---|---|---|
| | 0.57 | 0.67 | 0.63 | 0.55 | 0.52 | 0.76 | 0.68 | 0.09 | 0.14 | 0.5 | 0.94 | 0.92 | 0.92 | 0.79 | 0.69 | 0.69 | 0.75 | 0.17 | 0.53 | 0.47 | 0.19 | 0.73 | 0.58 | 0.72 | 0.87 | 0.76 |
| | 0.15 | 0.18 | 0.17 | 0.16 | 0.11 | 0.13 | 0.12 | 0.03 | 0.06 | 0.1 | 0.17 | 0.17 | 0.16 | 0.2 | 0.11 | 0.16 | 0.21 | 0.05 | 0.1 | 0.1 | 0.05 | 0.14 | 0.12 | 0.13 | 0.17 | 0.16 |
| $\Gamma_s$ | 1 | -0.5 | 0.3 | -0.3 | -0.03 | -0.04 | 0 | 0 | 0 | -0.01 | 0 | 0 | 0.02 | -0.01 | 0.06 | 0 | 0.02 | 0 | 0 | -0.02 | 0 | -0.02 | 0.1 | 0 | -0.02 | -0.02 |
| | 1 | 0.2 | 0.2 | 0.2 | 0.1 | 0.2 | 0 | 0 | 0 | 0.05 | 0 | 0 | 0.04 | 0.1 | 0.2 | 0 | 0.1 | 0 | 0 | 0.04 | 0 | 0.09 | 0.09 | 0 | 0.1 | 0.08 |
| $\Delta\Gamma_s$ | | 1 | -0.5 | 0.4 | 0 | 0.01 | 0 | 0 | 0 | 0 | 0 | 0 | 0 | 0 | 0.01 | 0.02 | 0 | 0 | 0 | 0 | 0 | 0.0 2 | -0.01 | -0.01 | 0.01 | 0 |
| | | 1 | 0.2 | 0.2 | 0 | 0.2 | 0 | 0 | 0 | 0 | 0 | 0 | 0 | 0 | 0.2 | 0.1 | 0 | 0 | 0 | 0 | 0 | 0.1 | 0.1 | 0.2 | 0.1 | 0 |
| $R_\perp$ | | | 1 | -0.4 | -0.2 | -0.06 | -0.01 | 0 | 0 | 0.03 | 0 | 0 | 0 | 0 | 0 | 0 | 0 | 0 | 0 | 0 | 0 | -0.02 | 0 | 0 | 0 | -0.01 |
| | | | 1 | 0.2 | 0.2 | 0.2 | 0.1 | 0 | 0 | 0.05 | 0 | 0 | 0 | 0 | 0.2 | 0 | 0 | 0 | 0 | 0 | 0 | 0.1 | 0 | 0 | 0 | 0.09 |
| $R_0$ | | | | 1 | -0.02 | 0 | 0 | 0 | 0 | -0.02 | 0.02 | -0.02 | -0.03 | 0 | 0 | 0.01 | 0 | 0 | 0 | 0 | 0 | 0.03 | -0.01 | 0 | 0 | 0.01 |
| | | | | 1 | 0.1 | 0 | 0 | 0 | 0 | 0.04 | 0.06 | 0.05 | 0.05 | 0 | 0 | 0.1 | 0 | 0 | 0 | 0 | 0 | 0.1 | 0.09 | 0 | 0 | 0.08 |
| $\delta_\parallel$ | | | | | 1 | 0.2 | 0 | 0 | 0 | -0.02 | -0.01 | 0 | 0.02 | 0 | 0.02 | 0 | 0 | 0 | 0 | 0 | 0 | -0.03 | 0 | 0.01 | 0 | 0 |
| | | | | | 1 | 0.1 | 0 | 0 | 0 | 0.07 | 0.04 | 0 | 0.04 | 0 | 0.09 | 0 | 0 | 0 | 0 | 0 | 0 | 0.08 | 0 | 0.09 | 0 | 0 |
| $\delta_\perp$ | | | | | | 1 | 0.5 | 0 | 0 | 0.02 | 0 | 0 | 0 | 0.02 | -0.4 | -0.01 | -0.02 | 0 | 0 | 0.02 | 0 | 0 | 0 | 0.01 | 0.01 | 0 |
| | | | | | | 1 | 0.1 | 0 | 0 | 0.06 | 0 | 0 | 0 | 0.1 | 0.2 | 0.1 | 0.1 | 0 | 0 | 0.07 | 0 | 0 | 0 | 0.2 | 0.1 | 0 |
| $\Delta m_s$ | | | | | | | 1 | 0 | 0 | -0.02 | 0 | 0 | -0.01 | 0 | 0.05 | 0.01 | 0 | 0 | 0 | 0 | 0 | 0 | 0 | 0 | 0 | 0 |
| | | | | | | | 1 | 0 | 0 | 0.07 | 0 | 0 | 0.06 | 0 | 0.1 | 0.1 | 0 | 0 | 0 | 0 | 0 | 0 | 0 | 0 | 0 | 0 |
| $M_{B_s^0}$ | | | | | | | | 1 | 0 | 0 | 0 | 0 | 0 | 0 | 0 | 0 | 0 | 0 | 0 | 0 | -0.05 | 0 | 0 | 0 | 0 | 0 |
| | | | | | | | | 1 | 0 | 0 | 0 | 0 | 0 | 0 | 0 | 0 | 0 | 0 | 0 | 0 | 0.02 | 0 | 0 | 0 | 0 | 0 |
| $\phi_s$ | | | | | | | | | 1 | -0.05 | 0 | 0 | 0 | 0 | 0 | 0 | 0 | 0 | 0 | 0 | 0 | 0 | 0 | 0 | 0 | 0 |
| | | | | | | | | | 1 | 0.04 | 0 | 0 | 0 | 0 | 0 | 0 | 0 | 0 | 0 | 0 | 0 | 0 | 0 | 0 | 0 | 0 |
| $\omega$ | | | | | | | | | | 1 | 0 | 0 | 0 | 0.2 | -0.06 | -0.2 | -0.02 | 0 | -0.07 | 0 | 0 | 0 | 0 | 0 | -0.02 | 0.01 |
| | | | | | | | | | | 1 | 0 | 0 | 0 | 0.2 | 0.08 | 0.2 | 0.1 | 0 | 0.09 | 0 | 0 | 0 | 0 | 0 | 0.05 | 0.05 |
| $f_{m,1}^s$ | | | | | | | | | | | 1 | -0.9 | -0.9 | 0 | 0.02 | 0 | -0.01 | 0 | 0 | 0 | 0.01 | 0.09 | 0.06 | 0.01 | -0.02 | -0.01 |
| | | | | | | | | | | | 1 | 0.2 | 0.2 | 0.3 | 0.07 | 0 | 0.3 | 0 | 0 | 0 | 0.02 | 0.1 | 0.05 | 0.08 | 0.2 | 0.2 |
| $\sigma_{m,1}$ | | | | | | | | | | | | 1 | 0.8 | -0.01 | -0.01 | 0 | 0.01 | 0 | 0 | 0 | -0.01 | -0.08 | -0.06 | 0 | 0.01 | 0 |
| | | | | | | | | | | | | 1 | 0.3 | 0.3 | 0.07 | 0 | 0.3 | 0 | 0 | 0 | 0.02 | 0.1 | 0.05 | 0 | 0.2 | 0 |
| $\sigma_{m,2}$ | | | | | | | | | | | | | 1 | 0 | -0.02 | 0 | 0.01 | 0 | 0 | 0 | -0.02 | -0.1 | -0.07 | -0.02 | 0.01 | 0.01 |
| | | | | | | | | | | | | | 1 | 0.3 | 0.07 | 0 | 0.3 | 0 | 0 | 0 | 0.02 | 0.1 | 0.05 | 0.08 | 0.2 | 0.1 |
| $f_{t,1}^s$ | | | | | | | | | | | | | | 1 | -0.01 | 0.4 | 0.5 | 0 | -0.06 | 0 | 0 | -0.02 | 0 | 0.02 | -0.03 | 0.01 |
| | | | | | | | | | | | | | | 1 | 0.1 | 0.3 | 0.4 | 0 | 0.1 | 0 | 0 | 0.2 | 0 | 0.2 | 0.3 | 0.2 |
| $\mu_t^s$ | | | | | | | | | | | | | | | 1 | 0.03 | 0.08 | 0.02 | 0.02 | -0.05 | 0 | -0.03 | 0 | -0.01 | -0.04 | -0.02 |
| | | | | | | | | | | | | | | | 1 | 0.2 | 0.2 | 0.02 | 0.06 | 0.06 | 0 | 0.1 | 0 | 0.1 | 0.1 | 0.09 |
| $\sigma_{t,1}^s$ | | | | | | | | | | | | | | | | 1 | 0.2 | 0 | 0.07 | 0 | 0 | -0.02 | 0 | 0.01 | -0.02 | 0 |
| | | | | | | | | | | | | | | | | 1 | 0.3 | 0 | 0.1 | 0 | 0 | 0.1 | 0 | 0.1 | 0.2 | 0 |
| $\sigma_{t,2}^s$ | | | | | | | | | | | | | | | | | 1 | 0 | -0.02 | 0 | 0 | 0 | 0 | 0 | 0.01 | -0.07 |
| | | | | | | | | | | | | | | | | | 1 | 0 | 0.1 | 0 | 0 | 0 | 0 | 0 | 0.3 | 0.2 |
| $\alpha_m^{Pr}$ | | | | | | | | | | | | | | | | | | 1 | -0.01 | -0.01 | -0.1 | -0.02 | 0 | 0 | -0.03 | -0.02 |
| | | | | | | | | | | | | | | | | | | 1 | 0.02 | 0.02 | 0.06 | 0.02 | 0 | 0 | 0.02 | 0.02 |
| $\mu_t^{Pr}$ | | | | | | | | | | | | | | | | | | | 1 | 0.2 | 0 | 0.1 | 0.02 | 0.02 | 0.2 | 0.4 |
| | | | | | | | | | | | | | | | | | | | 1 | 0.08 | 0 | 0.1 | 0.07 | 0.09 | 0.1 | 0.1 |
| $\sigma_t^{Pr}$ | | | | | | | | | | | | | | | | | | | | 1 | 0 | 0.2 | 0.03 | 0.03 | 0.3 | 0.2 |
| | | | | | | | | | | | | | | | | | | | | 1 | 0 | 0.1 | 0.07 | 0.09 | 0.1 | 0.1 |
| $\alpha_m^{LL}$ | | | | | | | | | | | | | | | | | | | | | 1 | 0 | 0 | 0 | 0 | 0 |
| | | | | | | | | | | | | | | | | | | | | | 1 | 0 | 0 | 0 | 0 | 0 |
| $f_{\tau_1}^{LL}$ | | | | | | | | | | | | | | | | | | | | | | 1 | -0.4 | -0.5 | 0.4 | 0.4 |
| | | | | | | | | | | | | | | | | | | | | | | 1 | 0.1 | 0.1 | 0.2 | 0.2 |
| $\tau_{t,1}^{LL}$ | | | | | | | | | | | | | | | | | | | | | | | 1 | 0.4 | -0.1 | -0.09 |
| | | | | | | | | | | | | | | | | | | | | | | | 1 | 0.1 | 0.08 | 0.08 |
| $\tau_{t,2}^{LL}$ | | | | | | | | | | | | | | | | | | | | | | | | 1 | -0.4 | -0.3 |
| | | | | | | | | | | | | | | | | | | | | | | | | 1 | 0.1 | 0.1 |
| $\mu_t^{LL}$ | | | | | | | | | | | | | | | | | | | | | | | | | 1 | 0.7 |
| | | | | | | | | | | | | | | | | | | | | | | | | | 1 | 0.2 |
| $\sigma_t^{LL}$ | | | | | | | | | | | | | | | | | | | | | | | | | | 1 |
| | | | | | | | | | | | | | | | | | | | | | | | | | | 1 |

Table 26: Mean and RMS of the correlation coefficient distributions over $\sim 300$ toy MC experiments for a fit to all parameters, with $2\,\mathrm{fb}^{-1}$; $\phi_s = -0.0368\,\mathrm{rad}$.

| | $\Gamma_s$ | $\Delta\Gamma_s$ | $R_\perp$ | $R_0$ | $\delta_\parallel$ | $\delta_\perp$ | $\Delta m_s$ | $M_{B_s^0}$ | $\phi_s$ | $\omega$ | $f^s_{m,1}$ | $\sigma_{m,1}$ | $\sigma_{m,2}$ | $f^s_{t,1}$ | $\mu^s_t$ | $\sigma^s_{t,1}$ | $\sigma^s_{t,2}$ | $\alpha^{Pr}_m$ | $\mu^{Pr}_t$ | $\sigma^{Pr}_t$ | $\alpha^{LL}_m$ | $f^{LL}_{\tau_1}$ | $\tau^{LL}_{t,1}$ | $\tau^{LL}_{t,2}$ | $\mu^{LL}_t$ | $\sigma^{LL}_t$ |
|---|---|---|---|---|---|---|---|---|---|---|---|---|---|---|---|---|---|---|---|---|---|---|---|---|---|---|
| | 0.48 | 0.58 | 0.6 | 0.51 | 0.52 | 0.47 | 0.55 | 0.09 | 0.62 | 0.74 | 0.95 | 0.93 | 0.93 | 0.83 | 0.62 | 0.78 | 0.78 | 0.17 | 0.54 | 0.5 | 0.2 | 0.76 | 0.59 | 0.74 | 0.91 | 0.8 |
| | 0.1 | 0.1 | 0.1 | 0.1 | 0.1 | 0.1 | 0.08 | 0.06 | 0.08 | 0.09 | 0.1 | 0.1 | 0.1 | 0.2 | 0.09 | 0.1 | 0.2 | 0.06 | 0.06 | 0.07 | 0.06 | 0.07 | 0.08 | 0.07 | 0.09 | 0.1 |
| $\Gamma_s$ | 1 | -0.3 | 0.2 | -0.2 | -0.02 | 0 | 0.01 | 0 | -0.2 | 0.04 | 0 | 0 | 0.02 | 0 | 0.04 | 0 | 0.02 | 0 | 0 | -0.01 | 0 | -0.01 | 0.1 | 0 | -0.01 | 0 |
| | 1 | 0.1 | 0.1 | 0.1 | 0.07 | 0 | 0.07 | 0 | 0.1 | 0.1 | 0 | 0 | 0.04 | 0 | 0.09 | 0 | 0.07 | 0 | 0 | 0.04 | 0 | 0.08 | 0.09 | 0 | 0.05 | 0 |
| $\Delta\Gamma_s$ | | 1 | -0.4 | 0.4 | 0 | -0.03 | 0 | 0 | 0.01 | 0.02 | 0 | 0 | 0 | -0.01 | 0.03 | 0 | 0 | 0 | 0 | 0 | 0 | 0.01 | -0.01 | 0 | 0 | 0 |
| | | 1 | 0.1 | 0.1 | 0 | 0.07 | 0 | 0 | 0.2 | 0.2 | 0 | 0 | 0 | 0.09 | 0.1 | 0 | 0 | 0 | 0 | 0 | 0 | 0.1 | 0.1 | 0 | 0 | 0 |
| $R_\perp$ | | | 1 | -0.4 | -0.2 | -0.03 | 0 | 0 | -0.2 | 0.03 | 0 | 0.01 | 0.01 | 0.01 | 0 | 0 | 0.01 | 0 | 0 | 0 | 0 | -0.02 | 0.01 | 0 | 0 | 0 |
| | | | 1 | 0.1 | 0.2 | 0.1 | 0 | 0 | 0.1 | 0.2 | 0 | 0.05 | 0.04 | 0.09 | 0 | 0 | 0.1 | 0 | 0 | 0 | 0 | 0.1 | 0.1 | 0 | 0 | 0 |
| $R_0$ | | | | 1 | -0.04 | -0.02 | 0.01 | 0 | 0.1 | -0.04 | 0.02 | -0.02 | -0.03 | 0 | 0 | 0 | 0 | 0 | 0 | 0 | 0 | 0.03 | 0 | 0 | 0 | 0 |
| | | | | 1 | 0.09 | 0.06 | 0.08 | 0 | 0.1 | 0.2 | 0.04 | 0.04 | 0.04 | 0 | 0 | 0 | 0 | 0 | 0 | 0 | 0 | 0.1 | 0 | 0 | 0 | 0 |
| $\delta_\parallel$ | | | | | 1 | 0.3 | -0.04 | 0 | 0.02 | -0.01 | 0 | 0 | 0.02 | 0 | 0.02 | 0 | 0 | 0 | 0 | 0 | 0 | 0 | -0.01 | 0 | 0 | 0 |
| | | | | | 1 | 0.1 | 0.09 | 0 | 0.08 | 0.1 | 0 | 0 | 0.05 | 0 | 0.09 | 0 | 0 | 0 | 0 | 0 | 0 | 0 | 0.08 | 0 | 0 | 0 |
| $\delta_\perp$ | | | | | | 1 | 0.1 | 0 | 0 | -0.04 | 0 | 0 | 0 | 0 | -0.2 | 0 | 0 | 0 | -0.01 | 0 | 0 | 0 | 0 | 0 | 0 | 0 |
| | | | | | | 1 | 0.07 | 0 | 0 | 0.08 | 0 | 0 | 0 | 0 | 0.07 | 0 | 0 | 0 | 0.05 | 0 | 0 | 0 | 0 | 0 | 0 | 0 |
| $\Delta m_s$ | | | | | | | 1 | 0 | 0.02 | -0.05 | 0.01 | -0.02 | 0 | -0.01 | 0.4 | 0 | 0.02 | 0 | 0.01 | -0.01 | 0 | -0.02 | 0.02 | 0.01 | 0 | 0 |
| | | | | | | | 1 | 0 | 0.09 | 0.1 | 0.04 | 0.05 | 0 | 0.08 | 0.1 | 0 | 0.08 | 0 | 0.06 | 0.05 | 0 | 0.08 | 0.09 | 0.1 | 0 | 0 |
| $M_{B_s^0}$ | | | | | | | | 1 | 0 | 0 | 0 | 0 | 0 | 0 | 0 | 0 | 0 | 0 | 0 | 0 | -0.05 | 0 | 0 | 0 | 0 | 0 |
| | | | | | | | | 1 | 0 | 0 | 0 | 0 | 0 | 0 | 0 | 0 | 0 | 0 | 0 | 0 | 0.02 | 0 | 0 | 0 | 0 | 0 |
| $\phi_s$ | | | | | | | | | 1 | -0.4 | 0 | 0 | 0 | 0.02 | -0.03 | -0.04 | -0.02 | 0 | 0 | 0 | 0 | 0 | 0 | 0 | 0 | 0.01 |
| | | | | | | | | | 1 | 0.1 | 0 | 0 | 0 | 0.08 | 0.1 | 0.1 | 0.08 | 0 | 0 | 0 | 0 | 0 | 0 | 0 | 0 | 0.06 |
| $\omega$ | | | | | | | | | | 1 | 0 | 0 | 0 | 0.2 | 0 | -0.2 | 0 | 0 | -0.09 | -0.01 | 0 | -0.02 | 0 | 0 | -0.04 | 0 |
| | | | | | | | | | | 1 | 0 | 0 | 0 | 0.2 | 0 | 0.2 | 0 | 0 | 0.08 | 0.07 | 0 | 0.1 | 0 | 0 | 0.08 | 0 |
| $f^s_{m,1}$ | | | | | | | | | | | 1 | -0.9 | -0.9 | 0.02 | 0.03 | 0 | 0.01 | 0 | 0 | 0 | 0.01 | 0.09 | 0.06 | 0 | 0 | 0 |
| | | | | | | | | | | | 1 | 0.2 | 0.2 | 0.3 | 0.06 | 0.2 | 0.3 | 0 | 0 | 0 | 0.02 | 0.1 | 0.05 | 0 | 0 | 0 |
| $\sigma_{m,1}$ | | | | | | | | | | | | 1 | 0.8 | -0.02 | -0.02 | 0 | -0.01 | 0 | 0 | 0 | -0.01 | -0.08 | -0.06 | 0 | 0 | 0 |
| | | | | | | | | | | | | 1 | 0.3 | 0.3 | 0.06 | 0.2 | 0.3 | 0 | 0 | 0 | 0.02 | 0.1 | 0.05 | 0 | 0 | 0 |
| $\sigma_{m,2}$ | | | | | | | | | | | | | 1 | -0.02 | -0.02 | 0 | -0.01 | 0 | 0 | 0 | -0.02 | -0.1 | -0.08 | -0.02 | 0 | 0 |
| | | | | | | | | | | | | | 1 | 0.3 | 0.06 | 0.2 | 0.3 | 0 | 0 | 0 | 0.01 | 0.09 | 0.05 | 0.06 | 0 | 0 |
| $f^s_{t,1}$ | | | | | | | | | | | | | | 1 | 0 | 0.4 | 0.5 | 0 | -0.05 | 0 | 0 | -0.02 | 0 | 0.03 | -0.05 | 0.01 |
| | | | | | | | | | | | | | | 1 | 0 | 0.4 | 0.5 | 0 | 0.1 | 0 | 0 | 0.1 | 0 | 0.1 | 0.2 | 0.2 |
| $\mu^s_t$ | | | | | | | | | | | | | | | 1 | 0 | 0.04 | 0.01 | 0 | -0.02 | 0 | -0.03 | 0.02 | 0 | -0.02 | 0 |
| | | | | | | | | | | | | | | | 1 | 0 | 0.1 | 0.02 | 0 | 0.06 | 0 | 0.1 | 0.1 | 0 | 0.09 | 0 |
| $\sigma^s_{t,1}$ | | | | | | | | | | | | | | | | 1 | 0.1 | 0 | 0.07 | 0.01 | 0 | 0 | 0 | 0.02 | -0.01 | 0 |
| | | | | | | | | | | | | | | | | 1 | 0.4 | 0 | 0.1 | 0.09 | 0 | 0 | 0 | 0.1 | 0.2 | 0 |
| $\sigma^s_{t,2}$ | | | | | | | | | | | | | | | | | 1 | 0 | -0.01 | -0.01 | 0 | -0.02 | 0.01 | 0.02 | 0 | -0.0 |
| | | | | | | | | | | | | | | | | | 1 | 0 | 0.1 | 0.1 | 0 | 0.2 | 0.1 | 0.1 | 0.2 | 0.2 |
| $\alpha^{Pr}_m$ | | | | | | | | | | | | | | | | | | 1 | -0.01 | -0.01 | -0.1 | -0.02 | 0 | 0 | -0.03 | -0.03 |
| | | | | | | | | | | | | | | | | | | 1 | 0.02 | 0.02 | 0.05 | 0.02 | 0 | 0 | 0.02 | 0.02 |
| $\mu^{Pr}_t$ | | | | | | | | | | | | | | | | | | | 1 | 0.2 | 0 | 0.2 | 0 | 0 | 0.3 | 0.4 |
| | | | | | | | | | | | | | | | | | | | 1 | 0.06 | 0 | 0.09 | 0 | 0 | 0.1 | 0.07 |
| $\sigma^{Pr}_t$ | | | | | | | | | | | | | | | | | | | | 1 | 0 | 0.2 | 0 | 0 | 0.3 | 0.2 |
| | | | | | | | | | | | | | | | | | | | | 1 | 0 | 0.09 | 0 | 0 | 0.09 | 0.1 |
| $\alpha^{LL}_m$ | | | | | | | | | | | | | | | | | | | | | 1 | 0 | 0 | 0 | 0 | 0 |
| | | | | | | | | | | | | | | | | | | | | | 1 | 0 | 0 | 0 | 0 | 0 |
| $f^{LL}_{\tau_1}$ | | | | | | | | | | | | | | | | | | | | | | 1 | -0.5 | -0.5 | 0.5 | 0.4 |
| | | | | | | | | | | | | | | | | | | | | | | 1 | 0.08 | 0.09 | 0.1 | 0.1 |
| $\tau^{LL}_{t,1}$ | | | | | | | | | | | | | | | | | | | | | | | 1 | 0.3 | -0.1 | -0.1 |
| | | | | | | | | | | | | | | | | | | | | | | | 1 | 0.1 | 0.07 | 0.07 |
| $\tau^{LL}_{t,2}$ | | | | | | | | | | | | | | | | | | | | | | | | 1 | -0.4 | -0.4 |
| | | | | | | | | | | | | | | | | | | | | | | | | 1 | 0.09 | 0.1 |
| $\mu^{LL}_t$ | | | | | | | | | | | | | | | | | | | | | | | | | 1 | 0.7 |
| | | | | | | | | | | | | | | | | | | | | | | | | | 1 | 0.1 |
| $\sigma^{LL}_t$ | | | | | | | | | | | | | | | | | | | | | | | | | | 1 |
| | | | | | | | | | | | | | | | | | | | | | | | | | | 1 |

Table 27: Mean and RMS of the correlation coefficient distributions over $\sim 300$ toy MC experiments for a fit to all parameters, with $2\,\text{fb}^{-1}$; $\phi_s = -0.736\,\text{rad}$.

| Relative bias in % with $\phi_s = -0.0368\,\text{rad}$ | | | | | | |
|---|---|---|---|---|---|---|
| | $\epsilon_{\cos\theta+}$ | $\epsilon_{\cos\theta-}$ | $\epsilon_{\varphi+}$ | $\epsilon_{\varphi-}$ | $\epsilon_{\cos\psi+}$ | $\epsilon_{\cos\psi-}$ |
| $\phi_s$ | $-4.62 \pm 4.09$ | $-6.78 \pm 3.96$ | $-3.41 \pm 4.10$ | $-4.75 \pm 4.00$ | $+4.29 \pm 4.18$ | $+0.52 \pm 3.96$ |
| $\lvert A_0\rvert^2$ | $-1.17 \pm 0.03$ | $+1.00 \pm 0.03$ | $-0.01 \pm 0.02$ | $-0.02 \pm 0.02$ | $-4.04 \pm 0.02$ | $+3.95 \pm 0.03$ |
| $\lvert A_\perp\rvert^2$ | $+9.40 \pm 0.08$ | $-10.84 \pm 0.09$ | $-0.06 \pm 0.08$ | $+0.00 \pm 0.09$ | $+4.86 \pm 0.09$ | $-4.94 \pm 0.08$ |
| $\Gamma_s$ | $+0.24 \pm 0.02$ | $-0.26 \pm 0.02$ | $+0.02 \pm 0.02$ | $-0.00 \pm 0.02$ | $+0.10 \pm 0.02$ | $+0.02 \pm 0.02$ |
| $\Delta\Gamma_s$ | $-4.66 \pm 0.80$ | $+5.90 \pm 0.90$ | $+0.04 \pm 0.79$ | $-0.39 \pm 0.85$ | $-1.38 \pm 0.85$ | $-0.59 \pm 0.86$ |
| $\delta_\perp$ | $-2.48 \pm 0.17$ | $+3.37 \pm 0.16$ | $-0.37 \pm 0.19$ | $-0.09 \pm 0.18$ | $-0.14 \pm 0.19$ | $-0.49 \pm 0.18$ |
| $\delta_\parallel$ | $+5.37 \pm 0.06$ | $-6.39 \pm 0.07$ | $+0.53 \pm 0.16$ | $+0.50 \pm 0.15$ | $+0.26 \pm 0.15$ | $+0.52 \pm 0.16$ |
| $\Delta m_s$ | $-0.00 \pm 0.01$ | $+0.02 \pm 0.01$ | $-0.00 \pm 0.01$ | $-0.01 \pm 0.01$ | $+0.01 \pm 0.01$ | $-0.01 \pm 0.01$ |

| Relative bias in % with $\phi_s = -0.736\,\text{rad}$ | | | | | | |
|---|---|---|---|---|---|---|
| | $\epsilon_{\cos\theta+}$ | $\epsilon_{\cos\theta-}$ | $\epsilon_{\varphi+}$ | $\epsilon_{\varphi-}$ | $\epsilon_{\cos\psi+}$ | $\epsilon_{\cos\psi-}$ |
| $\phi_s$ | $+2.20 \pm 0.26$ | $-2.91 \pm 0.26$ | $-0.59 \pm 0.27$ | $-0.18 \pm 0.25$ | $+0.96 \pm 0.26$ | $-1.21 \pm 0.25$ |
| $\lvert A_0\rvert^2$ | $-1.09 \pm 0.02$ | $+1.00 \pm 0.02$ | $-0.01 \pm 0.02$ | $-0.02 \pm 0.02$ | $-3.98 \pm 0.02$ | $+3.97 \pm 0.02$ |
| $\lvert A_\perp\rvert^2$ | $+8.97 \pm 0.07$ | $-10.35 \pm 0.08$ | $+0.02 \pm 0.08$ | $+0.06 \pm 0.08$ | $+4.74 \pm 0.08$ | $-4.93 \pm 0.07$ |
| $\Gamma_s$ | $+0.15 \pm 0.02$ | $-0.18 \pm 0.02$ | $+0.04 \pm 0.02$ | $+0.06 \pm 0.02$ | $+0.07 \pm 0.02$ | $-0.04 \pm 0.02$ |
| $\Delta\Gamma_s$ | $-2.97 \pm 0.77$ | $+3.31 \pm 0.84$ | $-0.89 \pm 0.87$ | $-1.35 \pm 0.84$ | $-0.56 \pm 0.85$ | $+1.68 \pm 0.86$ |
| $\delta_\perp$ | $-2.34 \pm 0.12$ | $+2.54 \pm 0.13$ | $-0.04 \pm 0.13$ | $-0.04 \pm 0.13$ | $+0.22 \pm 0.13$ | $-0.37 \pm 0.14$ |
| $\delta_\parallel$ | $+5.39 \pm 0.07$ | $-6.08 \pm 0.07$ | $+0.27 \pm 0.14$ | $+0.40 \pm 0.15$ | $+0.43 \pm 0.15$ | $+0.76 \pm 0.16$ |
| $\Delta m_s$ | $+0.00 \pm 0.01$ | $-0.00 \pm 0.01$ | $+0.01 \pm 0.00$ | $-0.00 \pm 0.01$ | $+0.00 \pm 0.01$ | $+0.00 \pm 0.01$ |

Table 28: Relative bias on the physics parameters when using acceptance curves in the fit which systematically deviate from the "true" acceptances. To determine the bias 500 toy experiments with data corresponding to a luminosity of $2\,\text{fb}^{-1}$ have been performed. In every toy experiment, the nominal acceptance curves (i.e. the angular acceptance curves shown in Figure 26) have been used for the generation of the toy data. The acceptance curves used in the fit have been altered from nominal by applying the multiplicative factors detailed in Section 11.1.1. The column $\epsilon_{\cos\theta\pm}$ corresponds to a systematic change of the acceptance curves in $\cos\theta$, $\epsilon_{\varphi\pm}$ to a systematic change in $\varphi$ and $\epsilon_{\cos\psi\pm}$ corresponds to a systematic change in $\cos\psi$. The upper table shows the results when using the Standard model value of $\phi_s = -0.0368$, the lower table shows the results assuming a New Physics scenario with $\phi_s = -0.736$.

# Chapter 5

# Analysis of the decay $B_s^0 \to \mu^+\mu^-$

D. Martinez Santos, S. Amato, M.-O. Bettler, W. Bonivento, A. Buechler, A. Camboni, X. Cid Vidal, L. De Paula, F. Dettori, H. Dijkstra, M. Gandelman, J.A. Hernando Morata, W. Hulsbergen, G. Lanfranchi, E. Lopez, G. Mancinelli, W. Manner, A. Perez Calero Yzquierdo, E. Polycarpo, H. Ruiz Perez, A. Sarti, O. Schneider, N. Serra, J. Serrano, E. Simioni, B. Storaci, F. Teubert, N. Tuning, J. van Tilburg and R. Vazquez

**Abstract**

In this note we describe a strategy to calibrate all the steps needed to extract the $B_s^0 \to \mu^+\mu^-$branching ratio from LHCb data using control channels and not relying on the simulation.

The ratio of offline reconstruction efficiencies between signal ($B_s^0 \to \mu^+\mu^-$) and normalization channels ($B^+ \to J/\psi(\mu^+\mu^-)K^+$ and/or $B^0 \to K^+\pi^-$) can be extracted using the ratio of different control channels (for instance, $B^0 \to J/\psi(\mu^+\mu^-)K^{*0}(K^+\pi^-)$) with a few percent precision. The ratio of trigger efficiencies can be extracted using events triggered independently of the signal (TIS), which with enough luminosity will give a few percent precision. The invariant-mass and the geometrical properties can be extracted using $B^0_{(s)} \to h^+h^-$events as signal candidates, and the events in the sidebands of the mass distribution as background candidates without relying on the simulation. There are several good control channels (for instance $J/\psi \to \mu^+\mu^-$ and $\Lambda \to p\pi^-$) to be able to calibrate the muon identification efficiency and the muon misidentification probability.

This strategy will allow LHCb to perform a measurement of the $B_s^0 \to \mu^+\mu^-$branching ratio that should not depend on how well our simulation reproduces real data.

# Contents

# 1 Introduction

One of the most mysterious facts in particle physics today is that, on one hand, New Physics (NP) is expected in the TeV energy range to solve the hierarchy problem, but, on the other hand, no signal of NP has been detected through precision tests of the electroweak theory or through flavour-changing and/or CP-violating processes in K and B decays. In the last decade, the domain of precision experiments in flavour physics has been extended from the kaon sector to the richer and better computable realm of B decays. The main conclusion of the first generation of B-decay experiments can be expressed by saying that the Cabibbo-Kobayashi-Maskawa (CKM) description of flavour-changing processes has been confirmed in $b \to d$ transitions, and NP effects are bounded to corrections up to 30% in the neutral $B_d^0$ mixing. (see for instance Ref. [1]).

On the other hand, NP effects can still be large in flavour-changing $b \to s$ transitions. Within the Standard Model (SM) flavour changing neutral currents (FCNC) occur through loop corrections. Potential NP contributions are constrained by other precision low-energy measurements involving similar loop corrections. A prime examples of low-energy constraint is the measurement of the inclusive branching ratio in $b \to s\gamma$ decays [2, 3], or the measurement of the anomalous magnetic moment of the muon [4, 5], $a_\mu = (g_\mu - 2)/2$. Together with the constraint on the Higgs mass [6], these measurements are among the most important indirect constraints on extensions of the SM, such as the minimal supersymmetric extension (MSSM).

The decay $B_s^0 \to \mu^+\mu^-$ has been identified as a very interesting measurement that could show a clear indication of NP and/or constrain the parameter space of models describing physics beyond the SM [7]. The Tevatron collider at Fermilab has already determined an upper limit [8] on the branching ratio of $\mathcal{B}(B_s^0 \to \mu^+\mu^-) < 3.6 \times 10^{-8}$ at 90% CL. The SM prediction has been computed to be $\mathcal{B}(B_s^0 \to \mu^+\mu^-) = (3.35 \pm 0.32) \times 10^{-9}$ [9] using the measurement of the $B_s^0$ oscillation frequency at Tevatron ($\Delta M_s = 17.8 \pm 0.1\,\mathrm{ps}^{-1}$) [10], which significantly reduces the uncertainties in the SM prediction.

Within the framework of the MSSM, this branching ratio is known [11] to increase as the sixth power of the ratio of Higgs vacuum expectation values, $\tan\beta$. Any improvement on this limit is therefore particularly important for models with large $\tan\beta$. In Figure 1 one can see an example of a "Higgs Penguin" contribution that would modify the SM prediction and depends on $\tan^6\beta$.

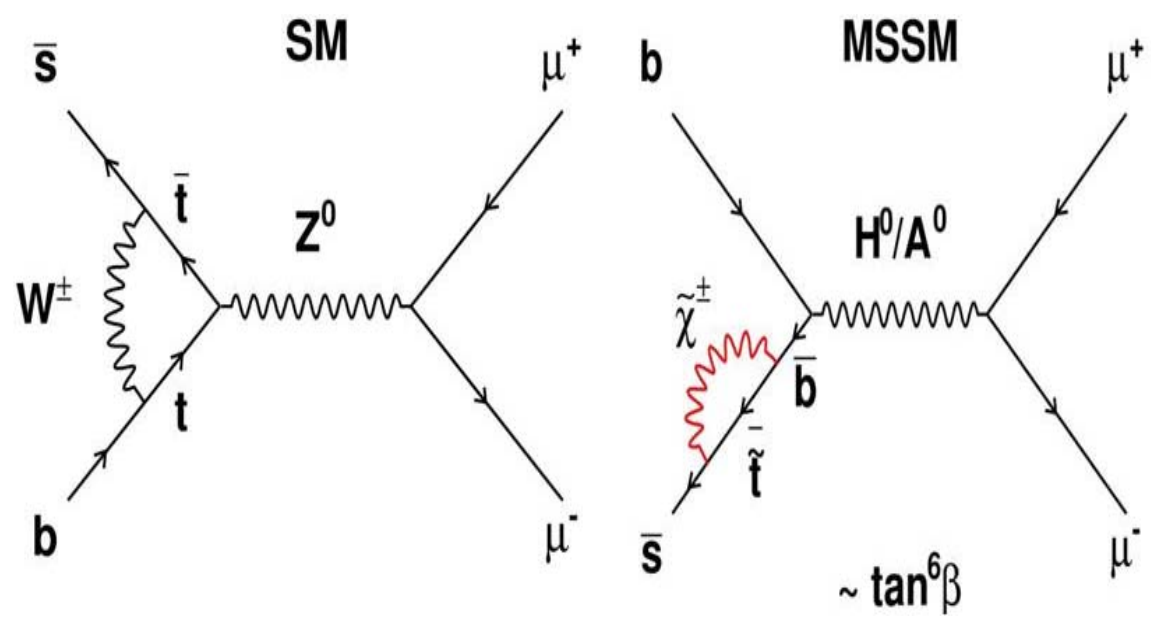

Figure 1: Examples of Feynman diagrams contributing to the decay $B_s^0 \to \mu^+\mu^-$ in the SM (left) and the MSSM (right).

The potential of LHCb to reach sensitivities of the order of $10^{-9}$ in the $B_s^0 \to \mu^+\mu^-$ decay was described in detail in Ref. [12]. The main conclusion in that note was that with little luminosity ($< 1\,\mathrm{fb}^{-1}$) LHCb can exclude any significant excess with respect to the SM.

Figure 2 shows the values of $\tan\beta$ and $M_A$, the mass of the CP-odd neutral Higgs, preferred by a global fit to several measurements within one particular realization (Non-Universal Higgs Masses, NUHM) of the MSSM [13], which is more general and includes the more popular "constrained" MSSM (cMSSM) realization. The best fit position is largely determined by the present $3.4\,\sigma$ discrepancy in the measured anomalous magnetic moment of the muon [5]. As a consequence, if such a discrepancy is not due to a statistical fluctuation, a sizable enhancement of the branching ratio is expected in this model, i.e. $\mathcal{B}(B_s^0 \to \mu^+\mu^-) \sim 10^{-8}$. LHCb could measure such a branching fraction with $5\,\sigma$ significance over the background with only $0.1\,\mathrm{fb}^{-1}$ of integrated luminosity.

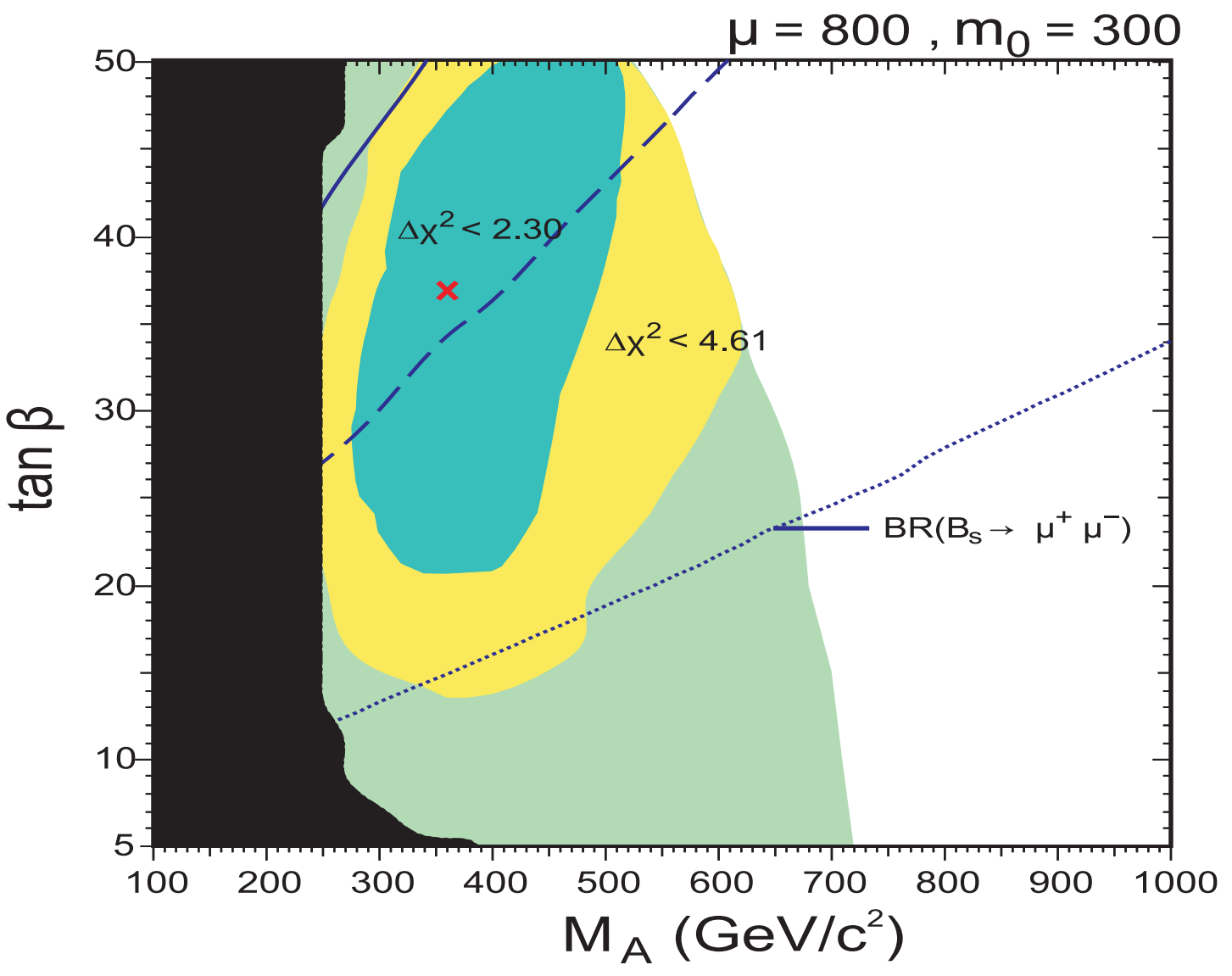

Figure 2: Best fit and $\chi^2$ contours in the plane $(M_A, \tan\beta)$ from the fit in Ref. [13] to several observables, including the anomalous magnetic moment of the muon. The dark area is excluded by previous measurements. The lines indicate the excluded region when $\mathcal{B}(B_s^0 \to \mu^+\mu^-) < 10^{-7}$ (continous), $2\times10^{-8}$(dashed), or $5\times10^{-9}$ (dotted).

In this note, we focus on the description of an analysis where the use of the simulation is minimized as compared to the analysis described in Ref. [12]. In Section 2 the software environment used to generate the signal, background and control channels samples is briefly described. In Section 3 a brief reminder of the analysis strategy is given. Sections 4 and 5 introduce the trigger and offline event selections. The offline selection has been modified with respect to Ref. [12] to have a selection as similar as possible between the signal and the control channels. In Section 6 the strategy to use these control channels to convert the observed number of candidates into a branching ratio is described. In Section 7 the $B^0_{(s)} \to h^+h^-$ control channels are used to calibrate the invariant mass and the geometrical likelihoods, while the calibration of the muon identification probability using $J/\psi \to \mu^+\mu^-$ and $\Lambda \to p\pi^-$ decays is described in Section 8. An update of LHCb's sensitivity to the $B^0_s \to \mu^+\mu^-$ search, based on the above strategies and using the updated simulation of the detector response, is presented in Section 9, including a cross-check with a less optimal but more robust analysis strategy. Finally, the main systematic limitation is discussed in Section 10 and conclusions are given in Section 11.

# 2 Signal, background and control channels MC samples

Pileup events are simulated[1] assuming an instantaneous nominal luminosity of $2 \times 10^{32}$ cm$^{-2}$s$^{-1}$ (and also for a maximum instantaneous luminosity of $5 \times 10^{32}$ cm$^{-2}$s$^{-1}$), an inelastic cross-section of 80 mb and a non-empty crossing rate of 30 MHz at the nominal luminosity. The total $b\bar{b}$ cross-section is assumed to be 500 $\mu$b and the probability to fragment into a $B^0_s$ meson is assumed to be 10%.

Given the low branching ratio of the signal, a detailed understanding of the background is crucial in this analysis. Several sources of background have been considered: combinatorial background (where two real muons in the event combine to form a signal candidate), misidentified hadrons and exclusive decays that could mimic the signal. The combinatorial background and the misidentification contribution to the $B^0_s \to \mu^+\mu^-$ analysis are studied using an inclusive sample of minimum bias events (with limited statistical significance), and a minimum bias sample containing a pair of $b$-quarks or $c$-quarks.

The inclusive $b\bar{b}$ ($c\bar{c}$) samples are obtained filtering a large dataset of minimum bias proton-proton interactions at $\sqrt{s} = 14$ TeV. In order to optimize the production, when a $b(c)$-quark is produced in the event, at least one is required to have a forward direction within 400 mrad of the beam axis. This cut reduces the probability that the decay products are outside the LHCb acceptance. The fraction of inclusive events with at least one of the two $b$ or $c$-quarks satisfying the cut is $(43.7 \pm 0.1)$% or $(42.3 \pm 0.1)$%.

The main background contribution turns out to be from events containing two real muons, see Ref. [12], therefore the combinatorial background can be studied in more detail using an inclusive sample of events containing at least one b-quark and two forward muons of opposite charge within 400 mrad of the beam axis; this is the so-called $b\bar{b} \to$ dimuon sample in Table 1. As shown in this table, the equivalent luminosity of the $b\bar{b} \to$ dimuon background sample is increased by a factor $\sim 70$ compared with the inclusive $b\bar{b}$ sample. The $b\bar{b} \to$ dimuon sample includes events where the two muons come from intermediate particles and events where the muon(s) are not related to the $b$-quark(s) in the event, hence is a more generic sample than the one used in Ref. [12]. However, the dominant background is still muons produced from the $b$-quarks as will be discussed in Section 5.

[1]All the Monte Carlo simulation samples used in this analysis were generated within the LHCb data challenge DC06, using the event generator PYTHIA [14], and the detector geometry as described by Dbase v30 and Gauss v25. The detector digitization is simulated using the program Boole v12 and the reconstruction is performed using the program Brunel v31.

Table 1: List of processes studied in this analysis, with their total cross-section, the cross-section after generator cuts, the number of events analyzed and the equivalent luminosity. In the case of $B_s^0 \to \pi^+\pi^-$ the 90% upper limit for the cross section is used.

| Process | Total $\sigma$ (pb) | MC $\sigma$ (pb) | # events | Luminosity ( $\mathrm{pb}^{-1}$) |
|---|---|---|---|---|
| Minimum Bias after L0 | $3.17 \times 10^{10}$ | $3.17 \times 10^{10}$ | 4.8M | $1.5 \times 10^{-4}$ |
| Inclusive $c\bar{c}$ | $3.6 \times 10^{9}$ | $1.54 \times 10^{9}$ | 0.2M | $1.3 \times 10^{-4}$ |
| Inclusive $b\bar{b}$ | $5 \times 10^{8}$ | $2.19 \times 10^{8}$ | 17.3M | 0.079 |
| $b\bar{b} \to$ dimuon | — | $4.84 \times 10^{6}$ | 25.7M | 5.31 |
| $B^+ \to J/\psi(\mu^+\mu^-)K^+$ | $2.42 \times 10^{4}$ | $4.31 \times 10^{3}$ | 467k | 108 |
| $B_d^0 \to J/\psi(\mu^+\mu^-)K^{*0}(K^+\pi^-)$ | $2.13 \times 10^{4}$ | $3.65 \times 10^{3}$ | 100k | 27 |
| $B_c^+ \to J/\psi(\mu^+\mu^-)\mu^+\nu_\mu$ | $1.19 \times 10^{3}$ | 246 | 59.7k | 242 |
| $B_s^0 \to K^+K^-$ | $1.85 \times 10^{3}$ | 635 | 629k | 991 |
| $B_s^0 \to \pi^+\pi^-$ | $< 1.7 \times 10^{2}$ | $< 34$ | 36.3k | $> 1.1 \times 10^{3}$ |
| $B_s^0 \to \pi^+K^-$ | 480 | 97.8 | 119k | $1.22 \times 10^{3}$ |
| $B_d^0 \to K^+\pi^-$ | $7.49 \times 10^{3}$ | $1.49 \times 10^{3}$ | 1.41M | 946 |
| $B_d^0 \to \pi^+\pi^-$ | $1.94 \times 10^{3}$ | 387 | 1.35M | $3.49 \times 10^{3}$ |
| $B_s^0 \to \mu^+\mu^-\gamma$ | 1.20 | 0.41 | 600 | $1.47 \times 10^{3}$ |
| $B_s^0 \to \mu^+\mu^-$ | 0.33 | 0.067 | 78k | 1.2M |

As in the case of the inclusive samples, the signal sample is obtained from a sample of minimum-bias events, including pileup, where a $b$-quark is produced. The hadronization process is repeated until the correct hadron type $B_s^0$ is produced. The $B_s^0$ meson is then forced to decay into two muons using the EvtGen software [15]. The number of events analyzed is shown in Table 1. In the case of exclusive $B$ decays, the decay products are required to be within 400 mrad of the beam axis, therefore the efficiency of the generator cuts depends on the $B$ decay. For instance, for $B_s^0 \to \mu^+\mu^-$ or $B_d^0 \to \pi^+\pi^-$ the efficiency of the generator cuts is $(20.2 \pm 0.1)\%$, while for decays like $B^+ \to J/\psi(\mu^+\mu^-)K^+$ it is $(17.9 \pm 0.1)\%$ as can be derived from the numbers in Table 1.

Several two-body decays, similar to the signal except for the particle identification (PID) likelihood, have been studied and are listed in Table 1. The branching ratio of the $B_s^0 \to \pi^+\pi^-$ decay is not known, hence the 90% upper limit is used throughout this paper. The process $B_c^+ \to J/\psi(\mu^+\mu^-)\mu^+\nu_\mu$, identified as a possible source of background given the fact that the mass of the $B_c^+$ ($6276 \pm 4$ MeV/$c^2$) is larger than that of the $B_s^0$ (5366.3 $\pm$ 0.6 MeV/$c^2$), has been studied in detail [16]. In addition, a possible contribution from the radiative decay $B_s^0 \to \mu^+\mu^-\gamma$ has been considered, as the branching ratio including Initial State Radiation (ISR) can be increased by an order of magnitude. The effect of Final State Radiation (FSR) is already included in the definition of the signal sample $B_s^0 \to \mu^+\mu^-$, therefore is not double counted here. The background from events with ISR photons is found to be negligible after the event selection. The $B^+ \to J/\psi(\mu^+\mu^-)K^+$ as well as the $B_d^0 \to J/\psi(\mu^+\mu^-)K^{*0}(K^+\pi^-)$ decays are important control channels, together with the two-body decays $B_{(s)}^0 \to h^+h^-$, as explained below.

# 3 Analysis strategy

The strategy to search for the $B_s^0 \to \mu^+\mu^-$ decay in LHCb was described in Ref. [12]. The basic concept is to apply a very efficient selection on signal events, removing obvious backgrounds to reduce the size of the data sample to be analyzed. Then each event is assigned a likelihood to be signal-like and a likelihood to be background-like. The likelihood is defined in terms of three quantities: Invariant Mass Likelihood (IML), the Particle Identification Likelihood (PIDL) and the Geometrical Likelihood (GL). The likelihood was built in this way to facilitate the calibration procedure explained in this note. For instance, we have been careful not to use explicitly the momentum information of the muon candidates in the definition of GL in order to avoid correlations between GL and IML or PIDL. We have checked that the remaining correlations are negligible.

Every event that has been triggered and selected offline is assigned to a 3D bin in the likelihood space described above. However, to relate the number of candidates in each bin with the total branching ratio an overall normalization factor is needed. The best strategy at LHC is to use another normalization channel with a precisely measured branching ratio. In this case, in order to compute the overall normalization factor, only the ratio of efficiencies between the signal and control channel needs to be evaluated, as explained in Section 6.

The IML assigns a probability that an event with a given invariant mass of the muon candidates is signal or background. As it will be described in Section 7, this probability can be obtained from $B_{(s)}^0 \to h^+h^-$ candidates, while the background probability can be obtained from the sidebands of the mass distribution.

The PIDL assigns a probability that the two muon candidates (defined by the requirement to have hits in the Muon Chambers applied at the selection level, see Section 5), are indeed muons and it is defined as an optimal combination of the difference in probabilities to be a muon compared with the probability to be a pion and with the probability to be a kaon. These probabilities use mainly information from the Muon Chambers, but also extra information from the Calorimeters and the RICH detectors. In Section 8 a procedure will be described to calibrate the information from the Muon Chambers using control channels.

The GL assigns a probability that the geometrical characteristics of the event are signal or background like. The geometrical variables included are:

- *Lifetime of the $B_s^0$ candidate.* This variable is computed using the distance between the reconstructed Secondary (SV) and Primary (PV) vertices, and the reconstructed momentum of the $B_s^0$ candidate. When more than one PV is reconstructed, the one that gives the minimum $B_s^0$ impact parameter is chosen.

- *Muon impact parameter significance.* This variable is the lowest impact parameter significance of the two muon candidates with respect to any of the primary vertices reconstructed in the event.

- *$B_s^0$ impact parameter.*

- *DOCA.* This variable is the distance of closest approach between the two muon candidates.

- *Isolation.* For each of the muon candidates, a search is performed for long tracks (traversing all tracking detectors and excluding the other muon candidate), that can make a "good" vertex with the muon candidate (i.e. DOCA $< 200\ \mu$m and the vertex coordinates along the beam axis should satisfy: 3 cm $> \mathrm{z}_{\mu+\mathrm{tr}} - \mathrm{z}_{\mathrm{PV}} > 0$). If we define $\alpha^{\mu+tr,PV}$ as the angle between the sum of the momentum of the muon and extra track and the direction defined by the PV and the vertex reconstructed using the muon and the extra track candidates, then the sum of the momenta is required to satisfy:

$$\frac{\left|\vec{P}_{\mu} + \vec{P}_{tr}\right| \cdot \alpha^{\mu+tr,PV}}{\left|\vec{P}_{\mu} + \vec{P}_{tr}\right| \cdot \alpha^{\mu+tr,PV} + P_{T\mu} + P_{Ttr}} < 0.4 \tag{1}$$

  where $P_{T\mu}$ and $P_{Ttr}$ are the transverse momentum (with respect to the beam line) of the muon candidate and the extra track. The number of tracks that satisfy these conditions is used as a discriminating variable for each of the muon candidates.

The five variables described above are combined using the mathematical method described in Ref. [12] to produce a uniform distribution between zero and one for signal candidates and peaked to zero values for background. As will be described in Section 7, this probability can be obtained from $B_{(s)}^0 \to h^+h^-$ candidates, while the background probability can be obtained from the invariant mass sidebands.

Once we know the signal and background likelihood for each event, we can test the compatibility of each branching ratio hypothesis with the distribution observed in these 3D bins, using the "modified frequentist" CLs method [17], and evaluate the exclusion limit or eventually the significance of the observation, as explained in Section 9.

# 4 Trigger selection

The LHCb trigger has two levels: L0 and HLT. The first level of triggering (L0), implemented on custom boards, will reduce the non-empty input rate from 30 to 1 MHz at a fixed latency of 4 $\mu$s. At this rate, events will be sent to a CPU farm with up to 2000 CPU boxes where several trigger algorithms will be executed. LHCb has the advantage, compared to the multipurpose experiments at LHC, to be able to trigger on muons with very low transverse momentum ($\sim$1 GeV/$c$) (relevant for the acceptance of signal and several control channels) and the possibility to trigger on hadrons at the lowest trigger level (relevant for the main control channels: $B^0_{(s)} \to h^+h^-$). The specific trigger cuts will need to be adapted to the real running conditions (the same is true for the offline cuts), but in the following we describe the trigger scenario used in this note.

Based on calorimeter and muon chamber information, the L0 trigger requires a muon, an electron or a hadron in the event with a transverse momentum ($P_{\mathrm{T}}$) or energy ($E_{\mathrm{T}}$) above some threshold. A pileup veto system is also foreseen: two dedicated silicon disks located upstream of the VELO will be used to reconstruct the longitudinal position of the interaction vertices and can be used to reject events with two or more such vertices. After the pileup selection, the event rate is approximately 7 MHz, with more than 90% single interactions at the nominal LHCb instantaneous luminosity of $2\times10^{32}\mathrm{cm}^{-2}\mathrm{s}^{-1}$ and the requirements on $P_{\mathrm{T}}$ ($E_{\mathrm{T}}$) can be relaxed. This selection together with the multiplicity measured in the Pileup detector and the Scintillator Pad Detector (SPD), and the total $E_{\mathrm{T}}$ measured in the calorimeter are referred to as Global Event Cuts (GEC) in the following. If the muon chambers detect a muon candidate in an event which satisfies the GEC and has $P_{\mathrm{T}}$ >1.3 GeV/$c$, or in an event which does not satisfy the GEC but has $P_{\mathrm{T}}$ >1.5 GeV/$c$, the event is triggered by L0 with an event rate of $\sim$200 kHz. Moreover, if two muon candidates are detected by L0, and the sum of their $P_{\mathrm{T}}$ >1.5 GeV/$c$, the event is triggered independently of the GEC decision with an event rate of $\sim$50 kHz for a total L0-muon rate of $\sim$220 kHz [18]. The efficiency of these L0 triggers on $B^0_s \to \mu^+\mu^-$ signal events selected offline is 97%, while for the $B^+ \to J/\psi(\mu^+\mu^-)K^+$ control channel it is 92%. Similarly, if a hadron with $E_{\mathrm{T}}$ >3.6 GeV is detected in the Hadron Calorimeter, the event will be triggered by L0 with an event rate of $\sim$700 kHz. The efficiency of this L0-hadron trigger on the $B^0_{(s)} \to h^+h^-$ control channels selected offline is $\sim$ 50%.

The second level of triggering (HLT) has access to all detector information, but given the limited amount of CPU power is divided in two steps: HLT1 and HLT2.

The HLT1 algorithms use only a partial reconstruction of the event seeded by the L0 information. Following the L0 decision, the HLT1 algorithms are separated in different alleys according to the L0 trigger decision. In the case of the muon-alley, the L0 trigger rate is further reduced to less than 100 kHz by confirming the L0-$\mu$ candidate(s) using the tracking stations. If the L0 single muon candidate is confirmed and has an impact parameter (IP) larger than 0.08 mm and a $P_{\mathrm{T}}$ >1.3 GeV/$c$ (*single muon trigger*) the event is accepted by HLT1 with a rate of $\sim$12 kHz and an efficiency of 94% on signal events selected offline and triggered by L0. If a second muon is found with a distance of closest approach (DOCA) less than 0.5 mm to the confirmed L0-$\mu$ and a $\mu\mu$ invariant mass larger than 2.5 $\mathrm{GeV}/c^2$ (*lifetime unbiased dimuon trigger*) the event is triggered at a rate of $\sim$4 kHz and an efficiency of 86% on signal events. Alternatively, if the invariant mass is larger than 0.5 $\mathrm{GeV}/c^2$ and the IP of both muons is greater than 0.15 mm (*lifetime biased dimuon trigger*) the event is accepted with a rate of $\sim$2.7 kHz and an efficiency of 81% on signal events. The total HLT1 muon alley rate is $\sim$17 kHz. It can be seen from the fact that the total rate is not very different from the sum of the individual rates that the *single* and *dimuon* triggers do not overlap too much. This is due to the relatively higher $P_{\mathrm{T}}$ cut needed in the case of the single muon trigger.

The HLT1 *lifetime unbiased dimuon trigger* would be the ideal trigger for this analysis as it is simple to understand and it should be very efficient. However, there is some inefficiency because the muon reconstruction at HLT1 has lower efficiency to reconstruct low momentum muons than the offline reconstruction. The fact that in HLT1 muons are reconstructed in the muon chambers before being extrapolated to the tracking stations implies a stringent requirement on the number of hits in the muon chambers and therefore a lower efficiency at low momentum. Moreover, as this trigger requires two online muons to be reconstructed, it is less robust against possibly worse online muon-id performance with real data. An interesting alternative is a hybrid of the *dimuon* and *single muon* triggers, called "*muon+track*", where two tracks that form a vertex in the VELO are extrapolated to the tracking stations and only one of them is required to be a muon in the muon chambers. This approach avoids the limited online reconstruction efficiency in the muon chambers, as it only requires one muon extrapolated from the VELO. The information provided by the extra track allows a more relaxed selection as compared to the single muon trigger. Selecting a muon with $P_{\mathrm{T}} > 1\,\mathrm{GeV}/c$ and IP $> 0.025$ mm, and an additional track with $P_{\mathrm{T}} > 0.6\,\mathrm{GeV}/c$ and IP $> 0.1$ mm, together with a DOCA between the muon and the extra track of less than 0.3 mm and a distance along the beam axis (DZ) greater than 1.5 mm with respect to the PV, gives a $\sim 12$ kHz trigger rate with an efficiency of $\sim 93$%.

The HLT1 hadron alley confirms the L0-hadron using the tracking stations, and if the track has an IP $> 0.1$ mm and a $P_{\mathrm{T}} > 6\,\mathrm{GeV}/c$ the HLT1 triggers at a rate of 4 kHz. Alternatively, if the confirmed track has IP $> 0.1$ mm and $P_{\mathrm{T}} > 2.5\,\mathrm{GeV}/c$, but a second track is found with DOCA<0.2 mm, IP $> 0.1$ mm and $P_{\mathrm{T}} > 1\,\mathrm{GeV}/c$, the event is also triggered with a rate of 6 kHz. The efficiency for the $B^0_{(s)} \to h^+h^-$ control channel is $\sim 87$% for a total HLT1 hadron alley rate of 9 kHz.

The total rate of HLT1 is $\sim$30 kHz including the rest of the alleys (ECAL alley, etc) not discussed in this note. At this rate, HLT2 performs the full event reconstruction, and several inclusive and exclusive selections are executed. The reconstruction at this trigger level is similar to the offline reconstruction. In particular, muons are reconstructed matching reconstructed tracks to the hits in the muon chambers avoiding the limitations of HLT1 discussed previously. It is clear from the above description of HLT1 that there are several inclusive triggers that should be able to trigger the signal and the $B^+ \to J/\psi(\mu^+\mu^-)K^+$ and $B^0_d \to J/\psi(\mu^+\mu^-)K^{*0}(K^+\pi^-)$ control channels very efficiently and in complementary ways:

Table 2: Trigger rates and efficiencies normalized to offline selected events for signal and control channels for the combined L0×HLT1×HLT2 triggers. The numbers in parentheses for the signal correspond to the efficiency in the sensitive region (GL > 0.5 and the mass difference with respect to the nominal $B_s^0$ mass should be within $\Delta M = \pm 60\,\mathrm{MeV}/c^2$).

| HLT2 trigger type | Rate (Hz) | $\epsilon(B_s^0 \to \mu^+\mu^-)$ | $\epsilon(B^+ \to J/\psi K^+)$ | $\epsilon(B_{(s)}^0 \to h^+h^-)$ |
|---|---:|---:|---:|---:|
| single muon | 550 | 53% (69%) | 29% | 0.8% |
| muon+track | 700 | 76% (90%) | 63% | 1.5% |
| dimuon | 460 | 89% (91%) | 69% | 0.3% |
| exclusive $B_{(s)}^0 \to h^+h^-$ | 30 | 70% (86%) | 0% | 30% |
| all triggers above | 1400 | 93% (94%) | 77% | 30% |

- **HLT2 Single muon:** If a muon candidate is reconstructed in HLT2 with an IP >0.175 mm and a $P_{\mathrm{T}}$ >3.5 GeV/$c$, the HLT2 triggers with a rate of ∼550 Hz and an efficiency of 73% on signal events selected offline and triggered by L0×HLT1 in the sensitive region.

- **HLT2 Muon+Track:** If a muon candidate is reconstructed in HLT2 with an IP >0.1 mm and a $P_{\mathrm{T}}$ >2 GeV/$c$, and a second track is found with $P_{\mathrm{T}} > 1\,\mathrm{GeV}$ and IP > 0.15 mm, together with a DOCA less than 0.15 mm and DZ greater than 1.5 mm, the HLT2 triggers with a rate of ∼ 700 Hz and an efficiency of 95% on signal events selected offline and triggered by L0×HLT1 in the sensitive region.

- **HLT2 Dimuon:** If two muon candidates are reconstructed in HLT2 with a vertex $\chi^2 < 20$ and an invariant mass larger than 2.9 GeV/$c^2$, or with an invariant mass larger than 0.5 GeV/$c^2$ and with the IP of both muons greater than 0.15 mm, the HLT2 triggers with a rate of ∼ 460 Hz and an efficiency of 96% on signal events selected offline and triggered by L0×HLT1 in the sensitive region.

On the other hand, the $B_s^0 \to \mu^+\mu^-$ signal candidates will also be triggered by an exclusive trigger common to the $B_{(s)}^0 \to h^+h^-$ decays and very similar to the offline selection described in Section 5 except for the requirement on the muon-id. The HLT2 exclusive selection efficiency to select $B_{(s)}^0 \to h^+h^-$ candidates was expected to be very efficient, as it is aligned to the offline selection. However, some inefficiency is expected due to the differences between the online and offline track reconstruction. The measured efficiency is ∼ 90% on signal events selected offline and triggered by L0×HLT1 in the sensitive region. In Table 2 one can see the performance and rates of the different triggers discussed in this section for signal and control channels. In the third column, the efficiency for signal is given for all selected events and in parenthesis for the events in the sensitive region (GL>0.5 and $\Delta M = \pm 60\,\mathrm{MeV}/c^2$). Trigger is understood to be the convolution of all the trigger steps, i.e. L0 × HLT1 × HLT2. As can be seen from the numbers quoted in Table 2, the trigger performance for signal is excellent and it is reasonable for control channels. The fact that the signal is triggered by several inclusive triggers gives a certain level of robustness to the analysis.

# 5 Event selection

The offline event selection has been modified with respect to Ref. [12]. The idea is still to use the selection just to reduce the size of the data sample to analyze, hence to keep most of the signal and remove "obvious" background. However, in order to keep the selection as similar as possible to the one for the control channels $B^0_{(s)} \to h^+h^-$, $B^+ \to J/\psi(\mu^+\mu^-)K^+$, and $B^0_d \to J/\psi(\mu^+\mu^-)K^{*0}(K^+\pi^-)$, some inefficiency on the signal has been allowed on events that in any case were shown not to have significant sensitivity in the final result as the ratio between signal and background was too small for them. The advantage of having a common selection is not only that the ratio of efficiencies is very close to one, hence the correction is smaller, but even more important is to select the $B$ candidates in the signal and control channels with similar phase space. This allows a simple calculation of the ratio of trigger efficiencies as will become clear in Section 6, and allows the use of $B^0_{(s)} \to h^+h^-$ events to calibrate the GL distribution, as will be explained in Section 7.

The common selection requires two tracks to satisfy the "isMuon" criteria (see Ref. [24]) with the exception of the $B^0_{(s)} \to h^+h^-$ decays, where all reconstructed tracks are assumed to be pions. The two tracks from the $B^0_{(s)}$ or $J/\psi$ decay are required to form a vertex with $\chi^2 < 14$, the impact parameter significance (with respect to the closest reconstructed PV) of the $B^0_{(s)}$ candidate must be IPS $< 6$, the distance of flight significance of the $B^0_{(s)}$ candidate must be $\mathrm{DOFS}(B^0_{(s)}|J/\psi) > 12$, and the transverse momentum of the $B^0_{(s)}$ candidate must be $P_{\mathrm{T}}(B^0_{(s)}) > 700\,\mathrm{MeV}/c$. In the case of $B^0_s \to \mu^+\mu^-$ and $B^0_{(s)} \to h^+h^-$ decays the minimum impact parameter significance of any of the two tracks with respect to the closest PV has to be $\mathrm{IPS}(\mu/h) > 3.5$, while for the $B^+ \to J/\psi(\mu^+\mu^-)K^+$ and $B^0_d \to J/\psi(\mu^+\mu^-)K^{*0}(K^+\pi^-)$ decays the IPS of the reconstructed kaon or $K^{*0}$ must be $\mathrm{IPS}(K^+/K^{*0}) > 3.5$. Table 3 shows the selection efficiencies for signal and control channels. The selection efficiencies (before the tight-mass window cuts are applied), for events reconstructed in the LHCb detector are:

$$
\begin{aligned}
\epsilon(B^0_s \to \mu^+\mu^-) &= 64.8\%,\\
\epsilon(B^0_{(s)} \to h^+h^-) &= 63.2\%,\\
\epsilon(B^+ \to J/\psi(\mu^+\mu^-)K^+) &= 62.5\%,\\
\epsilon(B^0_d \to J/\psi(\mu^+\mu^-)K^{*0}(K^+\pi^-)) &= 62.7\%.
\end{aligned}
$$

While these selection efficiencies are very similar, as was the goal in designing a common selection between signal and control channels, there is a partial cancellation between different cuts, in particular DOFS and IPS cuts. The effect of FSR (seen in Table 3 as the effect of the $\pm 60\,\mathrm{MeV}/c^2$ mass cut) is certainly different at the scale of the $B^0_s$ mass or at the scale of the $J/\psi$ mass. Moreover, in the case of the inclusive $B^0_{(s)} \to h^+h^-$ decays, where all particles are assumed to be pions, the invariant mass distribution develops a large tail at low masses due to the contribution of decays like $B_d^0 \to K^+\pi^-$ or $B_s^0 \to K^+K^-$, hence it does not make too much sense to apply the same tight invariant mass window cut.

Table 3: Efficiency of each individual offline cut (normalized to the sample selected by the previous cuts) on reconstructed events for signal ($B_s^0 \to \mu^+\mu^-$) and various control channels ($B_{(s)}^0 \to h^+h^-$, $B^+ \to J/\psi(\mu^+\mu^-)K^+$, $B_d^0 \to J/\psi(\mu^+\mu^-)K^{*0}(K^+\pi^-)$).

| Cut | $B_{(s)}^0$ final state $\mu^+\mu^-$ (%) | $h^+h^-$ (%) |
|---|---|---|
| $B_{(s)}^0$ mass ($\pm 600\,\mathrm{MeV}/c^2$) | 97.2 | 94.8 |
| $\chi^2(B_{(s)}^0) < 14$ | 98.3 | 97.9 |
| IPS($B_{(s)}^0$) $< 6$ | 98.6 | 98.3 |
| $P_\mathrm{T}(B_{(s)}^0) > 700\,\mathrm{MeV}/c$ | 97.8 | 97.9 |
| DOFS($B_{(s)}^0$) $> 12$ | 76.0 | 76.4 |
| minimum IPS($\mu/h$) $> 3.5$ | 92.6 | 92.6 |
| All cuts above | 64.8 | 63.2 |
| $B_{(s)}^0$ mass ($\pm 60\,\mathrm{MeV}/c^2$) | 90.6 | — |

| Cut | $B^{+,0}$ final state $J/\psi K^+$ (%) | $J/\psi K^{*0}$ (%) |
|---|---|---|
| $B^{+,0}$ mass ($\pm 500\,\mathrm{MeV}/c^2$) | 97.3 | 97.3 |
| $\chi^2(J/\psi) < 14$ | 98.3 | 98.3 |
| IPS($B^{+,0}$) $< 6$ | 97.4 | 97.5 |
| $P_\mathrm{T}(B^{+,0}) > 700\,\mathrm{MeV}/c$ | 98.4 | 98.9 |
| DOFS($J/\psi$) $> 12$ | 74.0 | 71.5 |
| IPS($K^+/K^{*0}$) $> 3.5$ | 92.1 | 95.1 |
| All cuts above | 62.5 | 62.7 |
| $J/\psi$ mass ($\pm 60\,\mathrm{MeV}/c^2$) | 94.2 | 93.7 |

Therefore, even if the ratio of efficiencies is not far from one in the MC simulation, one needs to cross-check each of the variables entering the selection between signal and calibration channels with real data to confirm that their ratios behave as expected.

The total efficiency on $B_s^0 \to \mu^+\mu^-$ signal events (including acceptance, trigger, reconstruction, PID and selection cuts) is 6.4%. The efficiencies for signal, control channels and the remaining background levels are given in Table 4, together with the expected number of events per $\mathrm{fb}^{-1}$. After this selection, the background composition is dominated by $b\bar{b}$ events, with about half of them containing two real muons. However, in the sensitive region (GL>0.5 and $\Delta M = \pm 60\,\mathrm{MeV}/c^2$) only $b\bar{b} \to$ dimuon events remain as irreducible background, completely dominated by semileptonic B decays. We expect to see about 10.4 signal events per fb$^{-1}$ in the sensitive region if the branching ratio has the SM value, while the expected background (dominated by the inclusive $b\bar{b} \to$ dimuon sample) is $90^{+70}_{-40}$ per fb$^{-1}$. Notice however, that all events selected are used in the evaluation of the sensitivity in Section 9, and the number of events in the sensitive region are given only for illustration as no additional selection cut is applied. It can also be seen in Table 4 that all the exclusive decay modes investigated so far do not contribute significantly to the total expected background.

Table 4: List of processes studied in this analysis, with their total selection efficiency (including acceptance, trigger, reconstruction, PID and selection cuts) and the number of events expected per fb$^{-1}$. The upper limit at 90% C.L. is also given when no event is left. In parenthesis are the number of signal and background events expected in the sensitive region (GL>0.5 and $\Delta M = \pm 60\,\mathrm{MeV}/c^2$). In the $b\bar{b} \to$ dimuon sample the total cross-section is not defined, hence the total efficiency is not quoted.

| Process | Total $\epsilon$ | $\mathrm{N_{exp}}/\,\mathrm{fb}^{-1}$ | 90% C.L. u.l. |
|---|---|---|---|
| Background | | | |
| L0-Minimum Bias | 0 | 0 (0) | < 1.5M |
| Inclusive $c\bar{c}$ | 0 | 0 (0) | < 57k |
| Inclusive $b\bar{b}$ | $1.3 \times 10^{-7}$ | 65k (0) | (< 2.3k) |
| $b\bar{b} \to$ dimuon | — | 33.2k (86) | — |
| $B_c^+ \to J/\psi(\mu^+\mu^-)\mu^+\nu_\mu$ | $1.49 \times 10^{-5}$ | 17.7 (0) | (< 9.5) |
| $B_s^0 \to K^+K^-$ | $1.09 \times 10^{-6}$ | 2.0 (2.0) | — |
| $B_s^0 \to \pi^+\pi^-$ | 0 | 0 (0) | < 2.1 |
| $B_s^0 \to \pi^+K^-$ | $3.40 \times 10^{-6}$ | 1.6 (0.8) | — |
| $B_d^0 \to K^+\pi^-$ | $4.23 \times 10^{-7}$ | 3.2 (1.1) | — |
| $B_d^0 \to \pi^+\pi^-$ | $1.48 \times 10^{-7}$ | 0.3 (0) | (< 0.7) |
| $B_s^0 \to \mu^+\mu^-\gamma$ | 0 | 0 (0) | < 1.6 |
| Control Channels | | | |
| $B^+ \to J/\psi(\mu^+\mu^-)K^+$ | 0.0326 | 788k | — |
| $B_d^0 \to J/\psi(\mu^+\mu^-)K^{*0}(K^+\pi^-)$ | 0.0202 | 428k | — |
| $B_{(s)}^0 \to h^+h^-$ (no $\mu$-ID) | 0.019 | 220k | — |
| Signal | | | |
| $B_s^0 \to \mu^+\mu^-$ | 0.0637 | 21.0(10.4) | — |

# 6 Normalization

The events left after the selection described in the previous section are distributed in bins of a three-dimensional space: IML, GL and PIDL as described in Section 3. The procedure to calibrate these probabilities without relying on the MC simulation will be explained in Sections 7 and 8. However, in order to translate the observed number of candidate events in each 3D bin into a measurement of the branching ratio, a global normalization factor is needed. If we decide to normalize to a calibration channel with a known branching ratio $\mathrm{BR}_{\mathrm{cal}}$, then the $B_s^0 \to \mu^+\mu^-$ branching ratio will be computed as

$$\mathrm{BR} = \mathrm{BR}_{\mathrm{cal}} \times \frac{\epsilon_{\mathrm{cal}}^{\mathrm{REC}} \epsilon_{\mathrm{cal}}^{\mathrm{SEL|REC}} \epsilon_{\mathrm{cal}}^{\mathrm{TRIG|SEL}}}{\epsilon_{\mathrm{sig}}^{\mathrm{REC}} \epsilon_{\mathrm{sig}}^{\mathrm{SEL|REC}} \epsilon_{\mathrm{sig}}^{\mathrm{TRIG|SEL}}} \times \frac{f_{\mathrm{cal}}}{f_{B_s^0}} \times \frac{N_{B_s^0 \to \mu^+\mu^-}}{N_{\mathrm{cal}}}\,, \tag{2}$$

where $f_{B_s^0}$ and $f_{\mathrm{cal}}$ refer to the probabilities that a $b$-quark "fragments" into a $B_s^0$ and into the $b$-hadron relevant for the chosen calibration mode. The ratio $f_{B_d^0}/f_{B_s^0} = f_{B^+}/f_{B_s^0}$ is known with an uncertainty of $\sim 13\%$ [19]. The total efficiency has been separated for convenience in three factors:

- $\epsilon^{\mathrm{REC}}$ is the efficiency to reconstruct all the tracks needed for the signal and calibration channel, including the effect of the limited acceptance of the detector;
- $\epsilon^{\mathrm{SEL|REC}}$ is the efficiency to select the events once they have been reconstructed;
- $\epsilon^{\mathrm{TRIG|SEL}}$ is the efficiency of the trigger on reconstructed and selected events.

The branching ratios $\mathcal{B}(B^+ \to J/\psi(\mu^+\mu^-)K^+) = (5.97 \pm 0.22) \times 10^{-5}$ and $\mathcal{B}(B_d^0 \to K^+\pi^-) = (1.88 \pm 0.07) \times 10^{-5}$ are both known [20] with a relatively good precision better than 4%, so they are good candidates as calibration channels. Moreover, they play complementary roles in terms of their differences with respect to the $B_s^0 \to \mu^+\mu^-$ signal: indeed, the $B^+ \to J/\psi(\mu^+\mu^-)K^+$ differs mainly because of the reconstruction of the extra kaon and the lower momentum muons, while $B_d^0 \to K^+\pi^-$ differs mainly because of the trigger and muon identification.

The normalization using $B_d^0 \to K^+\pi^-$ requires extracting the fraction of such events in the collected $B_{(s)}^0 \to h^+h^-$ sample, taking into account the imperfect PID performance. A possible strategy to obtain this fraction without relying on MC simulation is based on counting the number of candidates of each mode when hard PID cuts are applied. In this case the contamination from misidentification can be neglected, and only identification efficiencies play a relevant role. Hence, the true fraction of $B_d^0 \to K^+\pi^-$ plus $B_s^0 \to \pi^+K^-$ decays with respect to all the two-body $B$ meson decays ($f_{K\pi}$) can be expressed as a function of the observed fractions ($f'$) and the pion and kaon identification efficiencies as:

$$\begin{aligned} f_{K\pi} &= \frac{f'_{K\pi}/\epsilon_\pi\epsilon_K}{f'_{K\pi}/\epsilon_\pi\epsilon_K + f'_{KK}/\epsilon_K^2 + f'_{\pi\pi}/\epsilon_\pi^2} \\ &= \frac{f'_{K\pi}}{f'_{K\pi} + f'_{KK}\left(\frac{\epsilon_\pi}{\epsilon_K}\right) + f'_{\pi\pi}\left(\frac{\epsilon_K}{\epsilon_\pi}\right)} \end{aligned} \tag{3}$$

Pion and kaon identification efficiencies depend on the momentum of the particle. Here $\epsilon_\pi$ and $\epsilon_K$ refer to the overall efficiency corresponding to the momentum distribution of the hadrons in $B_{(s)}^0 \to h^+h^-$. The invariant mass distribution of these events allows the separation between the $B_d^0$ and $B_s^0$ components.

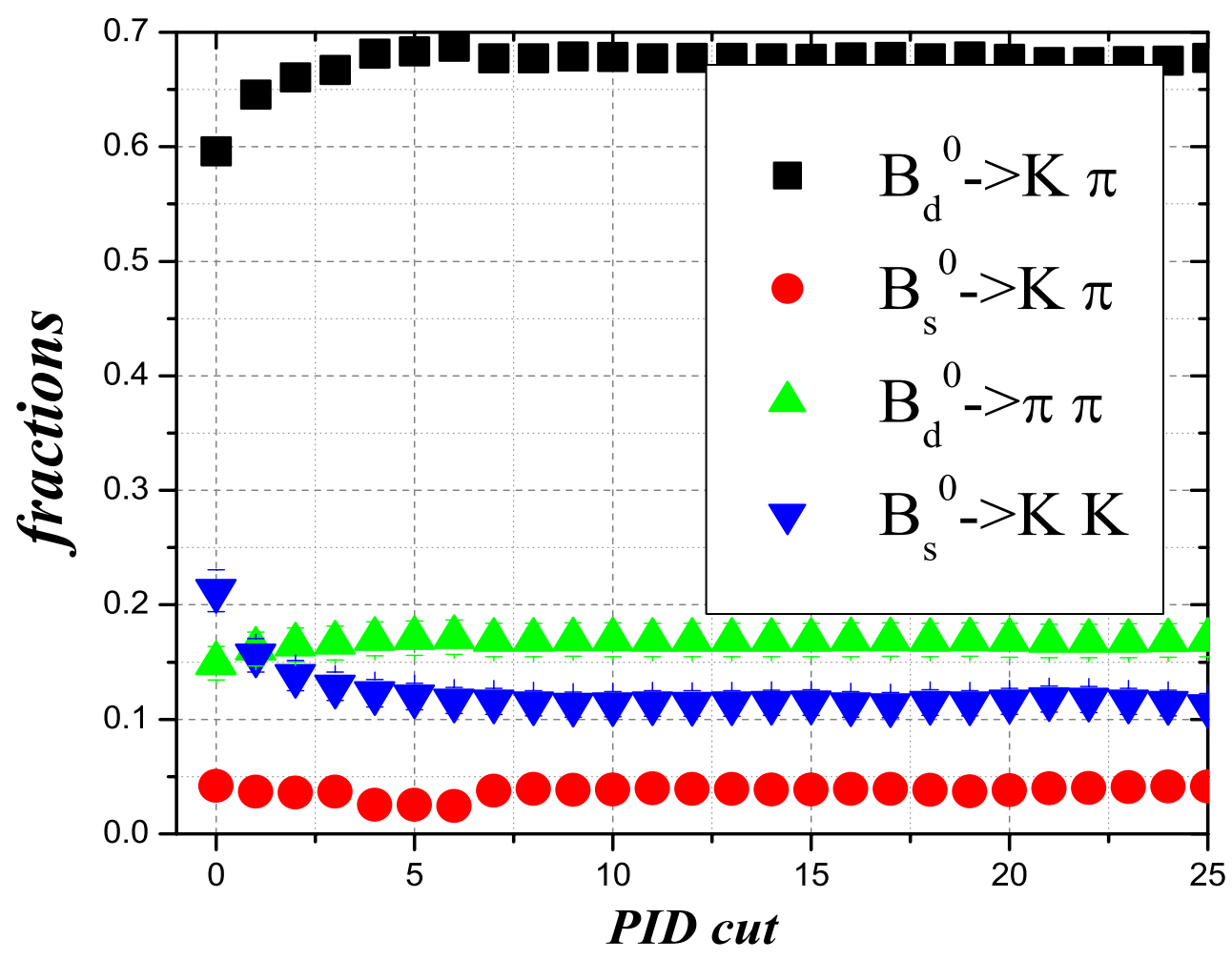

Figure 3: Measured fraction of $B_d^0 \to K^+\pi^-$, $B_d^0 \to \pi^+\pi^-$, $B_s^0 \to \pi^+K^-$ and $B_s^0 \to K^+K^-$ as a function of the cut in the difference in log-likelihood between the pion and kaon hypothesis.

The ratio between $\epsilon_\pi/\epsilon_K$ is accessible by counting the number of $B_d^0 \to K^+\pi^-$ and $B_d^0 \to \pi^+\pi^-$ when hard PID cuts are applied:

$$\frac{\epsilon_\pi}{\epsilon_K} = \frac{\mathcal{B}(B_d^0 \to K^+\pi^-)}{\mathcal{B}(B_d^0 \to \pi^+\pi^-)} \frac{N'_{\pi\pi}}{N'_{K\pi}} . \qquad (4)$$

Hence Eq. 3 can be computed to estimate the fraction of $B_d^0 \to K^+\pi^-$ in the original $B_{(s)}^0 \to h^+h^-$ sample. Figure 3 shows the measured fraction of the different $B_{(s)}^0 \to h^+h^-$ decay hypothesis as a function of the cut on the difference in log-likelihood between the pion and kaon hypothesis. A cut hard enough to ensure we are in the "plateau" will allow to compute the ratio between $\epsilon_\pi/\epsilon_K$.

Applying this method to $B_{(s)}^0 \to h^+h^-$ simulated events equivalent to $0.1\,\text{fb}^{-1}$ of data we obtain $f_{K\pi} = 0.68 \pm 0.03$, in agreement with the number expected from MC, $f_{K\pi} = 0.681$. The PID efficiencies will be calibrated by other means in LHCb [21], hence this method provides a useful independent cross-check. Once we know $f_{K\pi}$ we can use the number of $B_d^0 \to K^+\pi^-$ candidates to normalize the $B_s^0 \to \mu^+\mu^-$ branching ratio.

In Eq. 2 we have introduced the ratios of reconstruction efficiencies ($\epsilon^{\text{REC}}$), selection efficiencies ($\epsilon^{\text{SEL|REC}}$) and trigger efficiencies ($\epsilon^{\text{TRIG|SEL}}$). The ratio of selection efficiencies between $B_s^0 \to \mu^+\mu^-$ and $B^+ \to J/\psi(\mu^+\mu^-)K^+$ and $B_{(s)}^0 \to h^+h^-$ on reconstructed events has already been discussed in Section 5. This ratio is close to one in our current

MC simulation, but as explained in Section 5 this needs to be validated with real data. The other two factors are discussed in the following sub-sections, where the convenience of their definitions should become clear.

## 6.1 Ratio of reconstruction efficiencies

The difference in the reconstruction efficiency, $\epsilon^{\rm REC}$, between $B_s^0 \to \mu^+\mu^-$ and $B_d^0 \to K^+\pi^-$ is mainly due to the muon identification, as will be discussed in Section 8 and the different interactions with the material in the LHCb detector . We can avoid the need to compute the absolute muon-id efficiency if instead the $B^+ \to J/\psi(\mu^+\mu^-)K^+$ channel is used as a normalization mode; however, in this case, differences with respect to the signal occur because of the efficiency to reconstruct an extra charged track and, to a lesser extent, because of the different phase space of the muon pair. Notice that the effect of the detector acceptance is included in the definition of $\epsilon^{\rm REC}$. In this section, we describe the case when the normalization channel is $B^+ \to J/\psi(\mu^+\mu^-)K^+$.

The probability that all the tracks in the final state are within the LHCb acceptance depends on the phase space of these tracks. The charged particles may not traverse sufficient detector layers to be found by the track finding algorithms, mainly because the magnetic field introduces a cut-off at low momentum. The momentum spectrum of the $b$ hadrons produced in LHCb has considerable uncertainty. Therefore, until the spectrum has been properly calibrated, it will introduce a systematic uncertainty if the efficiencies are estimated using the simulation. Track finding algorithms are also sensitive to occupancy. Therefore, the reconstruction efficiency depends on how busy the events are. It is unlikely that the existing Monte Carlo simulation estimates the occupancy reliably. All these arguments justify the need for an alternative method to evaluate these efficiencies (or rather the ratio between signal and control channel efficiencies) without relying on the simulation. So we plan to pursue two approaches:

1) The efficiency ratio can be determined solely from the simulation, after the simulation has been either tuned or reweighted, using control samples, to properly reproduce the distributions to which the ratio is most sensitive, namely the momentum spectrum of the $B$ meson and the occupancy of the event. Remaining differences between data and simulation can then be used to assign a systematic uncertainty.

2) The efficiency ratio can be also estimated by considering another ratio of control channels, in order to probe explicitly the efficiency for reconstructing an extra track in the final state. A suitable control channel is $B_d^0 \to J/\psi K^{*0}$. If the tracks involved in the two control channels have similar phase space, one may expect:

$$\frac{\epsilon^{\rm REC}(B_s^0 \to \mu^+\mu^-)}{\epsilon^{\rm REC}(B^+ \to J/\psi K^+)} \simeq \frac{\epsilon^{\rm REC}(B^+ \to J/\psi K^+)}{\epsilon^{\rm REC}(B_d^0 \to J/\psi K^{*0})}\,. \tag{5}$$

The first method is being worked out in the context of the LHCb tracking working group, and is not discussed further here.

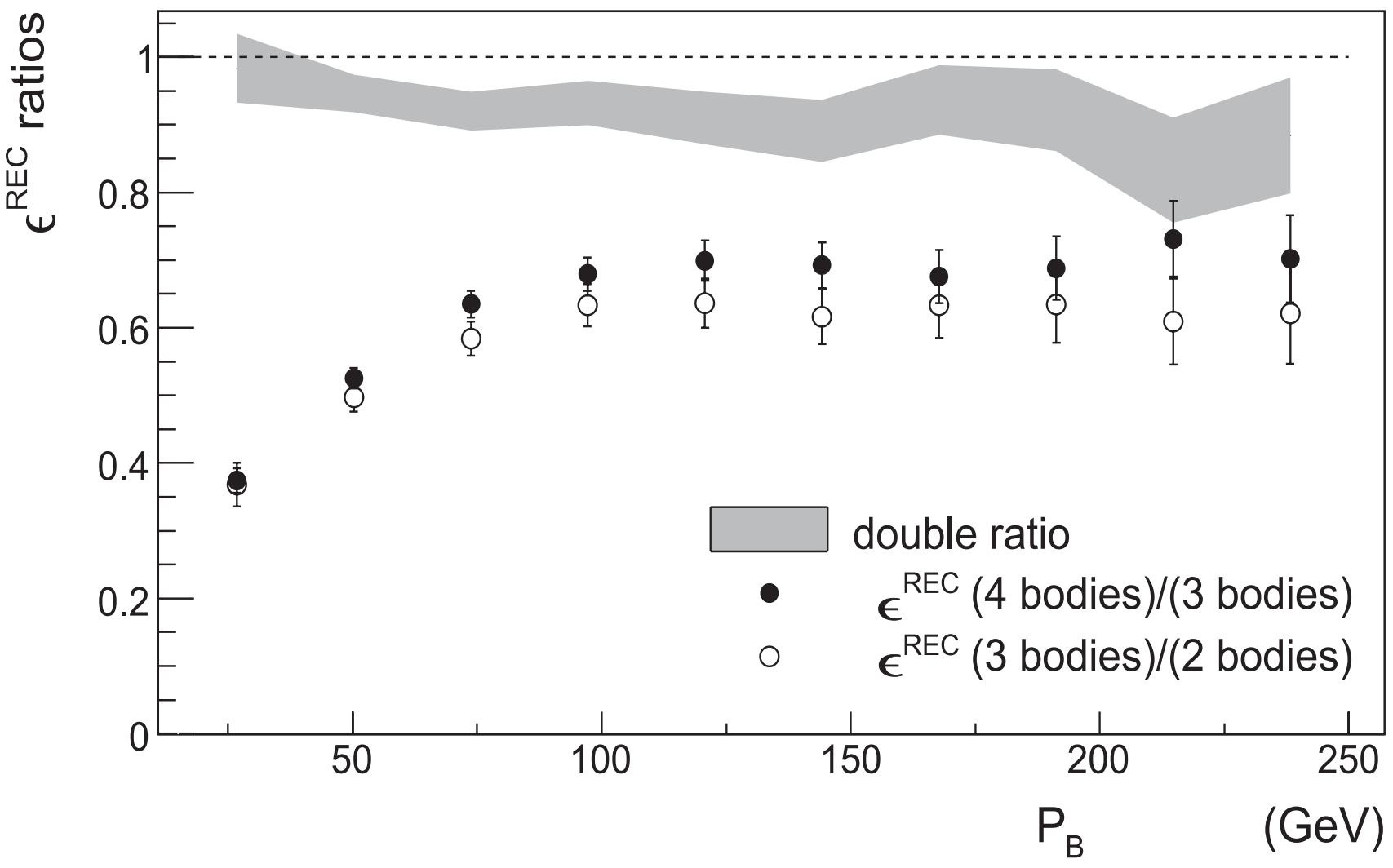

Figure 4: Ratio of reconstruction efficiencies between 4-body decays ($B_d^0 \to J/\psi(\mu^+\mu^-)K^{*0}(K^+\pi^-)$) and 3-body decays ($B^+ \to J/\psi(\mu^+\mu^-)K^+$) compared with 3-body decays vs 2-body decays ($B_s^0 \to \mu^+\mu^-$) as a function of the momentum of the $B$ candidate. The double ratio is shown as a shaded band.

The following example shows how the second method works. The branching ratios for these control channels are known with very good precision [20]: $\mathcal{B}(B^+ \to J/\psi(\mu^+\mu^-)K^+) = (5.97 \pm 0.22) \times 10^{-5}$ and $\mathcal{B}(B_d^0 \to J/\psi(\mu^+\mu^-)K^{*0}(K^+\pi^-)) = (5.36 \pm 0.25) \times 10^{-5}$. Hence, applying the common selection described in Section 5 and neglecting the effect of the trigger (which will be discussed in Section 6.2) one can obtain from the total numbers of selected $B^+ \to J/\psi(\mu^+\mu^-)K^+$ and $B_d^0 \to J/\psi(\mu^+\mu^-)K^{*0}(K^+\pi^-)$, denoted $N(B^+)$ and $N(B_d^0)$ respectively, the ratio[2]:

$$\frac{\epsilon_{\rm cal}^{\rm REC}\epsilon_{\rm cal}^{\rm SEL|REC}\epsilon_{\rm cal}^{\rm TRIG|SEL}}{\epsilon_{\rm sig}^{\rm REC}\epsilon_{\rm sig}^{\rm SEL|REC}\epsilon_{\rm sig}^{\rm TRIG|SEL}} = \frac{\mathcal{B}(B^+ \to J/\psi K^+)}{\mathcal{B}(B_d^0 \to J/\psi K^{*0})} \times \frac{N(B_d^0)}{N(B^+)}\,, \tag{6}$$

which is the first factor needed to compute the normalization if we assume the ratio of the trigger and selection efficiencies cancels out (in Section 5 the ratio of selection efficiencies was evaluated to be 0.997). With this approximation we can measure the

[2] $f_{B^+} = f_{B_d^0}$ is assumed.

ratio of reconstruction efficiencies using the equivalent of 0.1 $\text{fb}^{-1}$ of data to be

$$\frac{\epsilon_{\text{cal}}^{\text{REC}}}{\epsilon_{\text{sig}}^{\text{REC}}} = 0.627 \pm 0.006 \tag{7}$$

using the $B^+ \to J/\psi K^+$ and $B_d^0 \to J/\psi K^{*0}$ control channels. This result is found to be close to the ratio calculated using the MC simulation of $B_s^0 \to \mu^+\mu^-$ and $B^+ \to J/\psi K^+$ decays:

$$\left.\frac{\epsilon_{\text{cal}}^{\text{REC}}}{\epsilon_{\text{sig}}^{\text{REC}}}\right|_{\text{MC}} = 0.589 \pm 0.005 \tag{8}$$

The double ratio between the ratio using only control channels and the ratio using the $B_s^0 \to \mu^+\mu^-$ decays can be seen in Figure 4 and is close to one $(0.94 \pm 0.01)$ at the few percent level. This method relies on the approximations that the ratio between the two control channels is the same as the ratio between signal and the calibration channel, and that the efficiencies of the selections cancel out. If this is correct there is no other source of systematic uncertainties. However, given the fact that the acceptance, reconstruction and selection efficiencies affect differently the decays under study, we expect some differences at the few percent level as observed in the MC simulation. Hence, even if with very little luminosity the statistical uncertainty will be at the per mille level, the method is expected to have a precision of few percent due to the implicit approximations.

In Figure 4 the dependence on the momentum of the $B$ candidate is shown. It can be seen that both ratios are similar (within few percent) in all the phase space. This sort of check will be absolutely necessary when real data is in hand to understand if the approximations that seem to work fine in our simulation still hold with real data. There are other possible control channels to be studied such as the ratio between $B_{(s)}^0 \to h^+h^-$ and $B^\pm \to h^+h^-h^\pm$.

## 6.2 Ratio of trigger efficiencies

Calibrating a branching ratio measurement with control channels requires the determination of the corresponding trigger efficiencies, or at least the ratio between them. The method described in Section 6.1 for the determination of the reconstruction efficiencies requires in addition the ratio of trigger efficiencies between control channels themselves.

This section describes first a method for estimating from data the trigger efficiencies on channels with large event yield. Then, a modification of the method is proposed for its application to $B_s^0 \to \mu^+\mu^-$, where the event yield is small.

### 6.2.1 Estimation of trigger efficiencies for control channels

The first step in the procedure is to classify events in three categories according to the information required to trigger:

- **TIS** (Trigger Independent of Signal): the event is triggered without using any information associated with the signal under study.

- **TOS** (Trigger On Signal): the event is triggered using only information from the signal under study.
- **TOB** (Trigger On Both): both signal and non-signal information is needed to trigger.

Note that an event can be TIS and TOS simultaneously (TIS&TOS in the following), while a TOB event can be neither TOS nor TIS. The LHCb trigger system records all the information needed for such a classification.

As the ratios between reconstruction and selection efficiencies for different channels will be obtained using the method described in previous sections, all trigger efficiencies in this section are normalized to those events which would be both offline-reconstructed and selected. The overall trigger efficiency on such events can be expressed as:

$$\begin{aligned}\epsilon^{\mathrm{TRIG|SEL}} &= \frac{N^{\mathrm{TRIG|SEL}}}{N^{\mathrm{SEL}}} = \frac{N^{\mathrm{TIS|SEL}}}{N^{\mathrm{SEL}}}\frac{N^{\mathrm{TRIG|SEL}}}{N^{\mathrm{TIS|SEL}}} \\ &= \epsilon^{\mathrm{TIS|SEL}}\frac{N^{\mathrm{TRIG|SEL}}}{N^{\mathrm{TIS|SEL}}} \,. \end{aligned} \tag{9}$$

Both $N^{\mathrm{TRIG|SEL}}$ and $N^{\mathrm{TIS|SEL}}$ are observable quantities, provided enough event yield is available and that contributions from background can be subtracted. $\epsilon^{\mathrm{TIS|SEL}}$ is the efficiency of triggering without any information from the signal. This efficiency depends on the selection cuts applied on the signal, through the correlation between the momentum of the signal $B$ and the rest of the event (especially the other $b$ hadron). If the selections for the signal and control samples are similar enough so that the phase space of the $B$ is approximately the same then Eq. 9 can be used to calculate the ratio of trigger efficiencies as $\epsilon^{\mathrm{TIS|SEL}}$ approximately cancels in the ratio.

Table 5 shows the values of $\epsilon^{\mathrm{TIS|SEL}}$ for signal and control samples. The values are given after HLT1 and after HLT2, because even though the relevant numbers are after HLT2, the MC statistics used are not sufficient to make a precise enough comparison. This comparison can be done with more relative precision after HLT1 as the number of TIS events is an order of magnitude larger. Differences are small due to the special care taken in designing selections which are as similar as possible. The residual discrepancies can be traced back to the small remaining differences in the $P_{\mathrm{T}}$ distribution of the reconstructed $B$ candidate.

If the approximation that $\epsilon^{\mathrm{TIS|SEL}}$ cancels in the ratio is not good enough, then we can work in bins of the B phase space if enough statistics are available. Within a small enough bin of the momentum of the signal B, $\epsilon^{\mathrm{TIS|SEL}}$ does not depend on how the event has been selected. In particular it becomes observable as:

$$\epsilon_i^{\mathrm{TIS|SEL}} = \frac{N_i^{\mathrm{TIS\&TOS}}}{N_i^{\mathrm{TOS}}} \tag{10}$$

Once we know how to calculate $\epsilon^{\mathrm{TIS|SEL}}$, we can rewrite $N^{\mathrm{SEL}}$ in the denominator of Eq. 9 as:

Table 5: Trigger efficiencies for signal and control channels: total trigger efficiency on selected events ($\epsilon^{\mathrm{TRIG|SEL}}$) and probability to trigger independently of the signal ($\epsilon^{\mathrm{TIS|SEL}}$) computed from MC, fraction of TIS events in the triggered sample, and estimates of $\epsilon^{\mathrm{TIS|SEL}}$ and $\epsilon^{\mathrm{TOS|SEL}}$ obtained from the fraction of TIS&TOS events in the TOS and TIS samples, respectively. Results are given in % at two rates; ~30kHz after HLT1 and ~2kHz after HLT2.

| Channel | $\epsilon^{\mathrm{TRIG\|SEL}}$ (%) | $\epsilon^{\mathrm{TIS\|SEL}}$ (%) | $N^{\mathrm{TIS}}/N^{\mathrm{TRIG}}$ (%) | $N^{\mathrm{TIS\&TOS}}/N^{\mathrm{TOS}} \simeq \epsilon^{\mathrm{TIS\|SEL}}$(%) | $N^{\mathrm{TIS\&TOS}}/N^{\mathrm{TIS}} \simeq \epsilon^{\mathrm{TOS\|SEL}}$(%) |
|---|---|---|---|---|---|
| After L0+HLT1 | | | | | |
| $B_s^0 \to \mu^+\mu^-$ | $95.0 \pm 0.1$ | $4.4 \pm 0.1$ | $4.6 \pm 0.1$ | $4.4 \pm 0.1$ | $95.2 \pm 0.6$ |
| $B_{(s)}^0 \to h^+h^-$ | $36.5 \pm 0.2$ | $4.16 \pm 0.06$ | $11.4 \pm 0.2$ | $4.9 \pm 0.1$ | $39.4 \pm 0.8$ |
| $B^+ \to J/\psi K^+$ | $87.67 \pm 0.09$ | $4.40 \pm 0.06$ | $5.02 \pm 0.06$ | $4.50 \pm 0.06$ | $88.0 \pm 0.4$ |
| After L0+HLT1+HLT2 | | | | | |
| $B_s^0 \to \mu^+\mu^-$ | $93.2 \pm 0.1$ | $0.71 \pm 0.04$ | $0.76 \pm 0.05$ | $0.71 \pm 0.05$ | $91 \pm 2$ |
| $B_{(s)}^0 \to h^+h^-$ | $29.9 \pm 0.1$ | $0.72 \pm 0.03$ | $2.40 \pm 0.09$ | $0.76 \pm 0.05$ | $29 \pm 2$ |
| $B^+ \to J/\psi K^+$ | $76.6 \pm 0.1$ | $0.71 \pm 0.02$ | $0.93 \pm 0.03$ | $0.73 \pm 0.03$ | $76 \pm 1$ |

$$N^{\mathrm{SEL}} = \sum_i^{\mathrm{bins}} \frac{N_i^{\mathrm{TIS|SEL}}}{\epsilon_i^{\mathrm{TIS|SEL}}} \tag{11}$$

which implies the remarkable result that $\epsilon^{\mathrm{TRIG|SEL}}$ can be expressed in terms of fully observable quantities as:

$$\epsilon^{\mathrm{TRIG|SEL}} = \frac{N^{\mathrm{TRIG|SEL}}}{\sum_i^{\mathrm{bins}} \frac{N_i^{\mathrm{TIS|SEL}}}{\epsilon_i^{\mathrm{TIS|SEL}}}} , \tag{12}$$

Note also that $\epsilon_i^{\mathrm{TIS|SEL}}$ in each bin is independent of the decay considered, hence it can be extracted from any channel, or even from the combination of several channels to reduce statistical uncertainties.

The efficiency of triggering using only information from the signal candidate, $\epsilon_i^{\mathrm{TOS|SEL}}$, can be analogously measured by replacing $N^{\mathrm{TRIG|SEL}}$ by $N^{\mathrm{TOS|SEL}}$. The results of $N^{\mathrm{TIS\&TOS}}/N^{\mathrm{TOS}}$ together with the true value for $\epsilon^{\mathrm{TIS|SEL}}$ are shown in Table 5 for signal and control channels. The true value for $\epsilon^{\mathrm{TOS|SEL}}$ is quite close to $\epsilon^{\mathrm{TRIG|SEL}}$ shown in the second column in Table 5 as the contribution from non-TOS events is very small in the current design of the trigger (less than 1% for the decay $B_s^0 \to \mu^+\mu^-$).

Looking at the relatively more precise numbers after HLT1 it can be seen that there is good agreement except for the case of the $B^0_{(s)} \to h^+h^-$ control channels, where discrepancies are at the level of $\sim 20\%$. This is caused by the global event cuts (GECs) in the L0 hadron trigger. The GECs veto events with high multiplicity or with more than one primary vertex, in order to promptly remove complicated events from trigger processing. As a result, the whole event is used in the trigger decision and the TIS/TOS/TOB classification fails. A possible solution is to consider only $B^0_{(s)} \to h^+h^-$ events triggered through the muon lines, which do not use GECs, but that would mean reducing the available statistics of TIS events. An alternative solution being considered is removing GEC from the rest of the L0 lines.

As $\epsilon^{\mathrm{TIS|SEL}}$ is similar enough for signal and control channels, the ratio of trigger efficiencies between $B^+ \to J/\psi(\mu^+\mu^-)K^+$ and $B^0_d \to J/\psi(\mu^+\mu^-)K^{*0}(K^+\pi^-)$ needed for correcting the ratio of reconstruction efficiencies in Section 6.1 can be computed using Eq. 9. With 0.1 $\mathrm{fb}^{-1}$, the result is:

$$\frac{\epsilon^{\mathrm{TRIG|SEL}}_{B^0_d \to J/\psi K^{*0}}}{\epsilon^{\mathrm{TRIG|SEL}}_{B^+ \to J/\psi K^+}} = 1.00 \pm 0.03\,, \tag{13}$$

which is in agreement with the value expected from MC:

$$\left.\frac{\epsilon^{\mathrm{TRIG|SEL}}_{B^0_d \to J/\psi K^{*0}}}{\epsilon^{\mathrm{TRIG|SEL}}_{B^+ \to J/\psi K^+}}\right|_{\mathrm{MC}} = 1.000 \pm 0.004\,. \tag{14}$$

Similarly, the ratio of trigger efficiencies between $B^0_d \to K^+\pi^-$ and $B^+ \to J/\psi(\mu^+\mu^-)K^+$, can be calculated with $0.1\,\mathrm{fb}^{-1}$ as:

$$\frac{\epsilon^{\mathrm{TRIG|SEL}}_{B^+ \to J/\psi K^+}}{\epsilon^{\mathrm{TRIG|SEL}}_{B^0_d \to K^+\pi^-}} = 2.72 \pm 0.15\,, \tag{15}$$

which is again in agreement with the value expected from MC:

$$\left.\frac{\epsilon^{\mathrm{TRIG|SEL}}_{B^+ \to J/\psi K^+}}{\epsilon^{\mathrm{TRIG|SEL}}_{B^0_d \to K^+\pi^-}}\right|_{\mathrm{MC}} = 2.56 \pm 0.01\,. \tag{16}$$

This formalism based on binning on the momentum of the $B$ has an interesting application [22]. The inverse of the TIS efficiency in each bin can be used as a weighting factor on the TIS sample in order to recover the distributions of any property before the trigger. This is illustrated for the case of the kaon $P_{\mathrm{T}}$ distribution in $B^0_d \to K^+\pi^-$ events in Figure 5.

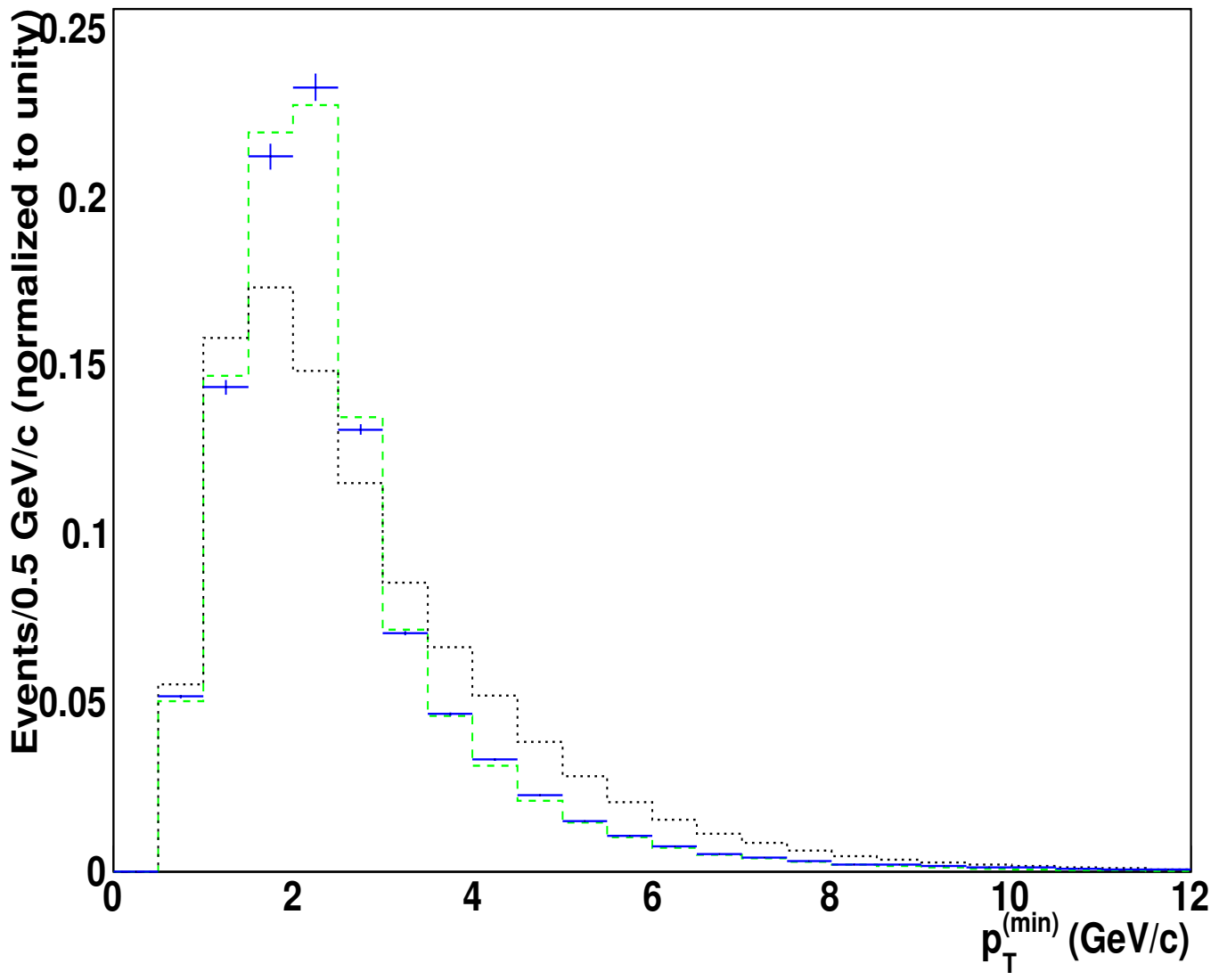

Figure 5: Minimum $P_{\mathrm{T}}$ (in GeV/$c$) distribution of the daughters measured in selected $B_d^0 \to K^+\pi^-$ events without trigger (green dashed line) compared with the distribution after trigger (black dotted) and computed using the weights described in the text (blue data points).

### 6.2.2 Evaluation of the trigger efficiency for $B_s^0 \to \mu^+\mu^-$

Because of the low event yield, Eq. 9 cannot be used directly to measure the trigger efficiency on $B_s^0 \to \mu^+\mu^-$. The method proposed here to obtain this efficiency without using the MC simulation consists of two steps:

1. The effect of the trigger on pairs of muons is parameterized using an independent sample of dimuon decays in events which would have triggered without using the information from the dimuon. There are plenty of possible sources of dimuon decays, but in this particular example the $B^+ \to J/\psi(\mu^+\mu^-)K^+$ control channel has been used.

2. The properties of the particles from $B_s^0$ decays before trigger for $B_s^0 \to \mu^+\mu^-$ are obtained by using the fact that the decay is kinematically similar to $B_{(s)}^0 \to h^+h^-$. The effect of the trigger could be unfolded by the reweighting method proposed above, but here we just assume that the properties of the particles before trigger are the same as those in TIS events.

For simplicity, the discussion is restricted to the single and dimuon triggers, which provide most of the efficiency for events in which both muons are offline reconstructed. For the emulation, the trigger is split into two complementary cases:

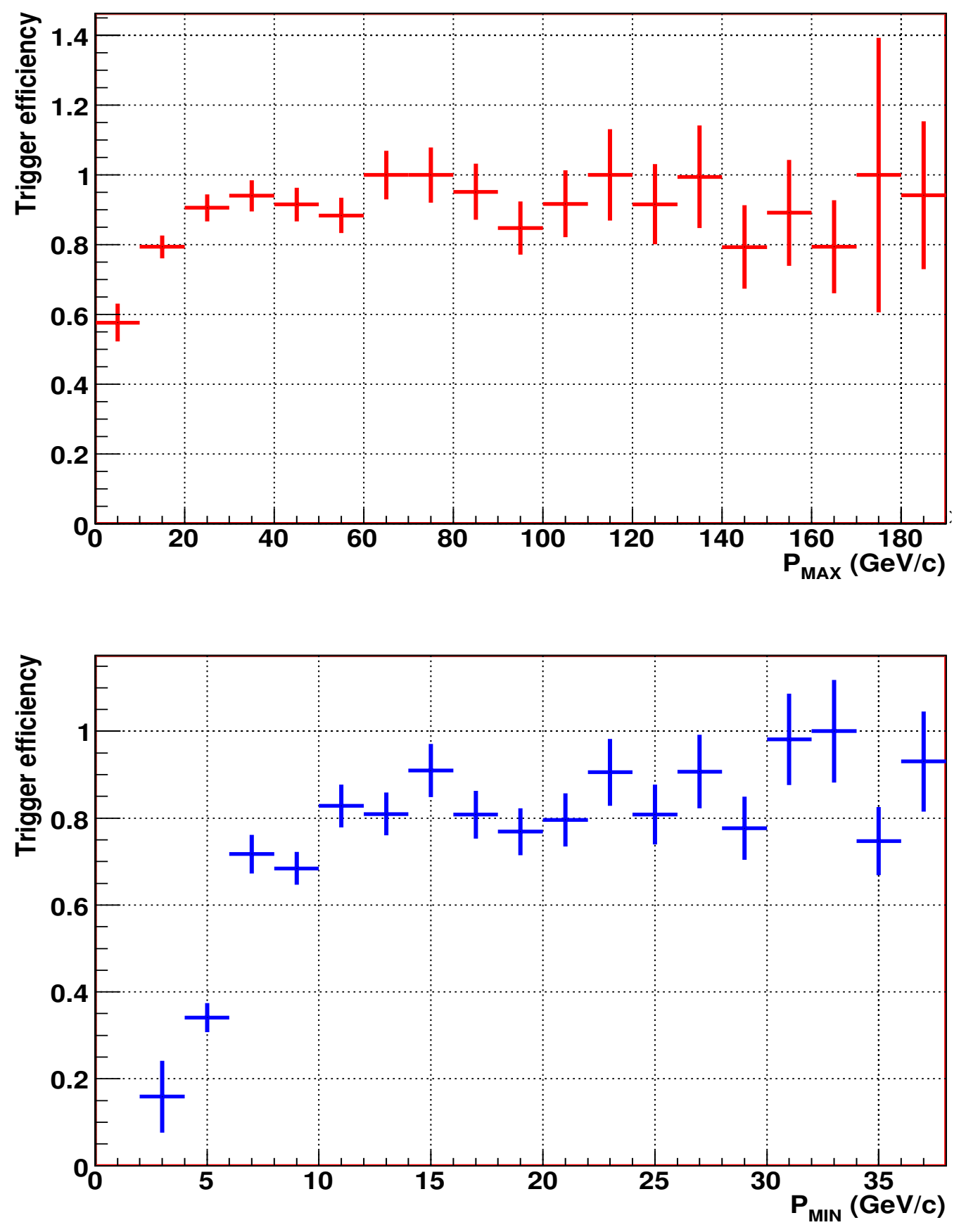

Figure 6: Efficiencies calculated using Eq. 9 for single (top) and dimuon (bottom) trigger lines in bins of the maximum (minimum) momentum of the muon candidates using selected $B^+ \to J/\psi(\mu^+\mu^-)K^+$.

a) One of the online muon tracks satisfies the IP and/or $P_{\mathrm{T}}$ cuts applied in the single muon line of the trigger.

b) Neither of the two online tracks satisfies the single muon line cuts. In this case the signal can only trigger on the dimuon line.

As the probability for an offline muon to be online reconstructed is strongly dependent on its momentum, the efficiency of the single muon line is parameterized as a function of the maximum momentum of the muon(s) which satisfy the single muon line cuts.

For the dimuon line, the efficiency is parameterized as a function of the minimum of both momenta, as both muons need to be identified. Other cuts, like invariant mass or vertex quality can be neglected as they are much softer than the ones applied offline.

Using $B^+ \to J/\psi(\mu^+\mu^-)K^+$ events the efficiency was computed using Eq. 9 in a given bin of the relevant parameter, i.e. maximum momentum in the single muon line and minimum momentum in the dimuon line. Notice that the trigger efficiency is only the efficiency of the particular trigger line under study. The results of these calculations can be seen in Figure 6. Notice that the normalization of Figure 6 has been done using $\epsilon^{\mathrm{TIS|SEL}}$ using Eq. 10. This is an overall normalization factor that will cancel in the ratio with the control channels.

In order to prevent distortions of the parameterizations due to the presence of background in the calibration sample, the offline selection can be tightened until the background level becomes negligible.

The last step is to use $B^0_{(s)} \to h^+h^-$ TIS events to determine the relevant phase space and weight the muon candidates accordingly in order to compute the trigger efficiency for signal.

So far, only the HLT1 trigger has been parameterized, hence the estimated HLT1 trigger efficiency on $B^0_s \to \mu^+\mu^-$ with $0.1\,\mathrm{fb}^{-1}$ after HLT1 is:

$$\epsilon^{\mathrm{TRIG|SEL}}_{B^0_s\to\mu^+\mu^-} = 0.92 \pm 0.07 \tag{17}$$

in good agreement with the MC expectation of 0.951. This same parameterization can be used to compute the fraction of signal TIS events in Equation 9. Using $0.1\,\mathrm{fb}^{-1}$ of data after HLT1, we obtain:

$$\frac{\epsilon^{\mathrm{TRIG|SEL}}_{B^+\to J/\psi K^+}}{\epsilon^{\mathrm{TRIG|SEL}}_{B^0_s\to\mu^+\mu^-}} = 0.95 \pm 0.02 \tag{18}$$

This result is in good agreement with the MC expectation:

$$\left.\frac{\epsilon^{\mathrm{TRIG|SEL}}_{B^+\to J/\psi K^+}}{\epsilon^{\mathrm{TRIG|SEL}}_{B^0_s\to\mu^+\mu^-}}\right|_{\mathrm{MC}} = 0.926 \pm 0.004\,. \tag{19}$$

Similarly, we can apply this procedure to the $B^0_d \to K^+\pi^-$ normalization channel for $0.1\,\mathrm{fb}^{-1}$ of data after HLT1 we expect

$$\frac{\epsilon^{\mathrm{TRIG/SEL}}_{B^0_d\to K^+\pi^-}}{\epsilon^{\mathrm{TRIG/SEL}}_{B^0_s\to\mu^+\mu^-}} = 0.39 \pm 0.02\,, \tag{20}$$

in good agreement with the MC expectation:

$$\left.\frac{\epsilon^{\mathrm{TRIG/SEL}}_{B^0_d\to K^+\pi^-}}{\epsilon^{\mathrm{TRIG/SEL}}_{B^0_s\to\mu^+\mu^-}}\right|_{\mathrm{MC}} = 0.372 \pm 0.001\,. \tag{21}$$

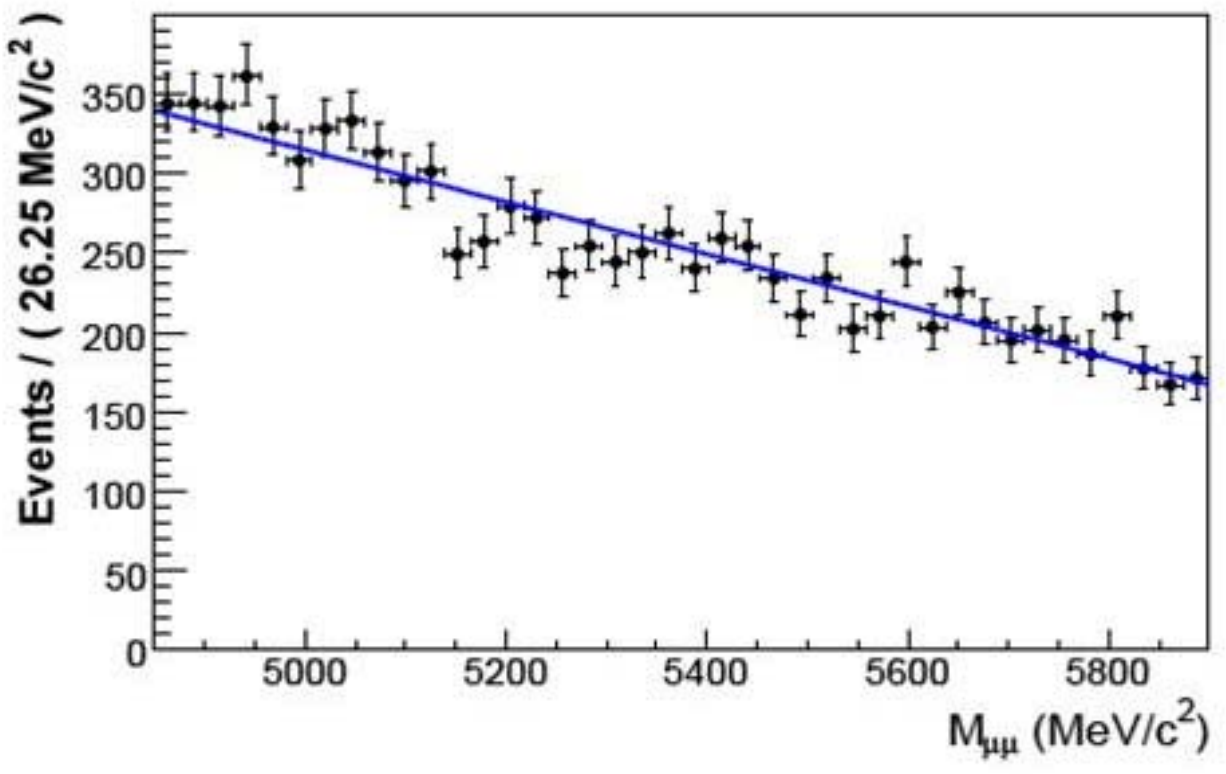

Figure 7: Invariant mass distribution for combinatorial background events in MeV/$c^2$ and the result of a linear fit (blue line).

# 7 Invariant mass and geometrical likelihood calibration

In the previous section we have described how to translate the number of signal candidates into a measurement of the branching ratio. This normalization factor multiplies the number of candidates in each bin of the likelihood function. In this section we will describe the procedure to calibrate the likelihood assigned to each candidate. The likelihood is composed of three independent probabilities: IML, GL and PIDL. The calibration of the PIDL will be discussed in Section 8; here we focus in the first two components, as both calibrations use the $B^0_{(s)} \to h^+h^-$ control channel.

## 7.1 Invariant mass likelihood calibration

The trigger and offline selections have a factor ten larger $B^0_s$ mass window ($\pm 600\,\mathrm{MeV}/c^2$) than the final cut applied in Section 5. This allows the use of events in the sidebands to estimate the combinatorial background in the signal region, without relying on any MC simulation. Figure 7 shows how a simple linear interpolation gives a good enough prediction; hence the invariant mass likelihood for a background event can be obtained from this simple parameterization.

The strategy to evaluate the invariant mass likelihood for a signal candidate is to use the decay $B^0_s \to K^+K^-$. This has the advantage of being kinematically very similar to the signal, and avoids tails in the invariant mass distribution due to reflections that may appear in decays like $B^0_d \to K^+\pi^-$ or $B^0_s \to \pi^+K^-$. However, in order to select this decay with enough purity, relatively strong cuts on the kaon identification are needed, which bias the momentum and hence change the invariant mass resolution we are trying to measure. The idea here is to evaluate the change in the measured resolution in the inclusive $B^0_{(s)} \to h^+h^-$ sample as a function of the cut in the kaon/pion ID, as can be seen in Figure 8. Notice that this $B^0_{(s)} \to h^+h^-$ sample is only used to evaluate the correction, and not the resolution itself. Figure 8 (top) shows the invariant mass assuming both the hadrons "h" are kaons. The distribution is fitted with a double Crystal-Ball parameterization [23] where the two components are forced to have the same mass resolution ($\sigma_{\rm fit} = 25.4\,{\rm MeV}/c^2$). This simplification does not introduce a sizable uncertainty in the evaluation of the correction as shown later in this section. Figure 8 (bottom) shows the invariant mass for well identified "h" as kaons or pions, i.e. with an absolute difference in the logarithm of the likelihood (DLL) larger than 20 ($\sigma_{\rm fit} = 21.5\,{\rm MeV}/c^2$). Notice, that the $B^0_d$ peak is not at the right position as the kaon hypothesis is used to compute the invariant mass. If the bias introduced by the cut on the kaon/pion-ID is the same for all $B^0_{(s)} \to h^+h^-$ decay modes, we can use the bias measured in the inclusive sample to correct the measurement of the resolution in the $B^0_s \to K^+K^-$ decay when strong kaon-ID cuts are applied. The study of MC samples with only $B^0_d \to K^+\pi^-$ or $B^0_s \to K^+K^-$ decays shows that the bias in the measured $\sigma$ using a Crystal-Ball parameterization as a function of the kaon/pion-ID cut is consistent with the measurements using the inclusive $B^0_{(s)} \to h^+h^-$ sample, see Figure 9, hence it validates the approach.

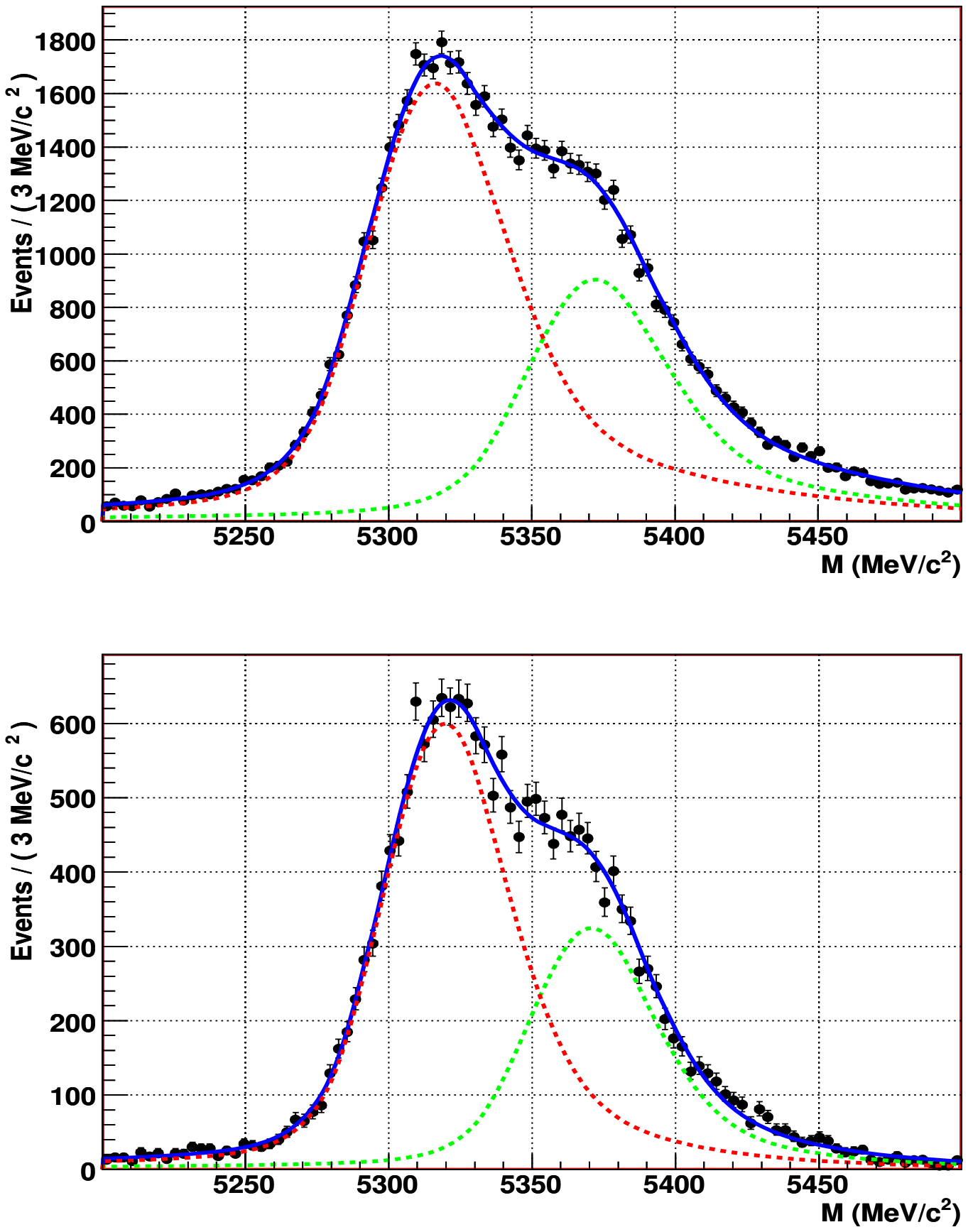

Figure 8: $K^+K^-$ invariant mass distribution for inclusive $B^0_{(s)} \to h^+h^-$ events, without cut on the kaon ID (top, $\sigma_{\rm fit} = 25.4\,{\rm MeV}/c^2$) and with a kaon and pion ID requirement $|{\rm DLL}| > 20$ (bottom, $\sigma_{\rm fit} = 21.5\,{\rm MeV}/c^2$). The two Crystal-Ball functions are forced to have the same width.

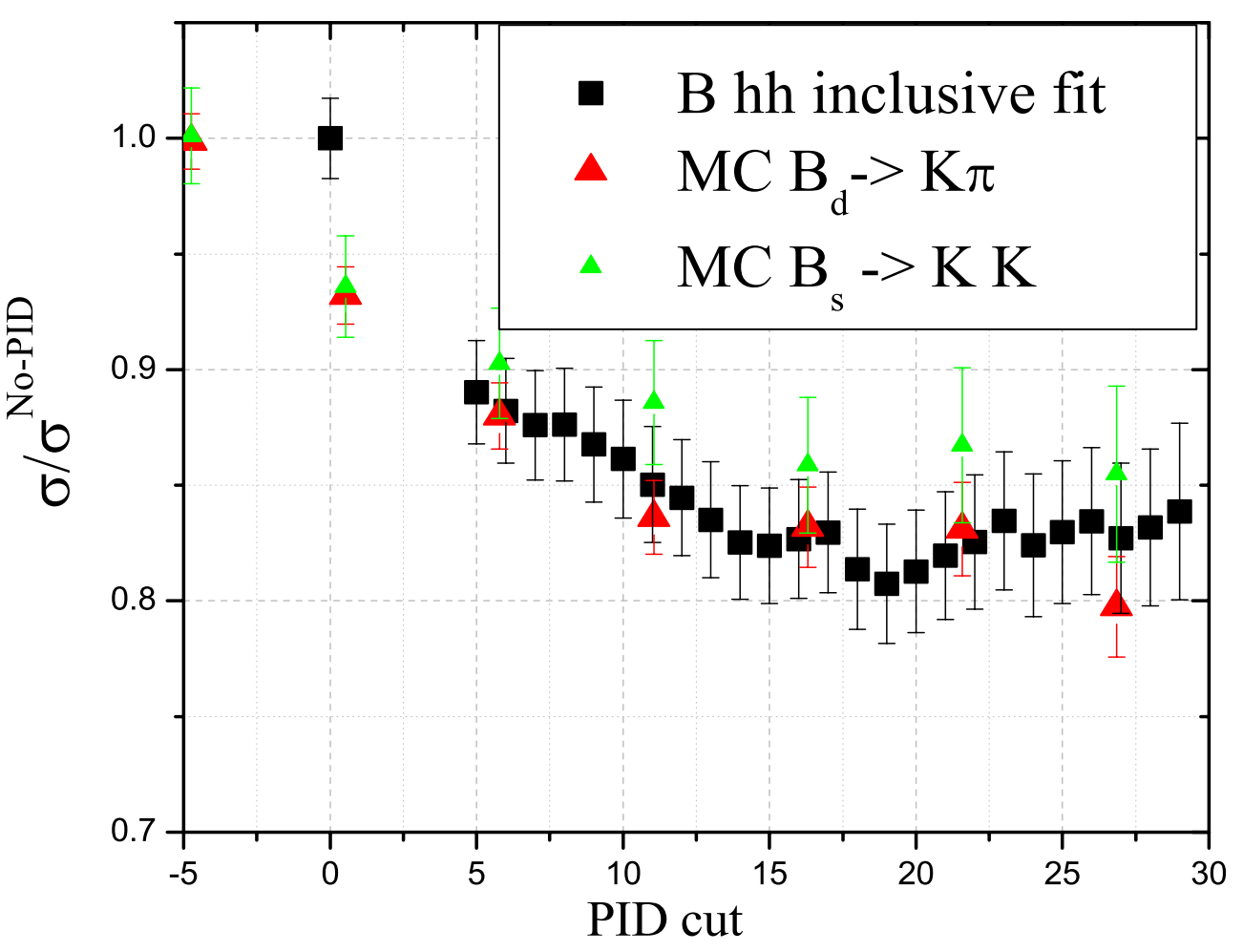

Figure 9: Shift of the invariant mass resolution ($\sigma/\sigma_0$) as a function of the PID cut measured in the inclusive $B^0_{(s)} \to h^+h^-$ sample (black squares), compared to the shift computed using samples of MC events with only $B^0_d \to K^+\pi^-$ (red triangles) and $B^0_s \to K^+K^-$ (green triangles) events.

Once the dependence of the mass resolution on the kaon ID cut is measured, then we only need to correct the measured resolution in the decay $B^0_s \to K^+K^-$ with the measured correction factor. As can be seen in Figures 9 and 10, this correction is not negligible ($\sim 20\%$), and once applied, the invariant mass resolution agrees nicely with the results of the fit using $B^0_s \to \mu^+\mu^-$ decays (see Figure 10).

## 7.2 Geometrical likelihood calibration

The mathematical method to build the geometrical likelihood is explained in Ref. [12]. It is a combination of six variables: the lifetime of the $B^0_s$ candidate, the lowest impact parameter significance of the two muon candidates and the impact parameter of the $B^0_s$ candidate w.r.t the closest primary vertex, the distance of closest approach of the two muon candidates, and the isolation of each of the muon candidates. The coefficients of the matrix that transforms these six variables into a single discriminating variable (named "geometrical likelihood", GL) distributed between zero and one, are computed from the MC simulation. In principle, one could imagine to redefine these coefficients using the $B^0_{(s)} \to h^+h^-$ sample and reoptimize the analysis, but our initial strategy is not to change the definition and just use the $B^0_{(s)} \to h^+h^-$ sample to measure the GL distribution with the coefficients defined "a priori". As a matter of fact, the definition of GL has not changed from the analysis in Ref. [12] in the evaluation of the sensitivity in Section 9. This turns out to be a good test of the calibration procedure, as the MC samples used to define GL are not completely compatible with the samples used in this note.

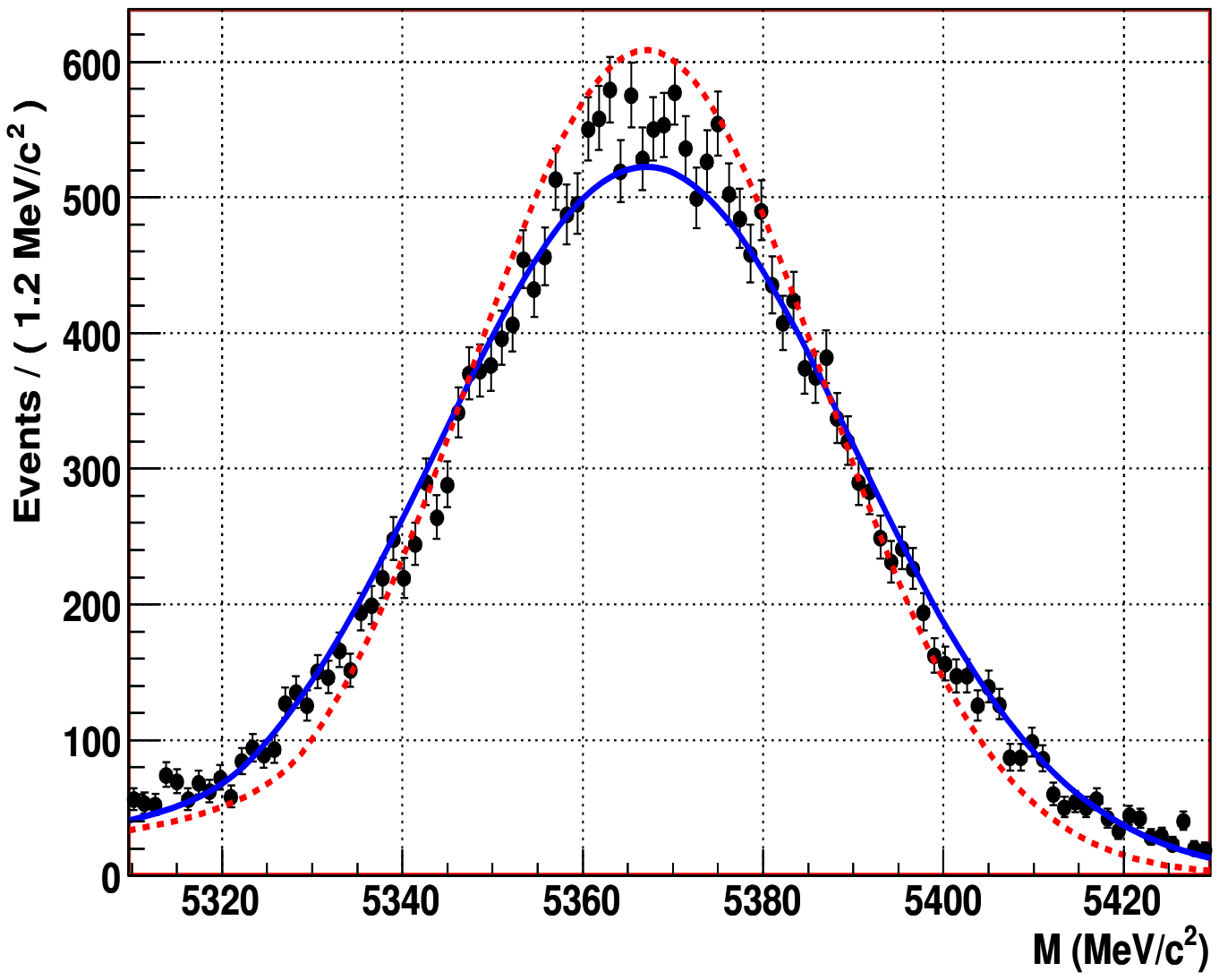

Figure 10: Invariant mass distribution for $B_s^0 \to \mu^+\mu^-$ candidates (data points) compared with the fit to the distribution from $B_s^0 \to K^+K^-$ candidates without (red dashed curve) and with (solid blue curve) the correction described in the text.

The strategy is to use the events in the sidebands of the invariant mass distribution to evaluate the GL distribution for background events.

The GL distribution for signal-like events can be evaluated in principle using $B^0_{(s)} \to h^+h^-$, however there are several issues to be considered. The GL distribution for $B_s^0 \to \mu^+\mu^-$ events is designed to be flat between zero and one, and identical to the one obtained from $B^0_{(s)} \to h^+h^-$ when no trigger is applied. Figure 11 shows the GL distribution computed from the simulation using $B_s^0 \to \mu^+\mu^-$ events separated according to the different triggers, but with arbitrary normalization. As can be seen, while the total distribution is almost flat as expected [3], because it is dominated by the dimuon trigger (see Section 4), the shape would be very different if the single muon trigger dominates. Hence, if for some reason in the real experiment the fraction of events selected by the single muon trigger is much higher, the shape would not be what is expected and will not agree with the one obtained from $B_{(s)}^0 \to h^+h^-$.

[3]It is not completely flat due to the differences between the MC used to define the GL (DC04), and the MC used to evaluate the performance (DC06).

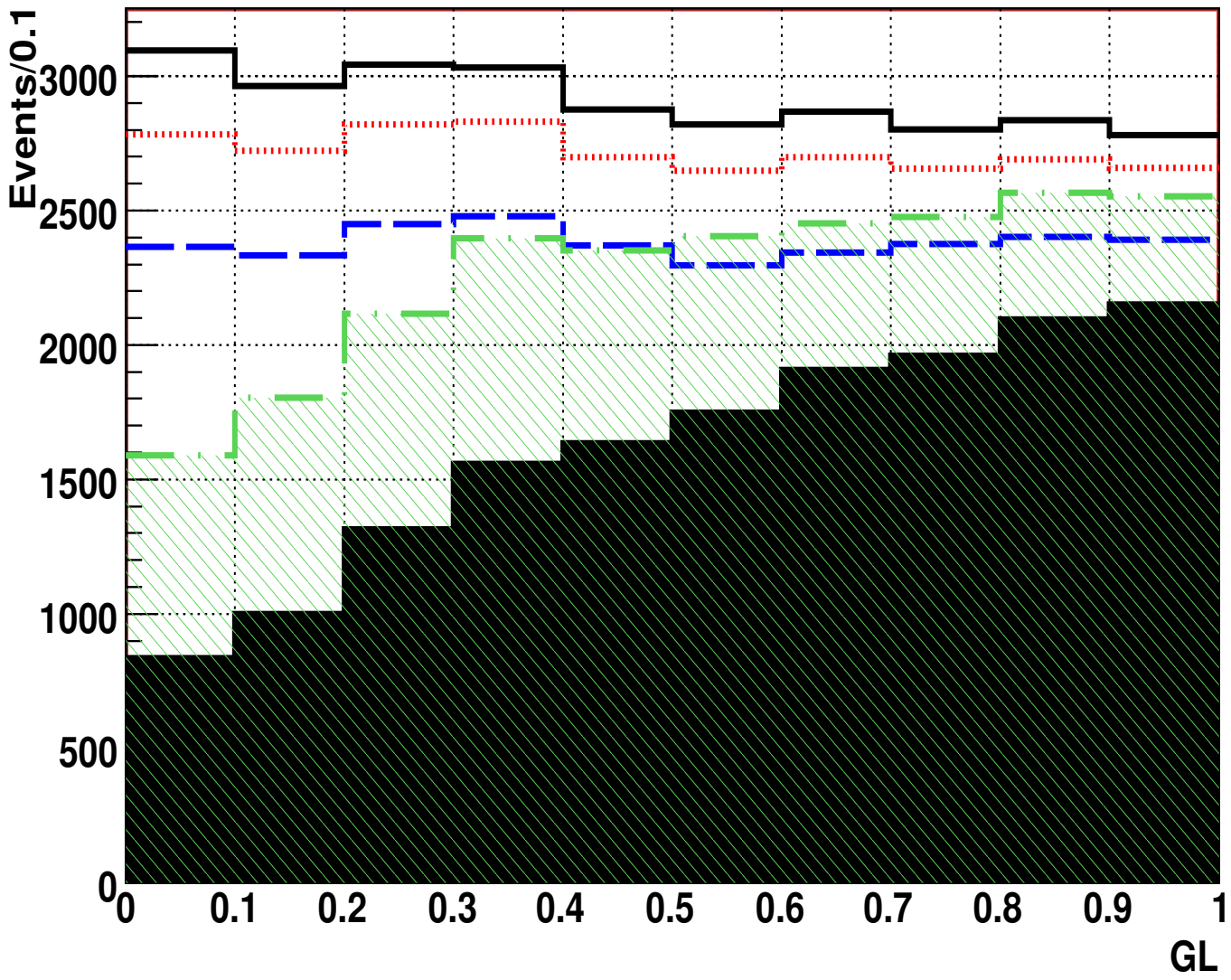

Figure 11: GL distribution computed for offline selected $B_s^0 \to \mu^+\mu^-$ candidates (black line), after trigger (red dotted line), after the dimuon trigger only (blue dashed), after muon plus track only (green dot-dashed with diagonal fill) and after the single-muon trigger only (black filled).

In order to avoid this situation, we need to evaluate the GL distribution for $B_s^0 \to \mu^+\mu^-$ events in the data and separate the different trigger contributions. As we do not have enough $B_s^0 \to \mu^+\mu^-$ candidates to do this, we need to use the same "trigger emulation" using control samples as described in Section 6.2.2.

The reason why the events triggered using the single muon trigger have such a different GL distribution compared with the events triggered with the dimuon trigger are the strong IP and $P_{\mathrm{T}}$ cuts. For the same reason, we expect that the shape obtained from triggered $B_{(s)}^0 \to h^+h^-$ candidates is going to be affected similarly, as this trigger relies strongly on the IP and $P_{\mathrm{T}}$ cuts. This is what is observed in Fig. 12 (top), where the GL distribution shows similarities with the one from the single muon trigger, but it is not the flat distribution that we would like to obtain. In this case the solution is to use again TIS events, as shown in Fig. 12 (bottom). In principle, using exclusive $B_{(s)}^0 \to h^+h^-$ decays would help to reduce the background, but this implies applying relatively strong PID cuts, that again modifies the shape of the GL distribution we are trying to obtain. The idea is then to use the inclusive $B_{(s)}^0 \to h^+h^-$ TIS events to determine the GL distribution for signal events. After background subtraction from the sidebands a precision of $\sim 6\%$ is expected in the sensitive region with $0.1\,\mathrm{fb}^{-1}$ of data.

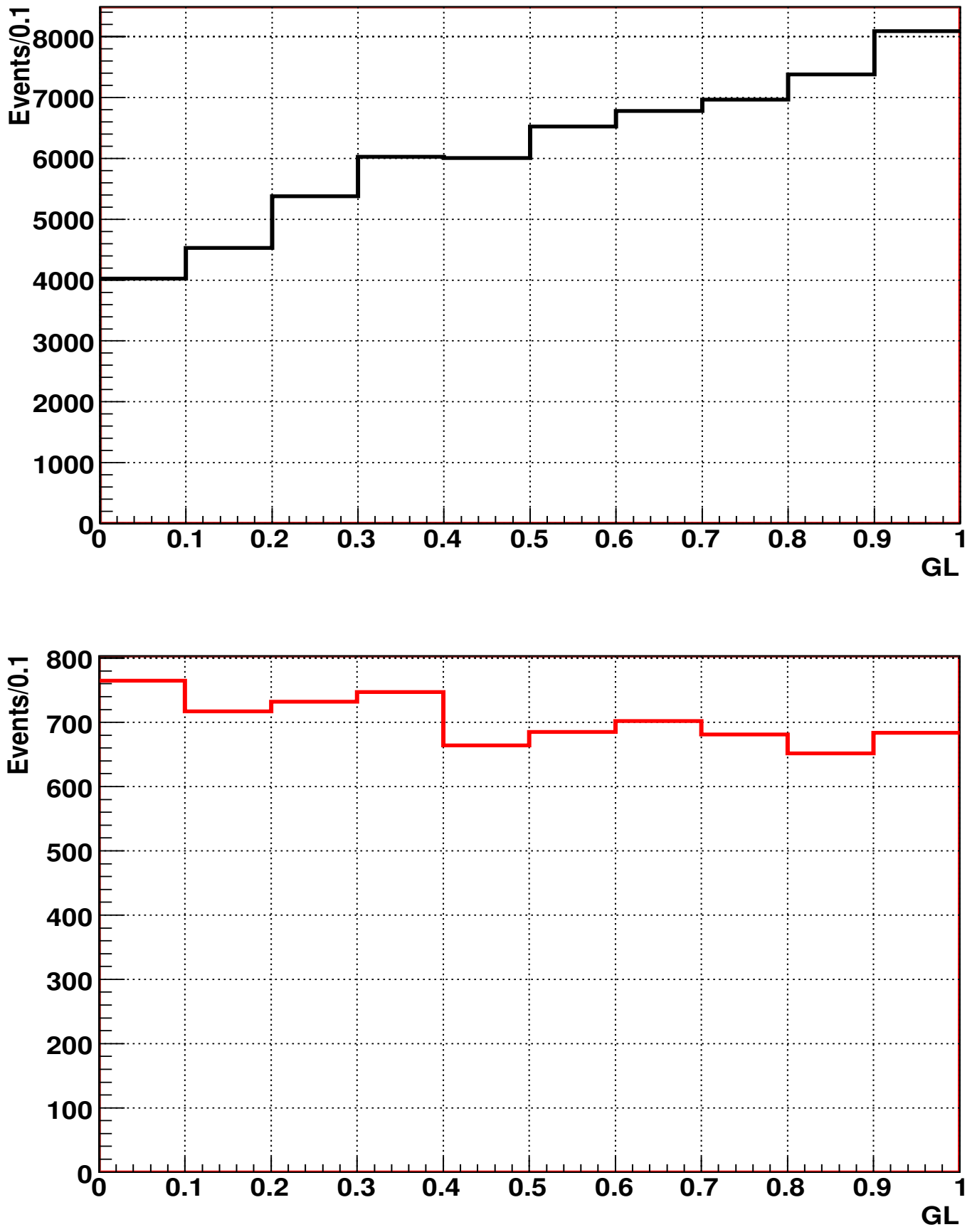

Figure 12: GL distribution computed for offline selected and triggered $B^0_{(s)} \to h^+h^-$ candidates (black histogram on top), and using only TIS events (red histogram on the bottom).

# 8 Muon ID calibration

The standard muon ID algorithm [24] is a two-step procedure. In the first step, a track is required to have a given number of muon hits inside a Field of Interest (FoI) to be considered as a muon. For example a muon with a momentum above 10 GeV/$c$ should have hits in all muons stations inside the FoI. In the second step, a discriminant variable is used to identify muons, constructed as the difference of likelihood between a track to be a muon or a non muon, according to the distribution of the distance (squared) of the hits inside the FoI, normalized by the pad side, to the point of the track extrapolation to the stations. This is called the delta log likelihood (DLL) and can be used according to the needs of the physics analysis. There are ongoing studies to improve this muon ID, in order to have a robust method against possible chamber inefficiencies, and especially in order to have a DLL independent of the muon momentum. This is particulary relevant for the calibration procedure, as the DLL obtained with the calibration sample of muons would then be valid for other samples, in particular for the muons of the $B_s^0 \to \mu^+\mu^-$ decay.

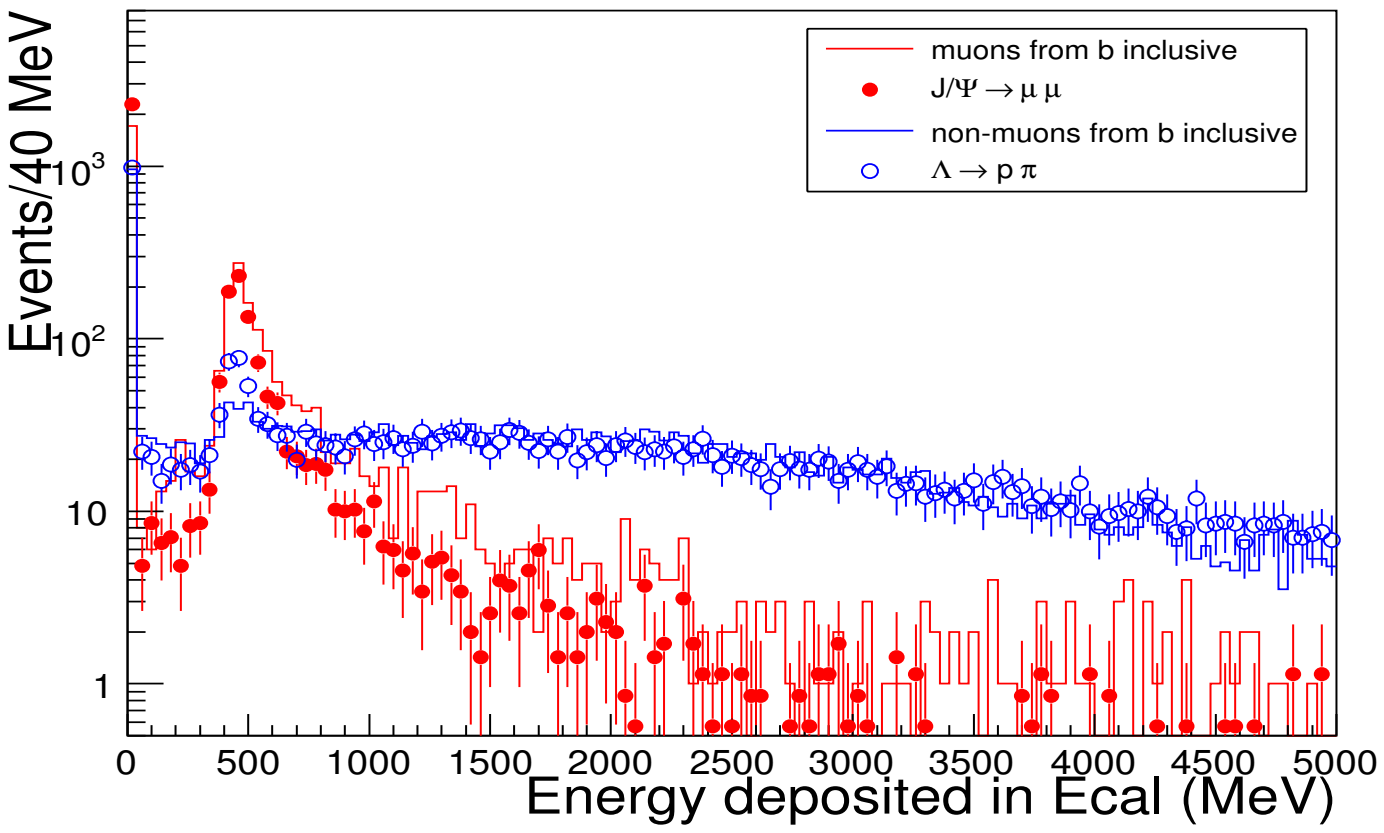

Figure 13: Energy deposited by MIPs in the ECAL ( MeV). The blue histogram corresponds to the energy distribution for non-muons while the red histogram is for muons. The blue (red) data points correspond to the energy distribution obtained from the calibration samples.

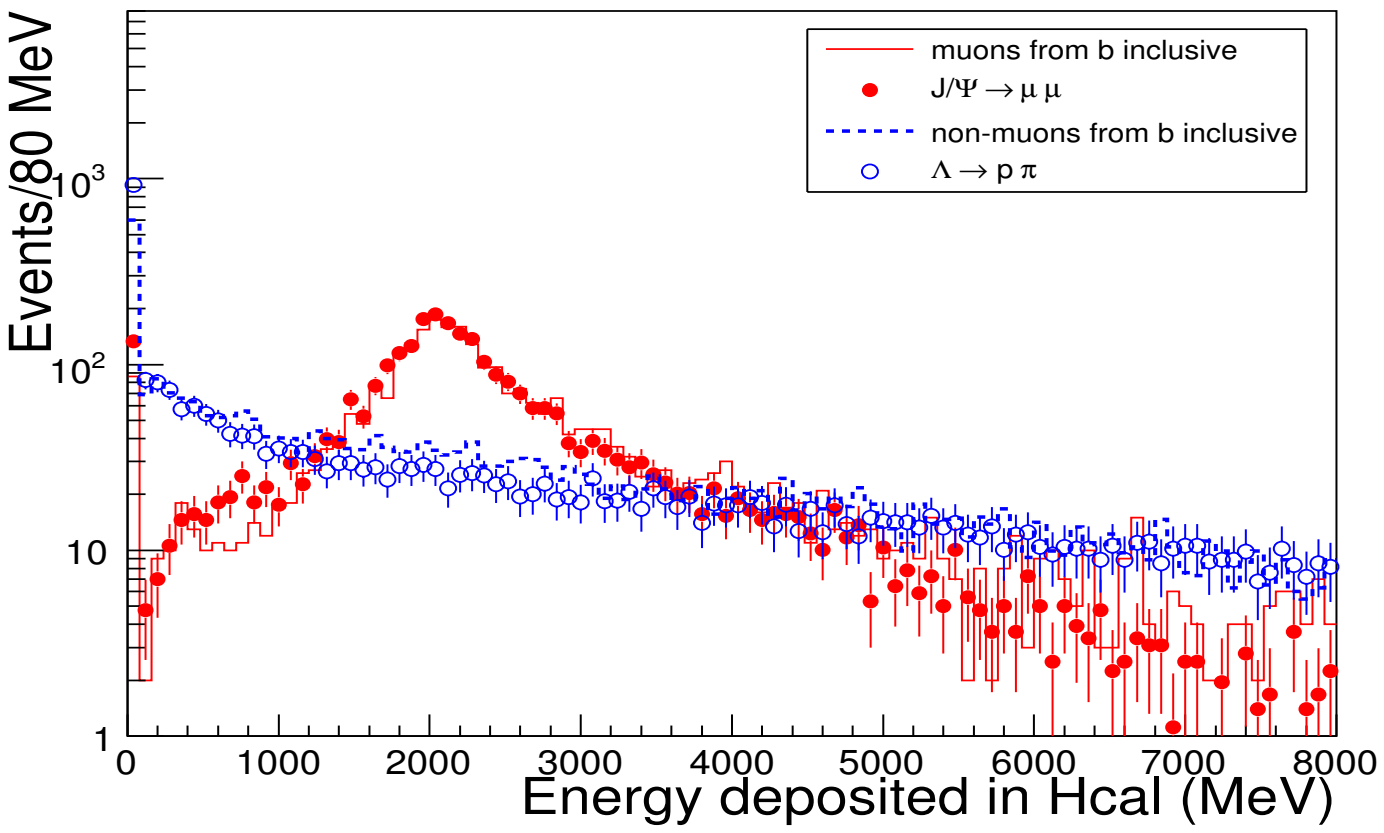

Figure 14: Energy deposited by MIPs in the HCAL ( MeV). The blue histogram corresponds to the energy distribution for non-muons while the red histogram is for muons. The blue (red) data points correspond to the energy distribution obtained from the calibration samples.

One possible strategy to calibrate the DLL is to obtain the distance squared variable distributions from a sample of muons from the decay $J/\psi \to \mu^+\mu^-$ and a sample of hadrons from the decay $\Lambda \to p\pi^-$ [25] as non muons. The $J/\psi \to \mu^+\mu^-$ decay is an abundant process at the LHC with a very clean signature. The inclusive $J/\psi$ cross section is $\sim 290\,\mu$b with 93% from prompt production and only 7% from $b$-hadron decays. Assuming the $\mathcal{B}(J/\psi \to \mu^+\mu^-) = (5.93 \pm 0.06)\%$ [20] and $\sim 10\%$ probability that both muons are within the LHCb acceptance, we expect an effective cross section of $1.7\,\mu$b which corresponds to an event yield of $1.7 \times 10^6$ per pb$^{-1}$. The strategy is to select $J/\psi$, in which one of the muons is identified without the information from the muon stations. Muons deposit in the calorimeter an energy compatible with a Minimum Ionizing Particle (MIP). Figures 13 and 14 show the energy distribution in ECAL and HCAL deposited by MIPs. The distributions obtained from the MC simulation show a clear peak which allows a clean selection.

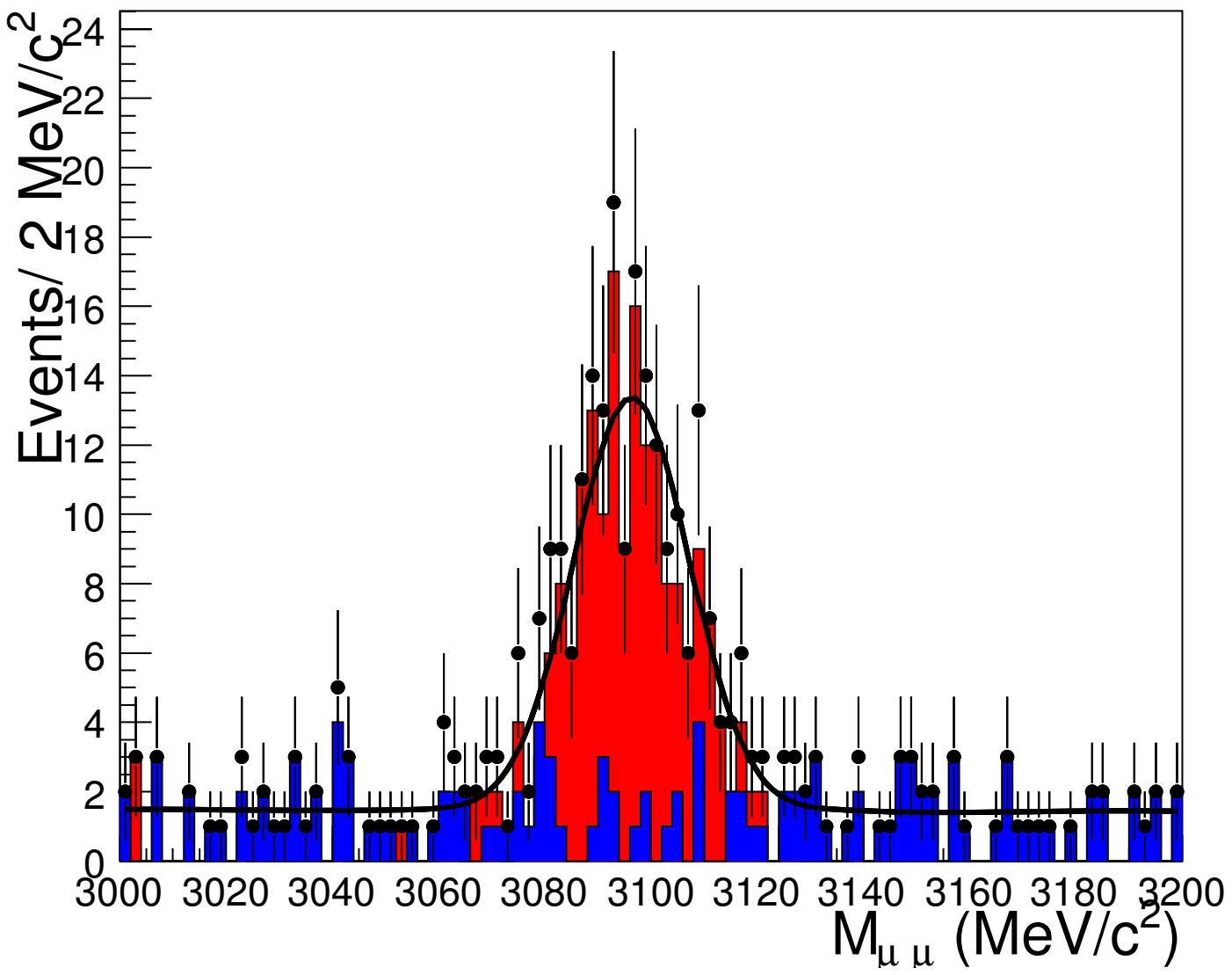

Figure 15: Invariant mass distribution of the events surviving the simple $J/\psi$ selection cuts in a sample of minimum bias events. The histogram in blue corresponds to the remaining combinatorial background.

With simple cuts on the vertex $\chi^2$ and the energy distribution in the calorimeters a sample of high purity muons can be obtained. To avoid bias introduced by the trigger, the muon is required not to be the only source of trigger in the event. Figure 15 shows the $J/\psi$ invariant mass distribution obtained from a sample of $\sim 4 \times 10^6$ minimum bias events after the L0 trigger.

In order to calibrate the non-muon distribution of the squared distance variable, the strategy is to use the decay $\Lambda \to p\pi^-$. This is an abundant process at LHC. The prompt $\Lambda$ and $\bar{\Lambda}$ production cross-section are $\sim 11\,$mb and $\sim 4.5\,$mb, respectively. The branching ratio into charged tracks is $\mathcal{B}(\Lambda \to p\pi^-) = (63.9 \pm 0.5)\%$. Assuming a visible proton-proton cross section of $\sim 64\,$mb, we expect a $\Lambda \to p\pi^-$ decay in every 6 minimum bias events produced. This channel is particularly interesting for studying separately the misidentification due to hadrons decaying and non-decaying in flight, due to the fact that it contains both protons and pions among the decay products. Moreover the $\Lambda$ is a very narrow resonance ($\sim 1\,\mathrm{MeV}/c^2$) and has a very long lifetime (corresponding to an average flight lenght of 7.9 cm in LHCb), therefore it can be easily selected without any particle ID using only tight cuts on impact parameters, flight distance and invariant mass.

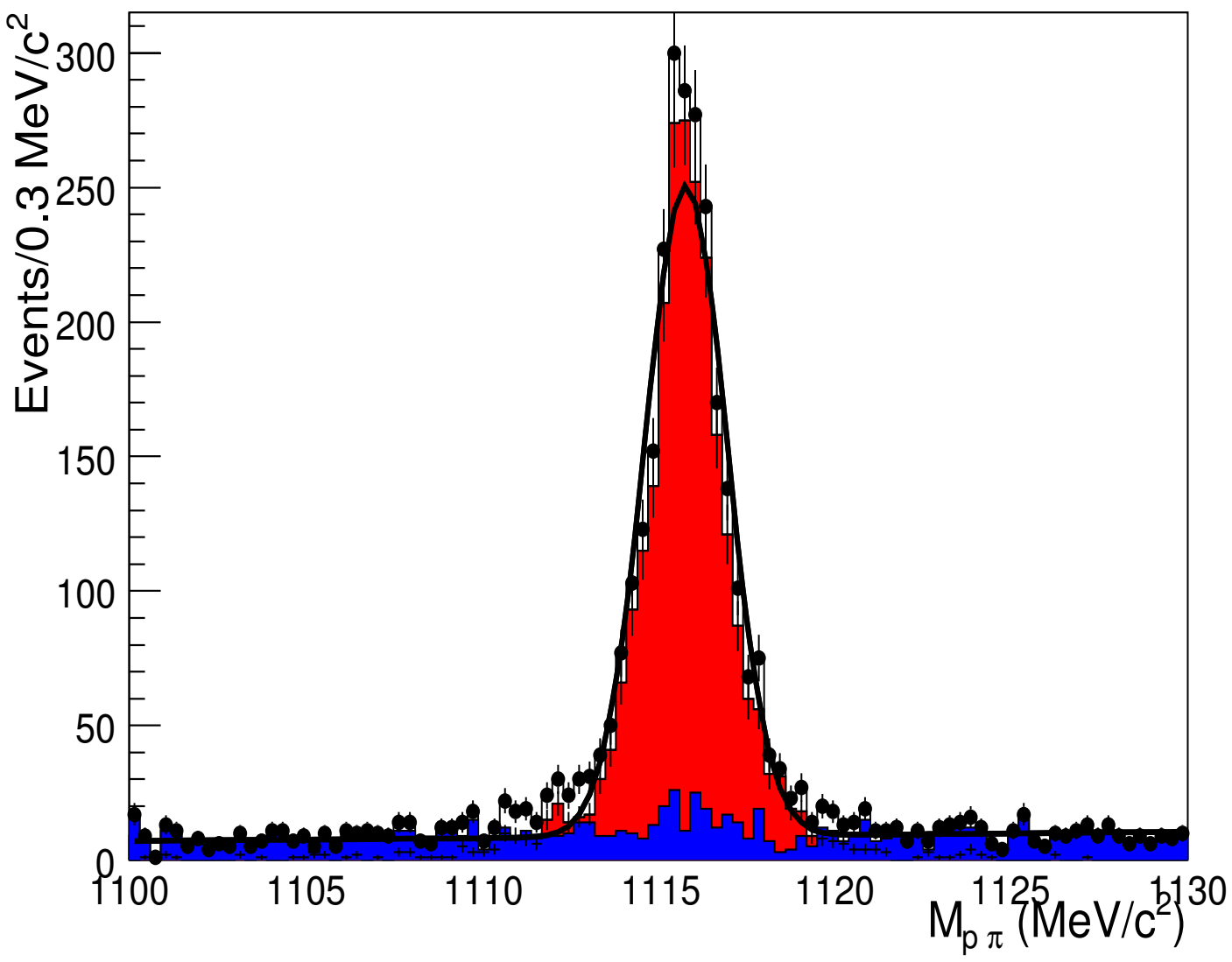

Figure 16: Invariant mass distribution of the events surviving the simple $\Lambda$ selection cuts in a sample of minimum bias events. The histogram in blue corresponds to the remaining combinatorial background

Figure 16 shows the $\Lambda$ mass peak emerging from a sample of $\sim 4 \times 10^6$ minimum bias events, with a purity larger than 95%.

As example of the use, of the $J/\psi \to \mu^+\mu^-$ and $\Lambda \to p\pi^-$ samples we can now extract from the data the muon ID DLL defined in Ref. [25]. Figure 17 shows the difference in muon ID DLL corresponding to muons and non-muons for an inclusive $b\bar{b}$ sample, together with the results obtained from the calibration samples described above. As can be seen, even if the agreement is not perfect, it indicates that this strategy is feasible.

# 9 Sensitivity to the $B_s^0 \to \mu^+\mu^-$ branching ratio

In this section we will apply the strategy discussed in the previous sections to the DC06 simulated data samples and update the LHCb potential from Ref. [12]. From the discussion in previous sections it should be clear that the systematic uncertainties only appear due to the approximations used to extrapolate the information obtained from the control channels into the phase space of the signal. However, these approximations are shown to be valid at the few percent level.The ratio of reconstruction efficiencies in section 6.1 was found to be correct within $\sim 6\%$, however this can be reduced if we use the MC infor-mation to correct for well understood effects. The biggest uncertainty found was $\sim 6\%$ in Section 7.2 due to the statistics of the background subtraction with $0.1\,\mathrm{fb}^{-1}$, but this uncertainty decreases with statistics and is not due to any approximation that introduces a systematic effect. Therefore, the dominant source of systematic uncertainties will be the $\sim 13\%$ uncertainty in the ratio of fragmentation functions when we normalize to the decays $B^+ \to J/\psi K^+$ or to $B^0_d \to K^+\pi^-$ control channels. These levels of systematic uncertainties do not have a sizable effect on the potential of LHCb to exclude a signal. However the degradation in the invariant mass resolution ($\sim 20\%$) due to a more realistic description of the beam pipe supports in the DC06 simulation, and more relevantly the inclusion of FSR, implies that we expect a slight degradation of the $B^0_s \to \mu^+\mu^-$ sensitivity as compared to that presented in Ref. [12].

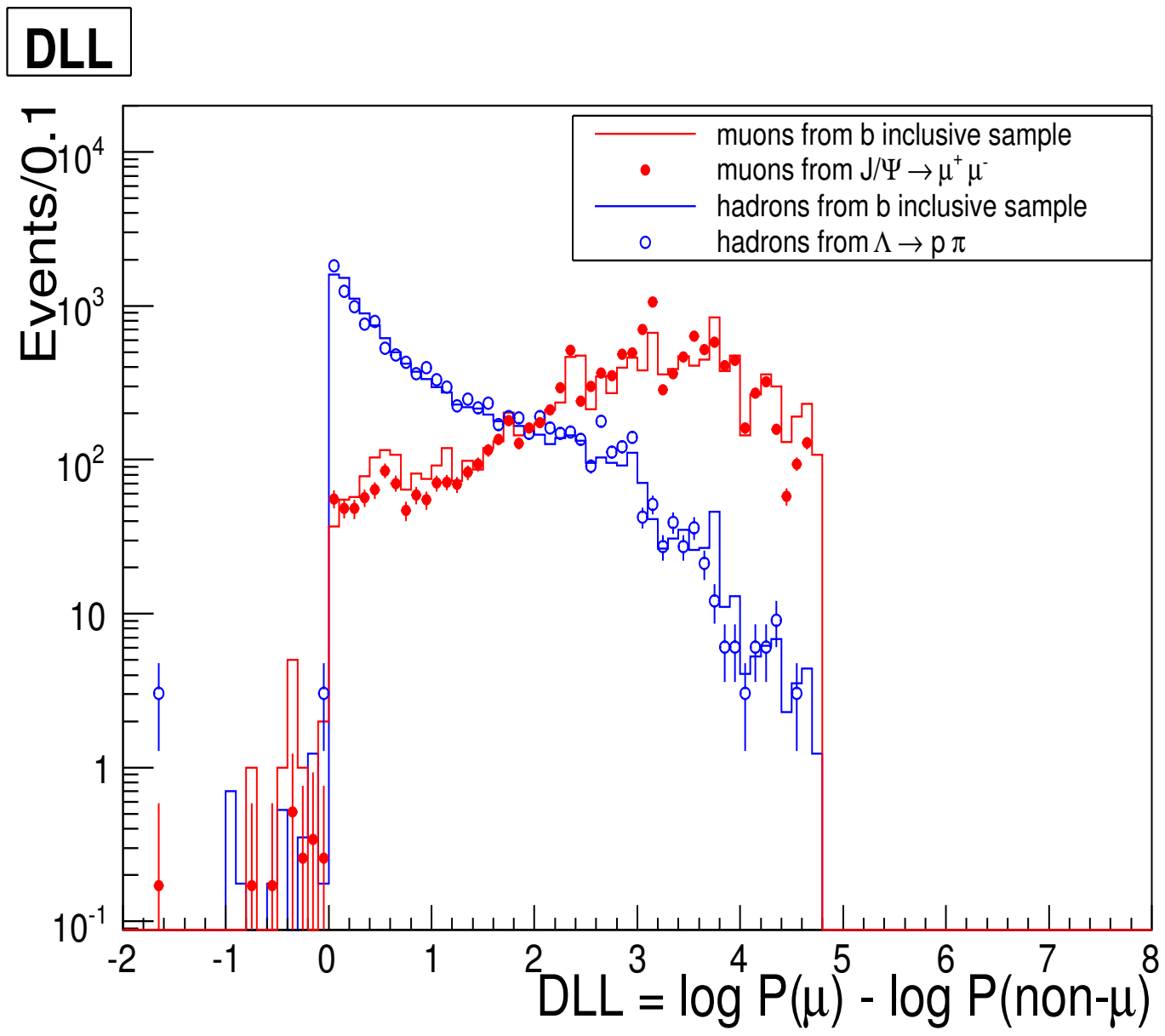

Figure 17: Muon ID DLL computed from an inclusive $b\bar{b}$ sample. The blue histogram corresponds to the expected DLL for non-muons while the red histogram is the expected DLL for muons. The blue (red) data points correspond to the DLL obtained from the calibration samples. $\Lambda \to p\pi^-$ ($J/\psi \to \mu^+\mu^-$).

## 9.1 Nominal sensitivity

After applying the offline selection described in Section 5 and the trigger described in Section 4, we expect 42 SM signal and 130k $b\bar{b}$ background events (from which 66k $b\bar{b} \to$ dimuon events) in the tight mass window for $2\,\mathrm{fb}^{-1}$ (nominal year). However, most of the background events fall in the region defined by GL $< 0.5$ (see Figure 18), leaving only $180^{+140}_{-80}$ $b\bar{b} \to$ dimuon background events compared with 21 SM signal events per nominal year in the sensitive region (GL $> 0.5$ and $\Delta M = \pm 60\,\mathrm{MeV}/c^2$). Notice that even if this is the most sensitive region all events passing the selection of section 5 in the tight mass window are used to compute the branching ratio. The signal and background annual yields in the most sensitive bins are shown in Table 6. Background yields are computed from the $b\bar{b} \to$ dimuon sample, where two events fall in the bin $0.5 < \mathrm{GL} < 0.65$ and two others in the bin $0.65 < \mathrm{GL} < 1.0$, and then extrapolated to an integrated luminosity of $2\,\mathrm{fb}^{-1}$. The IML and GL probalities were assumed uncorrelated for the background, which allowed increasing the effective luminosity of the simulated sample to $\sim 50\,\mathrm{pb}^{-1}$. From the expected number of signal and background events in each bin, the potential of the LHCb experiment on the exclusion or measurement of $\mathcal{B}(B_s^0 \to \mu^+\mu^-)$ can be extracted as in Ref. [12].

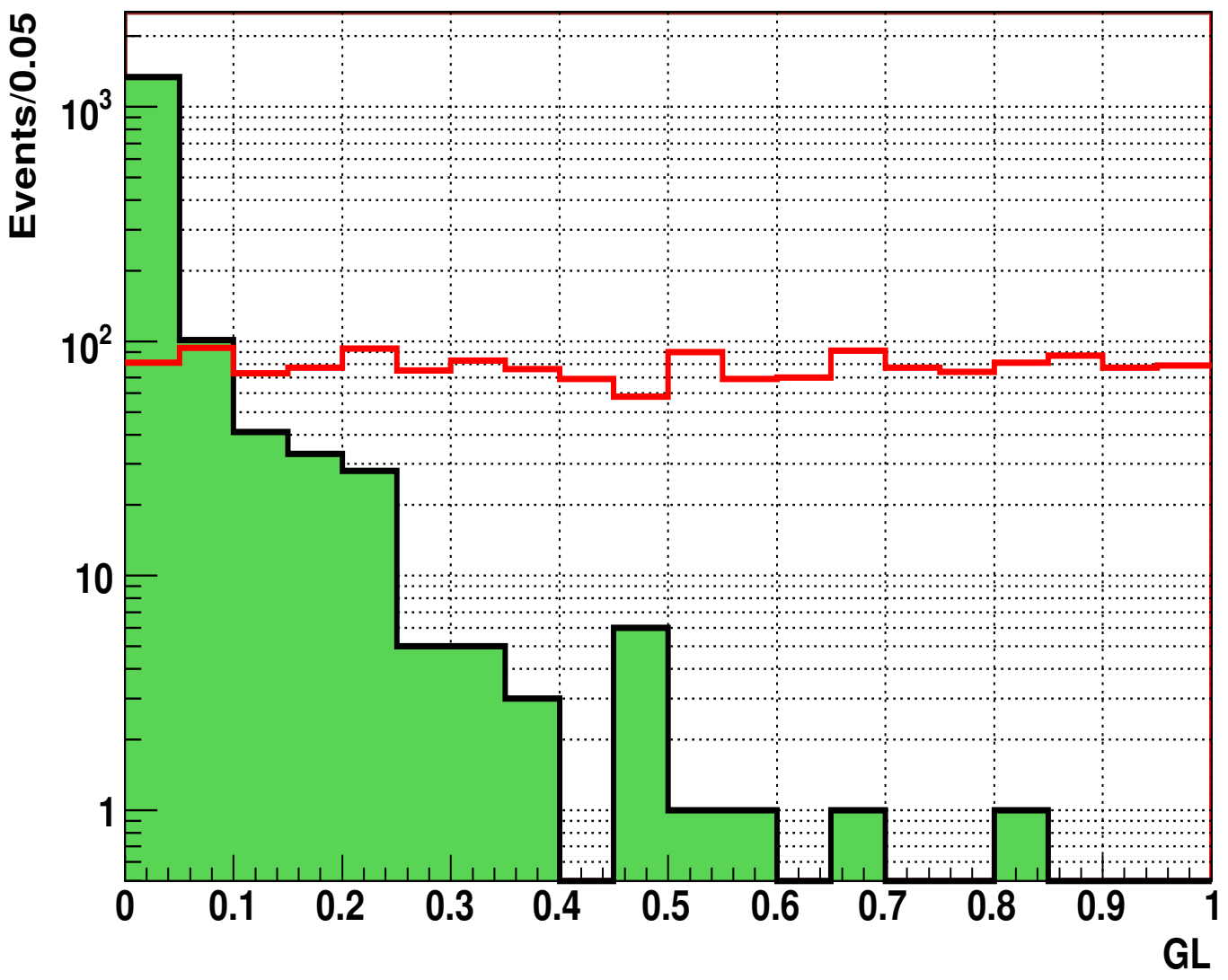

Figure 18: Distribution of the Geometrical Likelihood (GL) for signal (red open histogram) and the inclusive dimuon background (green filled histogram).

Figure 19 (top) shows the BR excluded at 90% CL as a function of the integrated luminosity up to $2\,\mathrm{fb}^{-1}$. The expected limit when Tevatron collects $8\,\mathrm{fb}^{-1}$ per experiment, $2\times10^{-8}$, is overtaken with less than $0.1\,\mathrm{fb}^{-1}$. With $\sim$$1\,\mathrm{fb}^{-1}$ limits up to the SM prediction can be set if no signal is present. In the case of presence of signal, the luminosity needed for the $3\,\sigma$ observation of a given BR is shown on Figure 19 (bottom). About $3\,\mathrm{fb}^{-1}$ are enough for a $3\,\sigma$ observation if the BR is the SM prediction. If the BR is $\sim 2 \times 10^{-8}$ as predicted in some MSSM scenarios (see Section 1), with very little luminosity ($< 0.5\,\mathrm{fb}^{-1}$) LHCb has the potential to claim a $5\sigma$ discovery.

Table 6: Expected numbers of SM signal and background events for $2\,\text{fb}^{-1}$ of luminosity in some bins of dimuon invariant mass and GL.

| | $0.5 < \text{GL} < 0.65$ | $0.65 < \text{GL} < 1$ |
|---|---|---|
| $m_{\mu\mu} \in [5406.6, 5429.6]$ MeV/$c^2$ | $S = 0.26\,,\ B = 16^{+21}_{-10}$ | $S = 0.61\,,\ B = 16^{+21}_{-10}$ |
| $m_{\mu\mu} \in [5384.1, 5406.6]$ MeV/$c^2$ | $S = 1.1\ \ ,\ B = 16^{+21}_{-10}$ | $S = 2.7\ \ ,\ B = 16^{+21}_{-10}$ |
| $m_{\mu\mu} \in [5353.4, 5384.1]$ MeV/$c^2$ | $S = 3.2\ \ ,\ B = 22^{+29}_{-14}$ | $S = 7.6\ \ ,\ B = 22^{+29}_{-14}$ |
| $m_{\mu\mu} \in [5331.5, 5353.4]$ MeV/$c^2$ | $S = 1.2\ \ ,\ B = 16^{+21}_{-10}$ | $S = 3.0\ \ ,\ B = 16^{+21}_{-10}$ |
| $m_{\mu\mu} \in [5309.6, 5331.5]$ MeV/$c^2$ | $S = 0.39\,,\ B = 16^{+21}_{-10}$ | $S = 0.91\,,\ B = 16^{+21}_{-10}$ |

## 9.2 Cross-check with a more robust analysis

The design of an alternative, robust analysis for the search of the decay $B^0_s \to \mu^+\mu^-$ has a two-fold motivation. First, the detector will not be completely understood during the initial phase of data taking. Uncertainties are difficult parameters to establish correctly. Thus, it is worth to search for the minimal set of simple variables to achieve a competitive sensitivity. In this section we will show the results of an analysis based on variables independent of the knowledge of the parameter uncertainties, but using the same statistical tools as in section 9.1.

Secondly, avoiding sophisticated statistical tools and using a simple procedure, such as a well established cut-and-count analysis may appear more convincing. In this section we will also show the results of such a simple cut-and-count analysis. In any case, alternative analyses obviously constitute an important cross-check of the results.

The robust analysis described below is based on variables that do not involve error estimates. Thus, the standard selection, described in Section 5 is not suitable for the robust analysis. A robust selection, based on similar parameters while avoiding the use of the parameter uncertainties, was designed. In this selection, the $\chi^2$ cut is replaced with a cut on the muons' distance of closest approach (DOCA), the $B^0_s$ IPS cut is replaced with a cut on the IP. A cut on the $B^0_s$ distance of flight (DOF) substitutes for the DOFS cut and a cut on the smallest muon IP with respect to any primary vertex substitutes for the cut on the muon IPS.

The values of the robust cuts are set such that the signal selection efficiency remains close to that of the standard selection (see Table 3 ). The cut values and their efficiencies on signal can be found in Table 7. The robust selection is found to be less effective in rejecting the dimuon background than the standard selection, the overall number of selected background events being larger by 31%.

Applying the same strategy as described in Ref. [12], one defines a likelihood function based on robust variables (lifetime and IP of the $B^0_s$, isolation and DOCA of the muons, and smallest muon IP). This GL is evaluated on independent MC samples. The samples

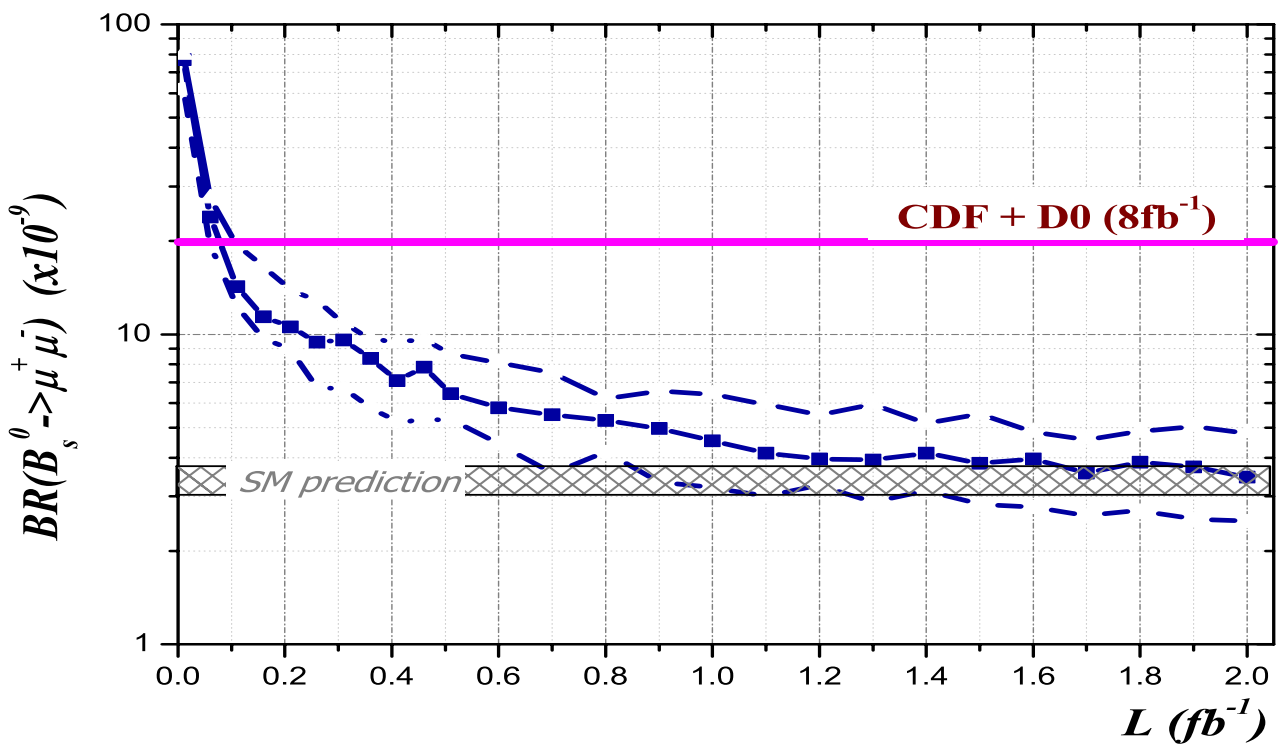


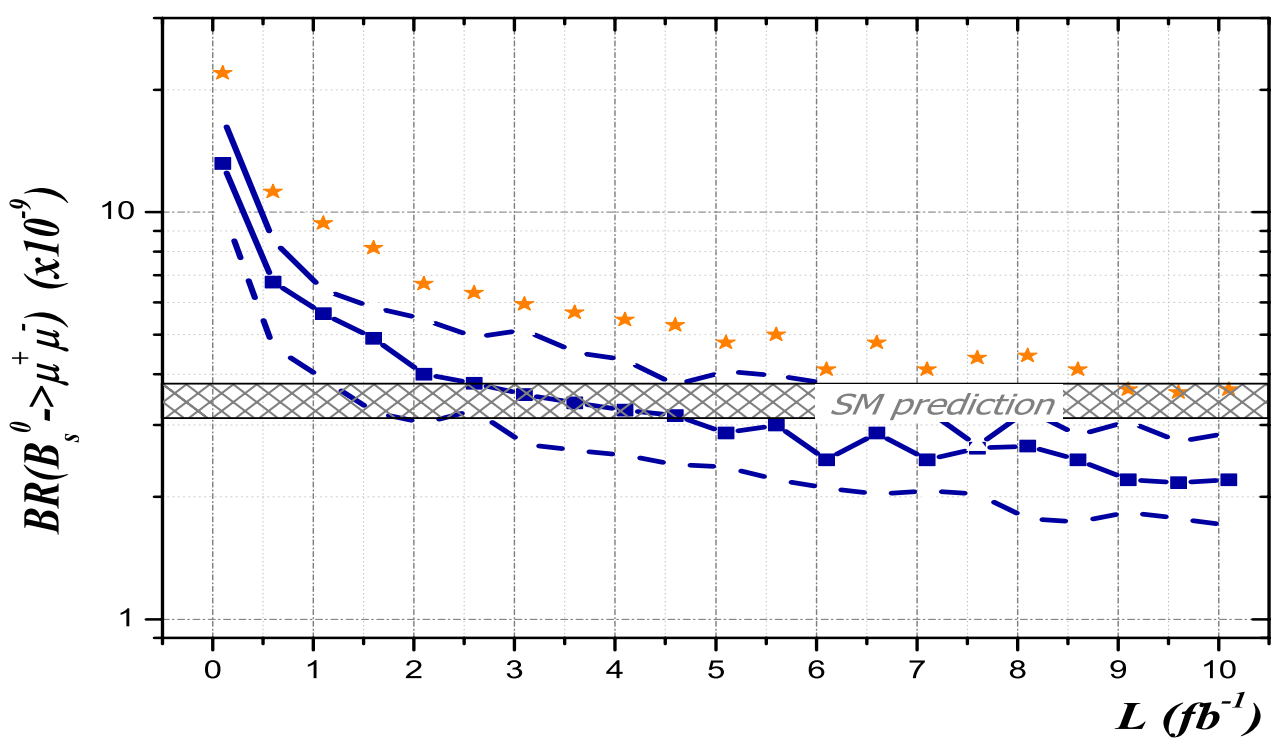


Figure 19: $B_s^0 \to \mu^+\mu^-$ BR excluded (if no signal is present) at 90% CL (top) and observed at $3\,\sigma$ (bottom) as a function of the integrated luminosity. Dashed lines define the 90% probability region due to the limited MC statistics used to evaluate the expected background. Orange stars in the bottom plot indicate the luminosity needed for a $5\,\sigma$ discovery.

used to define the likelihood consist of signal events and dimuon background events produced at an instantaneous luminosity of $2 \times 10^{32}\,\mathrm{cm}^{-2}\mathrm{s}^{-1}$ within the DC06 configuration and passing the selection cuts. No trigger is applied. The Geometry Likelihood is then applied to an independent set of signal and background samples generated at an instantaneous luminosity of $5 \times 10^{32}\,\mathrm{cm}^{-2}\mathrm{s}^{-1}$ within the DC06 configuration. The sensitivity does not seem to depend on the instantaneous luminosity within the uncertainties. The effective sizes of the MC samples are $1223\,\mathrm{fb}^{-1}$ and $0.04\,\mathrm{fb}^{-1}$ for the signal and the background respectively. The background efficiency is estimated under the assumption that the reconstructed $B_s^0$ mass and the GL are uncorrelated. The yields for $2\,\mathrm{fb}^{-1}$ of collected data in the sensitive region (GL $> 0.5$ and $\pm 60\,\mathrm{MeV}/c^2$ mass window) are detailed in Table 8. From the available MC background sample, 3 events fall in the $0.5 < \mathrm{GL} < 0.65$

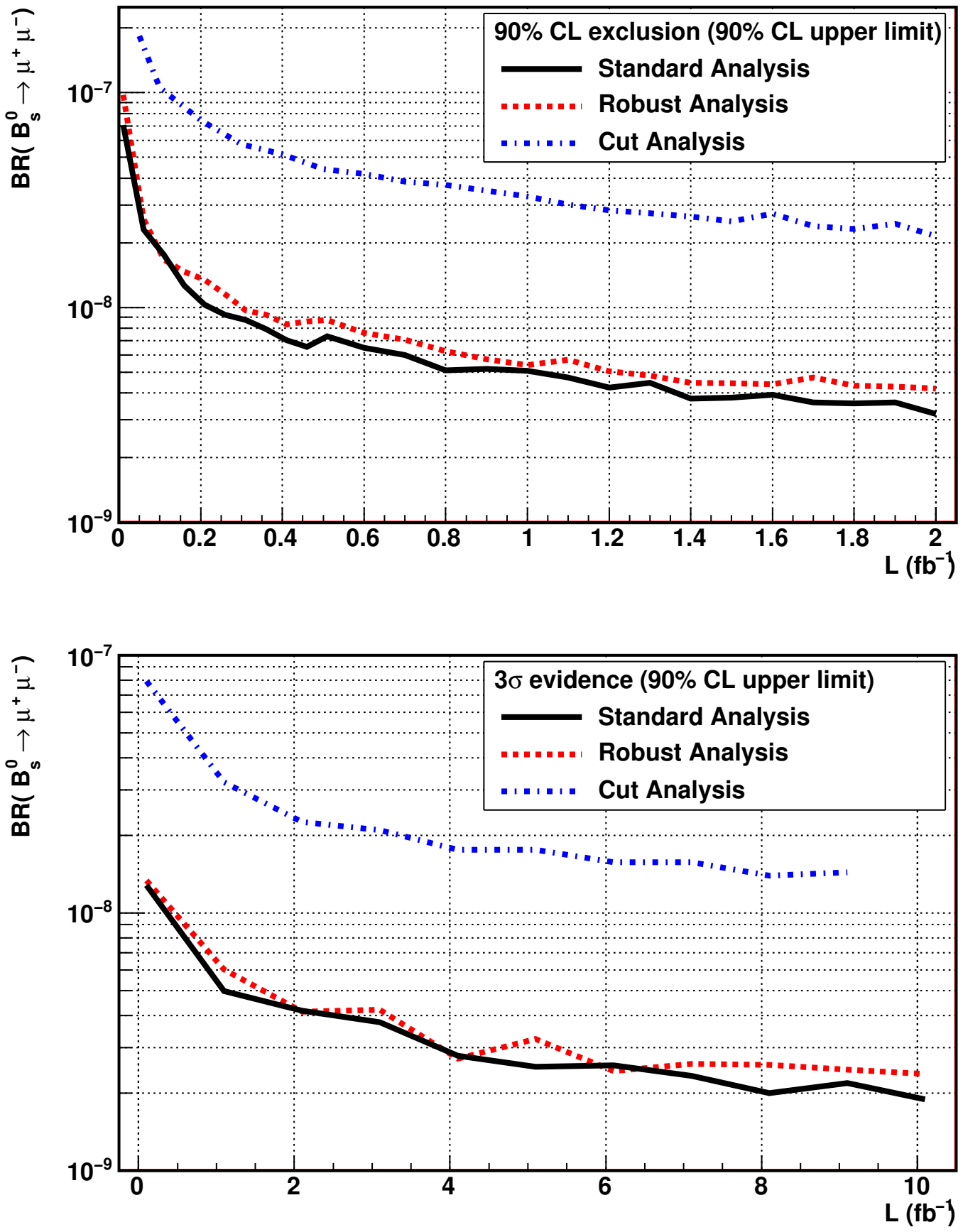


Figure 20: Expected 90% CL upper limit of $\mathcal{B}(B_s^0 \to \mu^+\mu^-)$ in absence of signal (top) and $\mathcal{B}(B_s^0 \to \mu^+\mu^-)$ at which a $3\,\sigma$ observation is expected (bottom) as a function of the integrated luminosity. In both cases, the background estimate has conservatively been set to its 90% CL upper limit. The black curve is the result of the analysis presented in Section 9.1, the red dashed curve is the result from the robust analysis, and the blue dashed-dotted curve is the result from the "cut analysis" (see text).

bin and only one in the GL $>$ 0.65 bin. These results are close to those obtained in Section 9.1.

Similarly to Section 9.1, one can derive the exclusion potential of the robust analysis if no signal is present. Figure 20 (top) shows the BR value excluded with 90% CL as a function of the integrated luminosity. The robust analysis is found to be slightly less sensitive than the analysis presented in Section 9.1. The BR value for which a $3\,\sigma$ observation is possible is shown as a function of the integrated luminosity in Fig. 20 (bottom).

The robust and standard analyses lead to compatible observation and exclusion sensitivities. Thus, the robust analysis constitutes a good cross-check for the standard one

Table 7: Efficiency of each individual offline cut (after all previous ones have been applied) on reconstructed $B_s^0 \to \mu^+\mu^-$ signal events in the case of the robust selection. The last two lines refer to the cuts applied in the "cut analysis" (see text).

| Cut | $\epsilon(B_s^0 \to \mu^+\mu^-)$ |
|---|---|
| $B_{(s)}^0$ mass ($\pm 600\,\mathrm{MeV}/c^2$) | 97.2% |
| DOCA $< 0.1\,\mathrm{mm}$ | 97.5% |
| $\mathrm{IP}(B_{(s)}^0) < 0.1\,\mathrm{mm}$ | 98.8% |
| $P_\mathrm{T}(B_{(s)}^0) > 700\,\mathrm{MeV}/c$ | 98.0% |
| $\mathrm{DOF}(B_{(s)}^0) > 1.7\,\mathrm{mm}$ | 76.2% |
| min. $\mathrm{IP}(\mu/h) > 0.06\,\mathrm{mm}$ | 91.2% |
| All cuts above | 63.7% |
| $B_{(s)}^0$ mass ($\pm 60\,\mathrm{MeV}/c^2$) | 90.6% |
| Further cuts for the cut analysis | |
| $B_{(s)}^0$ mass ($\pm 35\,\mathrm{MeV}/c^2$) | 79.2% |
| lifetime $B_{(s)}^0 > 4$ ps | 9.5% |
| All cuts above | 4.8% |

Table 8: Expected numbers of SM signal and background events for $2\,\mathrm{fb}^{-1}$ of luminosity in some bins of dimuon invariant mass and GL for the robust analysis.

| | $0.5 < \mathrm{GL} < 0.65$ | $0.65 < \mathrm{GL} < 1$ |
|---|---|---|
| $m_{\mu\mu} \in [5406.6, 5429.6]$ MeV$/c^2$ | $S = 0.23\,,\ B = 25^{+25}_{-14}$ | $S = 0.59\,,\ B = 8^{+24}_{-7}$ |
| $m_{\mu\mu} \in [5384.1, 5406.6]$ MeV$/c^2$ | $S = 1.14\,,\ B = 25^{+25}_{-14}$ | $S = 2.61\,,\ B = 8^{+24}_{-7}$ |
| $m_{\mu\mu} \in [5353.4, 5384.1]$ MeV$/c^2$ | $S = 3.45\,,\ B = 35^{+34}_{-19}$ | $S = 7.70\,,\ B = 12^{+23}_{-10}$ |
| $m_{\mu\mu} \in [5331.5, 5353.4]$ MeV$/c^2$ | $S = 1.35\,,\ B = 26^{+25}_{-14}$ | $S = 2.87\,,\ B = 9^{+24}_{-7}$ |
| $m_{\mu\mu} \in [5309.6, 5331.5]$ MeV$/c^2$ | $S = 0.41\,,\ B = 26^{+25}_{-14}$ | $S = 0.81\,,\ B = 9^{+24}_{-7}$ |

and should improve the confidence in the first result. This also implies that the standard analysis does not rely too heavily on error estimates.

Alternatively, a very simple cut analysis was designed applying, in addition to the robust preselection, a tighter cut on the $B_s^0$ mass and a cut on the $B_s^0$ lifetime. The mass cut window value is $\pm\sqrt{2}\ \sigma_\mathrm{mass}$ whereas the value of the $B_s^0$ lifetime cut optimizes the $\frac{N_S}{\sqrt{N_S+N_B}}$ ratio. Again, the cut optimisation was performed on $2 \times 10^{32}\,\mathrm{cm}^{-2}\mathrm{s}^{-1}$ DC06 configuration data for signal and background, resulting in the cut values and efficiency indicated in Table 7. The application of those cuts on the $5 \times 10^{32}\,\mathrm{cm}^{-2}\mathrm{s}^{-1}$ DC06 samples

for signal and background gives a yield of 3.3 SM signal events and no background left with 2 $\text{fb}^{-1}$ of data. The corresponding sensitivity for both exclusion and observation is shown in Fig. 20. The sensitivity of the cut analysis is significantly lower than those of the analyses involving the GL. More work is needed to improve the cut analysis such that it can become a useful cross-check.

# 10 Main systematic limitation

As explained in Section 6, the use of the $B^+ \to J/\psi(\mu^+\mu^-)K^+$ or $B^0 \to K^+\pi^-$ decay modes as normalization channels introduces a systematic uncertainty of $\sim 13\%$ due to our limited knowledge of the relative production rates of $B^0_s$ mesons compared to $B^+$ or $B^0$ mesons. With sufficient statistics, all other systematic uncertainties are expected to be eventually much smaller than that, because the whole analysis is based on a strategy where all the other ingredients are extracted from the data themselves. In this section, we first describe the limitations introduced by this "irreducible" normalization systematic uncertainty, before discussing possible prospects for improving the situation.

We note that the Tevatron experiments are currently using the $B^+ \to J/\psi(\mu^+\mu^-)K^+$ decay mode to normalize the result of their $B^0_s \to \mu^+\mu^-$ searches [8], which therefore also suffer from the same $\sim 13\%$ systematic uncertainty. However, this does not have important consequences at present, because their current upper limits on the $B^0_s \to \mu^+\mu^-$ branching fraction are still an order of magnitude larger than the Standard Model (SM) prediction. On the contrary, this may well become a serious limitation for LHCb, namely preventing the discovery of New Physics in case it only enhances the $B^0_s \to \mu^+\mu^-$ branching fraction by up to a factor 3 compared to the SM expectation (although in this case LHCb would still be able, with sufficient statistics, to make a $5\,\sigma$ observation of a non-zero signal as shown in Fig. 19). This is illustrated in Fig. 21, which compares, as a function of the true $B^0_s \to \mu^+\mu^-$ branching fraction and for different integrated luminosities, the expected statistical and total uncertainties with the precision needed to make a measurement inconsistent with the SM prediction at the level of $5\,\sigma$. For instance, with 10 $\text{fb}^{-1}$ of data the statistical uncertainty would allow us to claim a $\geq 5\,\sigma$ difference with respect to the SM if $\mathcal{B}(B^0_s \to \mu^+\mu^-) > 6 \times 10^{-9}$. However including the systematic uncertainty limits this ability to $\mathcal{B}(B^0_s \to \mu^+\mu^-) > 11 \times 10^{-9}$, even if a lot more statistics could be accumulated. If the total systematic uncertainty could be reduced to 10% (5%) the limitation would then be $\mathcal{B}(B^0_s \to \mu^+\mu^-) > 9\ (7) \times 10^{-9}$ at 10 $\text{fb}^{-1}$.

The currently assumed uncertainty of 13% is the quadratic combination of the error on the $B^+ \to J/\psi(\mu^+\mu^-)K^+$ or $B^0 \to K^+\pi^-$ branching fraction and of the error on the ratio of the production rates of the corresponding $B$ mesons:

$$\frac{f_{B^0_s}}{f_{B^+}} = \frac{f_{B^0_s}}{f_{B^0_d}} = 0.265 \pm 0.034\,. \qquad (22)$$

The above average, dominated by LEP results, has been obtained by HFAG [26], under the two assumptions that $f_{B^0_d} = f_{B^+}$ and that the production fractions are the same at

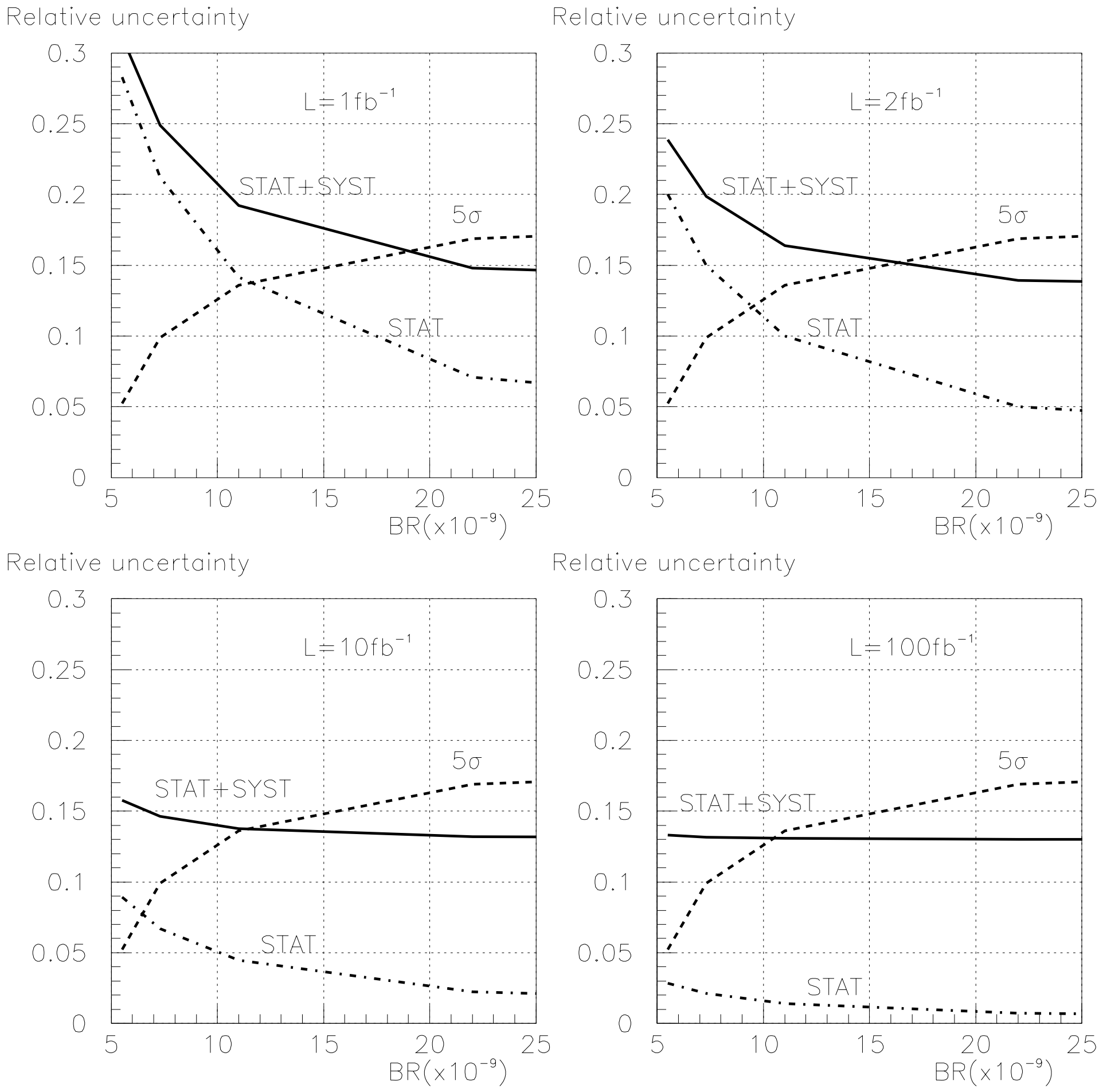


Figure 21: Relative uncertainty expected from LHCb on the measurement of the $B_s^0 \to \mu^+\mu^-$ branching fraction as a function of this branching fraction in units of $10^{-9}$, for four different integrated luminosities: 1 fb$^{-1}$, 2 fb$^{-1}$, 10 fb$^{-1}$ and 100 fb$^{-1}$. The solid curve shows the total (statistical and systematic) expected uncertainty, while the dash-dotted curve shows only the statistical uncertainty. A systematic uncertainty of 13% (due to the normalization to the $B^+ \to J/\psi K^+$ decay mode) is assumed, which becomes completely dominant after 10 fb$^{-1}$. The dashed curve (independent of the luminosity) is the uncertainty that needs to be achieved to be able to claim a $5\,\sigma$ discrepancy with respect to the SM prediction of $(3.35 \pm 0.32) \times 10^{-9}$ [9].

LEP and at the Tevatron, although the quoted uncertainty includes a scale factor taking into account a $1.8\,\sigma$ discrepancy in the LEP and Tevatron measurements of $\bar{\chi}$, the time-integrated mixing probability averaged over all $b$-hadron species. The second assumption

has also been questioned in a recent paper by the CDF collaboration, which presents a new measurement $f_{B_s^0}/f_{B^+} = 0.320 \pm 0.097$ [27] not included in the HFAG average. The prospects for clarifying the validity of the use of the HFAG average at LHCb or for improving significantly the knowledge of $f_{B_s^0}/f_{B^+}$ in the near future, either at the Tevatron or the LHC, are not too good. So it is likely that all future $B_s^0 \to \mu^+\mu^-$ results normalized against a $B^+$ or $B^0$ decay mode will be affected by a total normalization uncertainty of $\sim 13\%$.

An attractive possibility to avoid the uncertainty on $f_{B_s^0}/f_{B^+}$ would be to normalize against a $B_s^0$ branching fraction measured in $\Upsilon(5S)$ decays with a sufficiently good precision. The Belle collaboration has already collected 23.6 fb$^{-1}$ of data at the $\Upsilon(5S)$ resonance, which were recently used to perform the following branching fraction measurement [28]:

$$\begin{aligned}\mathcal{B}(B_s^0 \to D_s^-\pi^+) = \\ \big(3.67^{+0.35}_{-0.33}(\text{stat})^{+0.43}_{-0.42}(\text{syst}) \pm 0.49(f_s)\big) \times 10^{-3}\,.\end{aligned} \tag{23}$$

The total relative uncertainty is 20%, so this measurement is not yet competitive as a candidate to normalize the $B_s^0 \to \mu^+\mu^-$ results. The normalization uncertainty ($\sim 14\%$) is dominated by the knowledge of $f_s$, the fraction of events with $B_s^0$ mesons amongst all events with $B$ mesons of any type ($B^0$, $B^+$ and $B_s^0$), of which the latest world average is [26]

$$f_s = 0.194 \pm 0.029\,. \tag{24}$$

The error on $f_s$ is dominated by systematic uncertainties. For example, the best single measurement of $f_s$, performed by Belle from a measurement of the inclusive production of $D_s$ mesons with 1.9 fb$^{-1}$ of data at the $\Upsilon(5S)$ resonance, is $f_s = 0.179 \pm 0.014 \pm 0.041$ [29], where the second (and dominant) uncertainty is mainly due to assumptions on poorly known branching fractions, such as $\mathcal{B}(B_s^0 \to D_s^- X)$ and $\mathcal{B}(B \to D_s^- X)$. The other measurements of $f_s$, from CLEO and Belle, suffer from similar limitations. It has recently been claimed [30] that Belle could reach a precision of 6–8% on $f_s$, which would lead to a 8–9% normalization uncertainty. We also note that Belle has recently decided to take more data at the $\Upsilon(5S)$ resonance. A naive extrapolation seems to indicate that a 10% measurement of the $B_s^0 \to D_s^-\pi^+$ branching fraction should be possible with a sample of 100 fb$^{-1}$.

In case Belle would publish a $B_s^0$ branching fraction measurement with a total error of 10% or less, we should investigate the possibility to use this as the new normalization point for the $B_s^0 \to \mu^+\mu^-$ search. A Belle measurement of a two-body decay mode, such as $B_s^0 \to K^+K^-$, or a mode with two muons, such as $B_s^0 \to J/\psi\phi$, would be preferable because several other experimental systematics would cancel in the ratios $(B_s^0 \to \mu^+\mu^-)/(B_s^0 \to K^+K^-)$ and $(B_s^0 \to \mu^+\mu^-)/(B_s^0 \to J/\psi(\mu^+\mu^-)\phi)$. However, the branching fractions of the $B_s^0 \to K^+K^-$, and $B_s^0 \to J/\psi\phi \to \mu^+\mu^-K^+K^-$ decay chains are expected to be in the order of $10^{-5}$, which is almost an order of magnitude smaller than that of $B_s^0 \to D_s^-(K^-K^+\pi^-)\pi^+$. So a scenario where the best normalization channel becomes $B_s^0 \to D_s^-\pi^+$ rather than $B_s^0 \to K^+K^-$ or $B_s^0 \to J/\psi\phi$ is more likely.

# 11 Conclusion

In this note we have defined a strategy to calibrate all the steps needed to extract the $B_s^0 \to \mu^+\mu^-$ branching ratio using control channels and not relying on the simulation. The ratio of offline efficiencies between signal ($B_s^0 \to \mu^+\mu^-$) and control channels ($B^+ \to J/\psi(\mu^+\mu^-)K^+$ and/or $B_d^0 \to K^+\pi^-$) can be extracted using the ratio of different control channels (for instance, the ratio of $B^+ \to J/\psi(\mu^+\mu^-)K^+$ with $B_d^0 \to K^+\pi^-$ and/or $B_d^0 \to J/\psi(\mu^+\mu^-)K^{*0}(K^+\pi^-)$). The approximations inherent seem to be valid at the few percent level. The ratio of trigger efficiencies can be extracted using events triggered independently of the signal (TIS), which with enough luminosity will give a few percent precision. The invariant-mass likelihood and the geometrical likelihood can be extracted using $B_{(s)}^0 \to h^+h^-$ events as signal candidates, and the events in the sidebands of the mass distribution as background candidates without relying on the simulation. There are several good control channels (for instance $J/\psi \to \mu^+\mu^-$ and $\Lambda \to p\pi^-$) to be able to calibrate the muon identification efficiency and the muon misidentification probability.

This strategy will allow LHCb to have a measurement of the $B_s^0 \to \mu^+\mu^-$ branching ratio that should not depend on how well our simulation reproduces real data. The main systematic uncertainty arises from the use of the $B^+ \to J/\psi(\mu^+\mu^-)K^+$ or $B^0 \to K^+\pi^-$ decay modes as normalization channels, as it introduces a systematic uncertainty of $\sim 13\%$ due to our limited knowledge of the relative production rates of $B_s^0$ mesons compared to $B^+$ or $B^0$ mesons. This uncertainty may limit the LHCb potential in the long term to find new physics, hence any experimental determination of a $B_s^0$ decay with a 10% precision or better would be very useful.

On the other hand, the LHCb potential to find clear signs of new physics in the $B_s^0 \to \mu^+\mu^-$ decay with the initial LHC data has been confirmed in this note. The use of control channels described in this note give us confidence that systematic uncertainties will not be a limiting factor for the firsts years of LHC running.

# Chapter 6

# Analysis of the decay $B^0 \to K^{*0}\mu^+\mu^-$

T. Blake, J. Dickens, U. Egede, F. Jansen, F. Marinho, M. Patel,
A. Pérez-Calero Yzquierdo, W. Reece, N. Serra, H. Skottowe and G. Ybeles Smit

**Abstract**

The decay $B^0 \to K^{*0}\mu^+\mu^-$ is a promising channel in which to search for new physics at LHCb. The angular asymmetry $A_{FB}$ is one of a number of angular observables which are well calculable theoretically and offer discrimination between different new physics models.

The key challenges for this analysis are expected to be controlling the backgrounds and understanding the biases induced on angular observables, primarily by the geometric acceptance of the detector. Efficient on- and off-line selection criteria have been determined that allow signal events to be separated from the background and will give signal yields comparable to those from the B-factories with a few weeks of data-taking at nominal luminosity. These criteria are also designed to minimize the biases induced on the observables.

A number of methods to correct residual biases have been proposed and are under study. A range of methods to then extract the observables have been investigated. These will allow LHCb to make competitive measurements with the first data. Using a simple counting analysis, the zero-crossing point of $A_{FB}$ can be determined with a statistical uncertainty of $\pm 0.5\,\text{GeV}^2$ with a $2\,\text{fb}^{-1}$ integrated luminosity. As the dataset available increases, and understanding of the detector develops, this precision can be markedly improved by using more sophisticated analysis methods. Eventually, a full angular analysis will be used to yield the best precision and extract the complete information that is available from $B^0 \to K^{*0}\mu^+\mu^-$ decays.

# Contents

# 1 Introduction

The decay $B_d \to K^{*0}\mu^+\mu^-$ is a flavour-changing-neutral current which proceeds via a $b{\to}s$ transition through a loop diagram (Fig. 1). New physics processes can therefore enter at the same level as the Standard Model (SM) processes, making the decay a sensitive probe of new physics contributions. The partial rate as a function of the di-muon invariant mass squared ($q^2$) and the di-muon forward-backward asymmetry ($A_{\rm FB}$) can both be affected in many new physics scenarios [1]. The branching ratio has been measured to be $(9.8 \pm 2.1) \times 10^{-7}$ [2], [3], [4]. A large yield of $B_d \to K^{*0}\mu^+\mu^-$ events is therefore expected at LHCb and this decay is a promising channel in which to search for new physics. The $B_d \to K^{*0}e^+e^-$ is more challenging experimentally and has a slightly different physics interest. This decay is studied elsewhere [5].

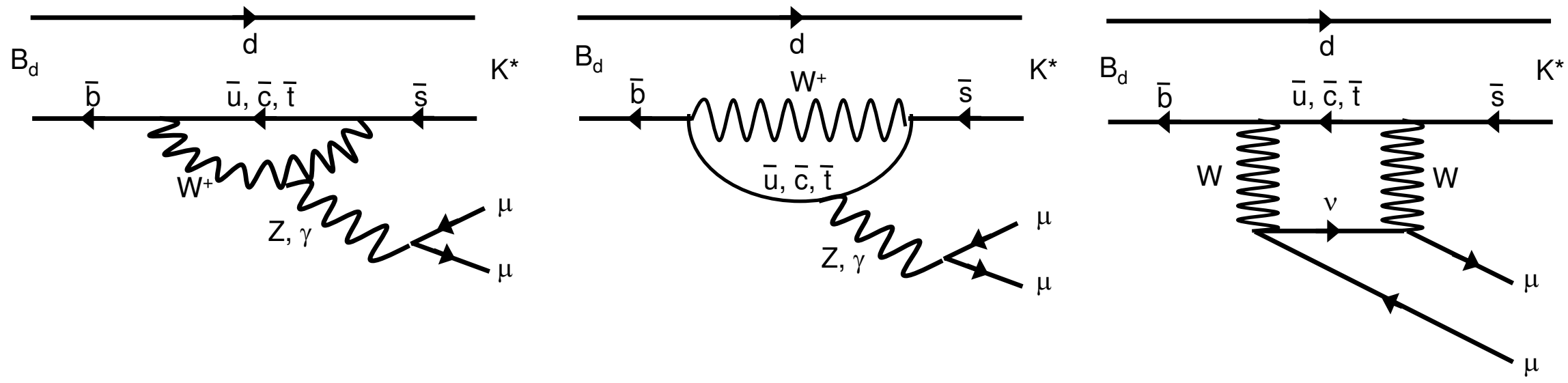


Figure 1: Dominant Standard Model Feynman diagrams for the $B_d \to K^{*0}\mu^+\mu^-$ decay.

For a $B_d$ ($\overline{B}_d$) decay, the forward-backward asymmetry $A_{\rm FB}$ is constructed from the number of forward- and backward-emitted positive (negative) muons in the di-muon rest frame according to Eq. 1.

$$A_{\rm FB}(q^2) = \frac{\int_0^1 \frac{\partial^2\Gamma}{\partial q^2 \partial \cos\theta_{\rm L}}\, {\rm d}\cos\theta_{\rm L} - \int_{-1}^0 \frac{\partial^2\Gamma}{\partial q^2 \partial \cos\theta_{\rm L}}\, {\rm d}\cos\theta_{\rm L}}{\int_0^1 \frac{\partial^2\Gamma}{\partial q^2 \partial \cos\theta_{\rm L}}\, {\rm d}\cos\theta_{\rm L} + \int_{-1}^0 \frac{\partial^2\Gamma}{\partial q^2 \partial \cos\theta_{\rm L}}\, {\rm d}\cos\theta_{\rm L}}\,. \tag{1}$$

Theory predictions for this asymmetry are well calculable in the range 1 GeV$^2$ $<q^2<$6 GeV$^2$ [6]. This paper focuses on this region but LHCb will be able to measure $A_{\rm FB}$ across the entire $q^2$ range. Although both the $q^2$ and $A_{\rm FB}$ spectra are sensitive to new physics, the zero-crossing point of the asymmetry has received particular theoretical attention, as the form factor ratio used in the calculation of this quantity is almost free of hadronic uncertainties. The position of this zero-crossing point is governed by the interference between the underlying vector and axial-vector currents. In the SM, the zero-crossing point is predicted to be at $q^2 = 4.36^{+0.33}_{-0.31}$ GeV [8].

At low $q^2$ the decay is dominated by the $C_7$ Wilson coefficient while, at high $q^2$, the behaviour is dictated by the $C_9$ and $C_{10}$ Wilson coefficients.

Theoretical predictions for the branching ratio from the SM and from various new physics scenarios are shown as a function of $q^2$ $(= s)$ in Fig. 2. The $A_{\rm FB}$ distributions in a number of new physics models are similarly shown in Fig. 3.

The $B_d \to K^{*0}\mu^+\mu^-$ decay can be uniquely described in terms of the kinematic variables $q^2$, $\theta_L$, $\theta_K$ and $\phi$ (illustrated in Fig. 4). For example, $\theta_L$ is the angle between the $B_d$ ($\overline{B}_d$) flight direction and the positive (negative) muon direction in the di-muon rest frame. Precise definitions of the angles, which can all be accurately measured by LHCb, are given in Ref. [10]. The partial rates $\frac{\partial^2\Gamma}{\partial\theta_L\,\partial q^2}$, $\frac{\partial^2\Gamma}{\partial\theta_K\,\partial q^2}$ and $\frac{\partial^2\Gamma}{\partial\phi\,\partial q^2}$ can be expressed in terms of $A_{\rm FB}$, the fraction of longitudinal $K^{*0}$ polarisation, $F_L$, the transverse asymmetry, $A_T^2$, and the observable, $A_{\rm Im}$, (Eqs 2, 3 and 4) [10]:

$$\frac{\partial^2\Gamma}{\partial\theta_L\,\partial q^2} = \left(\frac{3}{4}F_L\sin^2\theta_L + \frac{3}{8}(1-F_L)(1+\cos^2\theta_L) + A_{\rm FB}\cos\theta_L\right)\sin\theta_L. \quad (2)$$

$$\frac{\partial^2\Gamma}{\partial\theta_K\,\partial q^2} = \frac{3}{4}\sin\theta_K\left(2F_L\cos^2\theta_K + (1-F_L)\sin^2\theta_K\right). \quad (3)$$

$$\frac{\partial^2\Gamma}{\partial\phi\,\partial q^2} = \left(1 + \frac{1}{2}(1-F_L)A_T^2\cos 2\phi + A_{\rm Im}\sin 2\phi\right). \quad (4)$$

Rather than counting forward- and backward-events to determine $A_{\rm FB}$ alone, by fitting Eq. 2 (or Eqs 2 and 3) the shape information can be used, allowing $A_{\rm FB}$ and $F_L$ to be extracted.

The present measurements of the $A_{\rm FB}$ spectrum from the Babar and Belle experiments are shown in Figs 5(a) and 5(b) [2], [3]. The sign convention for $A_{\rm FB}$ used by these experiments is the opposite of that used by LHCb. Both experiments have observed $\mathcal{O}(100)$ $B_d \to K^{*0}\ell^+\ell^-$ events ($\ell = e, \mu$). In addition, the CDF experiment has observed $\sim 100$ $B_d \to K^{*0}\mu^+\mu^-$ events ($4.4\,{\rm fb}^{-1}$) [4].

Given projected total datasets for these experiments of $\sim 400$, $\sim 750$ and $\sim 8\,{\rm fb}^{-1}$ respectively, a total of $\sim 600$ events should be observed at all facilities. LHCb will collect a similar number of events with a few weeks of data-taking at luminosities $\mathcal{O}(10^{32}\ {\rm cm}^{-2}{\rm s}^{-1})$. A larger dataset will allow detailed measurements of the above parameters and, eventually, the measurement of the full angular distribution [11]. This will allow a considerably improved measurement of $A_{\rm FB}$. Given the background expectation, $A_{\rm FB}$ will be measured with a similar precision to the current B-factory data with a $0.3\,{\rm fb}^{-1}$ integrated luminosity.

The partial rate $\frac{\partial^3\Gamma}{\partial\theta_L\,\partial\theta_K\,\partial q^2}$ given in Eq. 5 shows that, while $A_{\rm FB}$ is weighted by $\sin^2\theta_K$, as $F_L$ is large[1], the majority of the rate is weighted by $\cos^2\theta_K$. Measuring the distribution of both angles simultaneously therefore gives enhanced sensitivity to $A_{\rm FB}$.

$$\frac{\partial^3\Gamma}{\partial\theta_L\,\partial\theta_K\,\partial q^2} = \left(\frac{9}{8}F_L\sin^2\theta_L\cos^2\theta_K + \frac{9}{32}(1-F_L)(1+\cos^2\theta_L)\sin^2\theta_K + \frac{3}{4}A_{\rm FB}\cos\theta_L\sin^2\theta_K\right). \quad (5)$$

[1]Ref. [3] cites $F_L = 0.67 \pm 0.23 \pm 0.05$ in the range 1 GeV$^2$ $<q^2<$6 GeV$^2$.

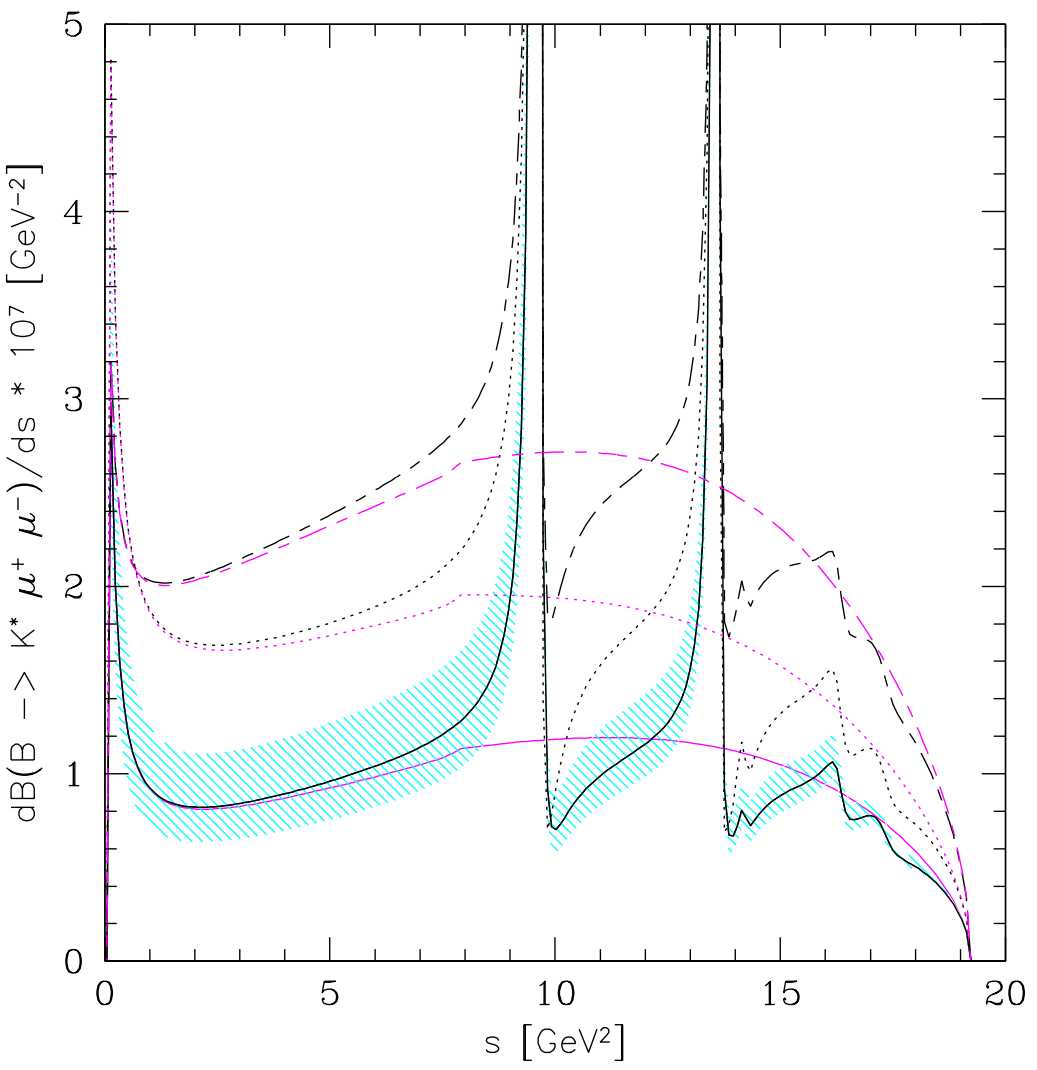


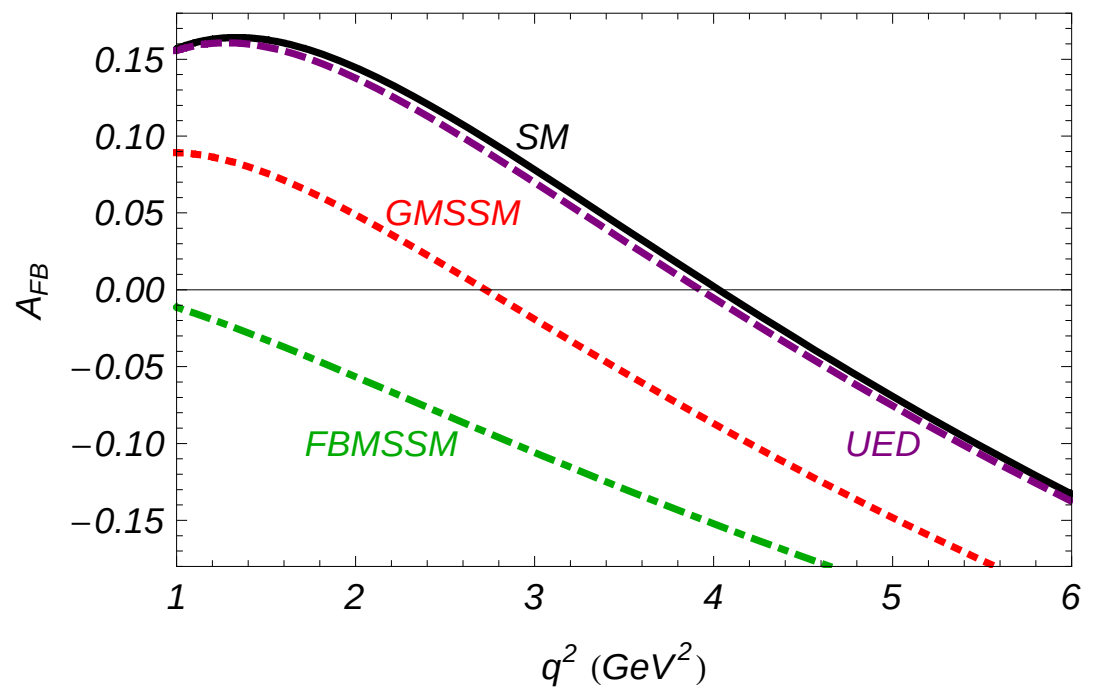


Figure 2: Theoretical $\mu^+\mu^-$ mass distributions in a number of models, taken from Ref. [7] (solid line – SM; dotted line and long-short dashed lines from new physics models described in Ref. [1]). The lower lines show the respective purely short-distance components. The shaded area around the SM line depicts the form factor-related uncertainties.

Figure 3: Theoretical $A_{\mathrm{FB}}$ distributions in a number of models. The solid line gives the SM prediction. The dashed lines show predictions from a universal extra dimensions (UED) model, a non-minimal flavour violating supersymmetric model (GMSSM) and a flavour blind supersymmetric model (FBMSSM). Details of these models can be found in Ref. [9].

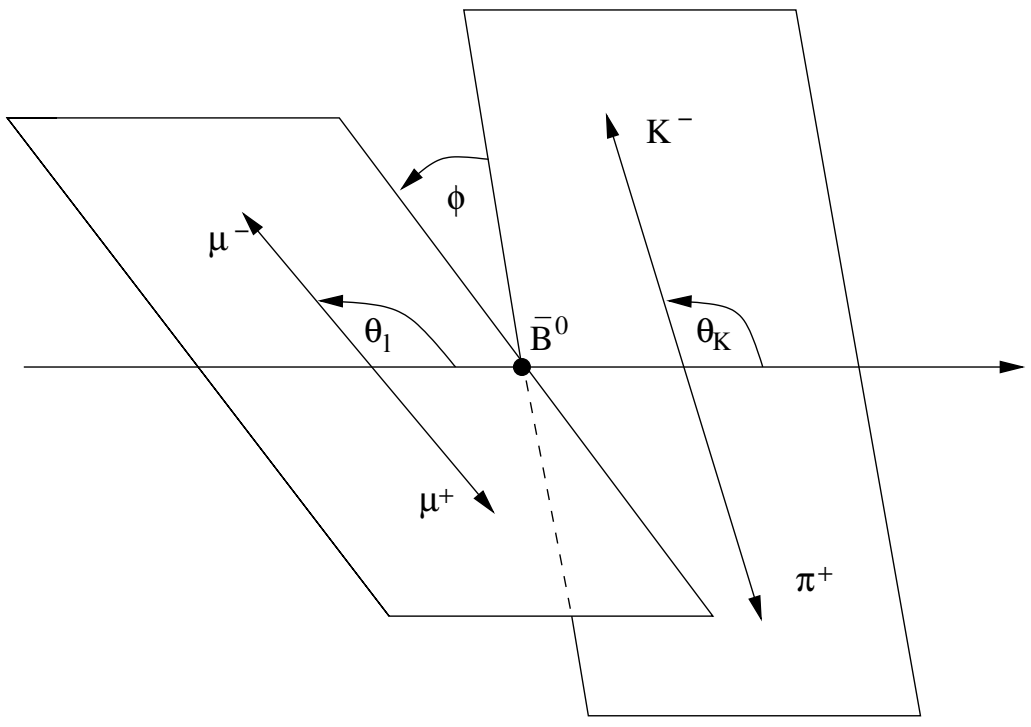


Figure 4: Definition of the angles $\theta_{\mathrm{L}}$, $\theta_{\mathrm{K}}$ and $\phi$ for the $\overline{\mathrm{B}}_{\mathrm{d}} \to \overline{\mathrm{K}}^{*0}\mu^+\mu^-$ decay. The horizontal line shows the $\overline{\mathrm{B}}_{\mathrm{d}}$ flight direction, the K and $\pi$ are shown in the $\overline{\mathrm{K}}^{*0}$ rest-frame and the $\mu^+\mu^-$ in the di-muon rest frame. The two planes that represent these frames are separated for illustration purposes.

In addition, a full angular analysis will allow new observables to be accessed, for example, additional transverse asymmetries which have different new physics sensitivities [12]. The $B_d \to K^{*0}\mu^+\mu^-$ decay will therefore be of interest throughout the lifetime of LHCb.

As discussed in the next section, a number of analysis issues will need to be treated in order to extract $A_{FB}$ from $B_d \to K^{*0}\mu^+\mu^-$ signal events. The remainder of this document describes the present understanding of how to acquire the data, perform the relevant corrections for the above effects and extract the observables. In Section 3, the way in which the events will be triggered is discussed. In Section 4, how to achieve the subsequent event filtering or 'stripping' is described. In Section 5, the offline event selection is summarised before, in Section 6, the manner in which the selected events will be corrected for detector and reconstruction biases is discussed. In Section 7, methods for extracting the observables are described and, finally, in Section 8, conclusions are presented.

# 2 Analysis Issues

An investigation of the event selection required to isolate $B_d \to K^{*0}\mu^+\mu^-$ candidates has been performed with Monte Carlo simulations of the signal and $b\bar{b}\to\mu^+\mu^- X$ background events [13], [14]. Standard LHCb tools were used to reconstruct candidates for the K and $\pi$ from the $K^{*0}$ decay and the two muons. The $K^{*0}$ and di-muon candidates were then combined to form $B_d$ candidates.

The kaon from the $K^{*0}$ decay is identified primarily using information from the two RICH detectors [15]. A requirement is made on the particle identification likelihood determined from these detectors. Averaged over momentum, kaons are identified with 95% efficiency, with 5% of pions misidentified as kaons [16].

Tracks extrapolated to the muon stations are identified as muons on the basis of sufficient hits being observed in the different stations, and on the basis of a likelihood

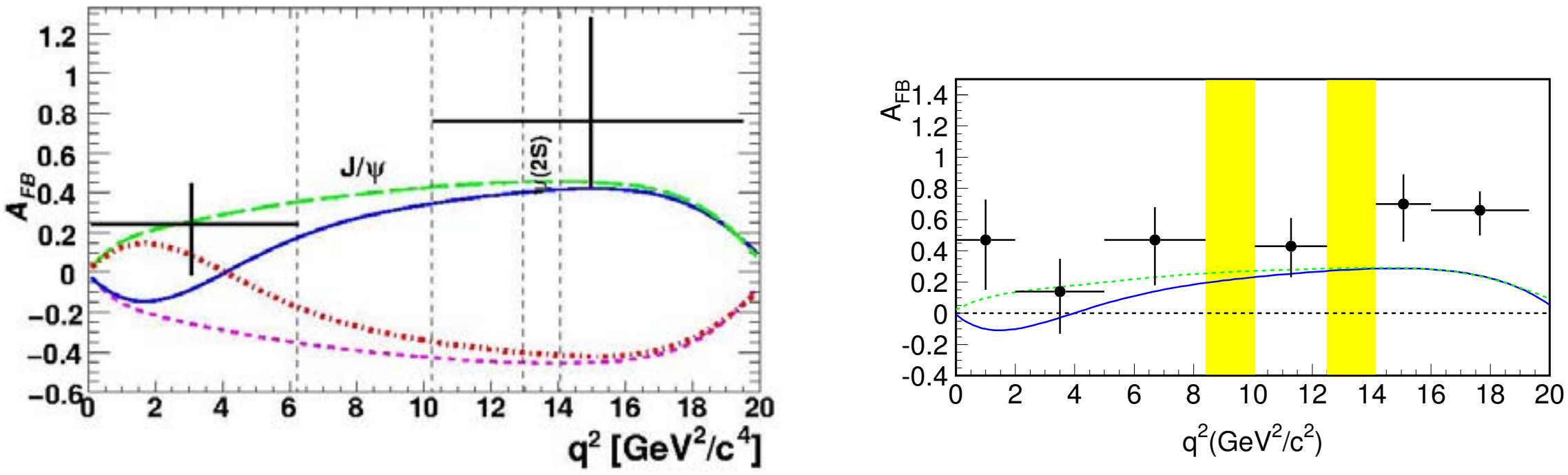


(a) (b)

Figure 5: The $A_{FB}$ spectrum measured by the BaBar (a) and Belle (b) experiments from 384 and 657 M $B\overline{B}$ pairs respectively. The solid line shows the SM expectation and the dashed lines the expectation from a range of supersymmetric models.

formed from the average distance between such hits and the relevant track [17]. Averaged over momentum, muons are identified with 93% efficiency, with 1% of pions misidentified as muons [16].

All four tracks are required to come from the same vertex and, as the analysis does not require any time dependence to be measured, the selection can use impact parameter and vertex displacement requirements to isolate the signal.

## 2.1 Acceptance Effects

In order to reconstruct $A_{\rm FB}$, knowledge of the absolute efficiency of the muon identification, trigger, and analysis requirements is not required. However, an efficiency that depends on $\theta_{\rm L}$ alters the measured value of $A_{\rm FB}$. Simulation studies indicate that the effect can be substantial – changing $A_{\rm FB}$ at the level of the statistical error expected from $2\,{\rm fb}^{-1}$ of data. It is therefore critical to prepare trigger and offline selections that minimise the variation of the angular efficiency as a function of $\theta_{\rm L}$.

Several effects which change the shape of the measured $\theta_{\rm L}$ distribution have been identified [18] and are discussed below. These effects are all observed to be symmetric about $\theta_{\rm L}=\pi/2$. While the $A_{\rm FB}$ value that is observed is scaled by such effects, there is no shift in the zero-point. However, the scaling of $A_{\rm FB}$ changes the gradient in this region, and hence does affect the experimental uncertainty on the zero-point. The gradient of $A_{\rm FB}$ is also dependent on the form-factors of the decay.

In order to traverse the LHCb dipole magnet and subsequently give hits in the muon detectors downstream of the magnetic spectrometer, muon candidates must have a certain minimum momentum. Requiring hits in the muon stations therefore removes muons with momenta $< 3\,{\rm GeV}$. Owing to the softer momentum spectrum of the signal muons in the low $q^2$ region, this is a larger effect for the $\rm B_d \to K^{*0}\mu^+\mu^-$ decay than for other B decays such as $\rm B_d \to J/\psi K^{*0}$ and $\rm B_s \to \mu^+\mu^-$. The present muon identification algorithms make requirements on the number of different muon stations with hits, depending on the momentum of the muon. For example, for muons with momenta $< 6\,{\rm GeV}$, hits are required in both the second and third muons stations, whereas for muons with momenta between 6 and 10 GeV, hits are required in the second, third and either of the fourth or fifth muon stations [17]. These criteria result in a muon acceptance that is a function of both $q^2$ and $\theta_{\rm L}$. Requiring that the muons be reconstructed therefore warps the momentum spectrum and changes the value of $A_{\rm FB}$ observed. This is illustrated in Fig. 6 where the effect of requiring $p > 3\,{\rm GeV}$ on the $\theta_{\rm L}$ efficiency is shown. Except where explicitly stated, this, and subsequent efficiency plots in this section, are made for events at the generation level i.e. without any effects from the detector and reconstruction. The correlation between $q^2$ and the muon momenta results in this efficiency also varying with $q^2$, with the largest effect at small $q^2$.

In addition, for forward-going $\rm B_d$ mesons, removing muons with low $p_{\rm T}$ removes events with $\theta_{\rm L} \sim 0, \pi$, again changing the $A_{\rm FB}$ value observed [18]. The effect of a $300\,{\rm MeV}$ $p_{\rm T}$ cut applied to both the muons is illustrated in Fig. 7. The plot in Fig. 8 shows the sensitivity that events at various $\theta_{\rm L}$ have to $A_{\rm FB}$, accounting for the number of events

produced with a given $\theta_L$, and demonstrates that the events removed are some of the most significant in the determination of this quantity. Applying requirements to just one of the muon $p_T$'s, to the di-muon $p_T$, or to the sum of the two muon $p_T$'s, has been shown to cause significantly less bias in the angular distributions.

Even after the momentum cuts have been applied, the $p_T$ cuts still have a large effect on the $\theta_L$ distribution. This is illustrated in Fig. 9 which shows the effect of the muon $p_T$ cut on the acceptance before any other cut is applied and after the application of the muon momentum cuts. It will therefore be necessary to understand the acceptance effect of both momentum and $p_T$ criteria.

A $p_T$ cut imposed on the K or the $\pi$ from the $K^{*0}$ decay is seen to have a similar effect on the $\theta_K$ efficiency. However, owing to the K,$\pi$ mass difference, the effect is not symmetric about $\theta_K = \pi/2$. Moreover, the correlation between the $\theta_L$ and $\theta_K$ variables is such that $\theta_K$ acceptance effects also warp the $\theta_L$ angular distribution.

The overall acceptance effect from the detector geometry and reconstruction is shown as a function of $\theta_L$ in Fig. 10. Effects of the type described above, caused by the cuts implicitly applied by the combination of the detector geometry and the reconstruction, can be seen. This plot is made for generated events that have been passed through the GEANT simulation of the detector. As shown in Fig. 8, the events most sensitive to $A_{FB}$ have $\theta_L \sim 0.3, 2.8$. At low $q^2$, events with these $\theta_L$ are retained with $\sim$ 4.5% efficiency, compared to $\sim$ 6.4% for events with $\theta_L \sim \pi/2$. A $p_T$ cut applied to both muons in any further event selection would remove a larger proportion of these sensitive events. The analogous efficiency as a function of $\theta_K$ is shown in Fig. 11.

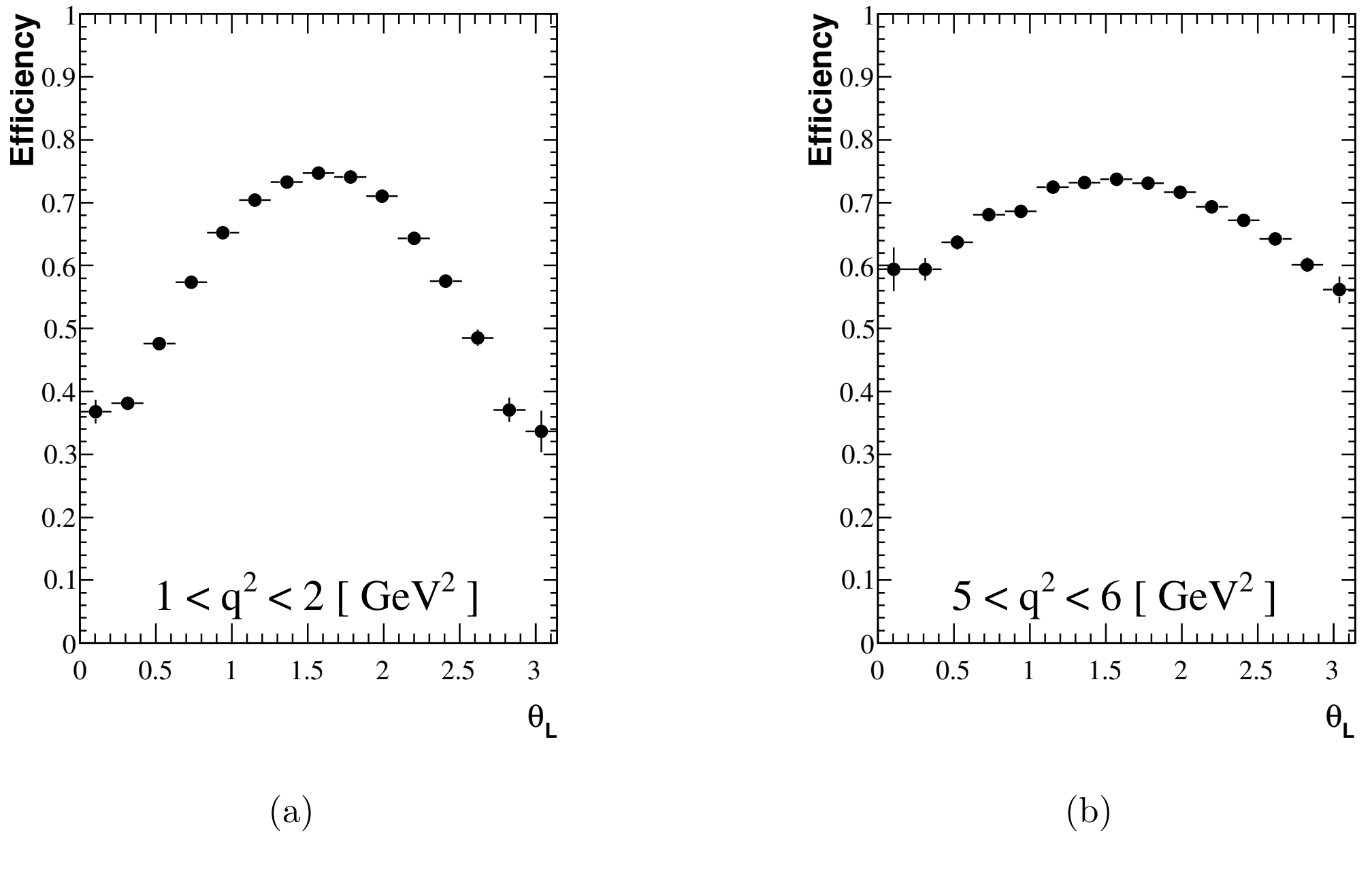


(a) (b)

Figure 6: The effect of a $\mu$ momentum >3 GeV cut on the signal efficiency as a function of $\theta_L$ for events with 1 GeV$^2$ $<q^2<$2 GeV$^2$ (a) and 5 GeV$^2$ $<q^2<$6 GeV$^2$ (b).

Correcting the acceptance effects from the detector geometry and reconstruction will be one of the central challenges for this analysis. While these angular biases are unavoidable, in order to get the best sensitivity to e.g. $A_{\mathrm{FB}}$, any additional biases induced, particularly those that remove very sensitive events, must be minimised. The angular efficiency of the trigger selections required to control the rate of events selected, and of the offline selection criteria needed to separate signal and background events, therefore

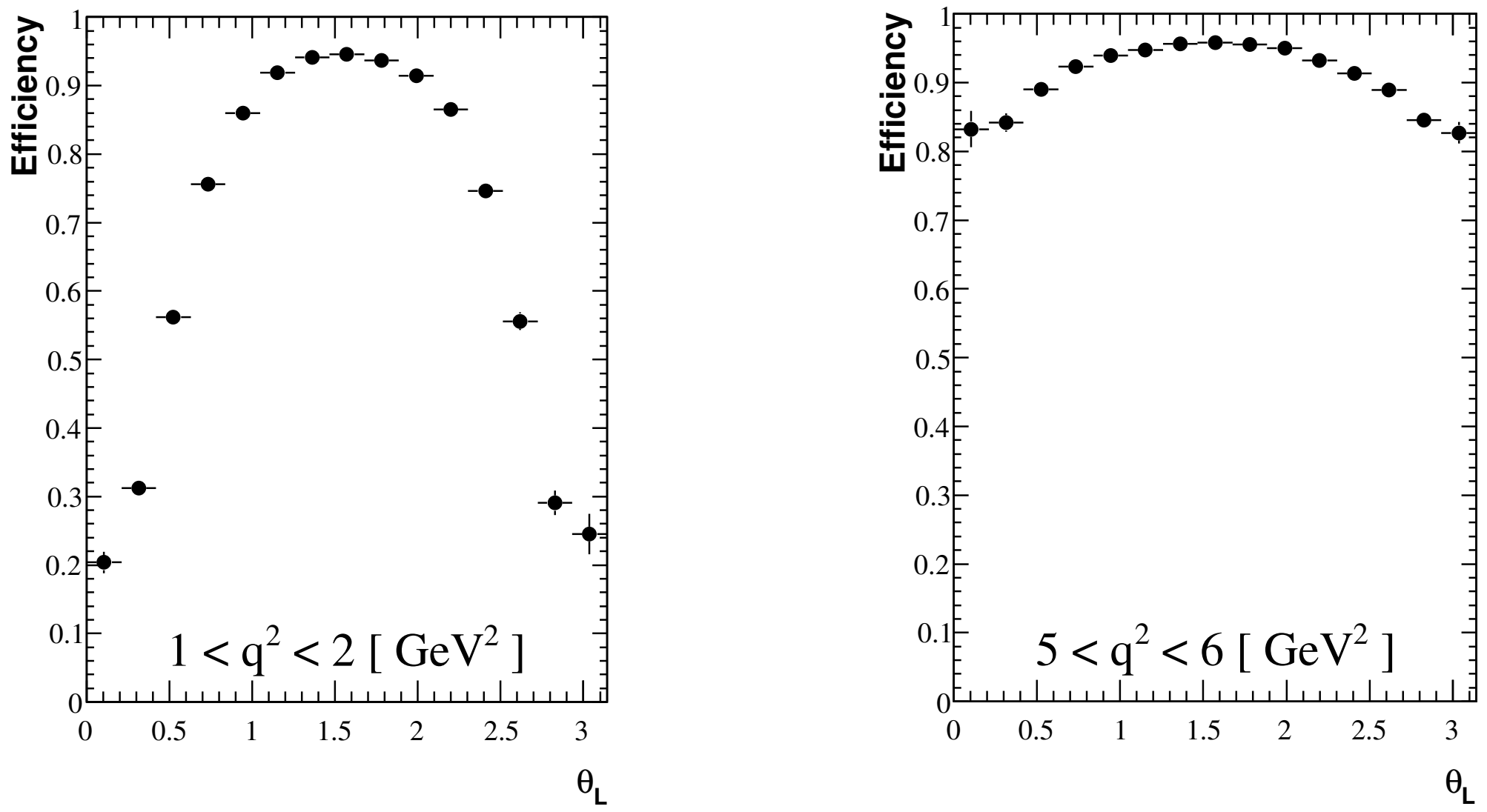


(a) (b)

Figure 7: The effect of a 300 MeV $\mu$ $p_{\mathrm{T}}$ cut on the signal efficiency as a function of $\theta_{\mathrm{L}}$ for events with 1 GeV$^2$ $<q^2<$2 GeV$^2$ (a) and 5 GeV$^2$ $<q^2<$6 GeV$^2$ (b).

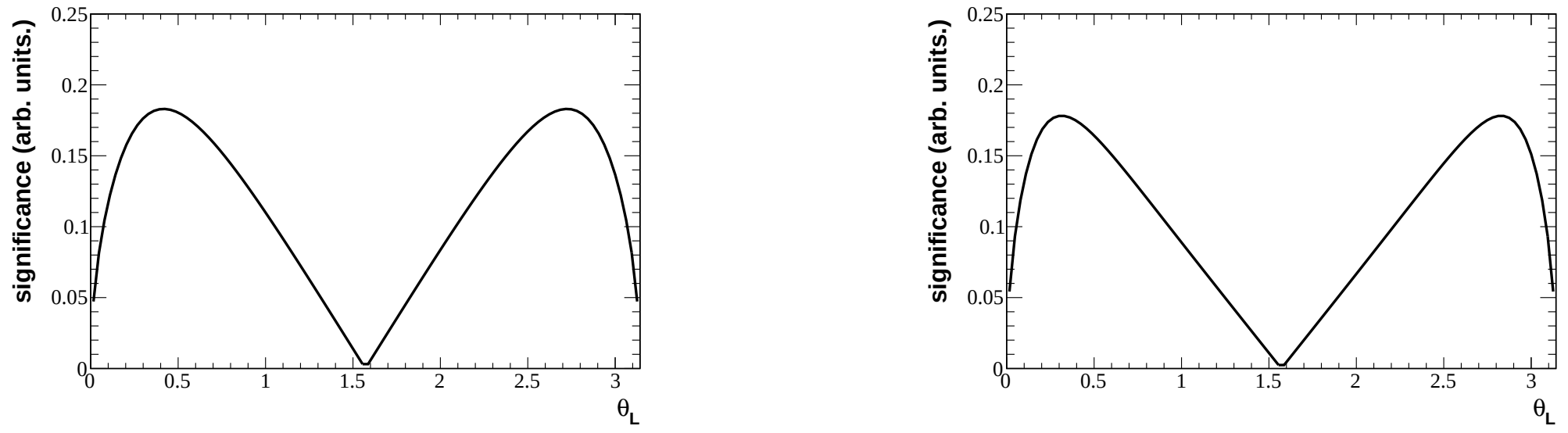


(a) (b)

Figure 8: The sensitivity to $A_{\mathrm{FB}}$ as a function of $\theta_{\mathrm{L}}$ for events with 1 GeV$^2$ $<q^2<$2 GeV$^2$ (a) and 5 GeV$^2$ $<q^2<$6 GeV$^2$ (b). The vertical axis shows the statistical significance of a bin in $\theta_{\mathrm{L}}$ for measuring a forward-backward asymmetry with $A_{\mathrm{FB}}$=0.10. This is defined as $\Delta N_{A_{\mathrm{FB}}}/\sqrt{N}$, where $\Delta N_{A_{\mathrm{FB}}}$ is the difference in the number of events expected at a given $\theta_{\mathrm{L}}$ with/without the forward-backward asymmetry, and $\sqrt{N}$ is the statistical uncertainty on the number of events. This is plotted in arbitrary units.

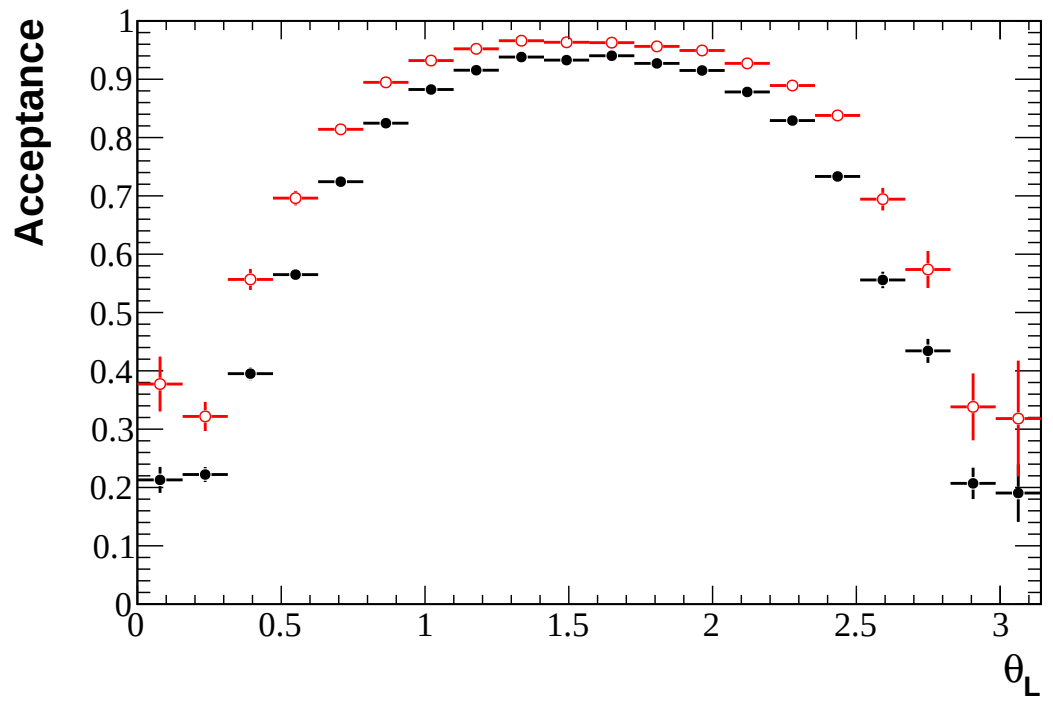


Figure 9: The effect of a 300 MeV $\mu$ $p_{\mathrm{T}}$ cut on the signal efficiency as a function of $\theta_{\mathrm{L}}$ for events with 1 GeV$^2$ $<q^2<$2 GeV$^2$ with (red open points) and without (black filled points) the application of the $\mu$ momentum >3 GeV cuts.

require careful consideration.

## 2.2 Background Effects

A second central issue will be understanding the distribution of the background. Present studies are limited by simulation statistics. However, it has been observed that the backgrounds fall into two classes [19].

Events in which the two muons selected originate in two different B decays give rise to symmetric $\theta_{\mathrm{L}}$ distributions. In a counting analysis (see Section 7.1), the shape of the $\theta_{\mathrm{L}}$ distribution of such 'symmetric' background events has no impact on $A_{\mathrm{FB}}$ but the number of such background events scales the $A_{\mathrm{FB}}$ value reconstructed. Such events cannot change

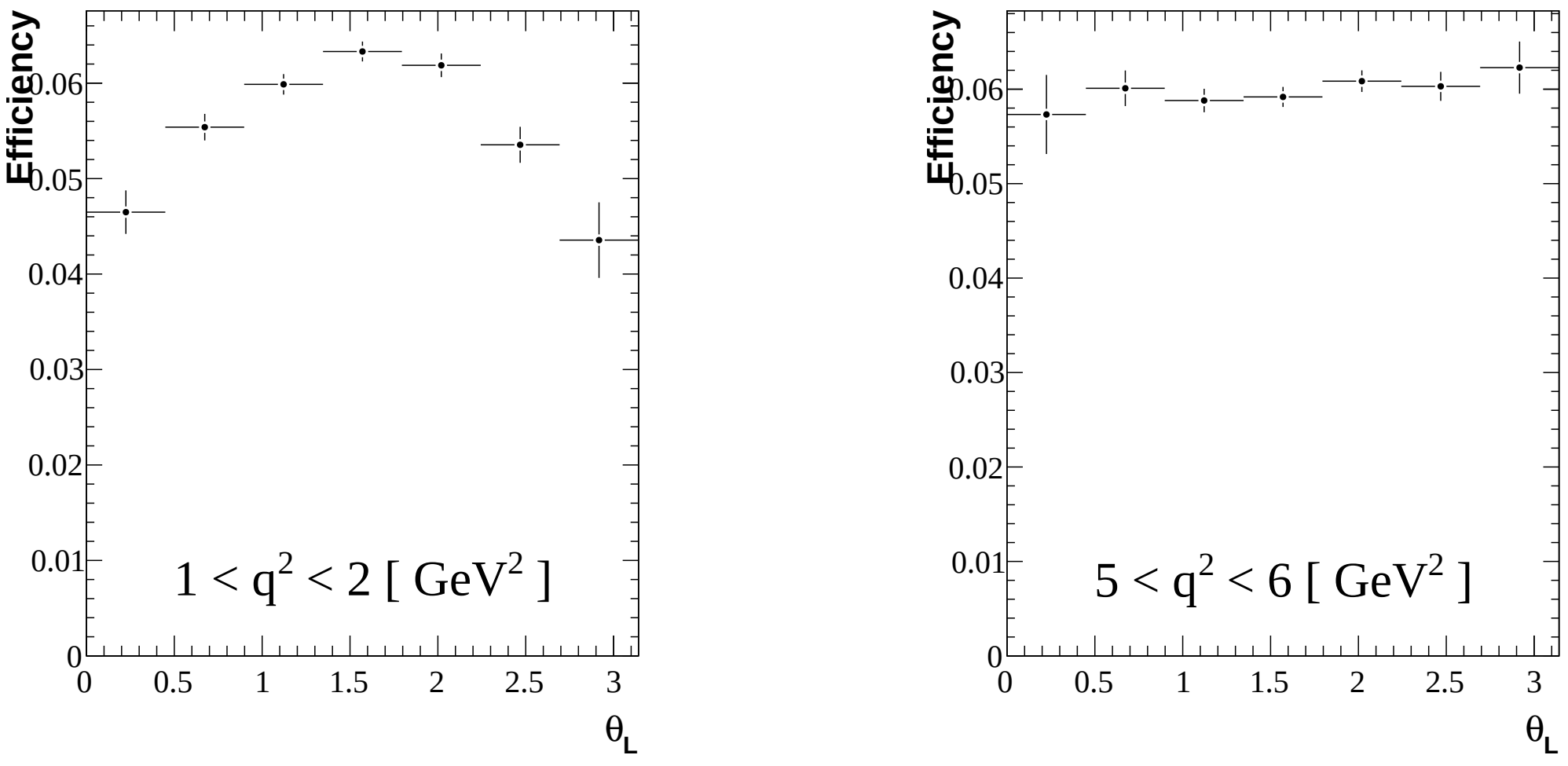


(a) (b)

Figure 10: The effect of the detector geometry and reconstruction on the signal efficiency as a function of $\theta_{\mathrm{L}}$ for events with 1 GeV$^2$ $<q^2<$2 GeV$^2$ (a) and 5 GeV$^2$ $<q^2<$6 GeV$^2$ (b).

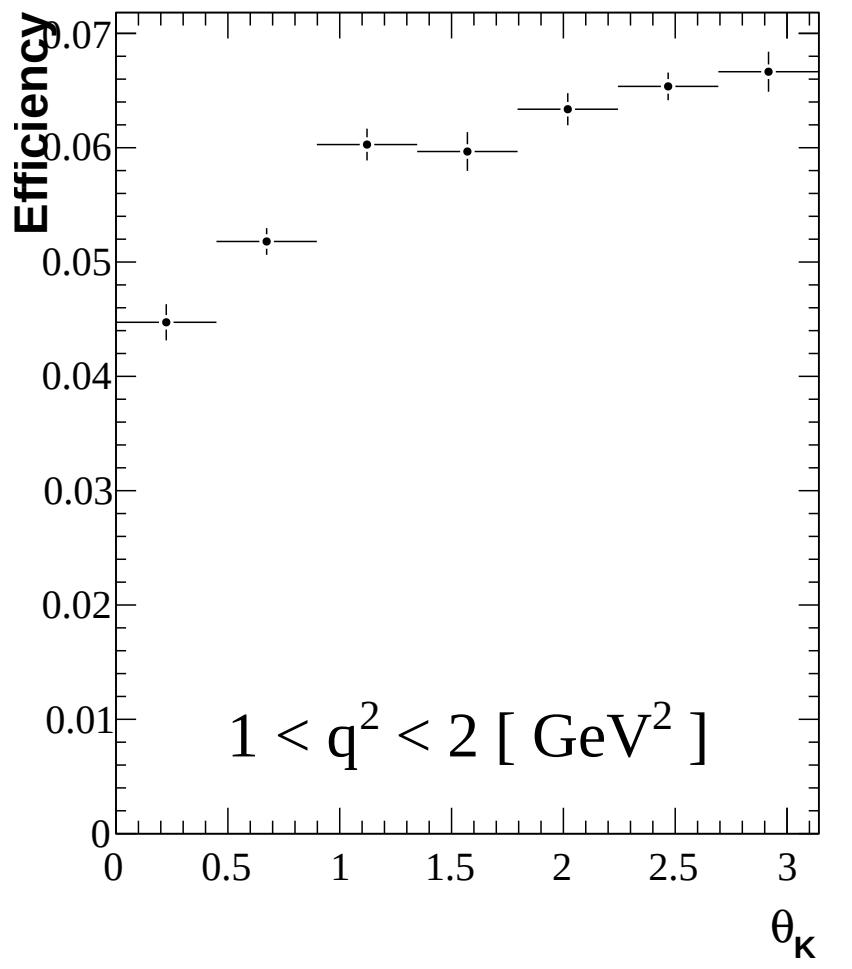


(a)

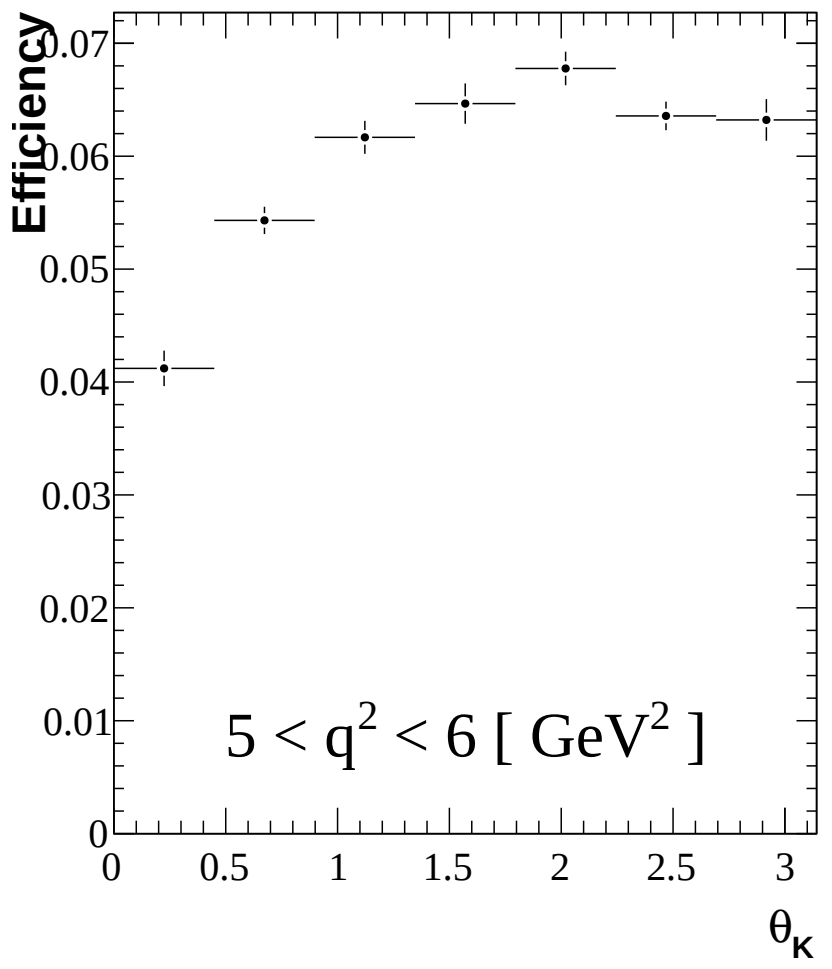


(b)

Figure 11: The effect of the detector geometry and reconstruction on the signal efficiency as a function of $\theta_{\mathrm{K}}$ for events with 1 GeV$^2$ $<q^2<$2 GeV$^2$ (a) and 5 GeV$^2$ $<q^2<$6 GeV$^2$ (b).

the zero-point observed but alter the precision with which the zero-point is determined.

Conversely, events where the two muons are selected from a single B decay chain, tend to give asymmetric $\theta_{\mathrm{L}}$ distributions. In these events, one muon may come from a semi-leptonic B decay, and the other from a subsequent semi-leptonic D decay. This results in one muon which tends to have a larger momentum than the other, and this larger momentum muon is likely to be reconstructed as forward-going. The shape of the $\theta_{\mathrm{L}}$ distribution of such 'asymmetric' background events dictates their effect on $A_{\mathrm{FB}}$ and such backgrounds can bias the zero-crossing point observed. The $\theta_{\mathrm{K}}$ distribution of the background events has not yet been studied but similar considerations will apply in terms of the origin of the kaon.

Non-resonant $\mathrm{B_d} \to \mathrm{K}^+\pi^-\mu^+\mu^-$ events also form a potential background. Using the data from e.g. Ref. [2], the branching ratio of non-resonant background can be limited to $\mathrm{BR}(\mathrm{B_d} \to \mathrm{K}^+\pi^-\mu^+\mu^-) < 4.0\times10^{-7}$ at 90% confidence level [18], comparable to that of the signal decay. In certain kinematic regions, these non-resonant events are expected to have the same $A_{\mathrm{FB}}$ distribution as the signal. However, isolating the relevant kinematic region has been found to be costly in signal efficiency [20]. The level of non-resonant background will therefore have to be established from the data, in order to understand whether it will have a significant bearing on the sensitivity. In addition, higher $\mathrm{K}^{*0}$ resonances will also need to be studied to ensure there is no contamination of the signal.

## 2.3 External Sources of Asymmetry

External sources of asymmetry that can bias the $A_{\mathrm{FB}}$ value observed will also have to be controlled. In order to cross-check asymmetries in the detector, the LHCb magnetic field will occasionally be reversed. Residual asymmetries will still need to be corrected,

as differing amounts of data will most probably be taken in the different magnetic field configurations. The precision with which such a correction can be determined is yet to be investigated. If CP observables are to be measured, the production asymmetry between $B_d$ and $\overline{B}_d$ decays will similarly have to be controlled. The production asymmetry is expected to be at the few percent level [21].

# 3 Trigger

The LHCb trigger consists of a hardware-based first-level (Level 0) and then a software-based "High Level Trigger" (HLT). Selections appropriate for the present analysis are described in the sections below. While a large number of different triggers are envisaged, many of which will trigger $B_d \to K^{*0}\mu^+\mu^-$ signal events, these lines will have different cuts and hence different acceptance effects. It will therefore be important to take the majority of $B_d \to K^{*0}\mu^+\mu^-$ events through just a few selections which will each have an acceptance correction determined for them (see Section 6). The trigger selection cuts have been chosen to try and minimise angular acceptance effects as much as possible.

The trigger algorithms were optimised with respect to offline selected signal events (see Section 5). The Level 0 trigger is well developed and the selections are detailed below. The HLT is still under development and the present form of the selections is outlined.

## 3.1 Level 0

The LHCb Level 0 (hardware) trigger is described in detail elsewhere [22]. The single muon selection requires muon candidates to have $p_T > 1.3$ GeV. The effect of this on the $\theta_L$ efficiency of offline selected $B_d \to K^{*0}\mu^+\mu^-$ events is illustrated in Fig. 12. This figure and other figures in this section show only the additional effect of the trigger on events which have already been reconstructed and offline selected.

The acceptance effect in Fig. 12 is markedly less severe than that from the detector geometry and reconstruction effects shown in Fig. 10. This owes both to the fact that the reconstruction already removes part of the events that would be cut by the above requirement but also to the kinematics.

For events with $\theta_L \sim 0, \pi$ the boost from the di-muon rest frame to the lab frame is likely to give one large and one small $p_T$ muon. While requiring either of the muons have large $p_T$ will result in such events being accepted, requiring both are large tends to result in such events being rejected. A OR-cut on the two muon $p_T$'s tends to keep events at $\theta_L \sim 0, \pi$, while, as seen in Fig. 7, an AND-cut removes such events. Conversely, for events with $\theta_L \sim \pi/2$, both muons are likely to have intermediate values of $p_T$. In this case an OR-cut will tend to remove events, as neither $p_T$ will be large.

The muon $p_T$ OR-cut used in the Level 0 single muon trigger therefore has high efficiency for events with $\theta_L \sim 0, \pi$, retaining the events with high sensitivity to $A_{FB}$, but lower efficiency around $\theta_L \sim \pi/2$. This is only the case at low $q^2$ (Fig. 12a), as at high $q^2$ the muon spectrum is harder and such effects play a lesser role (Fig. 12b).

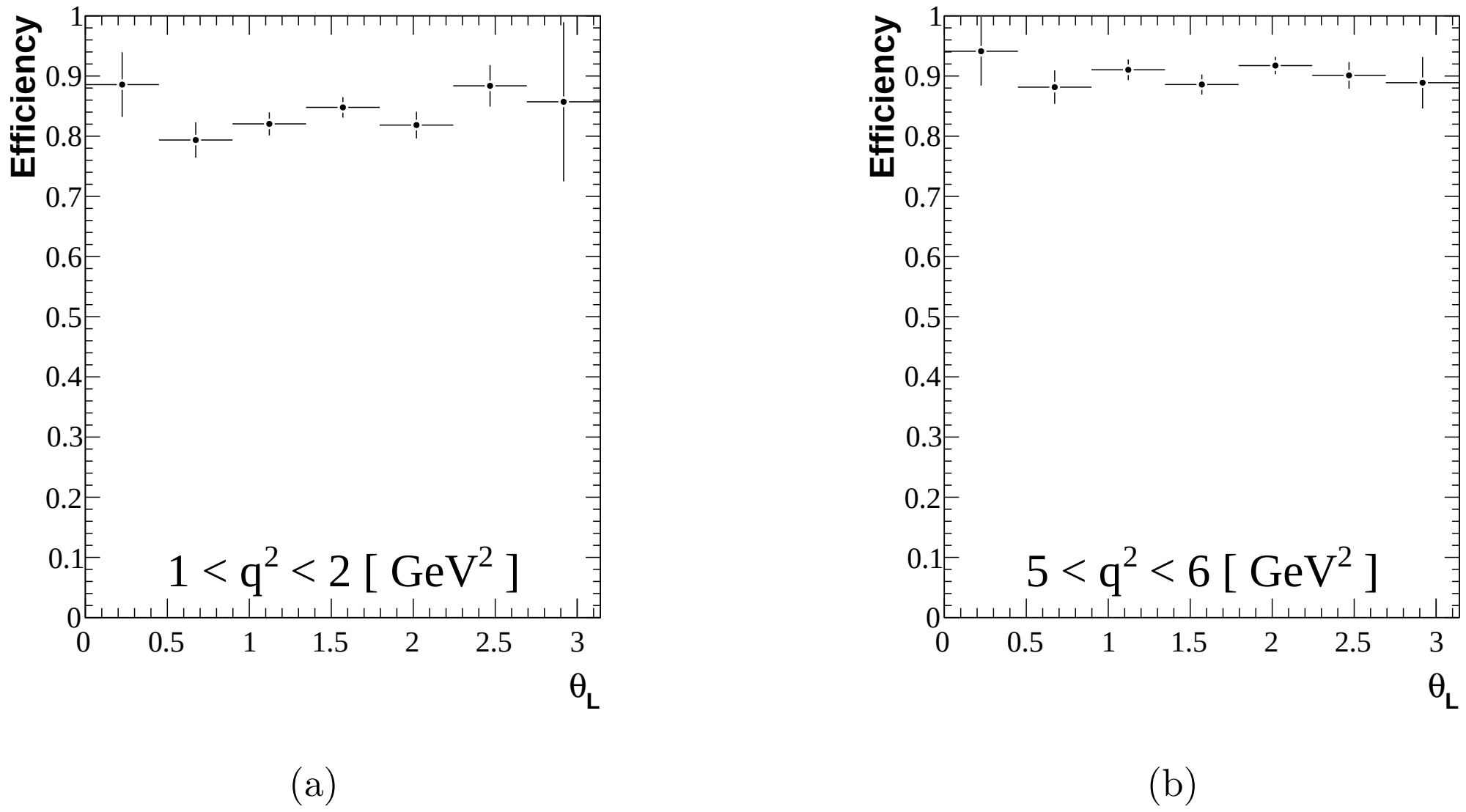


Figure 12: The efficiency of the Level 0 single muon trigger as a function of $\theta_{\rm L}$ for events with 1 GeV$^2$ $<q^2<$2 GeV$^2$ (a) and 5 GeV$^2$ $<q^2<$6 GeV$^2$ (b).

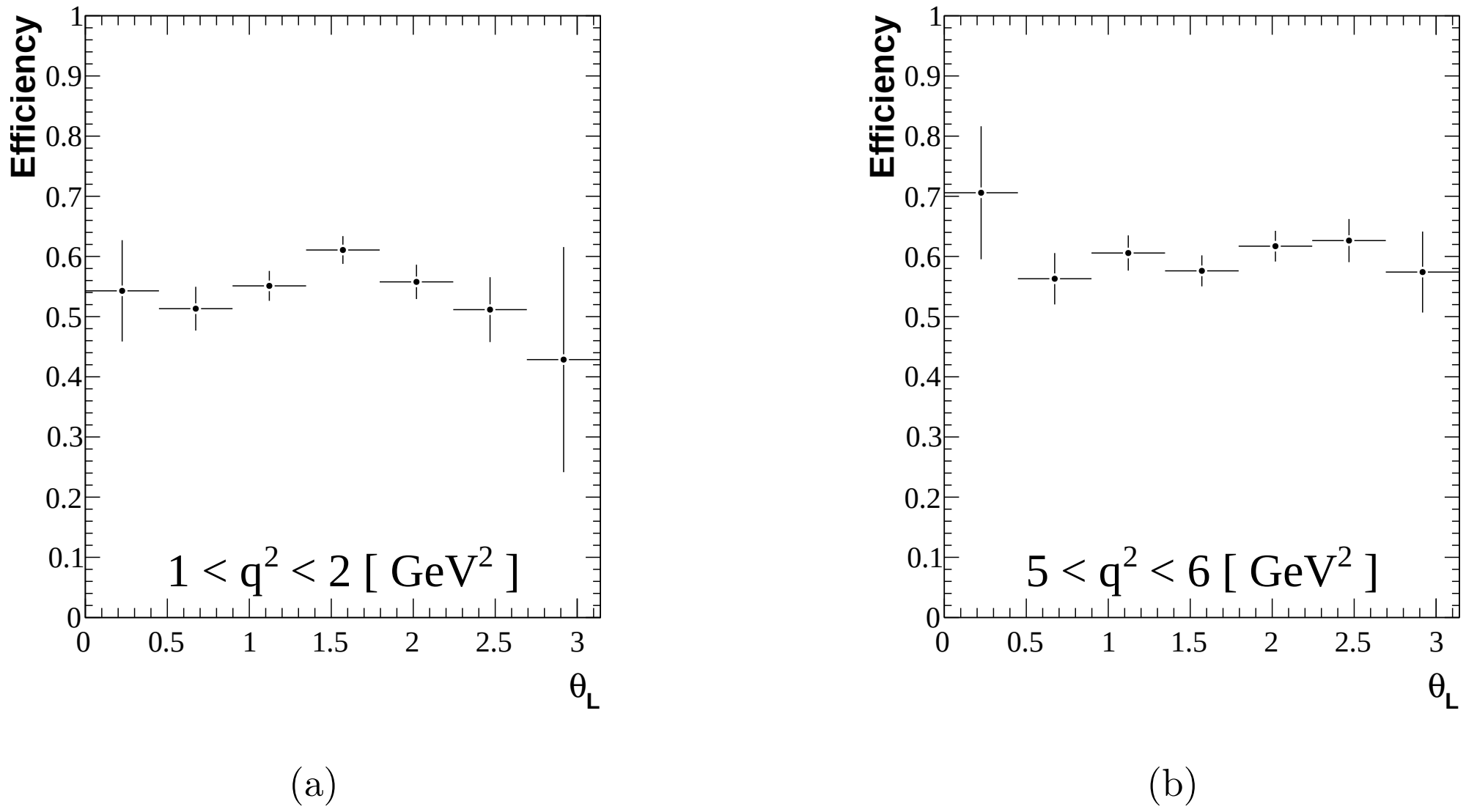


Figure 13: The efficiency of the Level 0 di-muon trigger as a function of $\theta_{\rm L}$ for events with 1 GeV$^2$ $<q^2<$2 GeV$^2$ (a) and 5 GeV$^2$ $<q^2<$6 GeV$^2$ (b).

The Level 0 di-muon selection requires that the muons have summed $p_{\mathrm{T}}$> 1.5 GeV and, in addition, each muon is required to have $p_{\mathrm{T}}$> 100 MeV. The effect of these requirements on the $\theta_{\mathrm{L}}$ efficiency is shown in Fig. 13. The shape is completely dominated by the 100 MeV requirement applied to both muons and, as might be expected from the above argument, the very sensitive events with $\theta_{\mathrm{L}}\sim 0, \pi$ are therefore depleted. In addition to the significantly lower efficiency, the acceptance effect is more severe than that seen for the Level 0 single muon trigger.

The efficiency of these selections with respect to offline selected $\mathrm{B_d} \to \mathrm{K}^{*0}\mu^+\mu^-$ events is given in Table 1. In addition, the efficiency of the Level 0 hadron line, which requires a cluster be found in the calorimeters with transverse energy $E_{\mathrm{T}} > 3.5$ GeV, is also shown. Other Level 0 trigger lines that make smaller contributions to the total efficiency are omitted. While the figures in this section show the efficiency in specific $q^2$ bins, this and other tables in this section, show the trigger efficiency averaged over the entire $q^2$ range. A total Level 0 trigger efficiency of $\sim$ 93% is expected. The vast majority of this efficiency comes from the single muon trigger.

| Level 0 Line | Signal Selection Efficiency (%) | Rate (kHz) |
|---|---|---|
| Single muon | 89.9±0.2 | 230 |
| Di-muon | 62.9±0.3 | 40 |
| Hadron | 27.0±0.3 | 720 |
| Total | 93.1±0.2 | 1000 |

Table 1: Level 0 trigger efficiencies with respect to offline selected $\mathrm{B_d} \to \mathrm{K}^{*0}\mu^+\mu^-$ events. The accept rate of each line is also shown.

The offline selection reduces the background expected to a level well below that of the signal (see Section 5), without imposing any $p_{\mathrm{T}}$ requirements. The use of $p_{\mathrm{T}}$ cuts is avoided, as such cuts remove events with high sensitivity to $A_{\mathrm{FB}}$ (see Section 2.1). However, the trigger, which is presently tuned on the offline selected events, must use $p_{\mathrm{T}}$ requirements to reduce the rate of minimum bias events selected. In the present trigger chain, it is not possible to form more discriminating quantities such as the vertex position. Retuning the offline selection after imposing a $p_{\mathrm{T}}$ cut equivalent to that which will, in any case, have already been made in the trigger, does not change the other selection criteria but gives a reduced efficiency [13]. Present simulation studies therefore suggest that, provided the background can be controlled to the extent that it does not dominate the sensitivity, the offline selection should avoid cutting significantly harder on $p_{\mathrm{T}}$ than the cuts used in the trigger.

## 3.2 High Level Trigger

The HLT of LHCb will also include single and di-muon selections. Owing to differences between the track reconstruction on- and off-line, not all muons that are found offline are reconstructed in the trigger. This is particularly the case for lower momentum muons.

In order to enhance the efficiency for channels where both muons may not have been reconstructed in the trigger, the HLT algorithms also include a so-called "muon+track" selection [23] where a single online muon is combined with another particle of any type (a "track"). In addition to $p_{\mathrm{T}}$ and impact parameter requirements on the muon and the track, this selection makes requirements on the displacement of the two particle vertex from the primary vertex, the vertex quality and the invariant mass of the muon+track combination.

The HLT will be divided into 'alleys' fed only by the relevant Level 0 trigger lines in the case of "HLT1" and a range of inclusive and exclusive selections in the case of the subsequent "HLT2" selections.

### 3.2.1 HLT1 Selection

The first level of the HLT (HLT1) first confirms the Level 0 trigger signal and then makes use of $p_{\mathrm{T}}$, impact parameter and, where appropriate, vertex position and quality requirements, in order to reduce the rate of minimum bias events to $\sim 40\,\mathrm{kHz}$ [16]. Applying the present HLT1 selections to offline and Level 0 selected events that pass the Level 0 single muon selection, the HLT1 muon+track and single muon selections are found to be the most efficient (Table 2). The total HLT1 efficiency (with respect to offline and Level 0 single muon selected events) is found to be 96%, with the muon+track selection contributing 87% of this and (with large overlap) the single muon selection 82%.

| HLT1 Line | Signal Selection Efficiency (%) | Rate (kHz) |
|---|---|---|
| Muon+track | 87.2±0.2 | 1.2 |
| Single muon | 82.0±0.3 | 9.0 |
| Di-muon | 52.3±0.3 | 8.0 |
| Total | 95.5±0.1 | 36.8 |

Table 2: HLT1 trigger efficiencies with respect to offline and Level 0 single muon selected $\mathrm{B_d} \to \mathrm{K}^{*0}\mu^+\mu^-$ events. The accept rate of each line is also shown.

The efficiency of the muon+track HLT1 selection as a function of $\theta_{\mathrm{L}}$ is shown in Fig. 14. As for the Level 0 trigger, after the application of the offline selection and the Level 0 single muon trigger selection, this HLT1 selection does not severely bias the $\theta_{\mathrm{L}}$ angular distribution. The efficiency as a function of $\theta_{\mathrm{K}}$ is also not significantly warped.

The efficiency of the single muon HLT1 selection as a function of $\theta_{\mathrm{L}}$ is shown in Fig. 15. As in the case of the Level 0 single muon trigger, at low $q^2$, a dip is seen in the efficiency for events with $\theta_{\mathrm{L}} \sim \pi/2$.

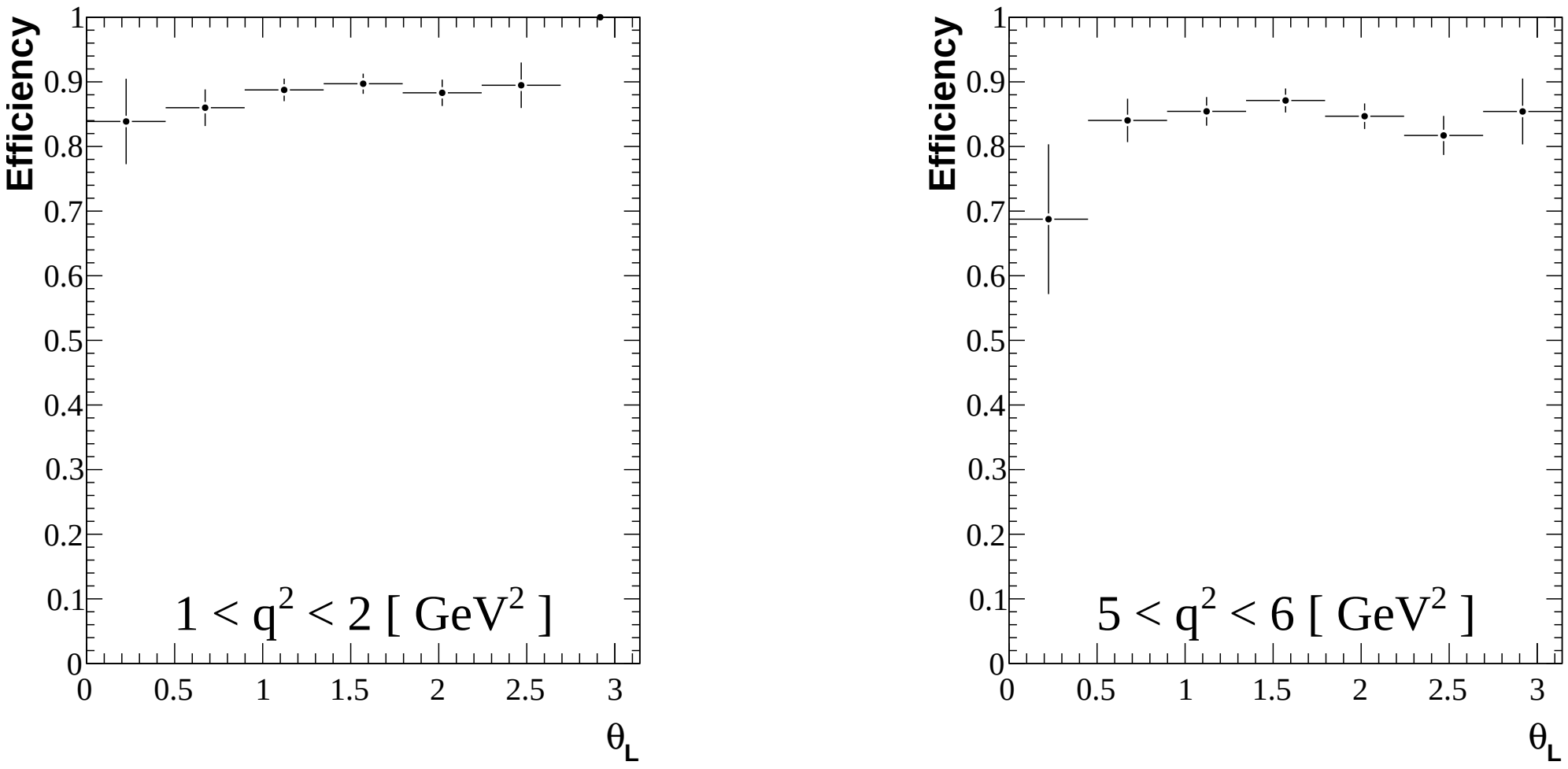


(a) (b)

Figure 14: The efficiency of the HLT1 muon+track trigger selection on offline and Level 0 single muon selected events. The efficiency is shown for events with 1 GeV$^2$ $<q^2<$2 GeV$^2$ (a) and 5 GeV$^2$ $<q^2<$6 GeV$^2$ (b).

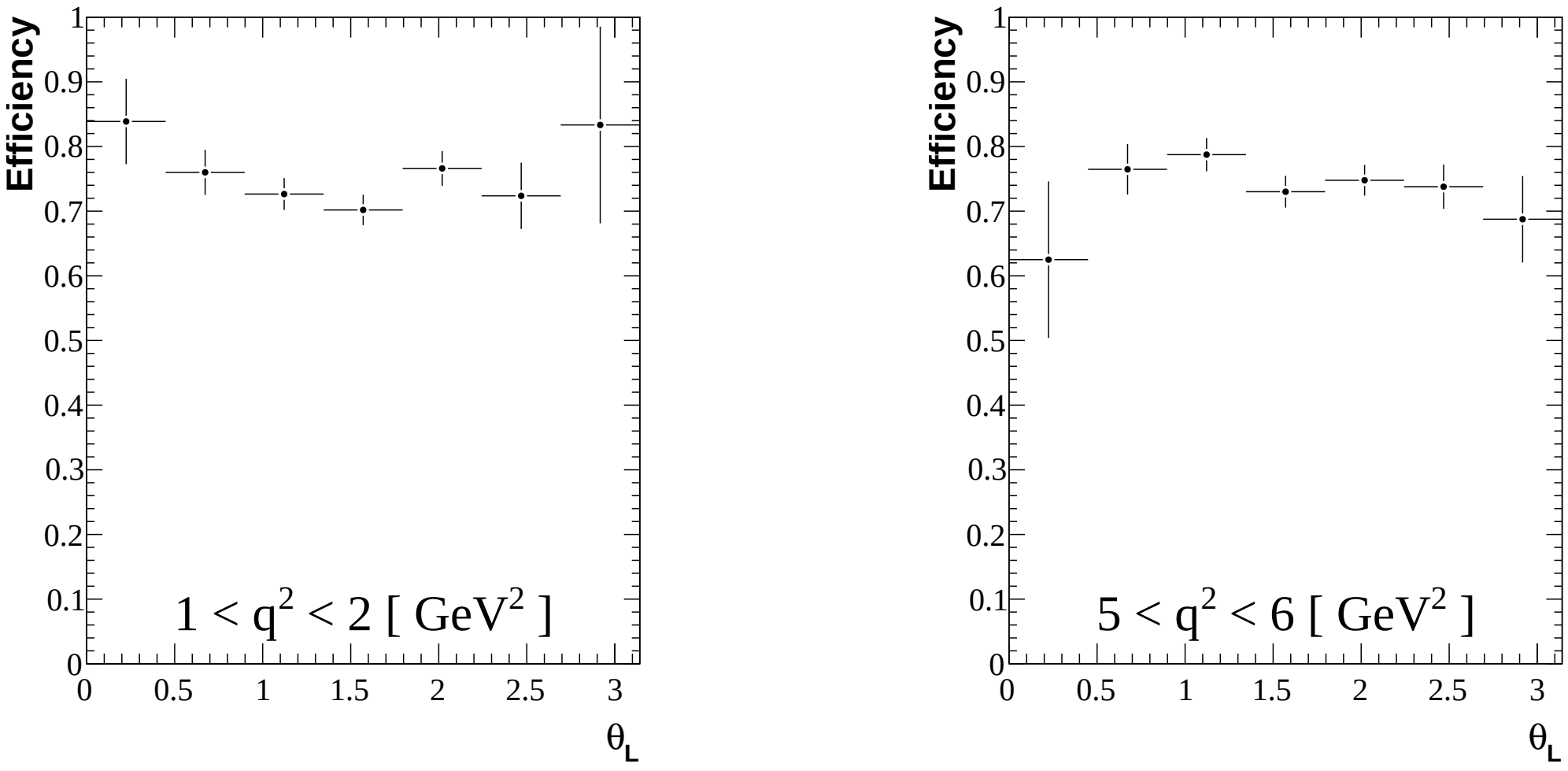


(a) (b)

Figure 15: The efficiency of the HLT1 single muon trigger selection on offline and Level 0 single muon selected events. The efficiency is shown for events with 1 GeV$^2$ $<q^2<$2 GeV$^2$ (a) and 5 GeV$^2$ $<q^2<$6 GeV$^2$ (b).

### 3.2.2 HLT2 Selection

Using a trigger scenario in which the bandwidth allocation is as shown in Table 3, the total efficiency of events passing the offline selection, Level 0 single muon selection, and HLT1 muon+track selection is found to be 97%. Again, a muon+track selection gives the bulk of this efficiency (95%), while one of the so-called "topological" selections, primarily designed to isolate hadronic decays by searching for a combination of three tracks with a high quality displaced vertex, gives a 74% efficiency.

At present the di-muon trigger contributes very little. A large difference between the on- and off-line reconstruction is observed at low $q^2$, which is under study. Improving the di-muon line should allow a further increase in the HLT2 efficiency. The hard $p_{\rm T}$ cut and downscaling used to give an HLT2 single muon line that meets the bandwidth constraints, and has the highest possible B purity, gives a low efficiency for $B_d \rightarrow K^{*0}\mu^+\mu^-$ events, where no $p_{\rm T}$ cut is applied offline.

The efficiency of the muon+track selection as a function of $\theta_L$ is shown in Fig. 16. As for the previous trigger levels, the selection induces little additional angular bias. The $\theta_K$ efficiency is similarly seen to have only a small additional angular bias.

## 3.3 Overall Trigger Efficiency

Given the above efficiencies for Level 0, HLT1 and HLT2 selections, the present simulation indicates that an overall trigger efficiency of 74% might be expected for the most efficient single trigger line (Level 0 single muon (90%), HLT1 muon+track (87%), HLT2 muon+track (95%)). With the present thresholds, the events would be selected by the trigger without severely biased the $\theta_L$ angular acceptance and, in particular, retaining events with high sensitivity to $A_{FB}$ (see Fig. 8, Section 2.1). The overall trigger efficiency as a function of $\theta_L$ is shown in Fig. 17. The $\theta_K$ efficiency is similarly not significantly warped (Fig. 18). The efficiency as a function of $q^2$ can be seen in Fig. 19. As might be expected, a drop is seen for events with low $q^2$ where the two muons are softer.

Adding other trigger lines, which have different acceptance effects, the overall efficiency can be increased to 87%.

The triggers outlined above are inclusive and should therefore also select other

| HLT2 Line | Signal Selection Efficiency (%) | Rate (kHz) |
|---|---|---|
| Muon+track | 94.7±0.2 | 0.64 |
| Topological | 73.5±0.3 | 0.39 |
| Single muon | 15.8±0.3 | 0.19 |
| Di-muon | 35.7±0.3 | 0.10 |
| Total | 97.5±0.1 | 2.00 |

Table 3: HLT2 trigger efficiencies with respect to offline, Level 0 single muon and HLT1 muon+track selected $B_d \rightarrow K^{*0}\mu^+\mu^-$ events. The accept rate of each line is also shown.

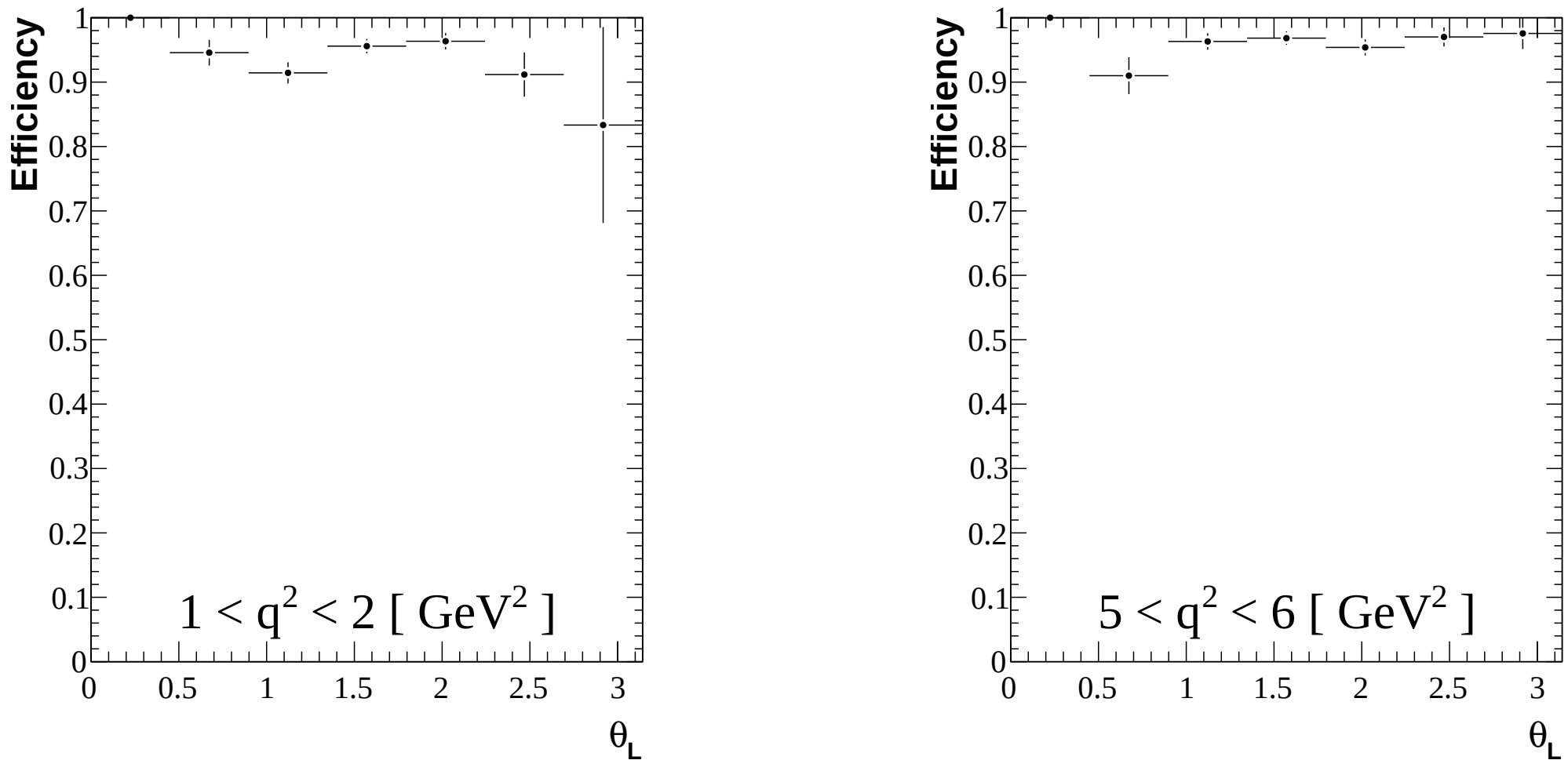


(a) (b)

Figure 16: The efficiency of the HLT2 muon+track trigger selection on events selected by the offline, Level 0 single muon and HLT1 muon+track selections. The efficiency is shown for events with 1 GeV$^2$ $<q^2<$2 GeV$^2$ (a) and 5 GeV$^2$ $<q^2<$6 GeV$^2$ (b).

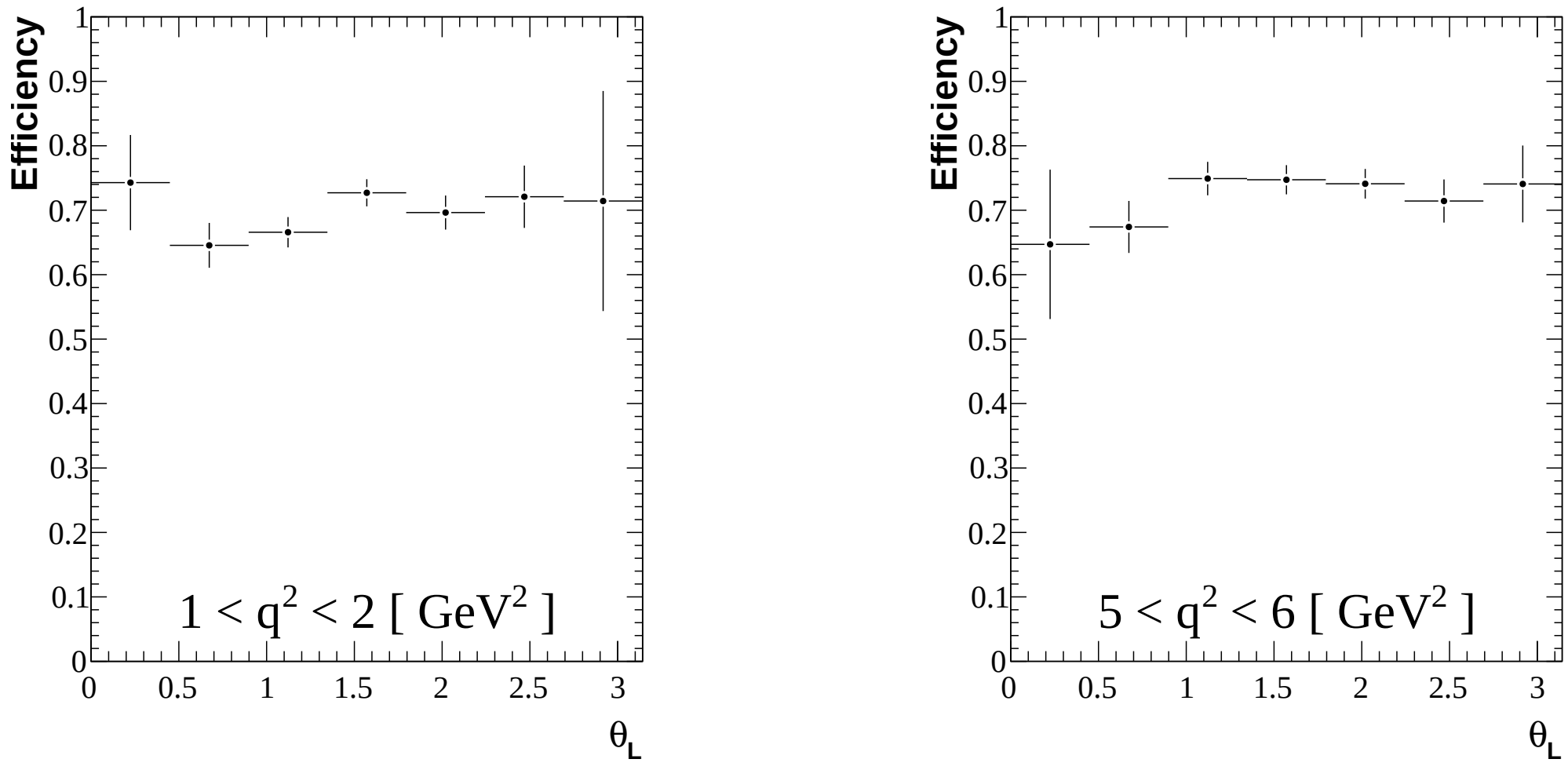


(a) (b)

Figure 17: The efficiency of the Level 0 single muon, HLT1 muon+track and HLT2 muon+track trigger as a function of $\theta_{\mathrm{L}}$. The efficiency is shown for events with 1 GeV$^2$ $<q^2<$2 GeV$^2$ (a) and 5 GeV$^2$ $<q^2<$6 GeV$^2$ (b).

$b \rightarrow s\ell^+\ell^-$ events which will form control channels for the $\mathrm{B_d} \rightarrow \mathrm{K}^{*0}\mu^+\mu^-$ analysis. The efficiency of these selections on the relevant control channels is yet to be established.

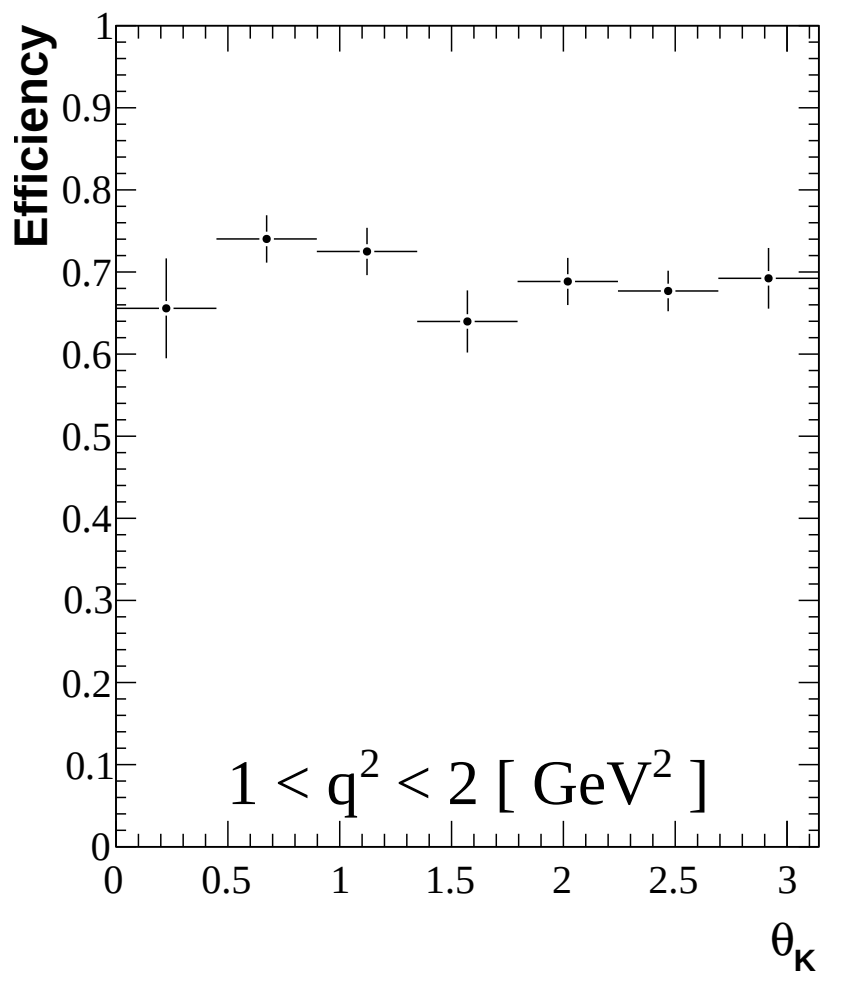


(a)

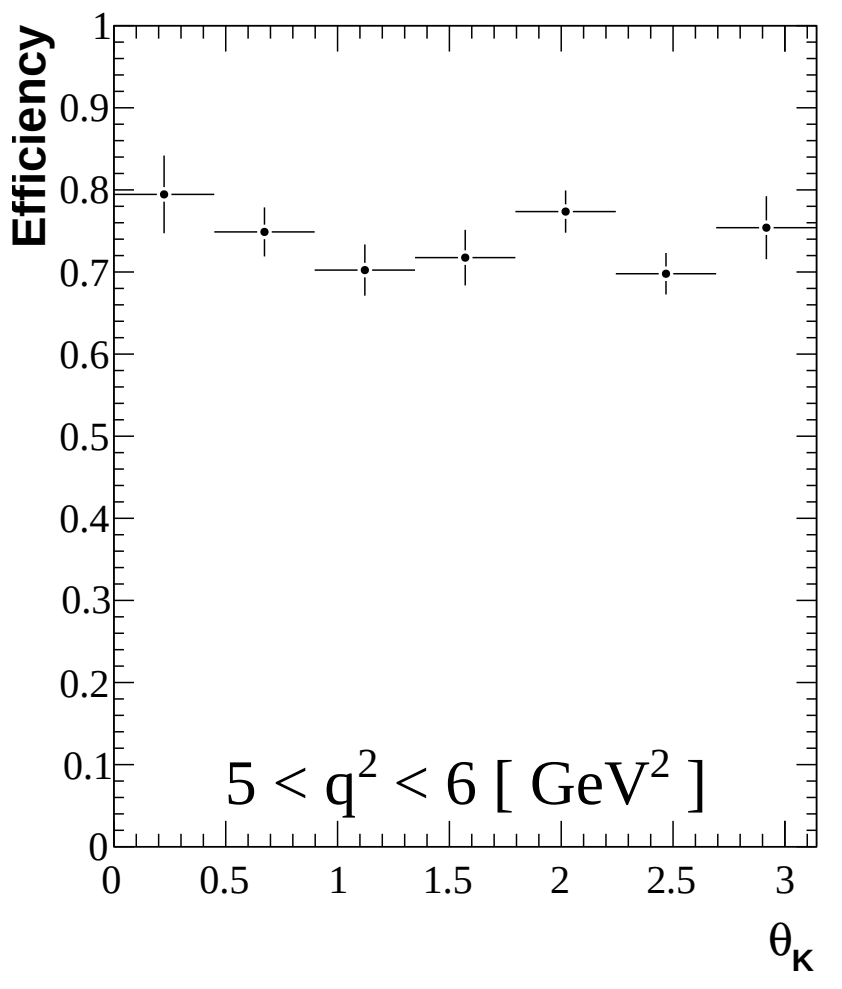


(b)

Figure 18: The efficiency of the Level 0 single muon, HLT1 muon+track and HLT2 muon+track trigger as a function of $\theta_{\mathrm{K}}$. The efficiency is shown for events with 1 GeV$^2$ $<q^2<$2 GeV$^2$ (a) and 5 GeV$^2$ $<q^2<$6 GeV$^2$ (b).

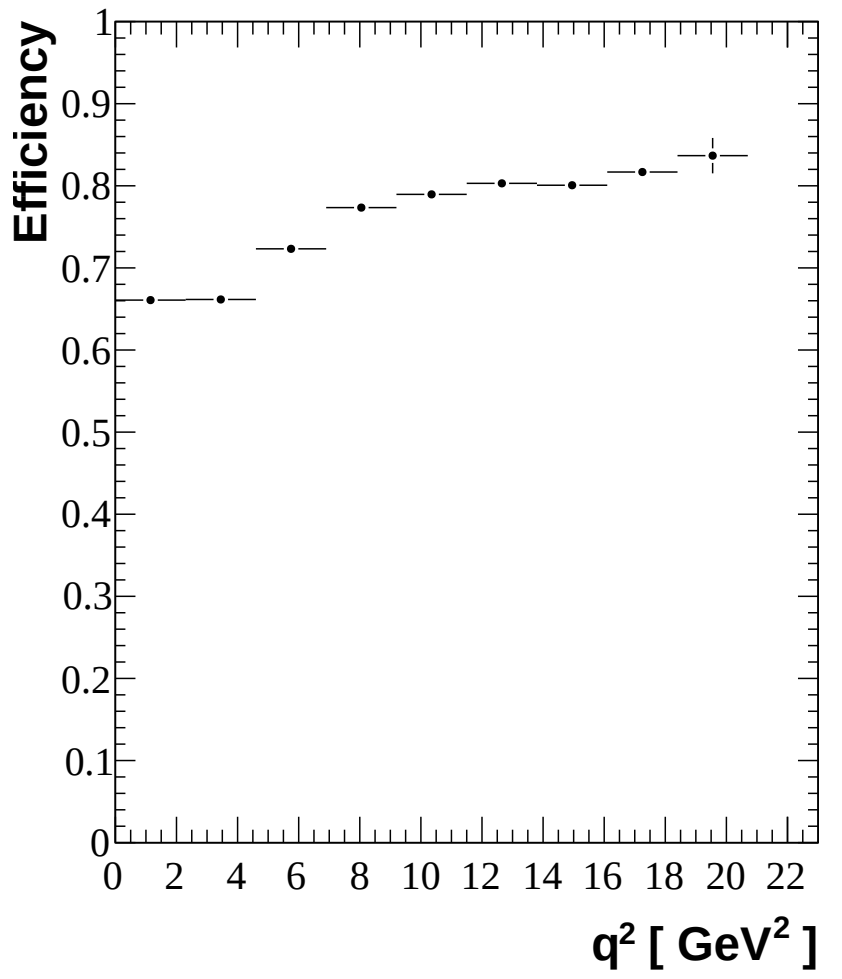


Figure 19: The efficiency of the Level 0 single muon, HLT1 muon+track and HLT2 muon+track trigger as a function of $q^2$.

# 4 Stripping

Triggered data will be fully reconstructed and ∼1 Hz of events from the signal selection will be written to a stream for further analysis. This procedure is known as “stripping”. The analysis of Ref. [13] suggests that this rate can be achieved, while retaining a $B_d$ mass sideband from 200 MeV below the nominal $B_d$ mass to 500 MeV above the nominal mass, and a $K^{*0}$ mass sideband of ±300 MeV around the nominal $K^{*0}$ mass. Such a selection would give $1.3 \times 10^4$ signal events/2 $fb^{-1}$, including 100% of those that would be selected by the cut-based offline selection described in Section 5.1.

As elsewhere, it is envisaged that quantities involving significances (estimate of a quantity divided by its error), which require detailed errors estimates, will be used only once the errors are suitably well modelled. The selections described in this and in the next sections use such quantities to demonstrate the eventual discrimination power that will be achieved.

The selection is performed by cutting on the $B_d$ flight distance, vertex $\chi^2$, impact parameter significance and pointing angle criteria (the angle between the reconstructed $B_d$ line of flight and the $B_d$ momentum vector), as well as the $K^{*0}$ daughter impact parameter significances and particle identification criteria. A full description of the selection and cut values can be found in Ref. [13].

Even if a multivariate offline selection is eventually used (see Section 5.2), it is envisaged that a stripping selection would use such cut-based criteria. Initially, such a selection would use only cuts on variables that do not require detailed error estimates. Once error estimates have been refined, the above quantities, some of which involve significances, should give the best discrimination power.

In order to estimate or limit the background from non-resonant $B_d \rightarrow K^+\pi^-\mu^+\mu^-$ events, a large range of $K\pi$ invariant masses will need to be selected. A separate selection with additional and/or tighter criteria will be required in order to meet the relevant rate requirements. Similarly, a stripping selection will be required for the $B^+ \rightarrow K^+\mu^+\mu^-$ events that will be used as a cross-check of the acceptance correction (see Section 6). Such selections are yet to be developed.

# 5 Offline Analysis

Investigation of the offline analysis requirements needed to separate the signal events from the background, and the resulting signal and background yields are described in detail elsewhere [13], [14]. Both cut-based and multivariate approaches have been studied. The latter will be particularly useful if, as is planned, it is possible to trigger and strip the data sample for offline analysis with relatively loose requirements.

Selection criteria will eventually be determined by optimising the sensitivity to e.g. $A_{\mathrm{FB}}$. As this sensitivity varies as a function of $\theta_{\mathrm{L}}$ (see Section 2.1), background events at different $\theta_{\mathrm{L}}$ will have a different effect on the overall sensitivity. Given the very limited simulation statistics available to study the background, it was impossible to account for this by studying the $\theta_{\mathrm{L}}$ distribution of background events. The selections detailed below

were therefore optimised by maximising the signal significance metric $S/\sqrt{S+B}$, where $S$ ($B$) is the number of signal (background) events expected from $2\,\mathrm{fb}^{-1}$ of data. This assumes that a background event impairs the measurement of $A_{\mathrm{FB}}$ by the same amount, regardless of the $\theta_{\mathrm{L}}$ value reconstructed, and that the signal events all have the same sensitivity to $A_{\mathrm{FB}}$. However, in both cut-based and multivariate selections, in order to keep the angular efficiency as flat as possible, selection criteria which badly warped the angular efficiency were avoided. Once it is possible to quantify systematic effects from e.g. the acceptance correction, and to study the background distribution in $\theta_{\mathrm{L}}$, the optimal selection may be different from that obtained by maximising $S/\sqrt{S+B}$.

The background sample used to determine selection requirements was composed of events in which $b\bar{b}$ quarks decayed to give two muons. Large samples of inclusive $b\bar{b}$ decays have indicated that this is the dominant source of background [20]. Other contributions, such as those from $c\bar{c}$ events, are neglected in the studies reported below.

## 5.1 Cut-Based Selection

The cut-based analysis investigated in Ref. [13] gives a signal selection efficiency of $0.71 \pm 0.02\%$. This efficiency is composed of $6.39 \pm 0.01\%$ from requiring that all the relevant particles are within the geometric acceptance of the detector; $91.94 \pm 0.06\%$ of events then have all the relevant tracks reconstructed; and the selection cuts then give an efficiency of $12.09 \pm 0.02\%$.

The 87% trigger efficiency that was discussed in Section 3.3 is used to compute the overall signal selection efficiency and the subsequent signal and background yields. The selection yields $4200^{+1100}_{-1000}$ signal events per $2\,\mathrm{fb}^{-1}$ of integrated luminosity. The background is estimated to be $200 \pm 140$ events, giving a background-to-signal ratio $B/S$=$0.05 \pm 0.04$ and a figure-of-merit for the selection $S/\sqrt{S+B} = 63^{+9}_{-8}$. The selection makes hard requirements on the $\mathrm{B_d}$ flight significance, impact parameter and pointing angle, requires large daughter impact parameter significances and suitable particle identification likelihoods for all particles. A table of cut values, taken from Ref. [13], is given in Appendix A.

## 5.2 Multivariate Selection

The multivariate analysis described in Ref. [14] uses a Fisher discriminant to separate signal and background events. The variables used to construct the discriminant were: the $\mathrm{B_d}$ pointing angle, vertex $\chi^2$, flight significance and $p_{\mathrm{T}}$; the $\mathrm{K}^{*0}$ flight significance, vertex $\chi^2$ and impact parameter significance; pion and kaon impact parameter significances and track $\chi^2$; kaon $\mathrm{K}-\pi$ and $\mathrm{K}-\mathrm{p}$ particle identification likelihoods; pion $\mathrm{K}-\pi$ and $\mu-\pi$ particle identification likelihoods; di-muon vertex $\chi^2$ and impact parameter significance; $\mu$ impact parameter significances, track $\chi^2$ and muon $\mathrm{K}-\pi$ and $\mu-\pi$ particle identification likelihoods. Of these, the $\mathrm{B_d}$ vertex $\chi^2$, flight and impact parameter significances and $p_{\mathrm{T}}$; kaon and pion $\mathrm{K}-\pi$ particle identification likelihoods and impact parameter significances were found to make the largest contribution to the discriminant. The Fisher coefficients that describe the contribution of each of these variables to the overall discriminant are

given in Appendix B.

The discriminant is shown for signal and background events in Fig. 20. Selecting events with discriminant values greater than 0.38 was found to give the largest $S/\sqrt{S+B}$. Again assuming an 87% trigger efficiency, the signal selection efficiency of $1.08 \pm 0.01\%$ then gives a signal expectation of $6200^{+1700}_{-1500}$ events per $2\,\mathrm{fb}^{-1}$. The residual background is estimated to be $1550 \pm 310$ events, giving a background-to-signal ratio $B/S$=$0.25 \pm 0.08$. The figure-of-merit for this selection is $S/\sqrt{S+B} = 71^{+11}_{-10}$. Accounting for the correlated component of the error on $S$, this is an improvement of $7 \pm 2$ over the figure-of-merit of the cut-based analysis.

The discriminant is shown for the residual background in Fig. 21. The background is dominated by events in which the particles are selected from fragments of two B decays (filled histogram in the figure, denoted '$b\bar{b}$' in the legend); the selected candidate includes a track that actually originated from the primary vertex but is mis-reconstructed to have a large impact parameter (filled histogram denoted 'from-PV'); or the candidate includes a 'ghost' track which does not match 70% of the hits from any single simulated particle (filled histogram denoted 'Ghost'). Partially reconstructed events (filled histogram denoted 'Part Reco') do not contribute significantly.

Having applied the above cut to the Fisher discriminant, it is observed that there still remain quantities on which a further cut would not remove any signal events, but would reduce the background e.g. the $\pi - \mu$ particle identification likelihood [14]. While the Fisher discriminant analysis already substantially improves $S/\sqrt{S+B}$ compared to

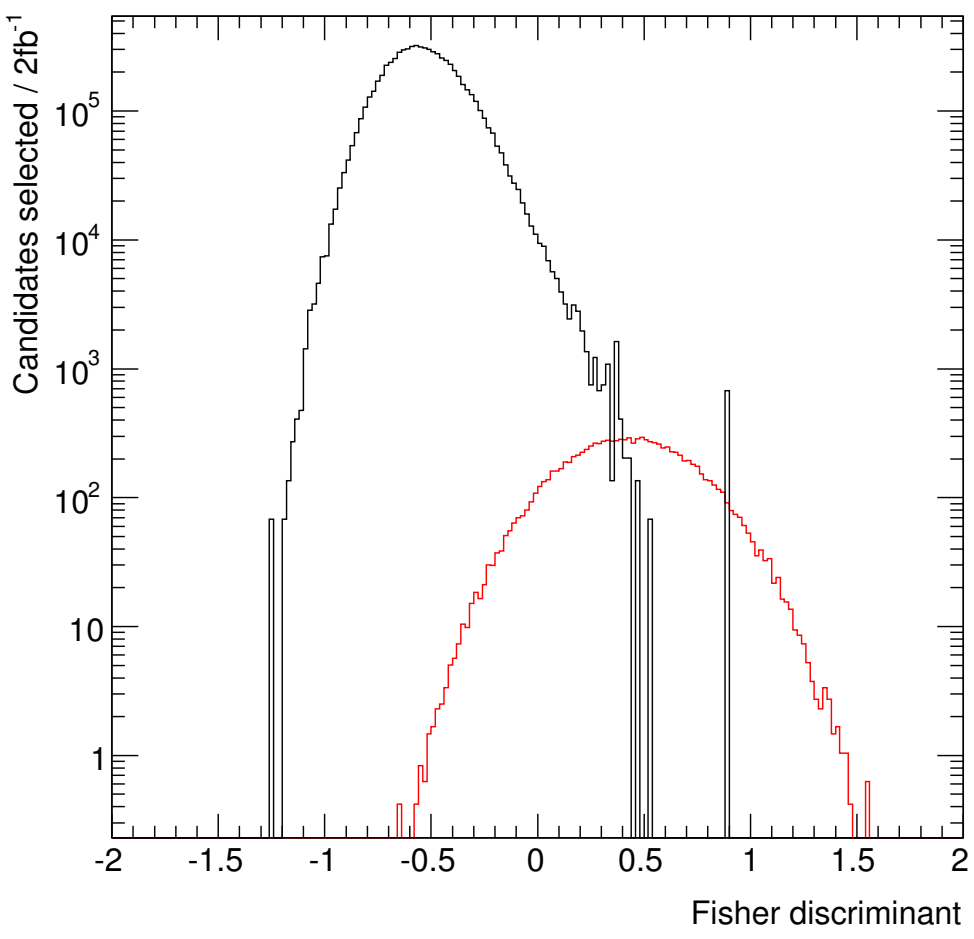


Figure 20: The Fisher discriminant of Ref. [14] for signal (red) and background (black) events.

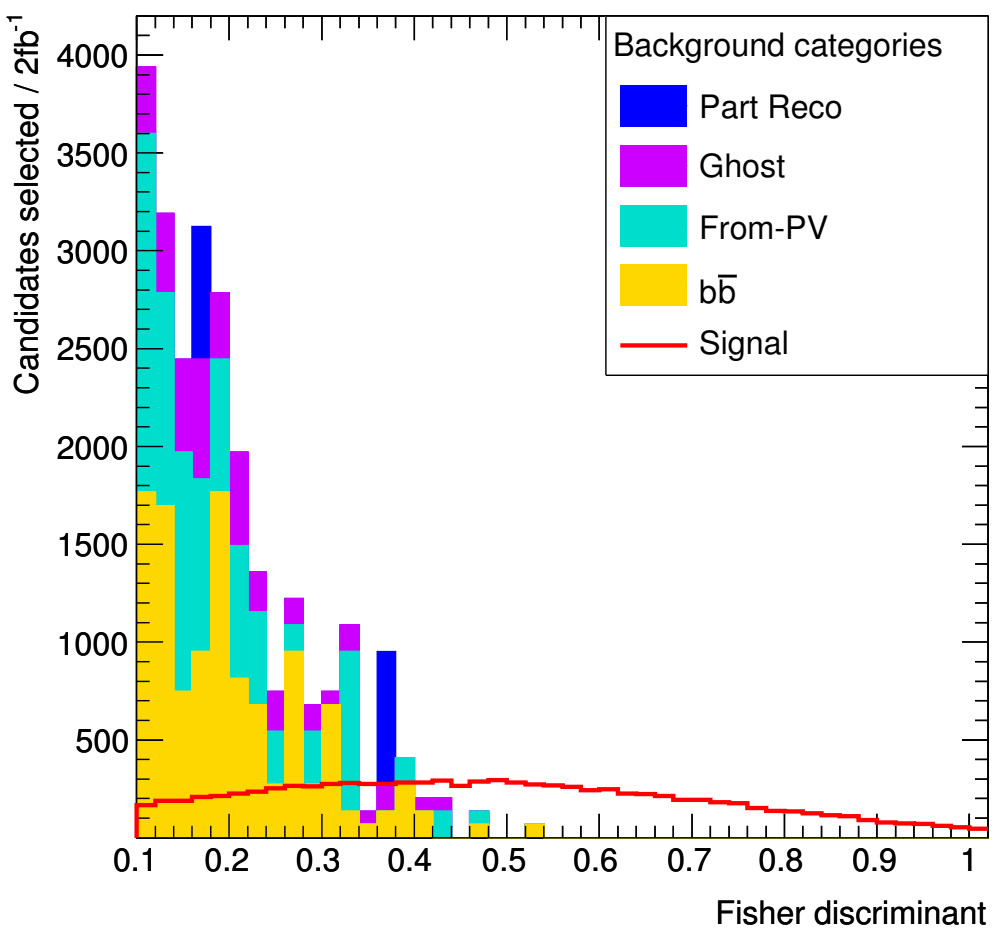


Figure 21: The tail of the Fisher discriminant distribution showing the signal events (open histogram) and background events (shaded histogram). The different shading indicates the composition of the background (see text).

a cut-based analysis, this indicates that it is still not optimal, and a further improvement in signal-background discrimination could be achieved by using a non-linear method such as a neural network.

## 5.3 Acceptance Effects

The $\theta_\mathrm{L}$ ($\theta_\mathrm{K}$) angular efficiency of the multivariate selection is shown in Fig. 22 (Fig. 23). As before, these plots are made for events that have already been reconstructed. At low $q^2$, the selection removes a larger number of events at $\theta_\mathrm{L} \sim 0, \pi$ than at $\theta_\mathrm{L} \sim \pi/2$ but the effect is smaller than that of the detector geometry and reconstruction (Fig. 10). As for the trigger selection (see Section 3), the offline selection therefore avoids introducing any strong additional angular efficiency.

Both the multivariate and cut-based selections use particle impact parameter significance rather than $p_\mathrm{T}$ information to separate signal and background. This is found to be both more discriminating against the background and reduces the angular acceptance effects described in Section 2. The impact parameter significance does still have a (lesser) effect on the angular acceptance, as the impact parameter error has a $p_\mathrm{T}$ dependence.

In the region in which theoretical predictions have the smallest uncertainties (1 GeV$^2$ $<q^2<$6 GeV$^2$), both of the proposed selections are observed to give a flat distribution of the ratio of the number of events selected to those generated as a function of $q^2$.

## 5.4 Performing the Analysis

### 5.4.1 Peaking Backgrounds

It will be necessary to try various candidate particle mass hypotheses in order to check for peaking sources of background. For example, a $\mathrm{B_d} \to \mathrm{J}/\psi\mathrm{K}^{*0}$ event where the pion is misidentified as a muon, and one of the muons as a pion, gives a $\mathrm{B_d} \to \mathrm{K}^{*0}\mu^+\mu^-$ candidate. By reversing the relevant mass hypotheses for the reconstructed tracks it will be possible to isolate a sample of such events. If the detector performs as expected, this particular background will be negligible.

One such peaking source of background that has already been identified is $\mathrm{B_s} \to \phi\mu^+\mu^-$ events where the one of the two kaons from the $\phi$ decay is misidentified as a pion and taken together with the genuine kaon to form a fake $\mathrm{K}^{*0}$. By selecting a suitable sample from the data, for example, by reversing particle identification requirements on the kaon, it is envisaged that the power of the $\mathrm{B_s}$ and $\phi$ mass veto employed to remove such background (see Ref. [14]), and the level of residual background expected, can be checked.

### 5.4.2 Forming a Signal-Background Discriminant

While a large number of correlated quantities are reconstructible, it is envisaged that, for simplicity, a relatively small number of such variables will eventually be used in the signal-background discriminant. Other variables that are uncorrelated with this discriminant,

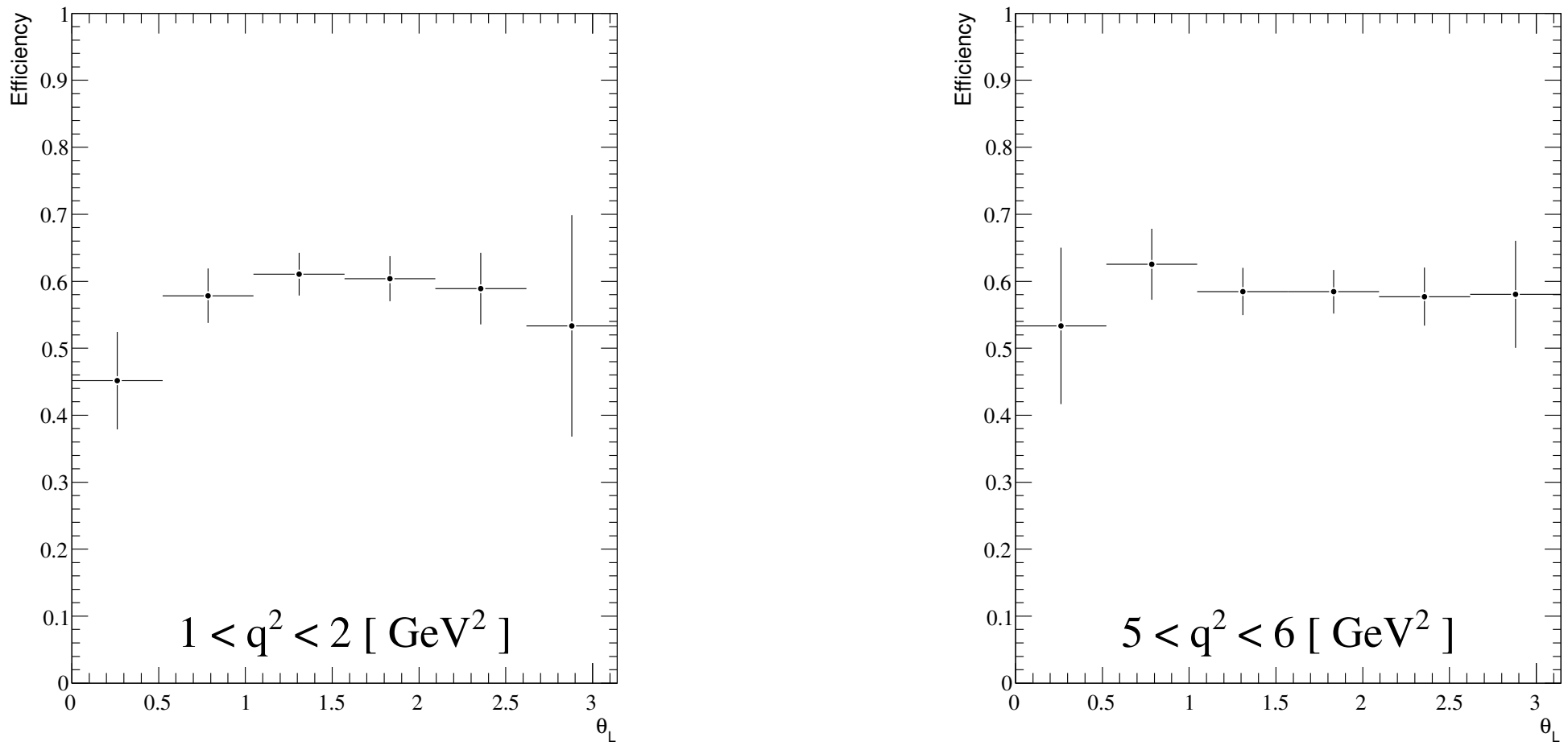


(a) (b)

Figure 22: The efficiency of the multivariate selection as a function of $\theta_{\mathrm{L}}$ for events with 1 GeV$^2$ $<q^2<$2 GeV$^2$ (a) and 5 GeV$^2$ $<q^2<$6 GeV$^2$ (b).

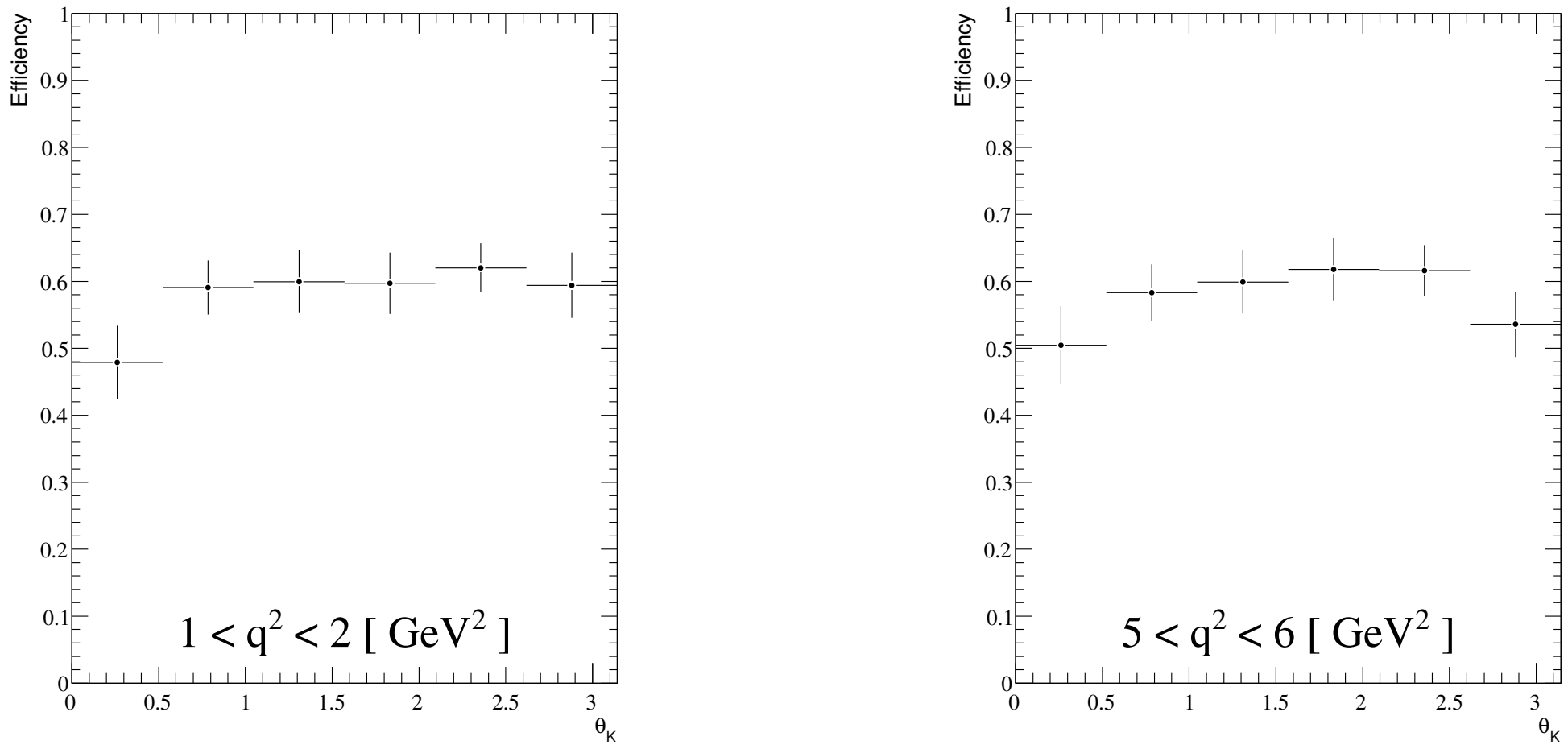


(a) (b)

Figure 23: The efficiency of the multivariate selection as a function of $\theta_{\mathrm{K}}$ for events with 1 GeV$^2$ $<q^2<$2 GeV$^2$ (a) and 5 GeV$^2$ $<q^2<$6 GeV$^2$ (b).

e.g. the $\mathrm{B_d}$ mass, will form separate selection criteria. Once the data are acquired, it will be necessary to understand which variables to use in the discriminant given the actual detector resolution. In the analysis described in Ref. [14], this was achieved by first forming a discriminant from a complete set of reconstructible quantities and then ordering these quantities according to their contribution to the discriminant. In order to remove the scale associated with each of the reconstructed quantities a transformation was applied that gave each quantity a Gaussian distribution with mean zero and unit

width. This also allowed the extent to which the variables were linearly correlated to be examined[2]. The performance of several candidate discriminants, formed selecting only those variables which made a contribution larger than some cut-off, was then determined, and the discriminant which gave the best performance selected. With the first LHCb data a trade-off will be required between achieving the best discrimination power and introducing a large number of variables which each needs to be understood.

The training and then performance evaluation of some discriminant, be it cut-based, Fisher, or some alternative, will require signal and background samples. It is envisaged that, for training, Monte Carlo simulated events will be used for the signal sample and a range of data sidebands for the background. Although differences between the simulation and the data will affect the former sample and make the discriminant sub-optimal, this will not affect the subsequent analysis, as the performance of the discriminant on background will be evaluated on data samples.

The choice of sideband for the background evaluation will be a source of systematic uncertainty, given that the background composition in the sideband will not entirely reflect the exact composition of the background in the signal region. This uncertainty will be evaluated by looking at the variation between different sideband samples and, assuming it is possible to understand the dominant background events, by using simulation to predict any variation in background composition. Simulation statistics are currently insufficient to estimate the potential size of this uncertainty.

# 6 Acceptance Correction

In order to extract the observables, the events passing the selection requirements must be corrected for the biases caused (primarily) by the detector geometry and reconstruction. Fig. 24 shows the acceptance model used to study the variation of the $A_{\mathrm{FB}}$ precision with the uncertainty on the acceptance correction. The analysis described in Section 7.3 was used to extract $A_{\mathrm{FB}}$. In order to simulate the acceptance variation in a given $2\,\mathrm{fb}^{-1}$ dataset, at various values of $\theta_{\mathrm{L}}$ an acceptance value was generated according to a Gaussian distribution. The generated acceptance values were then fitted with a parameterised curve. While the original acceptance distribution was applied to the simulated data, the fluctuated distribution was used to correct for the acceptance. The value of $A_{\mathrm{FB}}$ was then extracted from the corrected angular distribution. A large number of such fluctuated experiments indicated the average variation in $A_{\mathrm{FB}}$, given a certain Gaussian width used to generate the fluctuations. The process was repeated with a range of such widths. The resulting effect on the $A_{\mathrm{FB}}$ precision from a $2\,\mathrm{fb}^{-1}$ dataset is shown as a function of this width in Fig. 25. This assumes $A_{\mathrm{FB}}$=0.10. An acceptance determined at the $\sim$10% level has only a 5% impact on the measurement of $A_{\mathrm{FB}}$. An acceptance function which varies with both $\theta_{\mathrm{L}}$ and $\theta_{\mathrm{K}}$ has also been studied and results in a similar effect.

[2]A Fisher discriminant is optimal only in the case of linearly correlated Gaussian quantities. The level of non-linear correlations observed between different variables is discussed in Ref. [14].

At present, two methods are envisaged to determine the acceptance correction, both of which are under investigation:

- Use the simulation, tuned to reproduce the detector performance, to derive the acceptance correction;
- Reproduce the kinematics of the signal events with a control channel such as $B_d \to J/\psi K^{*0}$, triggered, stripped and selected with the same requirements as the signal, and derive a data-based acceptance correction.

These possibilities are discussed in the sub-sections below. In the case of the data-derived correction, several control channels, and/or some combination with simulated data, may be required in order to extract an overall correction. Assuming the feasibility of the data-derived method can be demonstrated, it is envisaged that both of the above methods will be employed to understand the acceptance correction and the differences between them used to understand systematic effects.

No forward-backward asymmetry is expected in the $B^+ \to K^+\mu^+\mu^-$ channel. While the lower branching ratio will give only around half as many events as the signal channel,

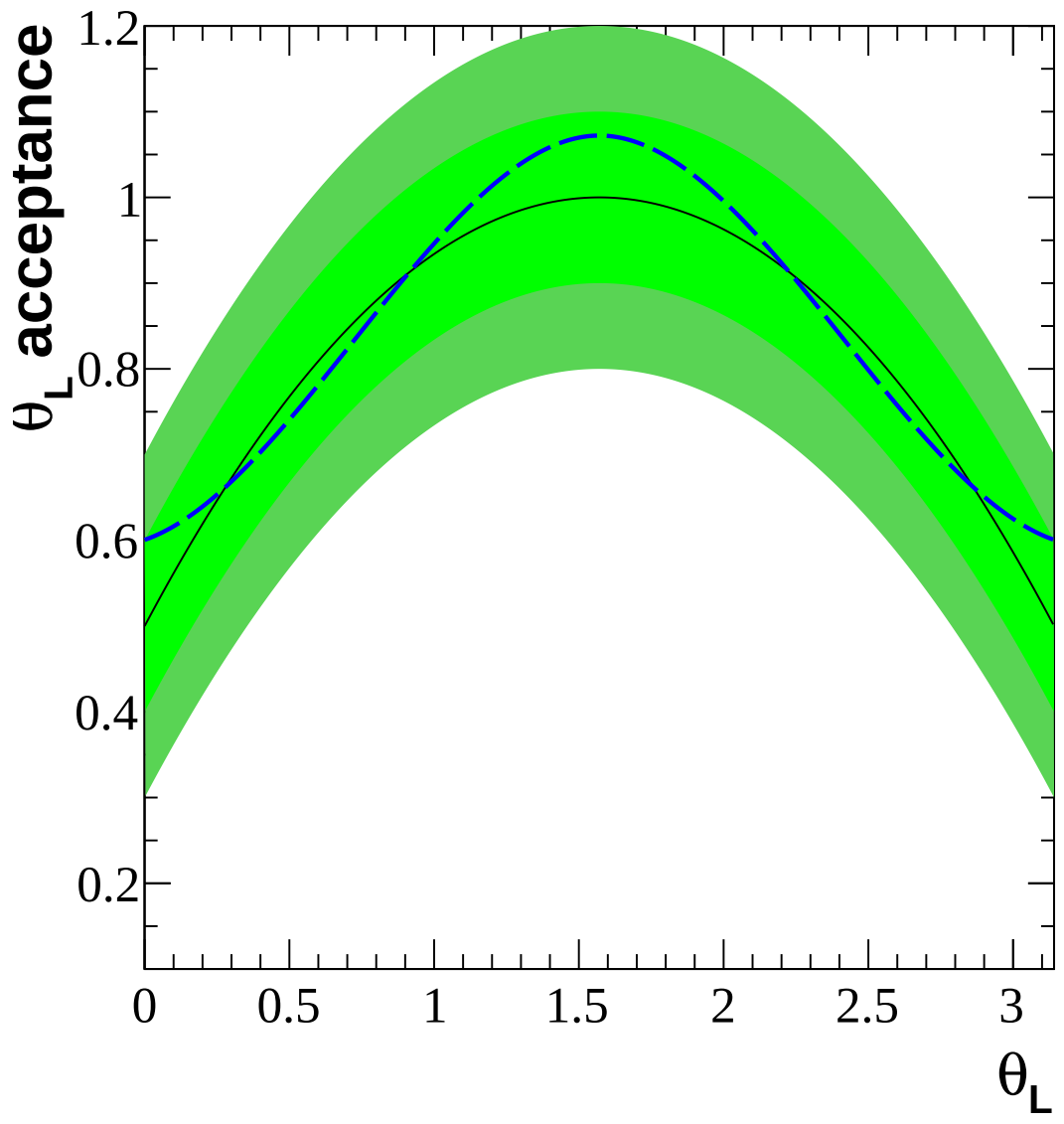


Figure 24: The acceptance model used to study the effect of the acceptance correction on the precision with which $A_{\rm FB}$ can be determined (solid line). The light (dark) green bands show the 1 (2)$\sigma$ Gaussian error bands within which the acceptance was fluctuated. In this example $\sigma = 0.1$. The dashed line shows an acceptance model derived from the fluctuated data that was then used in a single simulation.

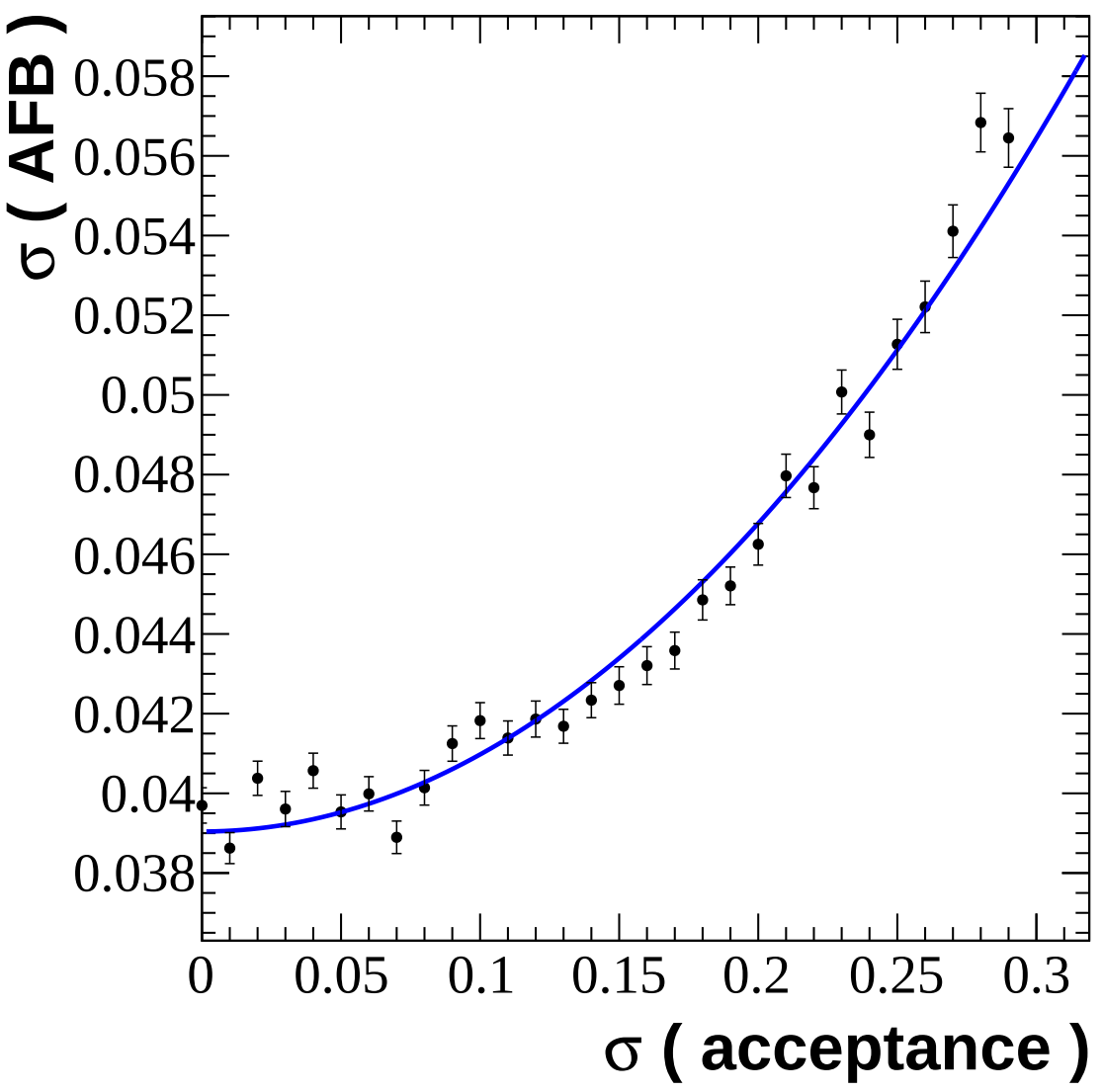


Figure 25: The variation in the $A_{\rm FB}$ precision from a $2\,{\rm fb}^{-1}$ dataset as a function of the uncertainty on the acceptance correction. A $\pm 0.1$ uncertainty on the acceptance gives rise to a 5% effect in the $A_{\rm FB}$ precision. A value of $A_{\rm FB}$=0.10 is assumed.

these data will allow an overall cross-check of the acceptance corrections derived from the above methods to be performed [24]. After making the relevant corrections, no forward-backward asymmetry should be observed. A selection for this channel that is as close as possible to the signal selection will be required.

## 6.1 Simulation Derived Acceptance Correction

A simulation derived acceptance correction will require the detector performance to be accurately reproduced. Several ideas exist for checking the performance of critical elements. It will be critical to ensure that the simulation reproduces the $B_d$ momentum spectrum determined from the data, as this will affect the momentum distribution of the decay products.

It is envisaged that the muon reconstruction can be calibrated using $J/\psi \rightarrow \mu^+\mu^-$ decays as source of genuine muons. By requiring one muon is detected in the muon chambers and a minimum ionising particle in the calorimeter for the second muon, the efficiency of finding the second muon can be mapped out as a function of momentum, position in the chambers etc. In order to remove the effect of any trigger bias, the muon that is identified in the muon chambers will also be required to have triggered the event. The study of $\mu^+$ separately from $\mu^-$ will also allow the extent of any residual detection asymmetry to be quantified. The single muon trigger line described above should give access to a very large sample of prompt $J/\psi$ events. The extent to which it will be possible to investigate the efficiency of muons with the relatively low momenta of interest to the present analysis is yet to be determined. Lambda decays will similarly provide an unambiguous source of pions for misidentification studies.

Calibration of the RICH detector will use $D^*$ decays to provide an unambiguous source of kaons and pions. It should therefore again be possible to compare the performance from real data with that from the simulation. Similarly, the tracking performance will be derived from both partially reconstructed, but kinematically over-constrained, decay modes and by comparing different parts of the tracking system against one another.

The fraction of generated signal events that survive the reconstruction and selection procedure as a function of the relevant variables e.g. $\theta_L$, $\theta_K$, $\phi$ and $q^2$, will give the signal acceptance function that can be used to weight the selected signal events. It is envisaged that systematic uncertainties on the acceptance correction will be determined by weighting the simulated events according to the uncertainties on the measured detector performance, and then recomputing the acceptance correction in order to determine the size of any change. For example, the vertex resolution will be determined from the data with some precision. Varying the resolution used in the simulation by that precision will give an estimate of the uncertainty on the acceptance correction coming from incorrect assumptions about vertex selection criteria.

## 6.2 Control Channel Derived Acceptance Correction

The use of $B_d \to J/\psi K^{*0}$ events to derive the acceptance correction is inhibited by the fact that the $q^2$ distribution of signal events and of $B_d \to J/\psi K^{*0}$ events are very different. The acceptance effects described in Section 2 are strong functions of $q^2$ and the largest effects occur when $q^2 \ll m_{J/\psi}{}^2$. The acceptance correction cannot therefore be simply derived from $B_d \to J/\psi K^{*0}$ events selected with the same offline, trigger and stripping requirements as the signal. However, the possibility of factorising the efficiencies in terms of the selected particles e.g. the $K^{*0}$ and each of the muons, is under investigation. Aspects of the offline/trigger/stripping selections which rely on more than one particle can clearly break this hypothesis of factorisation of the efficiencies. However, if the factorisation holds to a suitable extent, it would allow the efficiencies for $B_d \to K^{*0}\mu^+\mu^-$ signal events to be derived from $B_d \to J/\psi K^{*0}$ events.

In contrast to the $B_d \to K^{*0}\mu^+\mu^-$ signal events, the $p_T$'s of the two muons are strongly correlated in $B_d \to J/\psi K^{*0}$ events (Fig. 26). If the selection efficiency depends on the $p_T$ of both muons, or equivalently, on both $\theta_{\mu_1}$ and $\theta_{\mu_2}$, then the overall selection efficiency will not then factorise into a component from $\mu_1$ and a component from $\mu_2$.

If the efficiencies factorise then the number of selected $B_d \to J/\psi K^{*0}$ events with any $\mu^-$ $p_T$ can be used to compute the selection efficiency of a $\mu^+$ with, for example, $p_T$=1 GeV. For an event with these kinematics it will be necessary to account for the efficiency of generating a $\mu^-$ with some $p_T$ and a $\mu^+$ with $p_T$=1 GeV, as, while the selection efficiency may factorise, the distribution of the number of generated events as a function of the $p_T$'s does not: as demonstrated in Fig. 26, this distribution depends on both the $\mu^+$ and $\mu^-$ $p_T$'s. Despite the difference in the $q^2$ spectrum, the selection efficiency determined from $B_d \to J/\psi K^{*0}$ events could then be used to determine the efficiency in the kinematic region of interest in $B_d \to K^{*0}\mu^+\mu^-$ signal events.

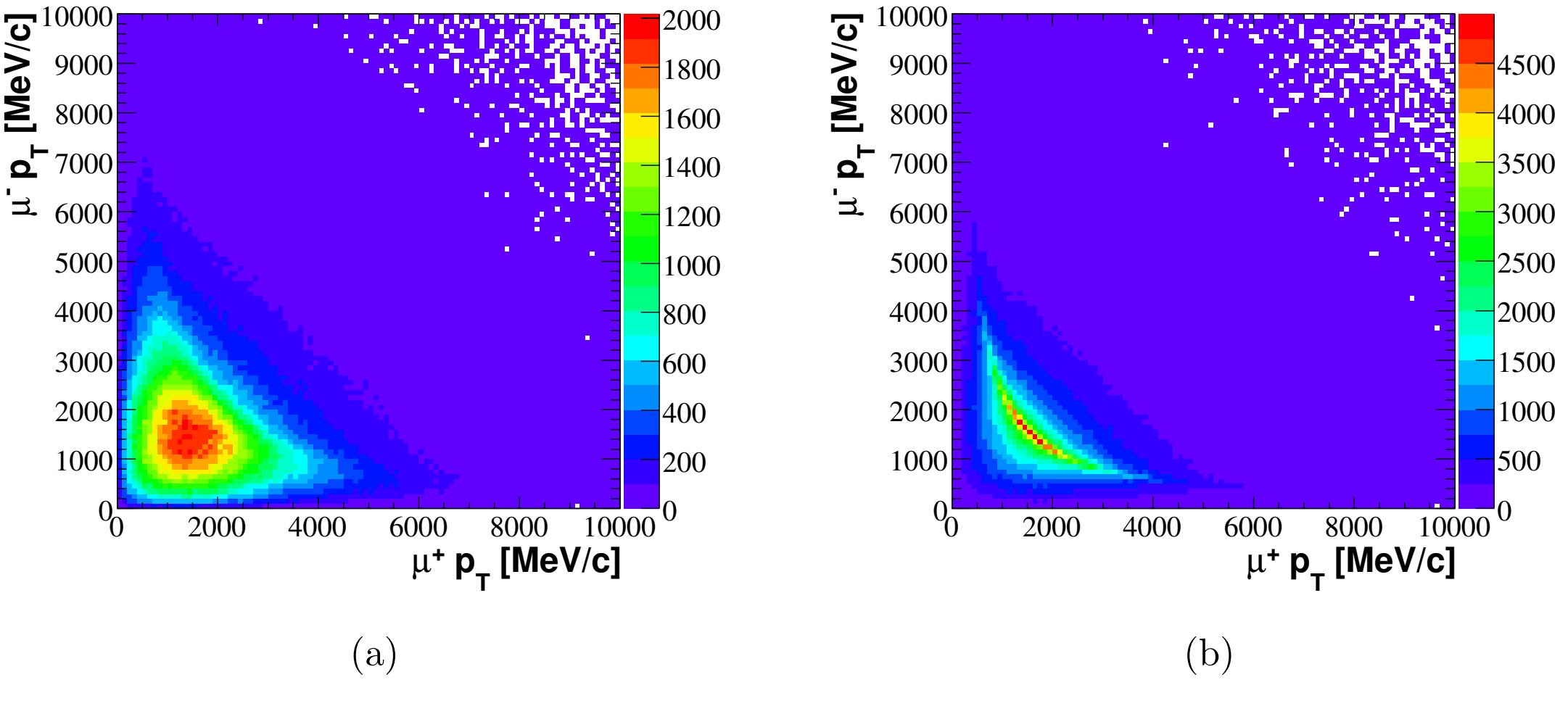


Figure 26: The $p_T$ distribution of the two muons for $B_d \to K^{*0}\mu^+\mu^-$ events (a) and $B_d \to J/\psi K^{*0}$ events (b).

To illustrate this, the single particle acceptances, $\epsilon_{\mathrm{X}}$, are assumed to be just a function of the relevant particle momenta, $p_{\mathrm{X}}$. The number of $\mathrm{B_d} \to \mathrm{J}/\psi\mathrm{K}^{*0}$ events reconstructed and selected, $N_{sel}$, can be written as a function of the generated distribution, $f$, and $p_{\mathrm{X}}$ as shown in Eqs 6–9.

$$N_{sel}(p_{\mathrm{K}^{*0}}) = \int\int \epsilon(p_{\mathrm{K}^{*0}}, p_{\mu_1}, p_{\mu_2}) f(p_{\mathrm{K}^{*0}}, p_{\mu_1}, p_{\mu_2}) dp_{\mu_1} dp_{\mu_2}, \quad (6)$$

$$\approx \int\int \epsilon_{\mathrm{K}^{*0}}(p_{\mathrm{K}^{*0}}) \epsilon_{\mu}(p_{\mu_1}) \epsilon_{\mu}(p_{\mu_2}) f(p_{\mathrm{K}^{*0}}, p_{\mu_1}, p_{\mu_2}) dp_{\mu_1} dp_{\mu_2}, \quad (7)$$

$$N_{sel}(p_{\mu_1}) \approx \int\int \epsilon_{\mathrm{K}^{*0}}(p_{\mathrm{K}^{*0}}) \epsilon_{\mu}(p_{\mu_1}) \epsilon_{\mu}(p_{\mu_2}) f(p_{\mathrm{K}^{*0}}, p_{\mu_1}, p_{\mu_2}) dp_{\mathrm{K}^{*0}} dp_{\mu_2}, \quad (8)$$

$$N_{sel}(p_{\mu_2}) \approx \int\int \epsilon_{\mathrm{K}^{*0}}(p_{\mathrm{K}^{*0}}) \epsilon_{\mu}(p_{\mu_1}) \epsilon_{\mu}(p_{\mu_2}) f(p_{\mathrm{K}^{*0}}, p_{\mu_1}, p_{\mu_2}) dp_{\mathrm{K}^{*0}} dp_{\mu_1}. \quad (9)$$

A simultaneous fit to Eqs 7–9 will allow the single particle efficiency functions, $\epsilon_{\mathrm{K}^{*0}}(p_{\mathrm{K}^{*0}})$ and $\epsilon_{\mu}(p_{\mu})$, to be determined as a function of the momenta, $p_{\mathrm{K}^{*0}}$, and, $p_{\mu}$, having accounted for the $\mathrm{B_d} \to \mathrm{J}/\psi\mathrm{K}^{*0}$ kinematics. To correct for the $\mathrm{B_d} \to \mathrm{K}^{*0}\mu^+\mu^-$ acceptance, an event-by-event weight could then be determined for signal events from the product of $\epsilon_{\mathrm{K}^{*0}}$, $\epsilon_{\mu_1}$ and $\epsilon_{\mu_2}$, given the $p_{\mathrm{K}^{*0}}$, $p_{\mu_1}$ and $p_{\mu_2}$ of a signal event.

In reality, these acceptances will also be a function of the polar angle of the particles, $\theta_{\mathrm{X}}$, and a fit as a function of both $p_{\mathrm{X}}$ and $\theta_{\mathrm{X}}$ will be required for each particle. Taking simulated $\mathrm{B_d} \to \mathrm{J}/\psi\mathrm{K}^{*0}$ events, and applying single particle efficiency functions which vary with both $p$ and $\theta$ to each particle, it has been shown that the efficiency functions can be recovered using an iterative procedure. Fig. 27 (a) and (b) show the input efficiency as a function of momentum (polar angle), in a single bin of polar angle (momentum). The data points indicate the efficiencies recovered by the iterative procedure. The error bars on these points reflect the statistical uncertainty expected from $8 \times 10^4$ $\mathrm{B_d} \to \mathrm{J}/\psi\mathrm{K}^{*0}$ events. This is $\sim 25\%$ of the number of such events expected from $2\,\mathrm{fb}^{-1}$ of data.

Potential sources of correlation between the particles that would break the factorisation include selection cuts depending on the vertexing. If the vertex were to be determined from just the two muon tracks, then the probability of passing vertex dependent cuts would depend on the $p_{\mathrm{T}}$ of both of the muons. For example, a cut on the vertex position will depend on the vertex resolution which will be more precise when both $p_{\mathrm{T}}(\mu_1)$ and $p_{\mathrm{T}}(\mu_2)$ are large. The inclusion of the $\mathrm{K}^{*0}$ in the vertex will reduce the effect but the extent to which the residual correlation will spoil the factorisation of the muon efficiencies is yet to be determined.

By considering the efficiency as a function of the above quantities, and performing an event-by-event weighting using the selected signal events, it will be possible to reduce the dependence on the decay model used. Both the correlations and potential differences between the signal and control samples will be sources of systematic uncertainty for the acceptance correction.

Although driven by the $\mathrm{B_d} \to \mathrm{J}/\psi\mathrm{K}^{*0}$ events that will be selected from the data, the

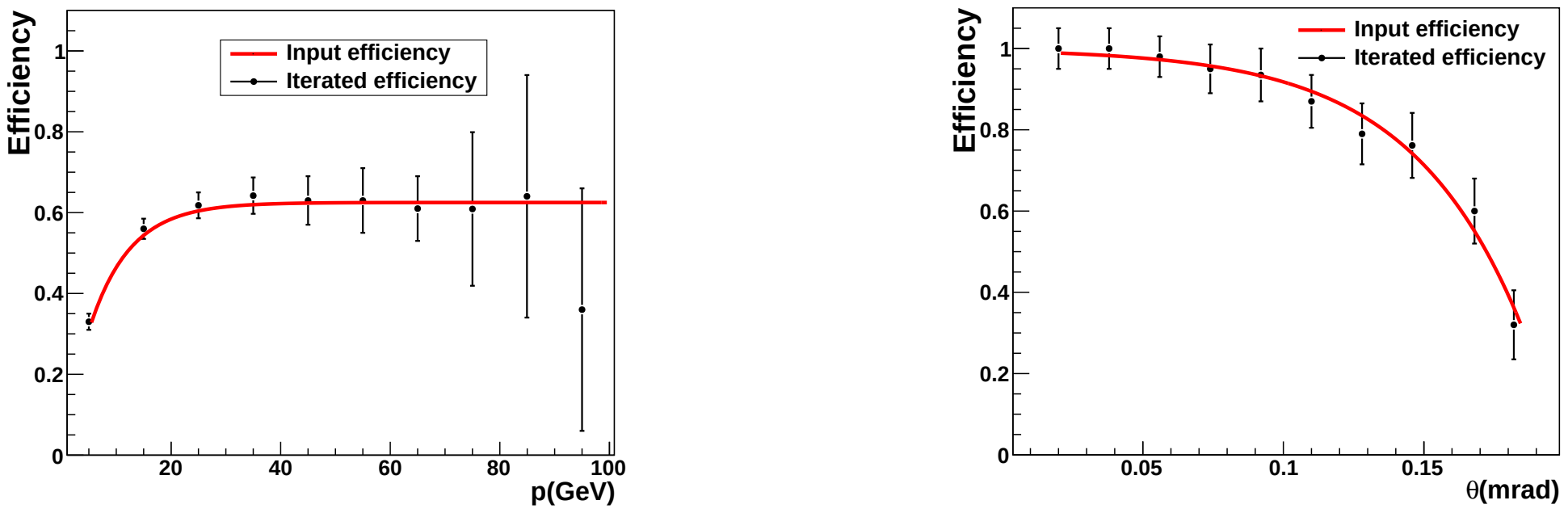


(a) (b)

Figure 27: The efficiency as a function of momentum (a) and polar angle (b) recovered from $B_d \to J/\psi K^{*0}$ events using an iterative procedure (points). The line indicates the input efficiency applied. The efficiency as a function of $p$ is shown for events with $110 < \theta < 128\,\mathrm{mrad}$ and the efficiency as a function of $\theta$ for events with $70 < p < 80\,\mathrm{GeV}$.

above procedure will also rely on an accurate model for the generated distribution, $f$. The $B_d \to J/\psi K^{*0}$ decay has been extensively studied at the B-factories and the decay model used in the LHCb simulation is derived from their data. The generated distribution for this decay is therefore thought to be reliable. In addition, as it is just the shape of the acceptance function that is important, the overall normalisation can be ignored.

However, as with a simulation derived correction, it will be necessary to tune or reweight the simulated $B_d \to J/\psi K^{*0}$ events to have the observed $B_d$ momentum spectrum. This can be achieved by recomputing $f$ for each $B_d \to J/\psi K^{*0}$ candidate observed, given the $B_d$ momentum of that candidate. The resulting single particle efficiency functions will then take into account the $B_d$ momentum spectrum observed in the data. A similar procedure will allow the effect of the complete removal of regions of the phase-space by selection requirements to be corrected for. While this event-by-event calculation of $f$ will be computationally intensive, the relevant calculation is independent of the fitted parameters and need not be repeated for each fit iteration.

Cross-checks of the acceptance correction will include comparing the shape of the efficiency function derived from the multiplication of K, $\mu^+$ and $\mu^-$ efficiencies with selected $B^+ \to J/\psi K^+$ and $B^+ \to K^+\mu^+\mu^-$ events. It should be possible to select these channels with similar selections to that of the signal.

Assuming the selection efficiency is identical to that of the signal, $\sim 3 \times 10^5$ $B_d \to J/\psi K^{*0}$ events should be triggered and selected in a $2\,\mathrm{fb}^{-1}$ dataset. The uncertainty in the acceptance correction that will result from such a sample remains to be investigated. The precision will depend both on the $B_d \to J/\psi K^{*0}$ statistics acquired and on the extent to which the factorisation of efficiencies breaks down.

If the factorisation of the efficiencies in $B_d \to J/\psi K^{*0}$ events fails owing to correlations between the two muons, then the selection efficiency could instead be determined from control channels that involve pions rather than muons. The effect of the muon identification could then be separately determined from $J/\psi \to \mu^+\mu^-$ decays (see Section 6.1

above). The relevant decay modes would again have to have well known angular distributions e.g. $D^0 \to K\pi$, $K_S \to \pi\pi$. Such decays would clearly have to be triggered through a separate line to the signal decays, and the effect of the trigger on the acceptance correction would therefore have to be determined separately. This can be achieved by using the ratio between events triggered independently of the signal and events triggered on the signal to weight e.g. the $\theta_L$ distribution of signal candidates. This method of removing the acceptance effect of the trigger is described elsewhere [25]. The trigger, muon identification and selection efficiency corrections could then be combined to form an overall acceptance correction. Cross-checks could again be performed with control channels such as $B^+ \to K^+\pi^+\pi^-$.

# 7 Extracting the Physics Parameters

Several methods of extracting the physics parameters from the angular distribution of the signal are available, with increasing levels of complexity. It is envisaged that the counting analysis detailed below will provide a robust first test of the analysis, and more refined methods, giving access to new observables, will be adopted as the data and the detector are better understood. Throughout this section the precisions cited are purely statistical. Additional uncertainties will arise from the acceptance correction and the limited knowledge of the relevant background distributions. As discussed in Section 2.1, if correctly accounted for, the acceptance correction will not modify the zero-crossing point, but will change the precision with which it can be determined. Understanding the background events with asymmetric $\theta_L$ distributions (see Section 2.2) will be particularly important, as such events can shift the value of $A_{FB}$ observed.

It is envisaged that the distribution of e.g $\theta_L$ and $q^2$ of background events will be obtained from the B mass sidebands. The subtraction of suitably normalised distributions or a simultaneous fit will allow the contribution of background events to be accounted for.

## 7.1 Binned Counting Analysis

A binned counting analysis groups together events in bins of $\theta_L$ and $q^2$ and, after correcting for the background and the acceptance, the number of forward- and backward-events are counted and used to form $A_{FB}$. Simulation studies indicate that, with $2\,\mathrm{fb}^{-1}$ of data, $A_{FB}$ can be measured with the precision shown in Fig. 28 [19]. It is estimated that a measurement of $A_{FB}$ with a comparable statistical uncertainty to that of present B-factory data will require $\sim 0.3\,\mathrm{fb}^{-1}$ of data. With $2\,\mathrm{fb}^{-1}$ of data, a linear fit allows the zero-crossing point to be determined with a statistical uncertainty of $\pm 0.5\mathrm{GeV}^2$.

While only two bins in $\theta_L$, one forward and one backward, are required to establish $A_{FB}$, an accurate acceptance correction will require a larger number of bins to model the shape in $\theta_L$. The background in each of the relevant bins will be estimated from $m_B$ mass sidebands. The level and distribution of this background is seen to have a large effect on the precision [19]. As this is the simplest way of extracting $A_{FB}$, it is envisaged that this

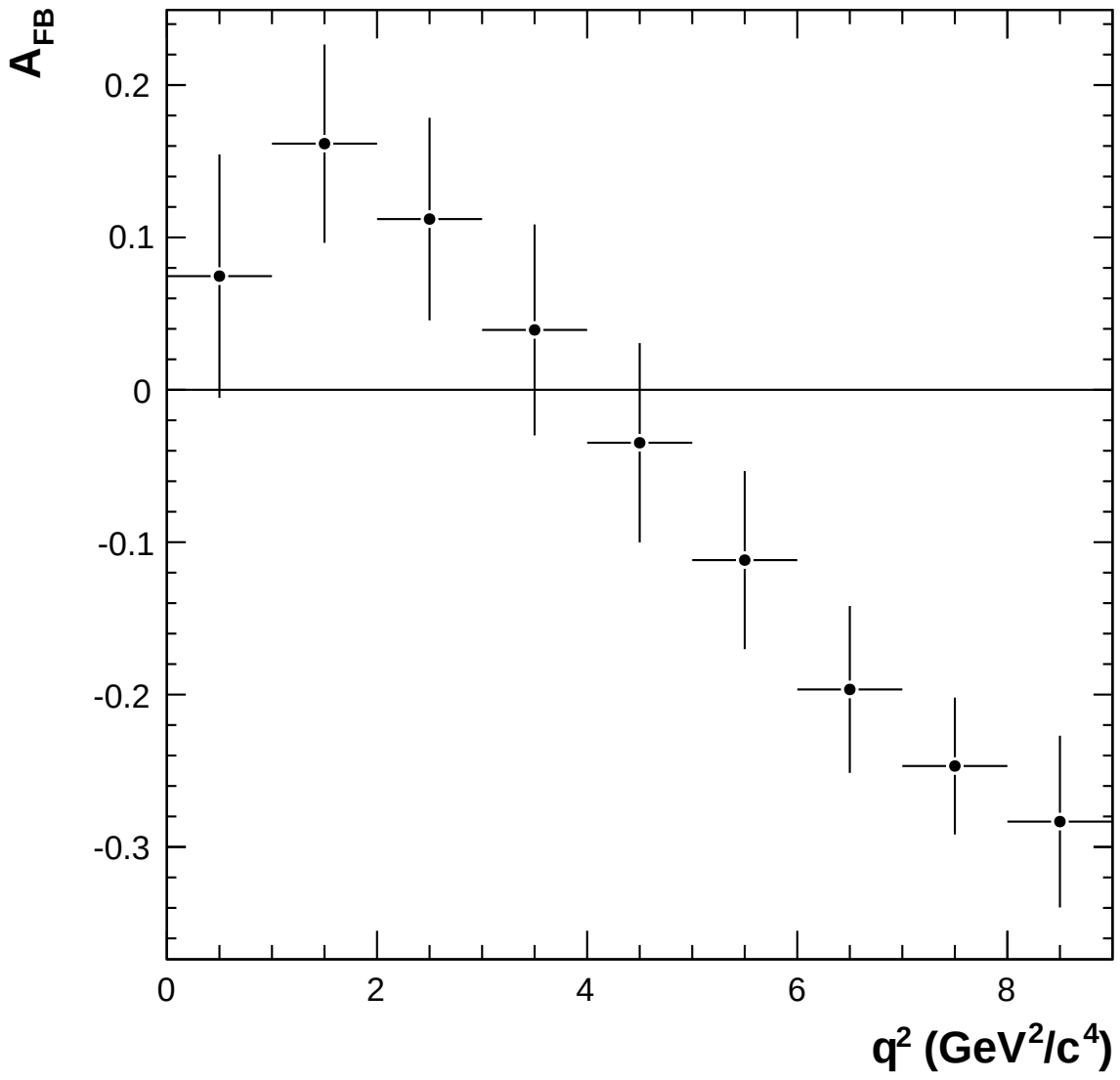


Figure 28: Precision with which $A_{\mathrm{FB}}$ can be determined from $2\,\mathrm{fb}^{-1}$ of data. Central values are taken from the Standard Model.

method will be applied to the first data. However, this does not give sensitivity to other angular observables.

## 7.2 Unbinned Counting Analysis

By binning together events in $\theta_{\mathrm{L}}$ which have varying sensitivities, some information is lost. An unbinned analysis allows the full exploitation of the available information. Unbinned fits of generic polynomials to the forward- and backward-going distributions as a function of $q^2$, will allow $A_{\mathrm{FB}}$ to be constructed "analytically" [26]. The zero-point will then be determined from the resulting polynomials. The order of the polynomials used will be a source of systematic bias. An alternative unbinned estimate can be made using a non-parametric approach which yields similar sensitivities [27].

## 7.3 Fit to Angular Projections

The above analyses do not exploit the information on the shape of the angular distribution(s). This information will be accessed by fitting the partial rate given in Eq. 2 in bins of $q^2$. This will allow the observables $A_{\mathrm{FB}}$ and $F_{\mathrm{L}}$ to be simultaneously determined as functions of $q^2$ [10][3]. In each bin of $q^2$, the background rate and the shape in $\theta_{\mathrm{L}}$ will be estimated from $m_{\mathrm{B}}$ mass sidebands. Simultaneously fitting the $\theta_{\mathrm{K}}$ distribution (Eq. 3), which also contains the observable $F_{\mathrm{L}}$, will improve the precision on $F_{\mathrm{L}}$ and hence on $A_{\mathrm{FB}}$.

[3] A fit that is also unbinned in $q^2$ would require $A_{\mathrm{FB}}$ and $F_{\mathrm{L}}$ to be described as functions of $q^2$ with some arbitrary polynomials, the parameters of which would need to be extracted from the fit. The form of these polynomials would be highly dependent on the underlying physics model.

Similarly, including the $\phi$ angular distribution will give a further handle on $F_{\rm L}$ (Eq. 4) and will introduce the additional observables $A_{\rm T}^2$ and $A_{\rm Im}$. As above, the precision is heavily dependent on the level and the distribution of the background and this will need to be studied in the data.

Fitting Eqs 2– 4, the precision on $A_{\rm FB}$ can be improved by a factor $\sim 2$ over the above counting methods. The precision with which the additional observables are determined is given in Ref. [11].

## 7.4 Likelihood Subtraction

In order to get the maximum information from the signal events, the correlation between $\theta_{\rm L}$ and $\theta_{\rm K}$ in a given event must be exploited. This requires a fit to the full two-dimensional $\theta_{\rm L}$, $\theta_{\rm K}$ distribution rather than to the projected distributions as above. As before, making such a fit unbinned avoids the sensitivity loss from grouping together events with different sensitivities.

It is eventually intended to perform an effective background subtraction in an unbinned fit using a likelihood subtraction method. This method is described in detail in Ref. [24], only a brief summary is provided here. The likelihood is calculated for background events in the sidebands, compared to the signal hypothesis, weighted appropriately, and then subtracted from the likelihood calculated in the signal region, where there is a mixture of background and genuine signal candidates. The fit then minimises,

$$-2\left(\sum_{i=1}^{N}\ln\left(P(\vec{x}_i,\vec{a})\right)-\frac{\widetilde{n}_B}{N_B}\sum_{j=1}^{N_B}\ln\left(P(\vec{x}_j,\vec{a})\right)\right), \qquad (10)$$

where $P(\vec{x}_j,\vec{a})$ is the probability to get an event with $\vec{x}=(\theta_{\rm L},\theta_{\rm K})$, given an underlying description of the signal with parameters $\vec{a}=(A_{\rm FB},F_L)$; $N$ is the total number of events in the signal region; $N_B$ is the number of events in the sidebands; and $\widetilde{n}_B$ is an estimator for the number of background events under the signal (extracted from a fit to the $m_{\rm B}$ distribution). The subtraction will use the sidebands to estimate the impact on the likelihood of having a mixture of signal and background under the peak rather than just purely signal. After the subtraction the subtracted-likelihood is an estimate for the likelihood associated with just the signal candidates. This is valid only if the shape of the background in the sidebands is the same as the shape of the background under the signal region.

This method has been implemented and an unbinned fit in $\theta_{\rm L}$ and $\theta_{\rm K}$ in bins of $q^2$ performed [24]. With $2\,{\rm fb}^{-1}$ of data the precision will not be greatly improved over the above fit of the $\theta_{\rm L}$ and $\theta_{\rm K}$ projected distributions. However, once enough statistics are accumulated, this technique could be used to perform an unbinned fit over the full three angles, providing access to new observables and leading to improvements in the precision on $A_{\rm FB}$, $F_{\rm L}$, and $A_{\rm T}^2$ [11]. In the case of $A_{\rm FB}$, a full angular fit could improve the precision with respect to the projected distributions method by $\sim 30\%$.

# 8 Conclusions

The decay $B_d \to K^{*0}\mu^+\mu^-$ is a promising channel in which to search for new physics at LHCb. The angular asymmetry in the $\theta_L$ angle, $A_{FB}$, is well calculable theoretically and will allow discrimination between different new physics models. Additional observables can be accessed using different angles. In some cases, these observables have very different new physics sensitivities.

The key challenges for this analysis are expected to be understanding the biases on angular observables induced by the geometric acceptance of the detector, reconstruction and signal selection requirements, and understanding the background.

The final state muons will provide a robust and discriminating signature that will allow the signal events to be triggered. The investigation of single muon, di-muon and muon+track variants for this online selection is on-going. Further event filtering or "stripping" is envisaged and present studies indicate the rate requirements can be fulfilled without any loss of signal efficiency.

An efficient offline selection can be formed using multivariate techniques which will give signal yields comparable to those from the B-factories with a few weeks of data-taking at nominal luminosity. A measurement of $A_{FB}$ with a comparable statistical uncertainty to that of present B-factory data will require $\sim 0.3\,\text{fb}^{-1}$ of data. Both data-derived and simulation-based acceptance corrections for effects induced by the above selections are under investigation. The uncertainties arising from these corrections and also from the lack of knowledge of the background composition are expected to give the dominant systematic effects.

Various methods of extracting the physics parameters have been investigated. It is envisaged that simple counting analyses will be employed initially, and more sophisticated methods adopted as the understanding of the analysis and the size of the dataset increase. Eventually, a full angular analysis will be used to yield the best precision and extract the complete information that is available from $B_d \to K^{*0}\mu^+\mu^-$ decays. Using the simple counting method, with $2\,\text{fb}^{-1}$ of data, the zero-crossing point of $A_{FB}$ can be determined with a statistical uncertainty of $\pm 0.5\,\text{GeV}^2$. A factor $\sim 2$ improvement in precision can be achieved with more complex analyses.

# A Cut-based Analysis Selection Criteria

| | |
|---|---|
| $B_d$: | Vertex $\chi^2 < 16$<br>$\cos\theta > 0.99995$<br>Flight significance $> 30$<br>sIPS $< 3$ |
| $K^{*0}$: | Flight significance $> 0.5$<br>Vertex $\chi^2 < 25$ |
| K: | $p > 1000\,\mathrm{MeV}$<br>$p_T > 200\,\mathrm{MeV}$<br>sIPS $> 4$<br>Track $\chi^2/\mathrm{dof} < 2$<br>DLL(K$-\pi$) $> -1$<br>DLL(K$-$p) $> -1$ |
| $\pi$: | $p > 1000\,\mathrm{MeV}$<br>$p_T > 200\,\mathrm{MeV}$<br>sIPS $> 4$<br>Track $\chi^2/\mathrm{dof} < 2$<br>DLL(K$-\pi$) $< 25$ |
| $\mu$: | $p > 3000\,\mathrm{MeV}$<br>sIPS $> 3$<br>Track $\chi^2/\mathrm{dof} < 2$<br>DLL($\mu - \pi$) $> -5$ |
| $\mu\mu$: | Vertex $\chi^2 < 25$ |

Table 4: Selection cuts used in the cut-based analysis. Taken from Ref. [13]. The meaning of the variables are described in detail in this reference.

# B Multivariate Analysis Fisher Coefficients

| Particle | Variable | Fisher coefficient | Ranking |
|---|---|---|---|
| $B_d$ | Vertex $\chi^2$ | $-1.133 \times 10^{-1}$ | $2.118 \times 10^{-1}$ |
| $B_d$ | $p_T$ | $+9.177 \times 10^{-2}$ | $8.915 \times 10^{-2}$ |
| $K^{*0}$ | $\sqrt{\text{flight distance } \chi^2}$ | $+7.173 \times 10^{-2}$ | $2.610 \times 10^{-1}$ |
| K | DLL K$-\pi$ | $+7.077 \times 10^{-2}$ | $2.117 \times 10^{-1}$ |
| $\mu\mu$ | $p_T$ | $-6.199 \times 10^{-2}$ | $8.280 \times 10^{-3}$ |
| $\pi$ | DLL K$-\pi$ | $-4.838 \times 10^{-2}$ | $1.331 \times 10^{-1}$ |
| $B_d$ | sIPS | $-4.319 \times 10^{-2}$ | $5.777 \times 10^{-2}$ |
| $\pi$ | $\sqrt{\text{sIP}\chi^2}$ | $+4.298 \times 10^{-2}$ | $2.406 \times 10^{-1}$ |
| $\mu$ 1 | $\sqrt{\text{sIP}\chi^2}$ | $+4.091 \times 10^{-2}$ | $2.116 \times 10^{-1}$ |
| $B_d$ | $\sqrt{\text{flight distance } \chi^2}$ | $+3.731 \times 10^{-2}$ | $2.395 \times 10^{-1}$ |
| K | $\sqrt{\text{sIP}\chi^2}$ | $+3.723 \times 10^{-2}$ | $2.492 \times 10^{-1}$ |
| $\mu$ 2 | $\sqrt{\text{sIP}\chi^2}$ | $+3.634 \times 10^{-2}$ | $2.097 \times 10^{-1}$ |
| K | DLL K$-$p | $+3.633 \times 10^{-2}$ | $1.644 \times 10^{-1}$ |
| $\mu$ 2 | DLL K$-\pi$ | $-3.112 \times 10^{-2}$ | $5.872 \times 10^{-2}$ |
| $\mu\mu$ | Vertex $\chi^2$ | $-3.013 \times 10^{-2}$ | $4.558 \times 10^{-2}$ |
| $\pi$ | Track ghost probability | $-2.830 \times 10^{-2}$ | $9.635 \times 10^{-2}$ |
| K | Track ghost probability | $-2.827 \times 10^{-2}$ | $1.024 \times 10^{-1}$ |
| $\mu$ 1 | DLL K$-\pi$ | $-2.722 \times 10^{-2}$ | $5.515 \times 10^{-2}$ |
| $\mu\mu$ | $\sqrt{\text{sIP}\chi^2}$ | $-2.595 \times 10^{-2}$ | $1.916 \times 10^{-1}$ |
| $\mu$ 2 | DLL $\mu - \pi$ | $+2.223 \times 10^{-2}$ | $6.481 \times 10^{-2}$ |
| $\mu$ 1 | DLL $\mu - \pi$ | $+2.120 \times 10^{-2}$ | $6.496 \times 10^{-2}$ |
| $B_d$ | $\cos\theta$ | $+1.432 \times 10^{-2}$ | $2.279 \times 10^{-1}$ |
| $K^{*0}$ | Vertex $\chi^2$ | $+1.198 \times 10^{-2}$ | $5.885 \times 10^{-2}$ |
| $\mu$ 2 | Track ghost probability | $-8.160 \times 10^{-3}$ | $1.510 \times 10^{-2}$ |
| $\mu$ 1 | Track ghost probability | $-6.802 \times 10^{-3}$ | $1.420 \times 10^{-2}$ |
| $\mu$ 1 | Track $\chi^2$/dof | $-5.170 \times 10^{-3}$ | $1.413 \times 10^{-2}$ |
| $\mu$ 2 | Track $\chi^2$/dof | $-5.037 \times 10^{-3}$ | $1.381 \times 10^{-2}$ |
| K | Track $\chi^2$/dof | $+4.044 \times 10^{-3}$ | $3.127 \times 10^{-2}$ |
| $\pi$ | Track $\chi^2$/dof | $-3.529 \times 10^{-3}$ | $2.756 \times 10^{-2}$ |
| K | DLL $\mu - \pi$ | $+2.358 \times 10^{-3}$ | $5.501 \times 10^{-3}$ |
| $K^{*0}$ | $\sqrt{\text{sIP}\chi^2}$ | $+2.147 \times 10^{-3}$ | $2.523 \times 10^{-1}$ |
| $\pi$ | DLL $\mu - \pi$ | $+6.709 \times 10^{-4}$ | $4.589 \times 10^{-3}$ |
| $\mu\mu$ | $\sqrt{\text{flight distance } \chi^2}$ | $-2.578 \times 10^{-4}$ | $2.263 \times 10^{-1}$ |

Table 5: Fisher coefficient calculated for each variable used in the Fisher discriminant. Taken from Ref. [14]. The meaning of the variables and the ranking is described in detail in this reference. As all variables are normalised to the same scale, the absolute value of each coefficient gives an indication of the discriminating power of the variable.

# Chapter 7

# Analysis of $B_s^0 \to \phi\gamma$ and other radiative B decays

S. Barsuk, I. Belyaev, A. Bret, M. Calvo Gomez, O. Deschamps, V. Egorychev, A. Golutvin, R. Graciani Diaz, M. Knecht, A. Kozlinskiy, F. Legger, F. Machefert, I. Machikhiliyan, M-N. Minard, G. Pakhlova, A. Puig Navarro, G. Rospabe, D. Savrina, O. Schneider, K. Senderowska, L. Shchutska, V. Shevchenko, F. Soomro, R. Vazquez Gomez, K. Voronchev, M. Witek, Y. Xie and O. Yushchenko

**Abstract**

This note presents the roadmap towards the first measurements with radiative modes $B_d^0 \to K^*\gamma$, $B_s^0 \to \phi\gamma$ and $B^+ \to \phi K^+\gamma$ at LHCb. The basic steps toward the first measurements of the radiative decays are presented. We have concentrated on the probing of photon polarization in $B_s^0 \to \phi\gamma$ decay. Based on our Monte Carlo simulation for 2 fb$^{-1}$ we expect to reach the uncertainty in the measurement of the photon polarization parameter $\mathcal{A}^{\Delta}$, $\sigma_{\mathcal{A}^{\Delta}} \approx 0.22$. The capability of the LHCb experiment to measure this and other parameters has been discussed for the different luminosity sets. Several critical aspects and important prerequisites, like calorimeter calibration, implementation of the HLT, determination of proper time acceptance function and $\gamma/\pi^0$ separation at high-$E_{\mathrm{T}}$ have been addressed. The measurements which can be performed already with a data-set of an integrated luminosity of less than 500 pb$^{-1}$, such as the measurements of the ratio of branching fractions for $B_s^0 \to \phi\gamma$ and $B^0 \to K^{*0}\gamma$, and the measurement of the direct $\mathcal{CP}$-asymmetry for $B^0 \to K^{*0}\gamma$ decay, are also discussed.

# Contents

# Introduction

This note presents the roadmap towards the first measurements of radiative penguin decays of b-hadrons at LHCb using a data-set of an integrated luminosity of up to 2 $fb^{-1}$. Measurements which require significantly larger integrated luminosities, like, e.g., the measurement of the Cabibbo-suppressed decays $B^0 \to \rho^0\gamma$ and $B^0 \to \omega\gamma$, are not addressed. While the focus of this note is on the measurement of the photon polarisation in $B_s^0 \to \phi\gamma$ decays, one of the LHCb key measurements which requires the statistics of 2 $fb^{-1}$, the note also discusses "early" measurements which can be performed already with a data-set of an integrated luminosity of less than 500 $pb^{-1}$.

A list of interesting measurements with radiative decays of b hadrons includes:

1. the measurement of the photon polarisation and $\mathcal{CP}$-violation parameters $\mathcal{C}$ and $\mathcal{S}$ in the decay $B_s^0 \to \phi\gamma$ [1, 2, 3, 4],
2. the measurement of the photon polarisation in the decays of polarised beauty baryons $\Lambda_b \to \Lambda^0\gamma$ and $\Lambda_b \to (\Lambda^* \to pK^-)\,\gamma$ [5, 6, 7],
3. the measurement of the photon polarisation in the decay $B^+ \to \phi K^+\gamma$,
4. the measurement of direct $\mathcal{CP}$-violation in the decay $B^0 \to K^{*0}\gamma$,
5. the precise measurement of the ratio of branching fractions for $B_s^0 \to \phi\gamma$ and $B^0 \to K^{*0}\gamma$ decays.

Some of these items have been intensively studied in LHCb [1, 2, 3, 4, 5, 6, 7], while for other topics, e.g. for the decay $B^+ \to \phi K^+\gamma$, only the first studies of the event selection have been performed. Table 1 gives the expected nominal event yields for a sample corresponding to 2 $fb^{-1}$ of integrated luminosity and the estimates of the background to signal ratio ($\mathcal{B}_{b\bar{b}}/\mathcal{S}$) for the above mentioned radiative B decays. The measurement of the photon polarisation in the decay $B_s^0 \to \phi\gamma$ is discussed in Section 3. The measurement of $B^0 \to K^{*0}\gamma$ and the determination of branching fractions for $B_s^0 \to \phi\gamma$ and $B^+ \to \phi K^+\gamma$ are discussed in Section 4.2.

The most important ingredients and prerequisites which are necessary for the successful measurements of radiative decays and in particular the photon polarisation in $B_s^0 \to \phi\gamma$ decays include:

- High Level Trigger for events triggered by high energy photons [8];
- the determination of the proper time acceptance function and the resolution function from data;
- the calibration of the LHCb Electromagnetic Calorimeter [9];
- the separation of high energy photons from energetic "merged" $\pi^0$-mesons, misreconstructed as single photons.

Table 1: The expected yields, $\mathcal{Y}$, per 2 fb$^{-1}$ of integrated luminosity and background to signal ratios, $\mathcal{B}_{b\bar{b}}/\mathcal{S}$, for radiative decays of beauty hadrons. The quoted yields do not include the efficiency of the High Level Trigger, which is expected to be rather high, see Section 3.2. The backgrounds only include backgrounds from $b\bar{b}$-events. Upper limits are quoted at 90 % CL.

| Decay Mode | | $\mathcal{Y}$ | $\mathcal{B}_{b\bar{b}}/\mathcal{S}$ |
|---|---|---|---|
| $B^0 \to K^{*0}\gamma$ | [2] | $7 \times 10^4$ | $0.6 - 0.7$ |
| $B_s^0 \to \phi\gamma$ | [2] | $1.1 \times 10^4$ | $< 0.6 - 0.9$ |
| $\Lambda_b \to \Lambda^0\gamma$ | [5] | 750 | $< 42$ |
| $\Lambda_b \to \Lambda^0\,(1670)\,\gamma$ | [5] | $2.5 \times 10^3$ | $< 18$ |
| $\Lambda_b \to \Lambda^0\,(1690)\,\gamma$ | [6] | $2.5 \times 10^3$ | $< 18$ |
| $\Lambda_b \to \Lambda^0\,(1520)\,\gamma$ | [6] | $4.3 \times 10^3$ | $< 9$ |
| $B^+ \to \phi K^+\gamma$ | | $\sim 7 \times 10^3$ | $< 2 - 4$ |

Of course this list is not exhaustive and probably will evolve with time. Only important aspects specific for the radiative decays are listed here. The items of the general interest, like tracking, hadron identification and flavour tagging are discussed elsewhere [10, 11, 12, 13, 14, 15]

This document is organized as follows. In Sections 1 and 2 the current theoretical and experimental status is briefly described. Section 3 describes the measurement of the photon polarisation in $B_s^0 \to \phi\gamma$ decays. Expected sensitivities are given followed by the discussion of the main systematics. In Section 4 we discuss "early" measurements which on one hand lead already to interesting physics results and on the other hand allow checks of the detector, trigger and reconstruction performance. The important items for successful measurements are described in in detail in Sections 3.2 and 3.5, and Appendices C and D.

# 1 Radiative penguin decays

Radiative decays of b hadrons are an example of flavour changing neutral currents. They are of significant interest due to their sensitivity to new physics and experimental accessibility. The theoretical perspective is given in Section 1.1. The general form of the effective Hamiltonian is used to explain the new physics sensitivity and the dominant source of theoretical uncertainties. The polarisation of emitted photons in radiative decays is found to be particularly worthy of study. The experimental challenges in the study of these processes are described in Section 1.2. This section shows that particular empahsis should be placed on the study of the radiative decays of $B_s^0$ mesons into a $\mathcal{CP}$ eigenstate.

## 1.1 Theoretical overview

Radiative decays of b hadrons caused by $b \to s\gamma$ transition are an interesting example of flavour-changing neutral current processes. From the theoretical point of view the lowest contribution to the amplitude comes from the one-loop process, see Figure 1, and hence, as with any loop process in quantum field theory, the decay pattern may be sensitive to the structure of heavy degrees of freedom of the theory. Another important, while in a sense more technical, reason is that the weak, electromagnetic and perturbative strong parts of the physics behind radiative decays are well under theoretical control. It is also advantageous that one can formulate theoretical predictions for a variety of different observables, such as decay rates, $\mathcal{CP}$ and isospin asymmetries, angular distributions, and cross-check in this way the robustness of the Standard Model (SM) framework. All that places the radiative B decays in the position of "standard candles of flavour physics" [16].

The conventional starting point for the theoretical analysis of B-decays is the effective $\Delta B = 1$ Hamiltonian [17]

$$\mathcal{H}_{\text{eff}} = \frac{G_F}{\sqrt{2}} \sum_{p=u,c} \lambda_p^{(q)} \Big( \mathcal{C}_1(\mu)\mathcal{Q}_1^p(\mu) + \mathcal{C}_2(\mu)\mathcal{Q}_2^p(\mu) + \sum_{i=3}^{8} \mathcal{C}_i(\mu)\mathcal{Q}_i(\mu) \Big), \tag{1.1}$$

where Cabibbo-Kobayashi-Maskawa (CKM) factors are given by $\lambda_p^{(q)} = V_{pq}^{\dagger} V_{pb}$, and the unitarity relationship has been used. The Wilson coefficients $\mathcal{C}_i(\mu)$ encode physics at mass scales larger than $\mu$ (and hence carry information about heavy particles - SM as well as new physics (NP) ones), while matrix elements of hadronic operators $\mathcal{Q}_i(\mu)$ are responsible for long-distance physics dominated by nonperturbative strong interactions. Poor knowledge of these latter factors is the main source of uncertainty of theoretical predictions. At leading order the dominant contribution comes from the electromagnetic penguin operator

$$\mathcal{Q}_7 = -\frac{e}{8\pi^2} \bar{m}_b(\mu) \bar{q} \sigma^{\mu\nu} (1 + \gamma_5) b \; F_{\mu\nu} \tag{1.2}$$

where $q = d, s$ and $\bar{m}_b(\mu)$ is the $\overline{\text{MS}}$ mass of the b-quark.

The standard theoretical procedure used for evaluation of hadronic matrix elements is based on the QCD factorization idea, augmented by soft-collinear effective theory (for

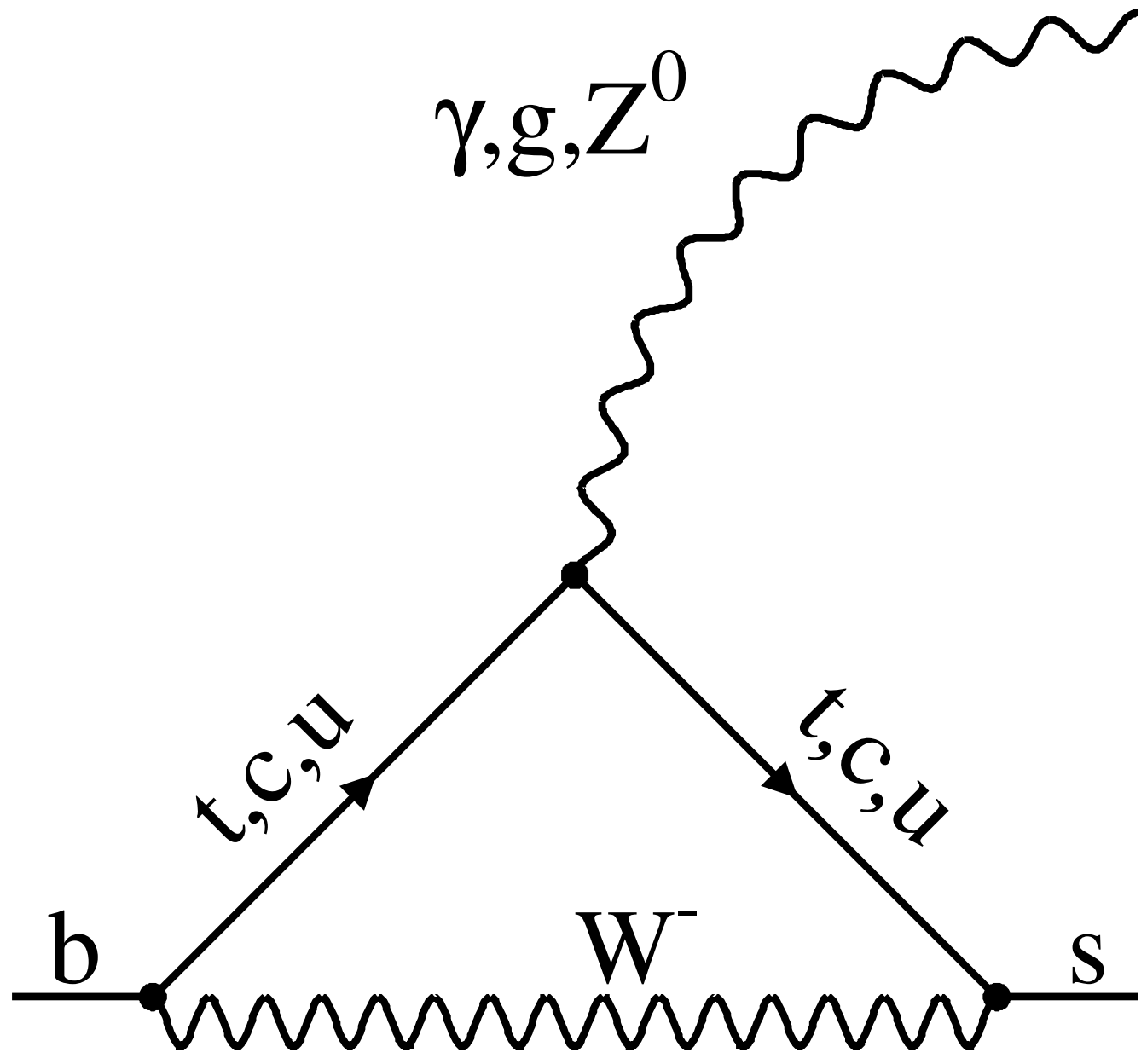


Figure 1: The penguin diagrams for b → s transitions.

recent review see [18] and references therein). The latter separates the matrix element of interest into non-perturbative but universal soft functions (form-factors, decay constants, light-cone distribution amplitudes) and hard scattering kernels calculated as perturbative series in $\alpha_s$. These calculations have been done in next-to-leading and partly in next-to-next-to-leading order [19]. However the whole factorization approach makes sense only in the leading order with respect to the small parameter $\Lambda_{QCD}/m_b$ and the question of a systematic construction of the $1/m_b$ expansion remains open (see recent discussion in [20]). Needless to say, having reliable SM theoretical predictions is a necessary prerequisite for addressing any NP scenario.

The state-of-the-art theoretical calculation [19] makes predictions for the exclusive branching ratio $B \to V\gamma$, listed in Table 2. A significant fraction of the uncertainty (about 1/4 of the total) in the branching ratio is due to the poor knowledge of the hadronic form-factor $T_1 (B \to K^*)$ at $q^2 = 0$, but uncertainties from other sources are

Table 2: The theoretical predictions for the branching fractions for exclusive $B \to V\gamma$ decays [19].

| Decay Mode | Branching ratio |
|---|---|
| $B^+ \to K^{*+}\gamma$ | $(4.6 \pm 1.4) \times 10^{-5}$ |
| $B^0 \to K^{*0}\gamma$ | $(4.3 \pm 1.4) \times 10^{-5}$ |
| $B^0_s \to \phi\gamma$ | $(4.3 \pm 1.4) \times 10^{-5}$ |

also important.

When discussing possible NP scenarios in the low energy effective Hamiltonian language, it is useful to distinguish those where only short-distance functions $\mathcal{C}_i(\mu)$ get modified with respect to the SM predictions and those where the structure of terms in the right hand side of Eq. (1.1) changes, for example, because new operators $\mathcal{Q}_{\rm i}^{\rm NP}$ (absent in the SM) appear. The models of the first sort are of "minimal flavour violating" type [21] since the only source of quark flavour mixing in this case is the CKM matrix of the SM. In a sense they are the most difficult to discover. The reason is clear: our ability to disentangle these scenarios crucially depends on the uncertainties in hadronic matrix elements, since the latter limits the accuracy of $\mathcal{C}_{\rm i}^{\rm exp}$ extracted from experimental data which we are interested in comparing with theoretical SM predictions of $\mathcal{C}_{\rm i}^{\rm th}$. The most prominent way of reducing this uncertainty seems to be further development of sophisticated numerical computations on the lattice, which is now the main tool of getting nonperturbative QCD results with the error budget under control. We refer interested readers to recent talks at the Lattice 2008 conference [22] where the current status of computations of hadronic matrix elements relevant for B decays is thoroughly reviewed. In short, we find ourselves in a peculiar situation: to discover minimal flavour violating NP in B decays of the discussed type we need to make progress in hardware and algorithms to get better knowledge of nonperturbative hadronic quantities (notably the form-factor $\rm T_1\,(B \to K^*)$) rather than to obtain more precise experimental results for exclusive branching ratios, since the error is already dominated by theory. But even with the lattice accuracy reached so far, experimental data on $\rm B \to K^*\gamma$ and other radiative penguin decays already put strong constraints on many models available on the theoretical market. The main qualitative message can be stated as follows: all observed branching ratios are generally consistent with each other and with the SM predictions and there is no doubt that the SM loop process depicted in Figure 1 is responsible for the main contribution to the radiative b decay.

It would however be wrong to conclude from the arguments presented above that rare radiative decays are not sensitive probes of NP and deserve no attention in this respect. The crucial point is that besides such a crude tool as branching ratios more complex information can be extracted from various $\mathcal{CP}$ asymmetries, time-dependent rates, angular distributions, isospin asymmetries and other quantities of analogous character. It is clear that if one considers non-minimally flavour violation scenarios with new operators and new complex phases, the choice of NP-sensitive observables becomes wider and typically for each particular beyond SM scenario one can distinguish observables most fit to search for it. For example, the theory prediction for the isospin asymmetry

$$\mathcal{A}^{\rm I}_{\rm B\to K^*\gamma} = \frac{\Gamma\left(\rm \bar{B}^0 \to \bar{K}^{*0}\gamma\right) - \Gamma\left(\rm B^- \to K^{*-}\gamma\right)}{\Gamma\left(\rm \bar{B}^0 \to \bar{K}^{*0}\gamma\right) + \Gamma\left(\rm B^- \to K^{*-}\gamma\right)}$$

is sensitive to the operator $\mathcal{Q}_6$ and hence provides constraints on corresponding NP [23]. In what follows we concentrate on another important class of observables related to the photon polarization, where a crucial contribution is expected from LHCb.

We can express the Hamiltonian for $b \to s\gamma$ in the following form:

$$\Delta\mathcal{H} = -\sqrt{8}G_F \frac{e\bar{m}_b}{16\pi^2}\mathcal{F}_{\mu\nu}\left[\mathcal{A}_L\bar{s}\sigma^{\mu\nu}\frac{1+\gamma_5}{2}b + \mathcal{A}_R\bar{s}\sigma^{\mu\nu}\frac{1-\gamma_5}{2}b\right] + \text{h.c.} \tag{1.3}$$

Here $\mathcal{A}_L(\mathcal{A}_R)$ corresponds to the amplitude for the emission of left (right) handed photons in the $b \to s_L\gamma_L(b \to s_R\gamma_R)$ decays. This can be easily seen by writing the electromagnetic field tensor for left (right) polarized photons as: $\mathcal{F}^{L,R}_{\mu\nu} = \frac{1}{2}(\mathcal{F}_{\mu\nu} \pm i\tilde{\mathcal{F}}_{\mu\nu})$, where $\tilde{\mathcal{F}}_{\mu\nu} = \frac{1}{2}\varepsilon_{\mu\nu\sigma\rho}\mathcal{F}^{\sigma\rho}$. Using the identity $\sigma_{\mu\nu}\gamma_5 = \frac{i}{2}\varepsilon_{\mu\nu\alpha\beta}\sigma^{\alpha\beta}$, one can see that only the $\mathcal{F}^L_{\mu\nu}$ part survives in the first term of the right-hand side of Eq. (1.3) and only $\mathcal{F}^R_{\mu\nu}$ in the second one. In the SM the amplitude ratio, representing the fraction of "wrong" helicity photons $\mathcal{A}_R/\mathcal{A}_L$ is proportional to the mass ratio $m_s/m_b$, since in the SM only the left-handed components of the external fermions couple to the W boson. Thus the leading contribution is given by the operator of Eq. (1.2). This naive $m_s/m_b$ scaling can however be destroyed by corrections, which take into account gluon emission. This effect may affect significantly the purity of the photon polarization. In papers [24] and [25] these contributions were estimated to be sufficiently large, about 10 %, however these results were based mainly on dimensional estimations. More precise calculations in perturbative QCD, taking into account the effects from hard gluon emission in the main part, give a contribution on the level of 3-4 % [26]. The nonperturbative corrections from the soft gluon emission via the c-quark loop, induced by the $\mathcal{O}_2$-operator, turn out to be about 1 %, nonperturbative contributions from the annihilation diagrams and other operators are of the same order or smaller as can be seen from the light cone sum rule method calculations [27]. Thus we conclude, following original arguments from [28], that the polarization of emitted photons in radiative decays is a good example of a nontrivial experimental observable sensitive to the Lorentz structure of effective Hamiltonian operator containing the photon emission vertex.

The admixture of photons with the "wrong" polarization may be rather large in some SM extensions like e.g. the Left Right Symmetric Model (LRSM). Here the enhancement of the right-handed photon fraction is due to $W_L - W_R$ mixing, and chirality flip along the internal t-quark line in the loop leads to a large factor $m_t/m_b$ in the amplitude for producing right-handed photons. It was shown that within the unconstrained minimal supersymmetric model (uMSSM) a strong enhancement of order $m_{\tilde{g}}/m_b$ is possible due to chirality flip along the gluino line and left-right squark mixing. In this case the degree of photon polarization, $\lambda_\gamma$, defined as

$$\lambda_\gamma = \frac{|\mathcal{A}_R|^2 - |\mathcal{A}_L|^2}{|\mathcal{A}_R|^2 + |\mathcal{A}_L|^2} \tag{1.4}$$

can take any value between $-1$ and 1 [29]. The model with anomalous right-handed top couplings [30] also allows sizeable contributions in $\mathcal{A}_R$, $-1 \leq \lambda_\gamma \leq -0.12$.

In models with non-supersymmetric extra dimensions there are also no reasons for the right-handed photon to be suppressed with respect to the left-handed one, so that in the general case left-handed and right-handed amplitudes are comparable and mixing-induced

$\mathcal{CP}$ asymmetries are of the order of one [31]. Thus the information one can get in this way is extremely interesting providing a typical example of a "null test" [32], since the photons are almost 100 % polarized in the SM.

## 1.2 Experimental Issues

The experimental challenge therefore is to measure the amplitude ratio $\left|\frac{\mathcal{A}(\mathrm{B}\to\Phi\gamma_\mathrm{R})}{\mathcal{A}(\mathrm{B}\to\Phi\gamma_\mathrm{L})}\right|$ where $\Phi$ represents some final hadronic state. There is no experimental possibility to measure photon polarization directly, but there are several indirect strategies. The first one is the study of angular distributions of the $\Phi$ decay products [33, 34]. In this way one is able to measure only the square of the amplitude ratio. Indeed, using the definition of the photon polarization parameter $\lambda_\gamma$ from Eq. (1.4) one has

$$\frac{\mathrm{d}\Gamma(\mathrm{B}\to\Phi\gamma)}{\mathrm{d}\Omega} \propto \left(|\mathcal{A}_\mathrm{R}|^2 + |\mathcal{A}_\mathrm{L}|^2\right) + \lambda_\gamma \left(|\mathcal{A}_\mathrm{R}|^2 - |\mathcal{A}_\mathrm{L}|^2\right). \tag{1.5}$$

It is worth noticing that the amplitudes $\mathcal{A}_\mathrm{R,L}$ corresponding to left-handed and right-handed photons do not interfere in this case since the polarization of the photon in the final state can (at least in principle) be measured independently. By studying the angular distribution one can extract $\lambda_\gamma$ from Eq. (1.5), in other words the method makes use of angular correlations among the decay products[1] in $\mathrm{B}\to[\Phi\to\mathrm{P_1P_2P_3}]\gamma$, where $\mathrm{P_i}$ is either a pion or a kaon. This technique was originally suggested in [33,34] and used for the decay $\mathrm{B}\to K\pi\pi\gamma$ with the sum over intermediate hadronic resonances. The radiative decay mode $\mathrm{B}\to[\varphi\to\mathrm{K^+K^-}]\mathrm{K}\gamma$ is considered in [35]. This mode is rather distinctive with many desirable features from the experimental point of view: the final state is a photon plus only charged mesons (for charged B mesons), the fact that $\varphi$ is narrow reduces the effects of intermediate resonances interference, etc. However the actual situation is rather involved. The possibility of measuring $\lambda_\gamma$ in this way depends on a delicate partial-wave interference pattern. The latter may be unfavourable and the asymmetry may escape detection [35].

Alternatively, one can study baryon decays $\Lambda_\mathrm{b}\to\Lambda^0\gamma\to\mathrm{p}\pi\gamma$ (or $\Lambda_\mathrm{b}\to\Lambda^*\gamma\to\mathrm{pK}\gamma$) and measure the photon polarization via the forward-backward asymmetry of the proton with respect to the $\Lambda_\mathrm{b}$ in the $\Lambda^0$ rest frame for polarized $\Lambda_\mathrm{b}$, (see [36, 37, 38] for details and references therein).

Aside from the experimental difficulties, the main problem of these two methods is the absence of interference between the amplitudes, corresponding to left- and right-handed photon emission, since they correspond to different and distinguishable (at least in principle) final states. Correspondingly they are sensitive only to the square of the amplitude ratio in the form of $\lambda_\gamma$, see Eq. (1.4). It would be advantageous to measure (the absolute value of) the amplitude ratio as it is. There are two possible ways to do that. The first one makes use of the fact that some photons convert in the detector material into electron-positron pairs. Thus it is possible to have the desired interference.

[1] Notice that there must be at least three particles in the final state.

It can be shown that for these processes the distribution in the angle $\phi$ between the $e^+e^-$ plane and the plane defined by the final state hadrons (e.g. $K\pi$ resulting from $K^*$ decay) should be isotropic for purely circular polarization, while the deviations from this isotropy includes the same parameter $\mathcal{A}_R/\mathcal{A}_L$, indicating the presence of right-handed photons [39, 40, 41, 42]. However multiple scattering does not allow to identify the decay plane for the low invariant mass $e^+e^-$-pair. This is not the case for pair creation from virtual photons where one can select pair masses above 30 MeV/$c^2$ without losing too much rate. However in this case other diagrams contribute with longitudinal virtual photons. The LHCb prospects for this measurement are discussed elsewhere [43].

Another way is to study the time evolution of $B^0_{(s)} \to \Phi^{\mathcal{CP}}\gamma$ decays, where $\Phi^{\mathcal{CP}}$ is some $\mathcal{CP}$-eigenstate. In this case the time-dependent decay rate can be conventionally parameterized as follows:

$$\Gamma_{B^0_{(s)}\to\Phi^{\mathcal{CP}}\gamma}(t) = |A|^2 e^{-\Gamma_{(s)}t}\Big(\cosh\frac{\Delta\Gamma_{(s)}t}{2} - \mathcal{A}^\Delta \sinh\frac{\Delta\Gamma_{(s)}t}{2} + \mathcal{C}\cos\Delta m_{(s)}t - \mathcal{S}\sin\Delta m_{(s)}t\Big) \quad (1.6a)$$

$$\Gamma_{\bar{B}^0_{(s)}\to\Phi^{\mathcal{CP}}\gamma}(t) = |A|^2 e^{-\Gamma_{(s)}t}\Big(\cosh\frac{\Delta\Gamma_{(s)}t}{2} - \mathcal{A}^\Delta \sinh\frac{\Delta\Gamma_{(s)}t}{2} - \mathcal{C}\cos\Delta m_{(s)}t + \mathcal{S}\sin\Delta m_{(s)}t\Big) \quad (1.6b)$$

Within the SM one has [44]:

$$\begin{aligned} \mathcal{C} &\approx 0 \\ \mathcal{S} &\approx \sin 2\psi \sin\varphi_{(s)} \\ \mathcal{A}^\Delta &\approx \sin 2\psi \cos\varphi_{(s)}, \end{aligned} \quad (1.7)$$

where $\psi$ is defined as

$$\tan\psi \equiv \left|\frac{\mathcal{A}\left(\bar{B}_{(s)} \to \Phi^{\mathcal{CP}}\gamma_R\right)}{\mathcal{A}\left(\bar{B}_{(s)} \to \Phi^{\mathcal{CP}}\gamma_L\right)}\right| \quad (1.8)$$

and related to the fraction of "wrongly"-polarized photons[2]; and $\varphi_{(s)}$ is the sum of $B^0_{(s)}$ mixing phase and $\mathcal{CP}$-odd weak phases for right $\mathcal{A}_R$ and left $\mathcal{A}_L$ amplitudes. From Eqs. (1.7) and (1.8) one can see that the measurement of $\mathcal{A}^\Delta$ and $\mathcal{S}$ directly determines the "wrongly"-polarized photon fraction [44].

For the $B^0$ system the parameter $\Delta\Gamma$ is negligible, and as a result the terms proportional to $\mathcal{A}^\Delta$ vanish:

$$\Gamma_{B^0\to\Phi^{\mathcal{CP}}\gamma}(t) \approx |A|^2 e^{-\Gamma t}\big(1 - \mathcal{S}\sin\Delta m t\big) \quad (1.9a)$$

$$\Gamma_{\bar{B}^0\to\Phi^{\mathcal{CP}}\gamma}(t) \approx |A|^2 e^{-\Gamma t}\big(1 + \mathcal{S}\sin\Delta m t\big) \quad (1.9b)$$

[2]Note that the parameter $\lambda_\gamma$, defined by Eq. (1.4) could be expressed as $\lambda_\gamma = \cos 2\psi$.

Also in the SM one expects $\sin\varphi = \sin(2\beta - \phi_{\mathrm{p}}) \approx \sin 2\beta$, where $\phi_{\mathrm{p}}$ is $\mathcal{CP}$-odd weak penguin phase. Therefore one gets:

$$\mathcal{S}_{\mathrm{B}^0} = \sin 2\psi \sin 2\beta. \tag{1.10}$$

On the contrary for the $\mathrm{B_s^0}$ system the parameter $\Delta\Gamma_{\mathrm{s}}$ is not negligible, providing a non-zero sensitivity to $\mathcal{A}^{\Delta}$. In the SM $\varphi_{\mathrm{s}}$ is expected to be small, $\sin\varphi_{\mathrm{s}} = \sin(2\beta_{\mathrm{s}} - \phi_{\mathrm{p}}) \approx 0$, thus the term with $\mathcal{S}$ vanishes:

$$\Gamma_{\mathrm{B_s^0}\to\Phi^{\mathcal{CP}}\gamma}(\mathrm{t}) \approx |\mathrm{A}|^2 \, \mathrm{e}^{-\Gamma_{\mathrm{s}}\mathrm{t}} \big(\cosh\frac{\Delta\Gamma_{\mathrm{s}}\mathrm{t}}{2} - \mathcal{A}^{\Delta}\sinh\frac{\Delta\Gamma_{\mathrm{s}}\mathrm{t}}{2}\big) \tag{1.11a}$$

$$\Gamma_{\mathrm{\bar{B}_s^0}\to\Phi^{\mathcal{CP}}\gamma}(\mathrm{t}) \approx \Gamma_{\mathrm{B_s^0}\to\Phi^{\mathcal{CP}}\gamma}(\mathrm{t}) \quad , \tag{1.11b}$$

and finally one gets:

$$\mathcal{A}^{\Delta}_{\mathrm{B_s}^0} \approx \sin 2\psi, \tag{1.12}$$

thus opening the possibility for the direct measurement of the photon polarization parameter $\sin 2\psi$ [45]. It is worth to stress here that for vanishing $\mathcal{S}$ and $\mathcal{C}$, both $\mathrm{B_s^0}$ and $\mathrm{\bar{B}_s^0}$ exibit the same decay-time evolution and therefore no flavour tagging is required for extraction of $\sin 2\psi$.

## 2 Current experimental status

Exclusive radiative B decays have been measured by the CLEO, BABAR, and Belle experiments. The first measurement of a radiative B decay was the measurement of the branching ratio of $\mathrm{B} \to \mathrm{K}^*\gamma$ by the CLEO collaboration [46]. Today the $\mathrm{B}^0 \to \mathrm{K}^{*0}\gamma$ signal can serve as a reference point for the measurement of other radiative decays and as a calibration signal. Its branching fraction is well measured by all three collaborations. The current measurements are summarized in Table 3. Within the SM the direct $\mathcal{CP}$-asymmetry for this decay is predicted to be less than 1 % [47]. No evidence of a deviation from this prediction has been seen in the measured charge asymmetry for the exclusive $\mathrm{B}^\pm$ and $\mathrm{B}^0$ decays to $\mathrm{K}^*\gamma$, see Table 3.

Table 3: The branching fraction for the decay $\mathrm{B}^0 \to \mathrm{K}^{*0}\gamma$ and the direct $\mathcal{CP}$-violation asymmetry for the decay $\mathrm{B}^0 \to \mathrm{K}^{*0}\gamma$ measured at the B factories.

| Experiment | | $\mathcal{B}(\mathrm{B}^0 \to \mathrm{K}^{*0}\gamma)$ $[10^{-6}]$ | $\mathcal{A}_{\mathcal{CP}}(\mathrm{B} \to \mathrm{K}^*\gamma)$ |
|---|---|---|---|
| BABAR | [48] | $39.2 \pm 2.0 \pm 2.4$ | $-0.013 \pm 0.036 \pm 0.010$ |
| Belle | [49] | $40.1 \pm 2.1 \pm 1.7$ | $-0.015 \pm 0.044 \pm 0.012$ |
| CLEO | [50] | $45.5^{+7.2}_{-6.8} \pm 3.4$ | $+0.08 \pm 0.13 \pm 0.03$ |
| PDG'08 & HFAG | [51, 52] | $40.1 \pm 2.0$ | $-0.010 \pm 0.028$ |

The $\mathrm{B}^+ \to \phi\mathrm{K}^+\gamma$ decay mode was observed first by the Belle collaboration [53] and the branching fraction has been measured both by the Belle and BABAR collaborations, see Table 4.

Table 4: Branching fractions for $\mathrm{B}^+ \to \phi\mathrm{K}^+\gamma$ and $\mathrm{B}_\mathrm{s}^0 \to \phi\gamma$ decays.

| Experiment | | $\mathcal{B}(\mathrm{B}^+ \to \phi\mathrm{K}^+\gamma)$ $[10^{-6}]$ | | $\mathcal{B}(\mathrm{B}_\mathrm{s}^0 \to \phi\gamma)$ $[10^{-6}]$ | |
|---|---|---|---|---|---|
| Belle | | $3.4 \pm 0.9 \pm 0.4$ | [53] | $57^{+18}_{-15}(\mathrm{stat})^{+12}_{-11}(\mathrm{syst})$ | [54] |
| BABAR | | $3.5 \pm 0.6 \pm 0.4$ | [55] | — | |
| PDG'08 | [51] | $3.5 \pm 0.6$ | | $< 120$@90 %CL | |
| HFAG | [52] | $3.5 \pm 0.6$ | | $57^{+21}_{-18}$ | |

The direct asymmetry for this decay is reported by BABAR collaboration to be $\mathcal{A}_{\mathcal{CP}} = -0.26 \pm 0.14 \pm 0.05$ [55]. Recently the Belle collaboration also observed for the first time a radiative penguin decay of the $\mathrm{B}_\mathrm{s}$ meson: $\mathrm{B}_\mathrm{s}^0 \to \phi\gamma$. The measured branching fraction was found to be in agreement with the SM expectations and the branching fraction for the $\mathrm{B}^0 \to \mathrm{K}^{*0}\gamma$ decay.

The measurement of the photon polarization in $\mathrm{b} \to \mathrm{s}\gamma$ transitions allows to search for right-handed FCNC and provides a powerful probe to test SM. The effects of photon polarization have been studied by BABAR and Belle through the measurement of time-dependent $\mathcal{CP}$-asymmetries for $\mathrm{B}^0 \to (\mathrm{K}^{*0} \to \mathrm{K}_\mathrm{S}^0\pi^0)\,\gamma$, $\mathrm{B}^0 \to \mathrm{K}_\mathrm{S}^0\pi^0\gamma$, $\mathrm{B}^0 \to \eta\mathrm{K}^0\gamma$, $\mathrm{B}^0 \to$

$K^0_S\rho^0\gamma$ and $B^0 \to \rho^0\gamma$ decays. The results for the parameters $\mathcal{C}$ and $\mathcal{S}$ defined in (1.9) for these decays are summarized in Table 5.

Table 5: The $\mathcal{CP}$-violation parameters $\mathcal{C}$ and $\mathcal{S}$ measured for various exclusive radiative decays.

| Experiment | | $\mathcal{C}(B^0 \to K^{*0}\gamma)$ | $\mathcal{S}(B^0 \to K^{*0}\gamma)$ |
|---|---|---|---|
| BABAR | [56] | $-0.14 \pm 0.16 \pm 0.03$ | $-0.03 \pm 0.29 \pm 0.03$ |
| Belle | [57] | $0.20 \pm 0.24 \pm 0.05$ | $-0.32^{+0.36}_{-0.33} \pm 0.05$ |
| | | $\mathcal{C}(B^0 \to K^0_S\pi^0\gamma)$ | $\mathcal{S}(B^0 \to K^0_S\pi^0\gamma)$ |
| BABAR | [56] | $0.36 \pm 0.33 \pm 0.04$ | $-0.78 \pm 0.59 \pm 0.09$ |
| Belle | [57] | $0.20 \pm 0.20 \pm 0.06$ | $-0.10 \pm 0.31 \pm 0.07$ |
| | | $\mathcal{C}(B^0 \to \eta K^0\gamma)$ | $\mathcal{S}(B^0 \to \eta K^0\gamma)$ |
| BABAR | [58] | $-0.32^{+0.40}_{-0.39} \pm 0.07$ | $-0.18^{+0.49}_{-0.46} \pm 0.12$ |
| | | $\mathcal{C}(B^0 \to K^0_S\rho^0\gamma)$ | $\mathcal{S}(B^0 \to K^0_S\rho^0\gamma)$ |
| Belle | [59] | $-0.05 \pm 0.18 \pm 0.06$ | $-0.11 \pm 0.33^{+0.05}_{-0.09}$ |
| | | $\mathcal{C}(B^0 \to \rho^0\gamma)$ | $\mathcal{S}(B^0 \to \rho^0\gamma)$ |
| Belle | [60] | $0.44 \pm 0.49 \pm 0.14$ | $-0.83 \pm 0.65 \pm 0.18$ |

Using Eq. (1.10) and taking $\sin 2\beta = 0.678 \pm 0.025$ [51], one finds for the current precision of the single most precise measurement of $\sin 2\psi$ in $B^0 \to K^{*0}\gamma$ decay:

$$\sigma^{B^0\to K^{*0}\gamma}_{\sin 2\psi} = 0.43, \tag{2.1}$$

where $\psi$ is defined through the ratio of amplitudes by Eq. (1.8).

# 3 Measurement of the photon polarisation in $B_s^0 \to \phi\gamma$

The measurement of the photon polarisation in $B_s^0 \to \phi\gamma$ events can be done either through the measurement of the $\mathcal{A}^{\Delta} \sinh \frac{\Delta\Gamma_s t}{2}$ or the $\mathcal{S} \sin \Delta m_s t$-terms, of Eq. (1.6) (see also Appendix A). The measurement of the amplitude of $\sinh \frac{\Delta\Gamma_s t}{2}$-terms can be done with untagged events. It is practically insensitive to the proper time resolution [3], but requires the knowledge of the proper time acceptance function, see Section 3.5. In contrast, the measurement of the amplitude of the fast oscillating $\sin \Delta m_s t$-term requires flavour tagging and is limited mainly by the proper time resolution [3] and has only a modest dependency on the uncertainty of the proper time acceptance function.

The relative amplitude of the oscillating term with respect to the term proportional to $\sinh \frac{\Delta\Gamma_s t}{2}$ is about $1.5\,(1-2\omega)\,\varepsilon^{\mathrm{tag}} \tan\varphi_s$ at $\tau = \tau^0_{B_s}$ and drops quickly to about $0.1\,(1-2\omega)\,\varepsilon^{\mathrm{tag}} \tan\varphi_s$ at $\tau = 3\tau^0_{B_s}$, where $\tau^0_{B_s}$ is the nominal lifetime of ${B_s}^0$ meson [51], $\varepsilon^{\mathrm{tag}}$ is the flavour tagging efficiency, $\omega$ is mistag rate, and $\varphi_s$ is the difference between the $B_s^0$ mixing phase and the weak phase of penguin amplitude, see Eqs. (1.6) and (A.1).

Taking this into account we concentrate on the measurement of photon polarisation through the measurement of $\mathcal{A}^{\Delta}$.

## 3.1 Event selection

The event selection criteria described in detail elsewhere [2], have been chosen to maximize the ratio $\frac{\mathcal{S}}{\sqrt{\mathcal{S}+\mathcal{B}}}$. The reconstruction efficiency $\varepsilon_{\mathrm{rec}}$, selection efficiency for reconstructed events $\varepsilon_{\mathrm{sel/rec}}$, the L0-trigger efficiency for selected events $\varepsilon_{\mathrm{L0/sel}}$ and the total efficiency $\varepsilon_{\mathrm{tot}}$ are summarized in Table 6.

Assuming the nominal detector performance we expect to observe $B_s^0 \to \phi\gamma$ signal mass peak with effective resolution of 98 MeV/$c^2$, see Figure 2 [2,3]. The width is dominated by the resolution of the Electromagnetic calorimeter [1, 2, 9], see Appendix C. The expected statistics for $B^0 \to K^{*0}\gamma$ and $B_s^0 \to \phi\gamma$ decays for the accumulated luminosity of 2 fb$^{-1}$ (see Table 1) will allow us to perform the precise measurement of the actual signal shape from data.

Table 6: The reconstruction efficiency $\varepsilon_{\mathrm{rec}}$, selection efficiency for reconstructed events $\varepsilon_{\mathrm{sel/rec}}$ and the L0-trigger efficiency for selected $B_s^0 \to \phi\gamma$ events $\varepsilon_{\mathrm{L0/sel}}$ [2].

| Efficiency | $B_s^0 \to \phi\gamma$ |
|---|---|
| $\varepsilon_{\mathrm{rec}}$ | 1.9% |
| $\varepsilon_{\mathrm{sel/rec}}$ | 11.7% |
| $\varepsilon_{\mathrm{L0/sel}}$ | 44.1% |
| $\varepsilon_{\mathrm{tot}}$ | 0.10% |

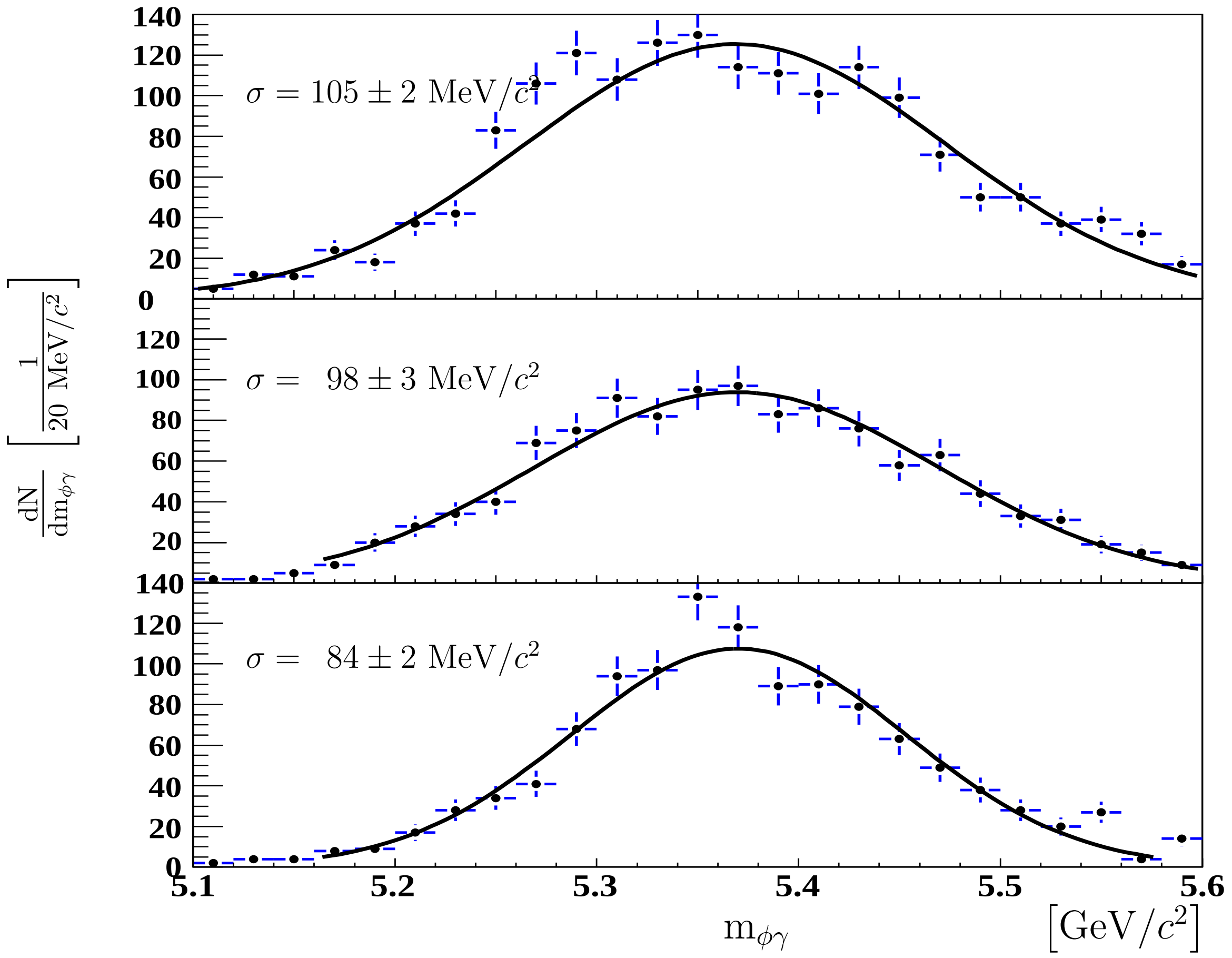


Figure 2: Signal invariant mass distribution after selection cuts for photons, detected in inner, middle and outer (from top to bottom) zones of the Electromagnetic Calorimeter. The curves represent the fit with a Gaussian signal function. The effective resolution for all three zones together is $98 \pm 2$ MeV/$c^2$.

The proper time resolution for the signal events can be described by a sum of two Gaussian functions[3] with the resolutions of $52 \pm 5$ and $114 \pm 7$ fs, and the fraction of narrow component of $51 \pm 9$ %, see Figure 3.

### 3.2 High Level Trigger

The structure of the High Level Trigger line for events triggered by high transverse energy photons is described in detail elsewhere [8]. In brief, the algorithm can be sketched as follows. The sequence of the photon HLT line begins with the confirmation of the L0

[3] The $B_s^0$ proper time resolution is dominated by the $\phi$-vertex resolution. The vertex is reconstructed with two kaon tracks and the error on its position depends strongly on the opening angle between two kaons. In particular it increases with the decreasing of the angle between the $\phi$ and the $B_s^0$ flight direction in the $B_s^0$ rest frame [1,3].

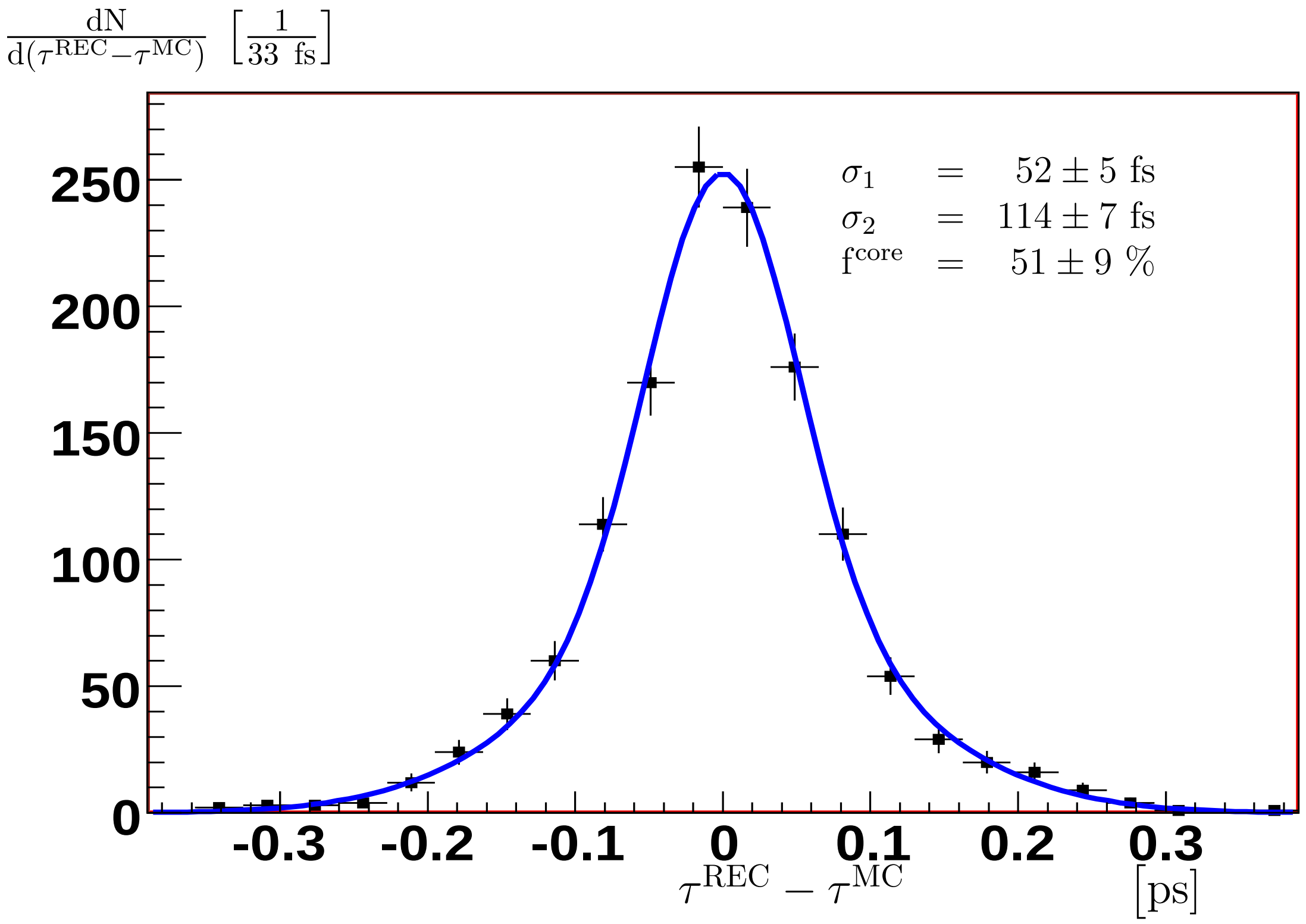


Figure 3: The proper time resolution for the $B_s^0 \to \phi\gamma$ events passing all selection criteria [3]. The curve represents the fit with a double Gaussian function. The effective resolutions are $52 \pm 5$ and $114 \pm 7$ fs and the fraction of the narrow component is $51 \pm 9$ %.

photon candidate. A fast clusterization is performed in the Ecal region close to the L0 photon candidate. The closest Ecal cluster with compatible energy is selected for subsequent analysis. A set of cluster shape variables is calculated to remove clusters which come from merged $\pi^0$. This reduces the minimum bias rate by a factor of 2 while preserving 90 % of the signal events. Then the 2D-Velo reconstruction is launched and 2D-Velo tracks with 2D-impact parameters in excess of 100 $\mu$m are selected for 3D-Velo reconstruction. The 3D-tracks with impact parameter in excess of 150 $\mu$m are reconstructed in the tracker stations behind the magnet and at least one track is required to have large transverse momentum $p_T > 700$ MeV/$c$. The reduced bandwidth after this step allows a full 3D-Velo reconstruction. The companion track is searched among all 3D-Velo tracks. The ones with distance of closest approach less then 200 $\mu$m are reconstructed in the tracker stations and kept if their transverse momentum is larger than 700 MeV/$c$. Then the two track vertex is formed, and a requirement on vertex quality is imposed. In the last step the photon is added to the vertex without additional cuts. Figure 4 illustrates the performance of the HLT1 line for $B_s^0 \to \phi\gamma$ events.

The efficiency of the first step of the High Level Trigger $\varepsilon_{HLT1}$ has been estimated to be around 70 % at 10 kHz minimum bias rate. It is worth to note here that for the HLT1 the option *"photon + single track"* instead of (or, in addition to) the option *"photon +*

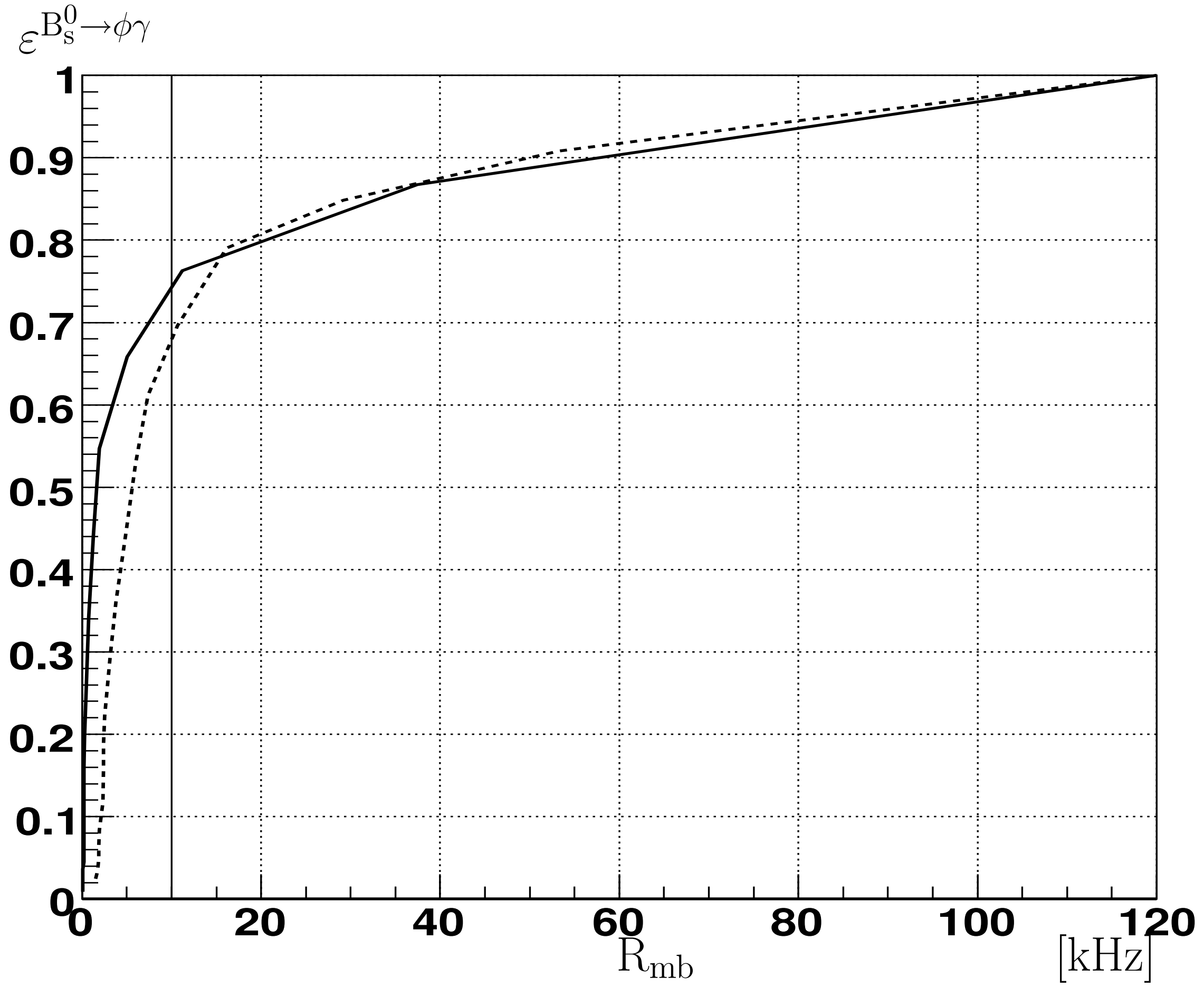


Figure 4: HLT1 efficiency $\varepsilon^{{B_s}^0 \to \phi\gamma}$ for selected events versus minimum bias rate $R_{mb}$ for two HLT1 strategies: *"photon + two tracks"* (solid line) and *"photon + single track"* (dotted line).

*two tracks"* is still open.

For the next trigger step (HLT2) the full reconstruction of the $B_s^0 \to \phi\gamma$ candidate is performed with relaxed cuts, compared to those used in the final selection. An efficiency of 90% has been found for a minimum bias background rate of 16 Hz. Also, the "inclusive $\phi$" trigger line, which selects the detached $\phi \to K^+K^-$ vertices, has an efficiency of 77% for $B_s^0 \to \phi\gamma$ events.

## 3.3 The background

Possible backgrounds have been studied in detail both for $B_s^0 \to \phi\gamma$ and $B^0 \to K^{*0}\gamma$ modes using slightly relaxed selection criteria to increase the event statistics [2]. The major part of the background consists of the random combination of two tracks from the same secondary vertex together with a merged $\pi^0$-meson, misidentified as a single high energy photon. Using a dedicated $\pi^0/\gamma$-separation algorithm, see Appendix D, this major

fraction of background can be significantly reduced.

Peaking background from badly reconstructed decays $B_s^0 \to \phi\pi^0$ where the energetic "merged" $\pi^0$ is misreconstructed as a photon, is expected to be small, since this decay is highly suppressed[4]. Even without $\gamma/\pi^0$-separation the cut on the polarisation of the vector meson drastically reduces this potential feed-down [1]. Assuming $\mathcal{B}(B_s^0 \to \phi\pi^0) = \mathcal{B}(B^0 \to K^{*0}\pi^0)$, and using the recent upper limit $\mathcal{B}(B^0 \to K^{*0}\pi^0) < 3.5 \times 10^{-6}$ [51] the final contribution from this source is expected to be less than 0.4% at 90% confidence level[5].

Another background which potentially leads to a contribution in the signal mass region is misreconstructed events $B^0 \to K^{*0}\gamma$ where the pion from the $K^{*0} \to K^+\pi^-$ decay is misidentified as a kaon. For a small fraction of such events the invariant mass of the true and fake kaon falls into the narrow mass window around the nominal $\phi$ meson mass. A Monte Carlo study shows that for this background the central value of the reconstructed mass is shifted by $\sim$100 MeV/$c^2$ upwards with respect to the nominal mass of $B_s^0$ meson and its width is about 2-3 times wider. However, the absolute contribution is negligible and amounts to 0.3 % even in case of a 100 % misidentification of pions as kaons, see Figure 5.

### Parameterization of the background as function of reconstructed proper time and reconstructed mass

The shape of the background mass distribution obtained with the limited Monte Carlo statistics is consistent with an exponential behaviour $\propto e^{-\mu m_{\phi\gamma}}$ with the parameter $\mu = 0.8\ (\mathrm{GeV}/c^2)^{-1}$, see Figure 6 [3].

More generally the background as a function of the reconstructed invariant mass can be described as:

$$f_b(m_{\phi\gamma}) \propto e^{-\mu m_{\phi\gamma}} \times P(m_{\phi\gamma}), \tag{3.1}$$

where $P(m_{\phi\gamma})$ is a smooth function (e.g. low-order polynomial) of the reconstructed invariant mass.

A crucial point of the analysis is the ability to extract the shape of the background under the signal using the sidebands. It has been demonstrated that the simultaneous fit of left and right sidebands, defined as 4.4–5.1 GeV/$c^2$ and 5.7–6.4 GeV/$c^2$ correspondingly, allows the precise determination of the background parameters [3].

Important for the analysis is the parameterization of the background distribution as a function of the reconstructed proper time. This parameterization can be extracted from the invariant mass sidebands. The nature of the background events in the low and high mass sidebands is different and one therefore expects different reconstructed proper time distributions for events of the left and right sidebands. Using a very general ansatz the

[4]This isospin-violating decay proceeds though a gluonic penguin graph, which in turn preserves isospin.

[5]This contribution has been estimated to be less than 4 % in reference [1] using $\mathcal{B}(B^0 \to K^{*0}\pi^0) < 3.6 \times 10^{-5}$.

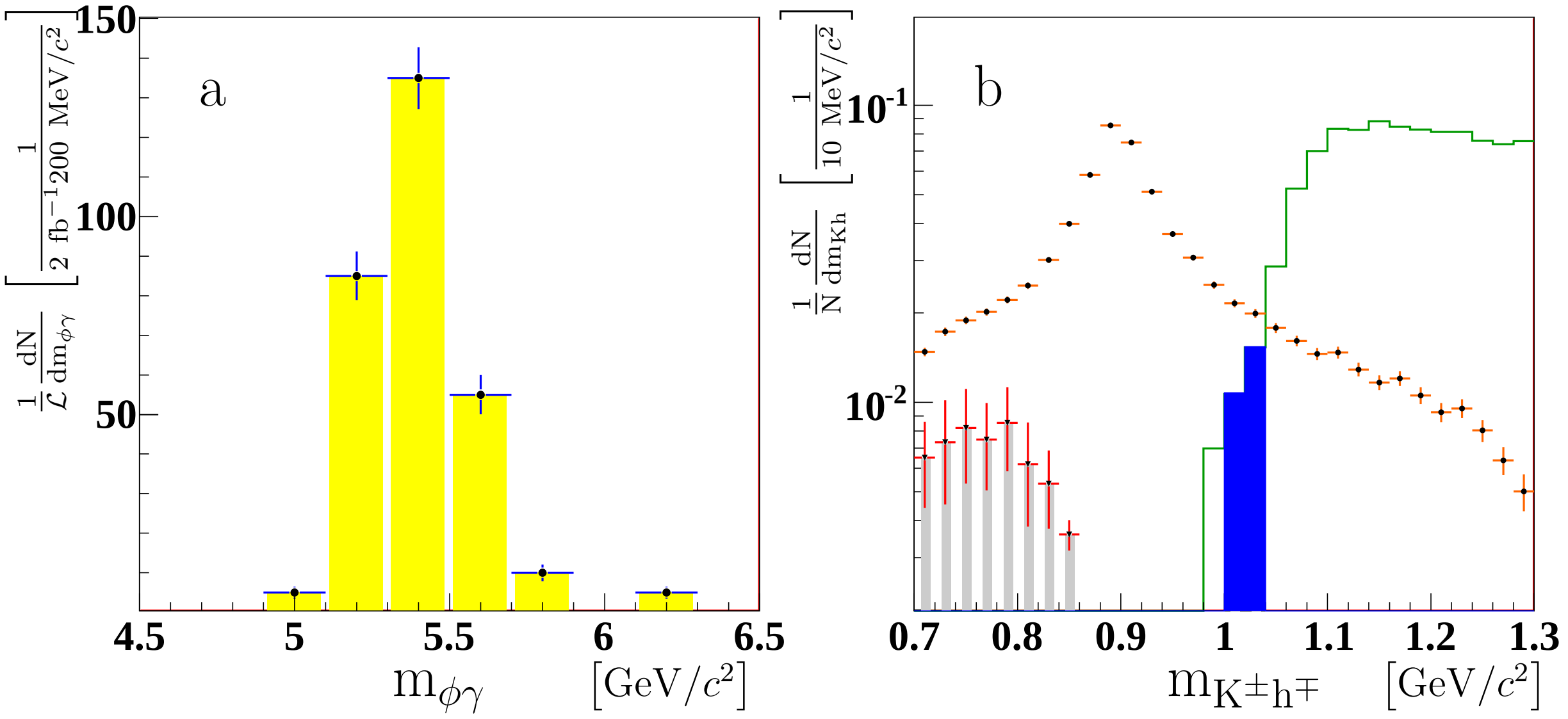


Figure 5: a) The absolute contribution of the feed-down from $B^0 \to K^{*0}\gamma$ decays misreconstructed as $B_s^0 \to \phi\gamma$ signal events for 2 $fb^{-1}$. b) The distribution of the $K\pi$ invariant mass from $B^0 \to K^{*0}\gamma$ decays (points with errors) and the corresponding mass distribution in case the pion is identified as second kaon (open histogram). The solid blue histogram shows the fraction of these misidentified $K^+K^-$ combinations, consistent with the mass window used for selecting $\phi$ candidates. The light grey histogram with error bars shows the original $K\pi$ invariant mass of the wrongly identified $\phi$ events. A 100 % misidentification of pions as kaons has been assumed for these plots.

background proper time distribution can be expressed as:

$$f_b(t) \propto \epsilon_{B_s^0 \to \phi\gamma}(t) \sum_i f_i e^{-t/\tau_i}, \tag{3.2}$$

where the common factor $\epsilon_{B_s^0 \to \phi\gamma}(t)$ is the proper time acceptance function for the signal decay $B_s^0 \to \phi\gamma$, which is assumed to be known in this section and is discussed in Section 3.5. A Monte Carlo study shows that the proper time acceptance can be parameterized as:

$$\epsilon_{B_s^0 \to \phi\gamma}(t) \propto \frac{(at)^c}{1 + (at)^c}. \tag{3.3}$$

The parameters have been found[6] to be [3]:

$$\begin{aligned} a &= 0.74 \pm 0.09\ ps^{-1} \\ c &= 1.86 \pm 0.15, \end{aligned}$$

see Figures 7 and 8.

[6] With a larger Monte Carlo sample more precise estimates for these parameters have been obtained, see (3.7).

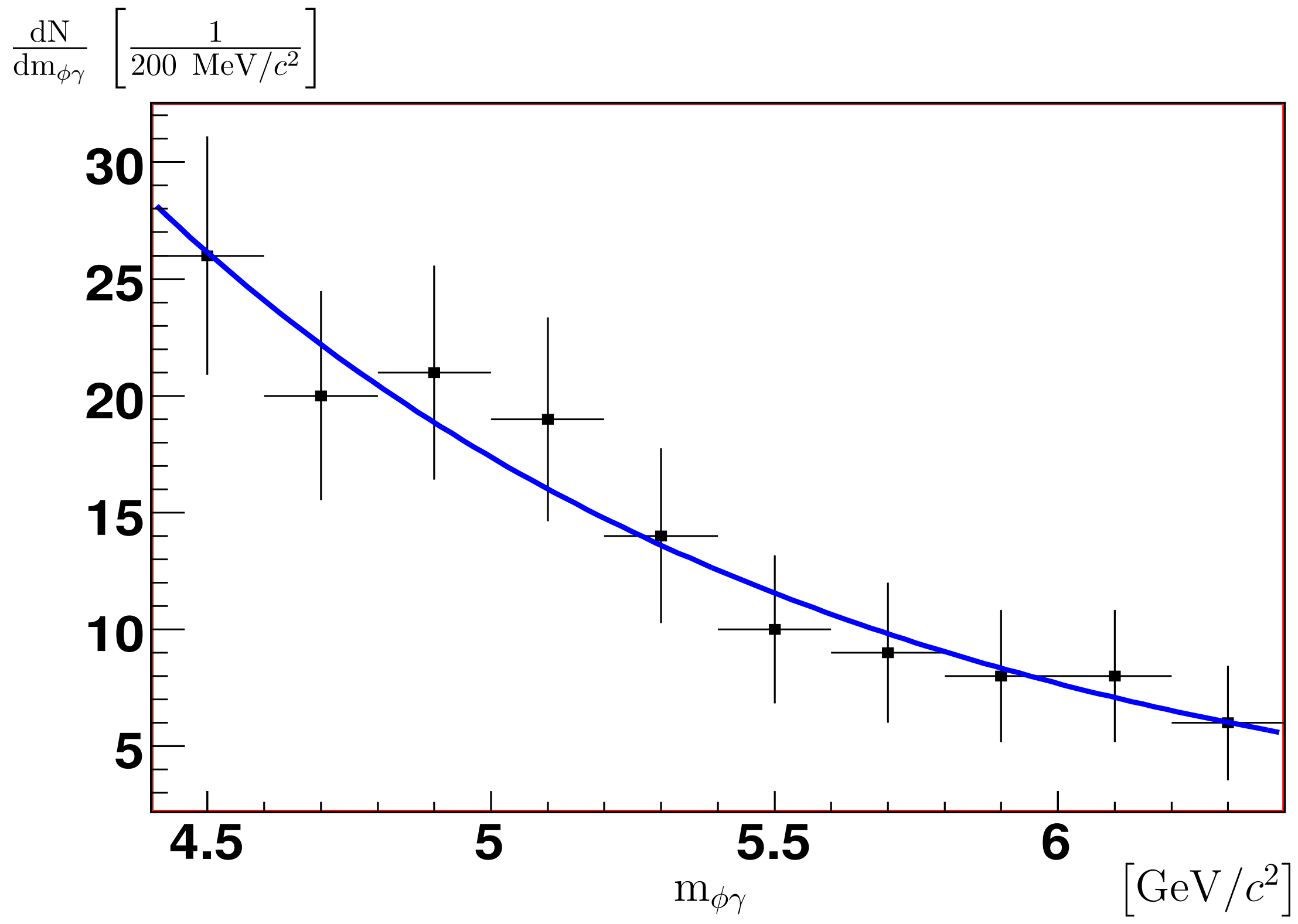


Figure 6: $b\bar{b}$-inclusive background mass distribution after relaxed selection for limited Monte Carlo statistics [3]. The curve represents the fit with an exponential function.

One can perform the fits of the left and right sidebands separately with the function (3.2) to determine the parameters $f_i$ and $\tau_i$ of the different background components. In practice it is more convenient to perform the fit of the sum of the left and right sidebands to determine the leading exponential factors $\tau_i$ and only then perform the separate fits of the left and right sidebands using the common set of $\tau_i$ to determine the relative composition $f_i$ for two sidebands separately ($f_i^L$ for the left sideband and $f_i^R$ for the right sideband respectively). For the limited Monte Carlo data samples it has been shown [3] that two exponential functions are enough to describe both left and right sidebands, see Figures 9 and 10.

The time parameters $\tau_i$ were determined from a fit to simulated background distributions (see Figures 9 and 10) and were found to be 0.45 ps and 8.7 ps. The decay time of short and long lived background component are significantly smaller and significantly larger than the nominal lifetime of $B_s$ meson [51] correspondingly[7].

The coherent description of the background events as a function of two variables, the reconstructed mass of the $B_s$ candidate and its proper time, is achieved by combining the separate parameterizations (3.1) and (3.2) assuming a smooth dependence of coefficients $f_i$ from (3.2) on the reconstructed mass $m_{\phi\gamma}$: $f_i \longrightarrow f_i\,(m_{\phi\gamma})$. With such a substitution

[7]Probably due to these large differences the resulting uncertainties on the physics parameters $\mathcal{A}^{\Delta}$, $\mathcal{C}$ and $\mathcal{S}$ were found to be almost insensitive with respect to the actual composition of the background [3].

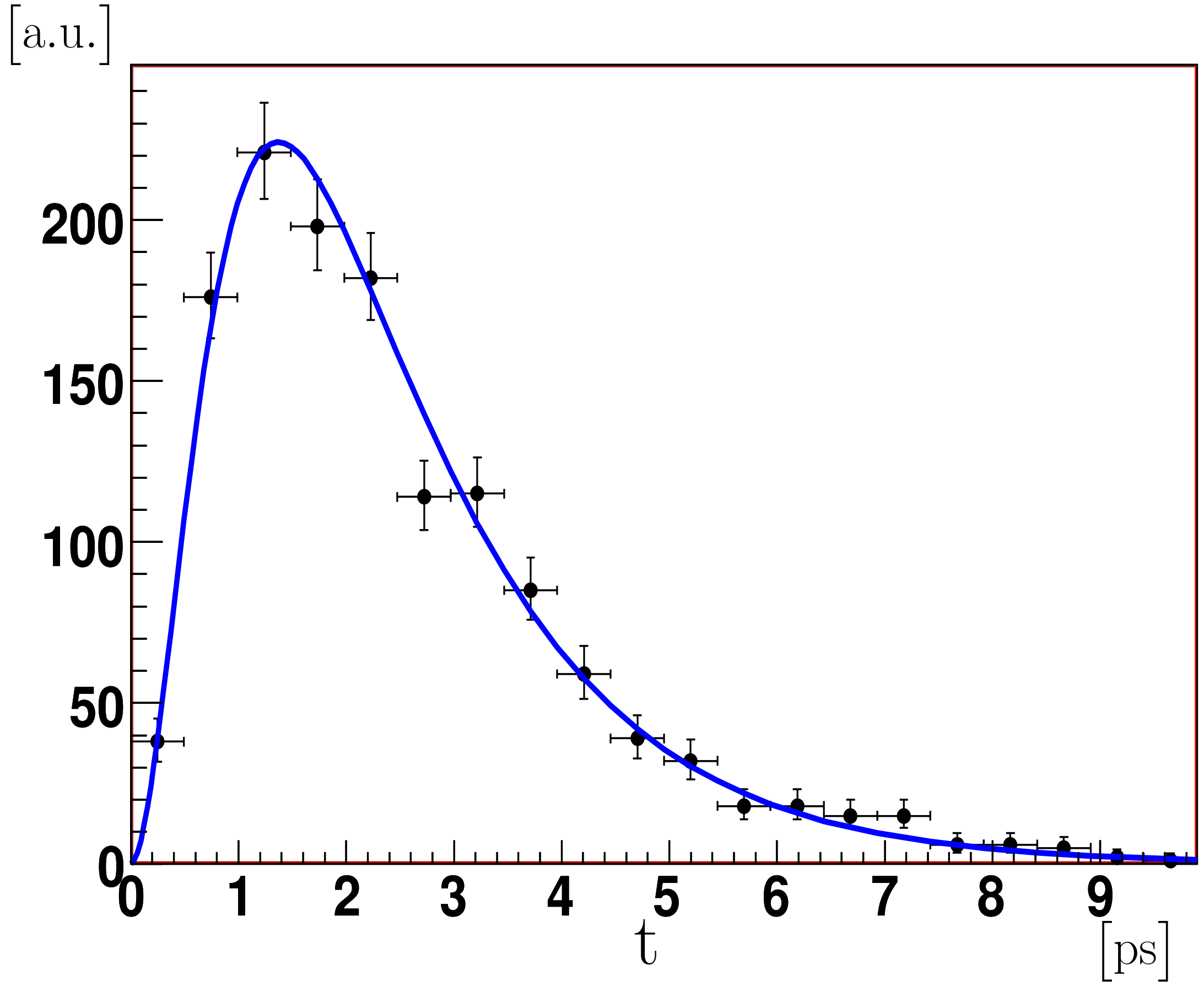


Figure 7: The proper time distribution for $B_s^0 \to \phi\gamma$ events [3] fitted with the function $f(t) = e^{-t/\tau}\epsilon_{B_s^0\to\phi\gamma}(t)$, where $\epsilon_{B_s^0\to\phi\gamma}(t)$ is defined by Eq. (3.3).

the distribution of the background as a function of two variables can be written as

$$f_b(m_{\phi\gamma}, t) \propto \left[e^{-\mu m_{\phi\gamma}} P(m_{\phi\gamma})\right] \times \left[\epsilon_{B_s^0\to\phi\gamma}(t) \sum_i f_i(m_{\phi\gamma}) e^{-t/\tau_i}\right], \tag{3.4}$$

where the first part is a function of reconstructed mass only, and describes the leading dependence of $f_b(m_{\phi\gamma}, t)$ on the reconstructed mass $m_{\phi\gamma}$, and the second part has a rather modest dependence on the reconstructed mass but is responsible to describe the dependency on the reconstructed proper time. A linear parameterization of $f_i(m_{\phi\gamma})$ has been used:

$$f_i(m_{\phi\gamma}) = f_i^0 + \delta f_i (m_{\phi\gamma} - m_{B_s}) \tag{3.5}$$

The parameters $f_i^0$ and $\delta f_i$ could be calculated from the values $f_i^L$ and $f_i^R$ determined from the two sidebands separately. Instead of using the functions (3.4) and (3.5) a simultaneous fit to the low-mass and high-mass sideband regions in the $(m_{\phi\gamma}, t)$-plane has been performed. It turned out that choosing the simplest expression for $P(m_{\phi\gamma}) = 1$ and the linear parameterization of $f_i$ a sufficiently good description of the time behaviour

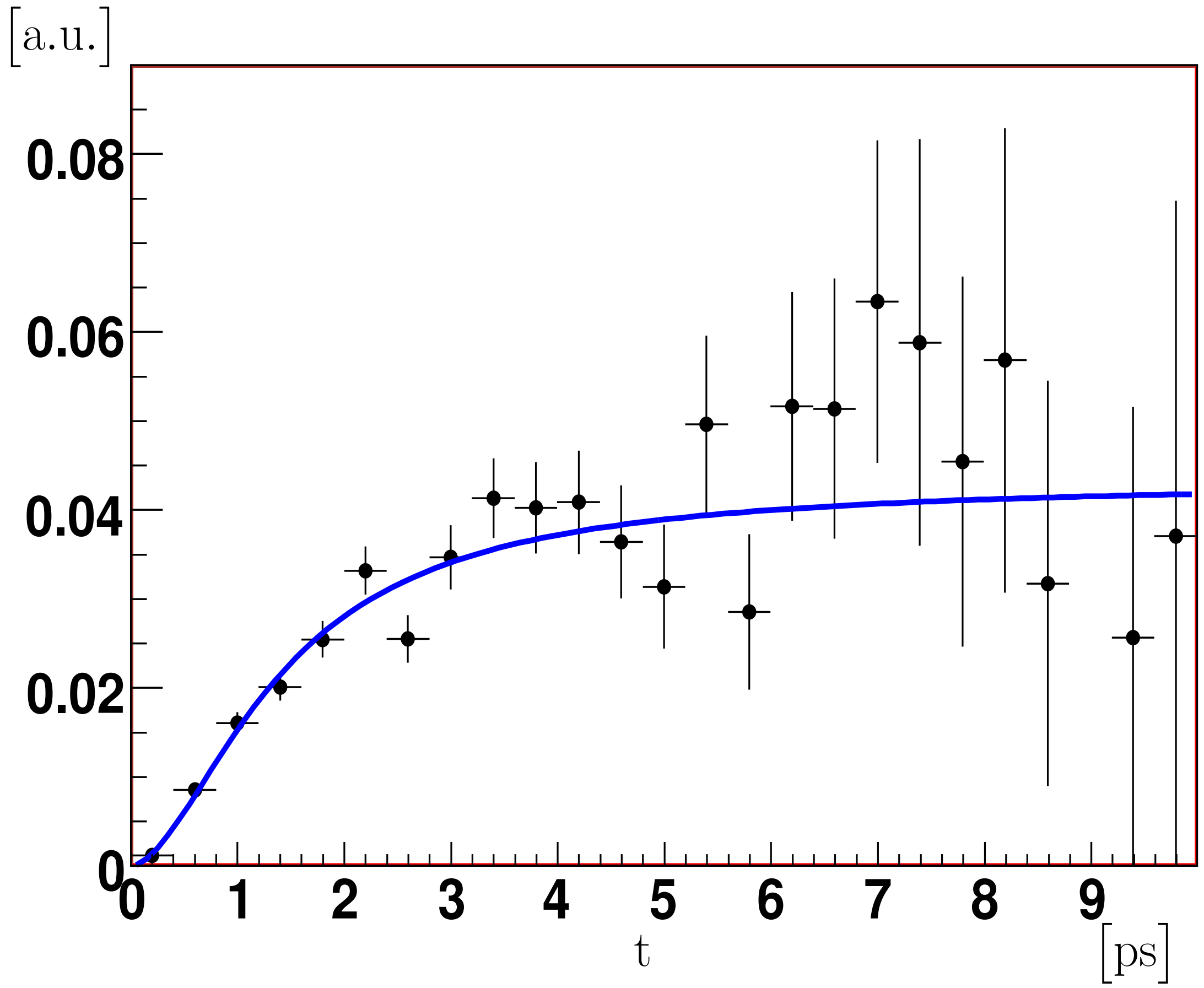


Figure 8: The proper time acceptance $\epsilon_{B_s^0\to\phi\gamma}(t)$ for $B_s^0 \to \phi\gamma$ events fitted with the function (3.3) [3].

is achieved. As shown in Table 7 all necessary background parameters are well determined [3].

The convergence of the fit can be improved using proper initial values, determined from the previous one-dimensional fits. However if some parameter from the fit where all parameters are left free tends to differ from the initial (and expected) value, this case requires a separate detailed investigation, as well as the case of a bad fit quality when all parameters fixed at their expected values. This could be a signature of some of the basic assumptions not being valid. For example it could indicate that the simple linear model (3.5) is not valid and one needs to use more complicated dependencies, e.g. a second order polynomial:

$$f_i\left(m_{\phi\gamma}\right) = f_i^0 + \delta f_i\left(m_{\phi\gamma} - m_{B_s}\right) + f_i''\left(m_{\phi\gamma} - m_{B_s}\right)^2 .$$

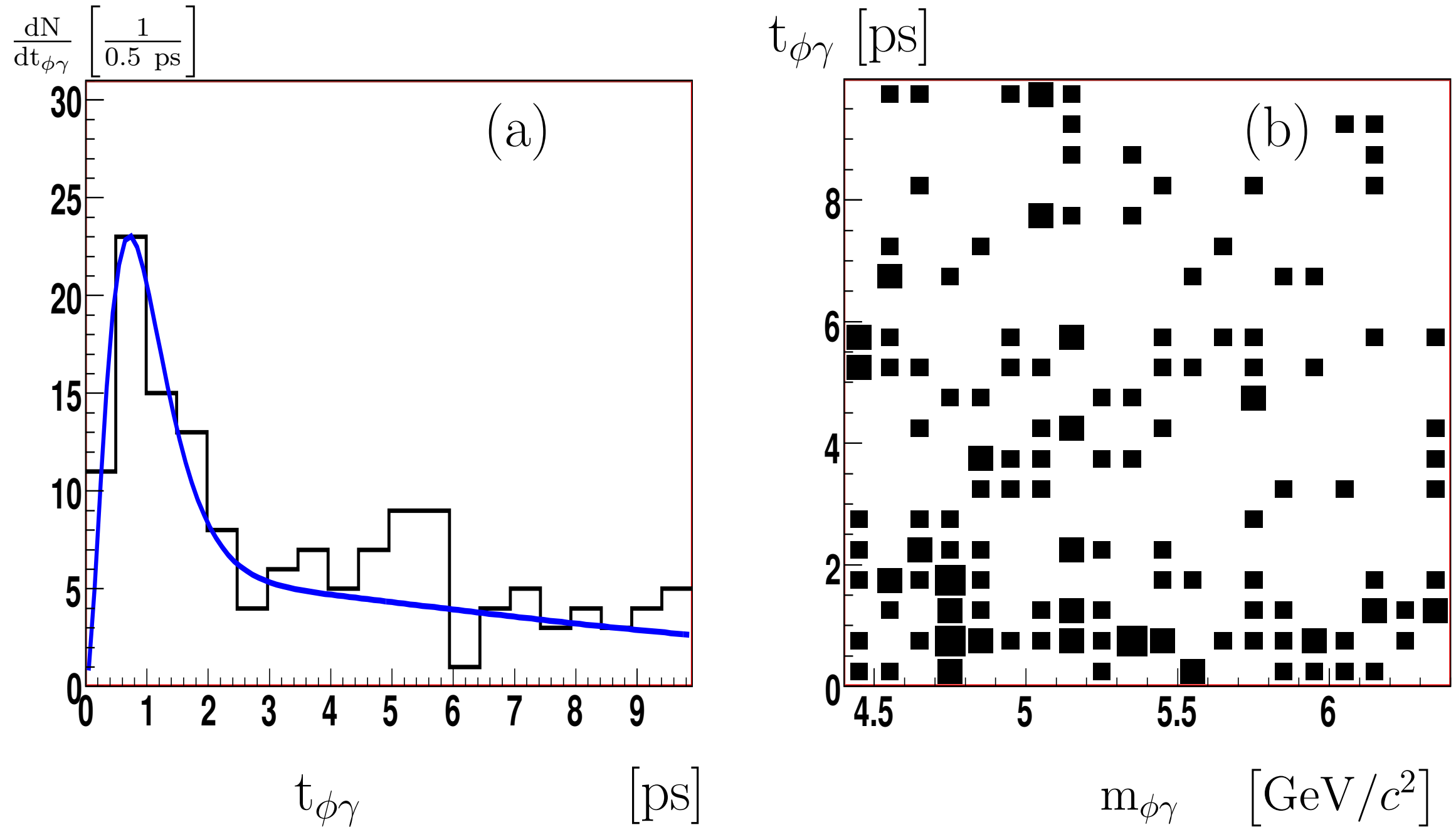


Figure 9: a) Proper time distribution of background events; b) proper time vs. invariant mass distribution for $b\bar{b}$-inclusive events after the relaxed selection cuts.

### 3.4 Fit procedure and results

After a detailed study and verification of the background parameterization models (3.4), using the low-mass and high-mass sideband regions in the $(m_{\phi\gamma}, t)$-plane, one can start the usual procedure of the unbinned maximum likelihood fit. The construction of the simplest likelihood function is described in Appendix B. We are planning to use the technique of blind analysis and instead of the physical parameter $\mathcal{A}^{\Delta}$ use some linear function $\mathcal{A}^{\Delta}_{\text{blind}} = \alpha\mathcal{A}^{\Delta} + \beta$, with "unknown" parameters $\alpha$ and $\beta$.

We are going to start with the fit ignoring the tagging information. All parameters, including the normalization of signal and background, except $\mathcal{A}^{\Delta}_{\text{blind}}$ could be fixed at their expected values. The result of this fit needs to be compared with the result of the fit where the background parameters are kept free[8].

Also, we are planning to perform series of fits where the generic signal parameters,

[8]Probably here one can also use free parameters for the signal normalization.

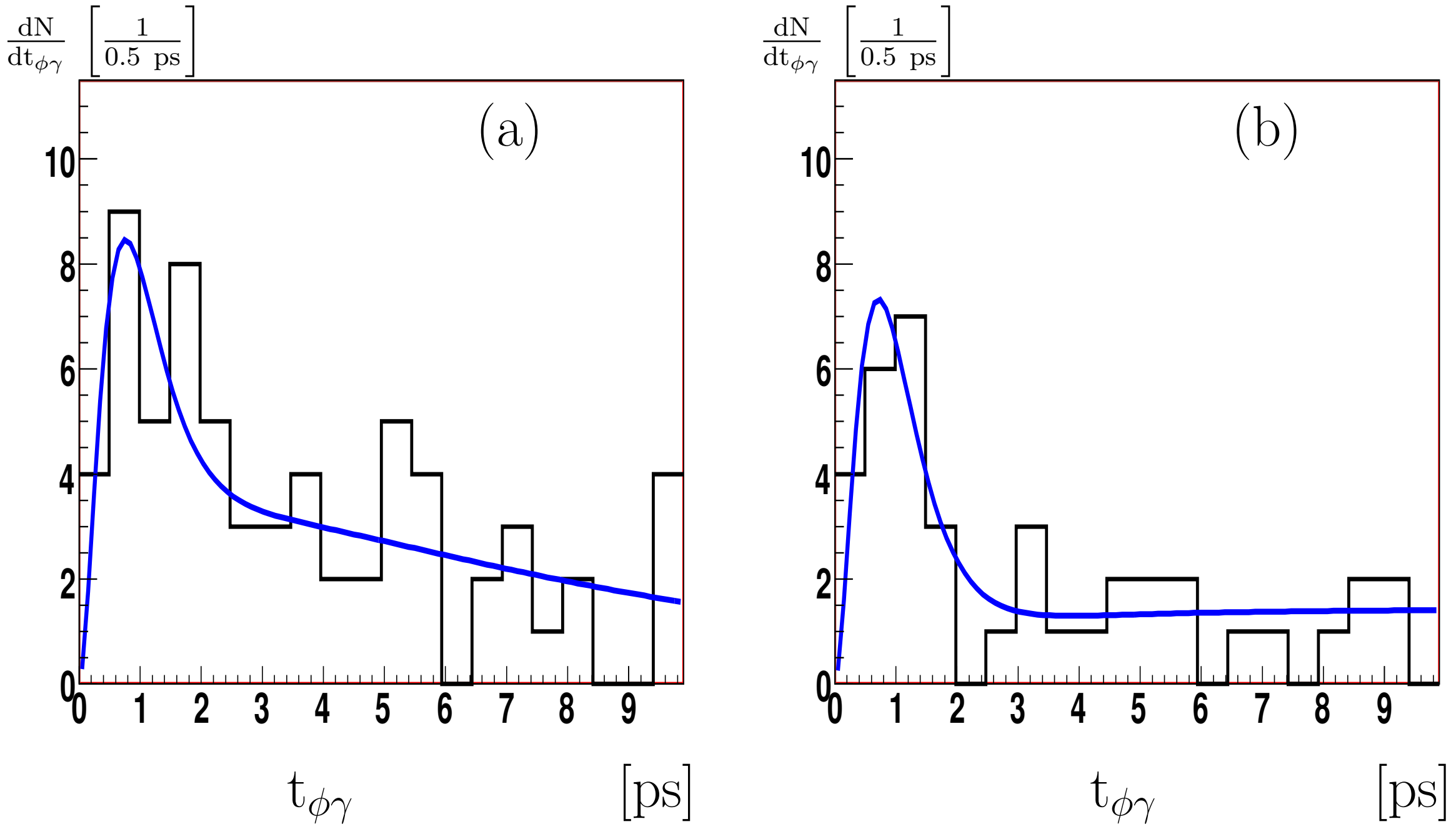


Figure 10: Proper time distribution of background events [3]: a) for left-sideband region with invariant mass 4.4–5.1 GeV/$c^2$; b) for right-sideband region with invariant mass 5.7–6.4 GeV/$c^2$.

like the nominal lifetime of the $B_s$ meson, and the value of $\Delta\Gamma_s$ are allowed to vary [9].

It is worth to note here that $\Delta\Gamma_s$ must be fixed or constrained in the fit using external measurements, e.g., the precise measurements from $B_s^0 \to \phi J/\psi$ channel, where $\Delta\Gamma_s$ is expected to be measured with 6-8 % precision [10].

For fits of the considered proper lifetime range and the loose constraint on $\Delta\Gamma_s$ possible variations in the nominal lifetime of the $B_s$ meson are absorbed in $\Delta\Gamma_s$ and $\mathcal{A}^{\Delta}_{\rm blind}$, without affecting the background parameterizations. In a similar way variations of $\Delta\Gamma_s$, either as a fixed or constraint parameter, are absorbed by the released lifetime of $B_s$ meson and $\mathcal{A}^{\Delta}_{\rm blind}$, without affecting the shape of the background.

Toy Monte Carlo studies [3, 4] show that the uncertainty on parameter $\mathcal{A}^{\Delta}$ for the fit procedure that we are planning to use in the first measurement (not using the tagging information) is the same as in case of the combined analysis exploiting the flavour tagging

[9] Since for small $\Delta\Gamma_s$ the expression (1.11) looks as:

$$\Gamma_{B_s^0 \to \Phi^{\mathcal{CP}}\gamma}(t) \approx |A|^2 e^{-\Gamma_s t}\left[1 - \frac{\mathcal{A}^{\Delta}\Delta\Gamma_s t}{2} + \frac{1}{2}\left(\frac{\Delta\Gamma_s t}{2}\right)^2 + ...\right] \tag{3.6}$$

we are sensitive mainly to the product $\left(\mathcal{A}^{\Delta}\Delta\Gamma_s\right)$, and the obtained value of $\mathcal{A}^{\Delta}_{\rm blind}$ should have a very strong dependence on the value of $\Delta\Gamma_s$ used. However, for these series of fits we do not care about the value of $\mathcal{A}^{\Delta}_{\rm blind}$.

Table 7: The errors on the background shape parameters obtained from 2D simultaneous fit of the low-mass and high-mass sideband regions on the on the $(m_{\phi\gamma}, t)$-plane for statistics equivalent to 2 $fb^{-1}$ of integrated luminosity. The values of input parameters are specified in the column "Input value". The parameter $c_1^0 = 1$ is fixed.

| Parameter | Input/Expected Value | Error |
|---|---|---|
| $c_1^0$ | 1 | — |
| $\delta c_1$ | 0.041 | 0.20 |
| $1/\tau_1$ | 2.23 | 0.04 |
| $c_2^0$ | 0.025 | 0.0014 |
| $\delta c_2$ | -0.007 | 0.004 |
| $1/\tau_2$ | 0.118 | 0.005 |

information[10]. The distribution of the fitted values of $\mathcal{A}^{\Delta}$ for $\mathcal{O}(10^4)$ toy experiments, each of them equivalent to 2 $fb^{-1}$ of accumulated data is shown in Figure 11. The expected uncertainty has been estimated to be $\sim 0.22$ for a $\frac{\Delta\Gamma_s}{\Gamma_s}$ value of 0.12 [3].

The fitted values of $\mathcal{A}^{\Delta}$ are practically unbiased and the errors returned by the fit are reliable. Figure 12 shows the corresponding pull distribution.

The expected statistical uncertainty for $\mathcal{A}^{\Delta}$ is independent of the value of $\mathcal{A}^{\Delta}$ and scales with square root of the luminosity: $\sigma_{\mathcal{A}^{\Delta}} \propto \mathcal{L}^{-\frac{1}{2}}$ [4]. In case of a combined fit of tagged and untagged events the parameter $\mathcal{A}^{\Delta}$ is found to be independent of the parameters $\mathcal{S}$ and $\mathcal{C}$: the correlation coefficients do not exceed 2 % [3,4]. The uncertainties and corresponding pulls are summarized in Table 8.

The result of a toy experiment with $\mathcal{A}^{\Delta} = 0.4$ for a data set equivalent to 2 $fb^{-1}$ of accumulated data is shown in Figure 13.

[10] For flavour tagging the following parameters have been used: the tagging efficiency $\epsilon^{tag} = 0.610 \pm 0.002$ and the mistag rate $\omega = 0.30$ [3,4,15]. A single tagging category has been assumed here. Considering multiple tagging categories has the potential to improve the results on $\mathcal{S}$ and $\mathcal{C}$ parameters which depend on the tagging [15].

Table 8: The uncertainties for parameters $\mathcal{A}^{\Delta}$, $\mathcal{C}$, and $\mathcal{S}$ from the combined fit of tagged and untagged samples and the corresponding pull parameters, obtained for $\mathcal{O}(10^4)$ toy experiments, each of them equivalent to 2 $fb^{-1}$ of accumulated data.

| Parameter | Uncertainty | Pull mean | Pull sigma |
|---|---|---|---|
| $\mathcal{A}^{\Delta}$ | $0.217 \pm 0.002$ | $0.03 \pm 0.01$ | $1.03 \pm 0.01$ |
| $\mathcal{C}$ | $0.115 \pm 0.001$ | $0.01 \pm 0.01$ | $1.08 \pm 0.01$ |
| $\mathcal{S}$ | $0.114 \pm 0.001$ | $0.02 \pm 0.01$ | $1.00 \pm 0.01$ |

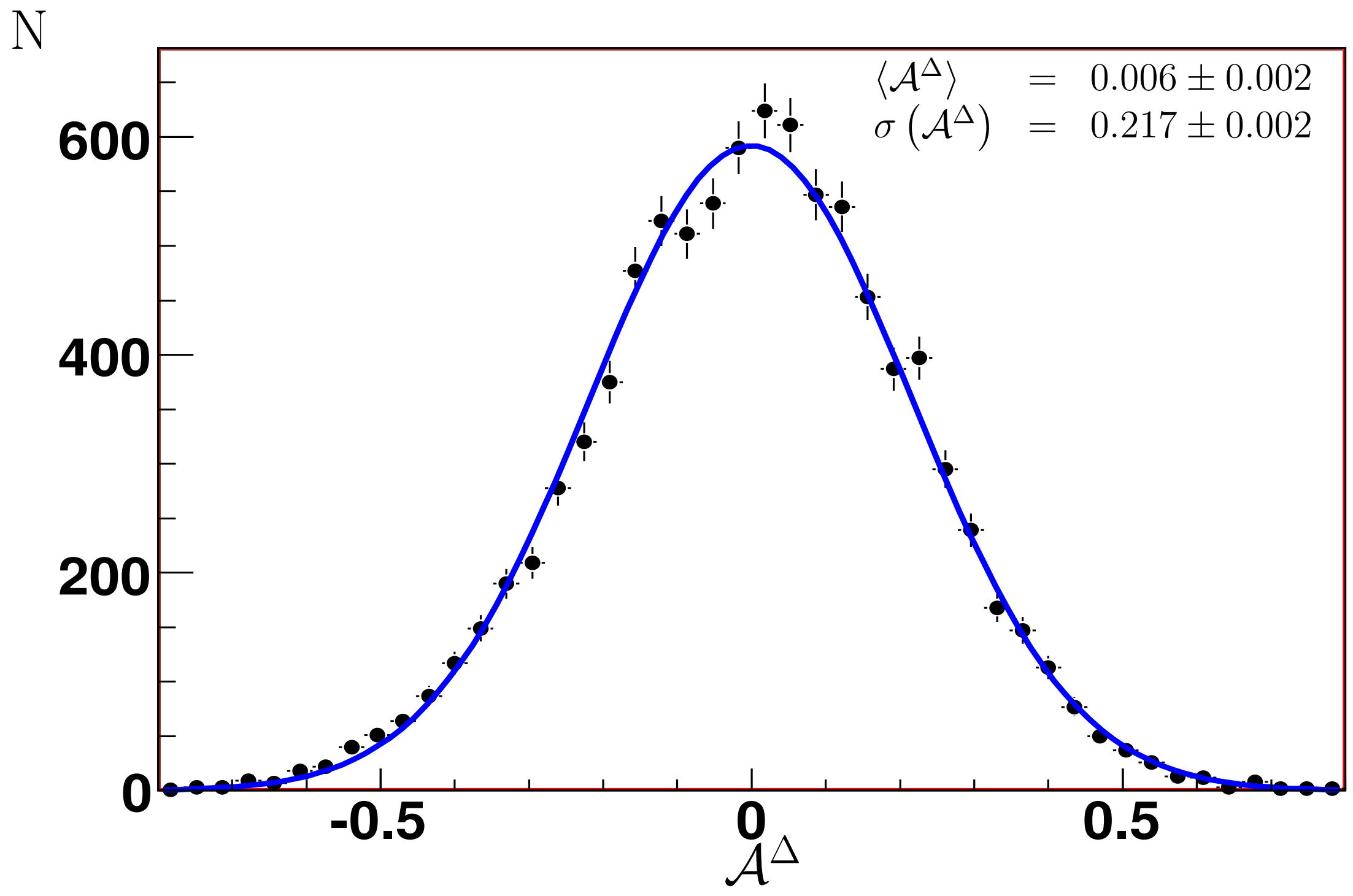


Figure 11: The distributions of fit results for $\mathcal{A}^{\Delta}$ for $\mathcal{O}(10^4)$ toy experiments, where $\mathcal{A}^{\Delta} = 0$ has been used as input value. The curve shows a Gaussian fit, with mean value $0.006 \pm 0.002$ and width $0.217 \pm 0.002$.

### 3.5 Proper time acceptance function

The determination of the proper time acceptance function is one of the most critical aspects of the whole analysis and probably the major source of systematical uncertainties. For small $\Delta\Gamma_s$ and $\mathcal{A}^{\Delta}$, neglecting the proper time resolution and using the over-simplified representation of the proper time acceptance function as the linear function $\epsilon(\tau) = \epsilon_0 (1 + \epsilon_1 \tau)$, one gets the observed distribution for signal events as function of the dimensionless variable $\mathrm{t}' = \Gamma_s \mathrm{t}$:

$$\frac{\mathrm{d}}{\mathrm{dt}'}\mathcal{N}_{\mathrm{B_s^0}\to\Phi^{\mathcal{CP}}\gamma} \propto \mathrm{e}^{-\mathrm{t}'}\left[1 - \left\{\mathcal{A}^{\Delta}\left(\frac{\Delta\Gamma_s}{2\Gamma_s}\right) + \epsilon_1\right\}\mathrm{t}' + \frac{1}{2}\left(\frac{\Delta\Gamma_s}{2\Gamma_s}\right)^2 (\mathrm{t}')^2 + ...\right].$$

Using the estimate of $\frac{\Delta\Gamma_s}{\Gamma_s} \approx 0.12$, one obtains that to achieve an acceptance related uncertainty $\sigma\left(\mathcal{A}^{\Delta}\right) \sim 0.1$, which corresponds to half of the expected statistical precision for 2 fb$^{-1}$ of accumulated data, one needs to know the slope of the proper time acceptance function with a precision of $0.006 \times \Gamma_s$.

Three methods to extract the acceptance function $\epsilon_{\mathrm{B_s^0}\to\phi\gamma}(\tau)$ for $\mathrm{B_s^0} \to \phi\gamma$ decay from the data are proposed. One can extract the proper time acceptance function using either the calibration channel $\mathrm{B^0} \to \mathrm{K^{*0}}\gamma$ or the calibration channel $\mathrm{B_s^0} \to \phi \mathrm{J}/\psi$. Note here that

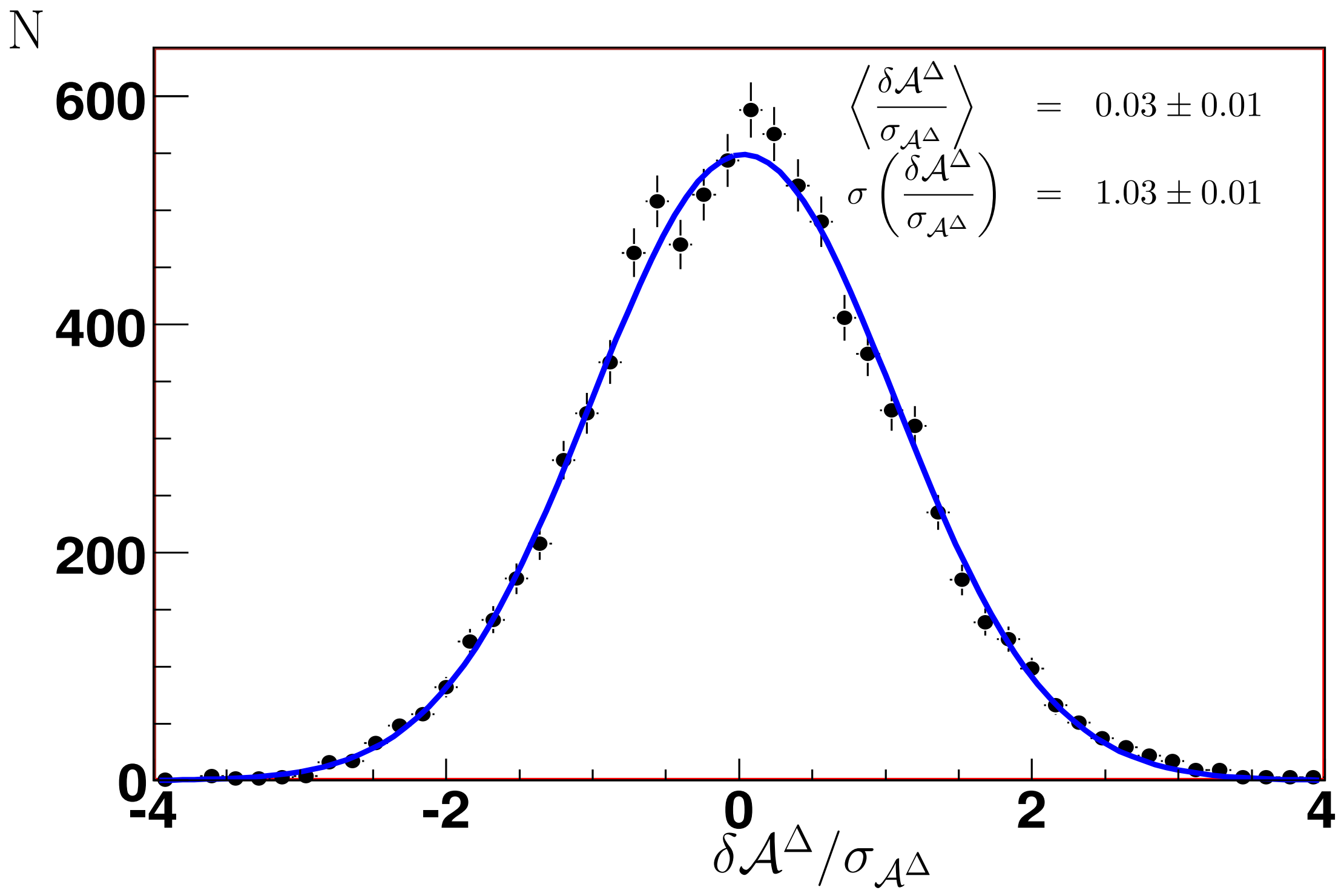


Figure 12: The pull distributions of $\mathcal{A}^{\Delta}$ for $\mathcal{O}(10^4)$ toy experiments. The curve shows a Gaussian fit, with mean value $0.03 \pm 0.01$ and width $1.03 \pm 0.01$.

to achieve the desired statistical precision in the knowledge of the proper time acceptance function, only calibration channels with large event yields can be used. Alternatively to the proposed calibration channels the proper time acceptance function can be extracted per event using a "lifetime swimming" method.

**Extraction of proper time acceptance using $\mathrm{B}^0 \to \mathrm{K}^{*0}\gamma$ decays**

The first method relies on the fact that for $\mathrm{B}^0 \to (\mathrm{K}^{*0} \to \mathrm{K}^+\pi^-)\,\gamma$ the theoretical lifetime distribution is simple and well known. For these decays the shape of the proper time acceptance function can easily be extracted from the measured proper time distribution. For $7 \times 10^4$ reconstructed $\mathrm{B}^0 \to \mathrm{K}^{*0}\gamma$ events, neglecting backgrounds, the parameters for the acceptance function $\epsilon_{\mathrm{B}^0\to\mathrm{K}^{*0}\gamma}(\tau)$ in Eq. (3.3) can be determined with the following precision:

$$\begin{aligned} \sigma_{\mathrm{a}} &= 0.01 \\ \sigma_{\mathrm{c}} &= 0.02 \quad . \end{aligned}$$

It has been shown that a bias $\delta\mathrm{a} = 0.04$ in the parameter a translates into a shift of the observed value of $\mathcal{A}^{\Delta}$, $\delta\mathcal{A}^{\Delta} \sim 0.2$. Under the assumption that the shape of the proper time acceptance function is the same for $\mathrm{B}^0_{\mathrm{s}} \to \phi\gamma$ and $\mathrm{B}^0 \to \mathrm{K}^{*0}\gamma$ this method provides

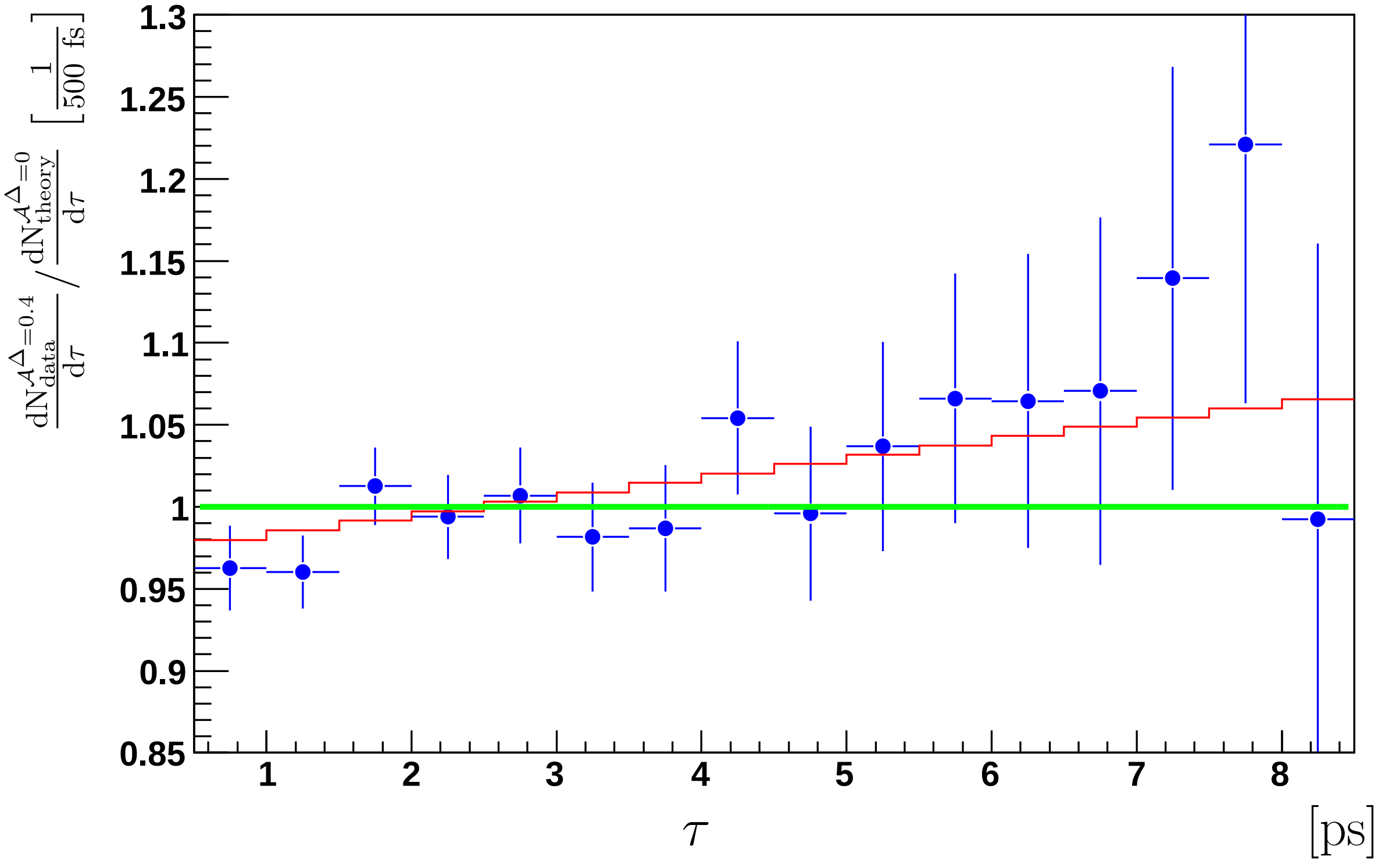


Figure 13: The ratio of Monte Carlo data for $B_s^0 \to \phi\gamma$, simulated with $\mathcal{A}^{\Delta} = 0.4$, to the theory expectation for $\mathcal{A}^{\Delta} = 0$, as a function of the $B_s^0$ proper lifetime for a toy experiment equivalent to 2 $fb^{-1}$ of accumulated data. The red histogram indicates the expected dependence.

an adequate statistical precision for the determination of the proper time acceptance function. The systematic uncertainty in $\mathcal{A}^{\Delta}$ related to the statistical errors of the above acceptance parameters corresponds to a quarter of the expected statistical error of $\mathcal{A}^{\Delta}$ for a nominal year of data taking.

The off-line and HLT selection criteria for $B_s^0 \to \phi\gamma$ and $B^0 \to K^{*0}\gamma$ events are very similar, and the L0 trigger condition are identical. However, due to the small energy release in the decay $\phi \to K^+K^-$ the vertex resolution for $B_s^0 \to \phi\gamma$ events is significantly worse than for $B^0 \to K^{*0}\gamma$ events, resulting in a difference in the proper time acceptance for both channels. A Monte Carlo study with large samples of $B_s^0 \to \phi\gamma$ and $B^0 \to K^{*0}\gamma$ clearly exhibits this difference:

$$\left.\begin{aligned} a &= 0.84 \pm 0.04\ \mathrm{ps}^{-1} \\ c &= 2.16 \pm 0.08 \end{aligned}\right\} \quad \text{for } B_s^0 \to \phi\gamma \tag{3.7a}$$

$$\left.\begin{aligned} a &= 1.00 \pm 0.04\ \mathrm{ps}^{-1} \\ c &= 2.20 \pm 0.09 \end{aligned}\right\} \quad \text{for } B^0 \to K^{*0}\gamma \tag{3.7b}$$

Here the uncertainties are statistical only. The Monte Carlo samples used for this

comparison are a factor 6(30) smaller than the the yield expected for data for the $B^0_s \to \phi\gamma(B^0 \to K^{*0}\gamma)$ channel correspondingly.

One clearly observes a significant difference in the parameter a. Without taking this difference into account one would get a large bias in the measurement of $\mathcal{A}^{\Delta}$. A proper re-weighting of the $B^0 \to K^{*0}\gamma$ events relying on Monte Carlo can be used to correct the difference.

### Extraction of proper time acceptance using "lifetime swimming" acceptance

A method to evaluate the per-event acceptance as a function of proper decay time by sliding the decay tree along the B flight direction is under investigation. For a given $B_{(s)}$ decay, the whole decay tree is moved along the $B_{(s)}$ flight direction, leaving the $B_{(s)}$ decay kinematics unchanged.

The effect on the proper time acceptance of an impact parameter cut on any final state particle or a cut on the $B_{(s)}$-meson direction angle can be evaluated event by event[11]. This method has been used by CDF to measure the $B^+$ lifetime in the decay mode $B^+ \to D^0\pi^+$ [61]. Its application by LHCb for the measurement of the $B_d$ lifetime in $B_d \to D^-\pi^+$ decays is described in Ref. [62].

As a proof of principle, this idea has also been applied to the decay mode $B^0 \to K^{*0}\gamma$. Fully simulated $B^0 \to K^{*0}\gamma$ events have been reconstructed using standard LHCb software and selected using particle identification, vertex information and invariant masses. In addition, the impact parameters of kaons and pions are required to be above 0.1 mm. The B direction angle is required to be less than 0.05 mrad. The last two requirements are expected to distort the proper time distribution. The before-acceptance proper time distribution, which is an exponential function convoluted by a resolution function, is assumed to be known. The proper time distribution for signal events is shown by the open histogram in Figure 14. Superimposed is the sum of the normalized per event probability density functions (pdf) of proper time, which is the product of the known before-acceptance proper time distribution and the per-event acceptance function. As expected, the two distributions match very well.

This demonstrates that the per-event pdf defined in this way is a good description of the proper time distribution and can be directly used in maximum likelihood fitting for the extraction of physical parameters.

Alternatively, we can evaluate the average acceptance function by dividing the sum of the per-event pdfs by the before-acceptance proper time distribution. Figure 15 shows the average acceptance functions estimated using truth information and using the per-event acceptance method. The ratio between them is shown in Figure 16, which shows no dependence on proper time above 0.3 ps.

While these preliminary results are encouraging, more work is needed to get this method to work for $B \to V\gamma$. In particular we need to understand how to use this method in the presence of background.

[11] The $B_{(s)}$ direction angle variable is defined as the angle between the the $B_{(s)}$ reconstructed momentum and the $B_{(s)}$ flight direction, evaluated as the vector from the primary vertex to the $B_{(s)}$ decay vertex.

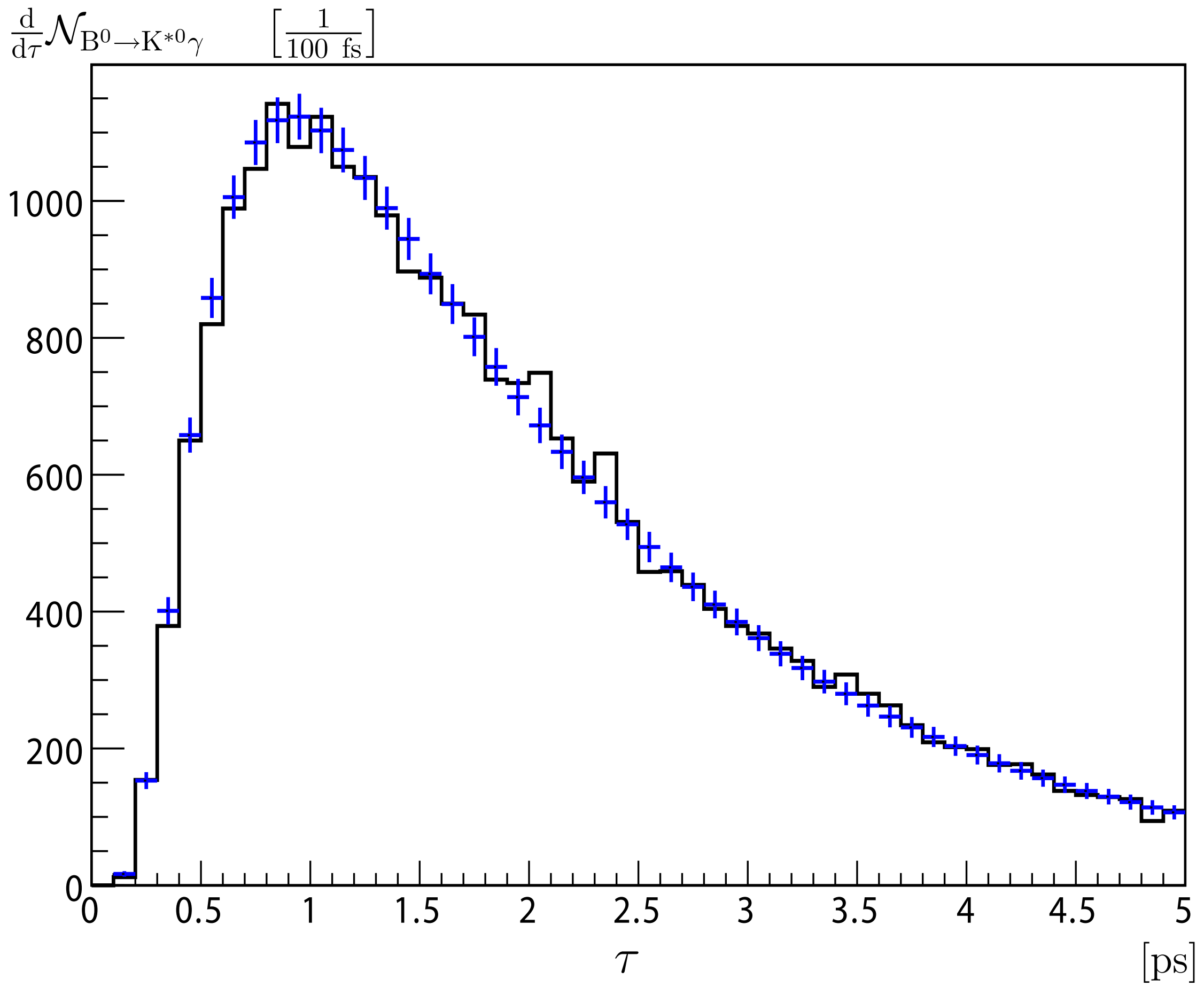


Figure 14: The proper time distribution of signal events $B^0 \to K^{*0}\gamma$ (open histogram), superimposed with the sum of normalized per event probability density functions (blue histogram with error bars).

**Extraction of proper time acceptance function from $B_s^0 \to \phi J/\psi$ decay**

The third approach relies on the possibility of selecting a large sample of *lifetime-unbiased* $B_s^0 \to \phi J/\psi$ events [10, 63]. *If* such selection can be done without cuts on the impact parameters of the kaons the events can be used for the determination of the acceptance $\epsilon_{B_s^0 \to \phi\gamma}(\tau)$. In this case the shape of the proper time acceptance function can be obtained using the relative acceptance of $B_s^0 \to \phi J/\psi$ events, where one applies the cuts, similar to the selection cuts for $B_s^0 \to \phi\gamma$ events, to the distribution without such cuts. Clearly one needs to mimic the decay $B_s^0 \to \phi\gamma$ as close as possible, and it is necessary to use $B_s$ vertices reconstructed from dikaon system only. Cuts on the $B_s$-vertex quality, direction angle, $\chi^2$ of the lifetime fit and other cuts need to be applied with respect to this "fake" $B_s$ vertex for the "lifetime-unbiased" selected $B_s^0 \to \phi J/\psi$ events. Using this technique one can obtain the relative (with respect to the *lifetime-unbiased* selection) $\epsilon_{B_s^0 \to \phi J/\psi}(\tau)$. E.g. applying the cut on the direction angle of $B_s$, one can evaluate its acceptance, as

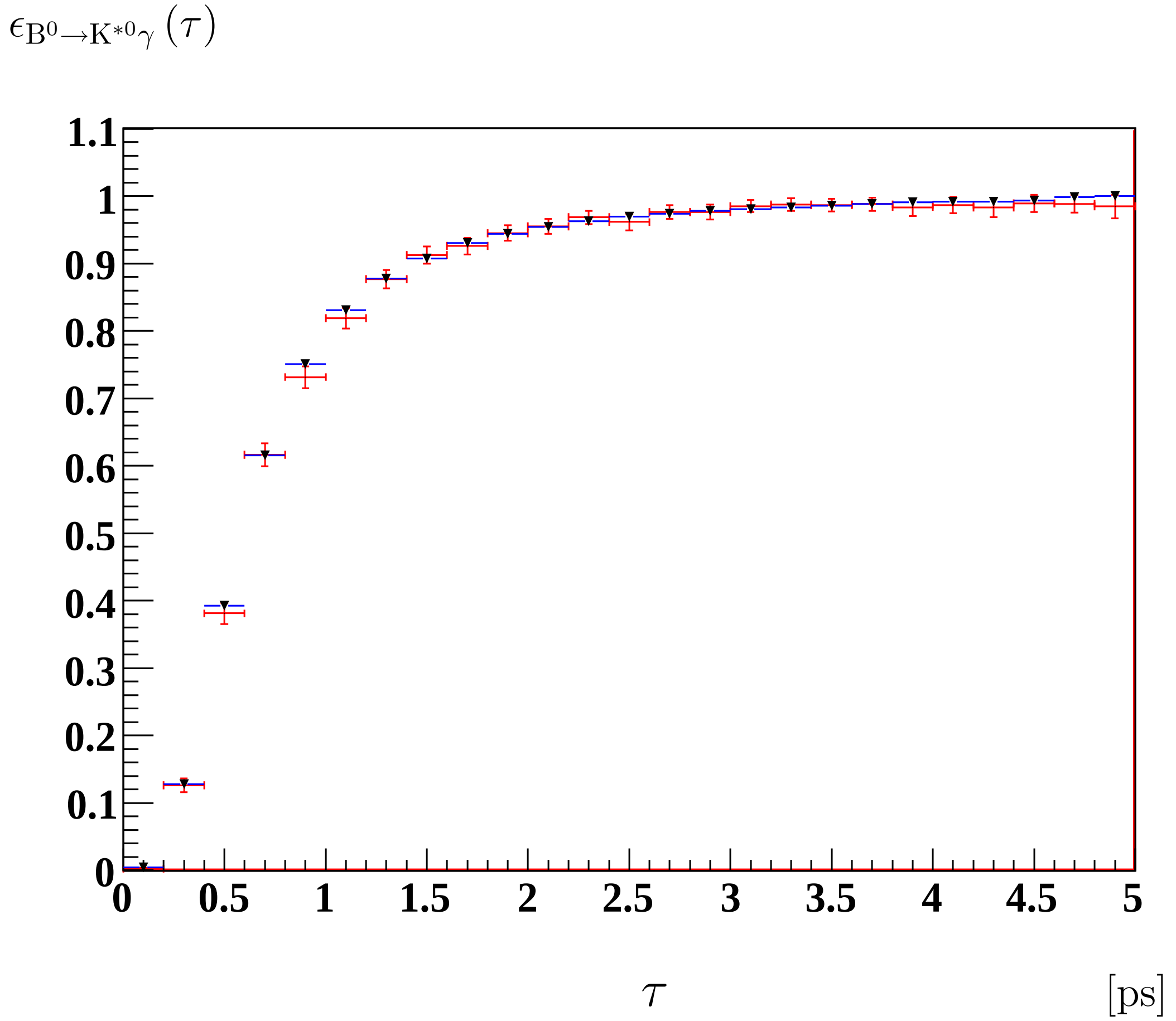


Figure 15: The proper time acceptance functions determined using Monte Carlo truth information $\epsilon^{MC}_{B^0 \to K^{*0}\gamma}(\tau)$ (red crosses) and "per-event-acceptance" method $\epsilon^{acc}_{B^0 \to K^{*0}\gamma}(\tau)$ (black triangles with blue error bars).

shown in Figure 17.

Due to the difference in the event kinematics, triggering, reconstruction and identification of $J/\psi \to \mu^+\mu^-$ with respect to energetic photon in $B^0_s \to \phi\gamma$, the acceptances $\epsilon_{B^0_s \to \phi J/\psi}(\tau)$ and $\epsilon_{B^0_s \to \phi\gamma}(\tau)$ will be different. However, we hope that with the proper re-weighting of $B^0_s \to \phi J/\psi$ events one could obtain a re-weighted acceptance function $\epsilon^*_{B^0_s \to \phi J/\psi}(\tau)$ for which the ratio of lifetime acceptance functions $r_\epsilon(\tau)$, defined as

$$r_\epsilon(\tau) \equiv \frac{\epsilon_{B^0_s \to \phi\gamma}(\tau)}{\epsilon^*_{B^0_s \to \phi J/\psi}(\tau)} \qquad (3.8)$$

is constant.

The available Monte Carlo event samples of $B^0_s \to \phi\gamma$ and $B^0_s \to \phi J/\psi$ events are limited and do not allow to prove or disprove this hypothesis. To get a feeling of the validity of this procedure the available event statistics were increased by releasing all but one selection cut.

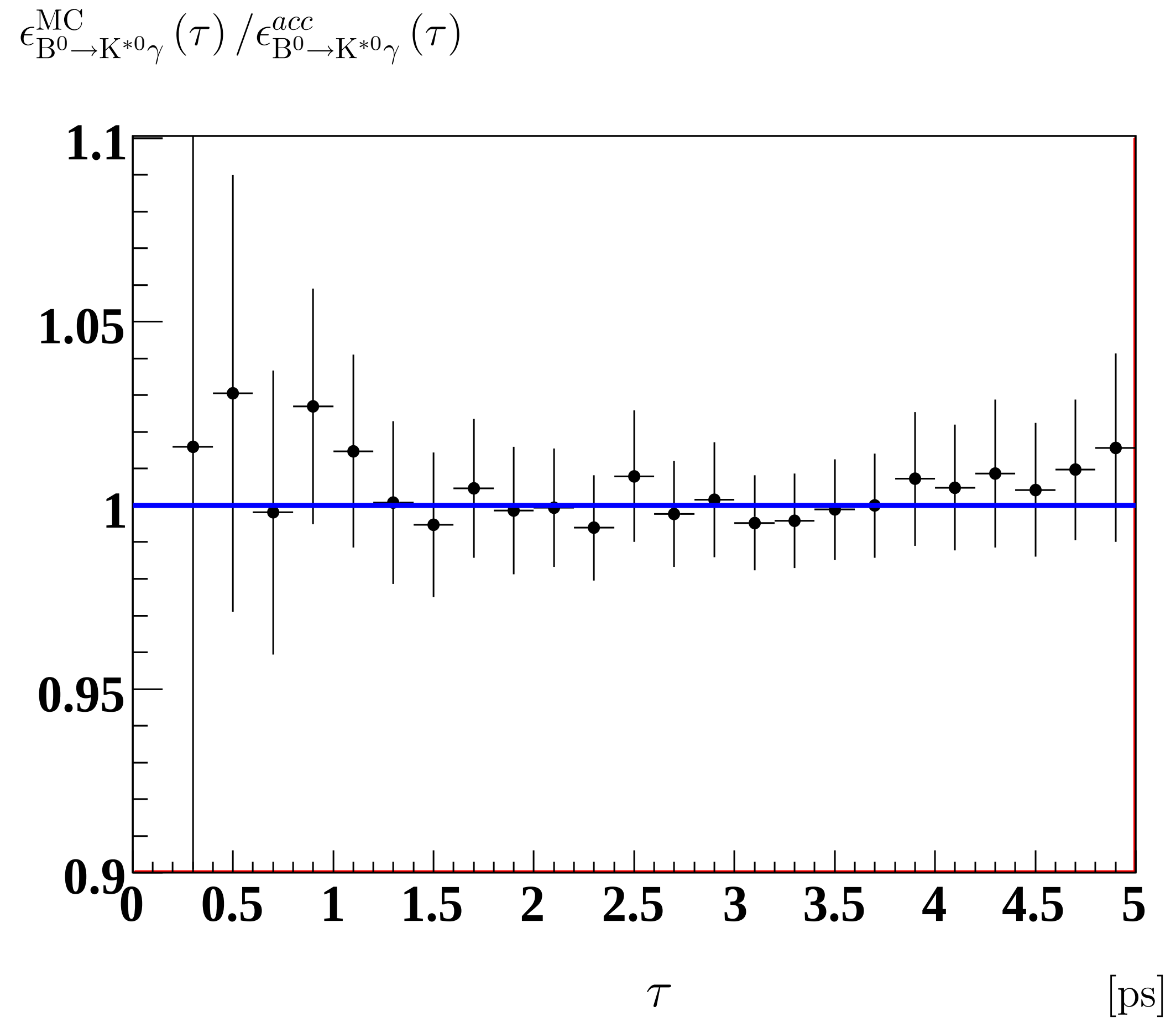


Figure 16: The ratio of proper time acceptance functions determined using Monte Carlo truth information $\epsilon^{\mathrm{MC}}_{\mathrm{B^0\to K^{*0}\gamma}}(\tau)$ and the "per-event-acceptance" method $\epsilon^{acc}_{\mathrm{B^0\to K^{*0}\gamma}}(\tau)$.

The cuts with the largest effect on the proper time acceptance are the cut on the minimum $\chi^2_{\mathrm{IP}}$ of impact parameter of the kaon with respect to the primary vertex, and the cut on the angle between the $\mathrm{B_s}$ momentum and the flight direction, evaluated as the vector from the primary vertex to the $\mathrm{B_s}$ decay vertex.

Figure 18 shows the ratio of proper time acceptance functions $\mathrm{r}_\epsilon(\tau)$ in the case one applies the cut on the direction angle of the $\mathrm{B_s}$ candidate. The minimal value of $\chi^2_{\mathrm{IP}}$ of impact parameter for the kaons have been used as re-weighting function. After the re-weighting the ratio of proper time acceptance function has only a modest dependence on the proper time for for $c\tau > 200\ \mu$m.

The usage of the decay angle of $\mathrm{B_s}$ as a re-weighting variable allows to decrease the dependence of the ratio $\mathrm{r}_\epsilon(\tau)$ on the proper time when a cut on the minimum value of $\chi^2_{\mathrm{IP}}$ value of the impact parameter of kaons is applied, as shown in Figure 19. The ratio of proper time acceptance functions $\mathrm{r}_\epsilon(\tau)$ is practically independent of the proper time for $c\tau > 200\ \mu$m within the statistical precision of the available Monte Carlo samples. For further studies the available Monte Carlo samples of $\mathrm{B^0_s} \to \phi\gamma$ would need to be increased

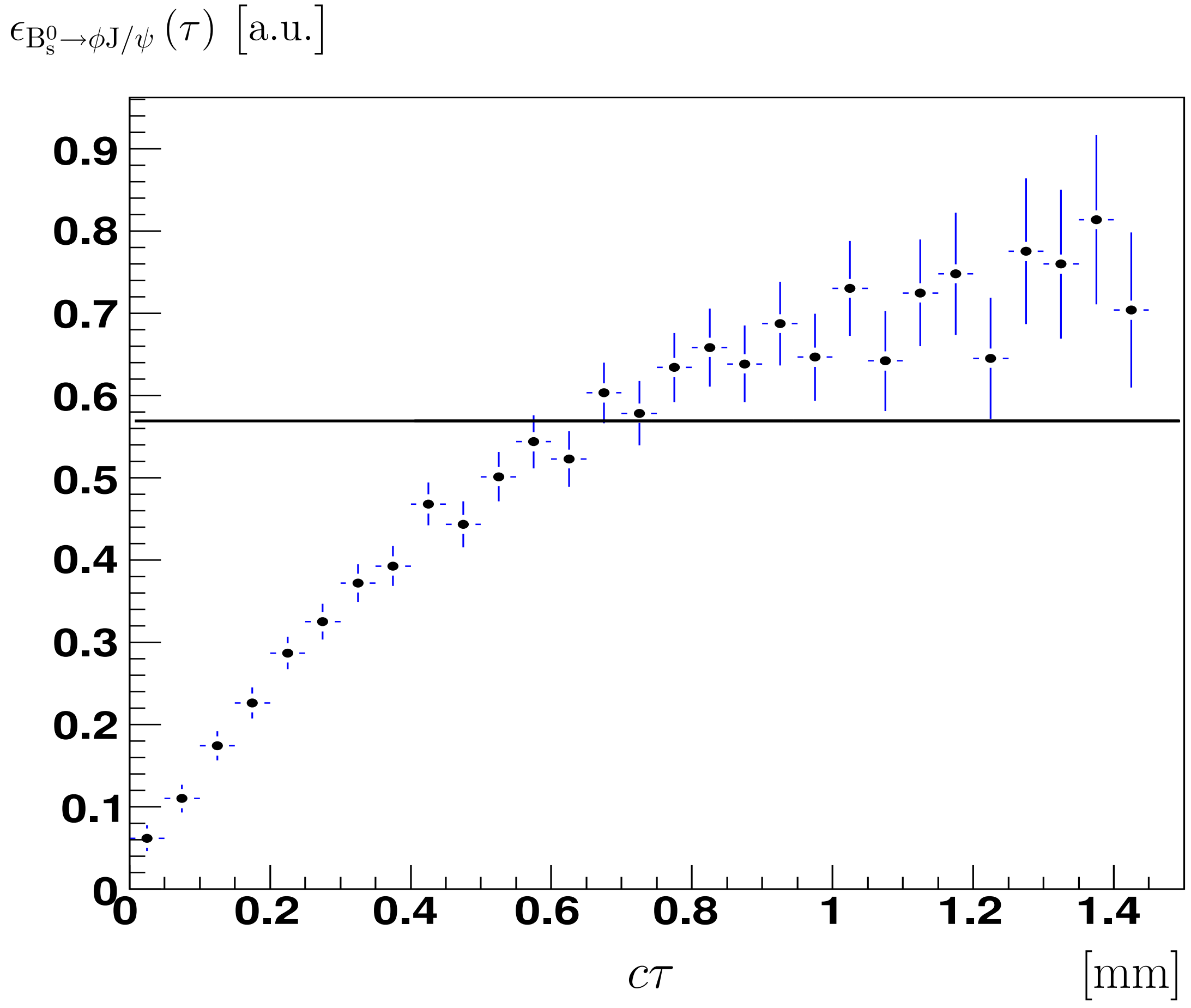


Figure 17: The relative lifetime acceptance function $\epsilon_{B_s^0\to\phi J/\psi}(\tau)$ for an applied cut on the direction angle of $B_s$-meson.

by a factor of 30 to reach the statistics equivalent to 2 $fb^{-1}$ of integrated luminosity.

It is possible that the combination of these re-weighting functions will eliminate all dependencies. However currently available samples of Monte Carlo events do not allow to estimate the precision of this method.

**Study of the proper time resolution and the possible proper time biases**

The technique of "fake" $B_s$-vertex reconstruction also allows to determine or calibrate the proper time resolution by comparison of the proper time determined for $B_s^0 \to \phi J/\psi$ events using the regular four-prong $B_s^0$ vertex and "fake" two-prong $B_s^0$ vertex. The difference is dominated by resolution in the reconstruction of the $\phi \to K^+K^-$ decay vertex. The possible biases in the determination of the proper lifetime also can be checked using this technique.

The effect of uncertainties of $\Gamma_s$ or $\Delta\Gamma_s$ on the measurement of $\mathcal{A}^{\Delta}$ has been studied using toy simulations corresponding to a scenario with $\Gamma_s = 0.6993$, $\Delta\Gamma_s = 0.084$ and

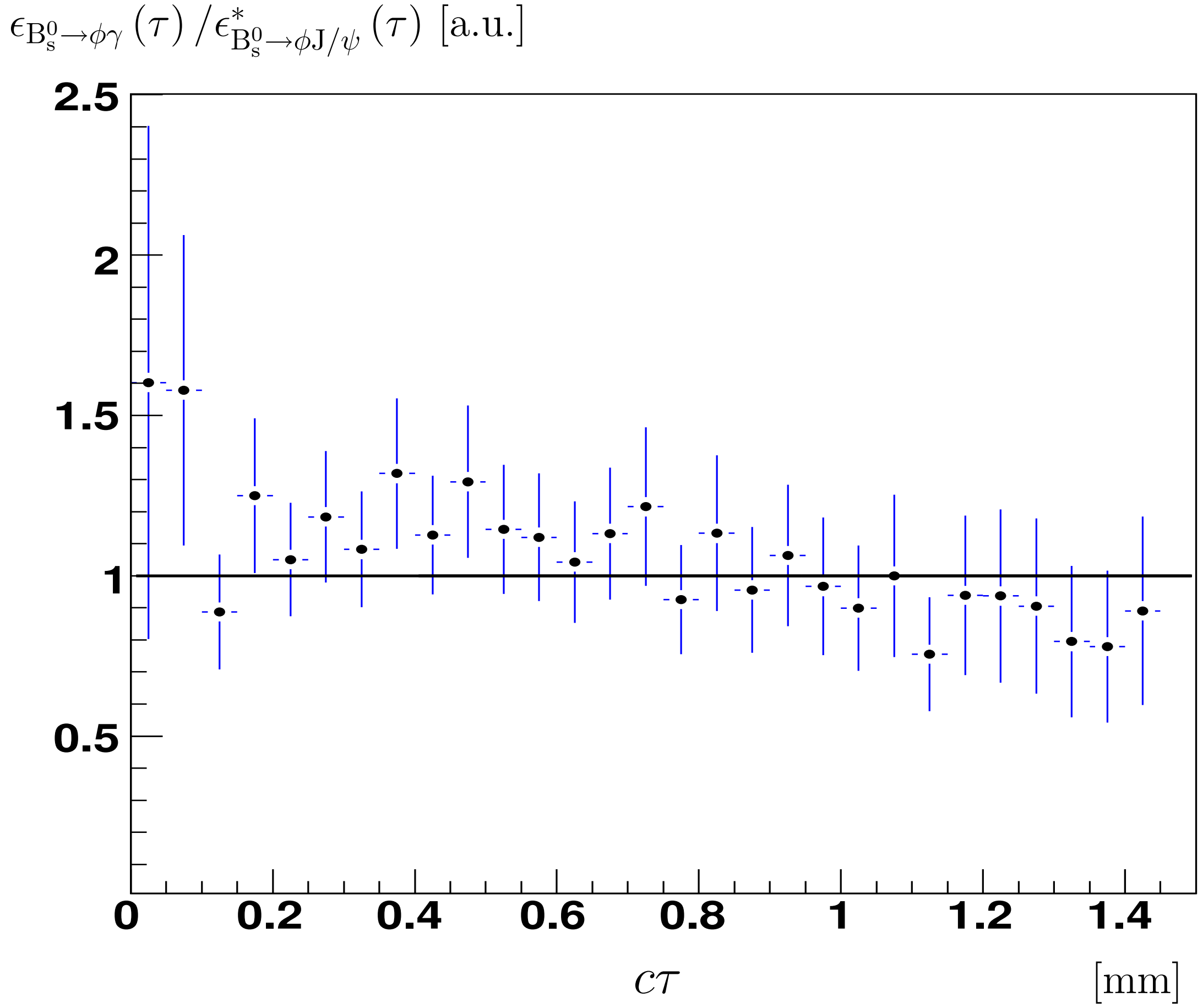


Figure 18: The ratio of lifetime acceptance functions $r_\epsilon(\tau)$ when a cut has been made on the direction angle, after re-weighting of $B_s^0 \to \phi J/\psi$ events with $\chi^2_{IP}$.

$\mathcal{A}^\Delta = 0.5$. The study shows that a relative shift of the value $\Gamma_s$ by 0.5% leads to a shift in $\mathcal{A}^\Delta$ by 0.2, independent of the value of $\mathcal{A}^\Delta$. If $\frac{\Delta\Gamma_s}{\Gamma_s}$ will be under-estimated by 10% (relative) $\mathcal{A}^\Delta$ will be over-estimated by 18%. If $\frac{\Delta\Gamma_s}{\Gamma_s}$ will be over-estimated by 10% (relative) $\mathcal{A}^\Delta$ will be under-estimated by 13%.

We will use the values of $\Gamma_s$ and $\frac{\Delta\Gamma_s}{\Gamma_s}$ measured at LHCb for the $B_s^0 \to \phi J/\psi$ decays, where these parameters will be measured with very small statistical error. It is expected that some of the systematic effects, such as wrong magnetic field, will affect $B_s^0 \to \phi J/\psi$ and $B_s^0 \to \phi\gamma$ in a similar way, therefore the bias in $\mathcal{A}^\Delta$ may be small if one uses the experimentally determined $B_s^0 \to \phi J/\psi$ lifetime even if it is significantly shifted compared to the true value. Other systematic effects, such as misalignment and proper time acceptance, may not completely cancel in the two decay modes for the measurement of $\mathcal{A}^\Delta$. This will be the subject of future studies.

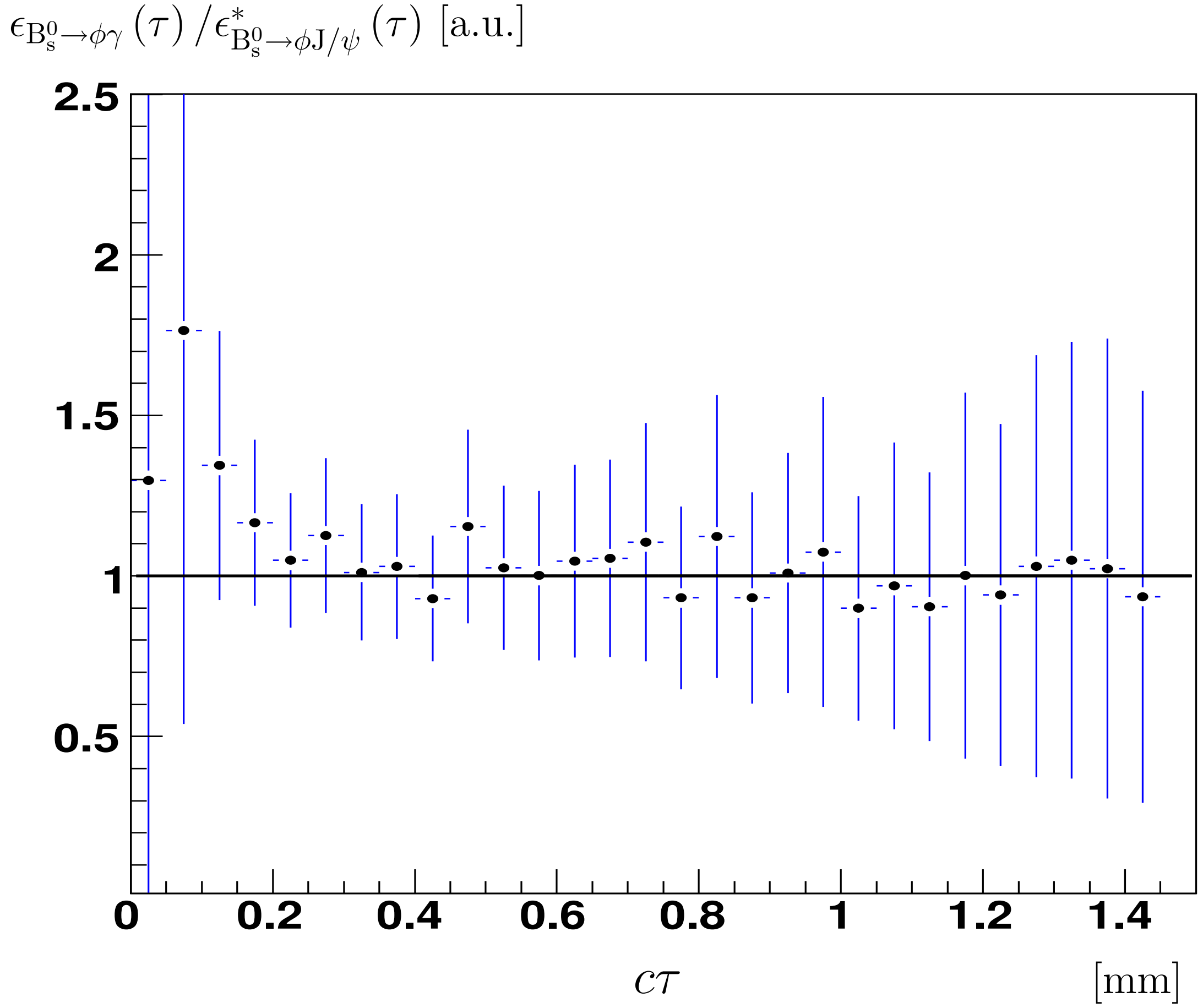


Figure 19: The ratio of lifetime acceptance functions $r_\epsilon(\tau)$ when a cut is applied on $\chi^2_{IP}$ for kaons, after re-weighting of $B_s^0 \to \phi J/\psi$ events with the decay angle of $B_s$-meson.

### The alternative selection of $B_s^0 \to \phi\gamma$

In order to minimize the systematic error related to the parameterization of the proper time acceptance function an alternative selection of $B_s^0 \to \phi\gamma$ events is under study. It is possible to replace the cut on the direction angle of the $B_s^0$-candidate with a cut on the reconstructed lifetime of the $B_s^0$-candidate. Such a selection allows to obtain a "step-like" shape of the proper time acceptance. Using such a selection one can achieve, for a well chosen value of the cut, the same efficiency for the signal events and the same suppression of minimum-bias background as the standard selection. Moreover with this approach one removes from the analysis the problematic area of small proper lifetimes, which causes problem for methods described above, see Figures 16 and 19. In addition it has been demonstrated that an HLT2 trigger selection without explicit cuts on the impact parameters of kaons is possible. However currently is not possible to check the performance of such a selection against the major expected background from $b\bar{b}$-events.

**The strategy for the proper time acceptance function**

All three methods described above will be used for the extraction of the proper time acceptance function, allowing to evaluate the systematic error.

Assuming that one can extract the actual proper time acceptance function for $B^0 \to K^{*0}\gamma$ or $B_s^0 \to \phi\gamma$ decays, different strategies for the use of this information are under consideration now. The actual strategy adopted will depend on the level of agreement between the shape of the acceptance function extracted for the calibration channel and the shape of the acceptance function expected from Monte Carlo simulation. In the case of good agreement one gets confidence in Monte Carlo simulation and in the predicted shape of the acceptance function for $B_s^0 \to \phi\gamma$ events. For the fit of the $B_s^0 \to \phi\gamma$ data one could, in this case, use the shape obtained from the Monte Carlo analysis. In the case of some discrepancy between the the shapes obtained from the Monte Carlo data and from the calibration channels the strategy could be twofold. First, obviously, one can try to tune the Monte Carlo simulation to reproduce the acceptance functions obtained for the calibration channels. Alternatively one can assume that the Monte Carlo simulation still reproduces reasonably the *ratio* of acceptance functions and one can use this hypothesis in the overall fit. Using the calibration channel $B^0 \to K^{*0}\gamma$ one can perform a common fit for the $B^0 \to K^{*0}\gamma$ and $B_s^0 \to \phi\gamma$ events and keep the parameters describing the shape of the acceptance function for $B^0 \to K^{*0}\gamma$ events free in the fit, while one fixes the shape of the ratio between the acceptance functions for $B_s^0 \to \phi\gamma$ and $B^0 \to K^{*0}\gamma$ decay modes from Monte Carlo simulation. The evaluation of the systematic errors and a possible bias requires special studies. The comparison of the result obtained by the first and second approaches will provide an estimate for the systematic error associated to the shape of the acceptance functions.

# 4 The early measurements

The early measurements are those that can be done with low integrated luminosity, significantly less than 2 $fb^{-1}$. In this section we discuss measurements with 10 $pb^{-1}$, 100 $pb^{-1}$ and 500 $pb^{-1}$. The expected yields are summarized in Table 9.

Table 9: The expected yields for data samples of 10 $pb^{-1}$, 100 $pb^{-1}$and 500 $pb^{-1}$ of integrated luminosity.

| Decay Mode | 10 $pb^{-1}$ | 100 $pb^{-1}$ | 500 $pb^{-1}$ |
|---|---|---|---|
| $B^0 \to K^{*0}\gamma$ | 370 | $4 \times 10^3$ | $2 \times 10^4$ |
| $B^0_s \to \phi\gamma$ | 55 | 550 | $3 \times 10^3$ |
| $B^+ \to \phi K^+\gamma$ | 35 | 350 | $2 \times 10^3$ |
| $\Lambda_b \to \Lambda^0\gamma$ | | | 200 |
| $\Lambda_b \to \Lambda^0\,(1520)\,\gamma$ | | 250 | $1.1 \times 10^3$ |
| $\Lambda_b \to \Lambda^0\,(1670)\,\gamma$ | | 130 | 650 |
| $\Lambda_b \to \Lambda^0\,(1690)\,\gamma$ | | 130 | 650 |

These estimates are obtained using a downscaling from 2 $fb^{-1}$, see Table 1, and assuming the nominal detector conditions and no inefficiency from the High Level Trigger (HLT). The numbers are probably optimistic, as there may be imperfections for early data, such as:

1. HLT (including HLT1 and HLT2) inefficiencies;
2. track reconstruction inefficiencies, and worsening of spatial or momentum resolution;
3. the absence or poorer hadron identification with RICH detectors;
4. inefficiency of calorimeter cells due to dead, masked and noisy channels;
5. imperfect calibration of calorimeter channels.

The effect of some of these points will be considered below.

## 4.1 The studies with 10 $pb^{-1}$

The number of detected signal events are summarized in the second column of Table 9. The study of background composition for $B^0 \to K^{*0}\gamma$ and $B^0_s \to \phi\gamma$ [2] shows that the background estimate is rather stable with respect to worse hadron identification - at most a doubling of the background level could be expected. The calorimeter commissioning is progressing well, and much less than 2 % of dead, noisy and masked cells in the calorimeters are expected. Using the estimates for the $B^0 \to K^{*0}\gamma$ decay, one can see that under the assumption that the L0 trigger [9] is operating properly, one can observe the $B^0 \to K^{*0}\gamma$ signal, even in the case of all conservative estimates. Even if the calorimeter is

miscalibrated up to 5%, which would correspond to an increase of the background level by a factor $\sim 2$, see Appendix C and Figure 20, the $B^0 \to K^{*0}\gamma$ signal should still be clearly visible. The position of the signal peak could in principle allow to estimate the global miscalibration factor for photons of very high energy, and the width will allow to estimate the channel to channel calorimeter miscalibration. Due to lower statistics the visibility of $B_s^0 \to \phi\gamma$ and $B^+ \to \phi K^+\gamma$ signals will not be evident. In summary one can conclude that with 10 $pb^{-1}$ of integrated luminosity, one should be able to see a $B^0 \to K^{*0}\gamma$ peak even in rather pessimistic scenarios.

## 4.2 The measurements with 100 $pb^{-1}$

The number of detected signal events are summarized in the third column of Table 9. With such an event statistics one can safely assume that the calorimeter (at least Ecal) is calibrated at the level of 1–1.5 %, and that the high energy calibration is also validated, see Appendix C.

We should be able to see $B_s^0 \to \phi\gamma$ and $B^+ \to \phi K^+\gamma$ signals. If the background to signal ratio, $\mathcal{B}/\mathcal{S}$, for $\Lambda_b \to \Lambda^0(1520)\gamma$, $\Lambda_b \to \Lambda^0(1670)\gamma$ and $\Lambda_b \to \Lambda^0(1690)\gamma$ is a somewhat lower compared to the upper limit that has been determined from the limited Monte Carlo statistics [5, 6, 7], than the observation of these channels can be possible[12].

The large statistics of $B^0 \to K^{*0}\gamma$ events allows us to further check and validate the calorimeter calibration at high energy, and to split the calorimeter into several regions to check the global miscalibration coefficients for these regions.

At this moment one can try to test the algorithms for $\pi^0/\gamma$-separation at very high energy, see Appendix D. Due to large statistics of $B^0 \to K^{*0}\gamma$ events the optimization of the algorithm could be performed using real data[13], directly analysing the behaviour of signal and background.

### The direct $\mathcal{CP}$-asymmetry for $B^0 \to K^{*0}\gamma$

With accumulated statistics around 100 $fb^{-1}$ one can make the measurement of the direct $\mathcal{CP}$ asymmetry for the $B^0 \to K^{*0}\gamma$ decay mode. The straightforward determination of this asymmetry is limited by the unknown or badly measured detector and production asymmetries. However using the calibration channel $B^0 \to K^{*0}J/\psi$, one can build the ratios $\mathcal{R}$ and $\bar{\mathcal{R}}$:

$$\mathcal{R} = \frac{\mathcal{N}_{B^0 \to K^{*0}\gamma}}{\mathcal{N}_{B^0 \to K^{*0}J/\psi}} \tag{4.1a}$$

$$\bar{\mathcal{R}} = \frac{\mathcal{N}_{\bar{B}^0 \to \bar{K}^{*0}\gamma}}{\mathcal{N}_{\bar{B}^0 \to \bar{K}^{*0}J/\psi}} \tag{4.1b}$$

[12] In general for these channels the background conditions are expected to be similar to the $B^0 \to K^{*0}\gamma$ mode.

[13] Note here that we do not need to know the exact value of the background suppression factor from this algorithm, we just need to know the signal efficiency.

From these ratios one can construct the asymmetry $\mathcal{A}_{\mathcal{R}}$:

$$\mathcal{A}_{\mathcal{R}} = \frac{\bar{\mathcal{R}} - \mathcal{R}}{\bar{\mathcal{R}} + \mathcal{R}}. \tag{4.2}$$

Clearly the production asymmetries and, to first order, the detector asymmetries cancel in the asymmetry $\mathcal{A}_{\mathcal{R}}$. As a result the asymmetry $\mathcal{A}_{\mathcal{R}}$ is a function of the direct $\mathcal{CP}$ asymmetries $\mathcal{A}^{\mathrm{dir}}_{\mathrm{K}^{*0}\gamma}$ and $\mathcal{A}^{\mathrm{dir}}_{\mathrm{K}^{*0}\mathrm{J}/\psi}$ only[14]:

$$\mathcal{A}_{\mathcal{R}} = \frac{\mathcal{A}^{\mathrm{dir}}_{\mathrm{K}^{*0}\gamma} - \mathcal{A}^{\mathrm{dir}}_{\mathrm{K}^{*0}\mathrm{J}/\psi}}{1 - \mathcal{A}^{\mathrm{dir}}_{\mathrm{K}^{*0}\gamma}\mathcal{A}^{\mathrm{dir}}_{\mathrm{K}^{*0}\mathrm{J}/\psi}} \quad , \tag{4.3}$$

where the direct $\mathcal{CP}$-asymmetries $\mathcal{A}^{\mathrm{dir}}_{\mathrm{K}^{*0}\gamma}$ and $\mathcal{A}^{\mathrm{dir}}_{\mathrm{K}^{*0}\mathrm{J}/\psi}$ are defined as:

$$\mathcal{A}^{\mathrm{dir}}_{\mathrm{K}^{*0}\gamma} = \frac{\mathcal{B}_{\bar{\mathrm{B}}^0\to\bar{\mathrm{K}}^{*0}\gamma} - \mathcal{B}_{\mathrm{B}^0\to\mathrm{K}^{*0}\gamma}}{\mathcal{B}_{\bar{\mathrm{B}}^0\to\bar{\mathrm{K}}^{*0}\gamma} + \mathcal{B}_{\mathrm{B}^0\to\mathrm{K}^{*0}\gamma}} \tag{4.4a}$$

$$\mathcal{A}^{\mathrm{dir}}_{\mathrm{K}^{*0}\mathrm{J}/\psi} = \frac{\mathcal{B}_{\bar{\mathrm{B}}^0\to\bar{\mathrm{K}}^{*0}\mathrm{J}/\psi} - \mathcal{B}_{\mathrm{B}^0\to\mathrm{K}^{*0}\mathrm{J}/\psi}}{\mathcal{B}_{\bar{\mathrm{B}}^0\to\bar{\mathrm{K}}^{*0}\mathrm{J}/\psi} + \mathcal{B}_{\mathrm{B}^0\to\mathrm{K}^{*0}\mathrm{J}/\psi}} \quad . \tag{4.4b}$$

The cancellation of reconstruction asymmetries from$\mathcal{A}_{\mathcal{R}}$ is exact if the selection criteria used for the $\mathrm{B}^0 \to \mathrm{K}^{*0}\gamma$ and the $\mathrm{B}^0 \to \mathrm{K}^{*0}\mathrm{J}/\psi$ selection do not result in a significant difference in the phase space for the important variables, or the reconstruction efficiency is constant. Otherwise the necessary corrections are of the order $\left(\mathcal{A}^{\mathrm{rec}}_{\mathrm{K}^{*0}\gamma} - \mathcal{A}^{\mathrm{rec}}_{\mathrm{K}^{*0}\mathrm{J}/\psi}\right)$, and clearly do not exceed $\frac{\int \varepsilon(\Phi)\left(\mathrm{p}^{\mathrm{s}}(\Phi) - \mathrm{p}^{\mathrm{c}}(\Phi)\right)\mathrm{d}\Phi}{\int \varepsilon(\Phi)\left(\mathrm{p}^{\mathrm{s}}(\Phi) + \mathrm{p}^{\mathrm{c}}(\Phi)\right)\mathrm{d}\Phi}$, where $\varepsilon(\Phi)$ is the reconstruction efficiency, $\mathrm{p}^{\mathrm{s}}(\Phi)$ and $\mathrm{p}^{\mathrm{c}}(\Phi)$ are probability distribution functions for the signal and calibration channel. These corrections can be estimated from Monte Carlo simulation. If they appear to be the limiting factors, the phase space can be subdivided into a few bins and the measurement of $\mathcal{A}_{\mathcal{R}}$ can be performed independently in each bin, and the results from the different bins can be averaged.

[14] The exact expression is:

$$\mathcal{A}_{\mathcal{R}} = \frac{\left(\mathcal{A}^{\mathrm{dir}}_{\mathrm{K}^{*0}\gamma} - \mathcal{A}^{\mathrm{dir}}_{\mathrm{K}^{*0}\mathrm{J}/\psi}\right) + \left(\mathcal{A}^{\mathrm{rec}}_{\mathrm{K}^{*0}\gamma} - \mathcal{A}^{\mathrm{rec}}_{\mathrm{K}^{*0}\mathrm{J}/\psi}\right)}{1 + \left(\mathcal{A}^{\mathrm{dir}}_{\mathrm{K}^{*0}\gamma} - \mathcal{A}^{\mathrm{dir}}_{\mathrm{K}^{*0}\mathrm{J}/\psi}\right)\left(\mathcal{A}^{\mathrm{rec}}_{\mathrm{K}^{*0}\gamma} - \mathcal{A}^{\mathrm{rec}}_{\mathrm{K}^{*0}\mathrm{J}/\psi}\right) - \mathcal{A}^{\mathrm{dir}}_{\mathrm{K}^{*0}\gamma}\mathcal{A}^{\mathrm{dir}}_{\mathrm{K}^{*0}\mathrm{J}/\psi} - \mathcal{A}^{\mathrm{rec}}_{\mathrm{K}^{*0}\gamma}\mathcal{A}^{\mathrm{rec}}_{\mathrm{K}^{*0}\mathrm{J}/\psi}} \quad ,$$

where the detector asymmetries $\mathcal{A}^{\mathrm{rec}}_{\mathrm{K}^{*0}\gamma}$ and $\mathcal{A}^{\mathrm{rec}}_{\mathrm{K}^{*0}\mathrm{J}/\psi}$ are defined as:

$$\mathcal{A}^{\mathrm{rec}}_{\mathrm{K}^{*0}\gamma} = \frac{\varepsilon_{\bar{\mathrm{B}}^0\to\bar{\mathrm{K}}^{*0}\gamma} - \varepsilon_{\mathrm{B}^0\to\mathrm{K}^{*0}\gamma}}{\varepsilon_{\bar{\mathrm{B}}^0\to\bar{\mathrm{K}}^{*0}\gamma} + \varepsilon_{\mathrm{B}^0\to\mathrm{K}^{*0}\gamma}}$$

$$\mathcal{A}^{\mathrm{rec}}_{\mathrm{K}^{*0}\mathrm{J}/\psi} = \frac{\varepsilon_{\bar{\mathrm{B}}^0\to\bar{\mathrm{K}}^{*0}\mathrm{J}/\psi} - \varepsilon_{\mathrm{B}^0\to\mathrm{K}^{*0}\mathrm{J}/\psi}}{\varepsilon_{\bar{\mathrm{B}}^0\to\bar{\mathrm{K}}^{*0}\mathrm{J}/\psi} + \varepsilon_{\mathrm{B}^0\to\mathrm{K}^{*0}\mathrm{J}/\psi}.} \quad .$$

One gets the explicit expression for $\mathcal{A}^{\rm dir}_{\rm K^{*0}\gamma}$:

$$\mathcal{A}^{\rm dir}_{\rm K^{*0}\gamma} = \frac{\mathcal{A}_{\mathcal{R}} + \mathcal{A}^{\rm dir}_{\rm K^{*0}J/\psi}}{1 + \mathcal{A}_{\mathcal{R}}\mathcal{A}^{\rm dir}_{\rm K^{*0}J/\psi}}. \tag{4.5}$$

The asymmetry $\mathcal{A}_{\mathcal{R}}$ could be measured with a statistical precision of 0.018 for 100 $\rm pb^{-1}$ of data due to the large statistics of $\rm B^0 \to K^{*0}\gamma$ and $\rm B^0 \to K^{*0}J/\psi$ events. The latter will be selected for calibration and normalization of other interesting decays of B hadrons [10,11]. The systematic error is expected to be rather small. As a result the final error on $\mathcal{A}^{\rm dir}_{\rm K^{*0}\gamma}$ will be dominated by the knowledge of $\mathcal{A}^{\rm dir}_{\rm K^{*0}J/\psi}$ from the B factories. It should be possible to measure the $\mathcal{A}^{\rm dir}_{\rm K^{*0}\gamma}$ with a precision which is better than the current combined precision from the B factories, $\sigma\left(\mathcal{A}^{\rm dir}_{\rm K^{*0}\gamma}\right) = 0.028$ [51].

### The ratio of $\rm B^0_s \to \phi\gamma$ to $\rm B^0 \to K^{*0}\gamma$ rates

Another interesting measurement could be the measurement of the ratio of the branching fractions for $\rm B^0 \to K^{*0}\gamma$ and $\rm B^0_s \to \phi\gamma$. Again, the direct determination of this ratio suffers both from the knowledge of detector and trigger efficiencies and knowledge of the production rates through the ratio of the fragmentation constants $\rm f_d/f_s$. However using the decay modes $\rm B^0 \to K^{*0}J/\psi$ and $\rm B_s{}^0 \to \phi J/\psi$ as normalization channels one can make a rather precise measurement of the double ratio $\mathcal{R}^{\rm B^0_s \to \phi\gamma}$:

$$\mathcal{R}^{\rm B^0_s \to \phi\gamma} = \frac{\mathcal{B}_{\rm B^0_s \to \phi\gamma}/\mathcal{B}_{\rm B_s{}^0 \to \phi J/\psi}}{\mathcal{B}_{\rm B^0 \to K^{*0}\gamma}/\mathcal{B}_{\rm B^0 \to K^{*0}J/\psi}} = \frac{\mathcal{B}_{\rm B^0_s \to \phi\gamma}/\mathcal{B}_{\rm B^0 \to K^{*0}\gamma}}{\mathcal{B}_{\rm B_s{}^0 \to \phi J/\psi}/\mathcal{B}_{\rm B^0 \to K^{*0}J/\psi}},$$

where again one expects the cancellation of most systematic uncertainties, resulting in a dominantly statistical error of less than 5% for 100 $\rm pb^{-1}$ of data.

### A first look into $\rm B^+ \to \phi K^+\gamma$ channel

The double ratio technique can be used again to measure, with reduced systematical error, the ratio of branching ratios $\rm B^+ \to \phi K^+\gamma$ and $\rm B^0 \to K^{*0}\gamma$:

$$\mathcal{R}_{\mathcal{D}}^{\rm B^+ \to \phi K^+\gamma} = \frac{\mathcal{B}_{\rm B^+ \to \phi K^+\gamma}/\mathcal{B}_{\rm B^0 \to K^{*0}\gamma}}{\mathcal{B}_{\rm B^+ \to K\pi\pi J/\psi}/\mathcal{B}_{\rm B^0 \to K^{*0}J/\psi}}.$$

The denominator of this double ratio is well known from the B factories [51]. However in this case not all particle identification efficiencies cancel, therefore additional corrections from Monte Carlo or other calibration channels are required[15]. But even in the case of large corrections the measurement could be the most precise determination of the $\rm B^+ \to \phi K^+\gamma$ branching ratio. Such statistics of $\rm B^+ \to \phi K^+\gamma$ events will allow a first look to be made

[15]Clearly if the the precise measurement of $\mathcal{B}_{\rm B^+ \to \phi K^+ J/\psi}$ will be available from the B factories, it would be very useful to decrease the systematic uncertainty.

at the Dalitz plots for three-kaon systems separately for the $B^+$ and $B^-$ cases. Also one can expect the first result on the resonant decomposition and the dominant partial waves. However, the results here depend crucially on the background level.

The clear peak of $B_s^0 \to \phi\gamma$ decay can be used for investigation of the background in this channel. In particular one needs to compare the proper time distribution in the left and right sidebands, and to check the stability and convergence of 2D fits for proper time versus the reconstructed mass for background events. Clearly the available statistics of background events will be much larger than any possible Monte Carlo event sample, allowing the detailed study of the proper time distributions separately for the left and the right sideband, which is a necessary prerequisite for the background parameterization in the whole 2D plane.

## 4.3 The measurements with 500 $pb^{-1}$ and beyond

The number of detected signal events are summarized in the last column of Table 9. Again these numbers do not include the efficiency of the High Level Trigger.

At this moment the calorimeter is assumed to be fully calibrated and understood. The large statistics of $B^0 \to K^{*0}\gamma$ events will allow this decay mode to be used widely as a normalization channel.

Assuming that with such statistics LHCb will be able to understand the trigger, particle reconstruction and identification, the scaling from Monte Carlo is a good estimate. However, for complicated items, like the flavour tagging, more experience may be required.

With this data set we will be able to commission the algorithm for the $\pi^0/\gamma$-separation at very high energy, see Appendix D. It is assumed that the algorithm has been tested and validated earlier. After the algorithm is commissioned we can expect an overall decrease in the level of background of a factor of 2 [2, 1].

Using 500 $pb^{-1}$ of data one can perform an unbinned maximum likelihood fit of $B_s^0 \to \phi\gamma$ events. The expected precision on $\mathcal{A}^{\Delta}$ for the untagged sample is about $\sim 0.4$, which is better than any current single measurement from B-factories. However, if the tagged analysis is possible, one can expect to obtain a precision for the measurement of the $\mathcal{CP}$ violating parameters $\mathcal{C}$ and $\mathcal{S}$ of $\sim 0.2$.

For the decay channel $B^+ \to \phi K^+\gamma$, one can start the detailed Dalitz analysis and the partial wave decomposition. In the case of a favourable contribution of partial waves and relative phases interesting results are expected.

If the background to signal ratio, $\mathcal{B}/\mathcal{S}$, for $\Lambda_b \to \Lambda^0(1520)\gamma$, $\Lambda_b \to \Lambda^0(1670)\gamma$ and $\Lambda_b \to \Lambda^0(1690)\gamma$ is somewhat lower than the upper limit that has been determined from the limited Monte Carlo statistics [5, 6, 7], then the observation of these channels should be possible. The possibility to observe the decay $\Lambda_b \to \Lambda^0\gamma$ also depends on the status of the High Level Trigger and analysis tools for selection of this challenging mode [5, 6, 7].

With 2 $fb^{-1}$ of accumulated luminosity the asymmetry $\mathcal{A}_{\mathcal{R}}$ defined by Eq. (4.2) and related to the direct $\mathcal{CP}$ asymmetry $\mathcal{A}^{\rm dir}_{K^{*0}\gamma}$ (see Eq. (4.3)), could be measured with a statistical precision better than 1 % due to the large statistics of $B^0 \to K^{*0}\gamma$ and $B^0 \to K^{*0}J/\psi$ events. The systematic error is expected to be rather small due to cancellation

of various systematic asymmetries in the $\mathcal{A}_{\mathcal{R}}$. However, further progress for $\mathcal{A}^{\mathrm{dir}}_{\mathrm{K}^{*0}\gamma}$ would definitely require more detailed understanding of the systematic uncertainties related to the detector and production asymmetries.

# Summary

The basic steps toward the first measurements of radiative B decays in LHCb have been presented. We have concentrated on the probing of photon polarisation in the $\mathrm{B}^0_\mathrm{s} \to \phi\gamma$ decay. Based on our Monte Carlo simulation for 2 $\mathrm{fb}^{-1}$ we expect to reach the uncertainty in the measurement of the parameter $\mathcal{A}^{\Delta}$, $\sigma_{\mathcal{A}^{\Delta}} \approx 0.22$. The capability of LHCb experiment to measure this and other parameters has been discussed for different luminosity sets.

The measurement of the direct $\mathcal{CP}$-asymmetry for $\mathrm{B}^0 \to \mathrm{K}^{*0}\gamma$ decay can be performed with statistical precision around 0.018 already with a data-set of an integrated luminosity around 100 $\mathrm{pb}^{-1}$, and the measurements of the ratio of branching fractions for $\mathrm{B}^0_\mathrm{s} \to \phi\gamma$ and $\mathrm{B}^0 \to \mathrm{K}^{*0}\gamma$ can be done with statistical precision less than 5%

Several critical aspects and important prerequisites, like calorimeter calibration, implementation of the HLT, determination of proper time acceptance function form data and $\gamma/\pi^0$ separation at high-$\mathrm{E_T}$ have been addressed.

# Acknowledgments

We would like to express our gratitude to our colleagues from the LHCb Rare Decays, Calorimeter and Trigger groups. Primarily we thank T. Skwarnicki for opening the door into the world of radiative penguins, A. Bondar, L. Camilleri, B. Delcourt, H. Dijkstra, U. Egede, R. Forty, J. Lefrancois, M. Merk, T. Nakada, N. Nikitin, P. Perret, B. Pietrzyk, G. Raven, M.-H. Schune, S. Stone and U. Uwer for their attention to this work and J. Amoraal, V. Breton, N. Brun, D. Golubkov, Yu. Guz, W. Hulsbergen, P. Koppenburg, G. Lanfranchi, O. Leroy, P. Pakhlov, J. Palacios, C. Parkes, T. du Pree, V. Romanovsky, P. Somogyi, and F. Teubert for the leading contributions, the great help, many useful suggestions and clarifications of various complicated aspects.

# Appendix A: Differential $\mathrm{B_s^0}$ decay rates and $\mathcal{CP}$-violation parameters

The observable rate for $\mathrm{B_s^0} \to \phi\gamma$ events is:

$$R(\mathrm{t}) = \varepsilon^{\mathrm{tag}} \int |\mathrm{A}|^2 \, \mathrm{e}^{-\Gamma\tau} \Big[ \cosh \frac{\Delta\Gamma\tau}{2} - \mathcal{A}^{\Delta} \sinh \frac{\Delta\Gamma\tau}{2} + (1-2\omega)\mathcal{C} \cos \Delta \mathrm{m_s}\tau - (1-2\omega)\mathcal{S} \sin \Delta \mathrm{m_s}\tau \Big] \mathcal{K}(\tau, \mathrm{t}) \, \mathrm{d}\tau \quad . \tag{A.1a}$$

The observed rate for $\mathrm{\bar{B}_s^0} \to \phi\gamma$ events is:

$$\bar{R}(\mathrm{t}) = \varepsilon^{\mathrm{tag}} \int |\mathrm{A}|^2 \, \mathrm{e}^{-\Gamma\tau} \Big[ \cosh \frac{\Delta\Gamma\tau}{2} - \mathcal{A}^{\Delta} \sinh \frac{\Delta\Gamma\tau}{2} - (1-2\omega)\mathcal{C} \cos \Delta \mathrm{m_s}\tau + (1-2\omega)\mathcal{S} \sin \Delta \mathrm{m_s}\tau \Big] \mathcal{K}(\tau, \mathrm{t}) \, \mathrm{d}\tau \quad . \tag{A.1b}$$

The rate of untagged events is:

$$R_{no-tag}(\mathrm{t}) = (1-\varepsilon_{tag}) \Big( \int |\mathrm{A}|^2 \, \mathrm{e}^{-\Gamma\tau} \Big[ \cosh \frac{\Delta\Gamma\tau}{2} - \mathcal{A}^{\Delta} \sinh \frac{\Delta\Gamma\tau}{2} \Big] \mathcal{K}(\tau, \mathrm{t}) \, \mathrm{d}\tau \Big) \quad , \tag{A.2}$$

where $\varepsilon^{\mathrm{tag}}$ is the tagging efficiency, $\omega$ is a mistag rate, and the kernel $\mathcal{K}(\tau, \mathrm{t}) = \epsilon(\mathrm{t}) \, \mathrm{G}(\tau - \mathrm{t})$ is parameterized as the product of the proper time acceptance function $\epsilon(\mathrm{t})$ and the Gaussian resolution function $\mathrm{G}(\tau - \mathrm{t})$. The parameters $\mathcal{C}$, $\mathcal{S}$ and $\mathcal{A}^{\Delta}$ are defined by (1.6). For equations (A.1) it has been assumed that the tagging efficiency for produced $\mathrm{B_s}$ and $\mathrm{\bar{B}_s}$ mesons are equal, and for equation (A.2) the production asymmetry also has been neglected.

# Appendix B: The likelihood function for the $B_s^0 \to \phi\gamma$ analysis

The probability distribution function (PDF) for the unbinned maximum likelihood fits for $B_s^0 \to \phi\gamma$ is constructed as follows. For each tagging category $\kappa$ the PDF can be written as:

$$P_\kappa(t, m) = f_s \frac{1}{\mathcal{N}_S}\left[e^{-\Gamma\tau}\big(I_+(\tau) + \kappa(1-2\omega)I_-(\tau)\big) \otimes \mathcal{K}(\tau, t)\right] \times G_m(m_{\phi\gamma}) + (1-f_s)\frac{1}{\mathcal{N}_B} f_b(m_{\phi\gamma}, t), \quad (B.1)$$

where $f_s$ is a fraction of signal events, $\kappa$ can possess three values: $\kappa = -1$ corresponding to $B_s$, $\kappa = 1$ to $\bar{B}_s$ and $\kappa = 0$ to untagged events, $f_s = \frac{S}{S+B}$ is the signal fraction, the kernel $\mathcal{K}(\tau, t) = \epsilon(t)\, G(\tau - t)$ is parameterized as the product of the proper time acceptance function $\epsilon(t)$ and the Gaussian resolution function $G(\tau - t)$, $G_m(m_{\phi\gamma})$ is the normalized mass PDF for the signal, which is parameterized by the simple Gaussian, background PDF $f_b(m_{\phi\gamma}, t)$ is defined by (3.4), and the helper functions $I_\pm(\tau)$ and the normalization factors $\mathcal{N}_S$ and $\mathcal{N}_B$ defined as:

$$\begin{aligned}
I_+(\tau) &= \cosh\frac{\Delta\Gamma\tau}{2} - \mathcal{A}^\Delta \sinh\frac{\Delta\Gamma\tau}{2}, \\
I_-(\tau) &= \mathcal{C}\cos\Delta m_s\tau - \mathcal{S}\sin\Delta m_s\tau \\
\mathcal{N}_S &= \int \left[e^{-\Gamma\tau}\big(I_+(\tau) + \kappa(1-2\omega)I_-(\tau)\big) \otimes \mathcal{K}(\tau, t)\right] dt \\
\mathcal{N}_B &= \iint f_b(m_{\phi\gamma}, t)\, dm_{\phi\gamma}\, dt.
\end{aligned}$$

The likelihood function is constructed as follows:

$$\mathcal{L}_0 = \left[\prod_{i=1}^{N_{ut}} P_0(m_i, t_i, \sigma_{t_i}),\right]\left[\prod_{i=1}^{N_{B_s}} P_{-1}(m_i, t_i, \sigma_{t_i})\right]\left[\prod_{i=1}^{N_{\bar{B}_s}} P_1(m_i, t_i, \sigma_{t_i})\right], \quad (B.2)$$

where $m_i, t_i, \sigma_{t_i}$ are the measured mass, proper time and proper time error for each event, correspondingly. For untagged analysis only one term in B.2 has to be considered.

The likelihood (B.2) function could be improved e.g. to take into account the overall integrated decay rates. Considering multiple tagging categories has the major potential to improve the results on $\mathcal{S}$ and $\mathcal{C}$ parameters which depend on the tagging. An improvement corresponding to 24% further statistics was seen by changing from a single to multiple tagging categories [15].

# Appendix C: Calorimeter Calibration

The importance of Ecal calibration is illustrated by Figure 20. For Ecal miscalibrated at around $\sim 3\%$ one gets a $\sim 20\%$ increase of the signal width, which can be interpreted as the effective $\sim 20\%$ increase of the combinatorial background in the signal mass interval. In addition the relative contribution to the background from incompletely reconstructed B-decays rises. We also believe that the separation of single photons from the high energy "merged" $\pi^0$-mesons degrades with degradation of Ecal calibration.

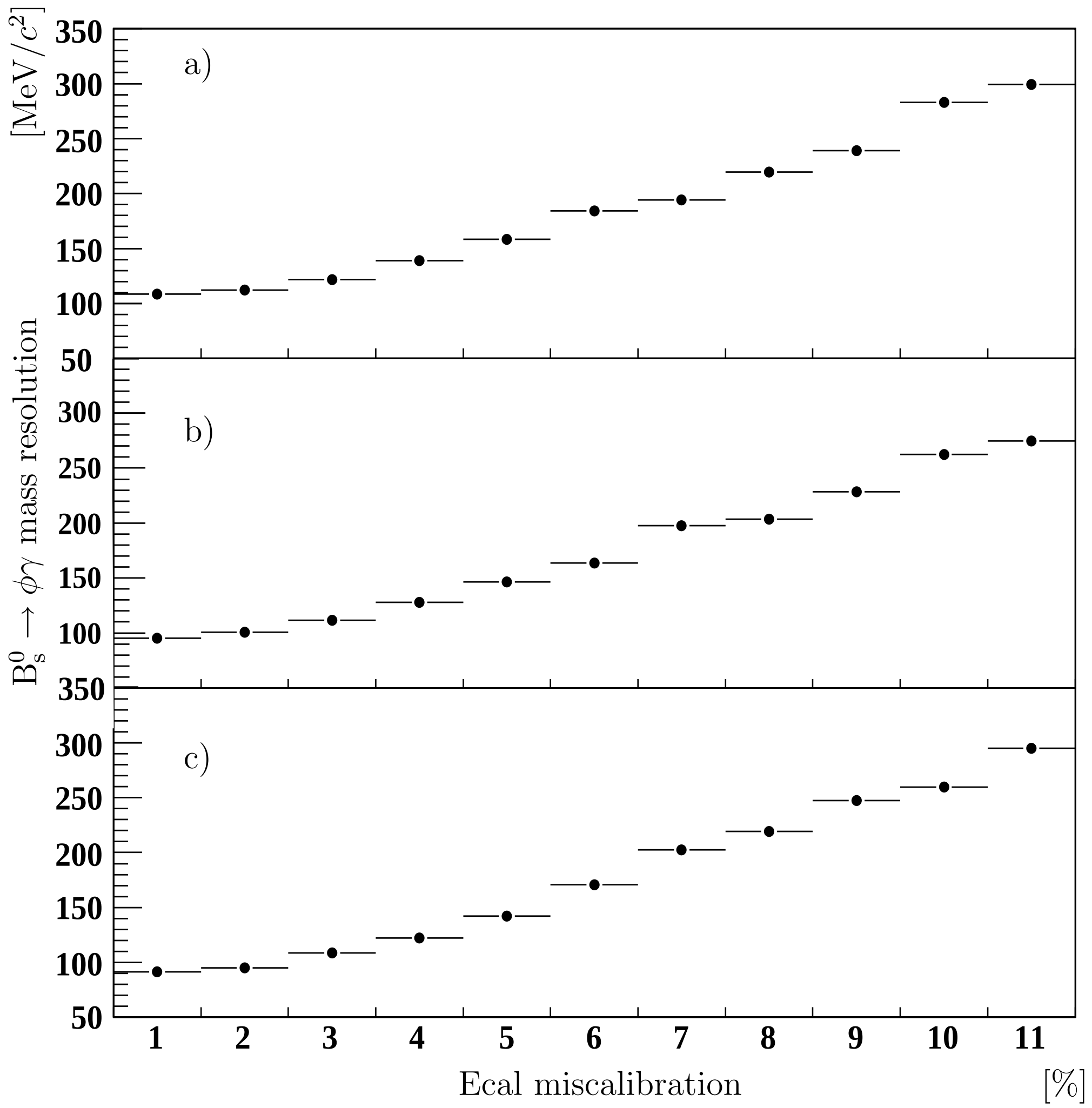


Figure 20: The mass resolution of $B_s^0 \to \phi\gamma$ peak as a function of Ecal miscalibrations for inner (a), middle (b) and outer (c) areas of the Calorimeter.

For the reconstruction of photon energy [64] both the energy deposited in electromagnetic calorimeter (Ecal) and the energy deposited in the Preshower detector (Prs) are used. Initial tests with cosmic particles of the individual modules before assembly of those modules in the Ecal shows that the initial light yield for different Ecal cells is

equalized within 4.3–8.0 % [65]. The phototube gains are remeasured in situ with LED's. Monte Carlo studies show that the equalization of module response using the energy flow technique effectively allows to reduce the miscalibration of the modules at the level 3.7–5% for 20% initial miscalibration of Ecal. Moreover for the bulk of the modules the miscalibration does not exceed 4% [66].

A few different techniques have been discussed for the fine calibrations of the Ecal and Prs detectors:

- calibration of Ecal using minimum ionizing particles [67],
- iterative calibration of Ecal with neutral pions with no energy deposition in Prs detector,
- simultaneous calibration of Ecal and Prs with neutral pions using a "Millipede-like" multidimensional minimization procedure,
- simultaneous calibration of Ecal and Prs using electrons [67].

Also the calibration of the Prs detector using the the clear signal from a pair of minimal ionizing particles from photon conversion for the reconstructed photon candidates has been discussed.

Each technique has its own advantages and disadvantages and therefore a unique application range. The calibration with minimum ionizing particles and the iterative calibration with $\pi^0$'s allow to calibrate the Ecal only. The calibration of the Preshower detector requires other techniques. The "Millipede-like" calibrations with $\pi^0$ and $e^{\pm}$ in theory both allow the simultaneous calibration of Ecal and Preshower detectors, but the procedures are more complicated and require the development of non-trivial code. Also from first principles one expects that the performance of these global methods to be very sensitive to the purity of the calibration sample and become rather fragile for samples with relatively low purity.

Also each calibration is efficient only for a certain energy range. For the selection of signal events a rather hard cut on the transverse energy of photons of the order of 2.8 GeV is usually applied [2, 5].

The calibration with minimum ionizing particles is rather simple but can be performed only for the middle and outer sections of the electromagnetic calorimeter [67]. The signals from minimum ionizing particles in the innermost part of the inner calorimeter are too small to be efficiently used for the calibration. The energy depositions from minimum ionizing particles are $\mathcal{O}\left(350\text{ MeV}\right)$ and the corresponding transverse energy is always less than 100 MeV, which is well outside of the most interesting energy region[16]. The calibration techniques based on $\pi^0 \to \gamma\gamma$ reconstruction are efficient for photons in the transverse energy range from $\sim$ 300 MeV to $\sim$ 1.5 GeV. The overlapping of the electromagnetic showers from two photons effectively sets the transverse energy cutoff. The

[16]Note that ADC calibration is 2.5 MeV of transverse energy per ADC channel, and the noise is $\sim$ 3 MeV per cell [9].

calibration techniques based on electrons are efficient in the energy range from $\sim 1$ GeV to 2–2.5 GeV. Here the transverse energy cutoff is set by the partial overlap of the electromagnetic shower from the initial electron and the shower from the bremstrahlung photon emitted before the magnet.

Therefore we have no adequate calibration techniques which directly provides us with the calibration which fits into the most interesting transverse energy range. Here we need to rely on phototube linearity and the LED-based monitoring system [9, 68]. A complementary approach is discussed below.

As concerns the calorimeter calibration, one also needs to determine the numerous correction factors [64, 69] which are related to the spatial resolution of the Ecal. The leading corrections in the first approximation can be extracted from a detailed Monte Carlo study, however, the subleading, but small corrections are very sensitive to the actual distribution of the passive material in front of the Ecal and in between the Preshower and Ecal detectors. Currently we do not see a way to extract these corrections for the photons directly from the data, however similar corrections for electrons and positrons probably can be extracted from data. And by comparison with Monte Carlo one can try to calibrate the correction factors for the photons. Since these correction factors essentially reflect the actual shape of the electromagnetic shower for the given distribution of the passive material, they are considered to be constants and they need to be determined from data only once. Last but not at least the covariance matrix for the photon momentum needs to be evaluated properly. Current evaluation of this matrix result in some underestimation of the errors [2]. Such aspects are important for valid constraints and for the correct evaluation of the errors for the proper lifetime. These aspects are not discussed further here.

## Iterative Ecal calibration with $\pi^0$

The iterative calibration of the electromagnetic calorimeter using the $\pi^0 \to \gamma\gamma$ peak in hadronic environment has been successfully used by the HERA-B Collaboration [70, 71] as the main calibration technique. It has been demonstrated that it gives a precise and robust Ecal calibration. This calibration does not rely on information from the other subdetectors, in particular it is tracking independent. It also has been shown that this calibration is fast and can be performed using very high rate of events available within the online environment.

The technique is based on the basic idea of the multiplicative energy correction coefficients and the direct relation between the energy shift of the photon and the shift of the visible position of $\pi^0$ peak:

$$\mathrm{m}^2_{\gamma\gamma} = 2E^\gamma_1 E^\gamma_2 \left(1 - \cos\theta_{\gamma\gamma}\right) \tag{C.1}$$

If one assumes that the shift in the reconstructed mass of $\pi^0$ is due to one photon only, e.g. the first photon, one gets:

$$\frac{\delta \mathrm{m}_{\gamma\gamma}}{\mathrm{m}_{\gamma\gamma}} = \frac{1}{2}\frac{\delta E^\gamma_1}{E^\gamma_1}.$$

From this identity one can get the calibration (correction) coefficient to be applied for the selected photon to get the right mass of $\pi^0$:

$$\lambda = \frac{E^{\mathrm{true}}}{E_{\gamma}^{\mathrm{rec}}} = 1 - 2\frac{\delta \mathrm{m}_{\gamma\gamma}}{\mathrm{m}_{\gamma\gamma}}. \tag{C.2}$$

The photon reconstruction procedure [64, 72] guarantees that the central cell of the cluster ("seed") always has the largest energy. Due to coarse structure of the Ecal, electromagnetic clusters rarely spread outside the direct neighbours of the "seed" cell. Moreover very often the seed cell contains the bulk of the energy of whole cluster, and one has the following relations:

$$E_{seed} > E_i \qquad \text{always} \tag{C.3a}$$
$$E_{seed} \gg \max E_i \qquad \text{very often} \tag{C.3b}$$
$$E_{seed} > \sum E_i \qquad \text{often} \tag{C.3c}$$
$$E_{seed} \gg \sum E_i \qquad \text{not rare,} \tag{C.3d}$$

where $E_{seed}$ represent the energy deposited in the seed cell of the cluster, $E_i$ stands for the energy deposited in non-"seed" cell, and the sum runs over all cells excluding the "seed" cell.

With a little simplification, the reconstructed energy of photon can be written as the sum of two terms [64, 73]:

$$E_{\gamma}^{\mathrm{rec}} = E_{Prs}\mathrm{f}_1\big(x_b, y_b\big) + \big(E_{seed} + \sum_{i \neq seed} E'_i\big)\mathrm{f}_2\big(x_b, y_b\big), \tag{C.4}$$

where $\mathrm{f}_1$ and $\mathrm{f}_2$ are some functions, $E_{Prs}$ - energy deposited in Preshower detector, $x_b$ and $y_s$ are x- and y-coordinates of the weighted barycenter of cluster, and $E'_i$ is the energy in the cell, corrected for the possible overlap with the neighbouring cluster, $E'_i \leq E_i$. The function $\mathrm{f}_2$ takes into account the transversal non-uniformity of the response, and in particular the small corrections for the transversal energy leakage and it is rather close to unity: $\mathrm{f}_2 = 1 + \mathcal{O}\,(2\ \%)$.

The property (C.3) and equation (C.4) imply that very often $E_{\gamma}^{\mathrm{rec}}$ is rather close to $E_{seed}$, and for the photons which do not deposit the energy in Preshower detector, $E_{Prs} = 0$, they are roughly proportional. As a result one can use value of calibration coefficient $\lambda$, obtained for the reconstructed photon as an estimate for the miscalibration of the "seed" cell. The actual procedure can works as follows.

For each Ecal cell, we select the reconstructed photons which has no (or very small) energy deposition in Preshower detector and which has this cell as "seed" cell. Clearly for the given event one has at most one such photon. Then we make the pairs from the selected photon and all other photons in the event and plot the invariant mass of the pair. It has been shown that if the initial miscalibration of Ecal does not exceed 15 % one should see the peak in the vicinity of the nominal mass of $\pi^0$ meson.The actual signal-to-background ratio, $\mathcal{S}/\mathcal{B}$, depends on the initial miscalibration, the simulated event types

and the selection criteria. The photon pairs are selected if they satisfy the following criteria:

- the transverse energy of the photon is required to be in excess of 300 MeV,
- the transverse energy of the photon pair is required to be in excess of 800 MeV,
- the energy deposited in the Preshower cells in front of the cluster is less than 10 MeV.

We have used minimum bias events, accepted by the L0 trigger [74] and photons have been selected using "no hits in the SPD detector" as neutrality criteria instead of matching of Ecal clusters with the reconstructed tracks. Therefore this kind of calibration is essentially *track independent*.

One can fit this distribution with a Gaussian function to describe the $\pi^0$-signal, and a polynomial to describe the background. Using the formula (C.2) one can extract the calibration coefficient from the deviation of the Gaussian mean from the nominal mass of the $\pi^0$-meson. The obtained calibration coefficient is ascribed to all photons from this class. After the repetition of the procedure for all Ecal cells one has the estimates for the calibration coefficients for all cells. The situation improves with more iterations and it stabilizes after 5-10 iterations.

It has been shown that if one defines the calibration coefficient according to equation (C.2) the convergence is worse and moreover it has a tendency to oscillate. The reasons for such overcalibration are rather clear - we have assumed that the whole energy of the photon is miscalibrated with the same coefficient as the energy of the seed cell. The alternative, improved formula:

$$\lambda' = \frac{E^{\mathrm{true}}}{E_\gamma^{\mathrm{rec}}} = 1 - \frac{\delta \mathrm{m}_{\gamma\gamma}}{\mathrm{m}_{\gamma\gamma}} \tag{C.5}$$

shows much better behaviour, more rapid convergence and it is more stable with respect to variation of $\mathcal{S}/\mathcal{B}$. Effectively this modified definition corresponds to prescription that only half of the shift in the invariant mass of di-photon is due to miscalibration of the "seed" cell for the one photon and only half of the energy of the photon is contained in "seed" cell. This redefinition looks rather natural and in practice drastically improves the convergence of the iterative procedure.

After 5-10 "primary" iterations the procedure stabilizes and no further improvement occurs. Then one needs to perform the reconstruction of the photons taking into account the new calibration constants and to reiterate again. It forms a set of "secondary" iterations. Essentially the purpose of this "secondary" iteration is to take into account the change of photon barycenter parameters $x_b$ and $y_b$, reevaluate the function $\mathrm{f}_s$ from equation (C.4) and the angle $\theta_{\gamma\gamma}$ from equation (C.1). The total convergence occurs after 2-3 "secondary" iterations. The final miscalibration level $\epsilon_{cal}$ after three "secondary iterations" with ten "primary iterations" each for the initial 10 % miscalibration can be parameterized as:

$$\epsilon_{cal} = \frac{4\ \%}{\sqrt{\mathrm{N}_{\pi^0}/100}}, \tag{C.6}$$

where $N_{\pi^0}$ is a number of reconstructed $\pi^0$ per cell. The miscalibration level $\epsilon_{cal}$ reaches $\sim 1$ % at $N_{\pi^0} \approx 2 \cdot 10^3$ and goes down to 0.4 % for $N_{\pi^0} = 10^4$.

With the described selection criteria the average number of reconstructed $\pi^0$ per minimum bias event is 0.023. The efficiency is so low mainly due to cut on Preshower energy. Combining this number with the equation (C.6), one gets the estimate for number of events $N_{\mathcal{L}}$ required to achieve a certain miscalibration level:

$$N_{\mathcal{L}} \sim \frac{5 \cdot 10^8}{\left(\epsilon_{cal}\,[\%]\right)^2} \tag{C.7a}$$

$$\epsilon_{cal} \sim \frac{2.2\ \%}{\sqrt{N_{\mathcal{L}}/10^8}}. \tag{C.7b}$$

However, the equations (C.7) should be used with the some care, since they assume some uniformity of $\pi^0$ distributions over the calorimeter surface. Clearly this distribution is not uniform and the good calibration of certain Ecal regions could be achieved with significantly smaller number of events. Moreover for the off-line environment, the available event sample is enriched by events with high activity and larger multiplicity of energetic $\pi^0$'s, therefore one can expect larger $\pi^0$ yields. The equations (C.7) could be rewritten in a term of data taking duration $t_{\mathcal{L}}$, assuming output rate from the on-line farm to be 2 kHz and the data taking efficiency $\epsilon_{\mathcal{L}}$:

$$t_{\mathcal{L}} \sim \frac{3\ \text{day}}{\epsilon_{\mathcal{L}}\left(\epsilon_{cal}\,[\%]\right)^2}$$

$$\epsilon_{cal} \sim \frac{1.7\ \%}{\sqrt{\epsilon_{\mathcal{L}} t_{\mathcal{L}}\,[\text{day}]}}.$$

Taking $\epsilon_{\mathcal{L}} \sim 50$ % one gets that one week of data taking is enough to achieve a miscalibration of around 1%.

It is worth to stress here that for precalibration one can also use photons with non-zero energy deposition in the Preshower detector. For this case one can accumulate significantly larger statistics of $\pi^0$-mesons, and for such conditions $\pi^0$ rides significantly lower background level. However for such sample of photons the multiplicative calibration does not hold anymore. As a result the procedure does not allow to reach the good calibration accuracy and suffers from the convergency to the right calibration constant. However, it allows rather fast and efficient precalibration of Ecal detector up to few percent level, thus providing a good starting point for the ultimate calibration.

It is possible that one could combine the best features of both approaches and perform the calibration with a sample of $\pi^0$-mesons, where only one photon has negligible energy deposition in the Preshower detector. This approach is under investigation now.

At high photon energy the two Ecal clusters from the $\pi^0 \rightarrow \gamma\gamma$ decay start to overlap and therefore one can not consider anymore both reconstructed photons to be independent. This effect of "overlapped" or "merged" $\pi^0$s efficiently sets the upper range for this type of calibration to be $\sim 1.5$ GeV of transverse energy [64, 9]. The lower range is set by the transverse energy cut to suppress the combinatorial background from soft photons and noisy clusters and is 300 MeV of transverse energy.

## Calorimeter calibration with electrons

A complementary approach to a combined Ecal and Preshower calibration is the calibration with electrons. The method relies heavily on the proper reconstruction of tracks and probably on the proper dE/dx-corrections for the track fit. The second vital prerequisite is the possibility to select a sample of clean electrons. It can be done either through usage of RICH detectors or through kinematic selection of electrons and positrons from $\gamma \to e^+e^-$ conversion. The method relies on the simultaneous "Millipede-like"-minimization of a global $\chi^2$ with respect to unknown calibration coefficients for Ecal and Preshower cells. The main advantage of this method consists in straightforward non-iterative calibration procedure which can be performed "on the fly" during normal data-taking or with a dedicated calibration run.

Requiring isolated tracks at the Ecal entrance and performing electron candidates selection we a get data set which can be used for direct calorimeter calibration (based on the $E_{CALO}/P_{track}$ ratio.

Requiring electron ID with RICH, R = 30 cm isolation from the nearest charged track, visible Prs energy ($E_{Prs} > 20$ MeV) and limited Hcal energy ($E_{Hcal} < 1.5$ GeV) for tracks with $p_{track} > 4$ GeV/$c$, we get a sample purity of $\sim 91$ %, see Figure 21.

With mean number of electrons per cell of $\sim 1000$, we can get a relative calibration precision of about $\sim 0.4$ % for all three Ecal regions. The electron candidates are distributed quite non-uniformly over the Ecal so we will need $\sim 150$M events (passed L0 trigger) to have at least 1000 electrons per cell even in regions with lowest occupancy.

The number of required events can be significantly reduced if we will accept possible calibration precision at the level of $\sim 1$ %. In this case the mean number of events per cell is around 100.

The selection of electrons without the RICH detector is possible by exploiting dE/dx in the VELO detector. This requirement significantly reduces statistics and, finally, we will get data set with worse purity, see Figure 22.

The relative statistics reduction (with respect to the first scenario) is a factor of 25. So, we will need around 400M events to get calibration precision at the level of around 1%.

More work is required to study the convergence of the method, its sensitivity to selection criteria, the sample purity and the shape of the background. The possibility to calibrate the Preshower detector needs to be verified.

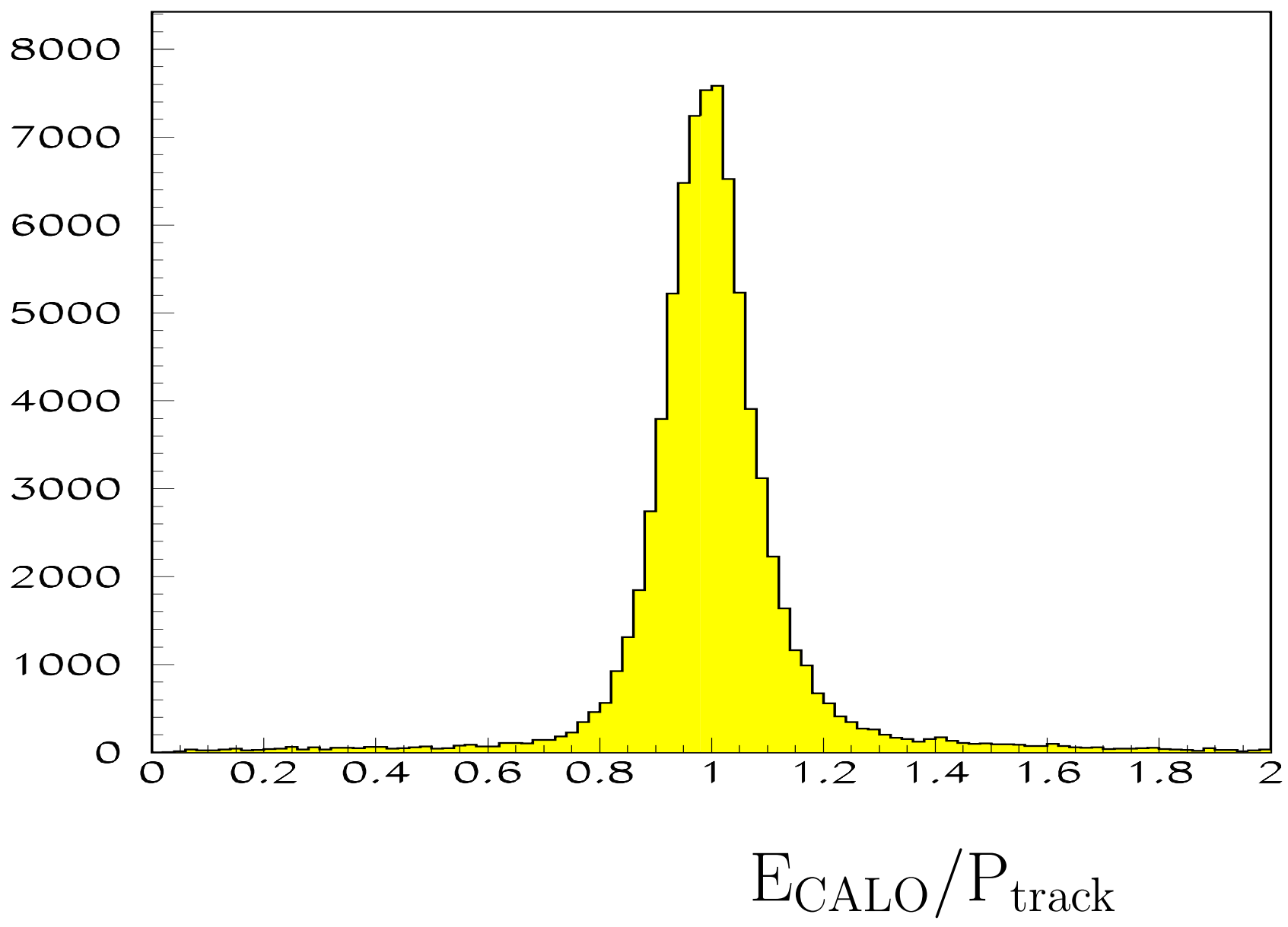

Figure 21: $E_{CALO}/P_{track}$ ratio for the electrons, selected using RICH information.

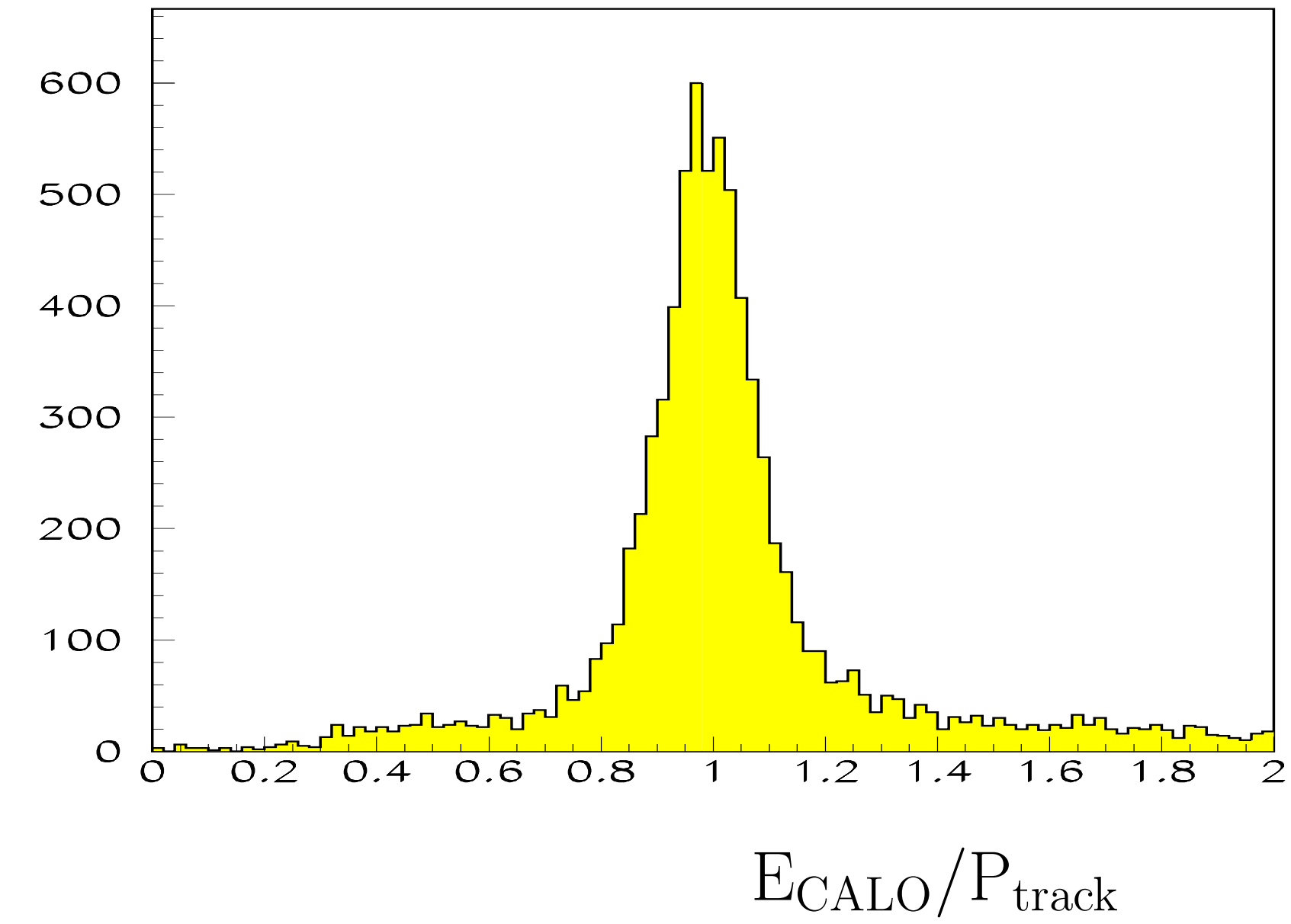

Figure 22: $E_{CALO}/P_{track}$ ratio for the electrons, selected using VELO information.

## Calorimeter calibration for high-$E_T$ photons

As has been mentioned in Section 4.3, each calibration method considered above has its own transverse energy range:

- the calibration with minimum ionizing particles allows to calibrate certain regions in Ecal for 15 MeV $\leq E_T{}^{\gamma} \leq$ 100 MeV,
- the iterative Ecal calibration with $\pi^0$ allows to calibrate Ecal in the transverse energy interval 300 MeV $\leq E_T{}^{\gamma} \leq$ 1.5 GeV,
- the calibration with $e^{\pm}$ allows to calibrate Ecal up to the transverse energy $2 -$ 2.5 GeV,

which are outside the most interesting region $E_T{}^{\gamma} \geq$ 2.5 GeV. To propagate the calibration constants obtained for lower energy one needs to rely on phototube linearity[17] and the LED-based monitoring system [9, 68]. But one needs to have an independent physical cross-check for the high energy calibration. The most appropriate candidate for such calibration channel is $\eta \to \gamma\gamma$. Due to the larger mass of $\eta$-meson compared to the $\pi^0$ the effective transverse energy cutoff occurs at significantly larger transverse energy, thus providing us access to isolated high energy photons. It has been shown that the peak $\eta \to \gamma\gamma$ is well visible, especially after vetoing $\pi^0 \to \gamma\gamma$ decays. The statistics of $\eta \to \gamma\gamma$ decays is more than order of magnitude less that the statistics of $\pi^0 \to \gamma\gamma$ events. In addition the $\eta \to \gamma\gamma$ peak sits on a larger combinatorial background due to larger available phase space for the diphoton system. These effects do not allow to use $\eta \to \gamma\gamma$ as a separate and efficient calibration channel. However, the position and the width of $\eta \to \gamma\gamma$ peak can be used for validation of Ecal calibration at a high transverse energy scale.

[17]The bulk of Ecal cells are equipped with phototubes with a nonlinearity term within ±1 % and the nonlinearity term does not exceed ±2 % for all phototubes [9].

# Appendix D: Separation of photons and $\pi^0$ at high-$\mathrm{E_T}$

The analysis of the remaining background events shows a large contribution from "merged" $\pi^0$ [2,1]. The development of an algorithm that allows to suppress this contribution would result in a lower level of background and therefore more precise determination of the physical parameters. Also it could provide some additional robustness of the results with respect to unexpected imperfections.

It has been shown that it is possible to separate "merged" $\pi^0$ from single photons using the detailed shower shape analysis with the following variables:

$\langle \mathrm{r}^2 \rangle = \mathrm{Tr}\,\mathcal{S}$ - the average energy weighted squared distance for the individual cells in the cluster, this variable is equal to the trace of the $2 \times 2$-matrix $\mathcal{S}$ of the second momenta of the cluster [64]. This variable characterizes the overall size of the cluster.

$1 - \dfrac{\langle \mathrm{r}^2 \rangle^2}{\langle \mathrm{r}^4 \rangle}$, where $\langle \mathrm{r}^4 \rangle$ is the average energy weighted $\mathrm{r}^4$ for the individual cells in the cluster. This variable characterizes the importance of the tails of the cluster and is analogous to a kind of 2D-kurtosis.

$\kappa = \sqrt{1 - 4\dfrac{\det \mathcal{S}}{\mathrm{Tr}^2\,\mathcal{S}}}$, where $\det \mathcal{S}$ is the determinant and $\mathrm{Tr}\,\mathcal{S}$ is the trace of the matrix $\mathcal{S}$. This variable characterizes the ratio of eigenvalues for the matrix $\mathcal{S}$ and therefore how stretched the cluster is.

$\alpha = \dfrac{\mathcal{S}_{xy}}{\sqrt{\mathcal{S}_{xx}\mathcal{S}_{yy}}}$ - this variable characterizes the orientation of the efficient ellipse, constructed from the second moments of the cluster, as well as how stretched the clusters are.

A likelihood function has been constructed from these variables properly taking into account their energy dependence, separately for all areas in Ecal.

For the testing sample of photons from $\mathrm{B}^0 \to \mathrm{K}^{*0}\gamma$ and merged $\pi^0$ from $\mathrm{B}^0 \to \mathrm{K}^{*0}\pi^0$ events, the achieved $\gamma/\pi^0$ separation is summarized in Table 10.

This technique could be further improved, e.g. for single photons the variable $\alpha$ should be slightly correlated with the azimuthal angle $\varphi_\gamma$ of the photon: $\alpha \propto \sin 2\varphi_\gamma$, while for $\pi^0$ no such correlation is expected. Also the "Bulos" variable or the effective cluster mass, determined from the specialized "merged" $\pi^0$ reconstruction algorithm [64] can be used for $\pi^0/\gamma$ separation.

Table 10: The efficiency of photon identification $\epsilon_\gamma$ and the probability of the "merged" $\pi^0$ misidentification as single energetic photon, $\epsilon_{\pi^0}$ for the different regions of the electromagnetic calorimeter.

| Ecal region | $\epsilon_\gamma$ | $\epsilon_{\pi^0}$ |
|---|---|---|
| Inner | 60 % | 1.4 % |
| Middle | 82 % | 2.3 % |
| Outer | 66 % | 1.1 % |

A slightly simplified version of this technique, based on a linear Fisher discriminant [75] instead of the original likelihood ratio approach, is currently used for the High Level Trigger, see Section 3.2 and Reference [8].